\documentclass[iop,numberedappendix]{emulateapj}
\pdfoutput=1
\usepackage{graphicx}
\usepackage{rotating}
\usepackage{threeparttablex}
\usepackage{verbatim}
\usepackage{amsmath, amsthm, amssymb}

\newcommand{\lapprox }{{\lower0.8ex\hbox{$\buildrel <\over\sim$}}}
\newcommand{\gapprox }{{\lower0.8ex\hbox{$\buildrel >\over\sim$}}}
\newcommand{\Msun}{\ensuremath{M_{\odot}}}

\newcommand{\LHa}{\ensuremath{L_{\rm{H}\alpha}}}
\newcommand{\LFUV}{\ensuremath{L_{\rm FUV}}}
\newcommand{\ICO}{\ensuremath{I_{\rm CO}}}

\newcommand{\aCO}{\ensuremath{\alpha_{\rm CO}}}
\newcommand{\tdep}{\ensuremath{\tau_{\rm dep}}}
\newcommand{\HII}{\ion{H}{2}}

\slugcomment{{\sc Accepted to ApJ:} May 5, 2014}
\shorttitle{Star Formation in NGC 300}
\shortauthors{Faesi et al.}

\begin{document}

\title{Molecular Cloud-scale Star Formation in NGC 300}
\author{Christopher M. Faesi\altaffilmark{1,2}, Charles J. Lada\altaffilmark{1}, Jan Forbrich\altaffilmark{1,3}, Karl M. Menten\altaffilmark{4}, Herv\'{e} Bouy\altaffilmark{5}}

\altaffiltext{1}{Harvard-Smithsonian Center for Astrophysics, 60 Garden Street, Cambridge, MA 02138}
\altaffiltext{2}{NSF Graduate Research Fellow}
\altaffiltext{3}{University of Vienna}
\altaffiltext{4}{Max Planck Institut f\"{u}r Radioastronomie}
\altaffiltext{5}{Centro de Astrobiolog\'{i}a, INTA-CSIC}

\begin{abstract}

We present the results of a galaxy-wide study of molecular gas and star formation in a sample of 76 {\HII} regions in the nearby spiral galaxy NGC~300. We have measured the molecular gas at 250~pc scales using pointed CO($J=2-1$) observations with the APEX telescope. We detect CO in 42 of our targets, deriving molecular gas masses ranging from our sensitivity limit of $\sim10^5$~{\Msun} to $7\times 10^5$~{\Msun}. We find a clear decline in the CO detection rate with galactocentric distance, which we attribute primarily to the decreasing radial metallicity gradient in NGC~300. We combine GALEX FUV, \textit{Spitzer} 24 $\mu$m, and H$\alpha$ narrowband imaging to measure the star formation activity in our sample. We have developed a new direct modeling approach for computing star formation rates  that utilizes these data and population synthesis models to derive the masses and ages of the young stellar clusters associated with each of our {\HII} region targets. We find a characteristic gas depletion time of 230~Myr at 250~pc scales in NGC~300, more similar to the results obtained for Milky Way GMCs than the longer ($>2$~Gyr) global depletion times derived for entire galaxies and kpc-sized regions within them. This difference is partially due to the fact that our study accounts for only the gas and stars within the youngest star forming regions. We also note a large scatter in the NGC~300 SFR-molecular gas mass scaling relation that is furthermore consistent with the Milky Way cloud results. This scatter likely represents real differences in giant molecular cloud physical properties such as the dense gas fraction.

\end{abstract}


\section{Introduction}
\label{sec:intro}

The physical process of star formation is of fundamental astrophysical importance over a wide range of interconnected scales. Individual stars and star clusters are observed to form within Giant Molecular Clouds (GMCs), which themselves form from the more diffuse interstellar medium (ISM) within galaxies. At the largest scales, the evolution of star formation across cosmic time plays a key role in galaxy evolution \citep[e.g.,][]{Daddi:2010go,Tacconi:2010cr}. An important first step toward a predictive theory of star formation is an understanding of the empirical relationship between star formation rates (SFRs) and interstellar gas. Half a century ago \cite{1959ApJ...129..243S} postulated that the star formation rate surface density in galaxies should scale as a power law with the surface density of gas, i.e. $\Sigma_{\rm SFR} \propto \Sigma_{\rm gas}^n$. More recently, \cite{1998ApJ...498..541K} demonstrated the existence of a nonlinear ($n\approx 1.4$) power law between galaxy-integrated $\Sigma_{\rm SFR}$ and $\Sigma_{\rm gas}$ across several orders of magnitude in spiral and starburst galaxies.

In the past decade, the rapid improvement of observational facilities has led to high-resolution, high-sensitivity multiwavelength probes of star formation and gas across larger samples of galaxies, as well as the first resolved studies within external galaxies \cite[see][and references therein for a comprehensive overview]{KennicuttJr:2012ey}. Investigations that separate the roles of atomic (\ion{H}{1}) and molecular (H$_2$) gas have shown that the more fundamental relation is the one between star formation and molecular gas \citep[e.g.,][]{2002ApJ...569..157W,Bigiel:2008bs}. Furthermore, multiple studies at kpc-resolution in both individual galaxies \citep[e.g.,][]{2007ApJ...671..333K,Blanc:2009hf}, and across samples of spiral galaxies \citep[e.g.,][]{Bigiel:2008bs,Rahman:2011bo,Leroy:2012ks} have found $n\approx 1$ -- a linear relation between $\Sigma_{\rm SFR}$ and the molecular gas surface density $\Sigma_{\rm mol}$. One interpretation of this linear slope is an approximately constant depletion time ($\tdep = \Sigma_{\rm mol} / \Sigma_{\rm SFR}$), the timescale for the molecular gas reservoir to be entirely converted into stars. Extragalactic studies consistently derive average depletion times of about 2~Gyr~\citep[e.g.,][]{Leroy:2013bf}.

Several factors complicate our understanding of the physical basis for the empirical power law relation between gas and SFR and the interpretation of {\tdep}. For example, the derived slope appears to depend on the physical scale sampled \citep{Schruba:2010hf,Calzetti:2012iz}, the choice of molecular gas tracer \citep{2007ApJ...669..289K,2008ApJ...684..996N}, and the fitting method used \citep{Shetty:2013jf}. Furthermore, while a linear slope (and constant {\tdep}) describes well the global average scaling relation in star-forming disk galaxies, individual galaxies show deviations from a single depletion time \citep[e.g.,][]{Saintonge:2011ey,Leroy:2013bf}. Second order effects such as variations in the ability of CO to trace the total molecular gas content \citep{Leroy:2011hc,Sandstrom:2013bw} may contribute to the scatter in {\tdep}.

Intriguing insights into the relation between star formation and molecular gas have come from studies of the nearest molecular clouds \citep[e.g.,][]{Heiderman:2010fz,Lada:2010cl}. For example, \cite{Lada:2010cl}, hereafter L10, demonstrated a tight linear correlation between the integrated star formation rate (SFR) and total mass in dense (i.e., gas with $n_{\rm{H}2} \gtrsim10^4$~cm$^{-3}$) molecular gas $M_{\rm dense}$ in a sample of ten well-resolved nearby molecular clouds. \cite{Lada:2012it} also showed that the same linear correlation exists between the SFR and \textit{total} gas mass $M_{\rm GMC}$, albeit with more scatter. They interpret the scatter as arising due to differences in the dense gas fraction $f_{\rm DG}=M_{\rm dense}/M_{\rm GMC}$, and posit that the fundamental scaling law in local clouds is of the form SFR~$\propto f_{\rm DG}M_{\rm GMC}$. The importance of the density structure of clouds in star formation has also been corroborated in studies of other galaxies; for example, \cite{2004ApJ...606..271G} find a linear relation between the total infrared luminosity (a tracer of dust-embedded star formation) and the luminosity in the dense gas tracer HCN in entire galaxies \citep[see also][]{2005ApJ...635L.173W}. Additionally, the \cite{Lada:2012it} correlation extends smoothly across more than five orders of magnitude and is consistent with the \cite{2004ApJ...606..271G} results to within a factor of three.

For their sample of Milky Way clouds, L10 derive a median {\tdep} of 180~Myr -- an order of magnitude shorter than the typical depletion times inferred for kpc-sized regions within disk galaxies. Does this difference reflect local environmental factors, the use of differing methods to derive SFRs and gas masses or densities, or the discrepant scales probed? To address this issue it is necessary to expand the local cloud sample to include a more heterogeneous set of galactic environments. Unfortunately, only the nearest clouds can be observed at a high enough resolution to be analyzed in a manner similar to that of L10. Furthermore, studies of more distant regions of the Milky Way are also complicated by uncertainties due to kinematic distance ambiguities and confusion along the line-of-sight. Compiling a set of measurements of star-forming regions within an external galaxy effectively places all regions at a single, consistent distance, avoiding the above issues. However, deriving star formation rates at GMC scales within other galaxies is not a trivial exercise, as integrated measurements of emission from young stellar populations and extrapolation of the Initial Mass Function (IMF) must be employed since individual stars cannot be resolved and the most massive stars dominate the luminosity of these populations. As a first step toward extending the study of star formation in GMCs within galaxies beyond the Milky Way, we have conducted a survey of the molecular ISM in the nearby spiral galaxy NGC~300, which we analyze alongside archival ultraviolet, infrared, and H$\alpha$ images in a manner as consistent as possible with the methodology used for the L10 sample.

NGC~300 is a southern-declination ($\delta=-37^{\circ}$), relatively small ($R_{25} \approx 5$~kpc), moderately inclined \citep[40$^{\circ}$;][]{1990AJ....100.1468P}, slightly subsolar metallicity spiral galaxy at a distance of 1.93~Mpc \citep{2004AJ....128.1167G}. It is forming stars at a global rate of about $0.15~\Msun$~yr$^{-1}$~\citep{2004ApJS..154..253H}. Previous observations have revealed large numbers of active {\HII} regions~\citep[e.g.,][]{Deharveng:1988wh} as well as supernova remnants \cite[e.g.,][]{Payne:2004cw}, suggesting that star formation has been proceeding actively throughout the galactic disk for several generations of stars. It is thus an ideal laboratory in which to study the relationship between molecular gas and the SFR in a large sample of individual star forming regions.

Our sample of star-forming regions in NGC~300 consists of 76 {\HII} regions from the catalog of \cite{Deharveng:1988wh}. We measure the molecular gas content on 250~pc scales using pointed $^{12}$CO($J=2-1$) observations. To infer star formation rates for these individual regions, we introduce a direct modeling approach in which we compare our multiwavelength observations to predictions from stellar population synthesis modeling.

The present paper is organized as follows. In Section~\ref{sec:data}, we present our APEX observations and ancillary data sets. Section~\ref{sec:phot} presents our measurements, while Section~\ref{sec:method} discusses how we perform physical parameter estimation from observables; in particular, Section~\ref{sec:models} introduces our direct modeling approach. In Section~\ref{sec:results}, we present our results, and finally discuss them in the context of both local and extragalactic studies in Section~\ref{sec:disc}. Appendix~\ref{sec:IMF} presents a new method for estimating uncertainties in the properties of stellar populations derived using population synthesis models, which we apply to our NGC~300 sample.

\section{Observations and Data}
\label{sec:data}

\begin{figure}[tb]
\includegraphics[width=\linewidth]{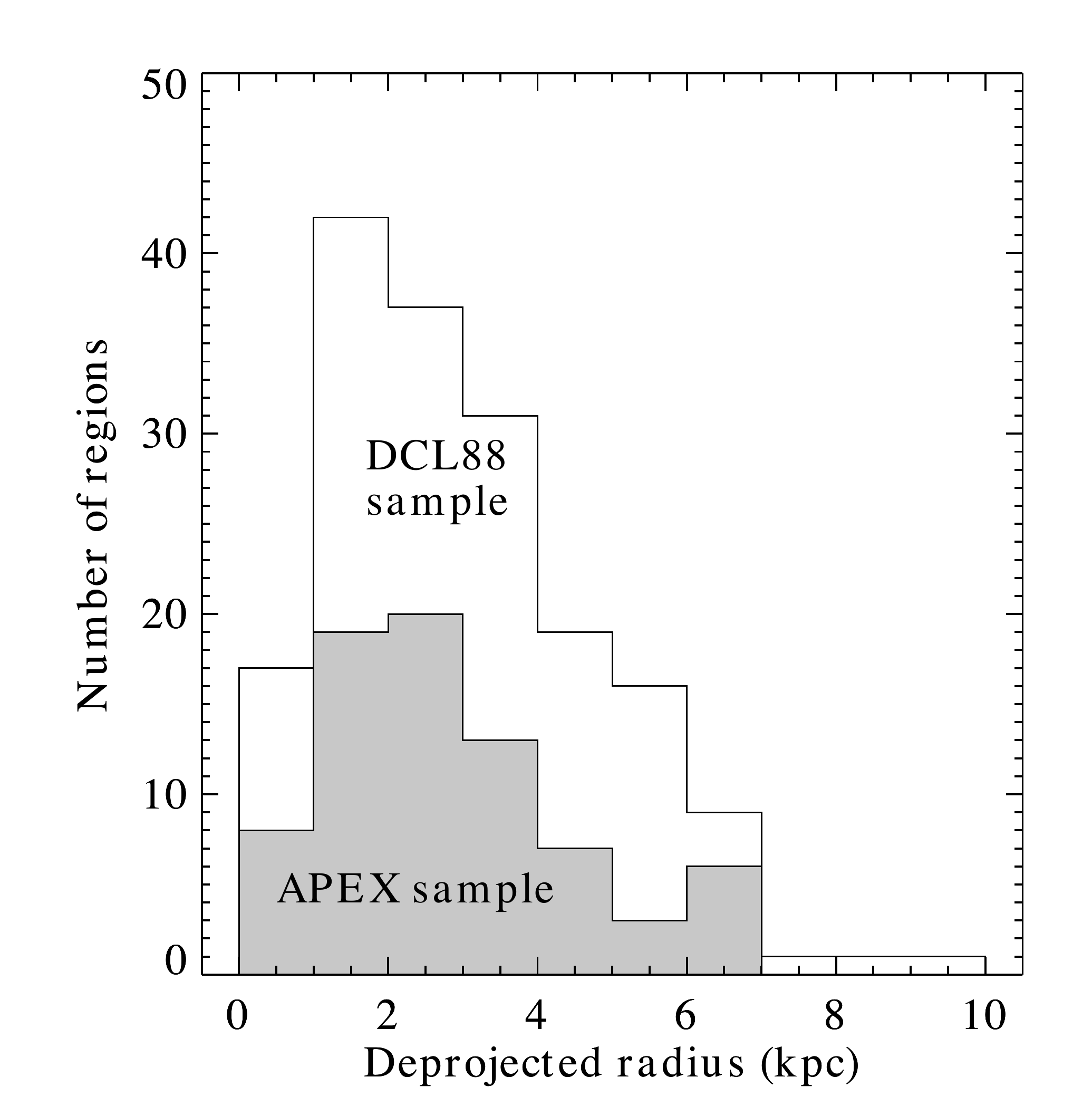}
\caption{\small{Histogram by deprojected radius of our APEX sample (shaded histogram) over plotted on the histogram of the full \cite{Deharveng:1988wh} catalog of {\HII} regions in NGC~300.}}
\label{fig:DCLhist}
\end{figure}

Our sample of star-forming regions in NGC~300 consists of 76 {\HII} regions from the catalog of \cite{Deharveng:1988wh}. We specifically selected regions from this catalog having a broad range of properties, including galactocentric radius, H$\alpha$ morphology, and luminosity in $24~\mu$m, FUV, and H$\alpha$. The histogram in Figure~\ref{fig:DCLhist} shows the distribution of galactocentric radii of both the full  \cite{Deharveng:1988wh} sample and our representative subsample of 76~regions. Table~\ref{tab:basicdata} summarizes the properties of NGC~300. In this section we describe the data sets we use to measure molecular gas masses and star formation rates in our sample.

\begin{center}
\begin{deluxetable}{l r}
\tablecaption{Basic properties of NGC 300 \label{tab:basicdata}}
\tablehead{}
\startdata
Morphological type &		SAs(d) \\
R.A. (J2000) &				$00^{\rm h}~54^{\rm m}~53.48^{\rm s}$ \\
Dec. (J2000) &				-37$^{\circ}$~41\arcmin~03.8\arcsec \\
Distance \citep{2004AJ....128.1167G} &		1.93 Mpc \\
Inclination &				39.8$^{\circ}$ \\
$r_{25}$ &				9.75\arcmin~(5.3 kpc) \\
Position angle of major axis &	114.3$^{\circ}$ \\
Helio. radial velocity &		144 km s$^{-1}$ \\
Metallicity \citep{Bresolin:2009hh} &	0.5--0.6 Z$_{\odot}$
\enddata
\tablecomments{\textit{\small{Data from HyperLeda database \citep{Paturel:2003jm}}}} \\
\end{deluxetable}
\end{center}

\subsection{APEX CO(2-1)}
\label{sec:APEX}

To measure the molecular gas content in our sample of 76 star-forming regions, we obtained CO($J=2-1$; rest frequency: 230.538~GHz) observations with the APEX-1 facility heterodyne receiver \citep{Vassilev:2008kg} on the Atacama Pathfinder Experiment  12 meter diameter submillimeter telescope \citep[APEX;][]{Gusten:2006ci}. Observations took place over several epochs between 2011 April 8 and December 16. We obtained over 100 hours of observation time divided amongst three project IDs: 13 hours between 2011 May 5 and May 12 under European Southern Observatory (ESO) project E-087.C-0507A-2011, and the remainder of the time under Max-Planck-Gesellschaft projects M-087.F-0033-2011 (2011 April 8--18 and June 4--August 7) and M-088.F-0022-2011 (2011 October 10--December 16; PI: J. Forbrich for all three projects). Scans were taken in on-off mode, with off positions chosen to be emission-free regions outside the galaxy disk. Each scan consisted of a total integration time (on+off) of 9 minutes. Calibration was performed using an extension of the chopper wheel method \citep[e.g.,][]{1976ApJS...30..247U} to set the absolute temperature scale and correct for spectral variations of the atmosphere. At least one calibration scan directly preceded each on-off scan. CO$(2-1)$ pointing was performed on strong point sources several times per night, and the pointing accuracy is estimated to be 1.5{\arcsec} on average. Focusing scans of $\sim3$ minutes per axis were obtained at least once per night. Uncertainties in measured intensity due to pointing and focusing errors are estimated to be less than 1\% at 230~GHz. The vast majority of our observations were obtained under favorable (precipitable water vapor levels between 0.2 and 4~mm) atmospheric conditions. We observed in single sideband mode using the XFFTS-2 spectral backend. This Fast Fourier Transform Spectrometer was configured to have 32768 76.3~kHz channels across two subbands, with a total bandpass of 2.5~GHz. All spectra were smoothed to a resolution of 1.1~MHz (1.39~km~s$^{-1}$) for analysis. Table~\ref{tab:obslog} provides a list of our APEX-observed sources, labeled by their original \cite{Deharveng:1988wh} number. The APEX-1 27'' (FWHM) beam size at 230~GHz corresponds to a physical scale of about 250~pc at the distance of NGC~300.

\begin{deluxetable*}{cccrrrrrc}
%\tablewidth{0pt}
\tablecaption{Summary of APEX observations\label{tab:obslog}}
\tablehead{
\colhead{DCL\#}	& \colhead{RA}	& \colhead{dec}		& \colhead{$T_{\rm RMS}$} &
\colhead{$v_{\rm HI}$} &	\colhead{$R_{\rm gal}$} &	\colhead{$I_{\rm CO}$} &	\colhead{Linewidth} &	\colhead{CO?} \\
& (J2000) & (J2000) &	\colhead{(mK)} &	\colhead{(km s$^{-1}$)} &	\colhead{(kpc)} &	\colhead{mK km s$^{-1}$)} 
& \colhead{(km s$^{-1}$)} \\
}
\startdata
  1 & 00:54:06.45 & -37:41:03.0 & 13.9 & 192.6 & 6.9 & $<$ 135 & \nodata & N \\
  2 & 00:54:08.53 & -37:37:56.1 & 12.6 & 196.5 & 6.5 & $<$ 122 & \nodata & N \\
  5 & 00:54:16.19 & -37:34:32.4 &  9.2 & 187.4 & 6.4 & $<$  89 & \nodata & N \\
  6 & 00:54:16.39 & -37:34:52.5 &  5.0 & 187.6 & 6.2 & $<$  48 & \nodata & N \\
  7 & 00:54:16.64 & -37:35:23.5 & 10.6 & 187.4 & 6.0 & $<$ 103 & \nodata & N \\
  9 & 00:54:17.62 & -37:35:07.4 & 18.6 & 185.7 & 6.0 & $<$ 180 & \nodata & N \\
 13 & 00:54:23.22 & -37:40:42.4 &  9.7 & 190.5 & 4.4 & $<$  94 & \nodata & N \\
 15 & 00:54:24.90 & -37:39:44.0 & 11.0 & 191.2 & 4.1 & $<$ 106 & \nodata & N \\
 17 & 00:54:25.40 & -37:39:06.0 & 11.1 & 191.2 & 4.1 & $<$ 107 & \nodata & N \\
 23 & 00:54:28.36 & -37:41:48.3 &  9.6 & 178.9 & 3.8 &  364$\pm$ 72 &  7.8 & D \\
 24 & 00:54:28.75 & -37:41:32.7 & 11.5 & 179.6 & 3.7 & $<$ 111 & \nodata & N \\
 29 & 00:54:31.57 & -37:38:15.2 & 12.8 & 190.0 & 3.4 & $<$ 124 & \nodata & N \\
 30 & 00:54:31.71 & -37:37:58.4 &  8.5 & 191.5 & 3.5 &  255$\pm$ 63 &  9.2 & D \\
 31 & 00:54:32.07 & -37:37:43.7 & 10.9 & 192.3 & 3.5 &  188$\pm$ 81\tablenotemark{a} &  3.8 & M \\
 34 & 00:54:32.70 & -37:38:42.0 &  5.9 & 188.4 & 3.2 &  313$\pm$ 44 &  7.0 & D \\
 37 & 00:54:35.38 & -37:39:32.4 & 12.3 & 185.2 & 2.6 &  431$\pm$ 92 & 10.1 & D \\
 41 & 00:54:38.75 & -37:41:23.5 &  7.0 & 174.1 & 2.2 &  899$\pm$ 52 & 14.3 & D \\
 43 & 00:54:39.54 & -37:42:30.2 & 14.0 & 167.4 & 2.4 & $<$ 136 & \nodata & N \\
 44 & 00:54:39.92 & -37:38:12.8 &  9.3 & 175.0 & 2.5 & $<$  90 & \nodata & N \\
 45 & 00:54:40.48 & -37:40:51.6 & 22.3 & 174.9 & 1.9 & $<$ 216 & \nodata & N \\
 46 & 00:54:40.56 & -37:43:00.5 &  6.7 & 164.2 & 2.5 &  322$\pm$ 50 &  9.4 & D \\
 49 & 00:54:42.15 & -37:39:02.8 &  9.5 & 172.8 & 1.9 &  346$\pm$ 71\tablenotemark{a} &  3.6 & D \\
 52 & 00:54:42.89 & -37:40:01.5 & 10.5 & 175.6 & 1.6 &  602$\pm$ 78 &  9.8 & D \\
 53C & 00:54:42.82 & -37:42:55.1 & 19.5 & 160.6 & 2.2 & $<$ 189 & \nodata & N \\
 54 & 00:54:43.67 & -37:39:46.3 & 12.2 & 172.7 & 1.5 & $<$ 118 & \nodata & N \\
 55 & 00:54:44.08 & -37:35:15.5 & 15.2 & 158.0 & 4.0 & $<$ 147 & \nodata & N \\
 56 & 00:54:44.27 & -37:40:23.1 &  9.1 & 170.7 & 1.3 &  194$\pm$ 68\tablenotemark{a} &  2.7 & M \\
 57 & 00:54:44.54 & -37:36:35.5 & 14.5 & 162.2 & 3.1 & $<$ 141 & \nodata & N \\
 61 & 00:54:45.39 & -37:38:44.0 &  7.8 & 162.4 & 1.8 &  251$\pm$ 58 &  6.8 & D \\
 63 & 00:54:45.60 & -37:37:53.0 &  5.7 & 156.4 & 2.3 &  178$\pm$ 42 &  5.1 & D \\
 64 & 00:54:46.40 & -37:40:20.0 & 11.4 & 165.2 & 1.1 & $<$ 110 & \nodata & N \\
 65 & 00:54:46.60 & -37:36:29.0 &  5.7 & 156.2 & 3.1 &  188$\pm$ 42 &  7.0 & D \\
 66 & 00:54:46.94 & -37:37:56.1 &  6.7 & 155.6 & 2.2 &  245$\pm$ 50 &  7.1 & D \\
 68 & 00:54:47.87 & -37:38:00.4 &  9.9 & 154.7 & 2.1 &  517$\pm$ 74 &  8.3 & D \\
 69 & 00:54:48.11 & -37:43:31.3 & 10.0 & 143.4 & 2.0 &  412$\pm$ 75 &  5.3 & D \\
 72 & 00:54:49.07 & -37:33:15.7 & 10.9 & 152.0 & 5.3 & $<$ 106 & \nodata & N \\
 76C & 00:54:50.89 & -37:40:23.6 &  8.3 & 151.5 & 0.5 &  643$\pm$ 62 & 14.1 & D \\
 79 & 00:54:51.15 & -37:38:22.8 &  6.1 & 150.8 & 1.8 & 1140$\pm$ 45 &  9.9 & D \\
 80 & 00:54:51.15 & -37:40:58.3 & 12.8 & 148.3 & 0.3 &  308$\pm$ 95\tablenotemark{a} &  4.6 & M \\
 81 & 00:54:51.38 & -37:41:41.8 &  7.0 & 144.6 & 0.6 &  201$\pm$ 52 &  5.7 & D \\
 85 & 00:54:51.97 & -37:41:35.1 & 14.5 & 142.5 & 0.5 &  284$\pm$108 &  5.6 & M \\
 86 & 00:54:52.44 & -37:40:36.4 &  8.7 & 145.1 & 0.3 &  353$\pm$ 65 &  9.9 & D \\
 88 & 00:54:53.12 & -37:43:44.0 &  8.3 & 133.8 & 1.9 &  591$\pm$ 62 & 13.0 & D \\
 89 & 00:54:53.47 & -37:41:00.2 &  9.5 & 140.6 & 0.0 & $<$  92 & \nodata & N \\
 93 & 00:54:54.83 & -37:43:41.5 &  7.1 & 127.9 & 1.8 &  319$\pm$ 53 &  8.5 & D \\
 98 & 00:54:56.38 & -37:40:28.1 & 12.5 & 132.3 & 0.6 &  471$\pm$ 93 & 17.1 & M \\
 99 & 00:54:56.40 & -37:39:37.0 & 14.6 & 135.7 & 1.2 & $<$ 141 & \nodata & N \\
100 & 00:54:56.40 & -37:41:10.0 &  7.1 & 127.7 & 0.4 &  283$\pm$ 53 & 10.4 & D \\
103 & 00:54:57.55 & -37:42:24.7 & 11.8 & 118.4 & 1.0 &  413$\pm$ 88 &  6.8 & D \\
109 & 00:55:00.37 & -37:40:33.0 &  9.0 & 119.1 & 1.1 &  626$\pm$ 67 & 14.0 & D \\
112 & 00:55:01.53 & -37:44:07.2 &  8.5 & 109.8 & 2.2 &  366$\pm$ 63 & 13.4 & M \\
114 & 00:55:02.20 & -37:39:42.0 &  8.7 & 119.3 & 1.7 &  971$\pm$ 65 & 16.1 & D \\
115 & 00:55:02.63 & -37:38:27.0 & 12.3 & 126.4 & 2.5 & $<$ 119 & \nodata & N \\
117 & 00:55:02.74 & -37:42:56.4 & 14.8 & 105.4 & 1.7 & $<$ 143 & \nodata & N \\
118B & 00:55:04.58 & -37:42:49.2 &  6.6 & 102.1 & 1.8 &  267$\pm$ 49 &  8.5 & D \\
119C & 00:55:02.87 & -37:43:13.2 &  8.9 & 105.6 & 1.8 &  336$\pm$ 66 &  7.2 & D \\
120 & 00:55:04.10 & -37:39:14.6 & 20.7 & 116.2 & 2.2 & $<$ 201 & \nodata & N \\
122 & 00:55:04.60 & -37:40:55.0 &  5.0 & 102.8 & 1.6 &  423$\pm$ 37 &  9.5 & D \\
124 & 00:55:05.35 & -37:41:19.4 & 14.5 &  99.8 & 1.7 & $<$ 141 & \nodata & N \\
126 & 00:55:07.45 & -37:41:04.1 & 11.5 &  99.3 & 2.0 &  370$\pm$ 86 &  4.9 & D \\
127 & 00:55:07.53 & -37:41:47.8 &  9.7 &  95.6 & 2.0 &  745$\pm$ 72 & 10.3 & D \\
129 & 00:55:08.85 & -37:39:27.4 & 13.8 & 110.8 & 2.7 &  251$\pm$103 &  5.1 & M \\
130 & 00:55:09.05 & -37:40:48.0 & 15.2 & 101.2 & 2.3 &  726$\pm$113 & 20.2 & M \\
133 & 00:55:09.95 & -37:47:53.1 & 12.2 & 102.3 & 4.9 & $<$ 118 & \nodata & N \\
136 & 00:55:12.20 & -37:39:07.0 & 12.4 & 110.5 & 3.3 & $<$ 120 & \nodata & N \\
137A & 00:55:12.79 & -37:41:37.0 & 12.5 &  91.1 & 2.8 &  324$\pm$ 93 &  8.2 & D \\
137B & 00:55:12.70 & -37:41:23.1 &  7.6 &  94.6 & 2.8 &  620$\pm$ 57 & 11.7 & D \\
137C & 00:55:13.86 & -37:41:36.9 &  5.7 &  91.2 & 2.9 &  735$\pm$ 42 &  9.4 & D \\
139 & 00:55:13.03 & -37:44:06.2 & 10.9 &  88.9 & 3.2 &  274$\pm$ 81 &  6.1 & D \\
140 & 00:55:14.96 & -37:44:14.7 &  8.1 &  87.3 & 3.5 &  312$\pm$ 60 &  8.4 & D \\
144 & 00:55:19.33 & -37:46:37.0 & 15.0 &  89.5 & 4.8 & $<$ 145 & \nodata & N \\
145 & 00:55:20.05 & -37:43:49.1 & 17.8 &  82.8 & 4.0 & $<$ 173 & \nodata & N \\
146 & 00:55:20.70 & -37:43:37.0 &  8.3 &  82.4 & 4.0 & $<$  80 & \nodata & N \\
147 & 00:55:24.34 & -37:39:33.8 & 11.2 &  97.1 & 4.8 & $<$ 108 & \nodata & N \\
150 & 00:55:28.00 & -37:44:17.0 &  7.5 &  78.8 & 5.1 & $<$  72 & \nodata & N \\
151 & 00:55:28.08 & -37:40:41.9 & 11.4 &  85.0 & 5.1 & $<$ 110 & \nodata & N
\enddata
\tablenotetext{a}{Gaussian-derived {\ICO}}
\end{deluxetable*}

\subsection{ESO/WFI H$\alpha$}

Our sample consists of {\HII} regions and thus our targets were selected based on the presence of H$\alpha$ recombination radiation. The H$\alpha$ luminosity of an {\HII} region is directly proportional to the number of ionizing photons emitted by the massive stars that ionize it~\citep[e.g.,][]{Osterbrock:2006ul}. The majority of these ionizing photons are emitted by stars with masses between 30 and 40~{\Msun} and thus lifetimes of 3-10~Myr~\cite[e.g.,][]{1998ARA&A..36..189K}. H$\alpha$ line emission is therefore a direct tracer of the most recent star formation activity.

NGC~300 was observed on 2000 August 5 with the Wide-Field Imager (WFI) on the Max Planck Gesellschaft (MPG)/European Southern Observatory (ESO) 2.2m telescope at La Silla, Chile, using the H$\alpha$ narrowband filter (central wavelength 6588.27~{\AA}, FWHM 74.31~{\AA}). The raw science and calibration images were downloaded from the ESO data archive. Airmasses ranged from 1.02 to 1.08 over the observation window. For the reduction procedure, we used a customized version of the ESO Multi-resolution Vision Module (MVM) image reduction system that has been updated from the official release version. This included flat-fielding, bias correction, and gain harmonization between the 8 separate imager chips. We then combined the seven 420-second images using SWarp\footnote{http://www.astromatic.net/software/swarp}. The final image has a pixel scale of 0\arcsec.23. A narrowband continuum image using a filter centered at $6655.61$~{\AA} with width 120.78~{\AA} was also obtained and reduced in a similar fashion. The continuum image was corrected for mosaicking artifacts by hand and then background-subtracted. We measured the PSFs of both images using several bright stars distributed at a range of positions on the detector chips, then convolved the H$\alpha$ image such that the two mean PSF resolutions matched. Finally, we subtracted a version of the continuum map, scaled by the ratio of filter widths and normalized to the same exposure times, from the H$\alpha$ map to obtain a line-only H$\alpha$ map. We corrected the final H$\alpha$ image for Galactic (Milky Way) extinction using the value of $A_R = 0.027$~mag from \cite{SF11} retrieved from the NASA/IPAC Extragalactic Database (NED). This was converted to H$\alpha$ nebular extinction by adopting a nebular-to-stellar extinction ratio of $A_{\rm{H}\alpha}/A_{\rm R} \approx 2$ \citep{1994ApJ...429..582C}. The top panel of Figure~\ref{fig:images} shows our ESO/WFI H$\alpha$ image of NGC 300 with our APEX-observed {\HII} region targets denoted as circles. The circle sizes represent both the APEX beam FWHM and the photometric aperture size used. Zoom-in images of individual regions are presented in Appendix~\ref{sec:stamps}.

\subsection{GALEX FUV}

The bulk of the emission at ultraviolet (UV) wavelengths longward of the Lyman continuum break in star-forming galaxies such as NGC~300 is direct photospheric emission from massive, luminous stars. Stars having a few tens of solar masses (and thus lifetimes $< 200$~Myr) dominate the integrated UV luminosity of a young stellar population, with shorter wavelengths probing progressively shorter timescales. The GALEX FUV band, centered at a wavelength of 1516~{\AA}, traces stars with characteristic lifetimes $< 100$~Myr~\citep[e.g.,][]{1998ARA&A..36..189K,2007ApJS..173..267S}. However, note that since our targets are selected based on the presence of significant H$\alpha$ emission, the stellar populations in our sample are almost certainly younger than about 10~Myr.

GALEX far-ultraviolet (FUV) images were downloaded from the MAST data archive using the GalexView tool. Observations were taken under the auspices of the Guest Investigator program GI1\_061002 between 2004 October 26 and December 15, with a total exposure time of 12988 seconds. All images went through the GALEX automated reduction pipeline \citep{2007ApJS..173..682M}. The FUV band data has an effective wavelength of 1516~{\AA} and resolution of 4.3{\arcsec}. The GALEX pointing uncertainty of 0.5{\arcsec} is insignificant compared to the 27{\arcsec} apertures over which we conduct photometry (see Section~\ref{sec:phot}). We corrected the GALEX image for Galactic extinction following a similar procedure to that described above for H$\alpha$, using the conversion $A_{\rm FUV}/E(B-V)=8.24$ from \cite{2007ApJS..173..293W} and the \cite{1989ApJ...345..245C} extinction law with a total-to-selective extinction of $R_V=3.1$. We present our GALEX FUV image in the bottom left panel of Figure~\ref{fig:images}. Zoom-in images of individual regions are presented in Appendix~\ref{sec:stamps}.

\subsection{\textit{Spitzer}/MIPS 24 $\mu$m}

Emission at 24~$\mu$m traces warm dust primarily heated by the intense UV fields of hot, young stars \citep[e.g.,][]{Draine:2003di}. Dust is ubiquitous in molecular clouds where these stars form, and thus much of the intrinsic UV emission does not escape the cloud and is instead absorbed and reradiated in the infrared.

NGC~300 was observed with the \textit{Spitzer} Space Telescope's Multiband Imaging Photometer for Spitzer \citep[MIPS;][]{2004ApJS..154...25R} on 2004 Oct 26. BCDs from AOR 6070016 (PI: G. Helou) were downloaded directly from the \textit{Spitzer} Heritage archive. Observations were taken in scan mode. These images underwent standard MIPS-24 reduction procedures as described in \cite{Gordon:2005jb}. We aligned and mosaicked the BCDs using the MOPEX\footnote{http://irsa.ipac.caltech.edu/data/SPITZER/ docs/dataanalysistools/tools/mopex/} image processing software, then rotated and regridded to the GALEX FUV pixel scale (1.5{\arcsec}) with MONTAGE\footnote{http://montage.ipac.caltech.edu/}. The MIPS 24~$\mu$m PSF is $\sim5.9\arcsec$. The 24~$\mu$m image is shown in the bottom right panel of Figure~\ref{fig:images}. Zoom-in images of individual regions are presented in Appendix~\ref{sec:stamps}.

\begin{figure*}[tb]
$\begin{array}{cc}
\multicolumn{2}{c}{\includegraphics[width=\linewidth]{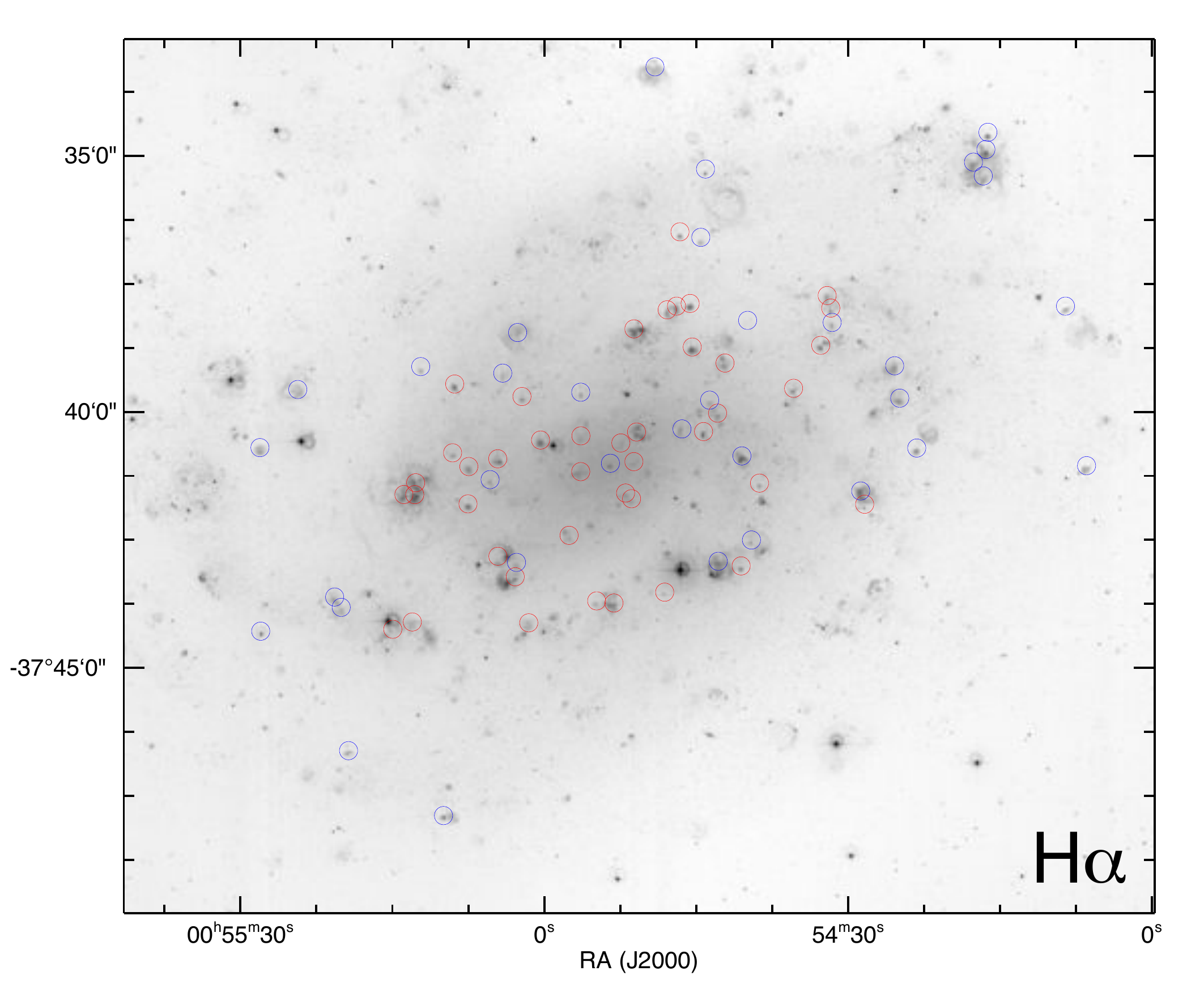}} \\
\includegraphics[width=0.5\linewidth]{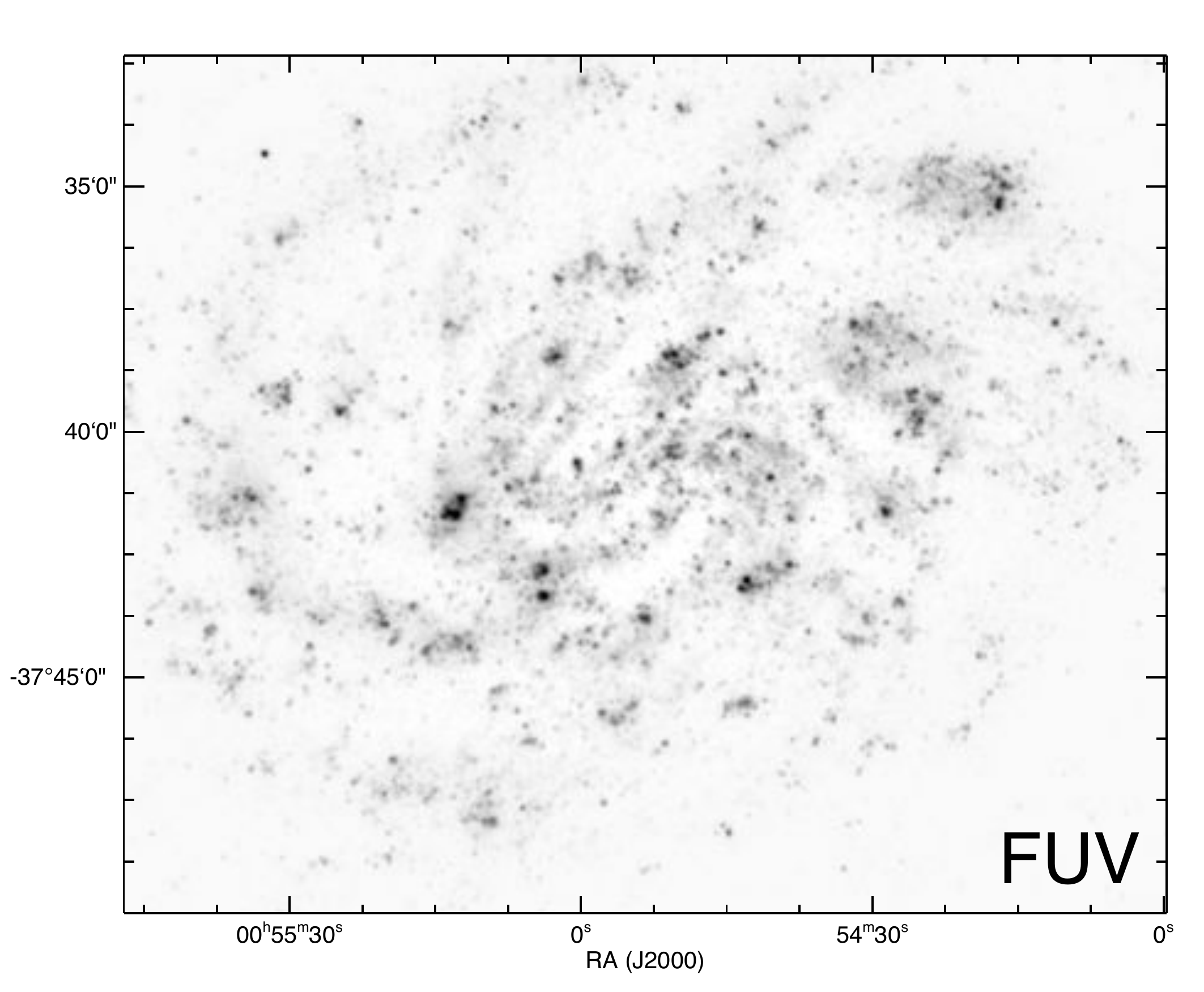} &
\includegraphics[width=0.5\linewidth]{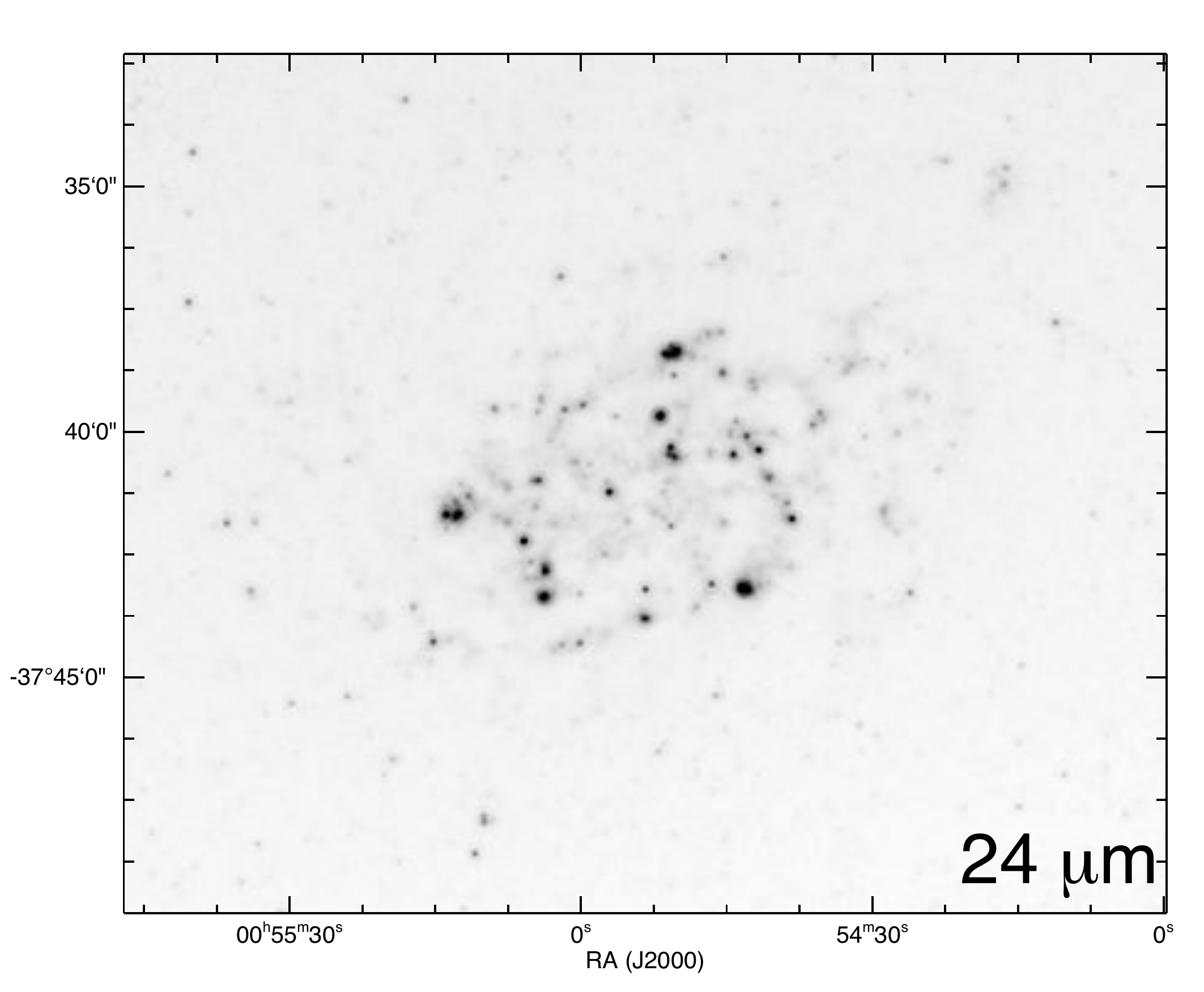} \\
\end{array}$
\caption{\small{ESO/WFI H$\alpha$ (top), GALEX FUV (bottom left), and \textit{Spitzer}/MIPS 24~$\mu$m (bottom right) images of NGC 300. All three images have been rescaled to a common field of view. Red circles on the H$\alpha$ image indicate APEX CO(2-1) detections, while blue circles show CO upper limits. The circle size (27{\arcsec}) corresponds to the APEX beam FWHM. See Appendix~\ref{sec:stamps} for zoom-in images at all three wavelengths.}}
\label{fig:images}
\end{figure*}

\section{Measurements and Photometry}
\label{sec:phot}

\subsection{APEX CO(2-1)}

We reduced our APEX CO data following standard procedures using the Gildas CLASS\footnote{http://www.iram.fr/IRAMFR/GILDAS/} software. To define the spectral window over which we search for CO emission, we derive \ion{H}{1} velocities for each source from the publicly available VLA first moment map of \cite{1990AJ....100.1468P} downloaded from NED. \ion{H}{1} velocities for our sources range from 78 to 197~km~s$^{-1}$. For each CO spectrum, we fitted and subtracted a polynomial baseline over a 300~km~s$^{-1}$ range approximately centered at the NGC~300 systemic velocity. We then combined all the spectra for each target and subtracted an additional first-order polynomial to account for any residual baseline slope. The spectral region within $\pm20$~km~s$^{-1}$ of the \ion{H}{1} velocity was not included in the baseline fit. Calibrated spectra are in units of corrected antenna temperatures ($T_A^*$), which we then converted to main beam brightness temperatures $T_{\rm MB}$ by dividing by the APEX-1 efficiency of $\eta_{\rm mb}=0.75$\footnote{http://apex-telescope.org/telescope/efficiency/}. We computed the integrated intensity $I_{\rm CO}=\int T_{\rm MB}(v)\, {\rm d}v$ for each target by integrating under the spectrum in a $\pm20$~km~s$^{-1}$ window centered on the \ion{H}{1} central velocity. In all cases this resulted in the inclusion of all significant emission. The formal uncertainty on $I_{\rm CO}$ is
\begin{equation}
\sigma_{I_{\rm CO}}=\sqrt{\Delta v \, w} \, \sigma_T,
\label{eqn:sigmaICO}
\end{equation}
where $\Delta v=1.39$~km~s$^{-1}$ is the spectral resolution, $w$ is the range over which the emission is measured (in km~s$^{-1}$), and $\sigma_T$ is the RMS noise of the spectrum (in K), as computed from line-free regions. As we have integrated all emission over the full $\pm 20$~km~s$^{-1}$ spectral window, we conservatively take $w$ to be 40~km~s$^{-1}$. We characterize the velocity extent of the CO emission using a parameter we label ``characteristic linewidth'', defined as $I_{\rm CO}$ divided by the peak temperature. The characteristic linewidth thus represents a relative measure of the velocity space spanned by the molecular gas without bias for any particular spectral morphology. Characteristic linewidths for our CO detections range from 5 to 16~km~s$^{-1}$, with a median of 8.5~km~s$^{-1}$.

Sources are considered to be a secure CO `detection' if both ${\ICO} > 3 \sigma_{I_{\rm CO}}$ and $T > 3\sigma_T$ in two or more consecutive channels near the \ion{H}{1} velocity. We classify sources fully satisfying one of the above criteria and meeting the other at least at the 2$\sigma$ level as `marginal detections'. All other sources are considered to be upper limits. We report upper limits as $2\sigma_{I_{\rm CO}}$, where $w$ in Equation (\ref{eqn:sigmaICO}) is now taken to be $17$~km~s$^{-1}$, twice the median linewidth of the detections. Since we have no direct information as to the spectral extent of potential CO lines below our detection threshold, this approximation provides the formal upper limit assuming that undetected lines are on average similar to detected ones. Following these criteria, our sample consists of 34 detections, 8 marginal detections, and 34 upper limits. The noise level is not uniform across the sample due to varying observing conditions and differing numbers of epochs for each source, and thus upper limits can be larger in magnitude than detections. For the remainder of this paper, we will use the terminology `CO detection' to refer to the aggregate of the secure and marginal detections.

As a cross-check, we also perform a three-parameter Gaussian fit to each spectrum in which CO is detected. This returns an additional estimate of {\ICO} as well as the CO central velocity and linewidth. The \ion{H}{1} and CO velocities are in very good agreement (centroids of both lines typically agree to within less than 4~km~s$^{-1}$, and at most 7~km~s$^{-1}$). The two measures of {\ICO} agree to within 2$\sigma_{I_{\rm CO}}$ for all 42 CO detections, and 32 out of 42 agree to within 1$\sigma_{I_{\rm CO}}$. However, the majority of the spectra were not particularly well-fit by a single Gaussian, and so we do not report the Gaussian fit parameters with the exception of the four sources discussed directly below.

Table~\ref{tab:obslog} lists quantities derived from the APEX data. We report the spectrally-integrated {\ICO} for the majority of sources; however, four CO detections exhibited obvious significant residual baseline errors within the integration window that downwardly biased this measure, and so we instead report the Gaussian-derived {\ICO} for these sources. We present the APEX spectra of our CO detections alongside zoom-in images in Appendix \ref{sec:stamps}.

\subsection{Photometry of ancillary data}
\label{sec:photometry}

We have used our APEX observations to probe the molecular (star-forming) ISM at 250~pc scales near {\HII} regions. To derive the star formation activity at these scales, we perform aperture photometry using apertures matched to this resolution. Since our pointed CO(2-1) observations provide no spatial information on the distribution of the gas at sub-resolution scales, we conduct photometry at the fixed scale of this resolution and centered on APEX targets, with only minor adjustments in some cases to incorporate emission from the entire {\HII}~region. We thus tacitly assume that, on average, the emission from young stars with which we trace the star formation rate is spatially coincident (to within 250~pc) with the CO-emitting clouds from which the stars presumably formed. In this section we describe our photometric procedure in detail.

We used the IDL \texttt{APER} task to perform background-subtracted aperture photometry on each of the GALEX FUV, \textit{Spitzer}/MIPS 24~$\mu$m, and H$\alpha$ images. For the majority of regions we used a 13.5{\arcsec} radius aperture, corresponding to the APEX CO(2-1) FWHM, centered on the APEX pointing position. In certain regions, the {\HII}~region clearly extended beyond the 13.5{\arcsec} aperture; in these cases, we adjusted the size and/or position of the aperture so as to encompass all contiguous emission while avoiding the addition of contaminating sources to the aperture. In most such regions the needed adjustments were slight; we report the aperture positions and sizes in Appendix~\ref{sec:stamps} for all regions for which changes were made. Background estimation was performed locally, though the choice of where to place the background annulus varied based on the structure of the emission as well as the level of source crowding. We performed the background selection procedure on a region-by-region basis using the H$\alpha$ image as reference. The same background regions were then used for the 24~$\mu$m and FUV images. For regions with compact emission and no other nearby sources, we utilized background annuli extending from 13.5{\arcsec} to 27{\arcsec} in radius. For sources with unassociated emission outside the aperture, we attempted to minimize contamination in the background annulus in the selection of the inner and outer radii. The outer radius was always chosen such that the annulus enclosed an area of at least three times the aperture area, when possible; in certain cases, reducing the outer radius was required to avoid contamination from other sources, but the annulus area was never less than twice that of the aperture. For sources in crowded regions, we placed much larger annuli such that a set of overlapping or nearby sources were all enclosed within the inner radius, and the annulus area was 2-3 times that of the standard 13.5{\arcsec} radius aperture. Inner radii chosen for crowded regions ranged from 16{\arcsec} to 36{\arcsec}, and outer radii from 28{\arcsec} to 72{\arcsec}. The background level per pixel for a given region was determined using the IDL task \texttt{MMM.PRO}, which uses a procedure that iteratively removes sources of emission within the annulus from the pixel distribution and returns the mode of the remaining pixels. Using our most crowded image, the GALEX FUV image, we verified that this procedure is robust even for annuli containing multiple sources by comparing average pixel values of nearby blank sky regions with the background levels found by our iterative algorithm.

The formal photometric uncertainty includes contributions from three sources, added in quadrature: noise within the aperture, as estimated by the scatter in pixel values in the annulus; uncertainty in the overall background level, as estimated by the error on the mean within the annulus; and, Poisson noise, which is directly estimated using the known gain values for each instrument. Poisson noise is insignificant ($< 0.1\%$) for H$\alpha$ and 24~$\mu$m, and contributes only a few percent at most to the uncertainty in the FUV. We have also included an additional uncertainty due to flux calibration: 5\% for GALEX FUV~\citep{2007ApJS..173..682M}, 4\% for \textit{Spitzer} 24~$\mu$m (MIPS Instrument Handbook), and 4\% for H$\alpha$ (see below). Finally, to incorporate any additional uncertainty due to narrowband continuum subtraction in the H$\alpha$ image, we used standard IRAF\footnote{IRAF is distributed by the National Optical Astronomy Observatory, which is operated by the Association of Universities for Research in Astronomy (AURA) under cooperative agreement with the National Science Foundation.} tasks to insert artificial point sources having a range of magnitudes into the final reduced and continuum-subtracted H$\alpha$ image and attempted to recover them. We treat the flux corresponding to the 3$\sigma$ limiting magnitude for a photometric detection over the background, scaled to the aperture size, as an additional uncertainty in our H$\alpha$ measurements. Regions for which the measured flux is less than three times the total photometric uncertainty are treated as upper limits, with 3$\sigma$ values reported. Table \ref{tab:results} lists the H$\alpha$, FUV, and 24~$\mu$m fluxes for each {\HII} region in our sample.

We validated our photometric method in two primary ways. First, for isolated point-like sources in the GALEX and \textit{Spitzer} images, we compared our $27\arcsec$ aperture measurements to photometry in which we vary the sizes of both the aperture and background annulus. We find that for apertures larger than twice the PSF, we recover at least 90\% of the flux we measure with the 27$\arcsec$ aperture. We also compared photometry computed using the mode background-subtraction method with iterative sigma-clipping and then taking the mean background, and find excellent agreement for the majority of sources in our sample.

We flux calibrated the H$\alpha$ image using observations of the standard star LTT17379 taken the same night as our NGC~300 observations. As an independent check on this calibration, we also conducted photometry on a number of previously measured {\HII} regions with measured fluxes published in \cite{2005ApJ...632..227R}. In particular, we measured the fluxes of 12 of their sources that are unambiguously uncrowded. We used the customized photometric aperture sizes specified in their Table~1 (9.6-36.0{\arcsec} in diameter), although the results change very little if we use 27{\arcsec} diameter apertures. We find a tight correlation between our photometry and their measured fluxes, and fit the relation using linear least-squares analysis to derive a calibration factor. This factor is within 4\% of the flux calibration determined from the standard star, and so we adopt 4\% as our flux calibration uncertainty, which we propagate into our final photometric errors. Note that \cite{2005ApJ...632..227R} report H$\alpha$ fluxes having accounted for \ion{N}{2} contamination of 10\% the total flux within the filter. Although in reality this number probably varies from source to source, the low average \ion{N}{2} fraction suggests that the effect on our results is likely minimal. Flux calibration is already applied to the Spitzer and GALEX images at the pre-download level, and we incorporate the reported calibration uncertainties into our results as discussed above. We applied standard aperture corrections of 1.047 and 1.16, respectively, for GALEX FUV and \textit{Spitzer} 24~$\mu$m measurements. These values were taken from the GALEX and \textit{Spitzer}/MIPS Instrument Handbooks and verified through tests on isolated point sources in our images.

We note that 17 of our photometric apertures overlap, a few significantly, with other apertures (see Appendix~\ref{sec:stamps} for zoom-in images of each region). In such cases the individual measurements are no longer entirely independent. We do not explicitly account for this effect in our analysis.

%% Table "tabulated photometry and results" placeholder
%% can also be placed at the end of the paper

\section{Methodology}
\label{sec:method}

In order to broadly contextualize our results, our goal is to measure molecular gas and star formation in a manner as consistent as possible with both the local sample of L10 and the external galaxies studied by, e.g., \cite{2004ApJ...606..271G}. $^{12}$CO becomes an effective tracer of molecular gas for gas with visual extinction $A_V \gtrsim 1-2$~magnitudes~\citep[e.g.,][]{Lombardi:2006fx}. Furthermore, \cite{Lada:2012it} demonstrated that deriving molecular cloud masses using either their $A_V \gtrsim 1$ extinction maps or $^{12}$CO images from \cite{2001ApJ...547..792D} results in masses that agree to within 12\% for the L10 local sample of molecular clouds. Thus our APEX $I_{\rm CO}$ measurements likely trace molecular clouds in a similar way to the $A_V \gtrsim 1$ ($A_K \gtrsim 0.1$) extinction maps of L10. In their sample of galaxies, \cite{2004ApJ...606..271G} also measure $I_{\rm CO}$, and thus the total molecular gas, just as we do in the present study. We introduce the term ``Giant Molecular Cloud Complex'' (GMCC) to indicate the (unresolved) molecular gas structure or structures measured within each 250~pc APEX beam, which may consist of one or more GMCs as well as potentially some amount of diffuse molecular gas. The mass of such an entity is hereafter designated $M_{\rm GMCC}$.

For estimating star formation rates, our multiwavelength data set provides a wealth of information on both the unobscured (H$\alpha$ and GALEX FUV) and embedded (\textit{Spitzer} 24~$\mu$m) star formation activity in NGC~300. Since we are targeting known {\HII} regions, we are probing the locations in which stars have formed very recently. Although we cannot resolve individual stars as L10 do in their sample of Milky Way star-forming regions, we can infer the integrated properties of the (young) stellar populations in each 250~pc sized region in our sample using population synthesis modeling. We introduce a new direct modeling procedure in which we compute the masses and ages of star-forming regions (\S~\ref{sec:sfrcalcs}). We then take the star formation rate to be the population mass divided by its age in a manner analogous to L10.

\subsection{Molecular gas mass}
\label{sec:mmol}

CO(2-1) integrated intensity was converted into CO line luminosity following \cite{1997ApJ...478..144S}:
\begin{equation}
\frac{L^{2-1}_{\rm CO}}{\rm{K~km~s}^{-1}~\rm{pc}^2} = 23.5 \left(\frac{\Omega_b}{\rm{arcsec}^2}\right) \left(\frac{D}{\rm{Mpc}}\right)^2 \left(\frac{I^{2-1}_{\rm CO}}{\rm{K~km~s}^{-1}}\right),
\end{equation}
where $\Omega_b$ is the beam solid angle, $D$ is the distance, and $I^{2-1}_{\rm CO}$ is the CO integrated intensity in the $J=2-1$ transition. We assume here implicitly that the beam size is larger than the source size, and so have replaced the solid angle of the source convolved with the beam with that of the beam. Given that GMCs in the Milky Way are typically $< 100$~pc in size, and that many of our sources show compact emission at other wavelengths, this assumption should be reasonable.

CO line luminosity is then converted into molecular gas mass $M_{\rm GMCC}$ using the CO-to-H$_2$ conversion factor, $\alpha_{\rm CO}$ as
\begin{equation}
\label{eqn:mmol}
\frac{M_{\rm GMCC}}{M_{\odot}} = 6.2 \left(\frac{0.7}{R_{21}}\right) \left(\frac{\alpha_{\rm CO}}{4.35}\right) \left(\frac{L^{2-1}_{\rm CO}}{\rm{K~km~s}^{-1}~\rm{pc}^2}\right)\, ,
\end{equation}
where $R_{21}$ is the CO(2-1)-to-CO(1-0) line ratio. This equation includes a factor of 1.36 to account for helium. $R_{21}$ depends on the excitation temperature and optical depth of the gas, and thus can vary within galaxy disks. Absent CO(1-0) data for NGC~300, we assume $R_{21}=0.7$, the average value found in the HERACLES survey of local disk galaxies by \cite{Leroy:2009di}, revised for the updated efficiency as discussed in \cite{Leroy:2013bf}. This choice is also similar to the average line ratio of $R_{21}=0.66$ in the Milky Way disk~\citep{Sakamoto:1995wq}. The uncertainty introduced by the assumption is likely small, but systematic; we note that enhancements of $R_{21}$ in spiral arms and in the inner regions of the Milky Way and other galaxies have been reported in the literature~\citep[e.g.,][]{1997ApJ...486..276S,Koda:2012ht}.

{\aCO} is well-known to exhibit variations both amongst and within galaxies~\citep[e.g.,][]{Bolatto:2013hl}. While the Milky Way value, $\alpha_{\rm CO}=4.35~M_{\odot}$~pc$^{-2}$~(K~km~s$^{-1}$)$^{-1}$, seems to describe well the average conversion factor in disk galaxies~\citep[][though see also \citealp{Sandstrom:2013bw}]{Leroy:2013bf}, our sample spans the entire disk of NGC~300 and potentially incorporates regions with differing physical conditions. In addition, NGC~300's metallicity is about one-half solar on average, and furthermore exhibits a decreasing trend with radius \citep{Deharveng:1988wh}. Thus we attempt to account for both the global deficiency of heavy elements and radial trend in our analysis.

\begin{figure}[tb]
\includegraphics[width=\linewidth]{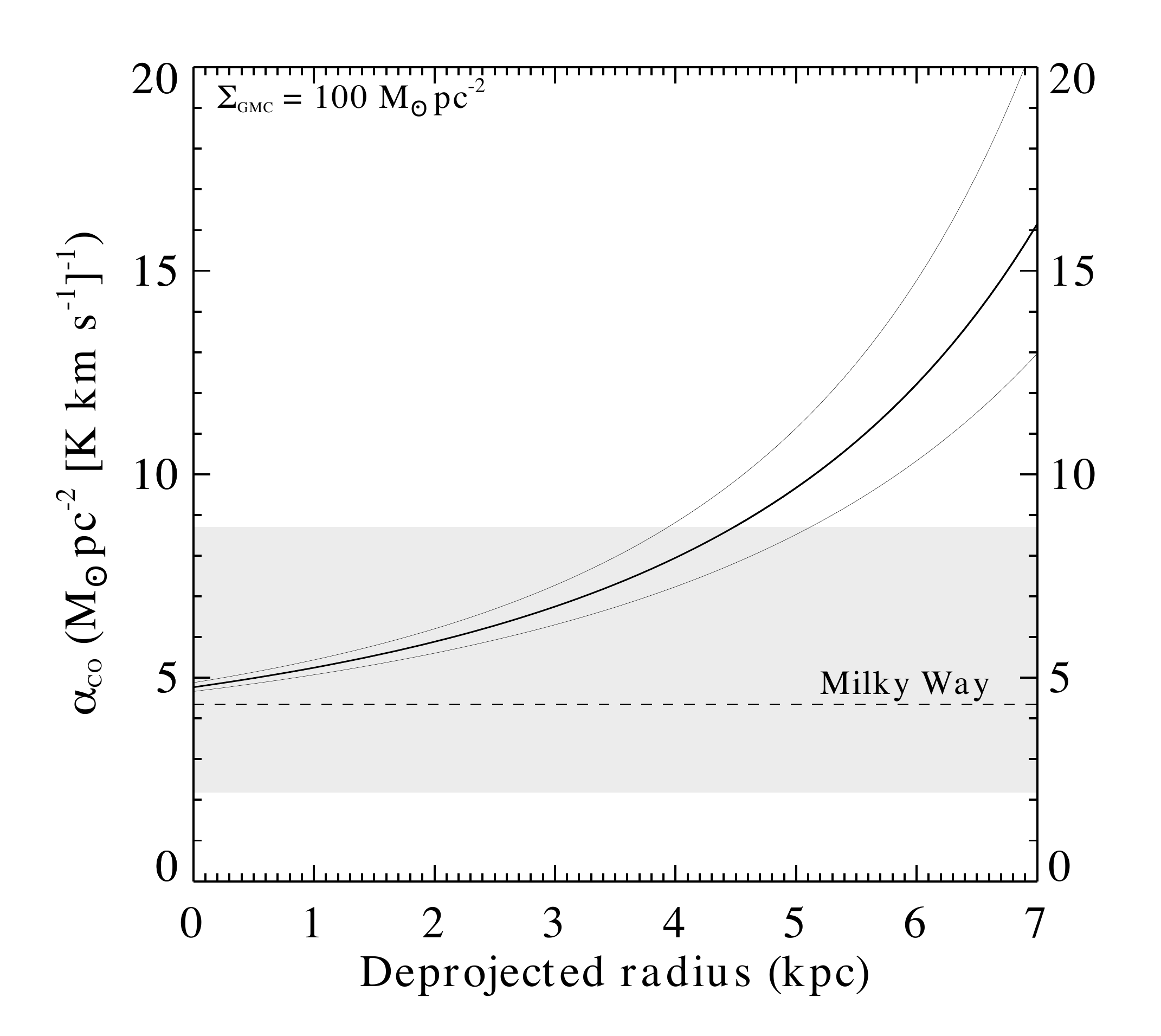}
\caption{\small{Radial variation of {\aCO} in NGC~300 due to decreasing metallicity with increasing galactocentric distance (dark line). The faint lines bracketing this relation indicate the range encompassed due to the uncertainty in the metallicity gradient. The Milky Way average value is shown as a horizontal dashed line, with the shaded region encompassing a factor of two variation. Our sample of {\HII} regions extends out to nearly 7~kpc.}}
\label{fig:aCOvar}
\end{figure}

Following \cite{Bolatto:2013hl}, we estimate the dependence of {\aCO} on metallicity as $\aCO \approx \alpha_{\rm CO, MW} f_{\rm COF}$, where $f_{\rm COF}$ is a correction that accounts for the H$_2$ gas in the outer layers of clouds where CO is mostly dissociated. It is a function of both the GMC surface density and the metallicity.  \cite{Bolatto:2013hl} suggest $f_{\rm COF}= 0.67 \exp{[0.4 / (\rm{Z}' \, \Sigma'_{\rm GMC})}]$, where $\rm{Z}'$ is the metallicity in units of the solar metallicity and $\Sigma'_{\rm GMC}$ is the characteristic surface density of molecular clouds in units of $100~\Msun$~pc$^{-2}$. As we have no resolved measurements from which to determine $\Sigma'_{\rm GMC}$ directly, we assume a value of unity (i.e. that GMCs have an average surface density of $100~\Msun$~pc$^{-2}$); studies of Milky Way GMCs find surface densities ranging from $\sim 40$ \citep[][Lada et al. 2013]{Heyer:2009ii} to $\sim 140~\Msun$~pc$^{-2}$ \citep{RomanDuval:2010gz}. The metallicity gradient in NGC~300 is given by $12+\log{\rm O/H} = 8.57 - 0.41 r/r_{25}$, where $r$ is the galactocentric radius and $r_{25}=5.3$~kpc is the radius at the $B$-band 25th magnitude isophote \citep{Deharveng:1988wh}. Taking the solar value to be $12+\log{\rm O/H}=8.66$ \citep{Bresolin:2009hh}, we derive the expression for the radial dependence of ${\rm Z}'$ in NGC~300:
\begin{equation}
{\rm Z}'(r) = 0.81 \times 10^{-0.41 r/r_{25}}.
\end{equation}
As Figure~\ref{fig:aCOvar} illustrates, the outwardly decreasing metallicity gradient in NGC~300 implies an increase in {\aCO} as a function of radius. Note that even if $\Sigma'_{\rm GMC}$ differs in NGC~300 from our Milky Way-motivated value of unity, the relative variation of {\aCO} with radius will be nearly preserved providing that $\Sigma'_{\rm GMC}$ is roughly constant throughout the disk. Furthermore, radial variation of $\Sigma'_{\rm GMC}$ within the disk would probably only exacerbate the change in {\aCO} with radius, as low surface density clouds are more likely to be found in the outer reaches of galaxies where the overall CO surface density is lower than at smaller radii~\citep[e.g.,][]{1998ApJS..115..241H,Schruba:2011em}.

\subsection{Star formation rates}

The size scales we investigate here ($\sim 250$~pc) are larger than the typical nearby massive GMCs (such as Orion) in the Milky Way, yet smaller than those probed by typical extragalactic studies that resolve $\sim$kpc-size regions and average over multiple stellar populations. In the Milky Way GMCs studied in L10, stars can be resolved, and thus star formation rates can be inferred by simply counting the total mass in young stars and dividing by their characteristic age. On the other hand, calculations of the star formation rate on kpc or galaxy-wide scales utilize integrated measures of unresolved stellar populations and population synthesis modeling. The models used typically assume that star formation has been proceeding continuously over $\sim100$~Myr timescales. While this number is appropriate when averaging over many stellar populations in various stages of evolution (as is the case at $\sim$kpc or larger scales), as the scale probed decreases, the assumption of continuous star formation begins to become less and less realistic. The limiting case is a single stellar population that formed over the course of only a few Myr -- practically instantaneous compared to the 100~Myr timescale applicable to larger regions. Assuming a 100~Myr timescale for a single region which is younger than an O-star lifetime (i.e., up to several Myr) has the effect of underestimating the true star formation rate, an effect which has been noted previously \citep{Chomiuk:2011iy}.

Are we targeting single stellar populations? Several of the sources we study here appear to contain only a single {\HII} region, as revealed by the presence of only a single structure in the H$\alpha$ line-only image, which has a resolution of 1.35{\arcsec} ($\sim$13~pc; FWHM of the measured PSF). Other sources exhibit secondary peaks within the 250~pc aperture, but the fact that both are H$\alpha$-bright at the same time suggests that these populations may be coeval and/or physically connected. We thus operate under the assumption that each of our 250~pc scale regions is well-approximated by a single stellar population that formed instantaneously, and use population synthesis models with the short timescales that reflect this assertion. To derive stellar masses and cluster ages for our sample, we compare our extinction-corrected FUV and H$\alpha$ observations to synthetic luminosity tracks from simulated instantaneous burst populations~\citep[e.g.,][]{Relano:2009jn}. Our \textit{Spitzer} 24~$\mu$m data are used to correct for extinction (Section~\ref{sec:ext}). We introduce and describe our direct population synthesis modeling technique in Section~\ref{sec:models}.

\subsubsection{Correcting for extinction}
\label{sec:ext}

While H$\alpha$ and FUV emission are both signposts for massive star formation, they can be significantly attenuated by dust. Since star forming regions tend to be rich in dust, this can leave only 20-40\% of the original emission at these wavelengths actually visible within some parts of a spiral galaxy \citep[e.g.,][]{1998ARA&A..36..189K}. The absorbed light is then re-radiated in the infrared as thermal emission from dust grains. Due to the intense UV radiation fields near young massive stars, dust can be significantly heated, and despite the fact that 24$\mu$m emission contributes only 5-10\% of the total infrared luminosity, it is considered a reasonable tracer of dust-obscured star formation in young star-forming regions~\citep[e.g.,][]{Leroy:2012ks}. The ratio of infrared to ultraviolet (or H$\alpha$, as a proxy) emission can therefore be used to correct for extinction. While not as direct as using the Balmer decrement or less attenuated recombination lines such as Pa$\alpha$, this method allows for region-by-region extinction estimates when more easily obtained infrared photometry is available. We are thus able to use our 24 $\mu$m photometry to account for the fact that extinction varies widely between the {\HII} regions in NGC~300 \citep{2005ApJ...632..227R}.

Typical multiwavelength prescriptions for computing SFRs in regions within galaxies linearly combine tracers of both unobscured and embedded star formation in the general form
\begin{equation}
\label{eqn:linearsfr}
\mathrm{SFR}=a L_{\rm vis} + b L_{\rm emb}
\end{equation}
to derive a complete census of star formation activity~\citep[e.g.,][]{Leroy:2008jk}. The first term in this expression -- the observed, or ``visible'' luminosity $L_{\rm vis}$ -- accounts for the UV or H$\alpha$ emission that escapes the local environment, while the second term -- the reradiated or embedded luminosity $L_{\rm emb}$ -- accounts for the remainder of the emission that is instead absorbed by dust and reradiated in the thermal infrared. Now if all the emission escaped, then instead we can write SFR$=aL_{\rm tot}$, with the same constant of proportionality as the first term in the previous expression since both utilize the same direct tracer. Since $L_{\rm tot}=L_{\rm vis} + L_{\rm emb}$, and the extinction is given by $A_{\lambda}=M_{\rm vis} - M_{\rm tot}$, where the $M$ are now in magnitudes, these equations can be algebraically combined to yield
\begin{equation}
A_{\lambda} = 2.5 \log \left(1 + \frac{b}{a} \frac{L_{\rm emb}}{L_{\rm vis}} \right)
\end{equation}
Taking 24~$\mu$m emission as our embedded tracer, we derive extinction corrections for our data based on Equation (7) of \cite{2007ApJ...666..870C} for H$\alpha$, and Equation (D10) of \cite{Leroy:2008jk} for FUV\footnote{The coefficient on L(24 $\mu$m) in \cite{Leroy:2008jk}, Equation (D10) is $10^{-43}$ (Leroy, private communication).}, both of which are of the form of Equation~(\ref{eqn:linearsfr}), whence
\begin{subequations}
\begin{align}
A_{\rm{H} \alpha} &= 2.5 \log \left(1 + 0.031 \frac{L[24 \mu \rm{m}]}{L[\rm{H}\alpha]}\right) \\
A_{\rm{FUV}} &= 2.5 \log \left(1 + 6.05 \frac{L[24 \mu \rm{m}]}{L[\rm FUV]}\right).
\end{align}
\label{eqn:ext}
\end{subequations}
Importantly, Equations~(\ref{eqn:ext}) were derived from calibrations applicable not to entire galaxies, but to individual regions within them. The dust heating in {\HII} regions is dominated by radiation from young stellar populations, while the average heating across a galaxy includes a nontrivial component from more evolved stars~\citep[e.g.,][]{Leroy:2012ks}. Using a calibration appropriate for integrated galaxies would thus underestimate the extinction and total SFRs in {\HII} regions \citep{Kennicutt:2009iv}.

\subsubsection{Models}
\label{sec:models}

We generate a grid of models using the publicly available \texttt{Starburst99} population synthesis code~\citep{1999ApJS..123....3L}\footnote{http://www.stsci.edu/science/starburst99/docs/default.htm}. The relevant parameter choices are listed in Table~\ref{tab:SB99}. The choice to use an instantaneous burst is motivated by the assumption that each of our 250~pc regions consists of a single primary stellar population that formed at roughly the same time, as discussed above. We choose a metallicity of 0.008 (0.4 Z$_{\odot}$), the closest option to NGC 300's characteristic metallicity of 0.56 Z$_{\odot}$ \citep{Bresolin:2009hh}. We use a Kroupa IMF, a two-part power law with slope $\alpha=-2.3$ above 0.5~{\Msun} up to a cutoff mass of 100~{\Msun}. See Section~\ref{sec:SFRunc} and Appendix~\ref{sec:IMF} for further motivation for the choice of cutoff mass and the implications of changing the IMF assumptions. The model outputs include H$\alpha$ line luminosities and low ($\sim 10$~{\AA}) resolution spectra, which we integrate over the GALEX FUV filter curve to derive FUV luminosities. The H$\alpha$ nebular continuum is insignificant ($\lesssim 1\%$) compared to the line emission over the full range of timescales we model.

\begin{center}
\begin{deluxetable}{lc}
\tablecaption{ \textit{Starburst99} Parameters \label{tab:SB99}}
\tablehead{
\colhead{Parameter}		& \colhead{Range}	 \\
}
Population Mass	& Fixed: $500$-$10^5$~{\Msun} \\
IMF				& Kroupa\tablenotemark{a} \\
$M_{\rm upper}$	& 100 \\
Stellar evolution tracks		& Geneva (high mass loss\tablenotemark{b}) \\
Metallicity			& 0.008 (0.4 $Z_{\odot}$) \\
Wind model		& Evolution \\
Mass interpolation	& Full isochrone \\
Time step			& 0.1 Myr \\
\tablenotetext{a}{Two-part power law over the ranges $0.1<M<0.5$ and $0.5<M<M_{\rm upper}$ (all masses in {\Msun}) with slopes $\alpha=-1.3$ and $-2.3$, respectively \citep{Kroupa:2002cm}.}
\tablenotetext{b}{\cite{1994A&AS..103...97M}}
\end{deluxetable}
\end{center}
\subsubsection{Deriving SFRs: A direct modeling approach}
\label{sec:sfrcalcs}

\begin{figure}[tb]
\includegraphics[width=\linewidth]{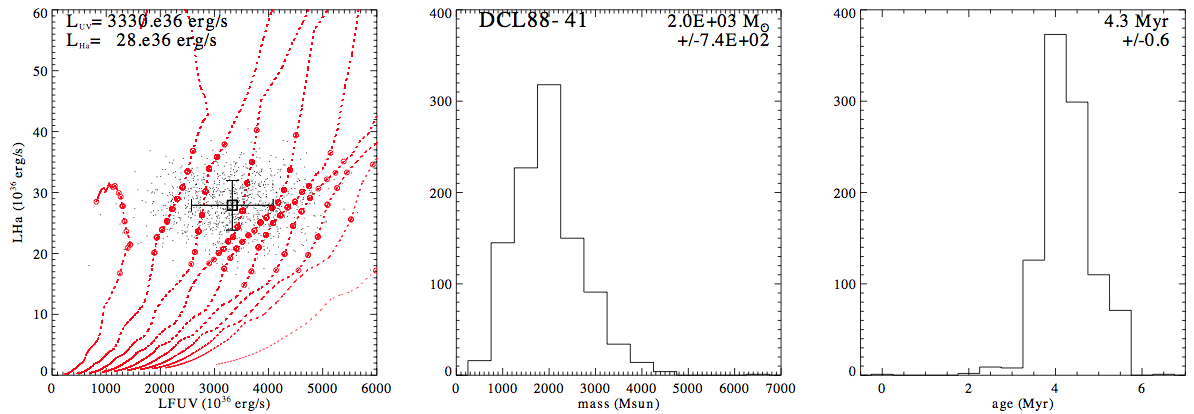}
\caption{\small{Graphical example of our Monte Carlo procedure for deriving stellar population masses ($M_*$) and ages ($t$), shown for the source DCL88-41 (left panel). The black square shows the measured H$\alpha$ and FUV luminosities and uncertainties, while the black dots indicate positions of the 1000 random draws from Gaussian luminosity distributions defined by these measurements (see the text). The red dotted lines show model tracks, and the red circles are individual best fit models; each corresponds to the closest grid point for a given black dot. The center and right panels show the posterior distributions for $M_*$ and $t$, with the best-fit values and uncertainties listed.}}
\label{fig:MC}
\end{figure}

\begin{figure*}[tb]
\includegraphics[width=\linewidth]{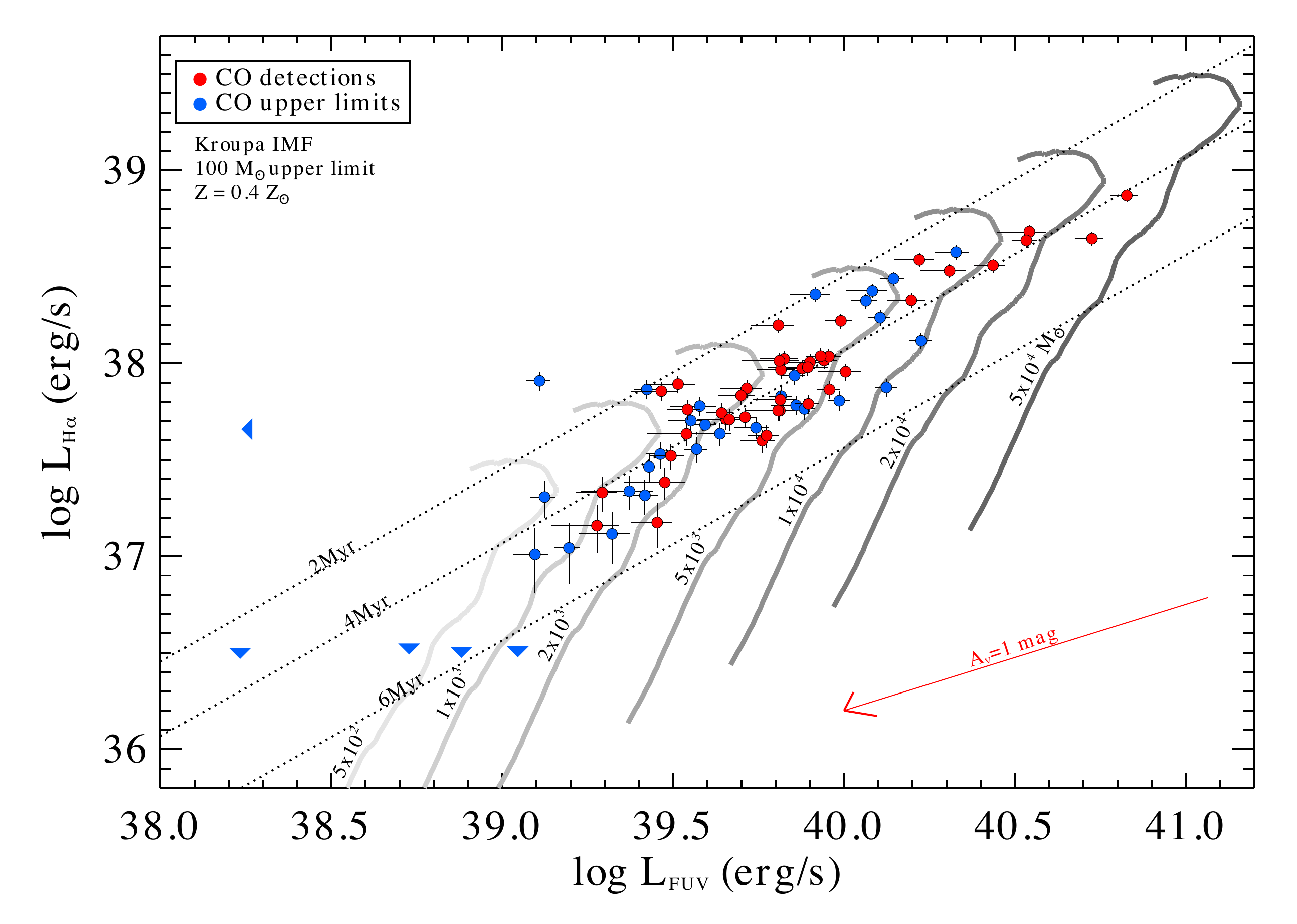}
\caption{\small{Extinction-corrected H$\alpha$ and FUV luminosities for our sample of 76 {\HII} regions in NGC300 (data points) plotted on \texttt{Starburst99} model tracks (gray lines). Red points show CO detections while blue points are CO upper limits. The models range in total stellar mass from 500 to $5\times10^4~\Msun$ and each model population evolves to the right then downward in the figure over 10~Myr. Lines of constant population age of 2, 4, and 6 Myr are shown to guide the eye. A unique mass and age for each region is derived by comparing the observed luminosities with those of model young stellar populations in ($L_{\rm FUV}$, $L_{\rm{H}\alpha}$) space. The reddening correction corresponding to one visual magnitude is shown with a red line.}}
\label{fig:HavsFUV}
\end{figure*}

To derive star formation rates on a region-by-region basis, we utilize our grid of \texttt{Starburst99} models with population mass (hereafter $M_*$) ranging from 500 to $10^5$~{\Msun} in steps of 500 {\Msun} and ages $t$ from 0.1 to 10~Myr in 0.1 Myr steps. Figure~\ref{fig:HavsFUV} shows extinction-corrected {\LHa} and {\LFUV} for our NGC~300 {\HII} regions overplotted on the model tracks. We then use an iterative Monte Carlo procedure to estimate the mass and age of each observed region as follows. For each iteration, we take a random draw from each of two independent Gaussian distributions corresponding to the observed H$\alpha$ and FUV luminosities. The H$\alpha$ distribution is defined to have mean equal to the measured $L_{\rm{H}\alpha}$ and standard deviation $\sigma_{L_{\rm{H}\alpha}}$, and the FUV distribution has mean equal to the measured $L_{\rm FUV}$ and standard deviation $\sigma_{L_{\rm FUV}}$ for that region. In other words, for each of FUV and H$\alpha$, we take a random draw from the distribution described by the measured luminosity and corresponding uncertainty. We then find the model stellar population with ($L_{\rm FUV}$,$L_{\rm{H}\alpha}$) closest to the Monte Carlo draws $L_{i, \rm{MC}}$ for that iteration by minimizing $\chi^2$ across the model grid, where
\begin{equation}
\chi^2 = \sum\limits_{i=1}^2 \frac{(L_{i,\rm{MC}} - L_{i, \rm{model}})^2}{\sigma_i^2}.
\end{equation}
Here $i$ runs from 1 to 2 corresponding to comparisons in H$\alpha$ and FUV. We then take the mass and age of this model population to be the best-fit values for that iteration. We repeat this procedure 1000 times for each source, then take the means of the distributions of these best-fit values as the final best-fit mass and age for the observed region. The standard deviations of the mass and age posterior distributions are taken to be the uncertainties on these parameters. Regions with photometric upper limits are treated somewhat differently. Instead of drawing from a Gaussian distribution for the wavelength at which the source is an upper limit, we draw from a random uniform distribution ranging from 0 to the upper limit value. Figure~\ref{fig:MC} illustrates our Monte Carlo procedure for the source DCL88-41 and shows the posterior distributions for $M_*$ and $t$ for this source. These distributions are generally centrally peaked and well-described by their first and second moments.

To derive the SFR for each {\HII} region in a manner analogous to that of L10, we divide the best-fit population mass by the population age:
\begin{equation}
\rm{SFR}=\frac{M_*}{t}.
\label{eqn:SFR}
\end{equation}
We then discard regions for which we derive a mass lower than 1000~{\Msun}, as these regions (1) partially scatter outside the parameter space spanned by our model grid, and (2) have highly uncertain masses and ages as a result of flat or double-peaked mass and age posterior distributions. We also discard regions with derived ages $< 1$~Myr, as all of these again lie outside our model parameter space. All seven of the regions discarded in this way are APEX CO upper limits.

\subsubsection{SFR uncertainties}
\label{sec:SFRunc}

In this section we discuss three possible contributions to the uncertainty in our modeling results: the assumption of instantaneous (vs. continuous) star formation, the form of the IMF, and the effects of stochastic sampling of the IMF.

One potential uncertainty in our derived star formation rates may be a result of our choice of an instantaneous burst model (in which we assume that the stellar population formed in a short time compared to its age) instead of a model in which star formation is continuous over a timescale comparable to the potential lifetime of a molecular cloud \citep[5--10~Myr; e.g.,][]{1989ApJS...70..731L}. To test the implications of changing this assumption, we compared the results we obtained from the models discussed in \S~\ref{sec:models} with those derived from continuous star formation models. In particular, we ran a series of \texttt{Starburst99} simulations with identical parameters to those described in Table~\ref{tab:SB99}, with the exception that we set SFRs to be continuous with rates ranging between $10^{-3}-10^{-2}~\Msun$~yr$^{-1}$. Initially, model tracks for the H$\alpha$ and FUV luminosities from a continuous star formation model evolve towards higher luminosities in both axes. However, H$\alpha$ remains constant after reaching a population age at which the birth rate of ionizing stars matches the death rate \citep[$\sim 8$~Myr;][]{Chomiuk:2011iy}, while FUV continues to increase until it reaches the corresponding FUV steady-state time ($\sim80$~Myr). We then compared our instantaneous burst SFRs to those derived from the continuous SFR models at specific timescales: 2~Myr, the luminosity-weighted average timescale for the emission of an H$\alpha$ photon in an instantaneous burst population \citep{Leroy:2012ks}; 3.7~Myr, the characteristic age of our stellar populations as computed from our instantaneous burst model grid; and, 10~Myr, the upper limit for the lifetime of a GMC~\citep{1989ApJS...70..731L}. For all three cases we derived SFRs within a factor of two of our previous results, though we note that using the 10~Myr timescale resulted in SFRs that were on average the most discrepant from our instantaneous burst results, again highlighting the importance of choosing model timescales judiciously. Thus we conclude that, provided that we use appropriate ($<10$~Myr) timescales for 250~pc star forming regions, a continuous star formation model is also a reasonable assumption. For our analysis, we prefer the instantaneous burst models, as they allow modeling of the stellar population masses and ages, and a computation of the SFR in a manner more similar to that of L10. Using a continuous SFR model would produce a systematic shift toward lower SFRs by a factor of on average about 40\%, the difference between the two sets of models when the continuous SFR tracks are read out at 3.5~Myr.

A second potential source of uncertainty is due to the fact that our analysis relies on tracers of only the most massive stars, which produce a majority of the luminosity but comprise a minority of the total stellar mass of the population. Thus there is some sensitivity of our modeling results to the functional form, slope, and upper mass cutoff of the IMF. Previous studies have investigated this issue in detail \citep[e.g.,][]{2007ApJ...666..870C,Leroy:2008jk,Chomiuk:2011iy}, finding that differences in derived SFRs using Salpeter and Kroupa-type IMFs amounted to factors of about 50\%. We thus caution that our results may suffer from additional systematic uncertainties at this level. However, assuming that the form of the IMF is similar across NGC~300's star clusters, this should not affect the relative difference in SFRs amongst regions, and so we do not include this systematic uncertainty in our reported SFRs. To illustrate the effects of changing the upper mass limit of the IMF, we present in Table~\ref{tab:varIMF} the median SFRs for our sample of {\HII} regions derived for a range of IMF upper limits from 50 to 120~{\Msun}. Over this entire range, SFRs change by a factor of at most 30\%. Note that upper mass limits lower than 50~{\Msun} lead to unphysical results in that the majority of the NGC~300 data points then lie outside the parameter space spanned by the tracks. We take the uncertainty on our derived SFRs to be $(+0.10,-0.02)$~dex, the relative difference in median SFR derived using our chosen 100~{\Msun} IMF upper mass limit as compared to that derived using 50 and 120~{\Msun} upper limits.

A third source of uncertainty is due to stochastic sampling of the upper end of the IMF, which is particularly problematic for low-mass stellar populations~\citep[e.g.,][]{2012ApJ...745..145D}. Since the SFR tracers we use here rely on the presence of massive stars, the uncertainty in the total derived population mass using these tracers (and thus the SFR) increases as the expected number of massive stars in the population declines. Previous studies have found that this effect is non-negligible for populations with fewer than about ten O-stars, or equivalently population masses $M_{\rm pop} \lesssim 3000~\Msun$~\citep{2003MNRAS.338..481C,Lee:2009bs}. To quantify and account for this effect, we have computed the ``most massive star'' (with mass $M_{\rm mm}$) expected in a stellar population as a function of the population mass for our fiducial Kroupa IMF by analytically integrating the upper end of the IMF and solving for the stellar mass above which only one star is expected (see Appendix \ref{sec:IMF} for details). Since the Poisson error on a histogram bin with value unity is also unity, the 1$\sigma$ range on the number of stars in the mass bin centered on $M_{\rm mm}$ extends from zero to two. Thus we take the FUV and H$\alpha$ luminosities emitted by a star with mass $M_{\rm mm}$ to be the stochastic contribution to the uncertainties on the FUV and H$\alpha$ luminosities of the population. This is propagated formally into uncertainties on $M_*$, $t$, and the SFR; see Appendix~\ref{sec:IMF} for a complete description of our procedure. The additional uncertainty from stochastic effects amounts to as much as a factor of 40\% in the low-mass ($M_*\sim1000~{\Msun})$ regime but becomes negligible ($<10\%$) above about $10^4$~{\Msun}.

\begin{center}
\begin{deluxetable}{cc}
\tablecaption{Effects of changing the IMF upper limit \label{tab:varIMF}}
\tablehead{
IMF upper limit	& Median SFR \\
 ({\Msun})	& ($10^{-3}~\Msun$~yr$^{-1}$)}
\startdata
50	& 1.08 \\
80	& 0.90 \\
100	& 0.86 \\
120	& 0.82
\enddata
\end{deluxetable}
\end{center}

\section{Results}
\label{sec:results}

\subsection{Molecular gas}

We detect CO$(J=2-1)$ in 42 of the 76 {\HII} regions surveyed (including marginal detections), following the criteria discussed in Section~\ref{sec:APEX}. Molecular gas masses $M_{\rm GMCC}$ in these regions range from 1$\times 10^5$ to $7\times 10^5~M_{\odot}$, extending the upper range of the L10 local cloud sample by almost an order of magnitude. In this section we examine the radial variation of the detection rate and the GMC Complex (GMCC) mass spectrum.

\subsubsection{Radial variation}
\label{sec:radvar}

\begin{figure}[tb]
\includegraphics[width=\linewidth]{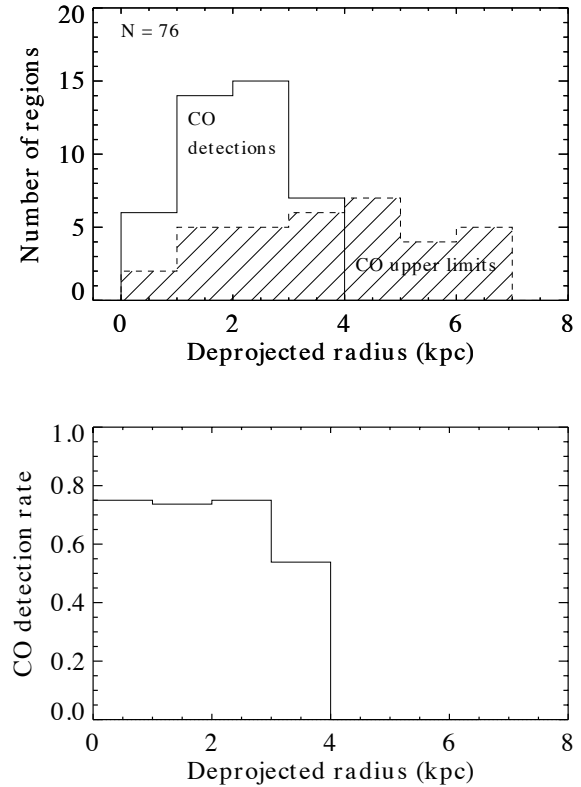}
\caption{\small{Histogram by galactocentric radius for the sample of NGC 300 {\HII} regions (top panel) detected in CO with APEX (open histogram) and those not detected in CO (hatched histogram). The bottom panel shows the detection rate as a function of radius, binned in 1 kpc bins. We do not detect CO beyond 4 kpc from the center of NGC~300, likely primarily due to the decreasing ability of CO to trace molecular gas in the outer, lower metallicity regions of the galaxy (see the text).}}
\label{fig:radialco}
\end{figure}

Our {\HII} region sample was chosen to cover a representative range in galactocentric radii of NGC~300's disk (Figure~\ref{fig:DCLhist}). To assess the distribution of gas-rich star forming regions within the galaxy, we plot histograms comparing the radial distributions of CO-detected sources and upper limits in Figure~\ref{fig:radialco}, binned by 1~kpc. In the bottom panel of the Figure we also show the CO detection rate as a function of galactocentric radius. The detection rate is simply defined to be the number of regions in a given radius bin in which CO is detected divided by the total number of observed regions in that radius bin. {\HII} region positions were deprojected using a custom IDL routine based on \texttt{im\_hiiregion\_deproject.pro} by J. Moustakas. The detection rate clearly decreases with increasing radius. The inner 3-4 kpc is dominated by CO detections, and we significantly detect CO in about 75\% of regions inside 3~kpc. However, despite the sample extending to $>6$~kpc, we do not detect CO in any regions with distances greater than 4 kpc. This trend is most likely a direct result of the fact that CO becomes a progressively poorer tracer of molecular gas toward the outer reaches of the galaxy due to the negative radial metallicity gradient. Since {\aCO} increases by up to a factor of three for the regions at largest galactocentric distance~(see \S~\ref{fig:aCOvar}), a cloud of a given mass in the outer reaches of NGC~300 would require a factor of nine longer integration time to detect in CO(2-1) than the same cloud near the galaxy center. We thus very likely miss such clouds in the outskirts despite being able to detect similar objects at smaller radii. The decrease in detection rate with radius may also be due in part to the fact that GMCs more distant from the galaxy's center are on average less massive than those within the disk. The average surface density of molecular gas is lower in the outer part of the Milky Way \citep[e.g.,][]{1998ApJS..115..241H}, and this may be the case for NGC~300 as well. If so, we may further be underestimating the masses of clouds at large galactocentric radii due to additional increase of {\aCO} in these regions caused by a lower $\Sigma'_{\rm GMC}$ than the Milky Way-average value we have assumed (see \S~\ref{sec:mmol}).

\subsubsection{GMC Complex mass spectrum}

\begin{figure}[tb]
\includegraphics[width=\linewidth]{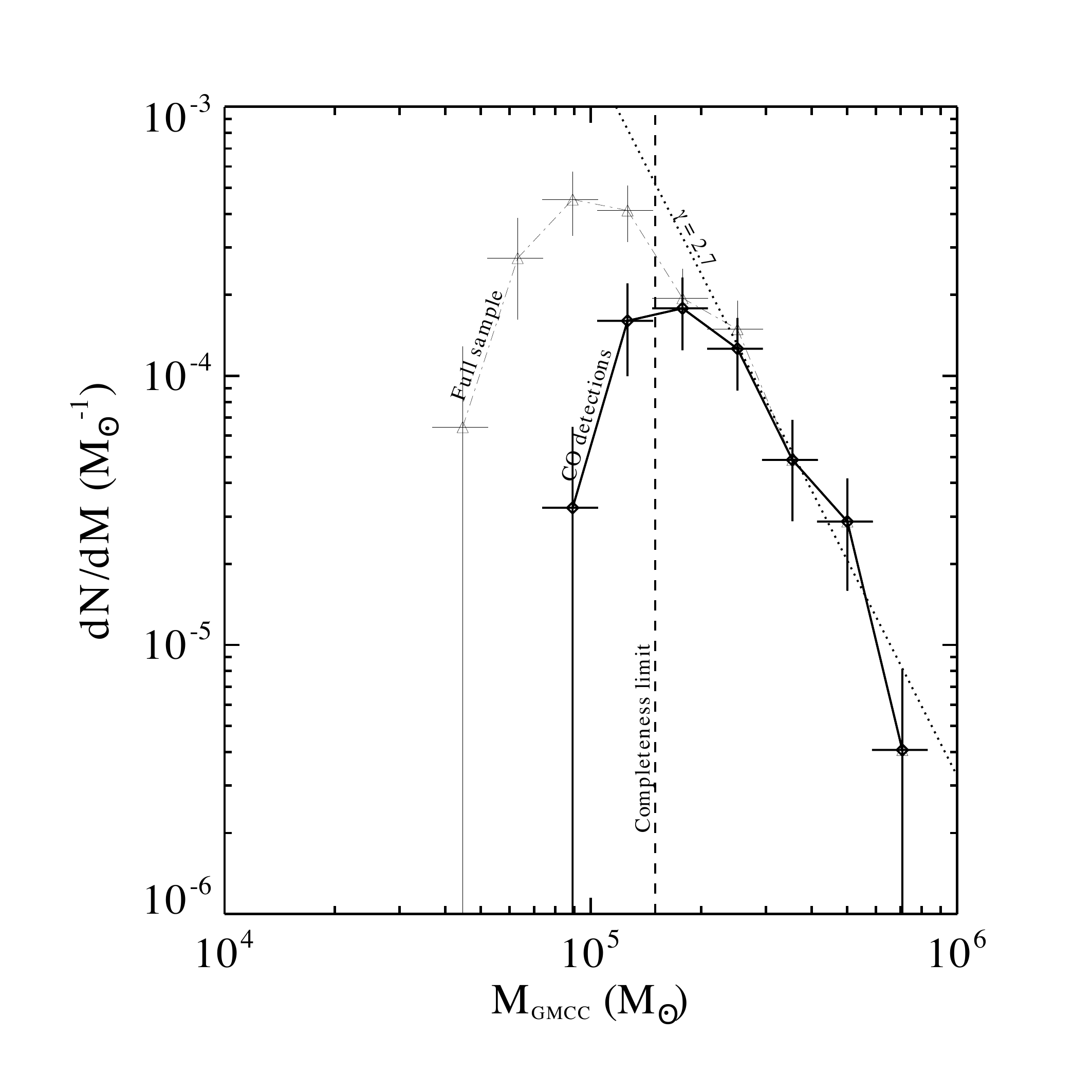}
\caption{\small{Differential mass spectrum $dN/dM$ of the CO-detected regions in our APEX sample (diamonds, solid line). Bins are 0.15~dex, corresponding to twice the typical measurement error on $M_{\rm GMCC}$. Triangles connected with a dot-dashed line show the spectrum of the full APEX sample (including CO upper limits). The vertical dashed line indicates our completeness limit, above which the majority of our mass measurements are CO detections. A power law fit to the bins above the completeness limit yields a best-fit slope of $\gamma=2.7 \pm 0.5$ (dotted line), similar to the slope of the cloud mass spectrum in the nearby spiral M33, which shares physical and morphological characteristics with NGC~300. The vertical error bars are Poisson errors on the histogram, while the horizontal bars indicate the bin sizes.}}
\label{fig:massspec}
\end{figure}

The mass spectrum of clouds (or other bound entities) is generally expressed in differential form and taken to be a power law with index $\gamma$, i.e.,
\begin{equation}
\frac{dN}{dM}=f(M)=M^{-\gamma}.
\end{equation}
We calculate the mass spectrum for our sample of 42 CO-detected sources in NGC~300 by separating the $M_{\rm GMCC}$ distribution into equally spaced logarithmic bins, where the bin size of 0.15~dex corresponds to twice the typical fractional uncertainty in $M_{\rm GMCC}$. We estimate $dN/dM$ for each bin as $N_{\rm bin}/\Delta M$, where $N_{\rm bin}$ is the number of sources in a bin, and $\Delta M$ is the (linear) width of that bin. We assume Poisson uncertainties on $dN/dM$, i.e. $\sigma = \sqrt{N}/dM$. We present the NGC~300 APEX GMCC mass spectrum in Figure~\ref{fig:massspec}. The vertical dashed line denotes the mass below which more than 20\% of our measurements are upper limits; we take this mass to be our completeness limit. Values of $dN/dM$ for the complete sample including CO upper limits are also shown to illustrate the completeness limit. Fitting a power law to the bins above the completeness limit using ordinary least squares fitting (accounting for Poisson errors), we find $\gamma=2.7\pm 0.5$ to be the best-fit slope. This fit is shown as a dotted line in the figure. Changing the bin size has only minor consequences on the best-fit $\gamma$: varying the bin size between 0.08 and 0.25~dex, the best-fit power law index ranges from 2.1 to 2.7.

While a GMCC mass spectrum does not necessarily map to a GMC mass spectrum with similar slope, it is nevertheless instructive to consider the distribution of GMCCs by mass in the context of cloud mass functions in other galaxies. Our derived slope of $\gamma=2.7$ is relatively steep compared to the power law indices computed for inner Milky Way clouds ($\gamma \approx 1.5$) by \cite{Rosolowsky:2005gt}. That same study found progressively steeper indices in the LMC ($\gamma=1.7$) and M33 ($\gamma=2.9$). They attribute differing power law indices amongst local group galaxies as being real differences between the GMC populations in these galaxies. In the case of M33, \cite{Rosolowsky:2005gt} suggest that the bottom-heavy mass function results from gravitational stability in the disk~\citep{2001ApJ...555..301M} that may inhibit the formation of massive GMCs there. Intriguingly, NGC~300 is much more similar to M33 than to the Milky Way in terms of metallicity (subsolar with a negative radial gradient), size, integrated luminosity (at several wavelengths), and morphology (late spiral). If NGC~300 also has a gravitationally stable disk, as does M33, this could potentially explain its apparently steep GMCC mass function.

However, we again caution that our sample represents a set of GMCCs -- $<250$~pc-scale conglomerations of CO-emitting gas near known {\HII} regions -- and is not a comprehensive catalog of GMCs in NGC~300. The number of clouds and fraction of gas organized in clouds may differ between GMCCs, and thus our GMCC mass function may or may not map to a GMC mass function with similar slope. Preliminary interferometric studies of a small subset of our sample do suggest, however, that a single discreet molecular structure generally dominates the CO luminosity within the 250~pc APEX beam (C. Faesi et al. 2014, in preparation).

\subsection{Star formation rates}
\label{sec:SFRs}

\begin{figure*}[tb]
$\begin{array}{cc}
\includegraphics[width=0.5\linewidth]{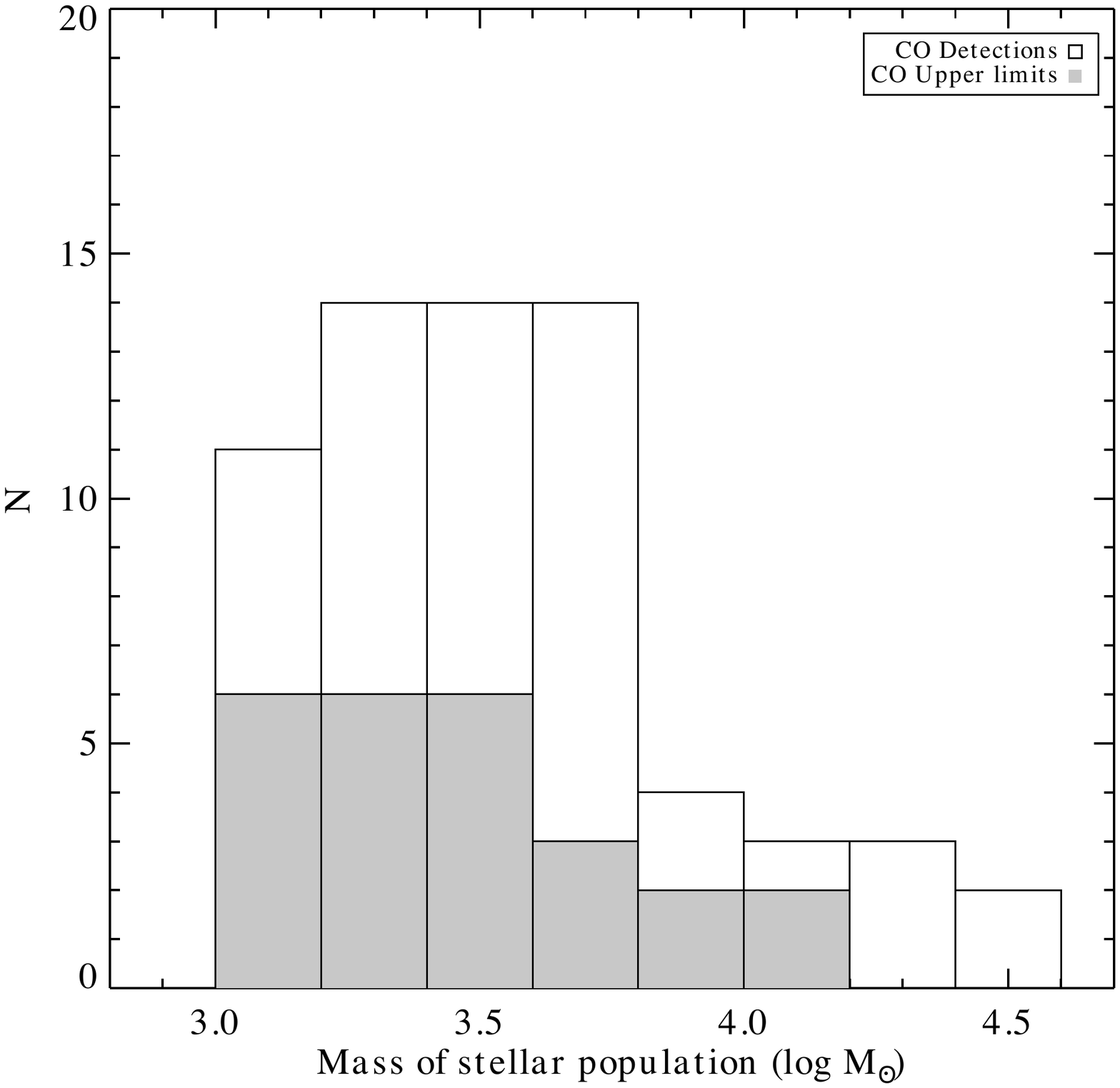} &
\includegraphics[width=0.5\linewidth]{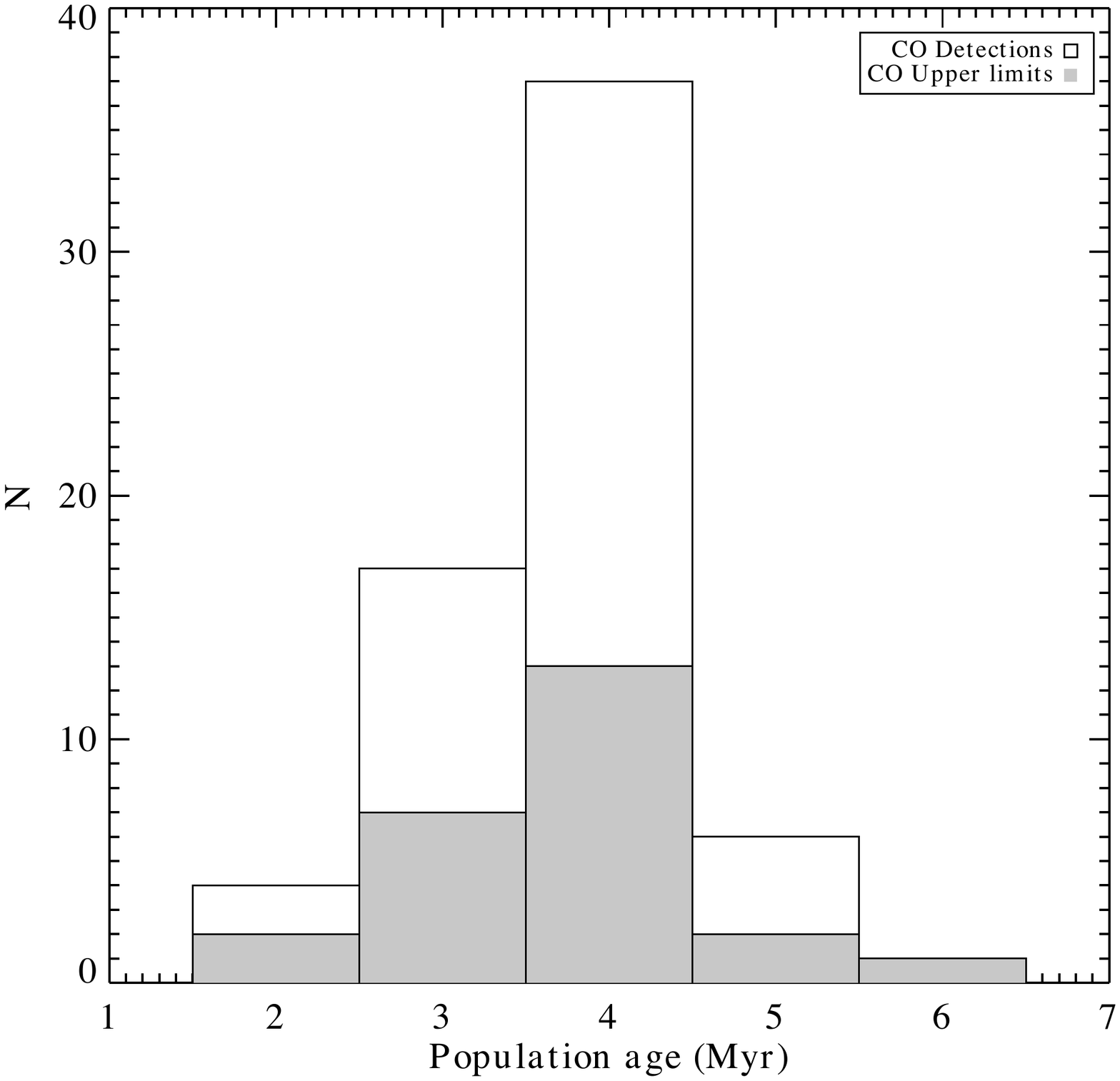} \\
\end{array}$
\caption{\small{Stacked histograms of the stellar mass (left) and age (right) of the best-fit model stellar populations to our sample of {\HII} regions. Regions in which CO is detected in our APEX survey are shown as open histograms, while those with CO upper limits are shown as shaded histograms.}}
\label{fig:hists}
\end{figure*}

Figure~\ref{fig:hists} presents the stellar mass ($M_*$) and age ($t$) histograms for our sample of NGC~300 HII regions computed using our new direct modeling approach, including both CO detections and upper limits (excluding those with $M_* < 1000~\Msun$ or $t<1$~Myr; \S~\ref{sec:sfrcalcs}). The $M_*$ distribution of the CO detections is centered at a higher mass (median 4000~{\Msun}) than that of the CO upper limits (median 2600~{\Msun}), suggesting that massive GMCs are preferentially associated with large stellar populations. The $M_*$ distribution of the CO detections is also much broader than that of the upper limits. These two differences evidently reflect that the most massive young clusters in our sample -- those with $M_* \gtrsim 1.5 \times10^4$~{\Msun} -- are always associated with CO clouds, while less massive clusters may or may not be. The age distributions are similar, both having a median of 3.7~Myr. A more quantitative interpretation is complicated by the varying CO detection threshold across our sample; it is possible that most (or all) of our {\HII} regions are associated with GMCs, and that many GMCs are simply undetected in our sensitivity-limited survey, particularly at large galactocentric radii where CO becomes a progressively less reliable tracer of molecular gas. In contrast to the decreasing CO detection rate with radius (\S~\ref{sec:radvar}), there is no significant trend in either $M_*$ or $t$ as a function of radius. There is no marked difference in the typical SFR between NGC~300 regions with APEX CO detections and those with upper limits (median $9.5 \times 10^{-4}$ and $9.1 \times 10^{-4}~\Msun$~yr$^{-1}$, respectively.

\begin{figure*}[tb]
\includegraphics[width=\linewidth]{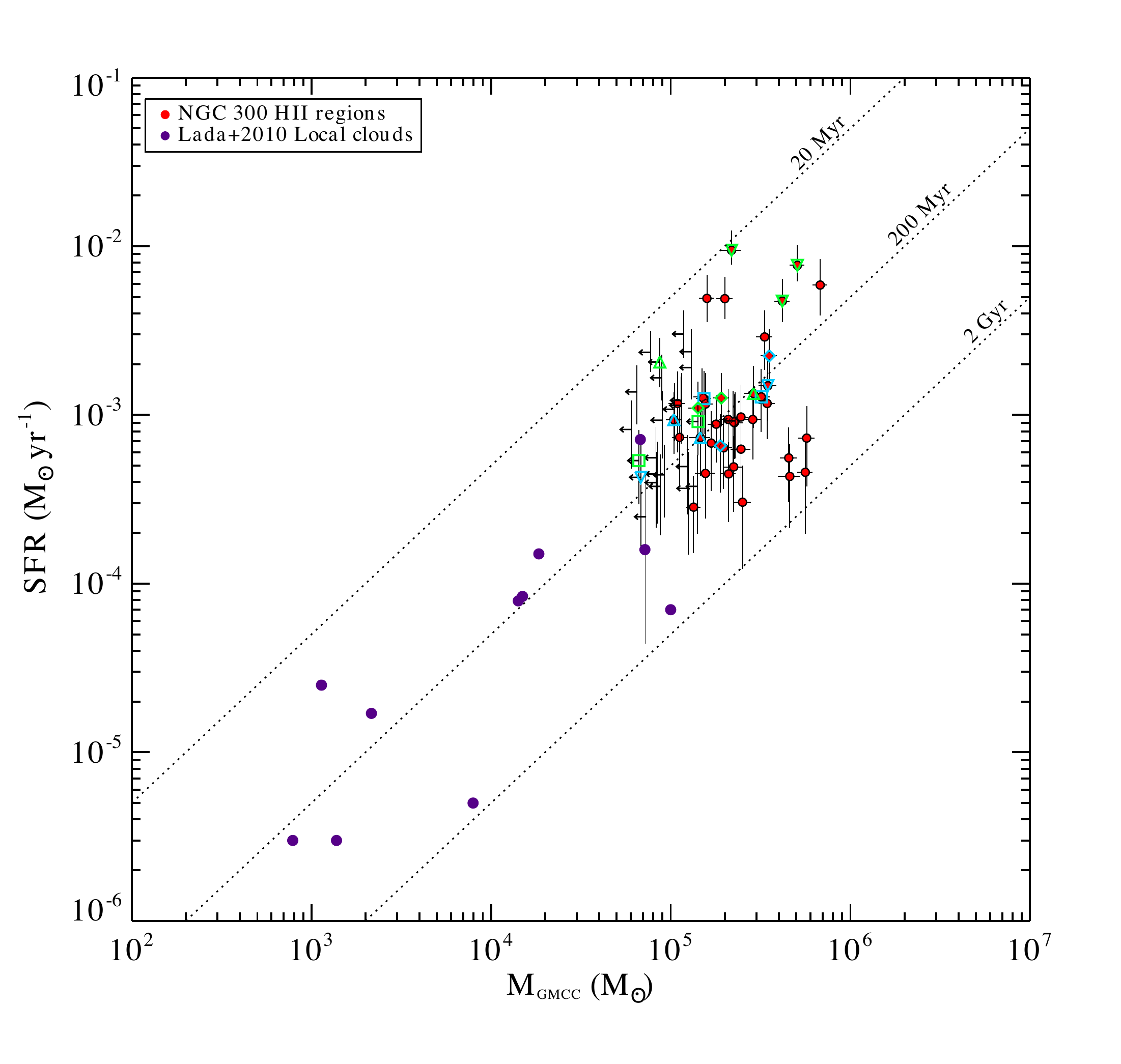}
\caption{\small{Star formation rate vs. molecular gas mass for the APEX CO-detected NGC 300 {\HII} regions (red circles) and upper limits plotted in the context of the L10 local Milky Way clouds (purple circles). Dotted lines indicate constant depletion times of 20 Myr, 200 Myr, and 2 Gyr. The NGC 300 clouds demonstrate a similar characteristic molecular gas depletion time ($\sim 230$ Myr) and level of scatter (0.4 dex) as the local clouds. Groups of sources for which apertures overlap are indicated with matching cyan or green triangles, diamonds, and squares.}}
\label{fig:sfrmmol}
\end{figure*}

Figure~\ref{fig:sfrmmol} shows SFR vs. $M_{\rm GMCC}$ derived for our sample of {\HII} regions using the direct modeling approach, computed as described in \S~\ref{sec:sfrcalcs}. These results are shown in the context of the L10 local Milky Way cloud sample. The scatter in the SFR-$M_{\rm GMCC}$ relation for the NGC~300 CO-detected regions is quite large (0.4~dex), approaching the dynamic range in $M_{\rm GMCC}$ (0.8~dex). While some of this scatter ($\sim0.2$ dex, on average) is due to uncertainties in our measurements and modeling procedure, the remainder is likely real, and may be driven by differing physical and/or evolutionary conditions across the sample. Groups of sources for which apertures overlap are indicated on the figure with matching colored symbols. The fact that $M_{\rm GMCC}$ often differs significantly between partially overlapping sources suggests that the underlying molecular gas distribution in some locations is clumpy on sub-resolution scales, a hypothesis that preliminary interferometric studies are beginning to confirm (C. Faesi et al., in preparation).

\section{Discussion}
\label{sec:disc}

\subsection{The star formation scaling relation in NGC 300}

From Figure~\ref{fig:sfrmmol}, it is clear that any potential power law trend in the scaling relation is masked by the large amount of scatter in the SFR-$M_{\rm GMCC}$ plane. We thus cannot directly address whether or not the SFR scaling relation in NGC~300 is linear at 250~pc scales, as it appears to be in $\sim$kpc-sized regions in other systems~\citep[e.g.,][]{Bigiel:2008bs,Leroy:2013bf}. A large amount of scatter might be expected at physical scales smaller than 1~kpc when sampling galaxy disks in an unbiased manner due to stochastic sampling of the galaxy's GMC mass function~\citep{Calzetti:2012iz}. However, we are specifically targeting regions with active star formation, and so our measurements are unlikely to be affected by this sampling effect. The majority of the scatter we recover most likely represents true physical differences between star-forming regions in NGC~300. In this section we place our sample in the context of the L10 local clouds and extragalactic studies, and discuss the implications of our results.

\subsection{Comparison with the Milky Way sample}

\begin{figure}[tb]
\includegraphics[width=\linewidth]{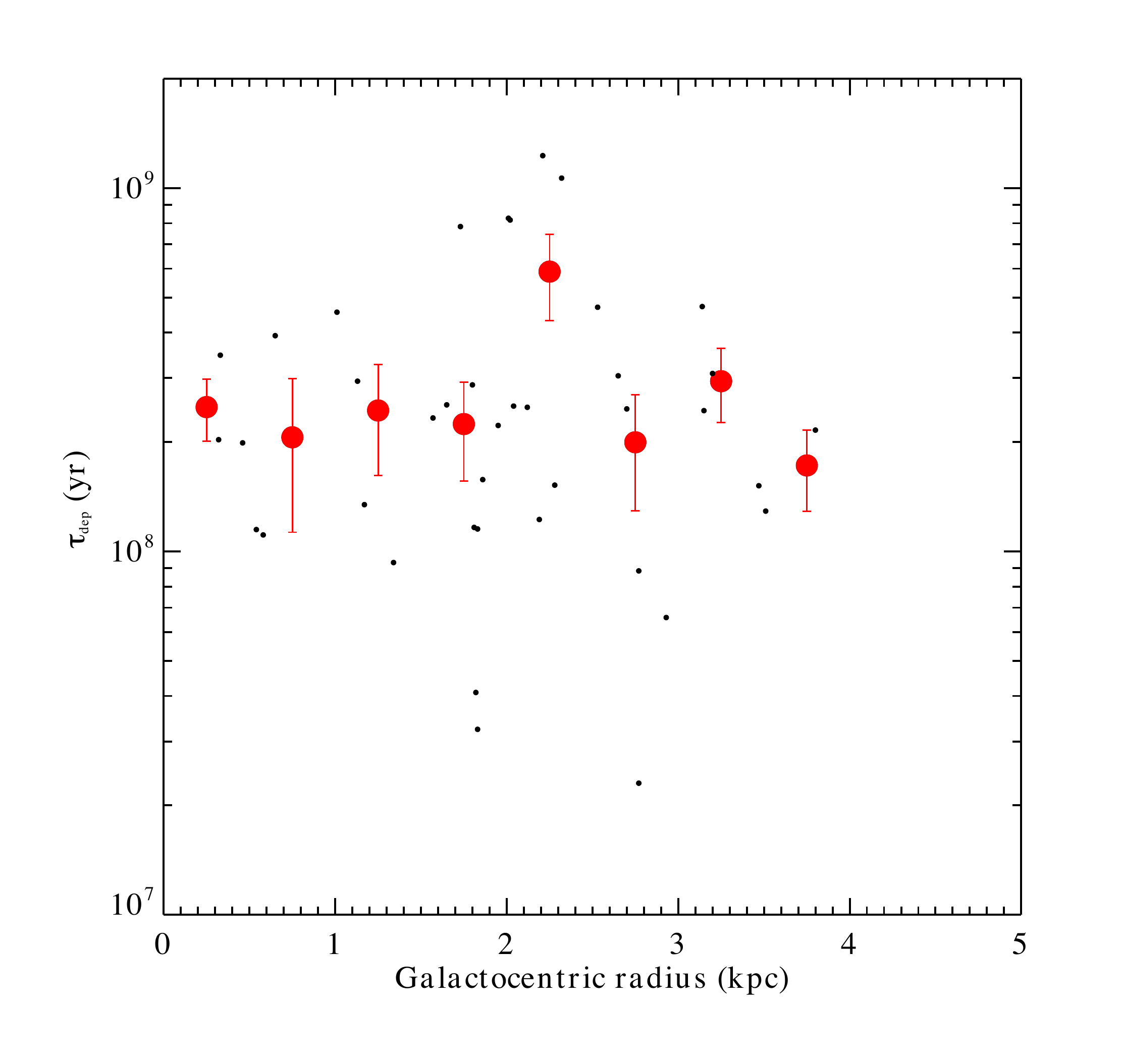}
\caption{\small{Molecular gas depletion time {\tdep} vs. galactocentric radius in the CO-detected regions in our NGC~300 sample. Black points show individual regions, while red points show the average {\tdep} computed in 0.5~kpc bins. The error bars on each point reflects the formal uncertainty on the mean in that bin. There is no obvious trend in {\tdep} with radius, although the scatter in {\tdep} seems to increase between $\sim 2$ and 3 kpc.}}
\label{fig:tdep_Rgal}
\end{figure}

From Figure~\ref{fig:sfrmmol}, it appears that the relation between star formation rate and molecular gas mass extends smoothly from local Milky Way clouds to NGC~300 GMCCs with masses of up to several times $10^5~M_{\odot}$, with a large amount of scatter present at all scales. This scatter corresponds to a wide range in star formation efficiency, or its inverse the molecular gas depletion time $\tau_{\rm dep} = M_{\rm GMC} /$SFR. Dotted lines in the Figure correspond to tracks of constant {\tdep} of 20~Myr, 200~Myr, and 2~Gyr. The median $\tau_{\rm dep}$ in NGC 300 is about 230~Myr (with a large amount of scatter), similar to the 180~Myr median depletion time in the L10 clouds. Amongst our CO-detected regions, we find a large range in {\tdep} from 23~Myr to 1.2~Gyr, similar to the 45~Myr to 1.6~Gyr range in the L10 sample. There is no particular trend in {\tdep} with galactocentric radius, as we demonstrate in Figure~\ref{fig:tdep_Rgal}. The red points in this figure show average depletion times binned by 0.5~kpc, and the error bars show the formal uncertainty on the mean for each bin. Interestingly, there is a notable increase in the scatter in {\tdep} at $\sim$2~kpc, followed by a decrease beyond $\sim$3~kpc.

Tracks of constant {\tdep} can also be interpreted as tracks of constant dense gas fraction according to the \cite{Lada:2012it} framework, which asserts that star formation occurs primarily in dense ($n \gtrsim 10^4$~cm$^{-3}$) gas. Clouds with high dense gas fractions (large dense gas masses $M_{\rm dense}$ as compared to total molecular gas masses $M_{\rm GMC}$) are rapidly turning that dense gas into stars, and thus exhibit a high efficiency (short {\tdep}). Conversely, clouds with low dense gas fractions are relatively inert, as the more diffuse gas is inefficient at directly forming stars (long {\tdep}), even if the small amount of dense gas that is present is actively star-forming. The 20~Myr {\tdep} track corresponds to a dense gas fraction of 100\% based on the linear fit between the SFR and mass in dense gas found by L10. If SFR$\propto f_{\rm DG} M_{\rm GMC}$, where $f_{\rm DG}=M_{\rm dense}/M_{\rm GMC}$ is the dense gas fraction, as proposed by \cite{Lada:2012it}, the 200~Myr and 2~Gyr tracks then represent 10\% and 1\% dense gas fractions, respectively. According to this interpretation, the large scatter we see in the NGC~300 clouds could be explained as variations in the dense gas fraction. Testing this scenario conclusively awaits data from dense gas tracers. We do note that the NGC~300 CO-detected regions show physically plausible dense gas fractions ranging between a few and nearly 100\%.

Differences in the evolutionary state of individual regions could provide an alternative explanation for the large scatter in the SFR-$M_{\rm GMCC}$ plane. It could be that clouds with short $\tau_{\rm dep}$ have used up some of the gas from which the current population of stars formed, and thus are shifted to the left in the diagram from the position they occupied with their original gas reservoir. In such a scenario clouds with long $\tau_{\rm dep}$ may simply be very young, and just beginning the process of star formation, i.e. shifted downward in SFR for a given $M_{\rm GMCC}$. However, we directly derive the presumptive ages of the regions we target, and we do not see any systematic trend in stellar population age $t$ with $\tau_{\rm dep}$. Evolutionary state in this sense thus does not seem to play a major role in explaining the scatter in {\tdep}. We do not, however, account for two potential additional evolutionary effects: (1) feedback from stellar populations onto their parent clouds, which may have decreased the molecular gas content from its original reservoir due to ionization and photodissociation~\citep[e.g.,][]{Dale:2012fg}, or (2) the possibility of continuing formation or accretion of molecular gas in GMCs~\citep[e.g.,][]{Burkert:2013hy}, which may act to increase the reservoir. Investigating these scenarios further is potentially very interesting but would require additional detailed modeling that is beyond the scope of this work.

\cite{Lada:2012it} also suggested that the linear relation between the SFR and GMC mass which describes the Milky Way sample and is consistent with our NGC~300 results also holds in entire galactic systems. This could be additional evidence pointing to a universal (if not necessarily surprising) linear relation between molecular gas mass and star formation rate. A potentially more interesting (and physically meaningful) relation is that between \textit{dense} gas and star formation, which \cite{Lada:2012it} show also extends from local clouds to entire galaxies studied in the dense gas tracer HCN. The scatter in the SFR-$M_{\rm dense}$ relation is also much lower than that in the SFR-$M_{\rm GMC}$ relation for the L10 sample. Future observations of the molecular gas associated with {\HII} regions in NGC~300 with, e.g., ALMA to trace the dense gas component will test the hypothesis that the amount of scatter decreases when the mass in only dense gas is considered, as is the case for the local clouds, and further illuminate the role of the dense gas fraction in star-forming GMCs.

\subsection{Comparison with standard extragalactic prescriptions}

\begin{figure}[tb]
\includegraphics[width=\linewidth]{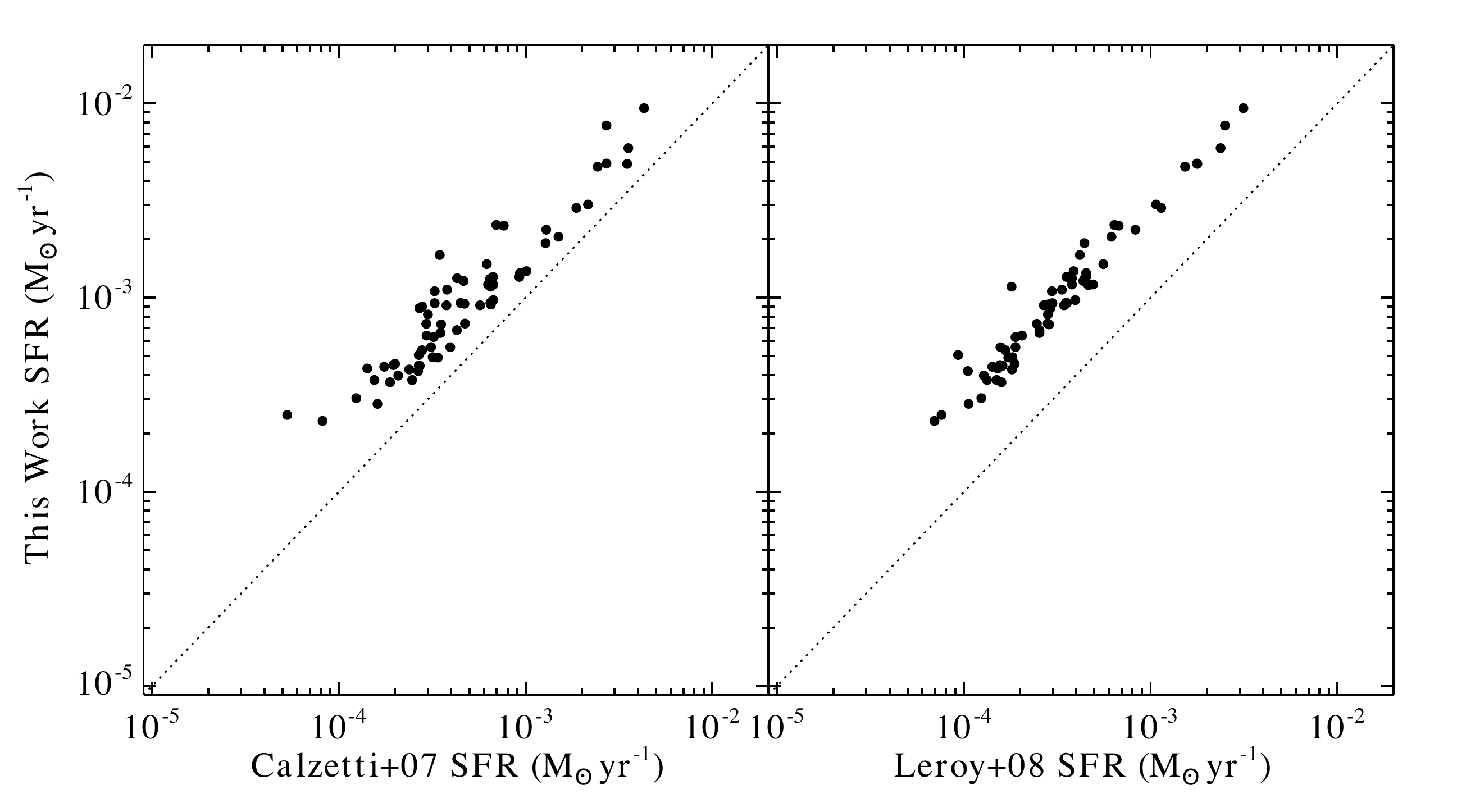}
\caption{\small{SFRs calculated using our direct modeling approach vs. SFRs computed from the literature prescriptions of \cite[][left panel]{2007ApJ...666..870C} and \cite[][right panel]{Leroy:2008jk}. We derive higher SFRs using our approach by a factor of about 2-3. The dotted lines show equality between the quantities. We use our data for all computations.}}
\label{fig:litcomp}
\end{figure}

As discussed in \S~\ref{sec:ext}, the population synthesis models used in standard prescriptions for estimating SFRs in galaxies and large regions within them typically utilize the continuous star formation approximation over 100~Myr timescales. These assumptions are appropriate for large regions of star-forming galaxies in which multiple stellar populations are in various stages of evolution such that the total star formation rate, averaged over sufficiently long timescales, appears continuous. However, these prescriptions may not be valid for use on smaller regions consisting of localized, instantaneous star-formation where the above assumptions begin to break down~\citep[e.g.,][]{Schruba:2010hf,Calzetti:2012iz}. To explore the effects of applying extragalactic prescriptions on {\HII} region-scales, we utilize our data to compute SFRs using two well-defined prescriptions from the literature and compare the results to those derived using our direct modeling approach. Figure~\ref{fig:litcomp} shows our SFRs plotted against those computed with the H$\alpha$+24$\mu$m calibration of \cite{2007ApJ...666..870C} and the FUV+24$\mu$m calibration of \cite{Leroy:2008jk}. We find an excellent correlation between SFRs derived using our method and each of these prescriptions, with Spearman rank correlation coefficients of 0.93 and 0.94, respectively. There is also a systematic offset such that our SFRs are higher by an average factor of 2.1 and 3.1 in relation to the \cite{2007ApJ...666..870C} and \cite{Leroy:2008jk} prescription-derived SFRs, respectively. The smaller offset with respect to the \cite{2007ApJ...666..870C} prescription may be because they targeted `{\HII} knots', centering their measurements on H$\alpha$ and 24 $\mu$m peaks (and thus peaks of star formation), albeit at larger scales than our study. In contrast, the \cite{Leroy:2008jk} calibration was derived for kpc-sized regions within galaxies, without specifically targeting star-forming regions directly. Furthermore, we recover higher SFRs than \cite{2007ApJ...666..870C} in the low-SFR regime. Since H$\alpha$ is heavily suppressed in lower mass star forming regions, our complementary use of FUV presumably allows us to recover low SFRs more accurately, although it is still subject to caveats regarding, e.g., stochastic sampling of the IMF at low cluster masses.

The most likely explanation for the systematic offset of our SFRs with respect to the extragalactic prescriptions has to do with the timescales used in the models. Unlike modeled emission from an instantaneous burst, FUV and H$\alpha$ luminosity \textit{increase} over time in a continuous SF model, eventually reaching a `steady-state' level where the stars that dominate the emission in a given tracer depart from the main sequence at the same rate that new stars replace them. The time to reach this steady-state condition is essentially the characteristic timescale over which the tracer probes star formation. Since slightly older (10-100~Myr) populations contribute significantly to FUV emission, the modeled FUV luminosity increases over time until about 100~Myr pass in the simulation. This has a significant effect on deriving SFRs from FUV observations: if a short timescale is assumed, a given FUV luminosity will correspond to a \textit{higher} SFR than it would for a long timescale. Thus continuous models based on 100~Myr timescales will \textit{underestimate} SFRs in regions where star formation has only been going on for a few to 10 Myr. {\HII} regions are typically younger than 10~Myr, and so the application of standard extragalactic SFR prescriptions to {\HII} regions at 250~pc scales results in an underprediction of the SFR, an effect we see in our results and that has been noted previously in the literature \citep[e.g.,][]{Chomiuk:2011iy,Lada:2012it}.

Many studies of statistically significant samples of regions within multiple galaxies have shown a linear or near-linear scaling of the surface density of star formation rate ($\Sigma_{\rm SFR}$) with that of molecular gas \citep[$\Sigma_{\rm mol}$; e.g.,][]{Bigiel:2008bs,Blanc:2009hf,Rahman:2011bo}. A linear scaling is equivalent to a constant average molecular gas depletion time, where here $\tau_{\rm dep}=\Sigma_{\rm mol}/\Sigma_{\rm SFR}$ is the ratio of surface densities instead of integrated quantities. When all regions within a galaxy (i.e. not just those identified as star-forming) are considered, these studies identify much longer depletion times than those we derive for our sample of {\HII} regions in NGC 300. For example, \cite{Leroy:2013bf} find a global molecular gas depletion time of $2.2$~Gyr (with about a factor of two scatter) in their comprehensive investigation of 30 nearby spiral galaxies. In contrast, we find a median depletion time of 230~Myr in our NGC~300 sample, about an order of magnitude lower. In the next section we explore some possible explanations for this difference.

\subsubsection{Integrated vs. Surface quantities}

Most extragalactic studies of star formation scaling relations derive surface (per area) quantities instead of integrated ones. Are there fundamental differences between this line of analysis and the integrated studies performed here and in most studies of local clouds? In our sample, we do not resolve individual molecular clouds with APEX, and so converting our integrated molecular gas measurements to gas surface densities simply involves dividing by the fixed area of the beam. For the SFRs, we assume that all emission in our tracers within the photometric aperture is from the young stellar population, and so converting SFR to SFR surface density implies dividing by the aperture area. Since, by design, we chose the aperture to be the same size as the APEX beam for the majority of sources, the relative position of data points in the $\Sigma_{\rm SFR}$--$\Sigma_{\rm GMCC}$ plane is very similar to that in the SFR--$M_{\rm GMCC}$ plane. For those sources with altered aperture sizes, the change in area is not large enough to shift the data points significantly. As a test, we have computed {\tdep} using surface quantities derived from our measurements and find a very similar median value of 220~Myr, as compared to 230~Myr from integrated measurements. Thus we conclude that the choice of integrated vs. surface measurements does not play a role in the $\tau_{\rm dep}$ difference between our results and kpc-scale extragalactic studies.

\subsubsection{Targeted vs. untargeted observations}

One crucial difference between the present study and those that sample SFR and molecular gas across the entirety of galactic disks is that we have purposefully biased our sample to target star-forming regions. If the star formation activity and molecular gas are both preferentially localized within such regions, this should have no effect on $\tau_{\rm dep}$, as the kpc (or larger) scale measurements will simply consist of several such regions added together (though with some dilution of signal due to a low filling factor). However, if there is a significant diffuse molecular component outside of star-forming regions, this could cause untargeted surveys to estimate a depletion time that refers to the time needed to consume \textit{all} the molecular gas, including that which is not actively star-forming. Cast in the context of the \cite{Lada:2012it} framework in which star formation activity is most closely connected to the dense gas fraction, the diffuse gas picked up in non-star-forming regions in untargeted surveys may have extremely low dense gas fractions, and thus low star formation rates. This could bias the depletion time to higher values and explain some of the discrepancy between our $\tau_{\rm dep}$ and that derived in untargeted surveys. However, the fact that we fail to detect CO in several of our {\HII} region targets places an approximate upper limit on the amount of molecular gas that might be missed in this way. Our average CO upper limit is $1.1\times 10^5~\Msun$, which corresponds to $2.0~\Msun$~pc$^{-2}$ over our 250~pc beam. Most extragalactic studies do not reach this level of sensitivity, and thus would also not pick up this quantity of gas in their kpc-scale beams \citep[though see also][]{Schruba:2011em}. However, the low gas densities we observe may be partially a result of ionization and photodissociation of the original molecular gas reservoir by the energetic stars in the nearby {\HII} regions, and so this upper limit may be an underestimate of the diffuse gas component in more quiescent regions of galaxies.

Furthermore, \cite{Schruba:2010hf} show that focusing measurements on peaks of star formation in M33 while using sub-kpc apertures leads to a factor of 2 to 3 higher SFRs and thus \textit{shorter} measured $\tau_{\rm dep}$ compared to values computed from observations that do not target star forming regions. This may partially, but not entirely, explain why the value of {\tdep} we derive is smaller than that computed over kpc-sized regions in galaxies. The remainder of the discrepancy may be due to a significant diffuse gas component, as discussed above. The contribution of large-scale, diffuse, non star-forming molecular gas to the total molecular mass budget of a galaxy is highly uncertain, but there is some recent evidence that it may be significant. For example, half the total emission in the spiral galaxy M51 appears to be coming from diffuse gas at  $>1$~kpc scales~\citep{Pety:2013fw}. If this is the case in other spiral galaxies (such as NGC~300), measurements on these large scales will naturally find longer {\tdep} than those that isolate star-forming regions. The depletion time measured over large scales in galaxies thus represents the time needed to consume the entire molecular gas reservoir, including that which is non star-forming. In contrast, the depletion time measured within star-forming regions reflects the time needed for conversion of gas primarily organized into molecular clouds into stars.

\section{Summary}

We have conducted a comprehensive multiwavelength survey of a sample of 76 {\HII} regions within the nearby ($d=1.93$~Mpc) spiral galaxy NGC~300 to assess the molecular gas and star formation activity at 250~pc scales. We measured $^{12}$CO ($J=2-1$) emission in targeted observations with APEX, and used this data to derive the CO luminosity and molecular gas mass. We utilized archival GALEX FUV and ESO/WFI H$\alpha$ maps to survey recent star formation, correcting for extinction with \textit{Spitzer}/MIPS 24$\mu$m data. We developed a new, direct modeling approach to calculating masses, ages, and star formation rates for single stellar populations, leveraging our multiwavelength dataset alongside customized \texttt{Starburst99} simulation runs (Figures~\ref{fig:HavsFUV}, \ref{fig:MC}). We summarize our main conclusions below.

\begin{enumerate}
\item{We detect $^{12}$CO(2-1) in 42 of the 76 {\HII} regions observed. The CO detection rate falls rapidly with radius, consistent with the previously measured metallicity gradient in NGC~300 and corresponding radial increase of {\aCO}, the CO-to-H$_2$ conversion factor (Figures~\ref{fig:aCOvar}, \ref{fig:radialco}).}
\item{We derive masses $M_*$ for stellar populations within our {\HII} region sample that range from $10^3$ to $3.8\times 10^4~M_{\odot}$ and population ages between 2 and 6~Myr (with a strong peak at 4 Myr; Figure~\ref{fig:hists}). There is no trend in any of $M_*$, age, or SFR with galactocentric radius out to 7~kpc.}
\item{We find that the mass function of GMC Complexes (GMCCs) associated with {\HII} regions in NGC~300 is steep (slope of -2.7), more similar to the GMC mass function in M33 than that of the Milky Way (Figure~\ref{fig:massspec}).}
\item{The NGC 300 sample extends the local cloud sample of L10 by about an order of magnitude in both mass and SFR. The scaling relation between SFR and molecular gas mass $M_{\rm GMCC}$ in NGC~300 is consistent with the relation found by L10 for the local clouds (Figure~\ref{fig:sfrmmol}).}
\item{The level of scatter in the SFR-$M_{\rm GMCC}$ relation in the NGC 300 sample is consistent with that in the local clouds, and is likely mostly due to intrinsic region-to-region differences in physical properties such as the dense gas fraction, as is the case in the local GMCs. The scatter corresponds to a range in molecular gas depletion times {\tdep} from 23~Myr to 1.2~Gyr for the NGC~300 sample, similar to the range of depletion times found in Milky Way clouds. There is no trend in {\tdep} with galactocentric radius in NGC~300 (Figure~\ref{fig:tdep_Rgal}).}
\item{We find a median {\tdep} of $230$~Myr at our resolution of 250~pc -- almost an order of magnitude shorter than that found in most extragalactic studies that resolve $\sim$~kpc scales. While extragalactic studies that sample kpc-scale regions in disks average over many stellar populations at different evolutionary states and can be modeled by a continuous star formation paradigm over 100~Myr timescales, the {\HII} regions we study here consist of individual young stellar populations better described by instantaneous burst models evolved over only a few Myr. The depletion times we measure correspond to the timescale for star formation to use up the gas reservoir in the star cluster's parent GMC, which may be the more relevant quantity in the context of GMC-regulated star formation in galaxies.}
\end{enumerate}

The method and analysis we present here for NGC~300 is general and applicable to other star-forming systems for which unresolved measurements of H$\alpha$, FUV, 24~$\mu$m, and some tracer of the molecular ISM are available. Similar studies in other galaxies will be important in elucidating any systemic differences based on galaxy properties such as metallicity, mass, or morphology, and in assessing the physical reason(s) for the scatter in {\tdep}.

\acknowledgements

We thank A. A. Goodman, D. Calzetti, A. Leroy, D. Wilner, R. Narayan, C. McKee, L. Blitz, L. Van Zee, and the participants of the Ringberg Conference ``Regulation of star formation in molecular gas: from galactic to sub-cloud scales'' for productive discussions that greatly improved this manuscript. We also wish to thank the anonymous referee for insightful comments that improved this manuscript.

This work is based in part on observations made with the \textit{Spitzer} Space Telescope, obtained from the NASA/IPAC Infrared Science Archive, both of which are operated by the Jet Propulsion Laboratory, California Institute of Technology under a contract with the National Aeronautics and Space Administration. This research has made use of the NASA/IPAC Extragalactic Database (NED) which is operated by the Jet Propulsion Laboratory, California Institute of Technology, under contract with the National Aeronautics and Space Administration. This research made use of Montage, funded by the National Aeronautics and Space Administration's Earth Science Technology Office, Computation Technologies Project, under Cooperative Agreement Number NCC5-626 between NASA and the California Institute of Technology. Montage is maintained by the NASA/IPAC Infrared Science Archive. C.M.F acknowledges support from a National Science Foundation Graduate Research Fellowship under Grant No. DGE-1144152. H. Bouy is funded by the Spanish Ram\'{o}n y Cajal fellowship program number RYC-2009-04497.

\pagebreak
\clearpage
\LongTables
\begin{turnpage}
\tabletypesize{\scriptsize}
%\begin{landscape}
%\centering
\begin{deluxetable*}{c rrrrrrrrr}
\tablecaption{Tabulated photometry and results \label{tab:results}}
\tablehead{
\multicolumn{1}{c}{DCL\#} &	\multicolumn{1}{c}{$F_{\rm{H}\alpha}$} &
\multicolumn{1}{c}{$F_{\rm FUV}$} &	\multicolumn{1}{c}{$F_{24}$} &
\multicolumn{1}{c}{$A_{{\rm H}\alpha}$} &	\multicolumn{1}{c}{$A_{\rm FUV}$} &
\multicolumn{1}{c}{$M_{\rm GMCC}$} &	\multicolumn{1}{c}{$M_*$} &
\multicolumn{1}{c}{age} &	\multicolumn{1}{c}{SFR} \\
&	\multicolumn{1}{c}{($10^{-14}$~erg~s$^{-1}$~cm$^{-2}$)} &	\multicolumn{1}{c}{$(\mu$Jy)} &	\multicolumn{1}{c}{(mJy)} &	\multicolumn{1}{c}{(mag)} &	\multicolumn{1}{c}{(mag)} &
\multicolumn{1}{c}{($10^5~\Msun$)} &	\multicolumn{1}{c}{$(10^3~\Msun)$} &	\multicolumn{1}{c}{(Myr)} &	\multicolumn{1}{c}{({\Msun}~yr$^{-1}$)} \\
%\endfirsthead
}
\hline
\multicolumn{10}{c}
{{\tablename\ \thetable{} -- continued from previous page}} \\
\hline
\multicolumn{1}{c}{DCL\#} &	\multicolumn{1}{c}{$F_{\rm{H}\alpha}$} &	\multicolumn{1}{c}{$F_{\rm FUV}$} &	\multicolumn{1}{c}{$F_{24}$} &	\multicolumn{1}{c}{$A_{{\rm H}\alpha}$} &	\multicolumn{1}{c}{$A_{\rm FUV}$} &
\multicolumn{1}{c}{$M_{\rm GMCC}$} &	\multicolumn{1}{c}{$M_*$} &	\multicolumn{1}{c}{age} &	\multicolumn{1}{c}{SFR} \\
&	\multicolumn{1}{c}{($10^{-14}$~erg~s$^{-1}$~cm$^{-2}$)} &	\multicolumn{1}{c}{$(\mu$Jy)} &	\multicolumn{1}{c}{(mJy)} &	\multicolumn{1}{c}{(mag)} &	\multicolumn{1}{c}{(mag)} &
\multicolumn{1}{c}{($10^5~\Msun$)} &	\multicolumn{1}{c}{$(10^3~\Msun)$} &	\multicolumn{1}{c}{(Myr)} &	\multicolumn{1}{c}{({\Msun}~yr$^{-1}$)} \\
\hline 
\endhead
\hline \multicolumn{10}{|r|}{{Continued on next page}} \\ \hline
\endfoot
\hline \hline
\endlastfoot
\hline
\multicolumn{10}{c}{CO detections} \\
\hline
 23 &  32.70~$\pm$ 1.32 &  855.0~$\pm$ 43.8 &  6.73~$\pm$ 0.43 & 0.08~$\pm$~0.03 & 0.29~$\pm$~0.08 &  2.89~$\pm$~0.32 &  4.0~$\pm$ 0.9 &  2.9~$\pm$~0.9 &  1.3$^{+0.6}_{-0.5} \times 10^{-3}$ \\
 30 &  14.60~$\pm$ 0.62 &  870.0~$\pm$ 45.1 &  3.70~$\pm$ 0.24 & 0.10~$\pm$~0.03 & 0.17~$\pm$~0.05 &  1.91~$\pm$~0.20 &  5.5~$\pm$ 1.1 &  4.3~$\pm$~1.1 &  1.3$^{+0.5}_{-0.4} \times 10^{-3}$ \\
 34 &  17.30~$\pm$ 0.72 &  459.0~$\pm$ 25.3 &  7.73~$\pm$ 0.49 & 0.17~$\pm$~0.05 & 0.55~$\pm$~0.14 &  2.23~$\pm$~0.21 &  3.1~$\pm$ 1.0 &  3.4~$\pm$~1.2 &  9.1$^{+4.8}_{-4.1} \times 10^{-4}$ \\
 37 &  11.60~$\pm$ 0.51 &  416.0~$\pm$ 22.0 &  8.80~$\pm$ 0.56 & 0.28~$\pm$~0.08 & 0.66~$\pm$~0.16 &  2.86~$\pm$~0.35 &  3.8~$\pm$ 1.1 &  4.0~$\pm$~1.3 &  9.4$^{+4.6}_{-4.0} \times 10^{-4}$ \\
 41 &   3.70~$\pm$ 0.25 &  127.0~$\pm$  8.6 &  5.84~$\pm$ 0.37 & 0.52~$\pm$~0.14 & 1.12~$\pm$~0.24 &  5.61~$\pm$~0.55 &  1.9~$\pm$ 0.9 &  4.2~$\pm$~1.6 &  4.6$^{+2.8}_{-2.6} \times 10^{-4}$ \\
 46 &   8.64~$\pm$ 0.40 &  254.2~$\pm$ 13.7 &  2.97~$\pm$ 0.19 & 0.14~$\pm$~0.04 & 0.41~$\pm$~0.11 &  2.10~$\pm$~0.20 &  1.5~$\pm$ 0.5 &  3.4~$\pm$~1.2 &  4.5$^{+2.4}_{-2.2} \times 10^{-4}$ \\
 49 &  20.20~$\pm$ 0.83 &  469.0~$\pm$ 24.9 &  7.90~$\pm$ 0.51 & 0.15~$\pm$~0.05 & 0.55~$\pm$~0.14 &  2.09~$\pm$~0.23 &  2.9~$\pm$ 0.9 &  3.1~$\pm$~1.1 &  9.4$^{+4.9}_{-4.2} \times 10^{-4}$ \\
 52 &  14.70~$\pm$ 0.62 &  641.0~$\pm$ 34.0 & 13.90~$\pm$ 0.89 & 0.34~$\pm$~0.09 & 0.67~$\pm$~0.17 &  3.47~$\pm$~0.40 &  6.3~$\pm$ 2.0 &  4.2~$\pm$~1.0 &  1.5$^{+0.7}_{-0.6} \times 10^{-3}$ \\
 61 &  30.80~$\pm$ 1.25 &  410.0~$\pm$ 21.6 &  8.81~$\pm$ 0.56 & 0.11~$\pm$~0.03 & 0.66~$\pm$~0.16 &  1.49~$\pm$~0.15 &  2.9~$\pm$ 0.6 &  2.2~$\pm$~0.9 &  1.3$^{+0.6}_{-0.5} \times 10^{-3}$ \\
 63 &  14.10~$\pm$ 0.60 &  344.0~$\pm$ 18.2 &  6.74~$\pm$ 0.43 & 0.18~$\pm$~0.05 & 0.62~$\pm$~0.16 &  1.12~$\pm$~0.10 &  2.4~$\pm$ 0.8 &  3.3~$\pm$~1.2 &  7.4$^{+4.0}_{-3.5} \times 10^{-4}$ \\
 65 &   4.67~$\pm$ 0.27 &  132.0~$\pm$  7.9 &  2.49~$\pm$ 0.16 & 0.20~$\pm$~0.06 & 0.60~$\pm$~0.16 &  1.34~$\pm$~0.12 &  1.0~$\pm$ 0.4 &  3.6~$\pm$~1.2 &  2.8$^{+1.5}_{-1.3} \times 10^{-4}$ \\
 66 &  19.50~$\pm$ 0.81 &  670.0~$\pm$ 34.4 &  8.86~$\pm$ 0.57 & 0.18~$\pm$~0.05 & 0.45~$\pm$~0.12 &  1.53~$\pm$~0.15 &  4.8~$\pm$ 1.3 &  3.8~$\pm$~1.1 &  1.2$^{+0.6}_{-0.5} \times 10^{-3}$ \\
 68 &  20.70~$\pm$ 0.85 &  714.0~$\pm$ 36.6 &  8.66~$\pm$ 0.56 & 0.16~$\pm$~0.05 & 0.42~$\pm$~0.12 &  3.19~$\pm$~0.35 &  4.8~$\pm$ 1.3 &  3.8~$\pm$~1.1 &  1.3$^{+0.6}_{-0.5} \times 10^{-3}$ \\
 69 &   1.92~$\pm$ 0.22 &   63.5~$\pm$  5.3 &  4.15~$\pm$ 0.27 & 0.66~$\pm$~0.20 & 1.38~$\pm$~0.28 &  2.51~$\pm$~0.28 &  1.4~$\pm$ 0.7 &  4.5~$\pm$~1.7 &  3.0$^{+2.0}_{-1.8} \times 10^{-4}$ \\
 76C &  53.50~$\pm$ 2.15 & 1250.0~$\pm$ 63.9 & 29.30~$\pm$ 1.87 & 0.21~$\pm$~0.06 & 0.71~$\pm$~0.17 &  3.33~$\pm$~0.34 &  9.6~$\pm$ 3.0 &  3.3~$\pm$~0.8 &  2.9$^{+1.3}_{-1.0} \times 10^{-3}$ \\
 79 &  95.11~$\pm$ 3.81 & 2295.7~$\pm$116.5 & 64.91~$\pm$ 4.16 & 0.25~$\pm$~0.07 & 0.81~$\pm$~0.19 &  6.79~$\pm$~0.64 & 20.6~$\pm$ 6.6 &  3.5~$\pm$~0.6 &  5.9$^{+2.5}_{-2.0} \times 10^{-3}$ \\
 81 &   9.92~$\pm$ 0.45 &  576.0~$\pm$ 30.6 &  4.34~$\pm$ 0.28 & 0.17~$\pm$~0.05 & 0.28~$\pm$~0.08 &  1.04~$\pm$~0.10 &  4.1~$\pm$ 1.0 &  4.4~$\pm$~1.3 &  9.4$^{+4.2}_{-3.5} \times 10^{-4}$ \\
 86 &   6.48~$\pm$ 0.33 &  437.0~$\pm$ 24.3 &  5.91~$\pm$ 0.38 & 0.33~$\pm$~0.09 & 0.46~$\pm$~0.13 &  1.79~$\pm$~0.19 &  4.3~$\pm$ 1.2 &  4.8~$\pm$~1.4 &  8.8$^{+4.3}_{-3.6} \times 10^{-4}$ \\
 88 &  38.30~$\pm$ 1.55 & 1120.0~$\pm$ 56.8 & 18.50~$\pm$ 1.18 & 0.19~$\pm$~0.05 & 0.54~$\pm$~0.14 &  3.53~$\pm$~0.37 &  8.0~$\pm$ 2.3 &  3.6~$\pm$~0.8 &  2.2$^{+1.0}_{-0.8} \times 10^{-3}$ \\
 93 &   9.35~$\pm$ 0.42 &  285.0~$\pm$ 16.0 &  6.41~$\pm$ 0.41 & 0.26~$\pm$~0.07 & 0.69~$\pm$~0.17 &  1.89~$\pm$~0.19 &  2.5~$\pm$ 0.8 &  3.8~$\pm$~1.3 &  6.6$^{+3.5}_{-3.1} \times 10^{-4}$ \\
100 &  17.10~$\pm$ 0.72 &  434.0~$\pm$ 23.7 & 12.90~$\pm$ 0.83 & 0.28~$\pm$~0.08 & 0.84~$\pm$~0.19 &  1.56~$\pm$~0.15 &  4.2~$\pm$ 1.5 &  3.6~$\pm$~1.1 &  1.2$^{+0.6}_{-0.5} \times 10^{-3}$ \\
103 &  11.20~$\pm$ 0.49 &  273.0~$\pm$ 15.4 &  3.34~$\pm$ 0.21 & 0.12~$\pm$~0.04 & 0.43~$\pm$~0.12 &  2.24~$\pm$~0.27 &  1.5~$\pm$ 0.5 &  3.1~$\pm$~1.1 &  4.9$^{+2.6}_{-2.2} \times 10^{-4}$ \\
109 &  21.90~$\pm$ 0.90 &  790.0~$\pm$ 40.5 &  4.76~$\pm$ 0.30 & 0.09~$\pm$~0.03 & 0.23~$\pm$~0.07 &  3.44~$\pm$~0.37 &  4.2~$\pm$ 1.1 &  3.6~$\pm$~1.1 &  1.2$^{+0.5}_{-0.5} \times 10^{-3}$ \\
114 &   8.47~$\pm$ 0.39 &  280.2~$\pm$ 15.9 &  7.77~$\pm$ 0.50 & 0.33~$\pm$~0.09 & 0.80~$\pm$~0.19 &  5.71~$\pm$~0.60 &  2.9~$\pm$ 1.0 &  4.0~$\pm$~1.4 &  7.3$^{+4.0}_{-3.5} \times 10^{-4}$ \\
118B &  81.66~$\pm$ 3.27 & 2519.6~$\pm$127.2 & 37.45~$\pm$ 2.40 & 0.18~$\pm$~0.05 & 0.50~$\pm$~0.13 &  1.59~$\pm$~0.15 & 17.9~$\pm$ 4.6 &  3.7~$\pm$~0.6 &  4.9$^{+1.9}_{-1.4} \times 10^{-3}$ \\
119C & 113.41~$\pm$ 4.54 & 2532.8~$\pm$127.7 & 37.67~$\pm$ 2.41 & 0.13~$\pm$~0.04 & 0.50~$\pm$~0.13 &  2.00~$\pm$~0.21 & 14.1~$\pm$ 2.8 &  2.9~$\pm$~0.5 &  4.9$^{+1.7}_{-1.2} \times 10^{-3}$ \\
122 &  18.40~$\pm$ 0.76 &  306.0~$\pm$ 17.5 & 11.80~$\pm$ 0.76 & 0.24~$\pm$~0.07 & 1.00~$\pm$~0.22 &  2.46~$\pm$~0.22 &  3.0~$\pm$ 1.1 &  3.1~$\pm$~1.2 &  1.0$^{+0.5}_{-0.5} \times 10^{-3}$ \\
126 &   7.46~$\pm$ 0.36 &  494.0~$\pm$ 27.1 &  5.12~$\pm$ 0.33 & 0.26~$\pm$~0.07 & 0.37~$\pm$~0.10 &  2.26~$\pm$~0.27 &  4.2~$\pm$ 1.1 &  4.7~$\pm$~1.4 &  9.0$^{+4.2}_{-3.5} \times 10^{-4}$ \\
127 &  12.79~$\pm$ 0.55 &  157.4~$\pm$  9.7 &  4.22~$\pm$ 0.27 & 0.13~$\pm$~0.04 & 0.78~$\pm$~0.19 &  4.54~$\pm$~0.50 &  1.2~$\pm$ 0.4 &  2.2~$\pm$~1.1 &  5.6$^{+2.9}_{-2.5} \times 10^{-4}$ \\
137A & 137.00~$\pm$ 5.49 & 5780.0~$\pm$290.0 & 49.40~$\pm$ 3.17 & 0.14~$\pm$~0.04 & 0.31~$\pm$~0.09 &  2.18~$\pm$~0.27 & 37.9~$\pm$ 6.3 &  4.0~$\pm$~0.3 &  9.5$^{+3.0}_{-1.7} \times 10^{-3}$ \\
137B &  83.00~$\pm$ 3.33 & 3110.0~$\pm$156.0 & 20.30~$\pm$ 1.30 & 0.10~$\pm$~0.03 & 0.25~$\pm$~0.07 &  4.18~$\pm$~0.42 & 17.5~$\pm$ 3.9 &  3.7~$\pm$~0.6 &  4.7$^{+1.7}_{-1.2} \times 10^{-3}$ \\
137C &  79.00~$\pm$ 3.17 & 4490.0~$\pm$225.0 & 41.00~$\pm$ 2.63 & 0.20~$\pm$~0.06 & 0.33~$\pm$~0.09 &  5.07~$\pm$~0.47 & 33.8~$\pm$ 6.1 &  4.4~$\pm$~0.4 &  7.7$^{+2.5}_{-1.5} \times 10^{-3}$ \\
139 &   9.86~$\pm$ 0.44 &  415.0~$\pm$ 22.3 &  2.74~$\pm$ 0.18 & 0.11~$\pm$~0.03 & 0.25~$\pm$~0.07 &  1.97~$\pm$~0.23 &  2.5~$\pm$ 0.6 &  3.9~$\pm$~1.3 &  6.4$^{+3.2}_{-2.7} \times 10^{-4}$ \\
140 & \nodata & $<$~  8.0 & $<$~ 0.01 & 0.03~$\pm$~0.08 & \nodata &  2.34~$\pm$~0.24 & \nodata &  9.9~$\pm$~1.7 & \nodata \\
\hline
\multicolumn{10}{c}{CO marginal detections} \\
\hline
 31 &  12.50~$\pm$ 0.54 &  728.0~$\pm$ 38.1 &  3.80~$\pm$ 0.24 & 0.12~$\pm$~0.04 & 0.20~$\pm$~0.06 &  1.42~$\pm$~0.17 &  4.8~$\pm$ 1.0 &  4.3~$\pm$~1.2 &  1.1$^{+0.5}_{-0.4} \times 10^{-3}$ \\
 56 &  15.60~$\pm$ 0.65 &  336.0~$\pm$ 20.9 & 15.30~$\pm$ 0.98 & 0.35~$\pm$~0.10 & 1.11~$\pm$~0.24 &  1.09~$\pm$~0.12 &  4.2~$\pm$ 1.7 &  3.6~$\pm$~1.2 &  1.2$^{+0.6}_{-0.6} \times 10^{-3}$ \\
 80 &   5.62~$\pm$ 0.30 &  242.0~$\pm$ 15.7 &  3.06~$\pm$ 0.20 & 0.21~$\pm$~0.06 & 0.44~$\pm$~0.12 &  1.56~$\pm$~0.20 &  1.8~$\pm$ 0.6 &  4.1~$\pm$~1.5 &  4.5$^{+2.4}_{-2.1} \times 10^{-4}$ \\
 85 &   8.77~$\pm$ 0.40 &  431.0~$\pm$ 23.7 &  4.16~$\pm$ 0.27 & 0.18~$\pm$~0.05 & 0.35~$\pm$~0.10 &  1.46~$\pm$~0.20 &  3.1~$\pm$ 0.7 &  4.2~$\pm$~1.4 &  7.3$^{+3.5}_{-3.0} \times 10^{-4}$ \\
 98 &  12.00~$\pm$ 0.52 &  472.0~$\pm$ 25.3 &  1.32~$\pm$ 0.08 & 0.05~$\pm$~0.01 & 0.11~$\pm$~0.04 &  2.46~$\pm$~0.31 &  2.3~$\pm$ 0.6 &  3.6~$\pm$~1.3 &  6.3$^{+3.2}_{-2.8} \times 10^{-4}$ \\
112 &   8.89~$\pm$ 0.41 &  133.8~$\pm$  8.2 &  2.47~$\pm$ 0.16 & 0.11~$\pm$~0.03 & 0.59~$\pm$~0.15 &  2.29~$\pm$~0.24 &  1.0~$\pm$ 0.2 &  2.4~$\pm$~1.2 &  4.2$^{+1.9}_{-1.6} \times 10^{-4}$ \\
129 &  13.20~$\pm$ 0.57 &  352.0~$\pm$ 19.6 &  5.55~$\pm$ 0.36 & 0.16~$\pm$~0.05 & 0.52~$\pm$~0.14 &  1.68~$\pm$~0.22 &  2.3~$\pm$ 0.8 &  3.3~$\pm$~1.2 &  6.8$^{+3.7}_{-3.3} \times 10^{-4}$ \\
130 &   3.04~$\pm$ 0.23 &  189.0~$\pm$ 13.8 &  3.61~$\pm$ 0.23 & 0.41~$\pm$~0.12 & 0.61~$\pm$~0.16 &  4.60~$\pm$~0.65 &  2.0~$\pm$ 0.7 &  4.7~$\pm$~1.7 &  4.3$^{+2.5}_{-2.2} \times 10^{-4}$ \\
\hline
\multicolumn{10}{c}{CO upper limits} \\
\hline
  1 &   4.70~$\pm$ 0.28 &  122.0~$\pm$  6.9 &  0.75~$\pm$ 0.05 & 0.07~$\pm$~0.02 & 0.24~$\pm$~0.07 & $<$ 2.17 & \nodata &  2.8~$\pm$~0.6 & \nodata \\
  2 &   2.66~$\pm$ 0.23 &  146.0~$\pm$  8.6 &  0.85~$\pm$ 0.05 & 0.13~$\pm$~0.04 & 0.22~$\pm$~0.07 & $<$ 1.77 &  1.0~$\pm$ 0.3 &  4.3~$\pm$~1.4 &  2.3$^{+1.2}_{-1.0} \times 10^{-4}$ \\
  5 &   9.92~$\pm$ 0.44 &  255.0~$\pm$ 15.4 &  3.91~$\pm$ 0.25 & 0.15~$\pm$~0.05 & 0.51~$\pm$~0.14 & $<$ 1.24 &  1.6~$\pm$ 0.5 &  3.3~$\pm$~1.2 &  4.9$^{+2.7}_{-2.4} \times 10^{-4}$ \\
  6 &  36.97~$\pm$ 1.49 &  797.2~$\pm$ 41.0 &  4.98~$\pm$ 0.32 & 0.06~$\pm$~0.02 & 0.24~$\pm$~0.07 & $<$ 0.65 &  3.5~$\pm$ 0.7 &  2.6~$\pm$~0.8 &  1.4$^{+0.6}_{-0.5} \times 10^{-3}$ \\
  7 &  27.00~$\pm$ 1.10 & 1830.0~$\pm$ 92.7 &  1.51~$\pm$ 0.10 & 0.02~$\pm$~0.01 & 0.03~$\pm$~0.01 & $<$ 1.30 & 10.6~$\pm$ 2.0 &  4.5~$\pm$~0.8 &  2.4$^{+0.9}_{-0.6} \times 10^{-3}$ \\
  9 &  24.13~$\pm$ 0.99 &  622.3~$\pm$ 32.5 &  3.04~$\pm$ 0.19 & 0.05~$\pm$~0.02 & 0.19~$\pm$~0.06 & $<$ 2.28 &  2.6~$\pm$ 0.6 &  2.8~$\pm$~1.0 &  9.2$^{+4.4}_{-3.7} \times 10^{-4}$ \\
 13 &   7.18~$\pm$ 0.35 &  266.0~$\pm$ 14.3 &  1.64~$\pm$ 0.10 & 0.09~$\pm$~0.03 & 0.24~$\pm$~0.07 & $<$ 0.83 &  1.5~$\pm$ 0.5 &  3.7~$\pm$~1.3 &  4.0$^{+2.1}_{-1.8} \times 10^{-4}$ \\
 15 &  13.30~$\pm$ 0.57 & 1193.8~$\pm$ 62.7 &  0.98~$\pm$ 0.06 & 0.03~$\pm$~0.01 & 0.03~$\pm$~0.01 & $<$ 0.89 &  8.7~$\pm$ 1.6 &  5.3~$\pm$~1.0 &  1.7$^{+0.6}_{-0.4} \times 10^{-3}$ \\
 17 &  16.64~$\pm$ 0.70 &  646.0~$\pm$ 34.2 &  3.18~$\pm$ 0.20 & 0.08~$\pm$~0.02 & 0.19~$\pm$~0.06 & $<$ 0.90 &  3.5~$\pm$ 0.9 &  3.7~$\pm$~1.2 &  9.3$^{+4.5}_{-3.8} \times 10^{-4}$ \\
 24 &  55.10~$\pm$ 2.21 & 1320.0~$\pm$ 66.8 &  7.25~$\pm$ 0.46 & 0.05~$\pm$~0.02 & 0.21~$\pm$~0.06 & $<$ 0.87 &  5.6~$\pm$ 0.9 &  2.7~$\pm$~0.7 &  2.1$^{+0.8}_{-0.6} \times 10^{-3}$ \\
 29 &   5.53~$\pm$ 0.30 &  271.8~$\pm$ 15.8 &  2.09~$\pm$ 0.13 & 0.15~$\pm$~0.05 & 0.29~$\pm$~0.08 & $<$ 0.92 &  1.8~$\pm$ 0.5 &  4.2~$\pm$~1.5 &  4.4$^{+2.2}_{-1.9} \times 10^{-4}$ \\
 43 &   6.97~$\pm$ 0.34 &  152.6~$\pm$  9.2 &  4.02~$\pm$ 0.26 & 0.22~$\pm$~0.07 & 0.77~$\pm$~0.19 & $<$ 0.88 &  1.3~$\pm$ 0.4 &  3.3~$\pm$~1.1 &  3.8$^{+2.1}_{-1.8} \times 10^{-4}$ \\
 44 & $<$~0.59 &    8.7~$\pm$  2.7 &  0.29~$\pm$ 0.02 & \nodata & 0.92~$\pm$~0.48 & $<$ 0.59 & \nodata &  9.4~$\pm$~1.8 & \nodata \\
 45 &  44.20~$\pm$ 1.78 &  555.0~$\pm$ 30.1 & 10.40~$\pm$ 0.67 & 0.09~$\pm$~0.03 & 0.60~$\pm$~0.15 & $<$ 1.30 &  3.8~$\pm$ 0.7 &  2.0~$\pm$~0.8 &  1.9$^{+0.8}_{-0.7} \times 10^{-3}$ \\
 53C &  71.10~$\pm$ 2.85 & 1630.0~$\pm$ 82.4 & 21.40~$\pm$ 1.37 & 0.12~$\pm$~0.04 & 0.45~$\pm$~0.12 & $<$ 1.18 &  8.8~$\pm$ 1.8 &  2.9~$\pm$~0.6 &  3.0$^{+1.2}_{-0.9} \times 10^{-3}$ \\
 54 &   5.05~$\pm$ 0.29 &   83.5~$\pm$ 10.5 &  6.16~$\pm$ 0.39 & 0.42~$\pm$~0.12 & 1.48~$\pm$~0.34 & $<$ 0.68 &  1.6~$\pm$ 0.8 &  3.6~$\pm$~1.4 &  4.3$^{+2.9}_{-2.7} \times 10^{-4}$ \\
 55 &   2.44~$\pm$ 0.22 &   92.5~$\pm$  6.4 &  1.28~$\pm$ 0.08 & 0.20~$\pm$~0.07 & 0.47~$\pm$~0.13 & $<$ 1.21 & \nodata &  3.8~$\pm$~1.0 & \nodata \\
 57 &   2.15~$\pm$ 0.22 &   61.1~$\pm$  4.4 &  1.29~$\pm$ 0.08 & 0.23~$\pm$~0.08 & 0.66~$\pm$~0.17 & $<$ 1.00 & \nodata &  3.8~$\pm$~0.7 & \nodata \\
 64 &   7.89~$\pm$ 0.37 &  425.0~$\pm$ 24.7 &  5.67~$\pm$ 0.36 & 0.27~$\pm$~0.08 & 0.46~$\pm$~0.13 & $<$ 0.60 &  3.6~$\pm$ 1.0 &  4.4~$\pm$~1.4 &  8.2$^{+4.0}_{-3.4} \times 10^{-4}$ \\
 72 &  24.69~$\pm$ 1.01 &  385.3~$\pm$ 20.0 &  2.12~$\pm$ 0.14 & 0.04~$\pm$~0.01 & 0.21~$\pm$~0.06 & $<$ 1.14 &  2.0~$\pm$ 0.5 &  1.8~$\pm$~0.9 &  1.1$^{+0.5}_{-0.5} \times 10^{-3}$ \\
 89 & $<$~0.59 &   44.5~$\pm$  7.1 & $<$~ 0.01 & \nodata & 0.01~$\pm$~0.03 & $<$ 0.46 & \nodata &  7.5~$\pm$~2.0 & \nodata \\
 99 & $<$~0.59 &   32.9~$\pm$  4.0 &  2.61~$\pm$ 0.17 & \nodata & 1.54~$\pm$~0.34 & $<$ 0.73 &  1.5~$\pm$ 1.1 &  6.2~$\pm$~3.8 &  2.5$^{+2.1}_{-2.1} \times 10^{-4}$ \\
115 &  27.73~$\pm$ 1.13 & 1717.8~$\pm$ 86.9 &  4.31~$\pm$ 0.28 & 0.06~$\pm$~0.02 & 0.10~$\pm$~0.03 & $<$ 0.77 & 10.2~$\pm$ 1.7 &  4.4~$\pm$~0.8 &  2.3$^{+0.8}_{-0.6} \times 10^{-3}$ \\
117 &   8.47~$\pm$ 0.39 &  238.3~$\pm$ 14.3 &  3.31~$\pm$ 0.21 & 0.15~$\pm$~0.05 & 0.47~$\pm$~0.13 & $<$ 0.84 &  1.5~$\pm$ 0.5 &  3.4~$\pm$~1.3 &  4.5$^{+2.5}_{-2.2} \times 10^{-4}$ \\
120 &   3.53~$\pm$ 0.25 &   70.8~$\pm$  6.1 &  5.45~$\pm$ 0.35 & 0.51~$\pm$~0.14 & 1.51~$\pm$~0.30 & $<$ 1.25 &  1.4~$\pm$ 0.7 &  3.9~$\pm$~1.5 &  3.7$^{+2.4}_{-2.2} \times 10^{-4}$ \\
124 &  10.39~$\pm$ 0.46 &  351.7~$\pm$ 20.6 &  2.94~$\pm$ 0.19 & 0.11~$\pm$~0.03 & 0.31~$\pm$~0.09 & $<$ 0.83 &  2.0~$\pm$ 0.6 &  3.6~$\pm$~1.3 &  5.6$^{+2.9}_{-2.6} \times 10^{-4}$ \\
133 &  11.71~$\pm$ 0.51 &  600.8~$\pm$ 31.5 &  9.55~$\pm$ 0.61 & 0.30~$\pm$~0.08 & 0.53~$\pm$~0.14 & $<$ 1.15 &  5.3~$\pm$ 1.5 &  4.4~$\pm$~1.2 &  1.2$^{+0.6}_{-0.5} \times 10^{-3}$ \\
136 &   1.26~$\pm$ 0.21 &   49.8~$\pm$  3.5 &  0.33~$\pm$ 0.02 & 0.11~$\pm$~0.05 & 0.25~$\pm$~0.08 & $<$ 0.88 & \nodata &  5.2~$\pm$~0.9 & \nodata \\
144 &   4.15~$\pm$ 0.26 &  193.0~$\pm$ 10.6 &  2.81~$\pm$ 0.18 & 0.25~$\pm$~0.08 & 0.49~$\pm$~0.13 & $<$ 1.41 &  1.6~$\pm$ 0.5 &  4.2~$\pm$~1.5 &  3.8$^{+2.0}_{-1.8} \times 10^{-4}$ \\
145 &  13.80~$\pm$ 0.59 &  670.0~$\pm$ 34.7 &  1.91~$\pm$ 0.12 & 0.06~$\pm$~0.02 & 0.12~$\pm$~0.04 & $<$ 1.42 &  3.7~$\pm$ 0.8 &  4.1~$\pm$~1.2 &  9.1$^{+4.1}_{-3.4} \times 10^{-4}$ \\
146 &  10.10~$\pm$ 0.45 &  388.0~$\pm$ 21.0 &  1.54~$\pm$ 0.10 & 0.06~$\pm$~0.02 & 0.16~$\pm$~0.05 & $<$ 0.67 &  2.0~$\pm$ 0.6 &  3.7~$\pm$~1.3 &  5.4$^{+2.8}_{-2.4} \times 10^{-4}$ \\
147 &  12.50~$\pm$ 0.54 &  828.0~$\pm$ 42.2 &  0.91~$\pm$ 0.06 & 0.03~$\pm$~0.01 & 0.05~$\pm$~0.01 & $<$ 1.05 &  4.8~$\pm$ 1.0 &  4.4~$\pm$~1.2 &  1.1$^{+0.5}_{-0.4} \times 10^{-3}$ \\
150 &  17.30~$\pm$ 0.72 &  129.0~$\pm$  7.5 &  0.43~$\pm$ 0.03 & 0.01~$\pm$~0.00 & 0.14~$\pm$~0.04 & $<$ 0.74 & \nodata & \nodata & \nodata \\
151 &   9.77~$\pm$ 0.44 &  176.0~$\pm$  9.6 &  1.41~$\pm$ 0.09 & 0.06~$\pm$~0.02 & 0.30~$\pm$~0.09 & $<$ 1.13 &  1.0~$\pm$ 0.2 &  2.0~$\pm$~1.3 &  5.1$^{+2.3}_{-1.9} \times 10^{-4}$
\end{deluxetable*}
\end{turnpage}

%%%%%%%%%%%%%%%%%%%%%%%%%%%%%%%%%%%
%%%%  APPENDIX A %%%%%
%%%%%%%%%%%%%%%%%%%%%%%%%%%%%%%%%%%
\clearpage
\appendix
\section{Modeling Uncertainties due to IMF Sampling Effects}
\label{sec:IMF}

In this Appendix we discuss in detail our method of calculating uncertainties in the direct modeling procedure we presented in \S~\ref{sec:models}. Our approach should be generally applicable to the estimation of uncertainties in population synthesis modeling which does not account for stochastic sampling of the IMF such as \texttt{Starburst99}~\citep[cf.][]{2012ApJ...745..145D}.

Measures of star formation that trace the upper end of the IMF, such as FUV and H$\alpha$ emission, are sensitive to both the form of the IMF as well as stochastic deviation from the expected distribution within a given population. The latter effect is particularly severe for low-mass stellar populations in which there may be very few massive stars, and so the addition or removal of even a single star can alter the population's FUV and ionizing photon output significantly. For this analysis we treat the IMF as a continuous probability distribution function (PDF) from which we draw each model population.

The \texttt{Starburst99} models we choose (Table~\ref{tab:SB99}) use the Kroupa IMF, which gives the number of stars $N$ in each mass bin as a two-part power law:
\begin{equation}
N(M) \, {\rm d}N = C M^{-\alpha} \, {\rm d}M,
\label{eqn:IMF}
\end{equation}
with
\begin{equation}
\alpha = \begin{cases}
& 1.3 < M/\Msun < 0.5 \\
& 2.3~{\rm if}~0.5 < M/\Msun < M_{\rm max}.
\end{cases}
\end{equation}
$C$ is a constant that sets the total population size (and mass), and $M_{\rm max}$ is the upper mass limit of the IMF. The default value in \texttt{Starburst99} is $M_{\rm max}=120~\Msun$, but here we leave $M_{\rm max}$ as a free parameter and explore the effects of changing it. Note that $C$ will take on different values for each segment of the broken power law.

If we are interested in knowing the number of stars of particular masses for a given population mass, we can solve for $C$ and then integrate Equation~(\ref{eqn:IMF}) over the desired range. The total population mass $M_*$ is computed by integrating the masses of all stars, i.e. it is the (un-normalized) first moment of the IMF:
\begin{eqnarray}
M_* &=& \int M N(M) \, dN \notag \\
&=& C_1 \int M^{1-\alpha_1} \, dM + C_2 \int M^{1-\alpha_2} \, dM \notag \\
&=& C_1 \int_{M_0}^{M_1} M^{1-\alpha_1} \, dM + C_2 \int_{M_1}^{M_{\rm max}} M^{1-\alpha_2} \, dM \notag \\
&=& C_1 \left[\frac{1}{2-\alpha_1}\left(M_1^{2-\alpha_1}-M_0^{2-\alpha_1}\right) + \frac{M_1^{\alpha_2-\alpha_1}}{2-\alpha_2}\left(M_{\rm max}^{2-\alpha_2} - M_1^{2-\alpha_2} \right) \right],
\end{eqnarray}

where we have enforced the fact that the IMF must be continuous at the point at which the slope changes to solve and substitute for $C_2$, i.e. $C_1 M^{\alpha_1} = C_2 M^{\alpha_2}$. Plugging in the Kroupa values ($\alpha_1=1.3,~\alpha_2=2.3,~M_0=0.1~\Msun,~M_1=0.5~\Msun$) results in

\begin{eqnarray}
M_*&=& C_1 \left(2.646 - \frac{5}{3} M_{\rm max}^{-0.3} \right) \notag \\
&=& C_1\gamma,
\end{eqnarray}
where the last expression defines the parameter $\gamma$, and thus $C_1=M_*/\gamma$. Numerical values of $\gamma$ for various choices of $M_{\rm max}$ are given in Table~\ref{tab:imf1}.

\begin{deluxetable}{cc}
\tablecaption{Numerical values of the parameter $\gamma$ \label{tab:imf1}
}
\tablehead{
$M_{\rm max}~(\Msun)$	& $\gamma$
}
$\infty$		& 2.646 \\
120			& 2.250 \\
100			& 2.227 \\
80			& 2.198 \\
50			& 2.131 \\
30			& 2.045 \\
20			& 1.968
\end{deluxetable}

\begin{figure}[tb]
\includegraphics[width=\linewidth]{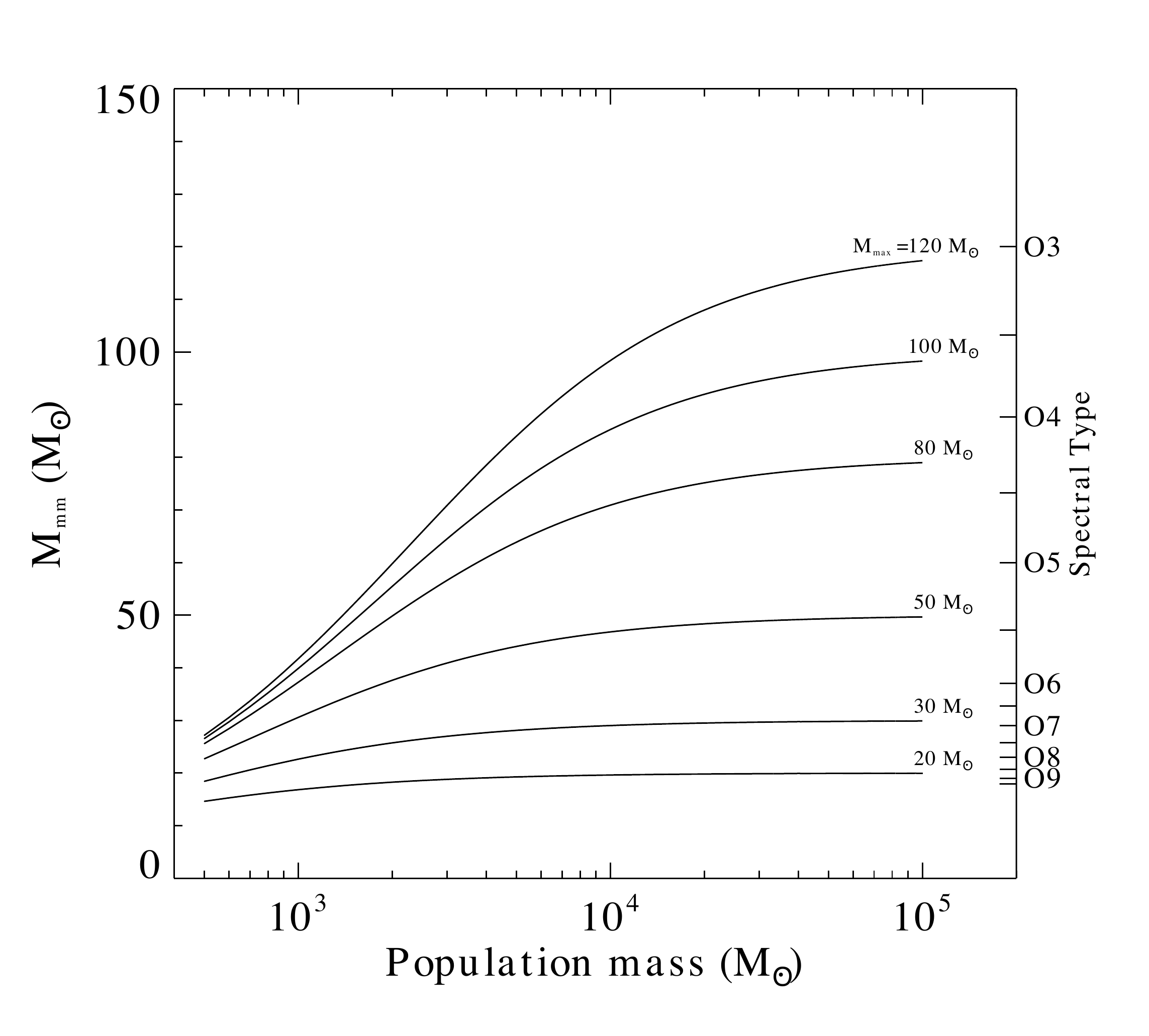}
\caption{\small{Mass of the most massive star $M_{\rm mm}$ in a newly formed stellar population generated using \texttt{Starburst99} as a function of the total stellar mass in the population. The separate curves show how this functionality differs for different IMF upper mass limits. $M_{\rm mm}$ was computed using Equation~\ref{eqn:Mmm}.}}
\label{fig:MmmvMpop}
\end{figure}

We can now estimate $M_{\rm mm}$, the stellar mass above which the IMF integrates to unity, which approximately corresponds to the highest mass a star will have in a newly formed stellar population of a given total mass $M_*$. We compute $M_{\rm mm}$ for each given $M_{\rm max}$ by setting $N=1$ in Equation~(\ref{eqn:IMF}) and integrating. Note that we only need to consider the second (higher mass) part of the power law.
\begin{equation}
N=1=C_2 \int_{M_{\rm mm}}^{M_{\max}} M^{-\alpha} \, dM = \frac{M_* M_1^{\alpha_2-\alpha_1}}{\gamma(\alpha-1)} \left(M_{\rm mm}^{1-\alpha}-M_{\rm max}^{1-\alpha} \right).
\end{equation}
Solving for $M_{\rm mm}$, we get
\begin{equation}
M_{\rm mm} = \left( \frac{2.6 \gamma}{M_*} + M_{\rm max}^{-1.3} \right)^{-1/1.3}.
\label{eqn:Mmm}
\end{equation}
Curves depicting $M_{\rm mm}$ as a function of $M_*$ for representative choices of $M_{\rm max}$ (and commensurate $\gamma$) are presented in Figure~\ref{fig:MmmvMpop}.

If the IMF is stochastically sampled according to Poisson statistics, the uncertainty on the highest mass bin (i.e. the one at $M=M_{\rm mm}$, which has only one star) is unity. If the UV luminosity of this star dominates the population's luminosity, then the Poisson uncertainties on the total FUV and H$\alpha$ luminosities of the population are simply the FUV and H$\alpha$ luminosities of a star of mass $M_{\rm mm}$. We compute these luminosities for a range of $M_{\rm mm}$ corresponding to the range of $M_*$ in our model grid using the same evolutionary tracks and atmospheric models chosen for our \texttt{Starburst99} runs, as follows.

For a given $M_*$ in our model grid, we first compute $M_{\rm mm}$ using Equation~(\ref{eqn:Mmm}). We then interpolate the Geneva high mass-loss evolutionary tracks~\citep{1994A&AS..103...97M} to estimate that star's mass $M$, effective temperature $T_{\rm eff}$, and bolometric luminosity $L_{\rm bol}$ at 2~Myr. We do this separately for the two tracks having initial masses bracketing $M_{\rm mm}$, then perform a weighted average to estimate each of the three quantities. Each weight is defined to be the normalized distance between $M_{\rm mm}$ and the initial mass of the track. We then calculate the star's radius from the definition of effective temperature, $R=\sqrt{L_{\rm bol} / (4\pi \sigma_{\rm SB} T_{\rm eff}^4)}$, where $\sigma_{\rm SB}$ is the Stefan-Boltzmann constant. We also compute its surface gravity as $\log{g}=\log{(GM/R^2)}$. We then select the two O-star atmospheric models from \cite{2002MNRAS.337.1309S} nearest in $T_{\rm eff}$ and $\log{g}$ to the values we derive to compute the FUV and H$\alpha$ luminosities as described below.

\begin{figure}[tb]
\includegraphics[width=\linewidth]{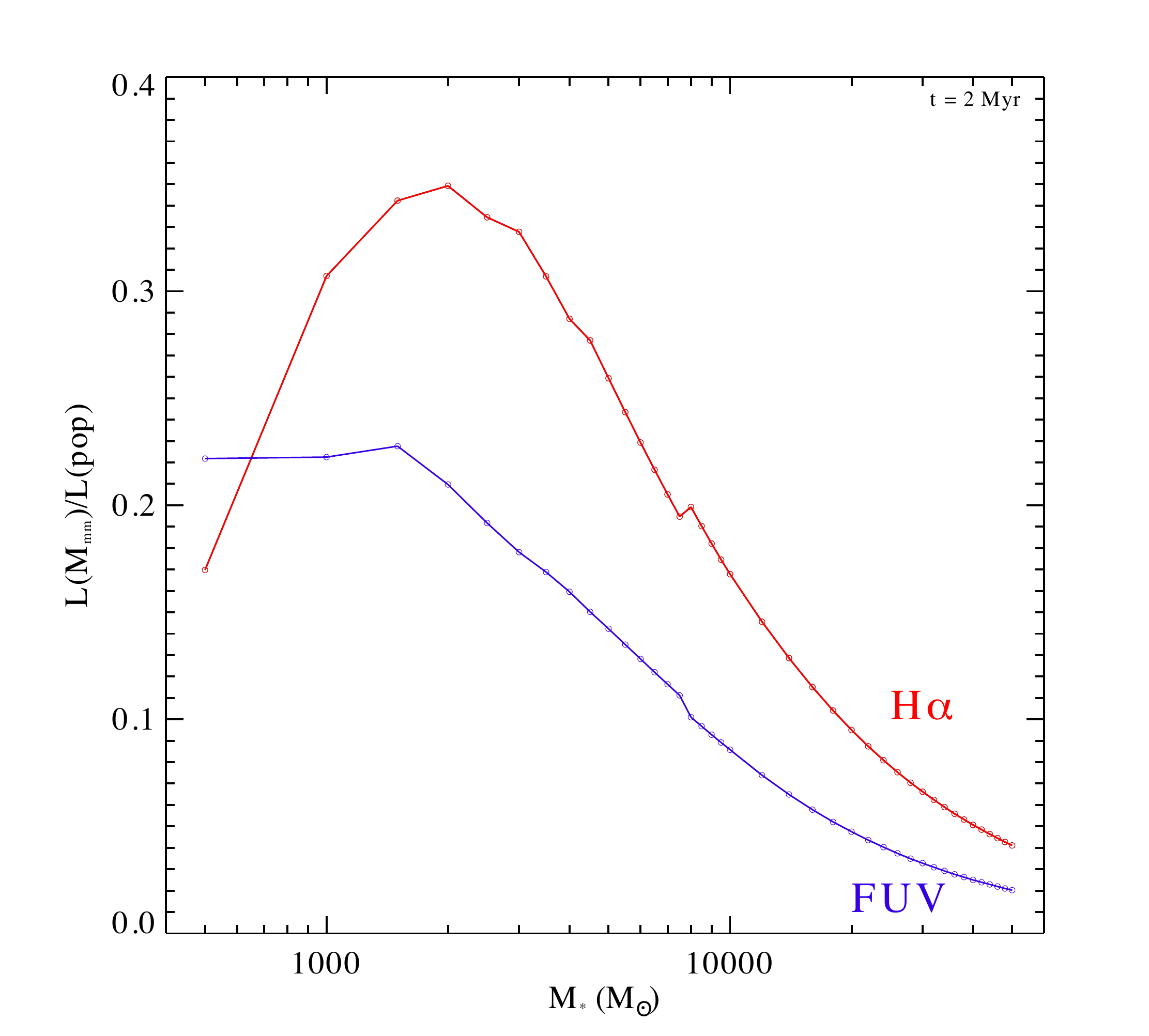}
\caption{Ratio of the luminosity of the most massive star in a population to the total luminosity of that population at 2~Myr after birth as a function of population mass. The blue curve shows this ratio for GALEX FUV luminosity, and the red curve shows the ratio for H$\alpha$, computed as described in the text. These ratios are taken to be the fractional uncertainties on the population mass and age, respectively, for our NGC~300 {\HII} regions.}
\label{fig:modeluncerts}
\end{figure}

To compute the FUV luminosity of a given model, we integrate the model spectrum over the GALEX FUV bandpass. To calculate the H$\alpha$ luminosity, we first integrate the model spectrum over the ionizing continuum to compute the ionizing photon rate $Q$(H$^0$), as
\begin{equation}
Q({\rm H}^0) = 4\pi R^2 \int_{\nu0}^\infty \frac{F_{\nu}}{h\nu} \rm{d}\nu,
\end{equation}
where $F_{\nu}$ is the flux density at frequency $\nu$, and $\nu0$ is the minimum frequency at which a photon can ionize hydrogen, i.e. $\nu=3.29 \times 10^{15}$~Hz, corresponding to $\lambda = 912$~{\AA}. The luminosity in a given recombination line is then linearly proportional to $Q$(H$^0$), assuming Case B recombination \citep[e.g.,][]{Osterbrock:2006ul}. The constant of proportionality is given by the ratio of the effective recombination rate coefficient of the line $\alpha^{\rm eff}_{\rm{H}\alpha}$ to the case B recombination rate coefficient $\alpha_{\rm B}$, i.e.
\begin{equation}
L(\rm{H}\alpha)=\left(\frac{\alpha^{\rm eff}_{\rm{H}\alpha}}{\alpha_{\rm{B}}}\right) h\nu_{\rm{H}\alpha} \, Q(\rm{H}^0).
\end{equation}
We use the above procedure to compute the FUV and H$\alpha$ luminosities for the two atmospheric models with $T_{\rm eff}$ and $\log{g}$ bracketing the values we derive above, then perform a weighted average using the normalized differences in $T_{\rm eff}$ between the derived value and model track as weights. Figure~\ref{fig:modeluncerts} shows the ratio of the luminosity of the most massive star in a population to that population's total luminosity at 2~Myr for both FUV and H$\alpha$. This represents an approximation of the stochastic uncertainty due to sampling the IMF. To apply this to our NGC~300 results, we take the ratio of FUV luminosities, i.e. $L_{\rm FUV}$($M_{\rm mm})/L_{\rm FUV}(\rm{pop})$, to be the fractional uncertainty on the derived mass, and the ratio of H$\alpha$ luminosities as the fractional uncertainty on the derived age. The uncertainty on the SFR is then computed by formally propagating through Equation~\ref{eqn:SFR}; fractional uncertainties in the SFR due to stochastic sampling range from 7 to 41\%. These uncertainties are added in quadrature with the other sources of uncertainty (see \S~\ref{sec:SFRunc}).

%%%%%%%%%%%%%%%%%%%
%% APPENDIX B %%
%%%%%%%%%%%%%%%%%%%%%

\clearpage
\section{Zoom-in images}
\label{sec:stamps}

Here we present zoom-in \textit{Spitzer}/MIPS 24~$\mu$m, GALEX FUV, and ESO/WFI H$\alpha$ images of all 76 of our  sources alongside their APEX spectra. The solid circles indicate the APEX 27{\arcsec} beam FWHM, which spans $\sim250$~pc at the 1.93~Mpc distance of NGC~300. Dotted circles show the positions of the photometric apertures for which we adjusted from the APEX pointing position. The full list of sources for which such an adjustment was made is shown in Table~\ref{tab:varphot}. Black circles indicate APEX CO detections and red circles indicate nondetections.

\begin{deluxetable}{crrrr}
\tablecaption{Adjusted photometric apertures \label{tab:varphot}}
\tablehead{
\colhead{DCL \#}	& \colhead{RA offset\tablenotemark{a}}	& \colhead{dec offset\tablenotemark{a}}
& \colhead{total offset\tablenotemark{a}}	& \colhead{Aperture} \\
	& ({\arcsec})	& ({\arcsec})	& ({\arcsec})	& size ({\arcsec}) \\
}
  6 &  -1.9 &  -0.7 &   2.0 &  11.0 \\
  9 &   6.2 &   1.1 &   6.3 &  12.0 \\
 15 &   0.0 &   0.0 &   0.0 &  15.5 \\
 17 &   2.8 &   2.1 &   3.5 &  13.5 \\
 29 &   0.9 &  -3.4 &   3.5 &  10.5 \\
 43 &   2.1 &   1.5 &   2.6 &  11.5 \\
 46 &  -1.9 &   0.7 &   2.0 &   9.0 \\
 57 &  -0.9 &  -5.2 &   5.3 &  11.0 \\
 72 &   2.4 &  -4.9 &   5.4 &  17.0 \\
 79 &  -4.5 &  -2.8 &   5.3 &  14.5 \\
 89 &  -1.7 &  -3.0 &   3.4 &  10.0 \\
112 &   0.0 &   0.0 &   0.0 &  10.5 \\
114 &  -7.6 &   3.4 &   8.3 &  13.5 \\
115 &   0.0 &   0.0 &   0.0 &  21.0 \\
117 &  -5.9 &  -2.8 &   6.6 &  10.5 \\
118B &  -4.7 &   2.1 &   5.2 &  17.5 \\
119C &  11.4 &  -3.7 &  12.0 &  13.5 \\
124 &  -2.1 &  -8.3 &   8.5 &  13.5 \\
127 &   0.0 &   0.0 &   0.0 &  10.5 \\
133 &  -3.6 &  -0.2 &   3.6 &  16.0 \\
\tablenotetext{a}{relative to the APEX pointing position as reported in Table~\ref{tab:obslog}}
\end{deluxetable}

% online-only figures for Faesi et al. manuscript
% Molecular Cloud-scale Star Formation in NGC 300
% all .pdf files referenced here should be in subfolder /online/

\begin{figure*}[b]
\begin{minipage}{0.24\linewidth}
\includegraphics[width=\linewidth]{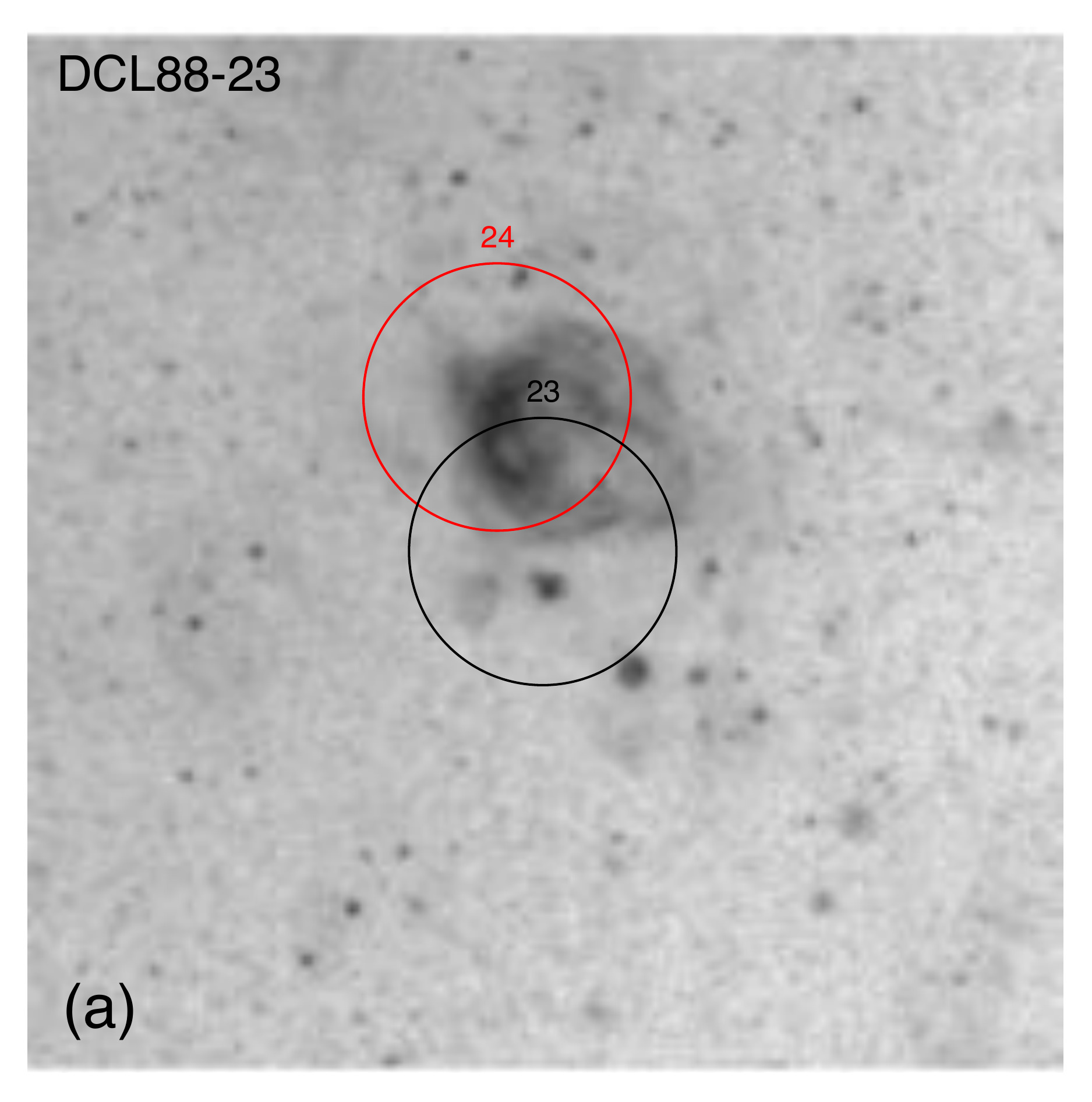}
\end{minipage}
\begin{minipage}{0.24\linewidth}
\includegraphics[width=\linewidth]{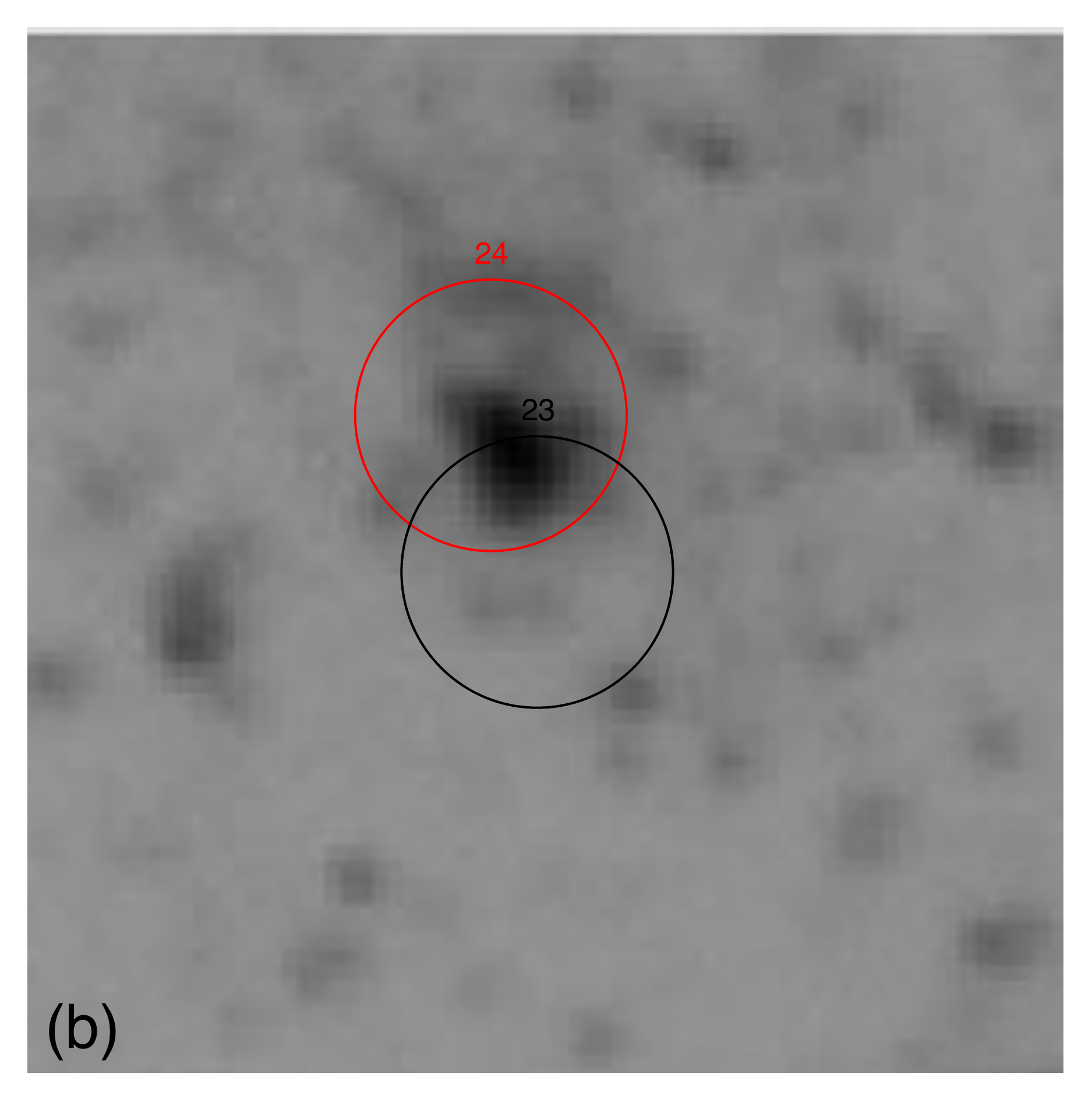}
\end{minipage}
\begin{minipage}{0.24\linewidth}
\includegraphics[width=\linewidth]{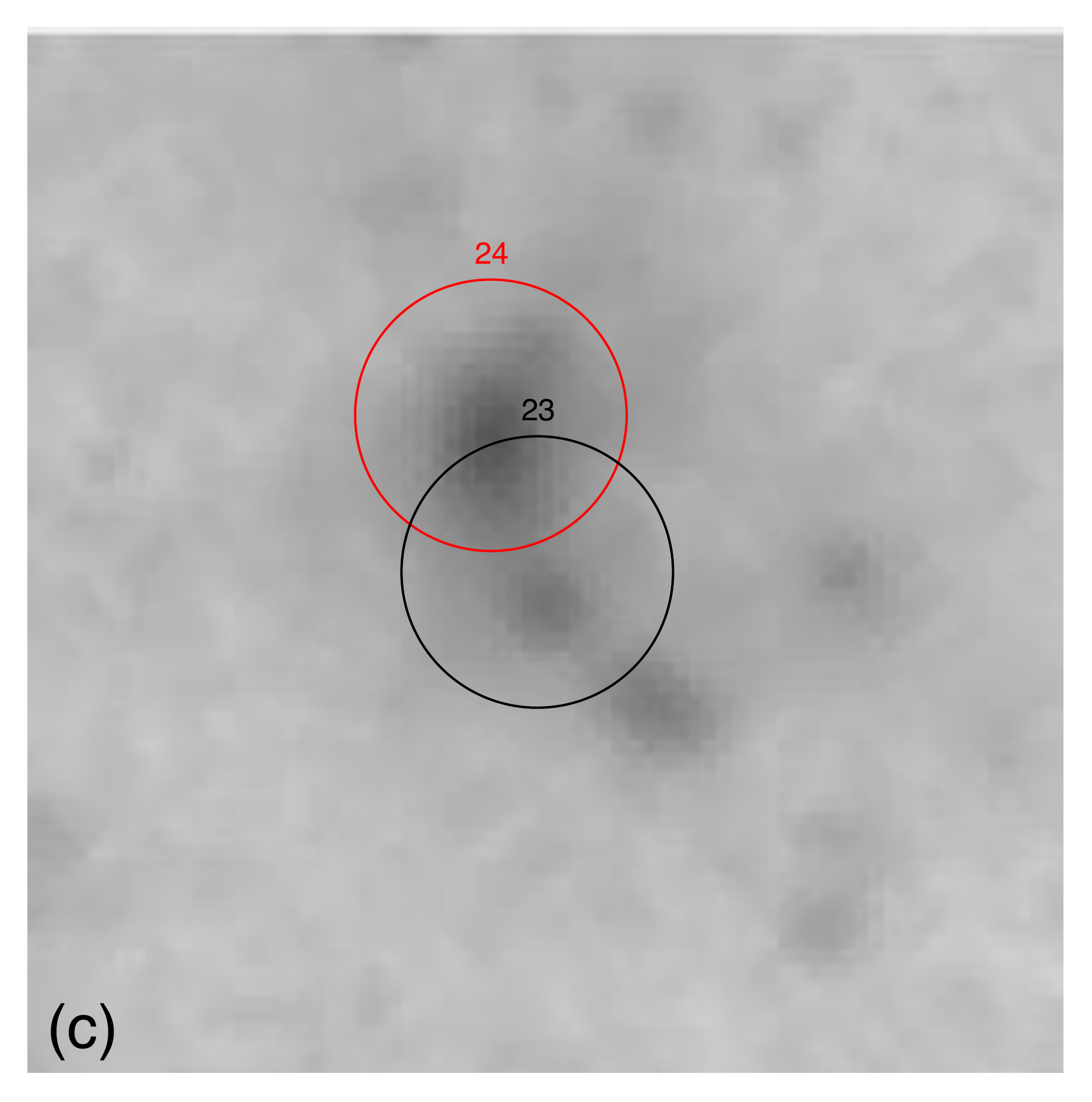}
\end{minipage}
\begin{minipage}{0.24\linewidth}
\includegraphics[width=\linewidth]{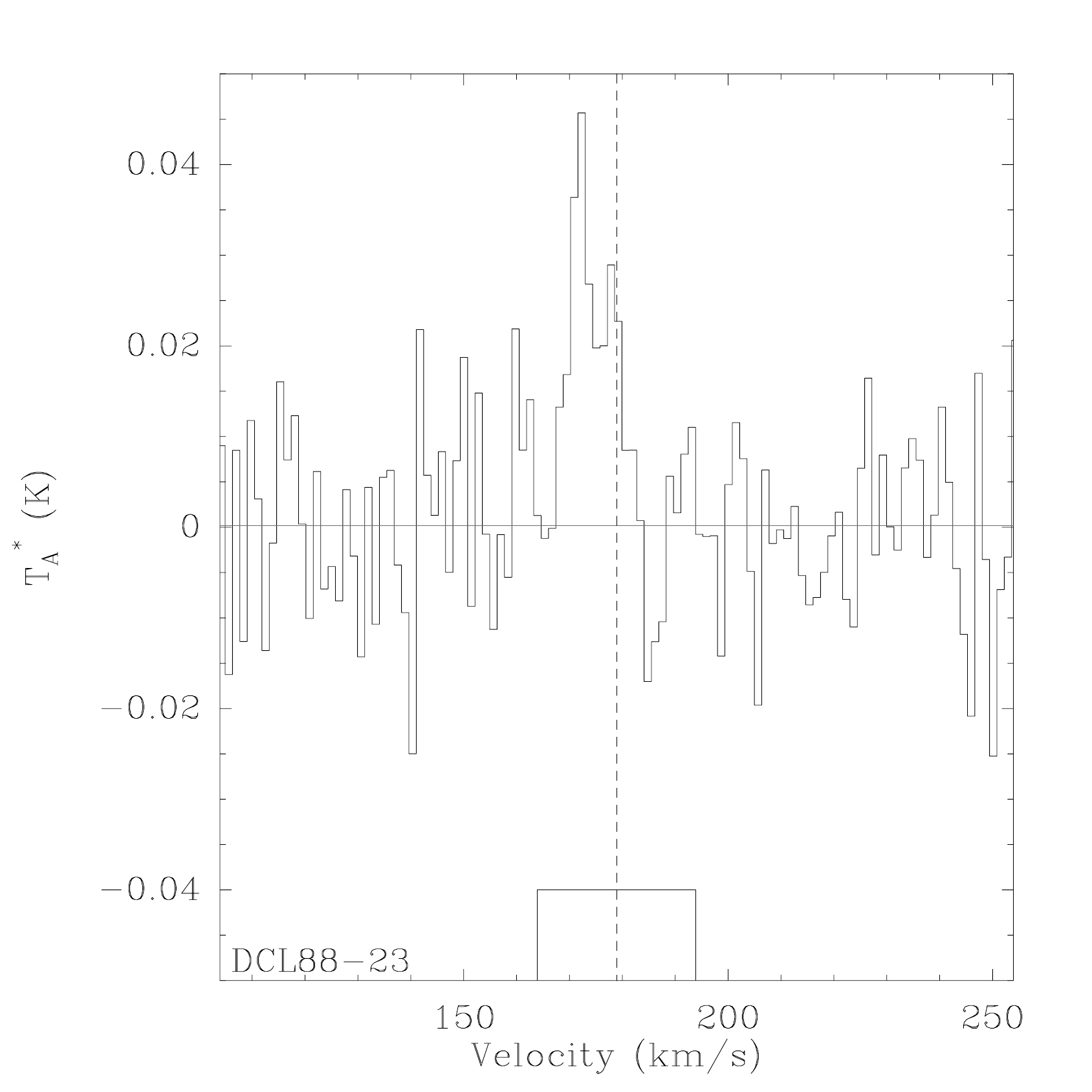}
\end{minipage}

\caption{\small{From left to right, ESO/WFI H$\alpha$ (a), GALEX FUV (b), and \textit{Spitzer}/MIPS 24~$\mu$m (c) images and APEX CO$(J=2-1)$ spectrum of our {\HII} region targets in NGC~300. The APEX 27{\arcsec} (250~pc) beam is indicated with a solid circle; this also denotes our photometric apertures for most sources. Dark circles indicate APEX CO detections while red circles show nondetections. Dotted circles show apertures which were moved from the APEX pointing position and/or altered in size (see Section~3.2). In each spectrum, the dashed line indicates the H~I velocity and the box shows the 30~km~s$^{-1}$ window over which we computed $I_{\rm CO}$. Note that the y-axis scale is different for each spectrum. The DCL source number of each source is listed on the H$\alpha$ image and the APEX spectrum; these numbers match those listed in Tables~\ref{tab:obslog} and~\ref{tab:results}. Right ascension increases to the left and declination to the top of the images. Note that the H$\alpha$ images shown here are not continuum-subtracted.}}
\label{fig:stamps}
\end{figure*}

\begin{figure*}
%\ContinuedFloat
\begin{minipage}{0.24\linewidth}
\includegraphics[width=\linewidth]{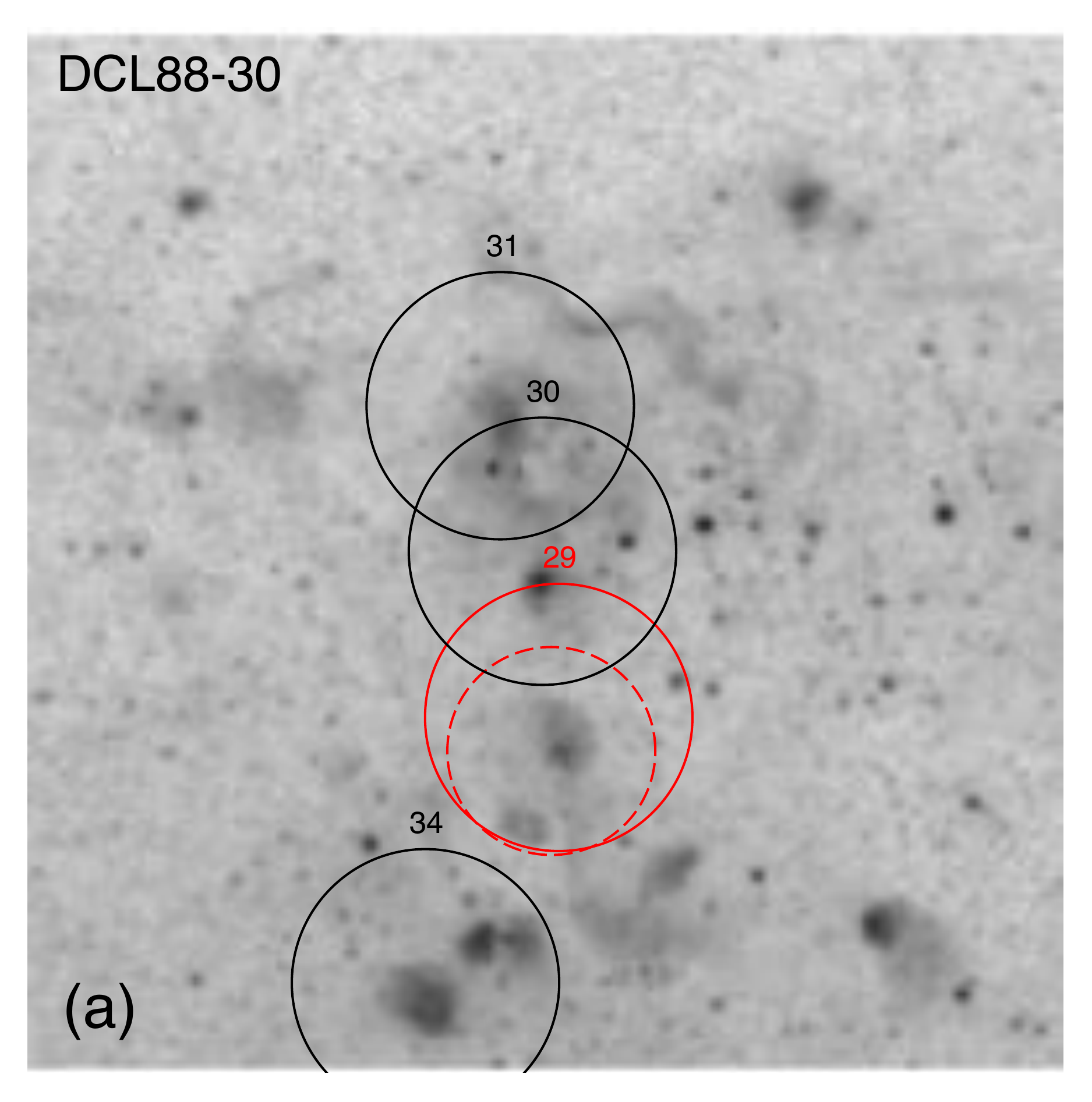}
\end{minipage}
\begin{minipage}{0.24\linewidth}
\includegraphics[width=\linewidth]{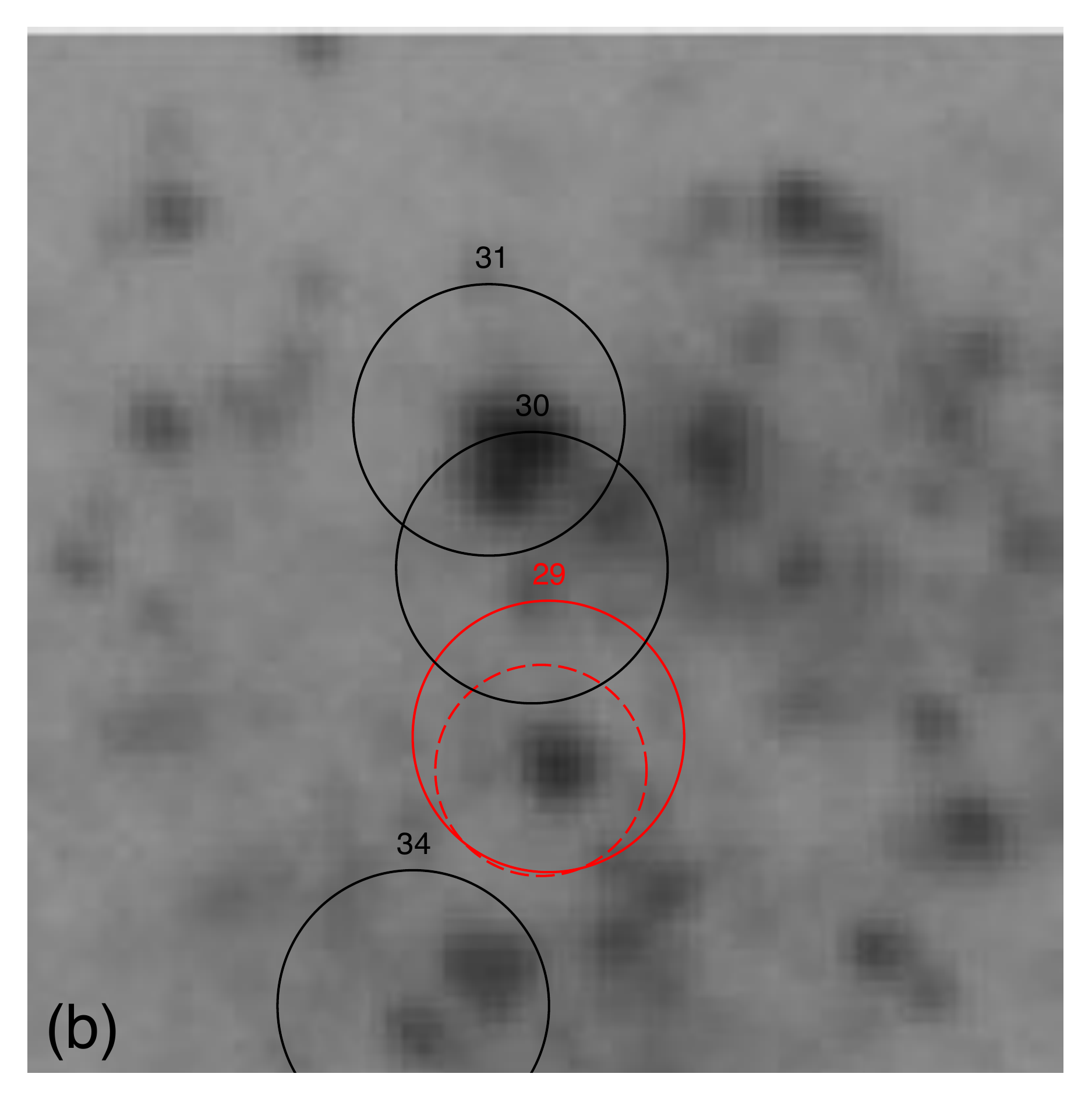}
\end{minipage}
\begin{minipage}{0.24\linewidth}
\includegraphics[width=\linewidth]{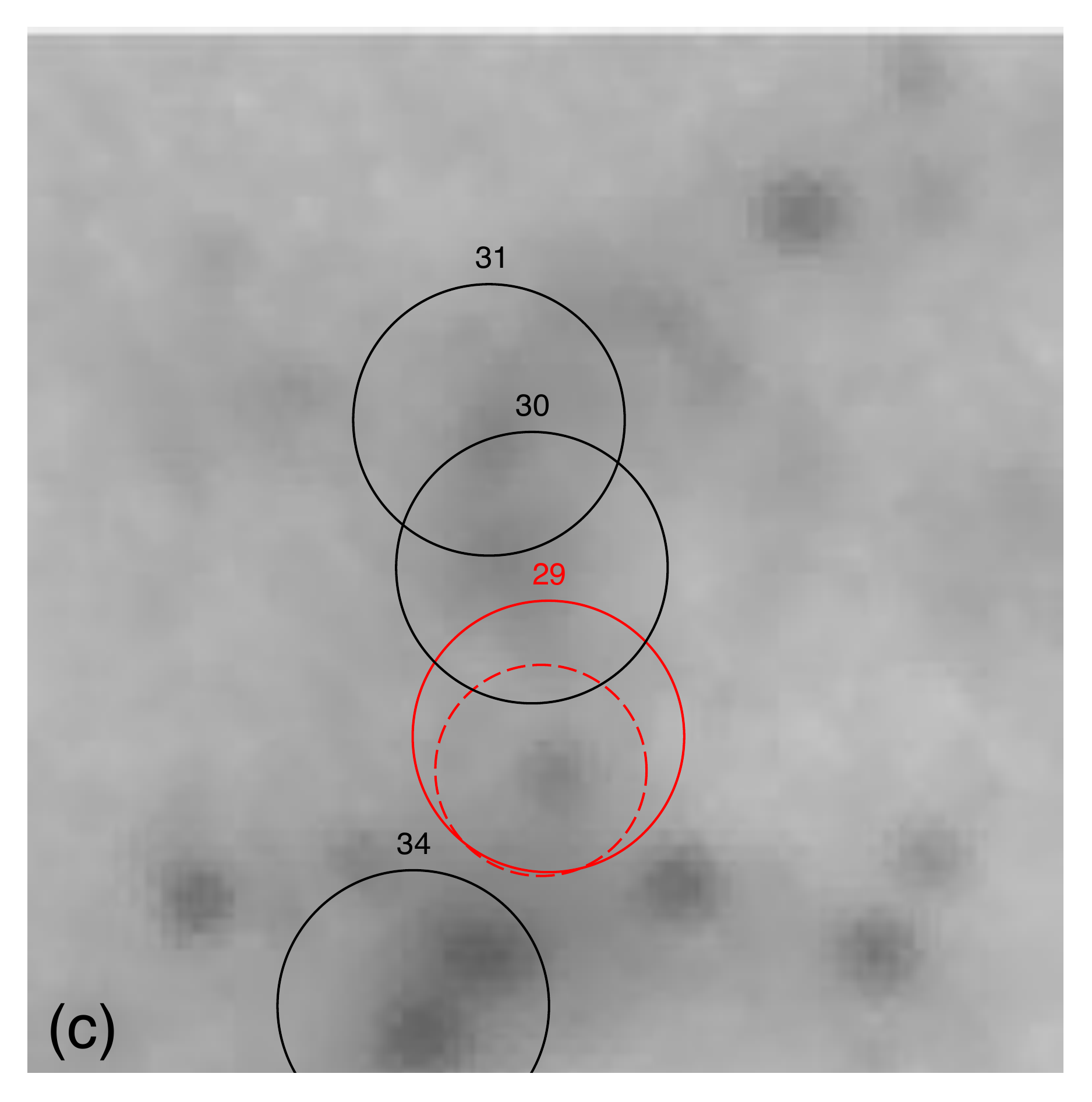}
\end{minipage}
\begin{minipage}{0.24\linewidth}
\includegraphics[width=\linewidth]{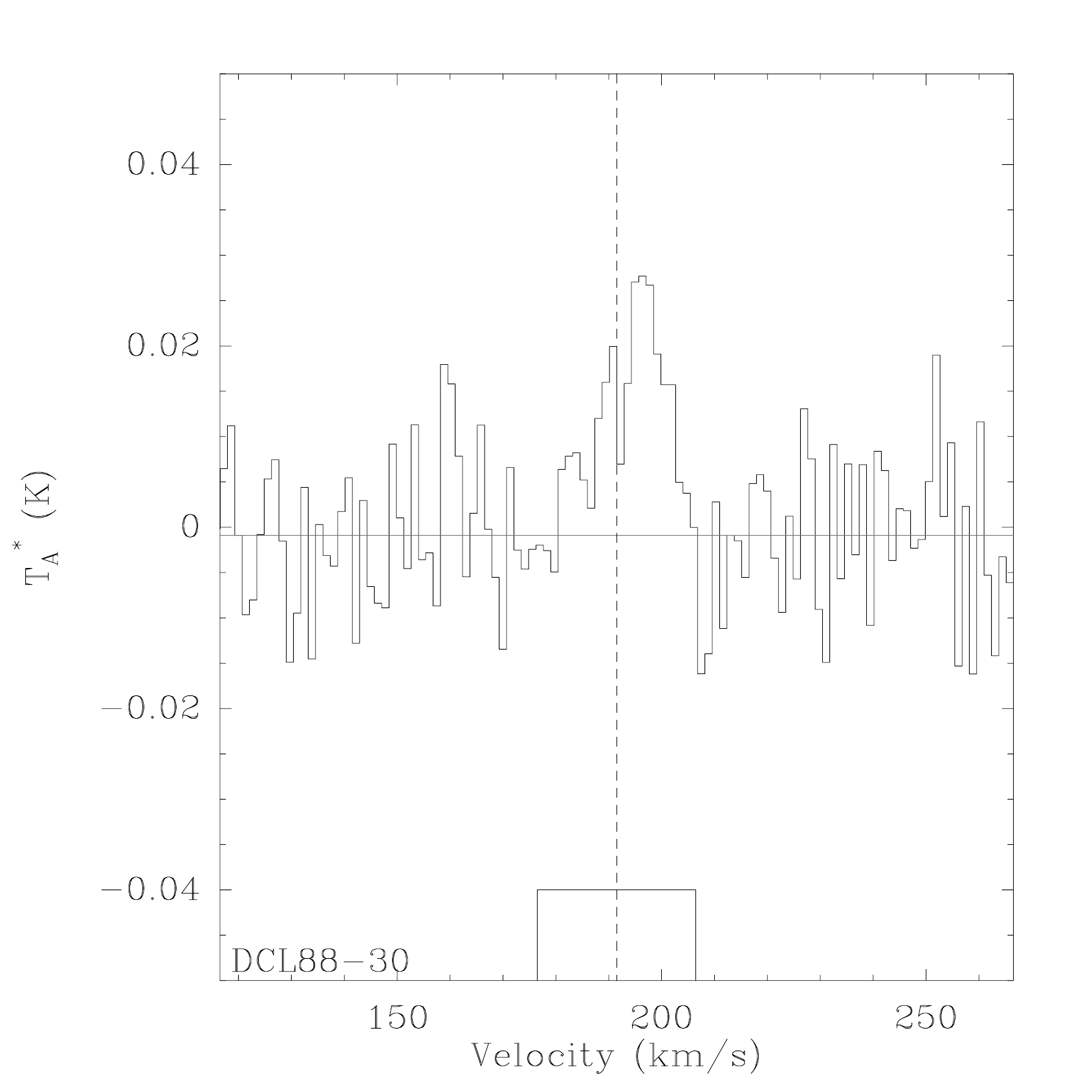}
\end{minipage}

\begin{minipage}{0.24\linewidth}
\includegraphics[width=\linewidth]{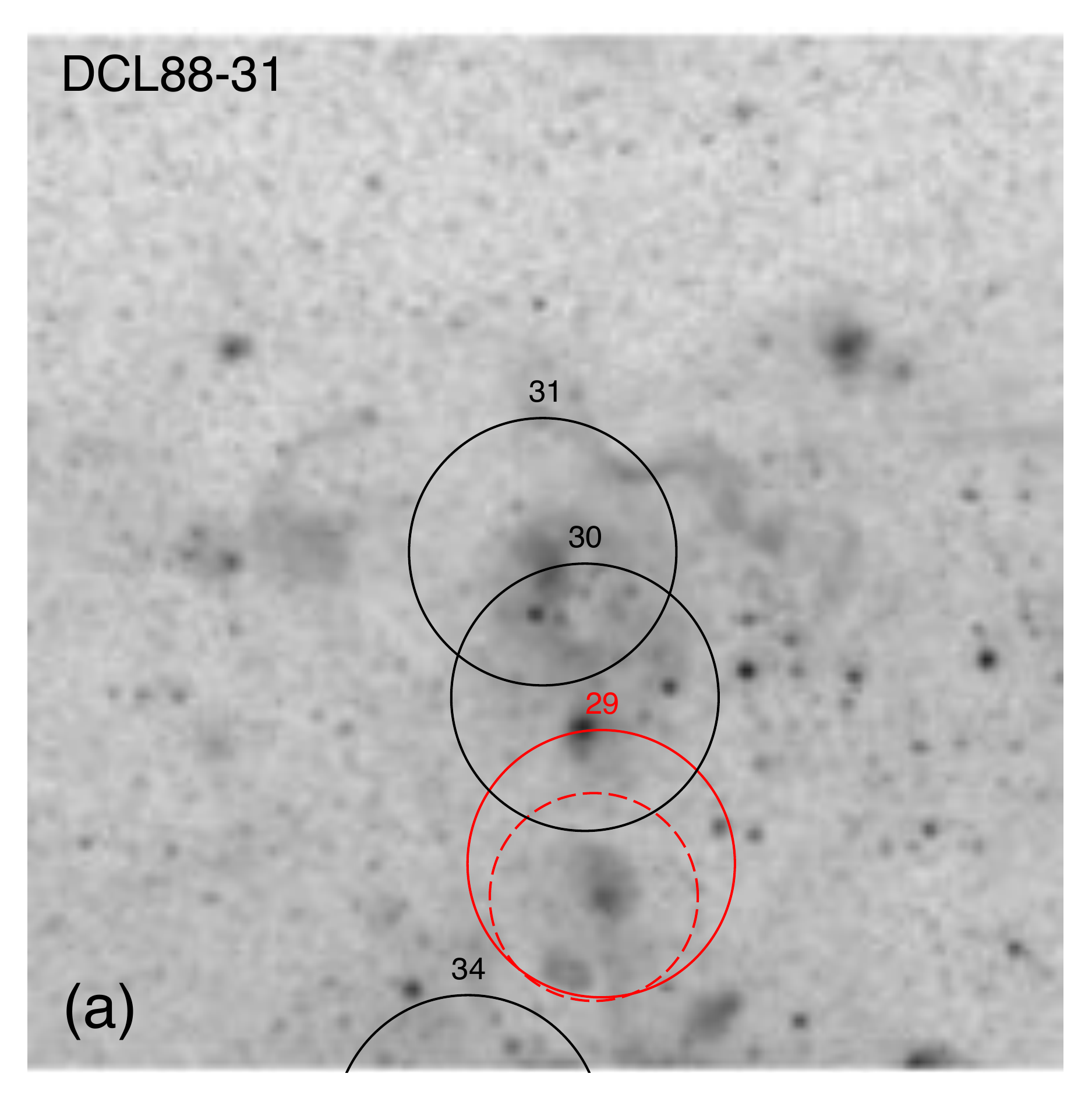}
\end{minipage}
\begin{minipage}{0.24\linewidth}
\includegraphics[width=\linewidth]{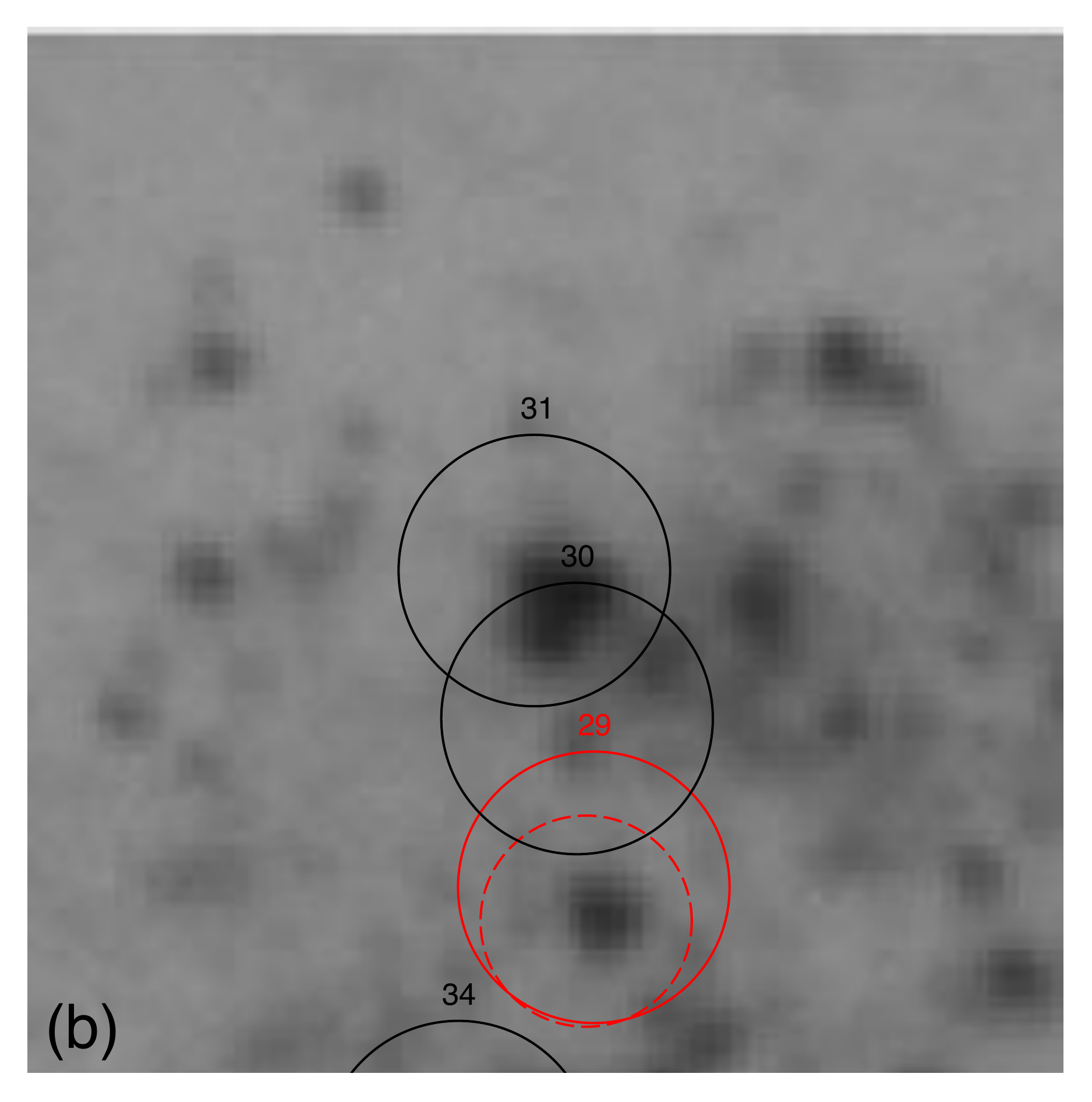}
\end{minipage}
\begin{minipage}{0.24\linewidth}
\includegraphics[width=\linewidth]{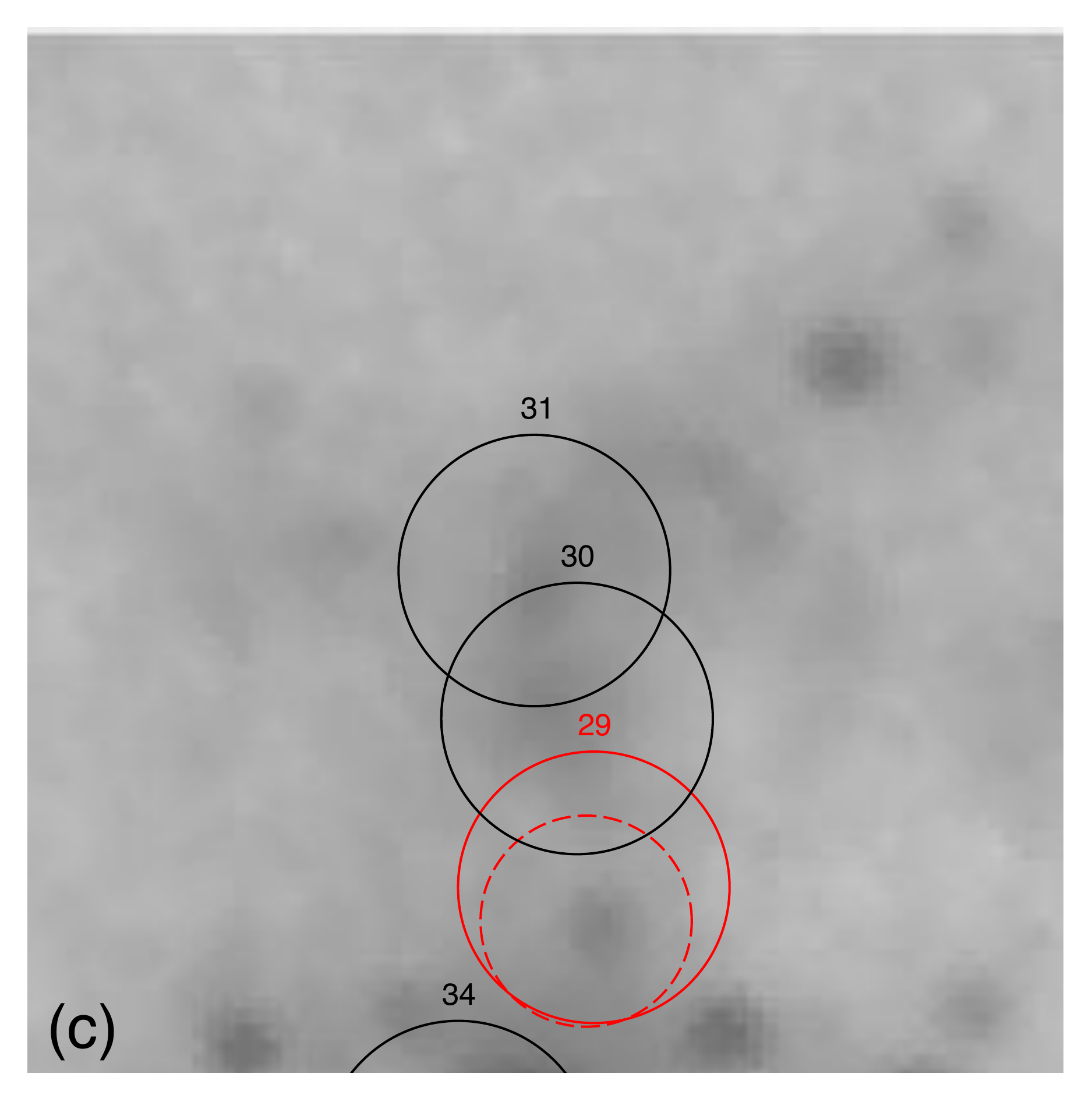}
\end{minipage}
\begin{minipage}{0.24\linewidth}
\includegraphics[width=\linewidth]{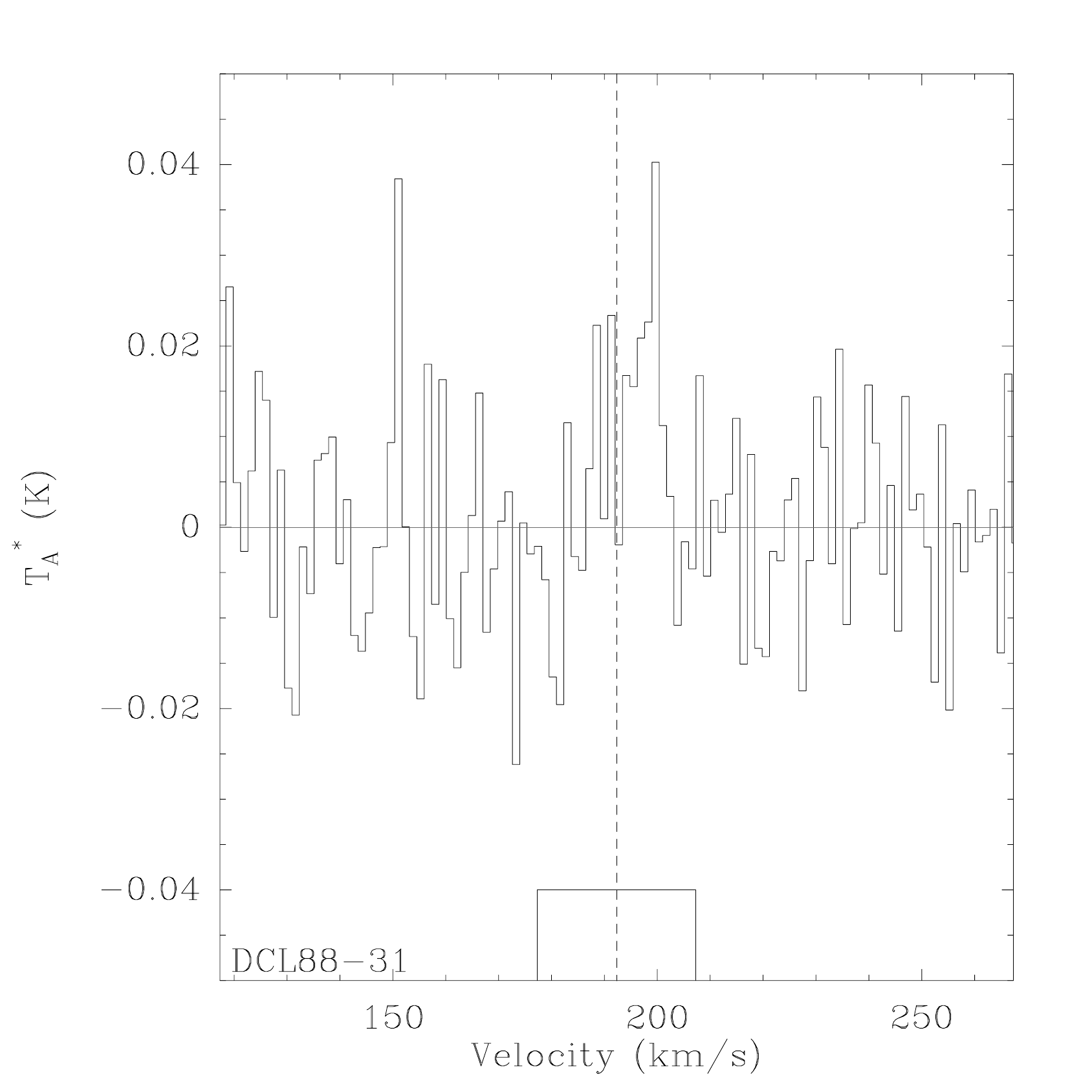}
\end{minipage}

\begin{minipage}{0.24\linewidth}
\includegraphics[width=\linewidth]{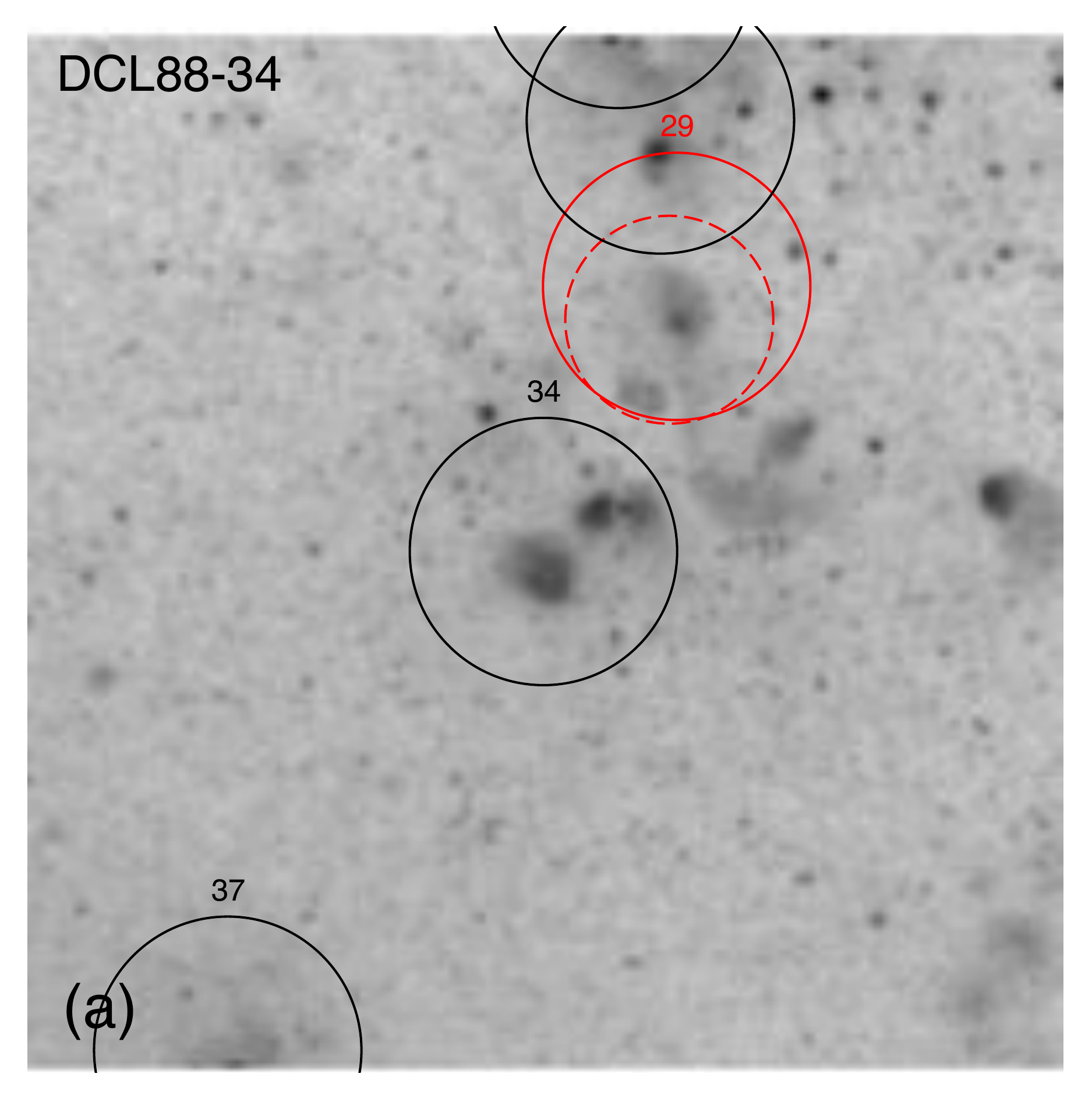}
\end{minipage}
\begin{minipage}{0.24\linewidth}
\includegraphics[width=\linewidth]{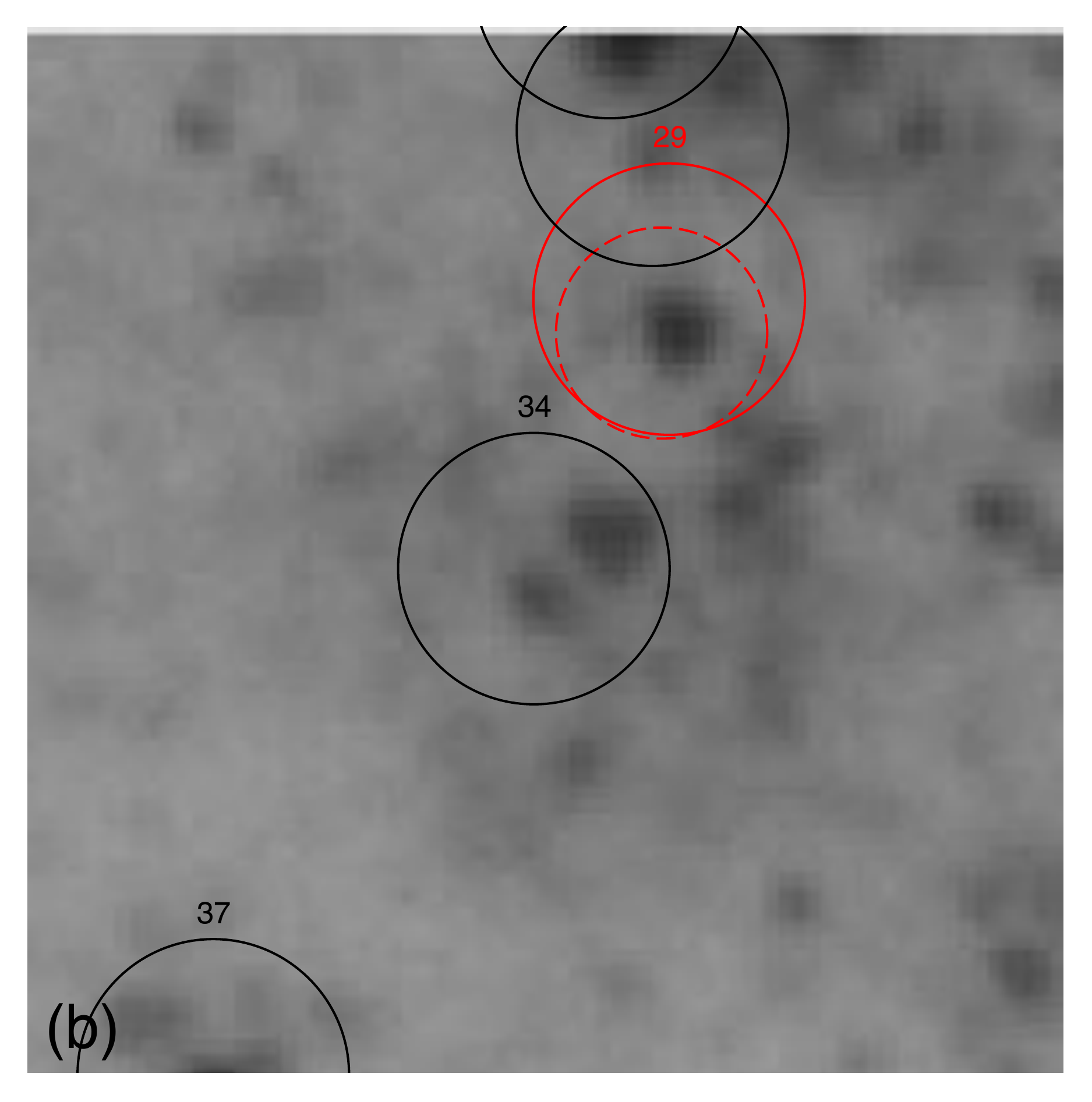}
\end{minipage}
\begin{minipage}{0.24\linewidth}
\includegraphics[width=\linewidth]{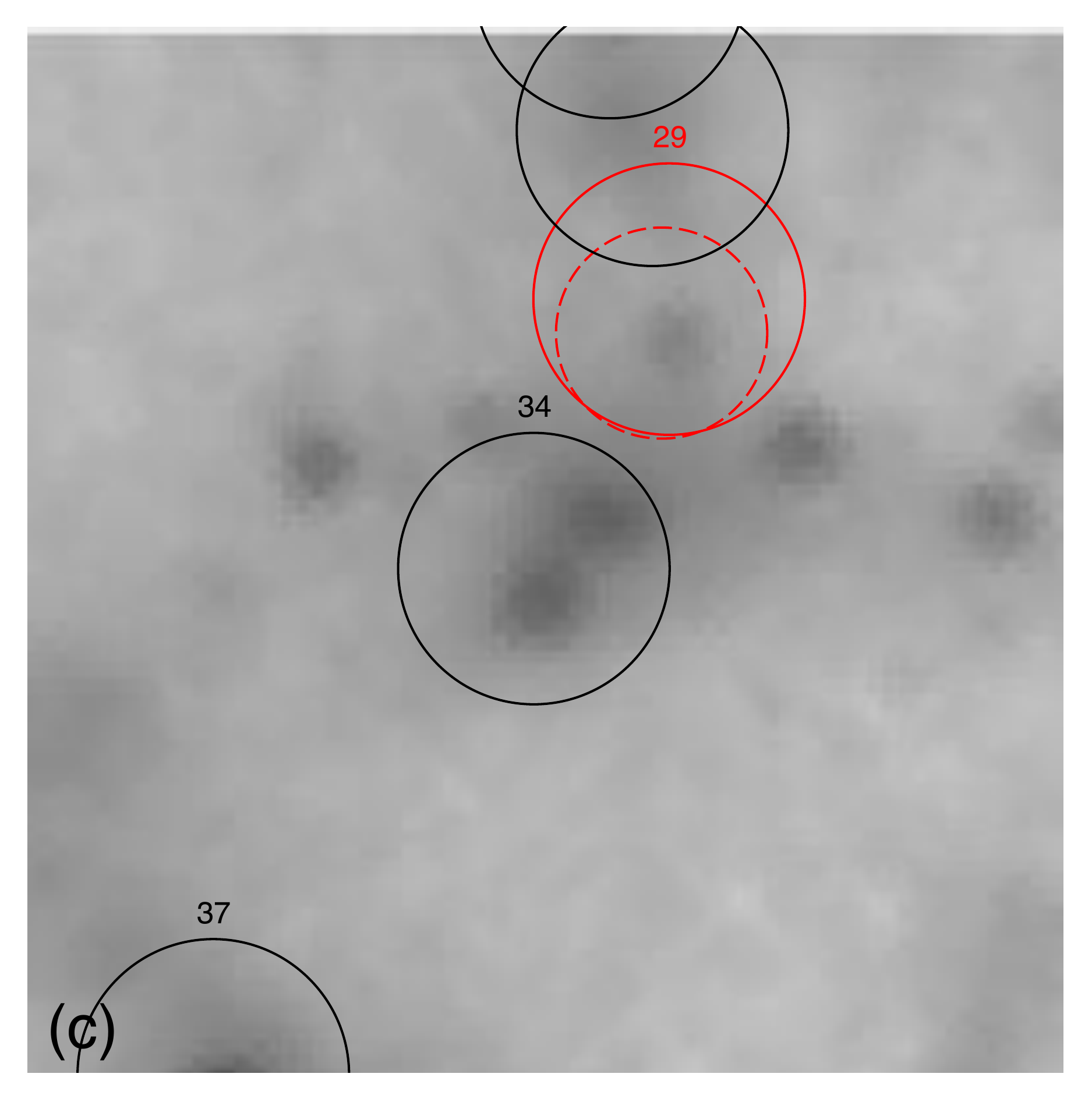}
\end{minipage}
\begin{minipage}{0.24\linewidth}
\includegraphics[width=\linewidth]{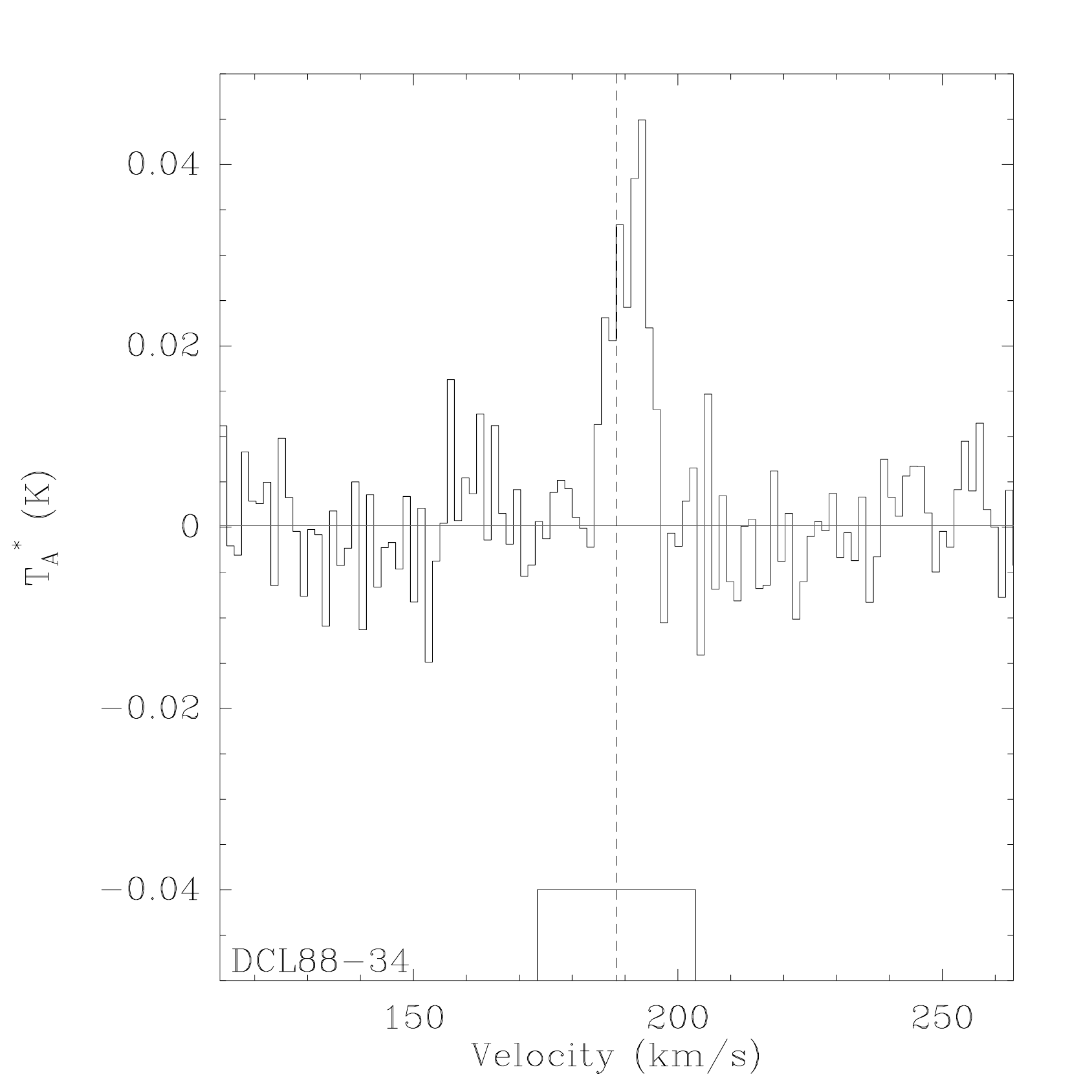}
\end{minipage}

\begin{minipage}{0.24\linewidth}
\includegraphics[width=\linewidth]{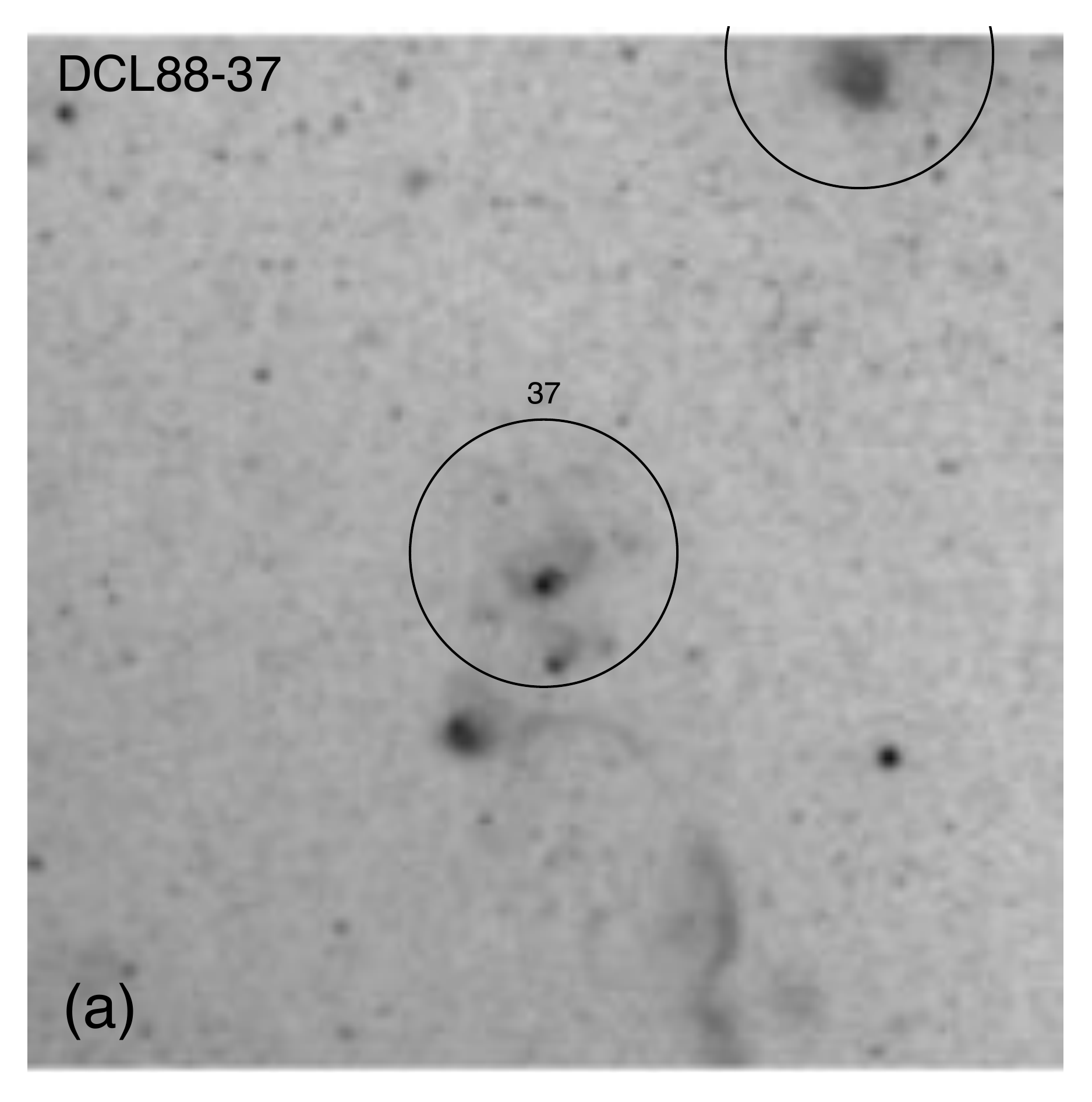}
\end{minipage}
\begin{minipage}{0.24\linewidth}
\includegraphics[width=\linewidth]{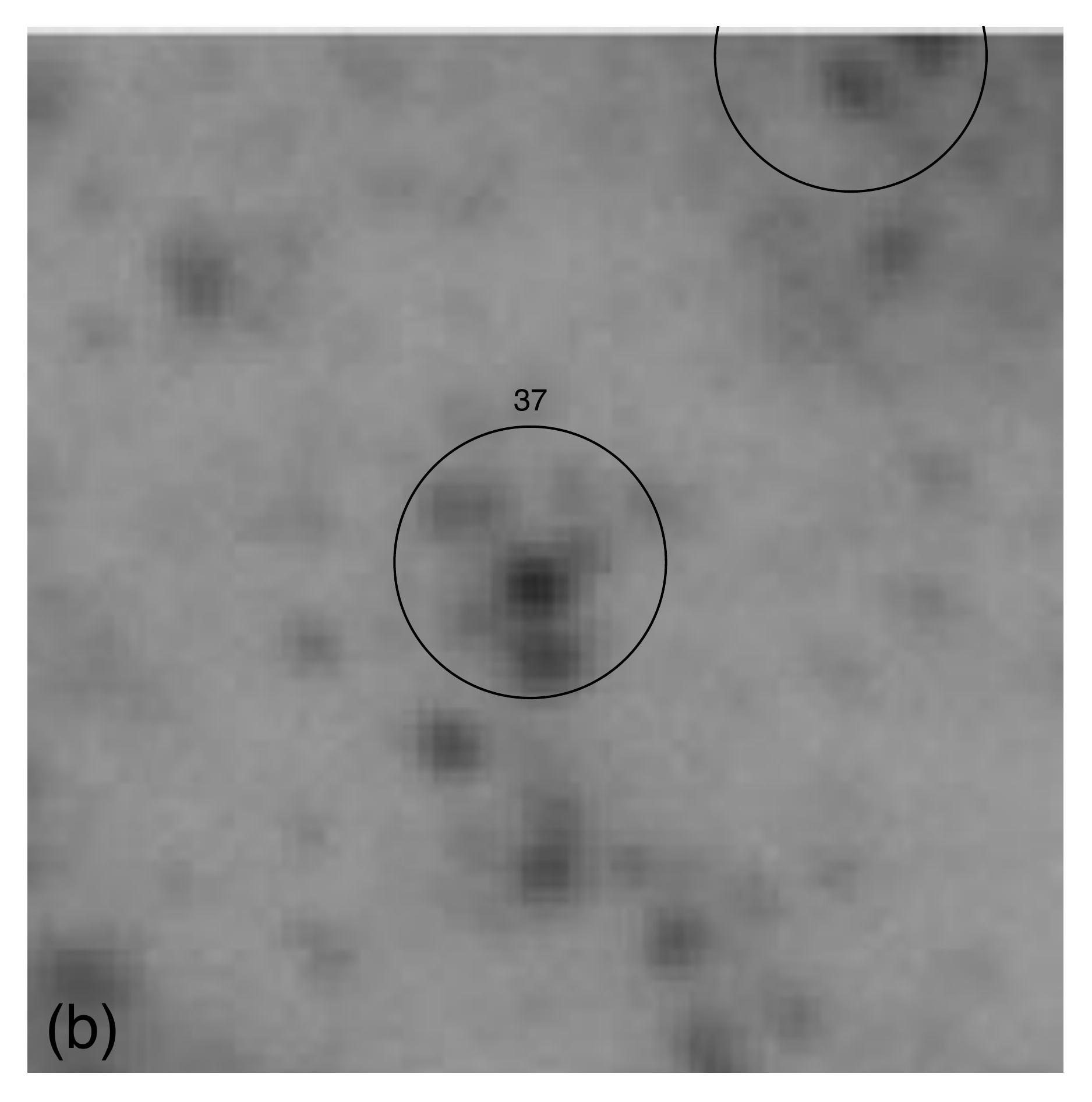}
\end{minipage}
\begin{minipage}{0.24\linewidth}
\includegraphics[width=\linewidth]{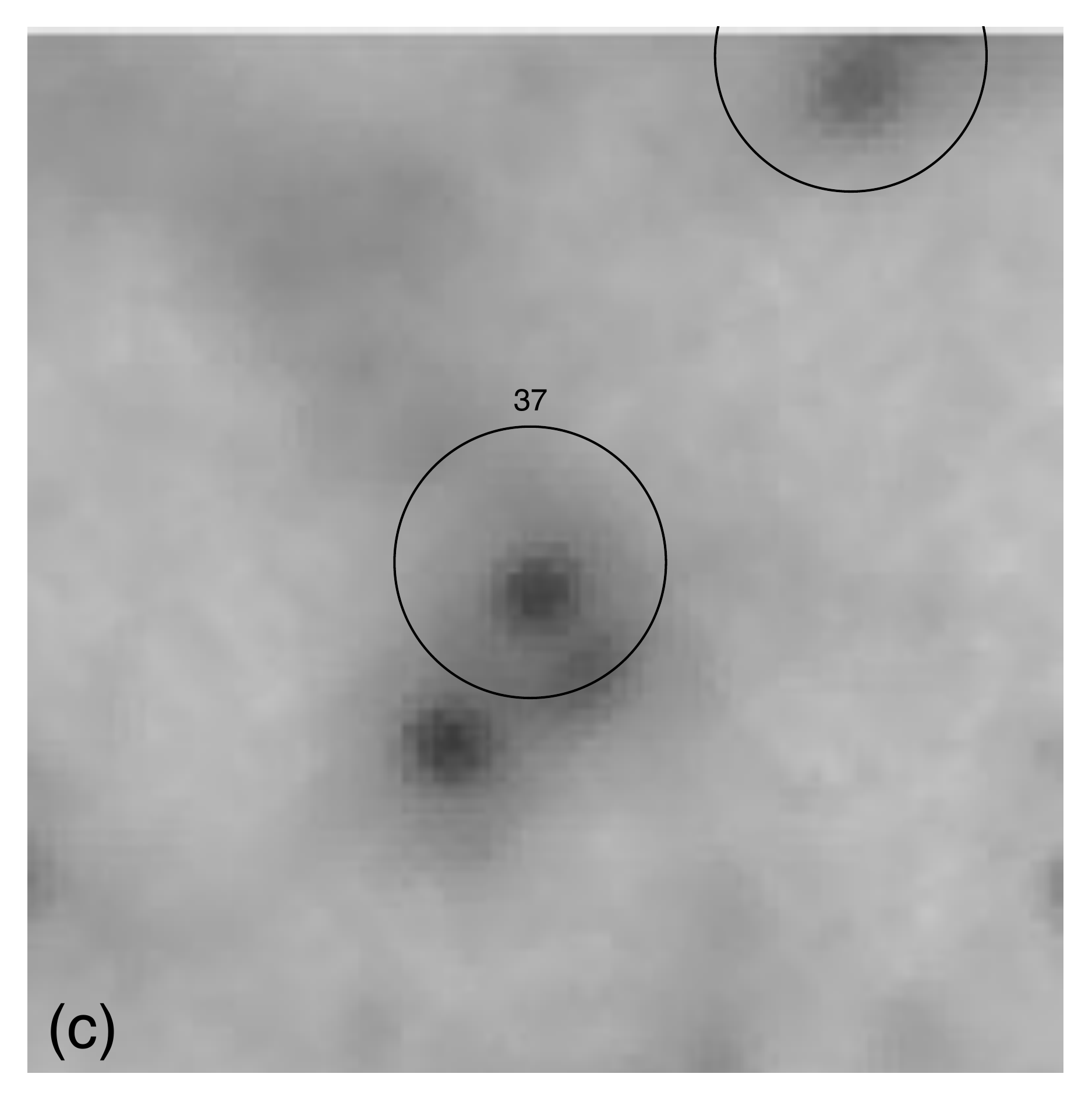}
\end{minipage}
\begin{minipage}{0.24\linewidth}
\includegraphics[width=\linewidth]{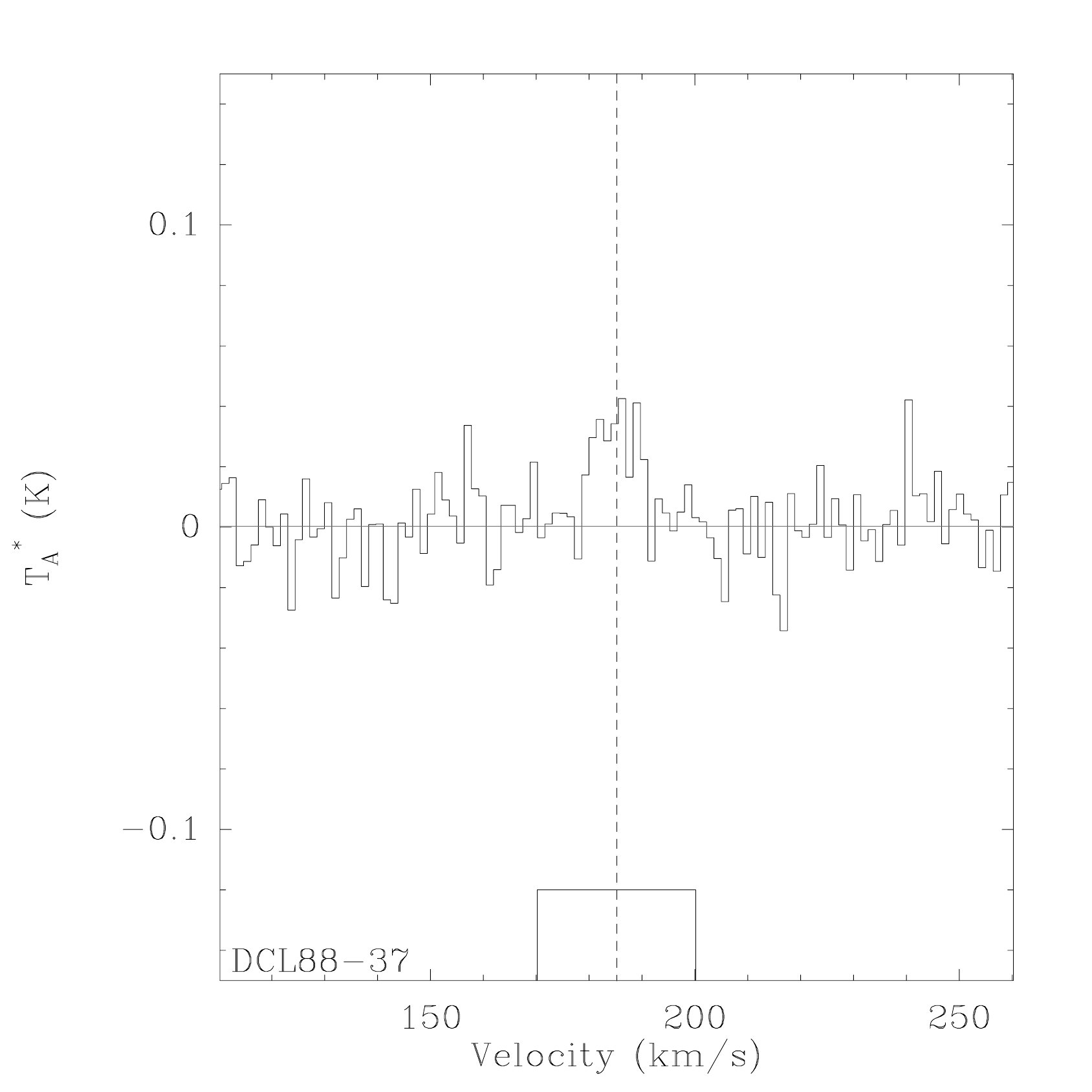}
\end{minipage}

\begin{minipage}{0.24\linewidth}
\includegraphics[width=\linewidth]{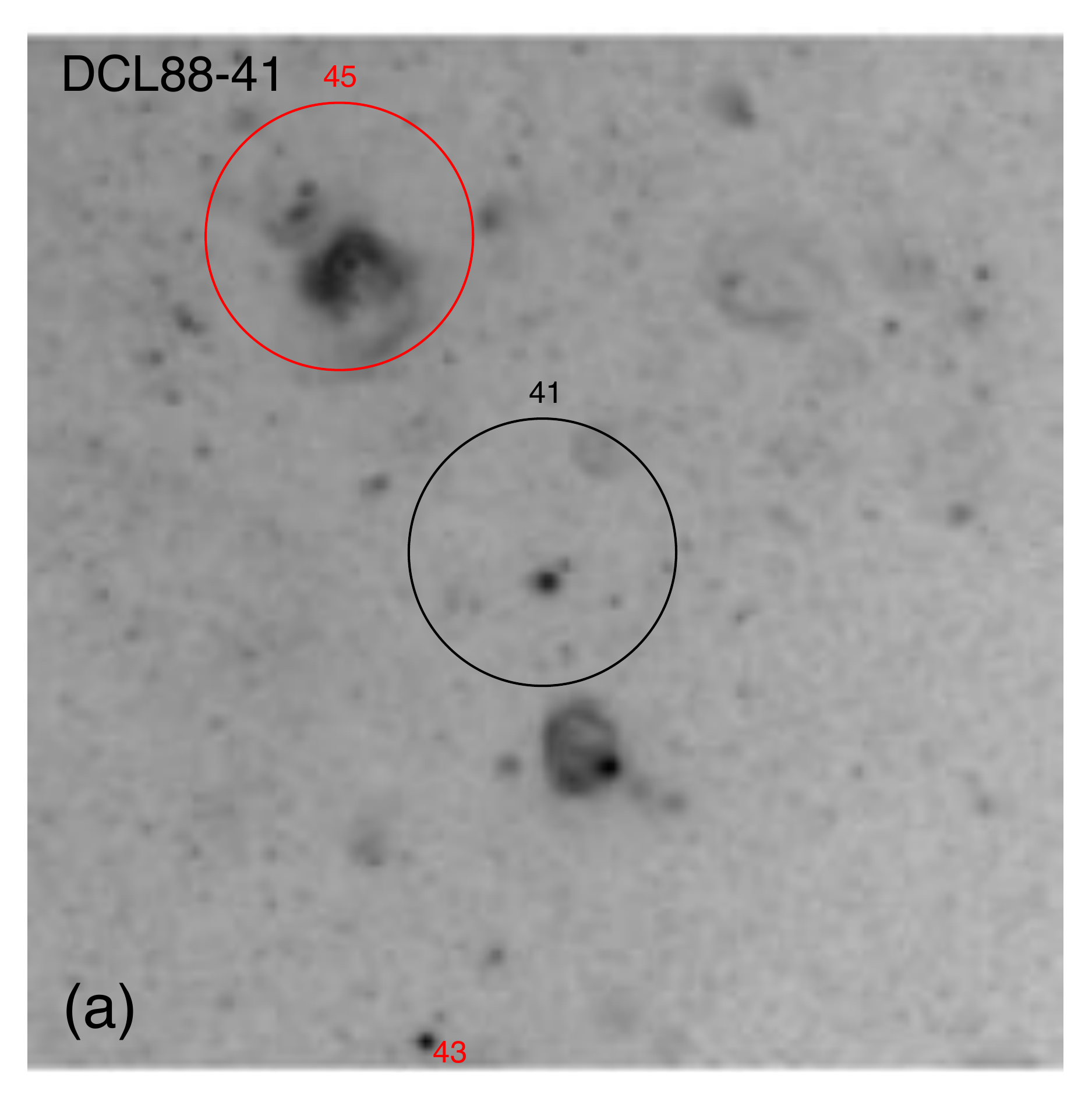}
\end{minipage}
\begin{minipage}{0.24\linewidth}
\includegraphics[width=\linewidth]{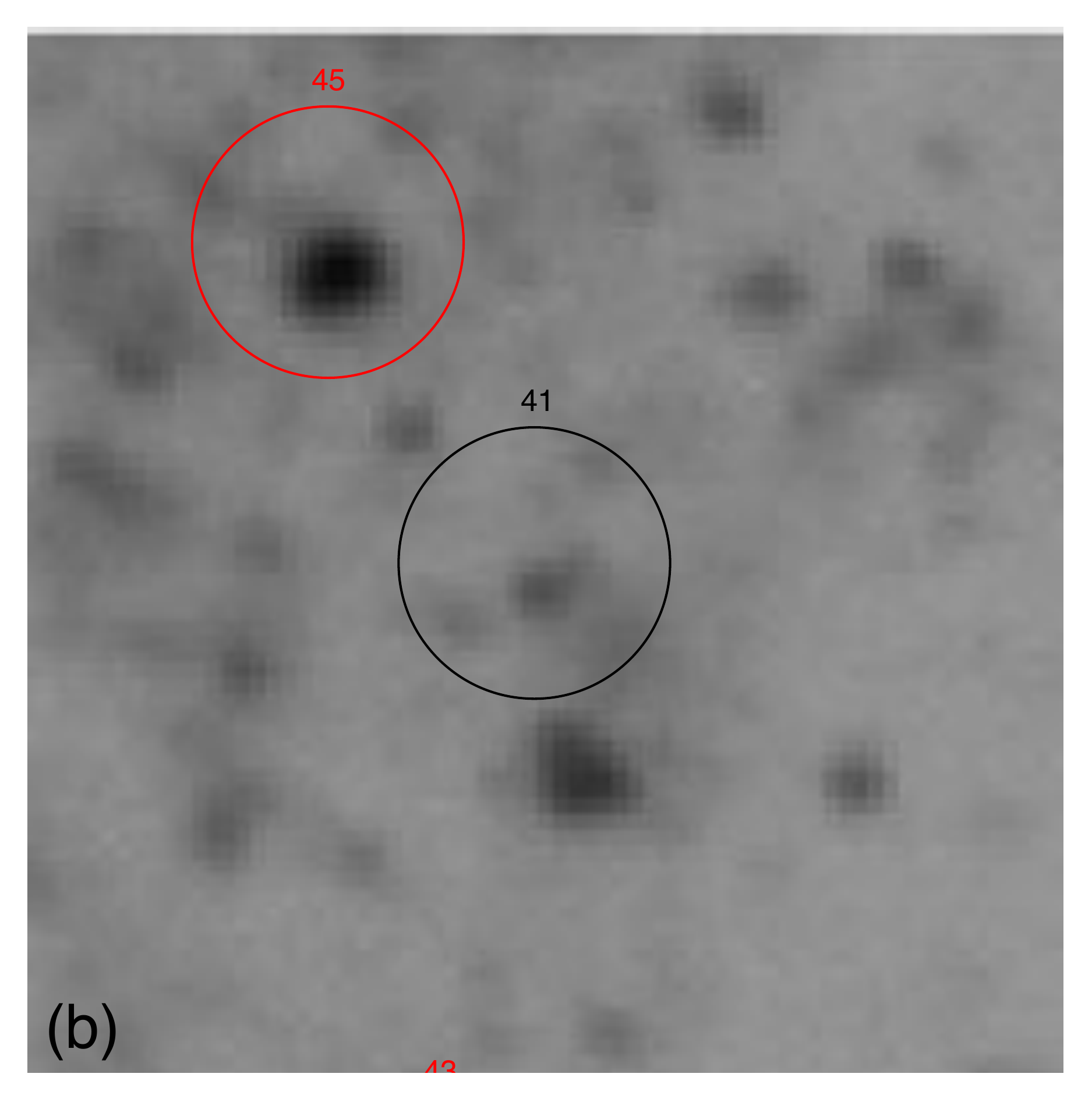}
\end{minipage}
\begin{minipage}{0.24\linewidth}
\includegraphics[width=\linewidth]{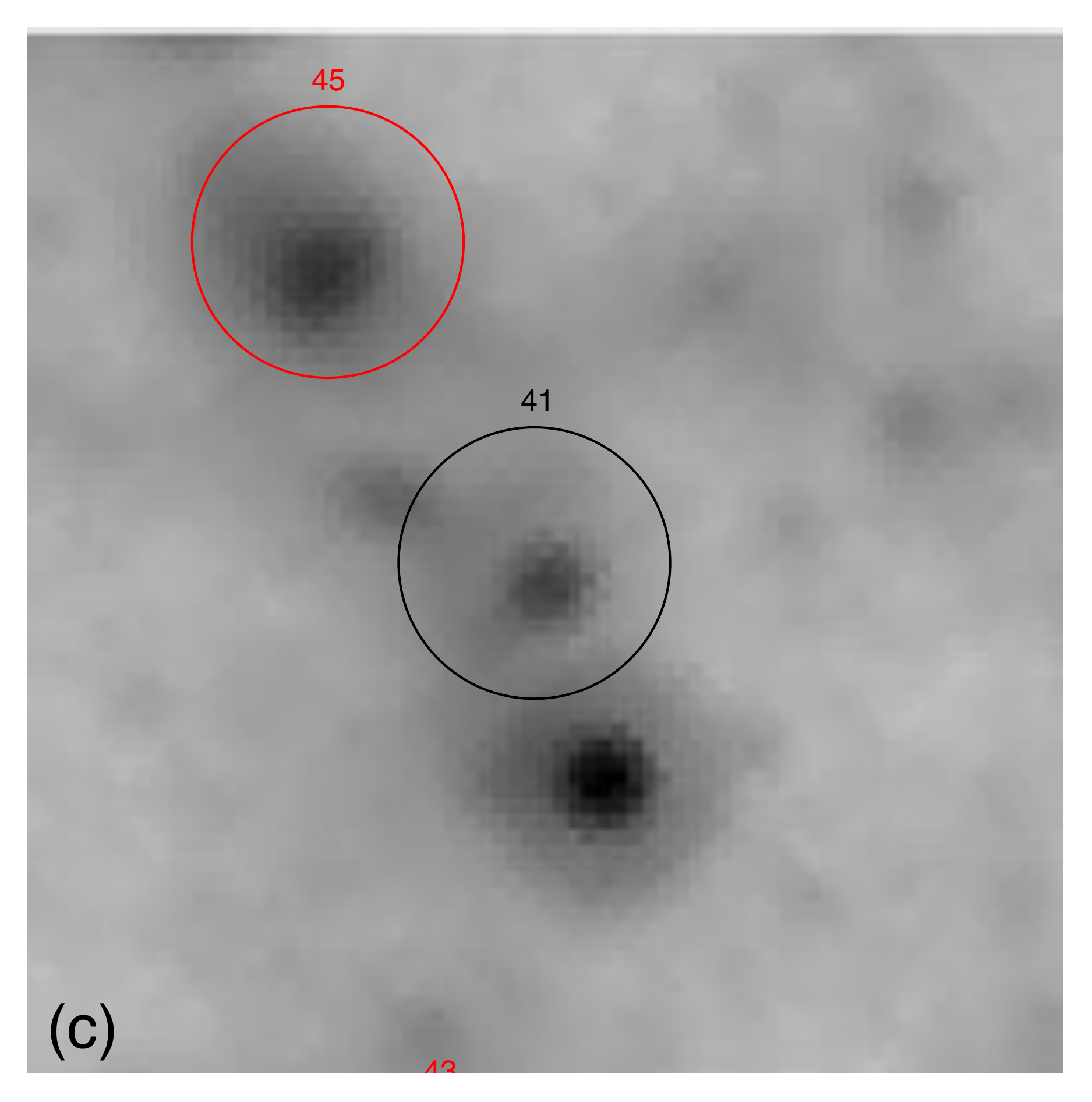}
\end{minipage}
\begin{minipage}{0.24\linewidth}
\includegraphics[width=\linewidth]{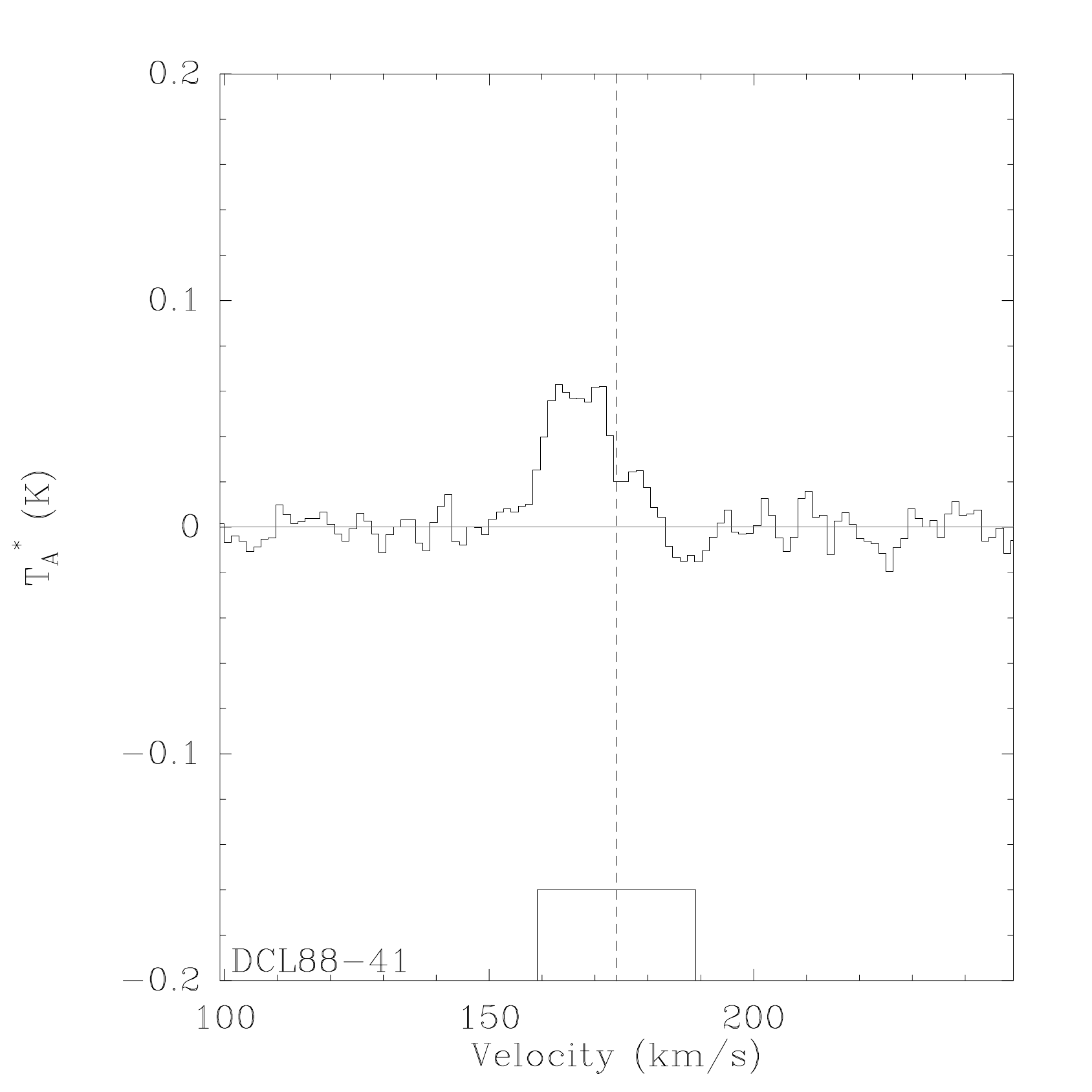}
\end{minipage}

\noindent\textbf{Figure~\ref{fig:stamps} -- continued.}
\label{fig:stamps2}
\end{figure*}

\begin{figure*}

\begin{minipage}{0.24\linewidth}
\includegraphics[width=\linewidth]{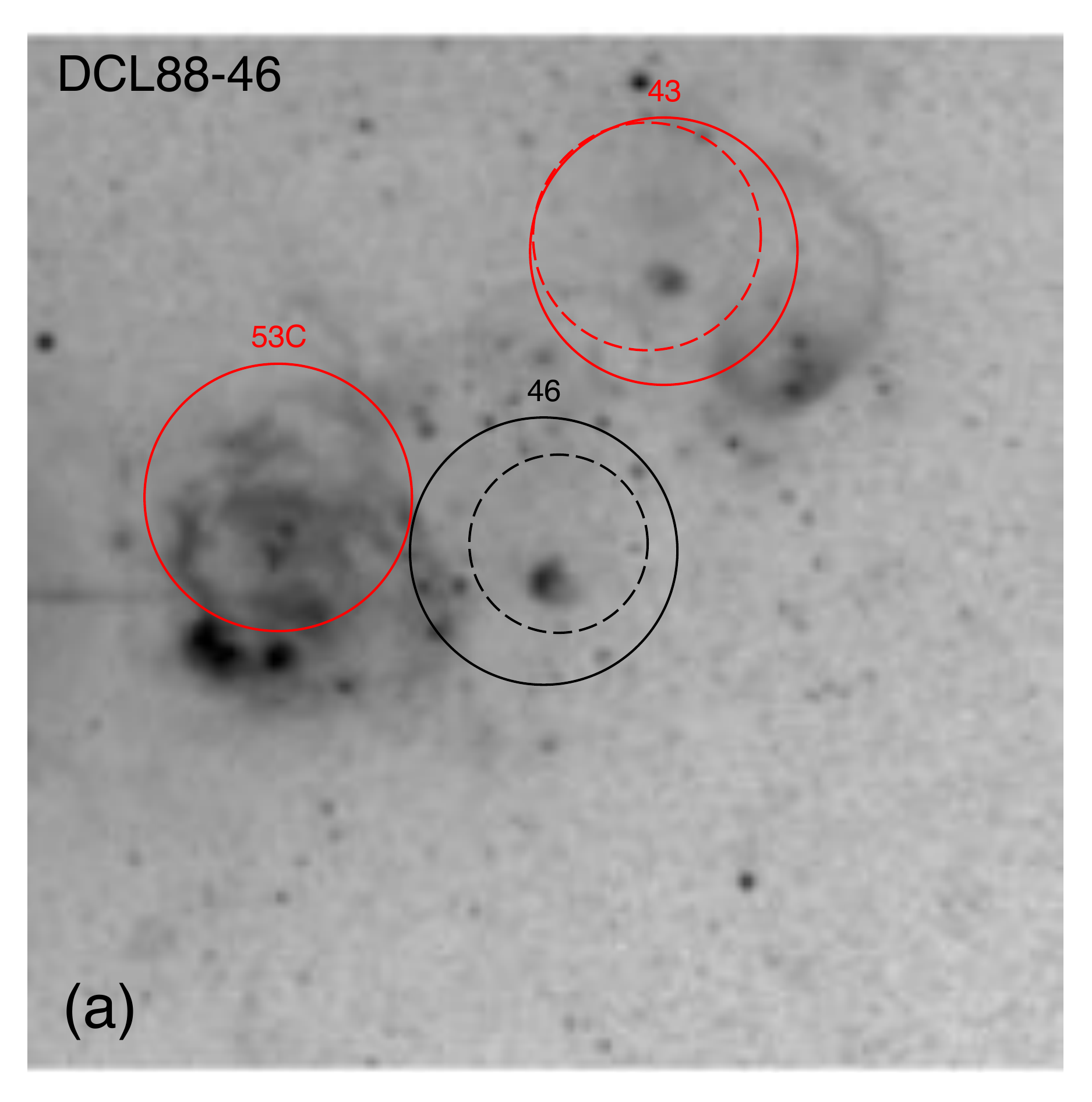}
\end{minipage}
\begin{minipage}{0.24\linewidth}
\includegraphics[width=\linewidth]{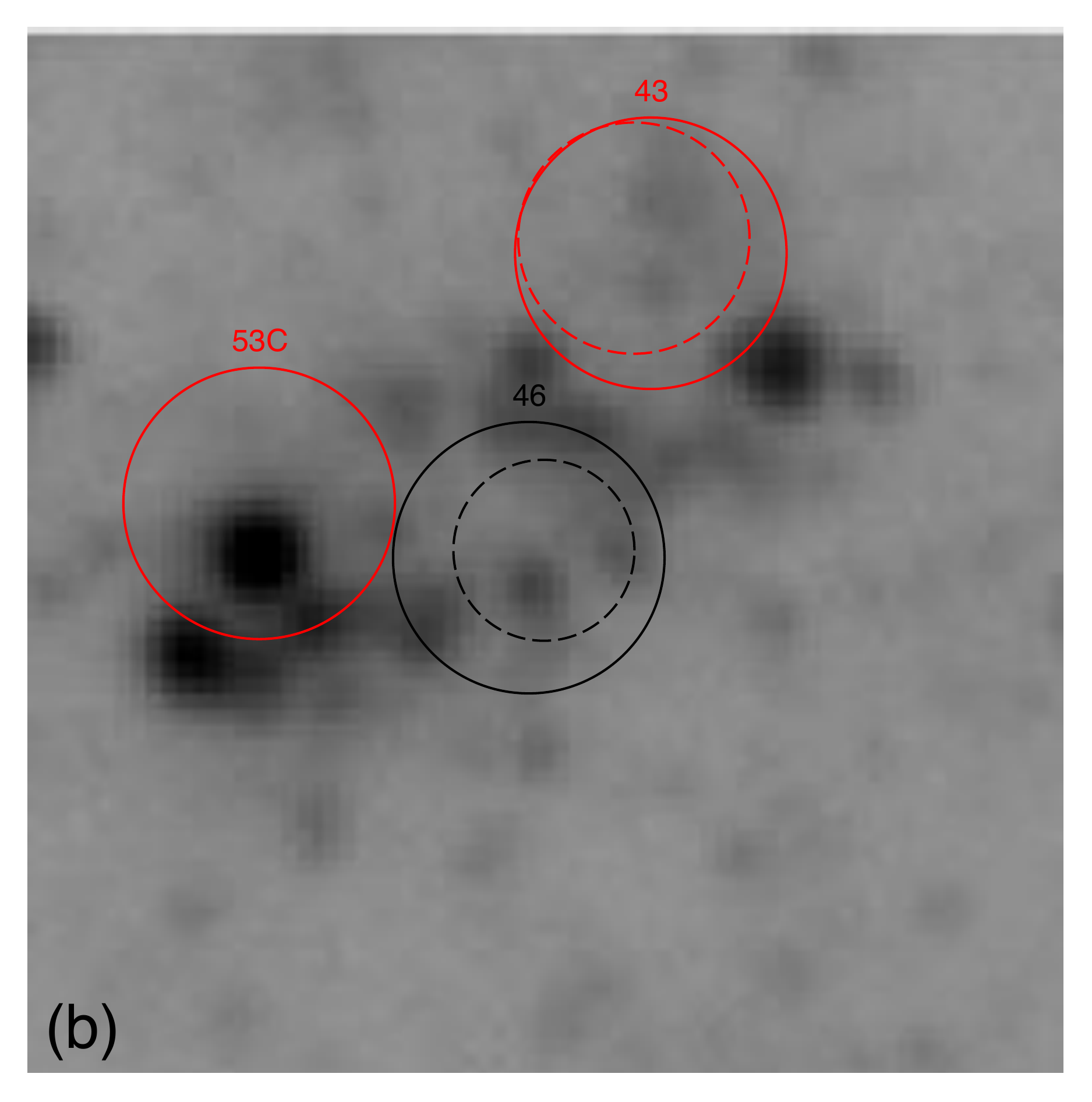}
\end{minipage}
\begin{minipage}{0.24\linewidth}
\includegraphics[width=\linewidth]{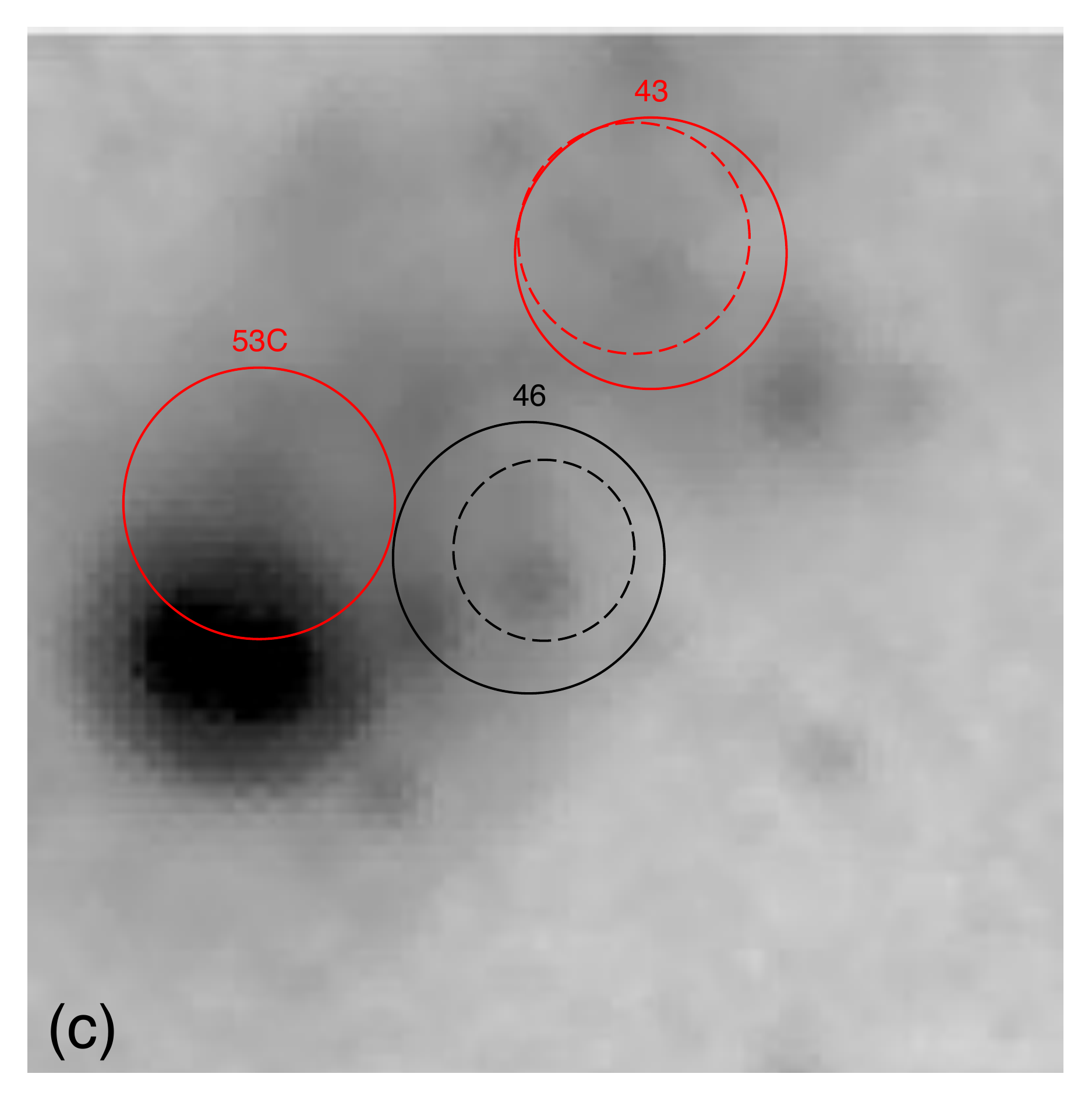}
\end{minipage}
\begin{minipage}{0.24\linewidth}
\includegraphics[width=\linewidth]{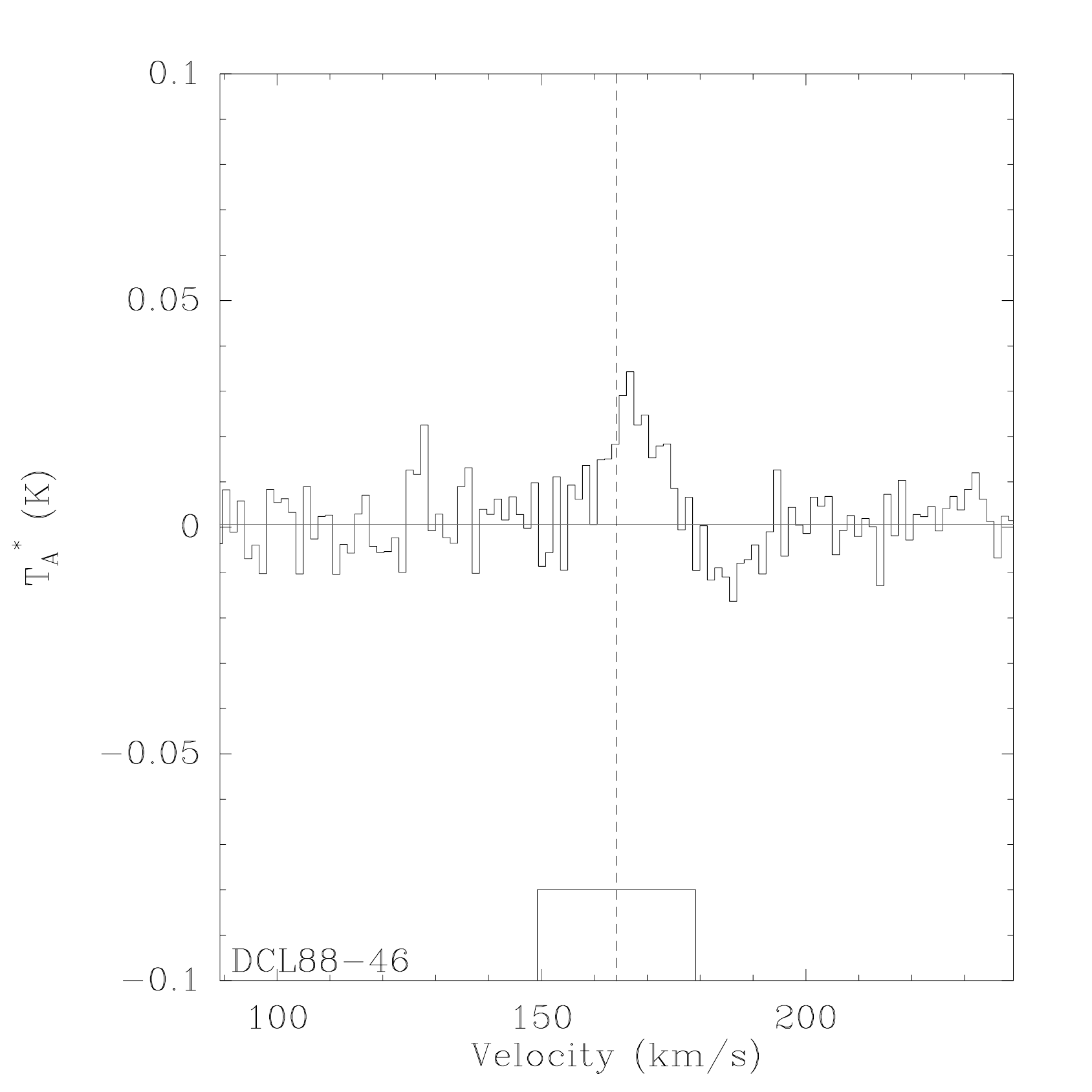}
\end{minipage}

\begin{minipage}{0.24\linewidth}
\includegraphics[width=\linewidth]{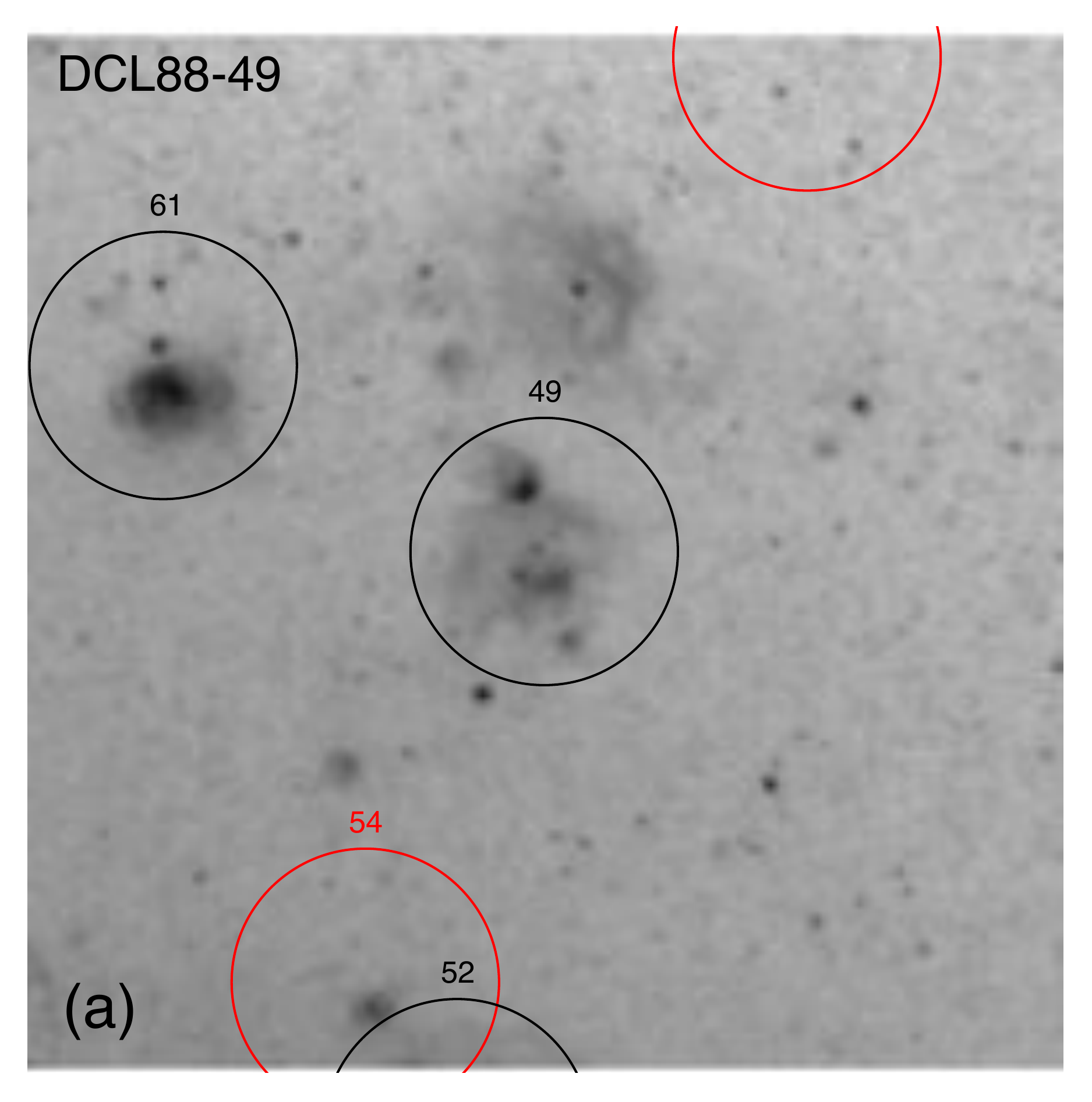}
\end{minipage}
\begin{minipage}{0.24\linewidth}
\includegraphics[width=\linewidth]{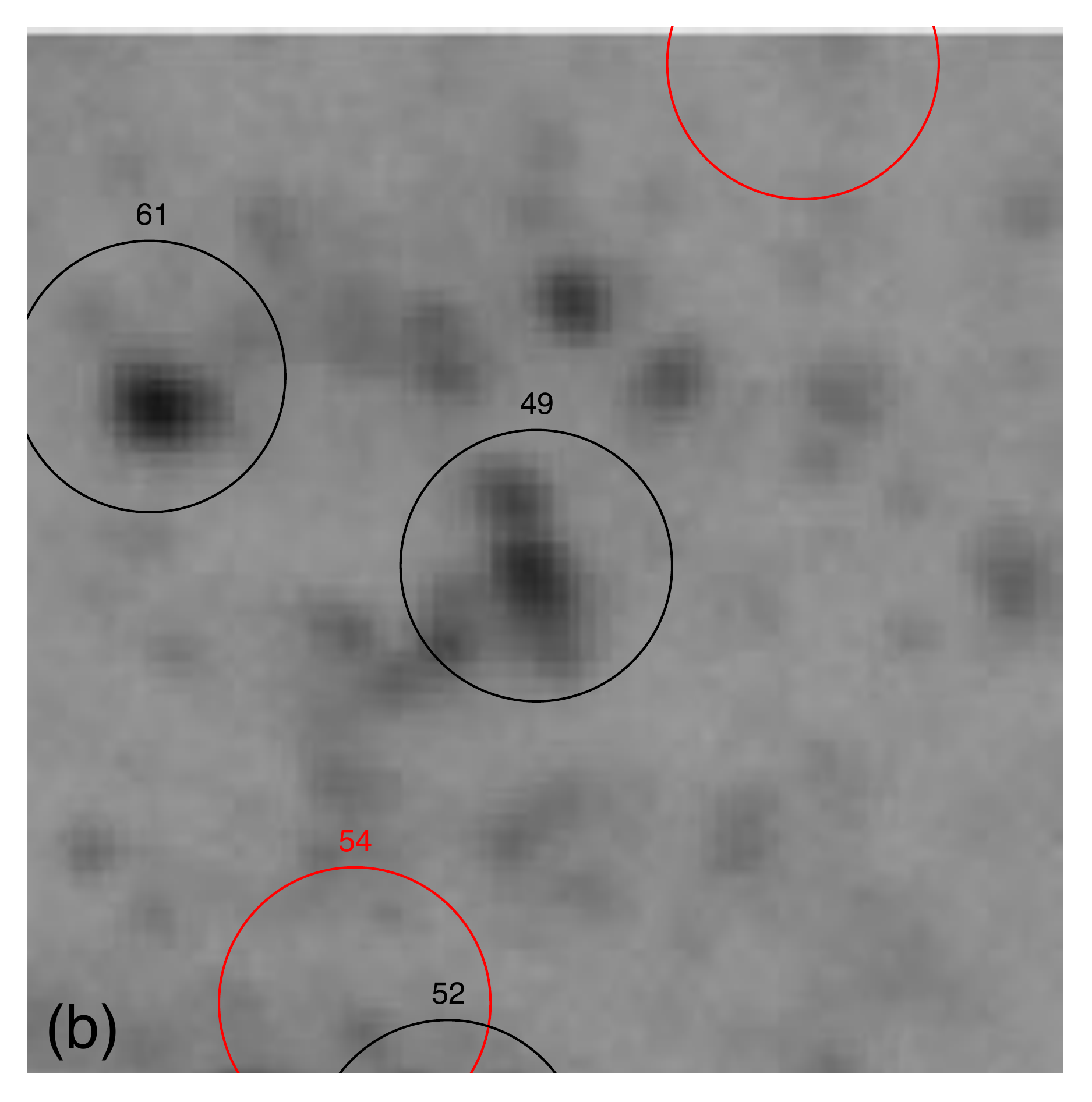}
\end{minipage}
\begin{minipage}{0.24\linewidth}
\includegraphics[width=\linewidth]{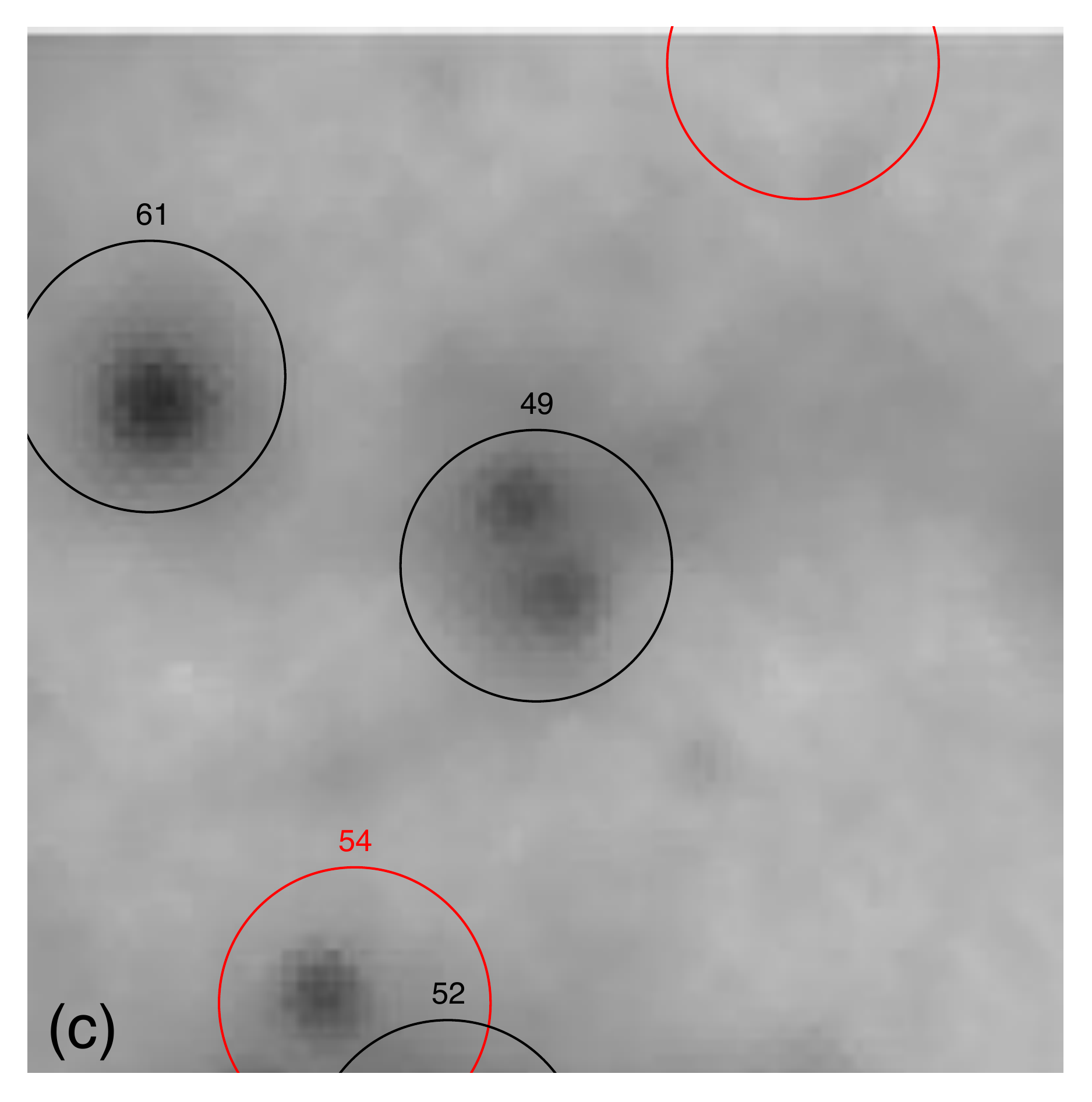}
\end{minipage}
\begin{minipage}{0.24\linewidth}
\includegraphics[width=\linewidth]{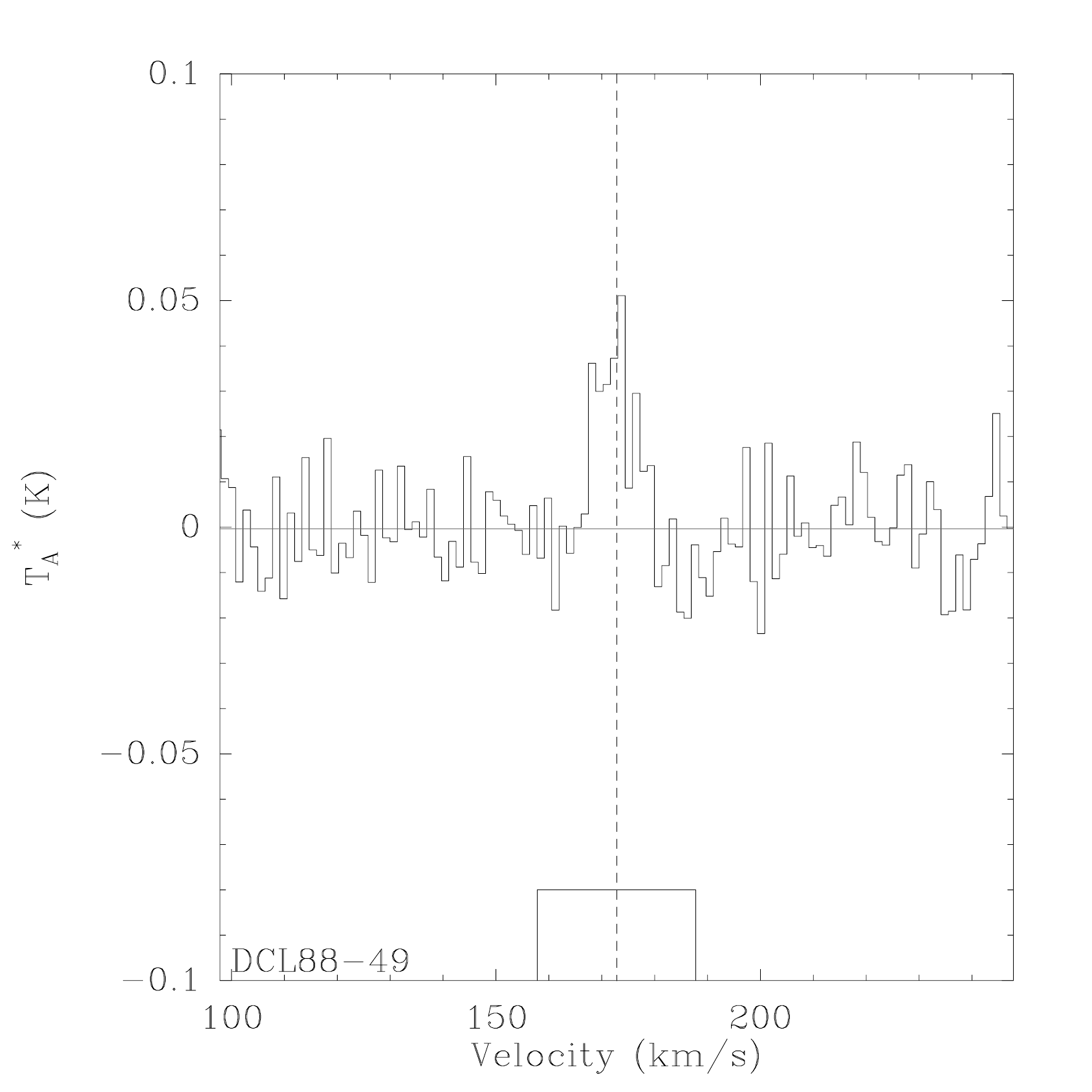}
\end{minipage}

\begin{minipage}{0.24\linewidth}
\includegraphics[width=\linewidth]{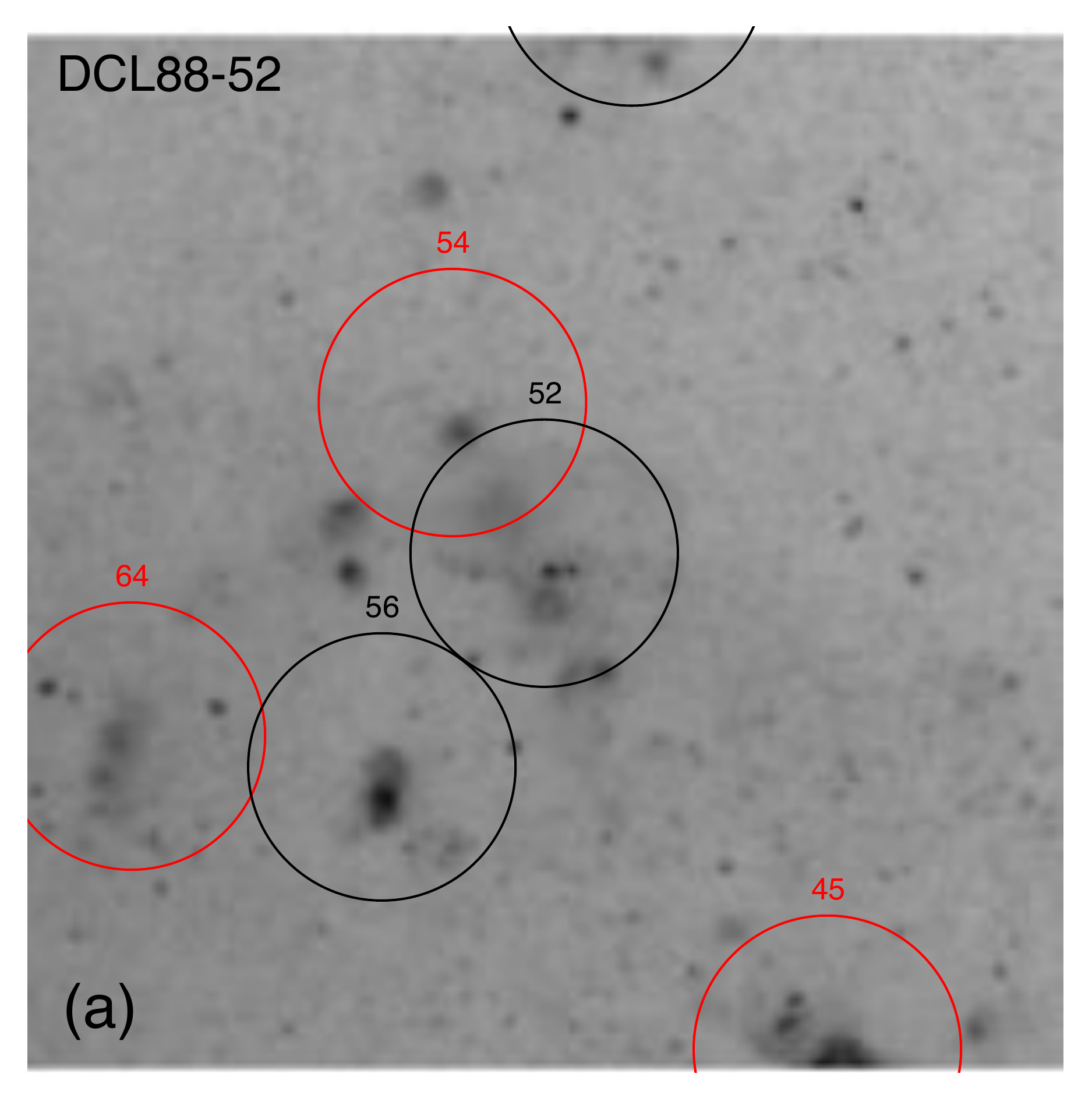}
\end{minipage}
\begin{minipage}{0.24\linewidth}
\includegraphics[width=\linewidth]{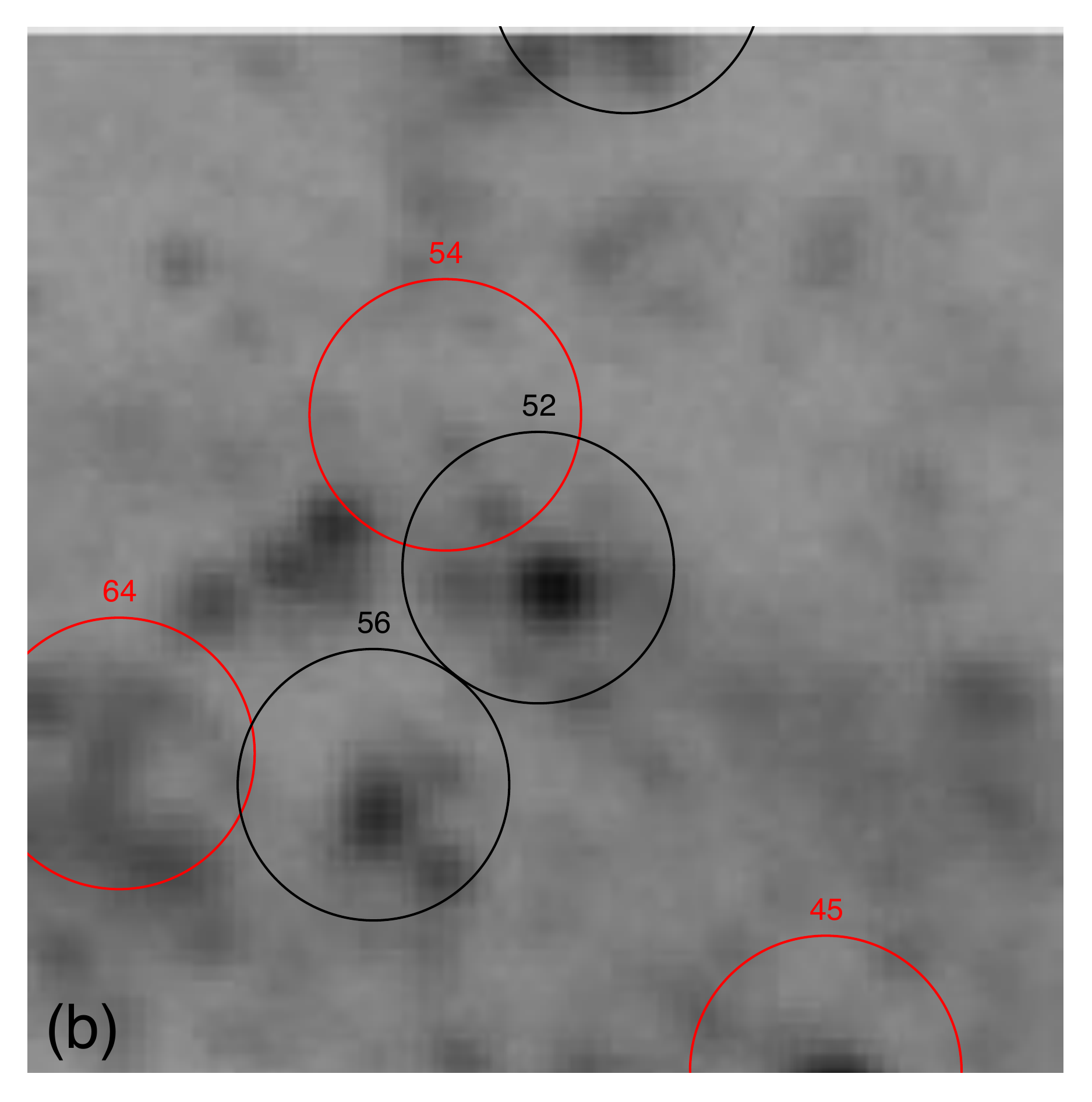}
\end{minipage}
\begin{minipage}{0.24\linewidth}
\includegraphics[width=\linewidth]{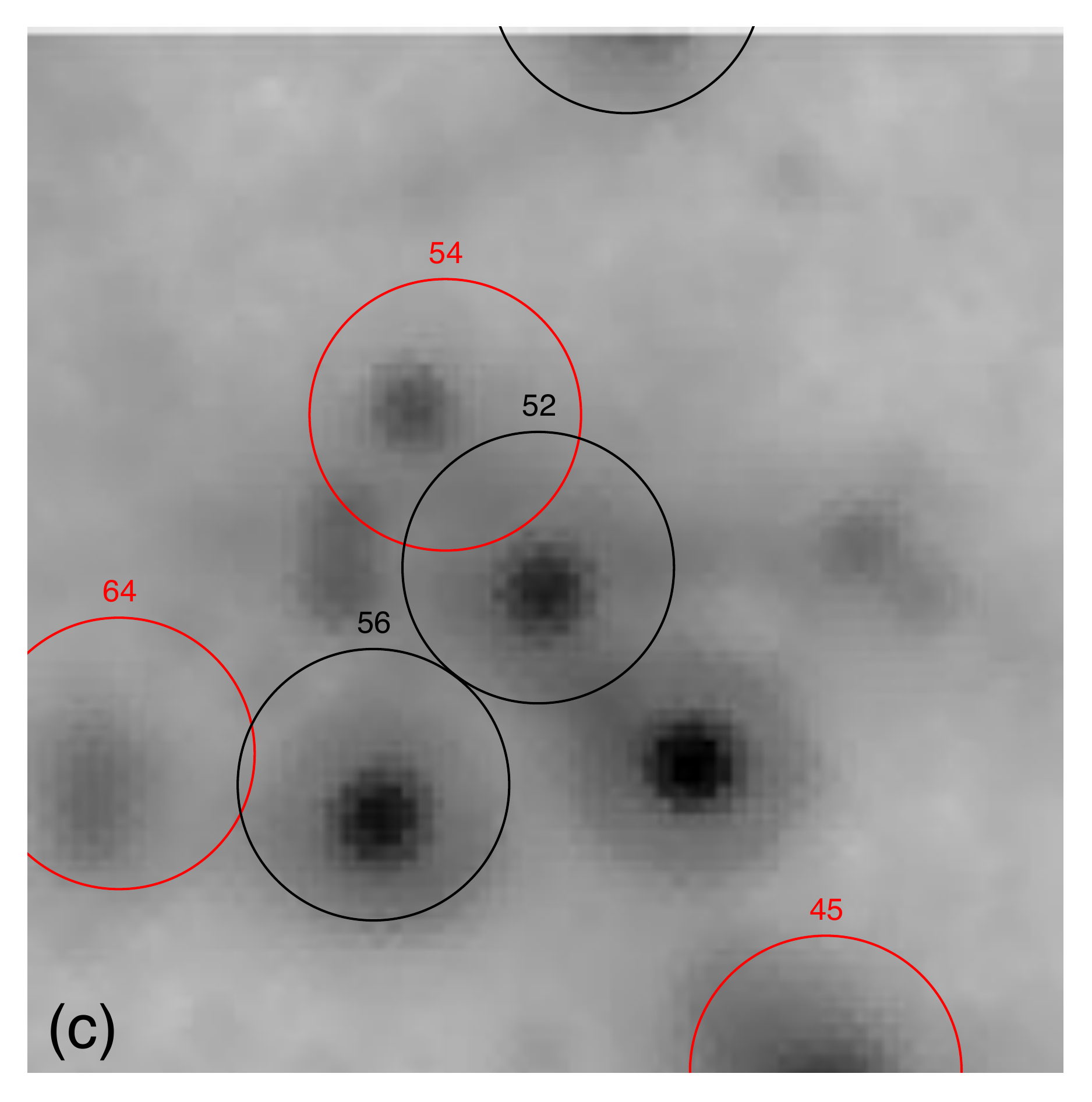}
\end{minipage}
\begin{minipage}{0.24\linewidth}
\includegraphics[width=\linewidth]{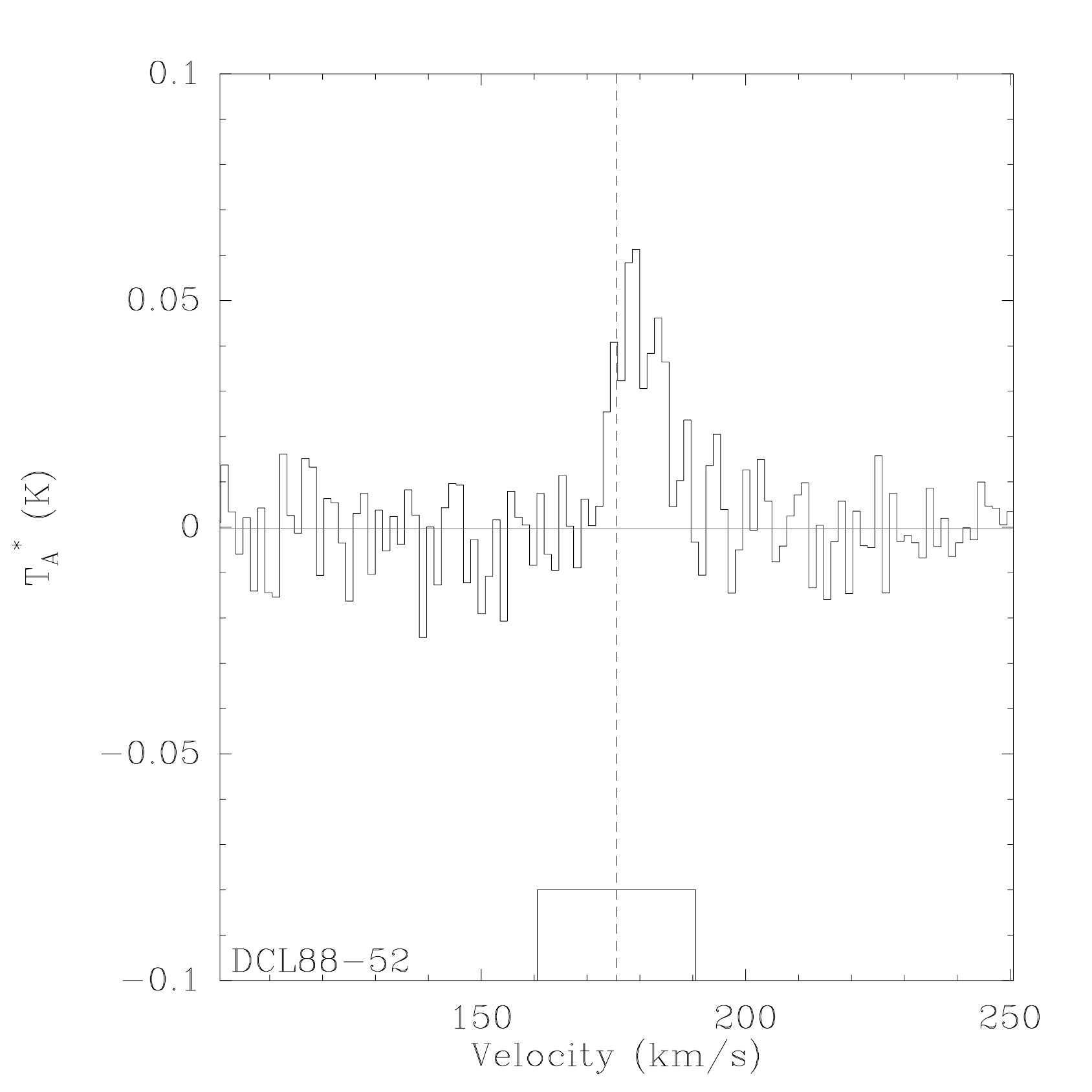}
\end{minipage}

\begin{minipage}{0.24\linewidth}
\includegraphics[width=\linewidth]{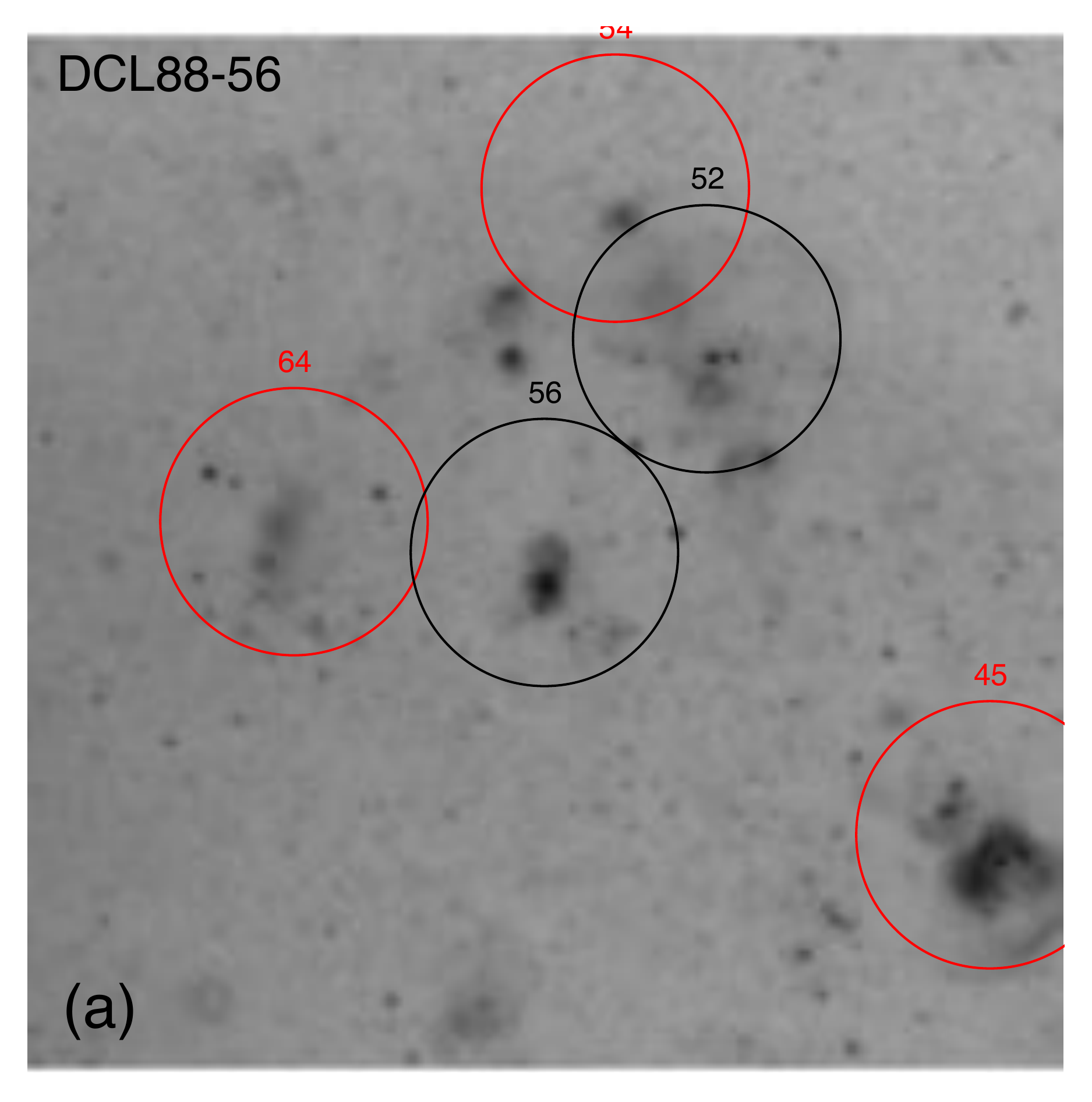}
\end{minipage}
\begin{minipage}{0.24\linewidth}
\includegraphics[width=\linewidth]{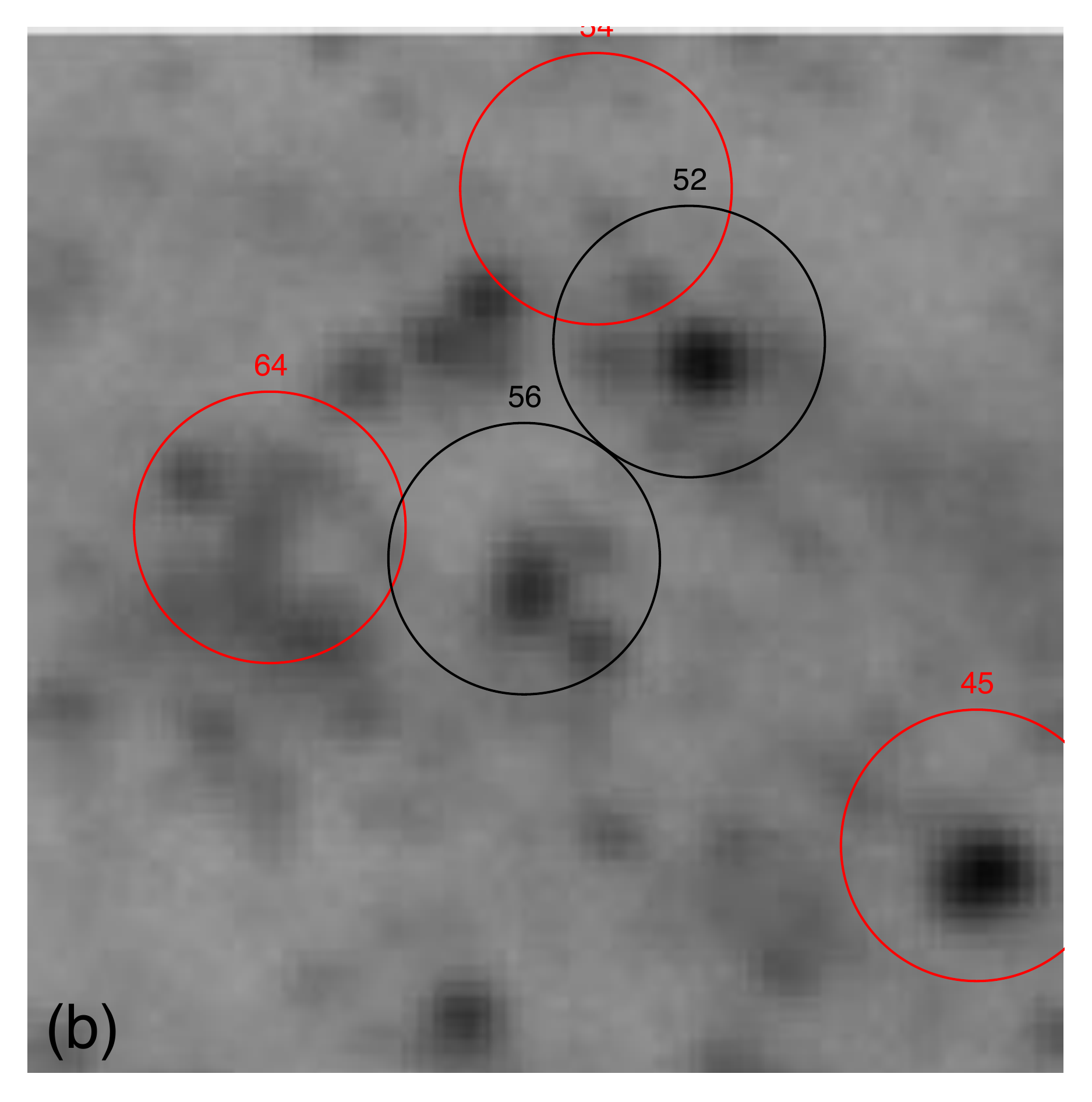}
\end{minipage}
\begin{minipage}{0.24\linewidth}
\includegraphics[width=\linewidth]{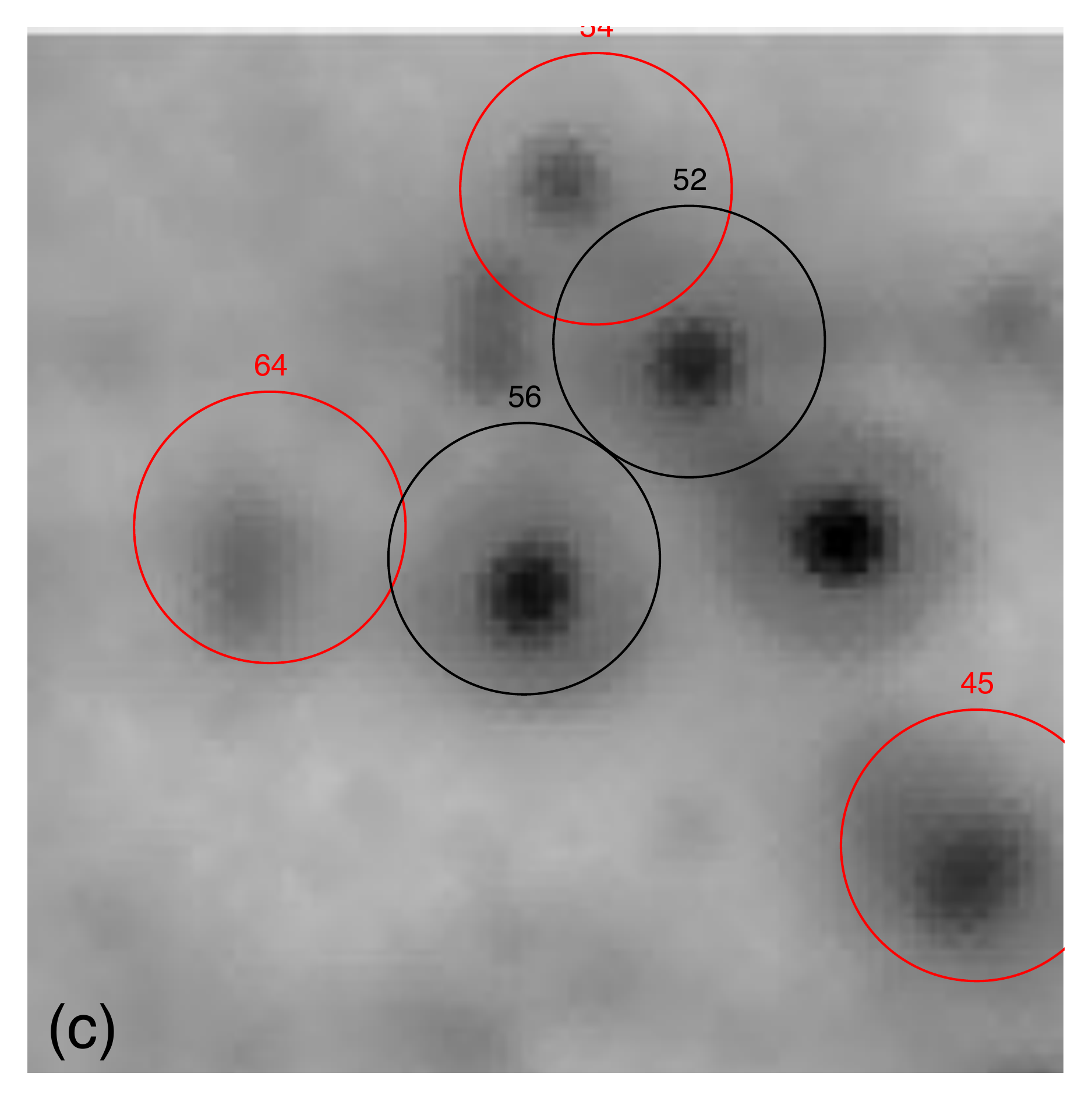}
\end{minipage}
\begin{minipage}{0.24\linewidth}
\includegraphics[width=\linewidth]{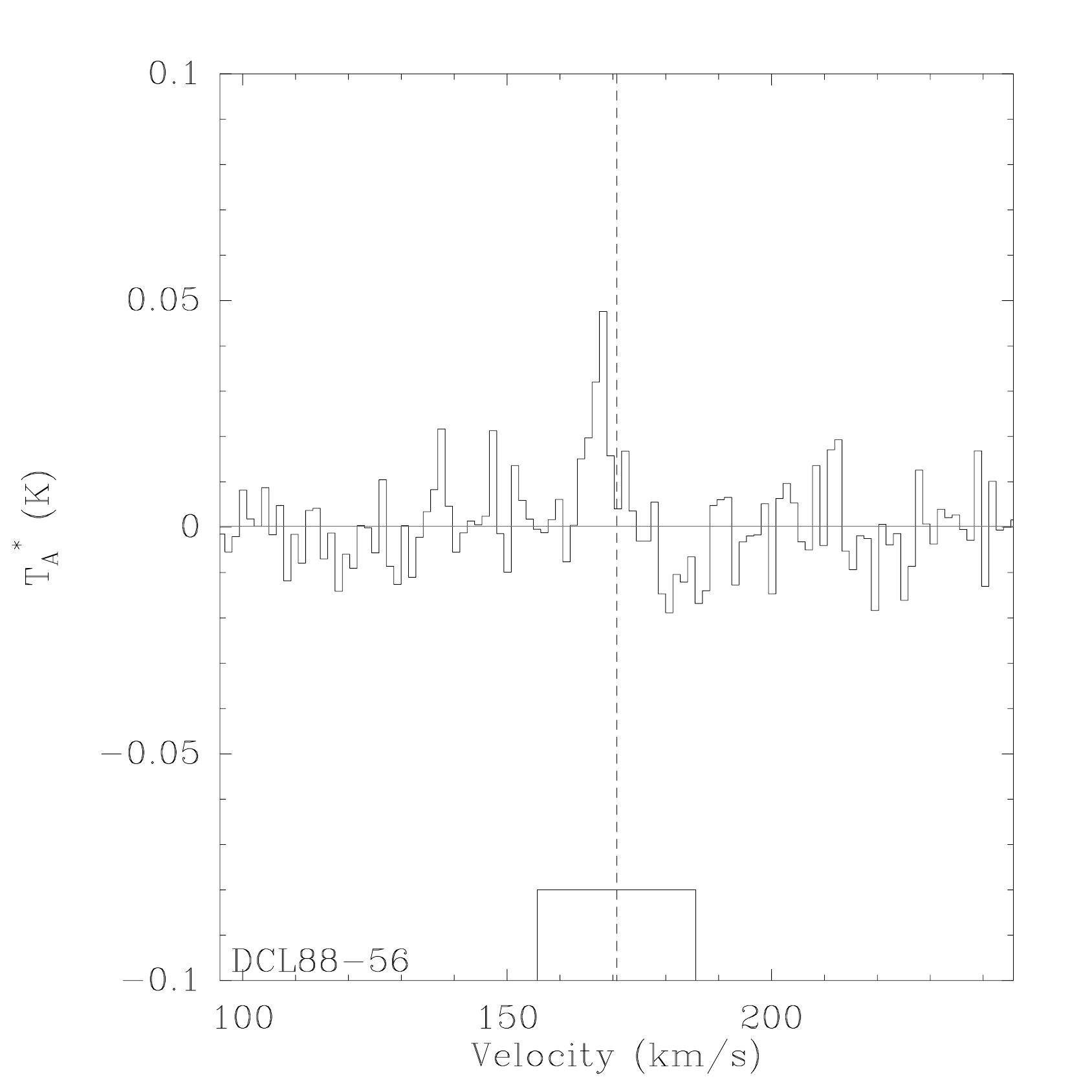}
\end{minipage}

\begin{minipage}{0.24\linewidth}
\includegraphics[width=\linewidth]{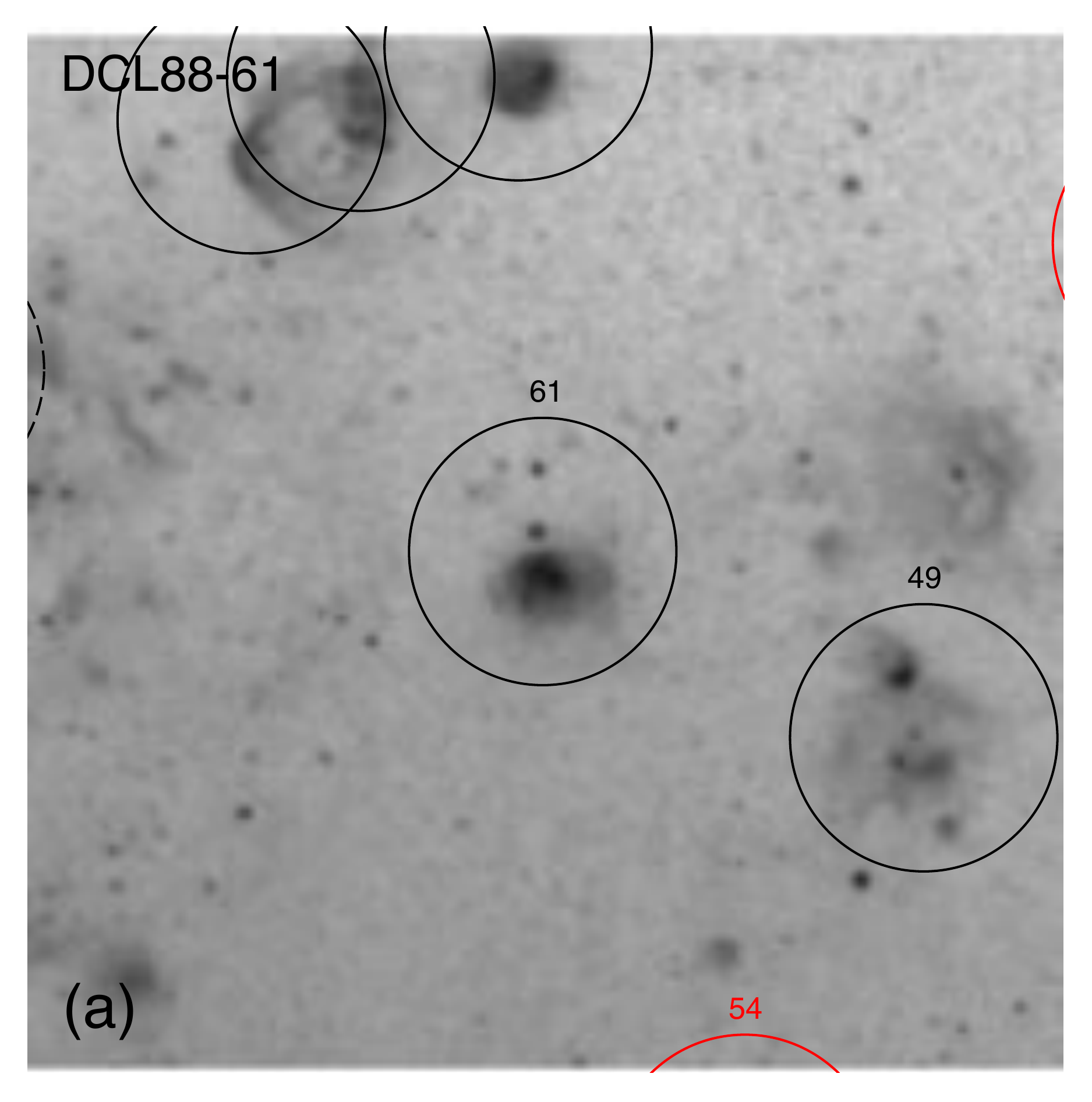}
\end{minipage}
\begin{minipage}{0.24\linewidth}
\includegraphics[width=\linewidth]{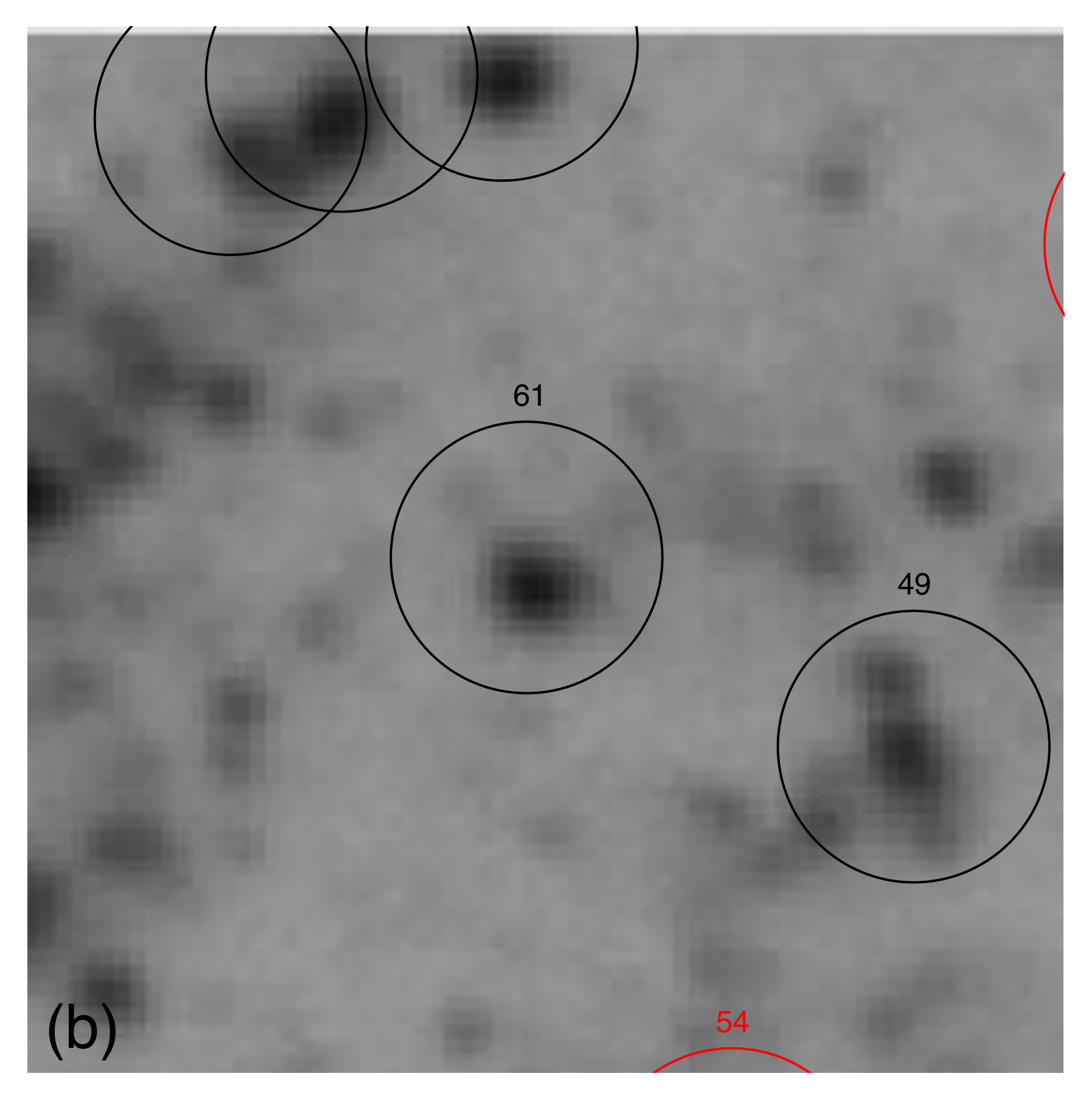}
\end{minipage}
\begin{minipage}{0.24\linewidth}
\includegraphics[width=\linewidth]{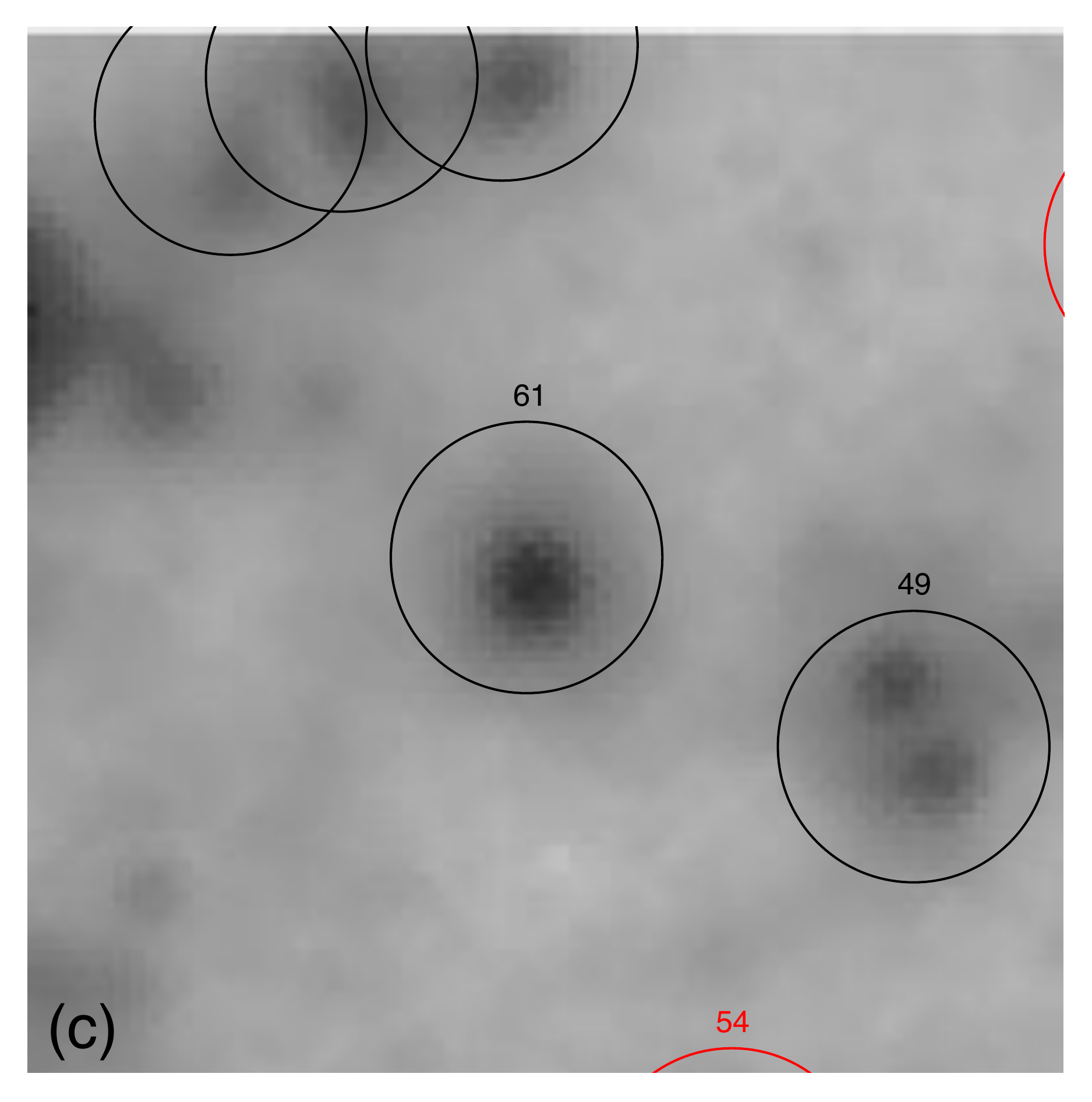}
\end{minipage}
\begin{minipage}{0.24\linewidth}
\includegraphics[width=\linewidth]{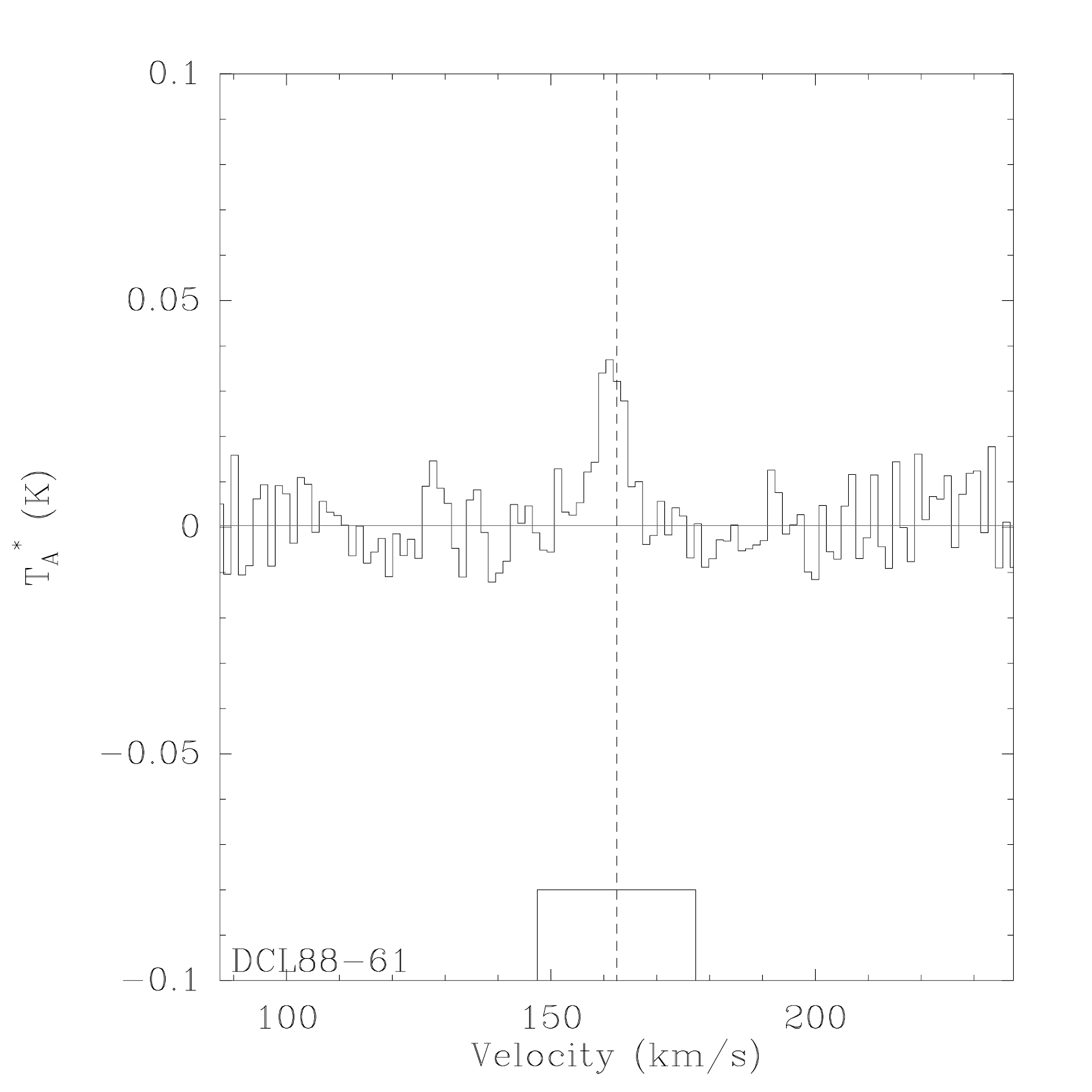}
\end{minipage}

\noindent\textbf{Figure~\ref{fig:stamps} -- continued.}
\end{figure*}

\begin{figure*}
%\ContinuedFloat

\begin{minipage}{0.24\linewidth}
\includegraphics[width=\linewidth]{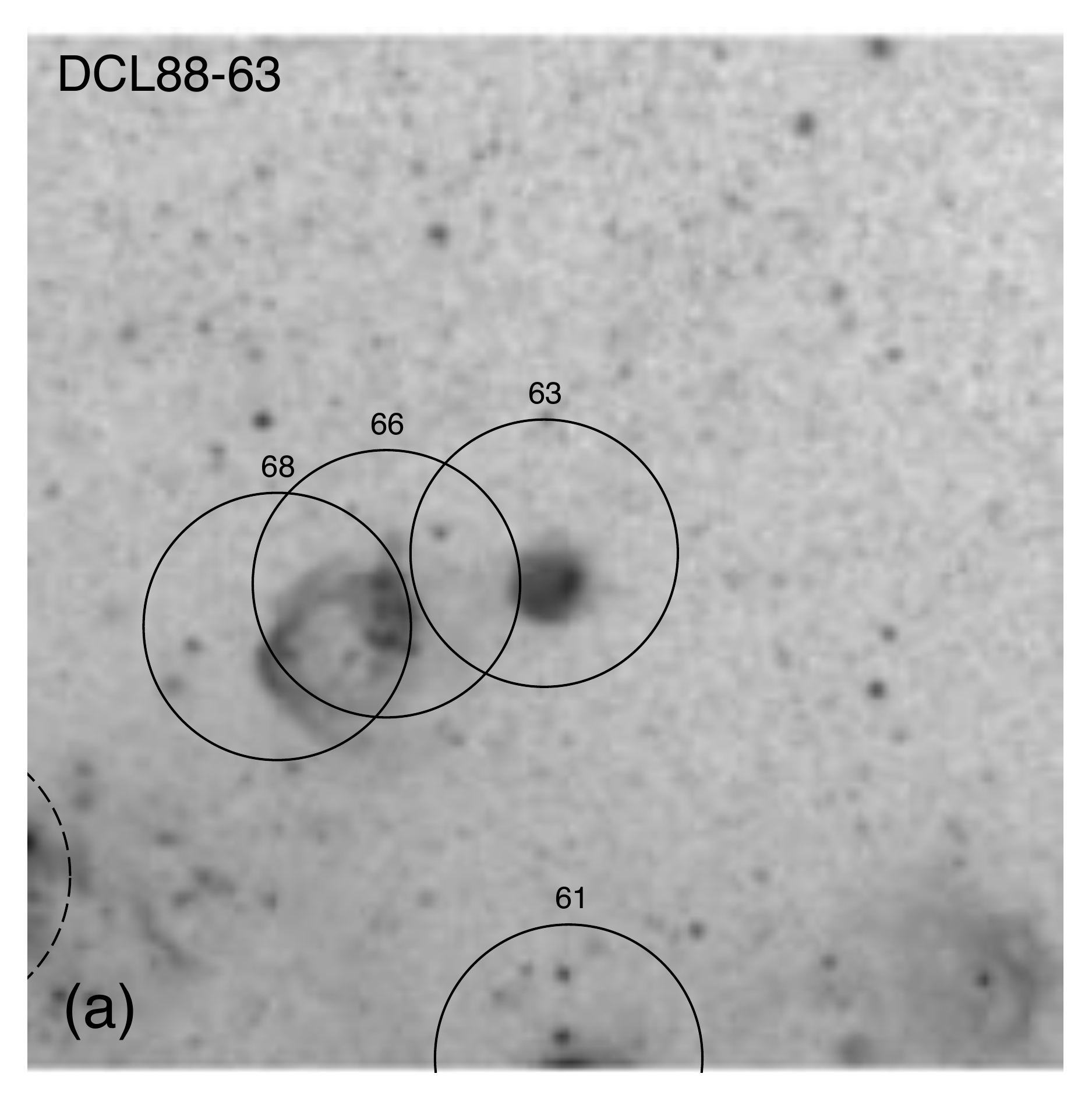}
\end{minipage}
\begin{minipage}{0.24\linewidth}
\includegraphics[width=\linewidth]{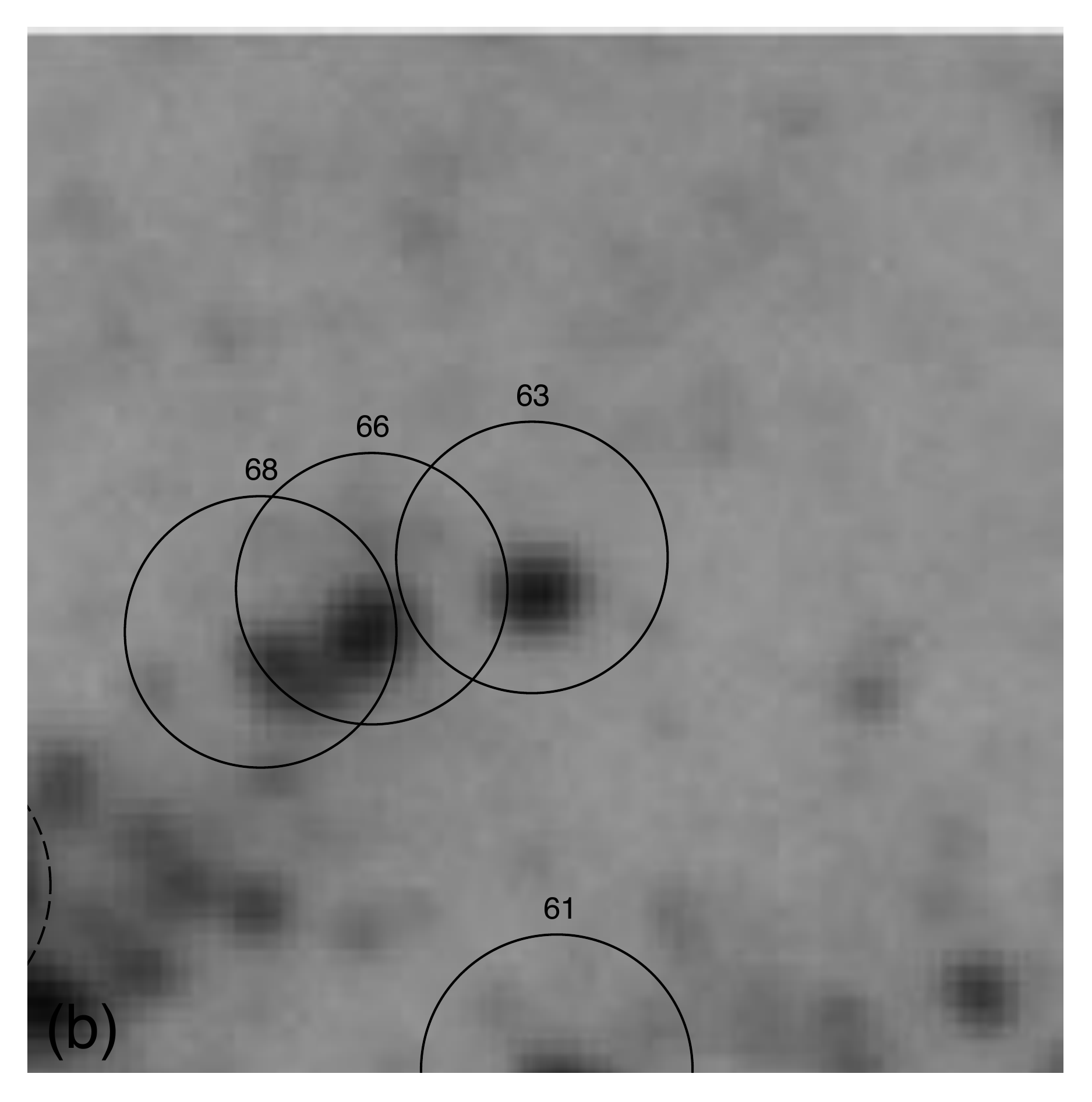}
\end{minipage}
\begin{minipage}{0.24\linewidth}
\includegraphics[width=\linewidth]{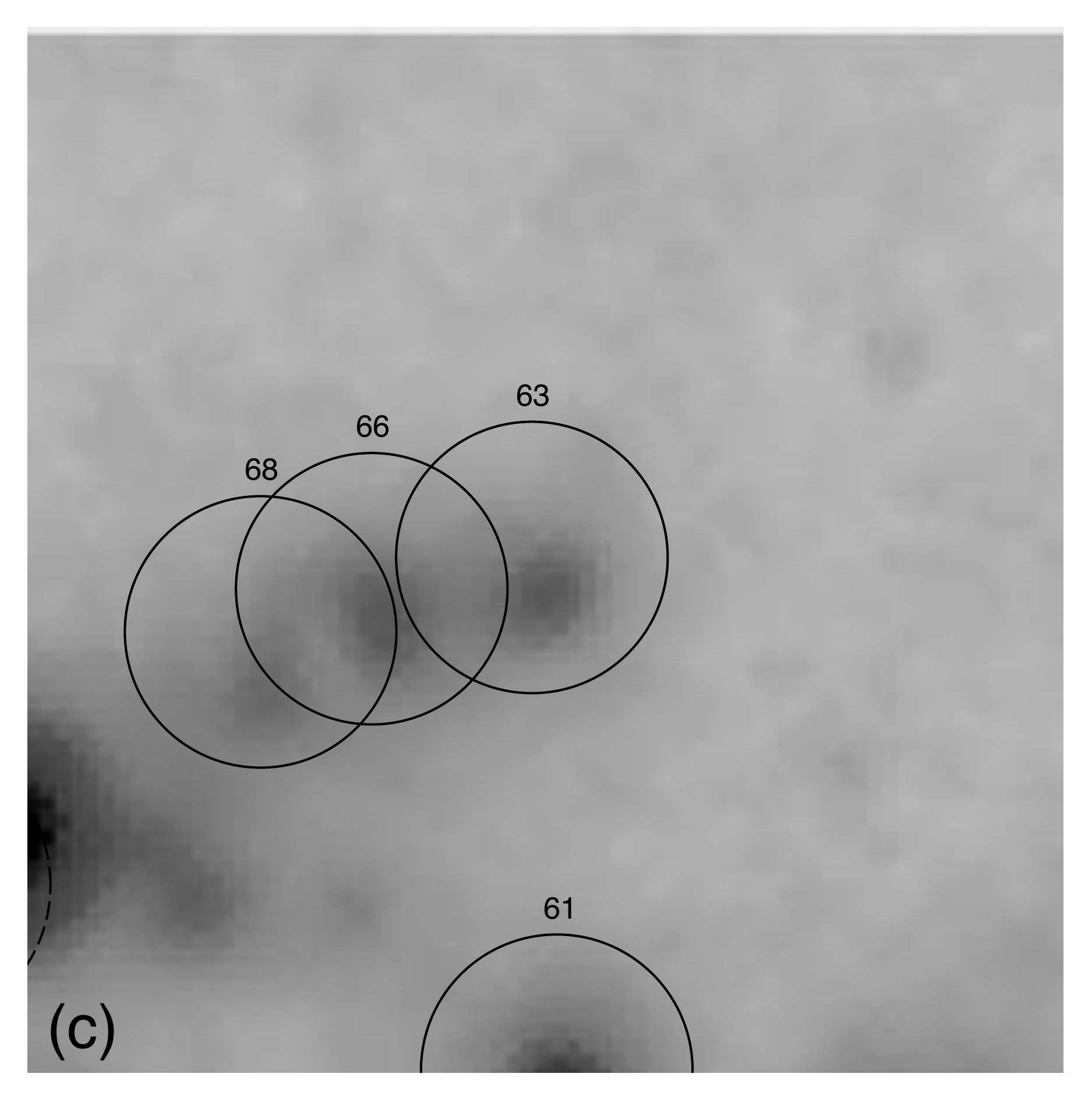}
\end{minipage}
\begin{minipage}{0.24\linewidth}
\includegraphics[width=\linewidth]{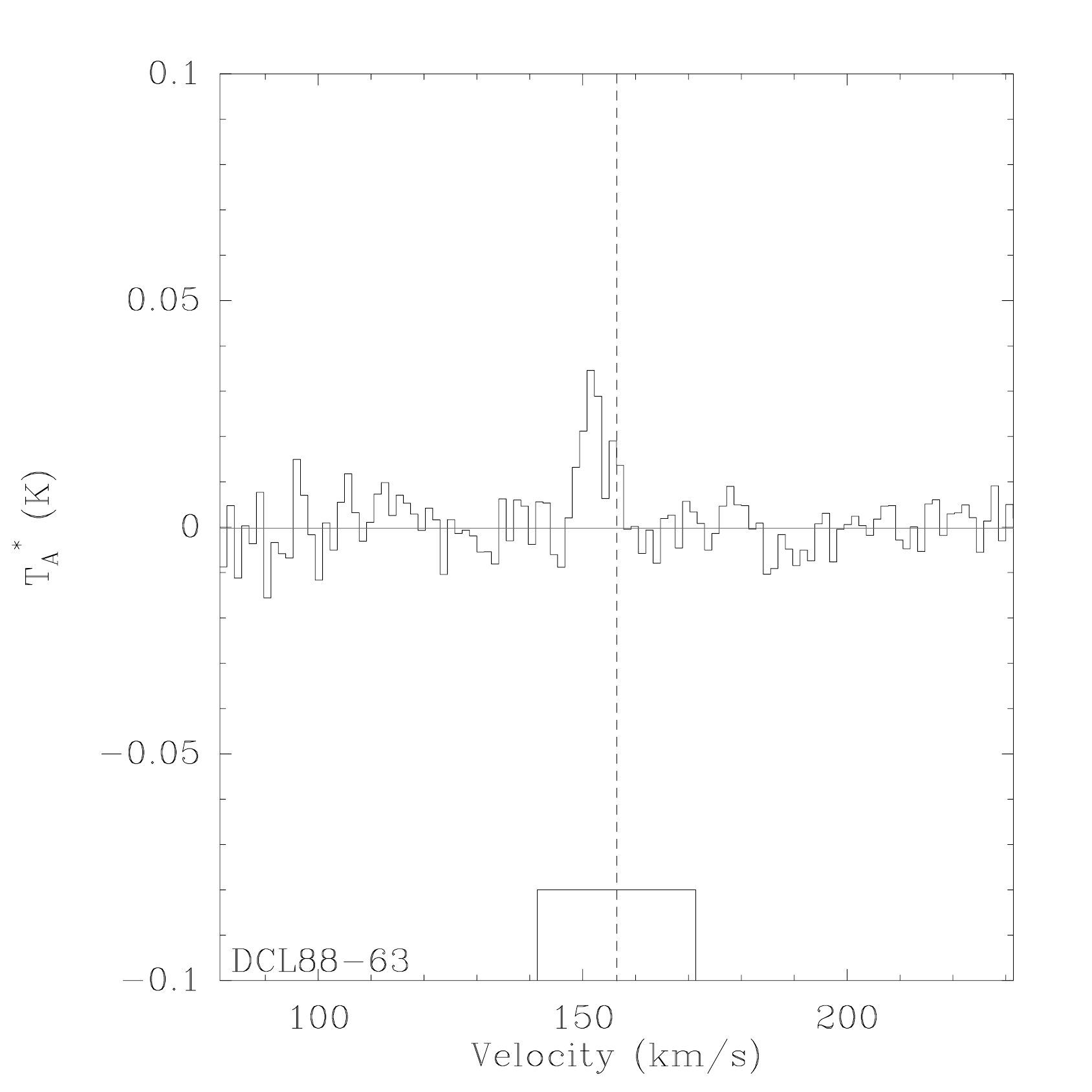}
\end{minipage}

\begin{minipage}{0.24\linewidth}
\includegraphics[width=\linewidth]{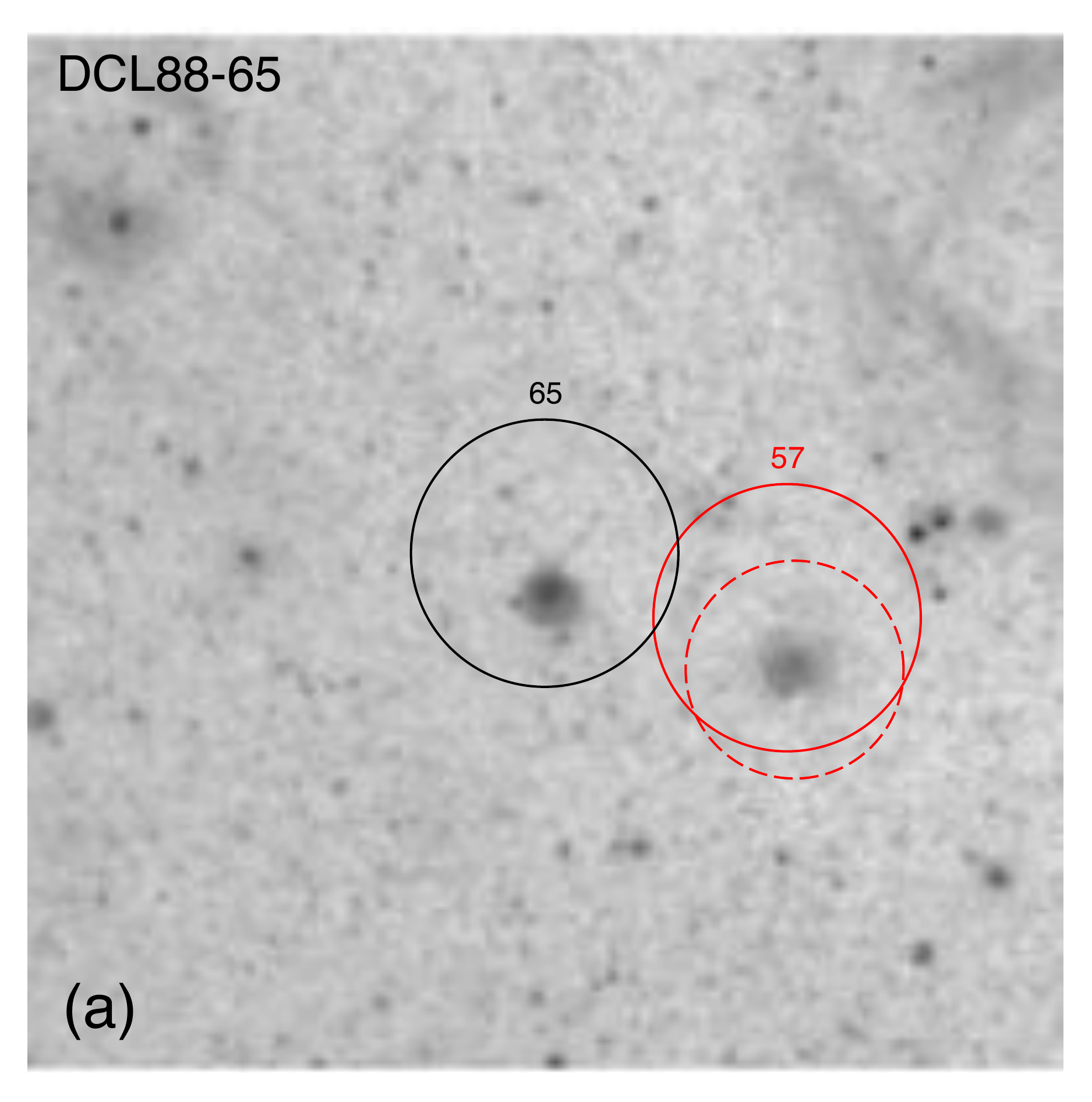}
\end{minipage}
\begin{minipage}{0.24\linewidth}
\includegraphics[width=\linewidth]{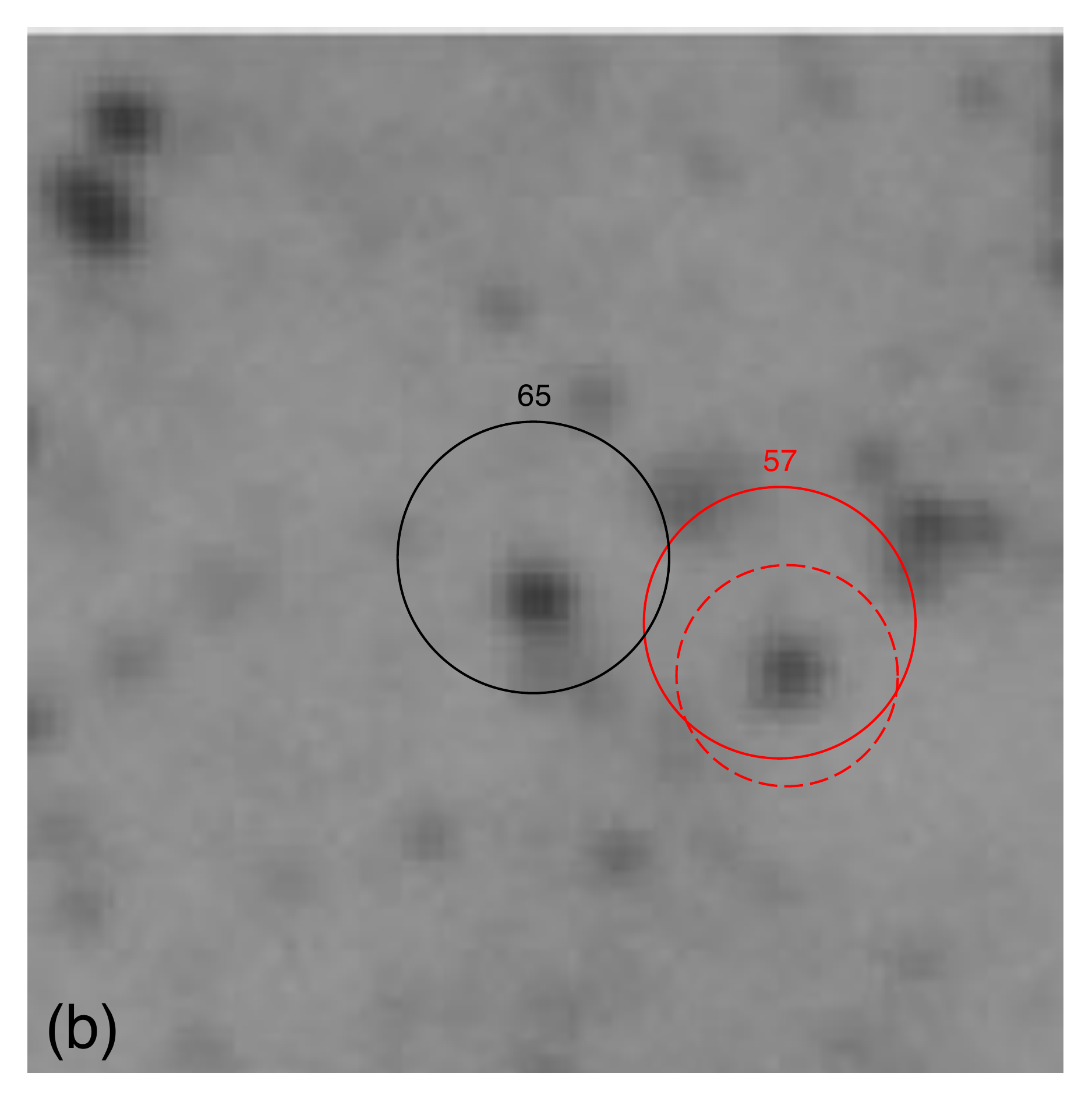}
\end{minipage}
\begin{minipage}{0.24\linewidth}
\includegraphics[width=\linewidth]{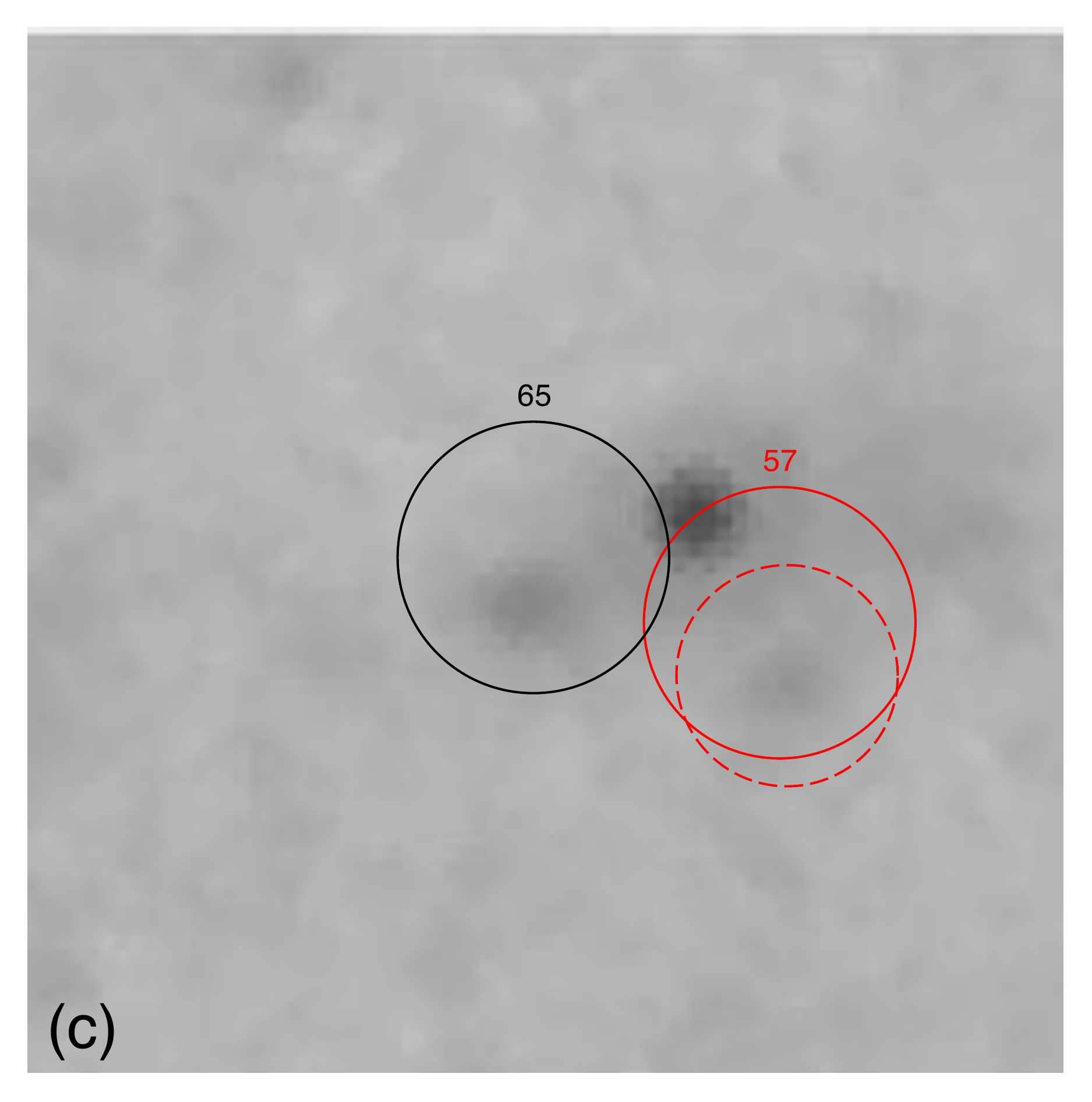}
\end{minipage}
\begin{minipage}{0.24\linewidth}
\includegraphics[width=\linewidth]{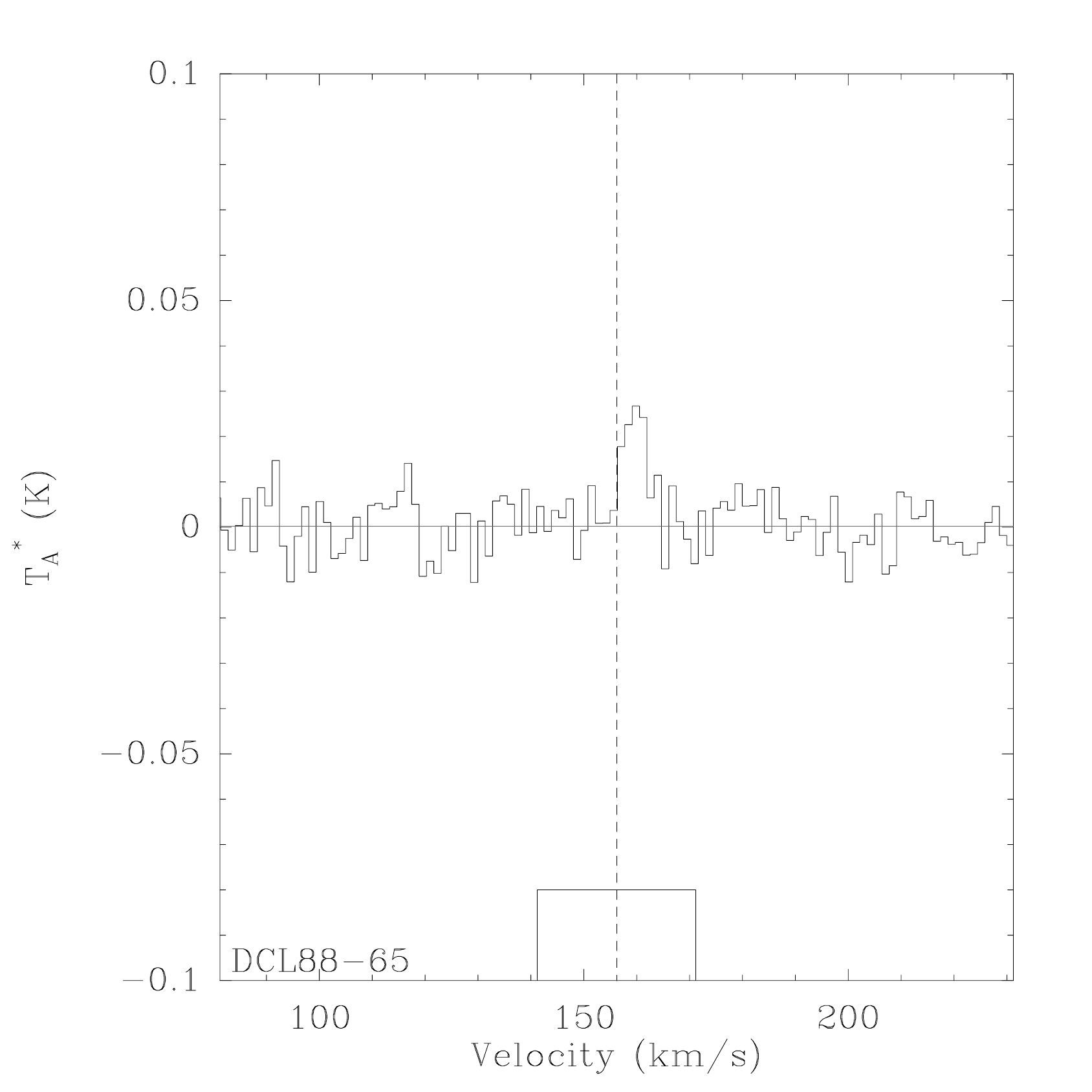}
\end{minipage}

\begin{minipage}{0.24\linewidth}
\includegraphics[width=\linewidth]{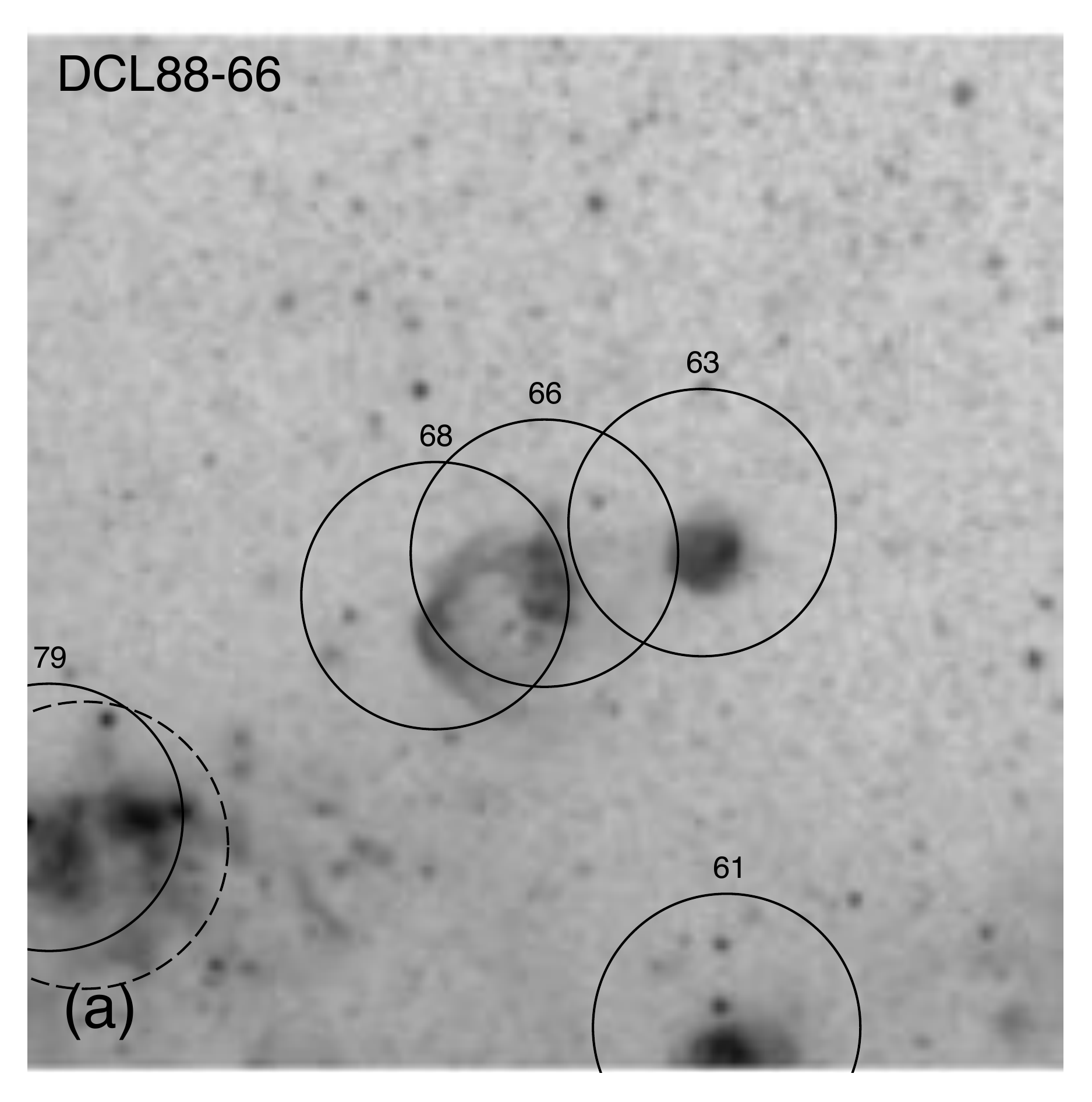}
\end{minipage}
\begin{minipage}{0.24\linewidth}
\includegraphics[width=\linewidth]{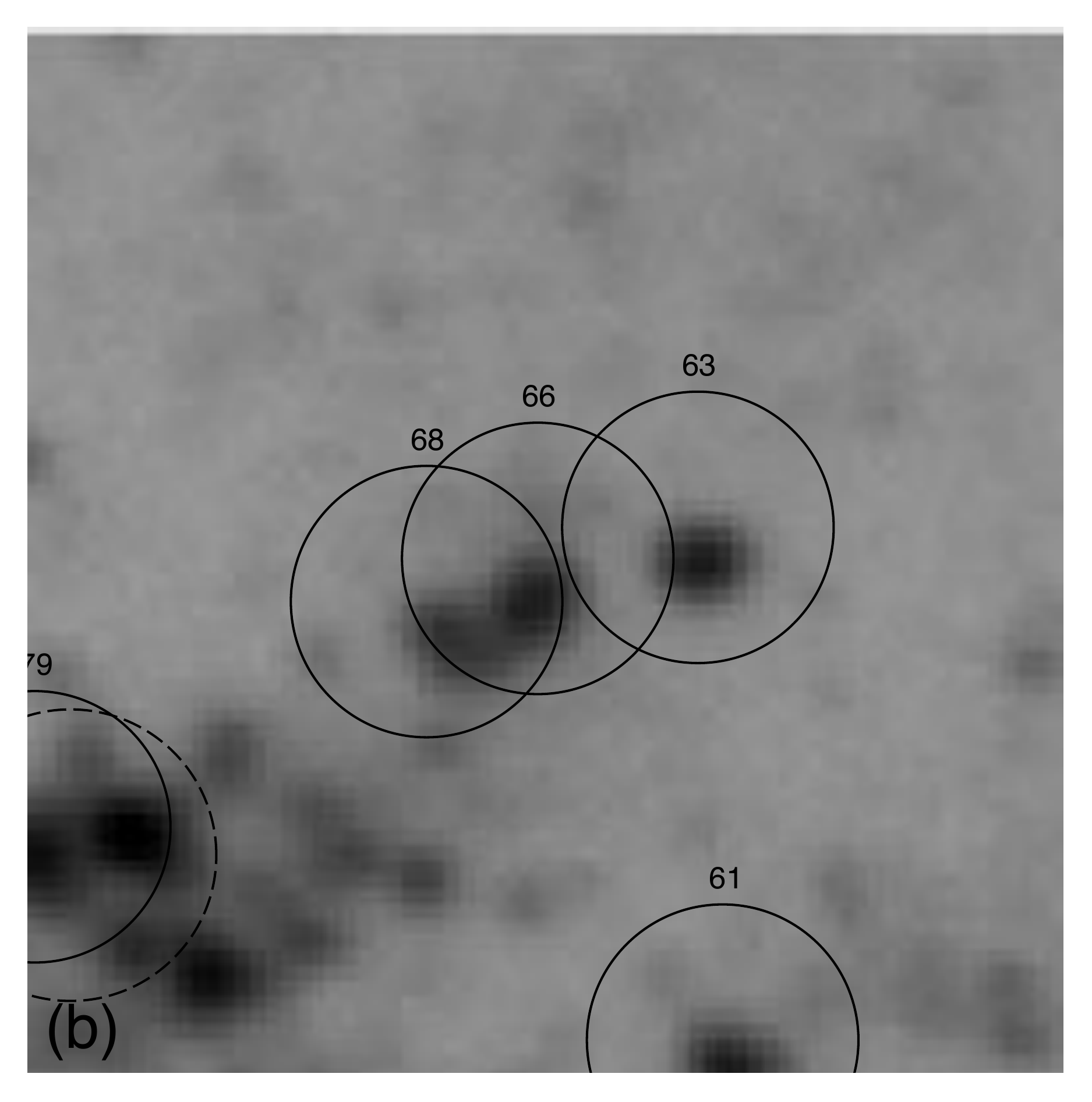}
\end{minipage}
\begin{minipage}{0.24\linewidth}
\includegraphics[width=\linewidth]{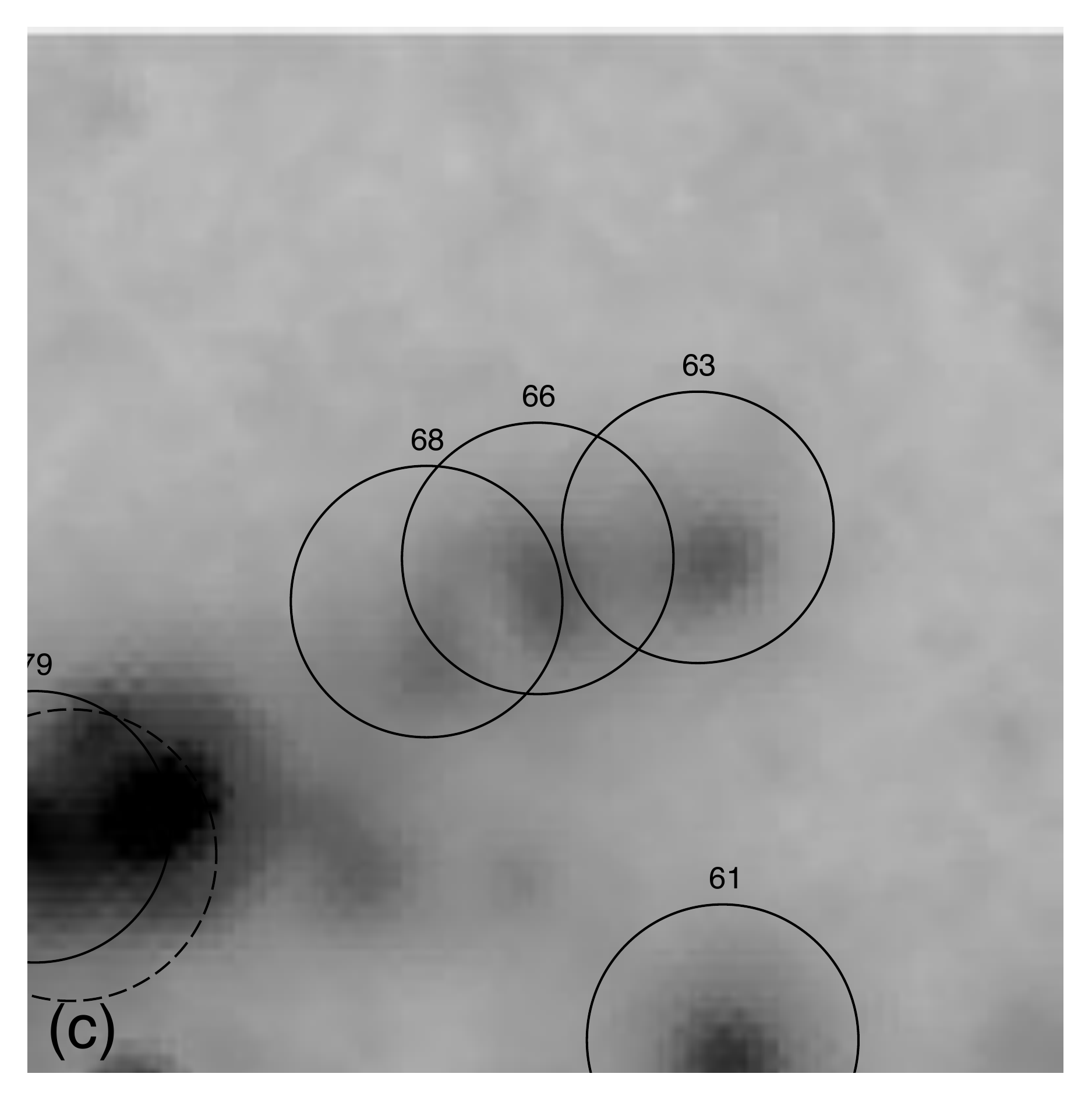}
\end{minipage}
\begin{minipage}{0.24\linewidth}
\includegraphics[width=\linewidth]{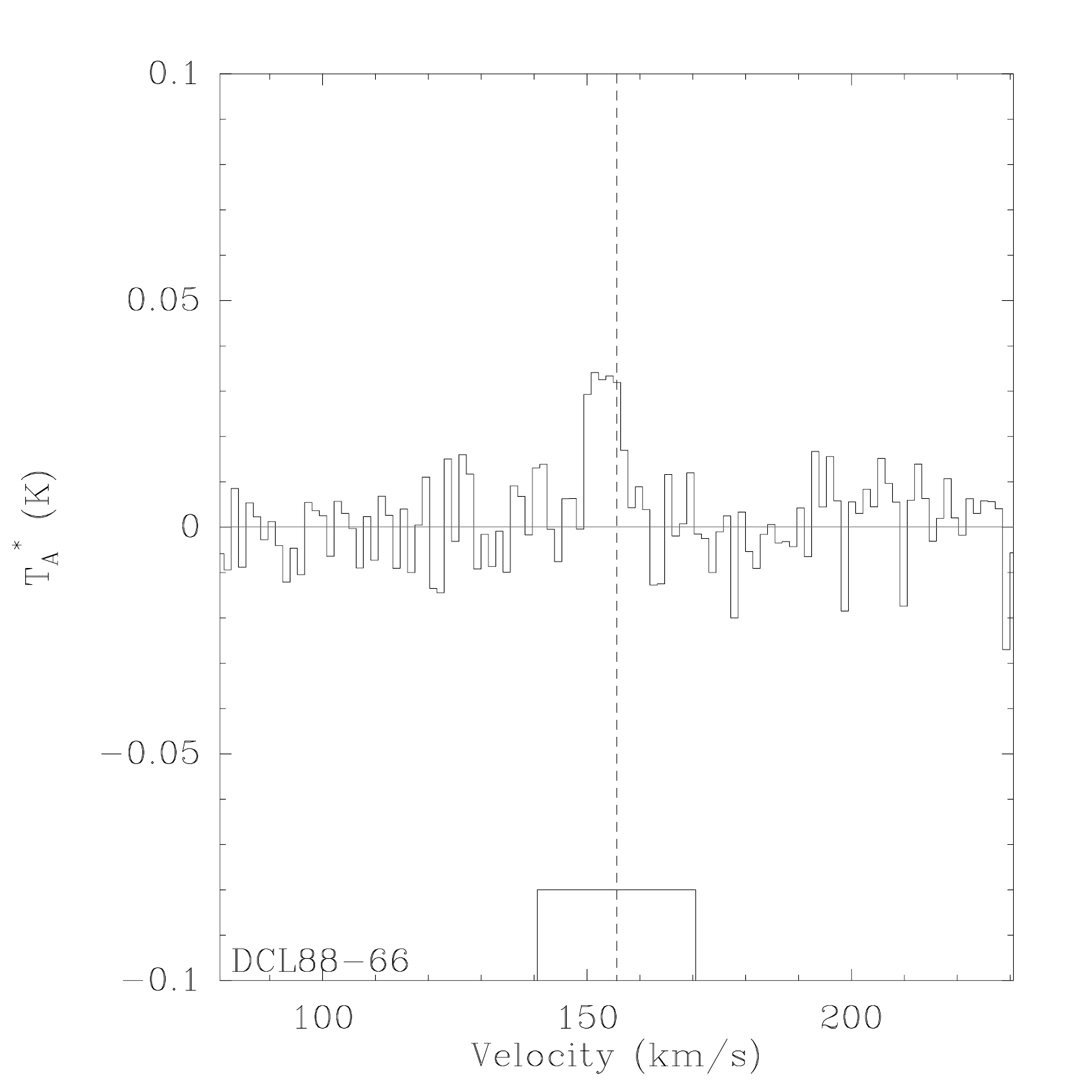}
\end{minipage}

\begin{minipage}{0.24\linewidth}
\includegraphics[width=\linewidth]{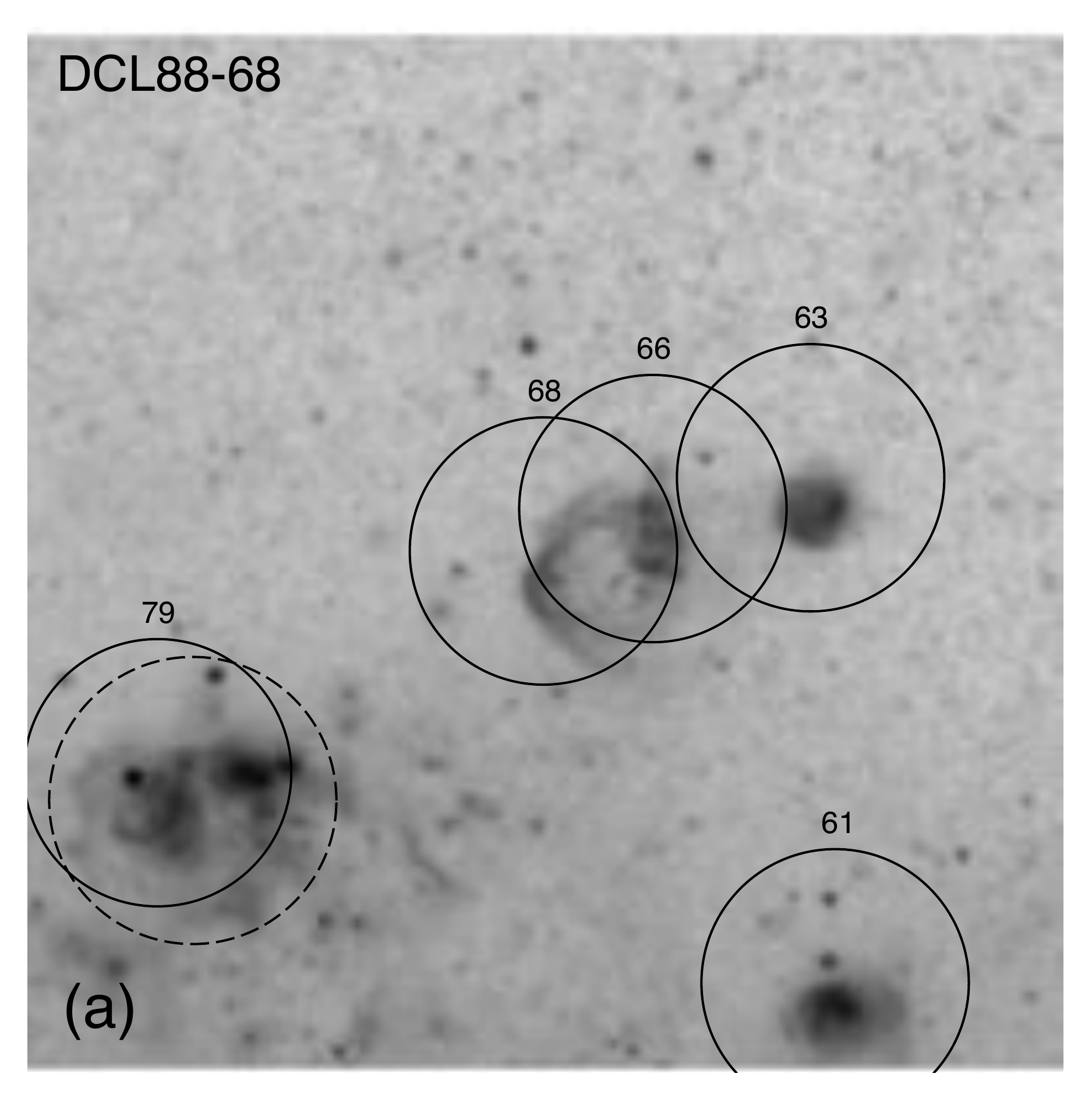}
\end{minipage}
\begin{minipage}{0.24\linewidth}
\includegraphics[width=\linewidth]{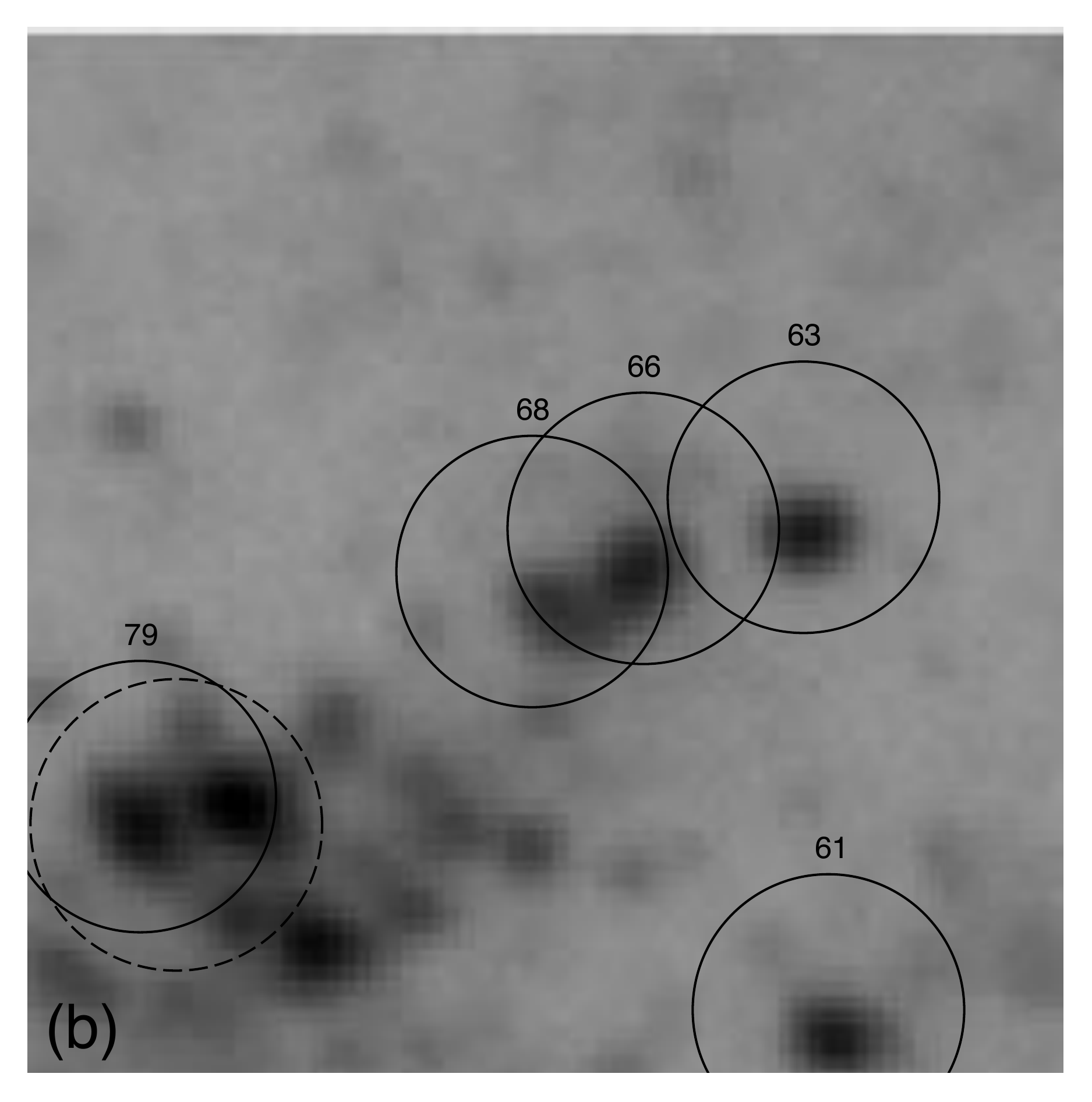}
\end{minipage}
\begin{minipage}{0.24\linewidth}
\includegraphics[width=\linewidth]{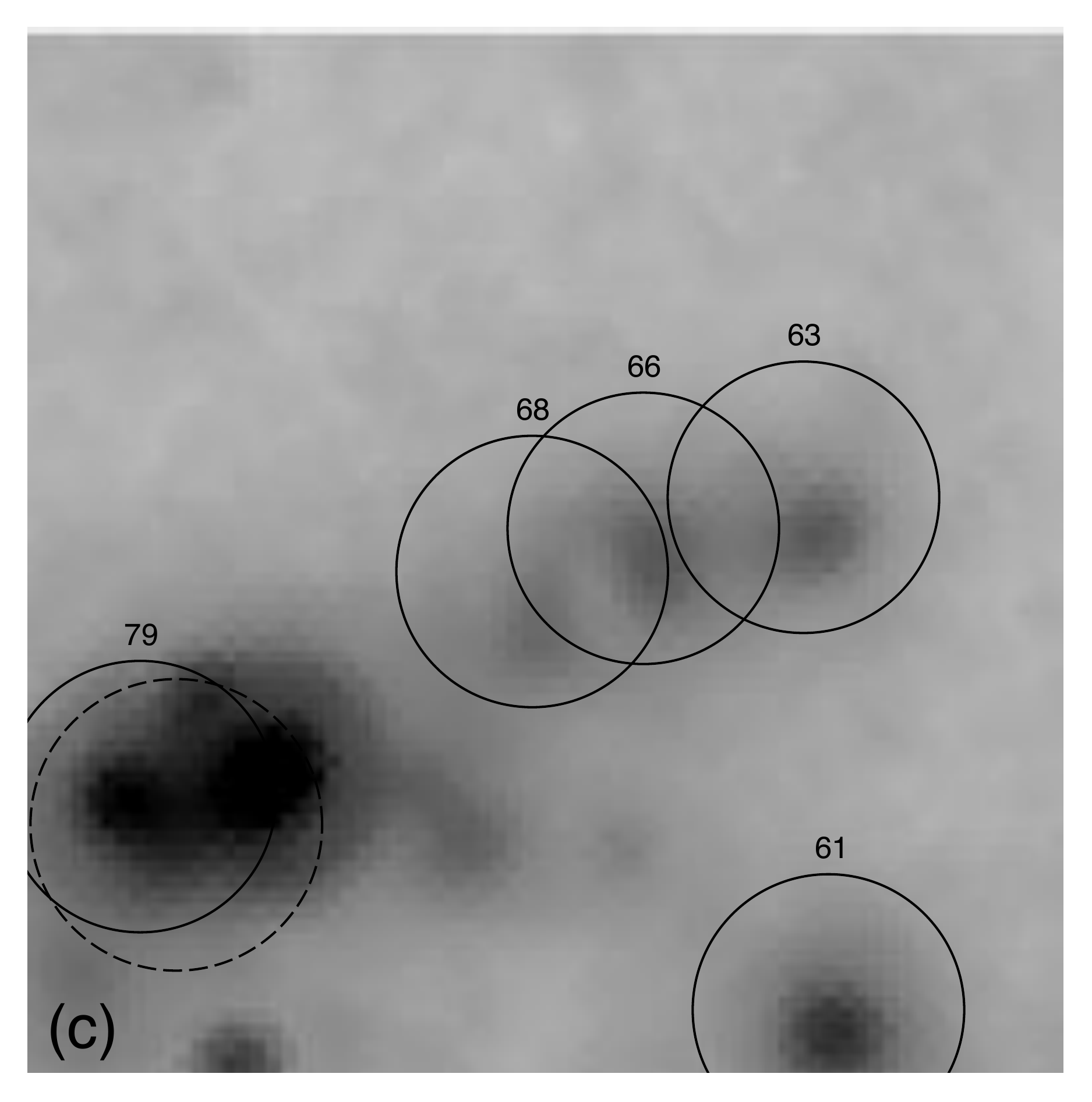}
\end{minipage}
\begin{minipage}{0.24\linewidth}
\includegraphics[width=\linewidth]{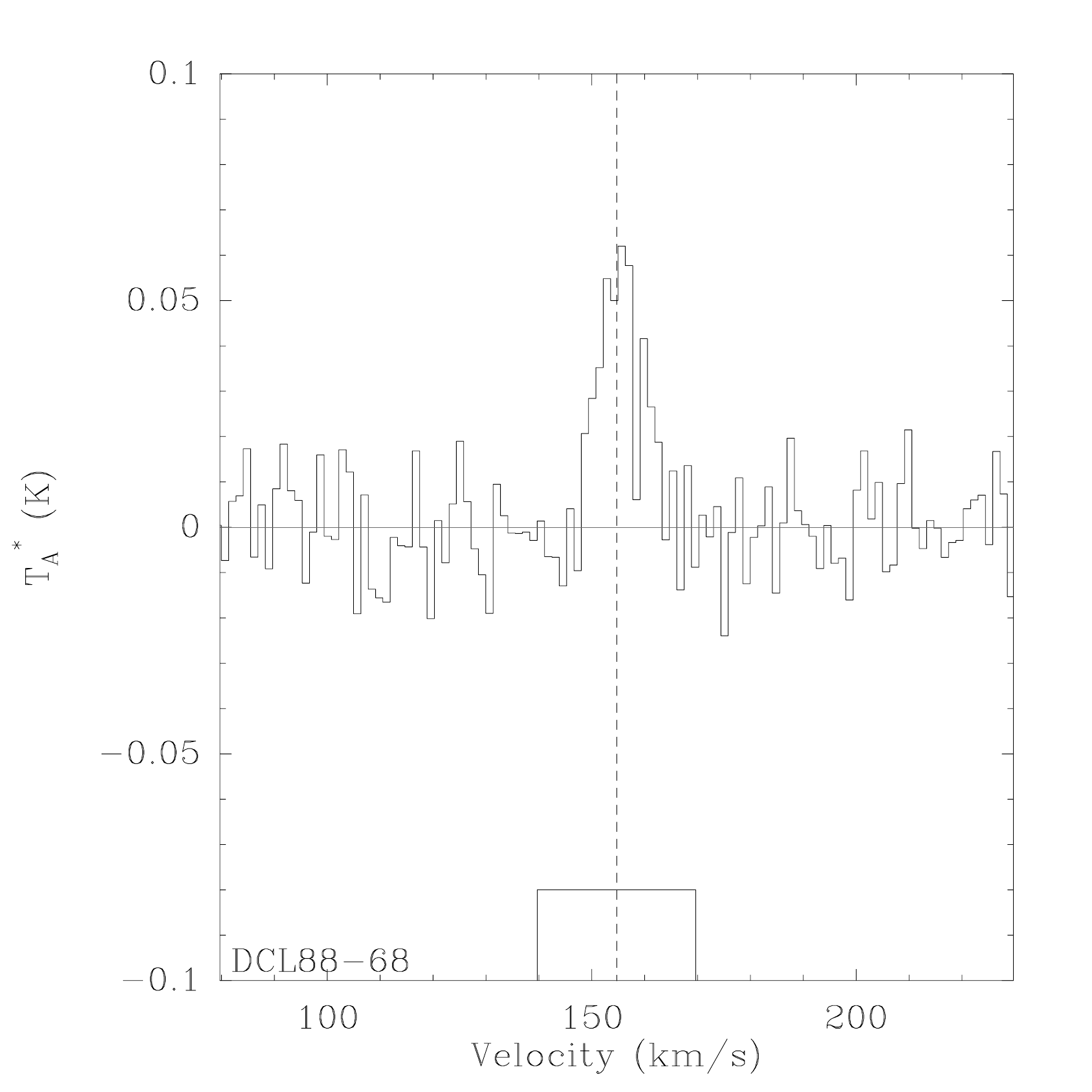}
\end{minipage}

\begin{minipage}{0.24\linewidth}
\includegraphics[width=\linewidth]{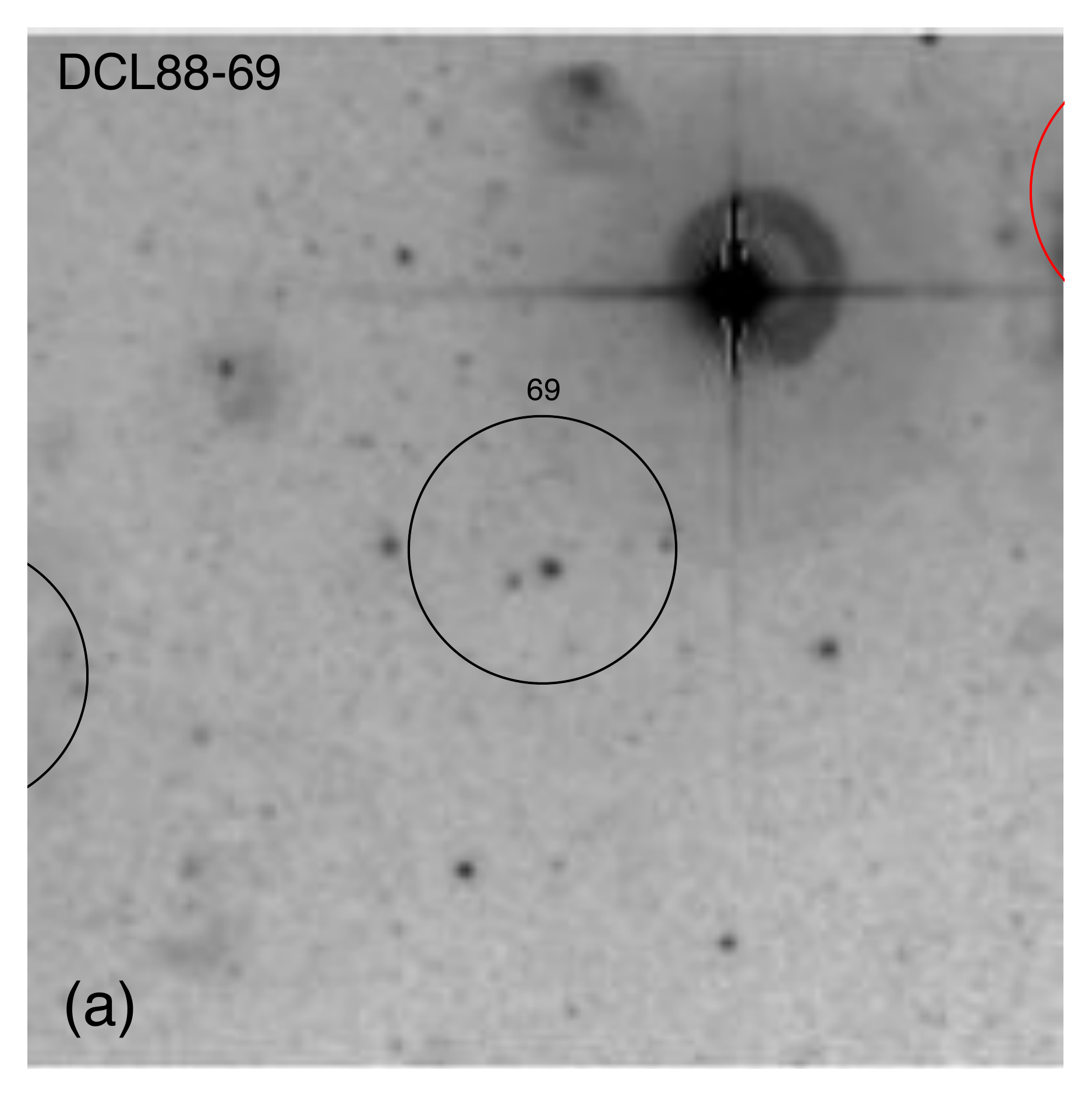}
\end{minipage}
\begin{minipage}{0.24\linewidth}
\includegraphics[width=\linewidth]{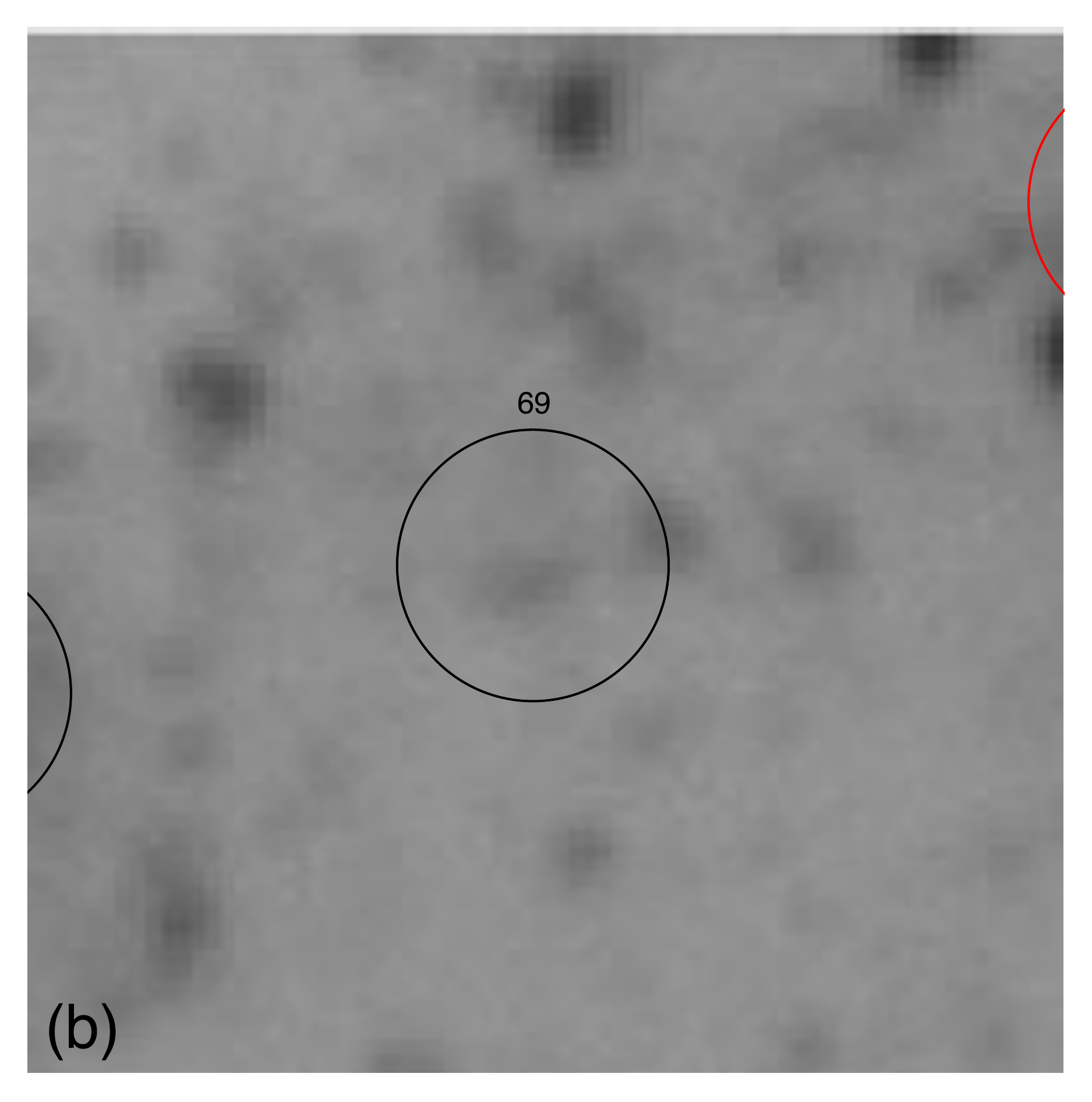}
\end{minipage}
\begin{minipage}{0.24\linewidth}
\includegraphics[width=\linewidth]{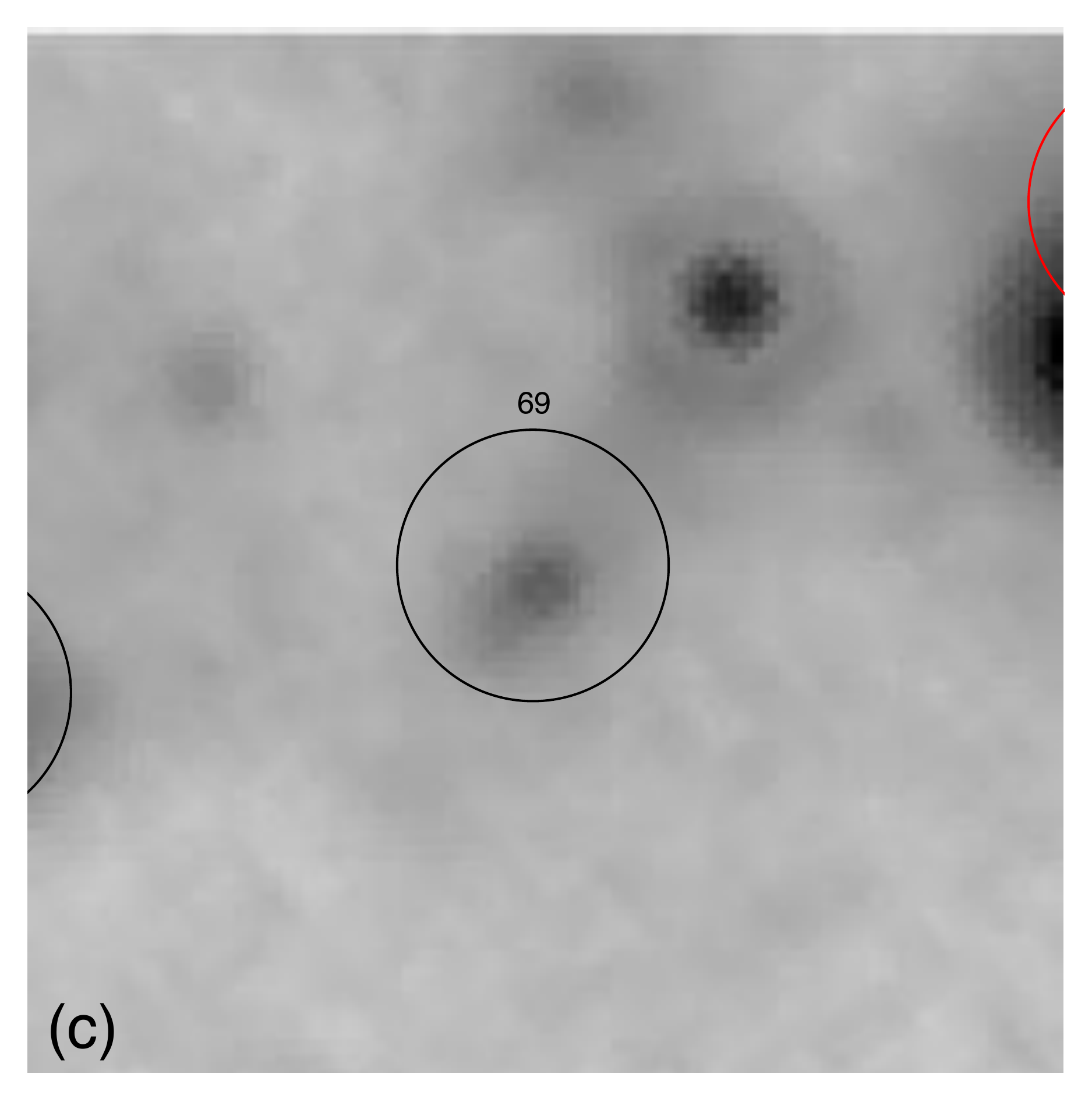}
\end{minipage}
\begin{minipage}{0.24\linewidth}
\includegraphics[width=\linewidth]{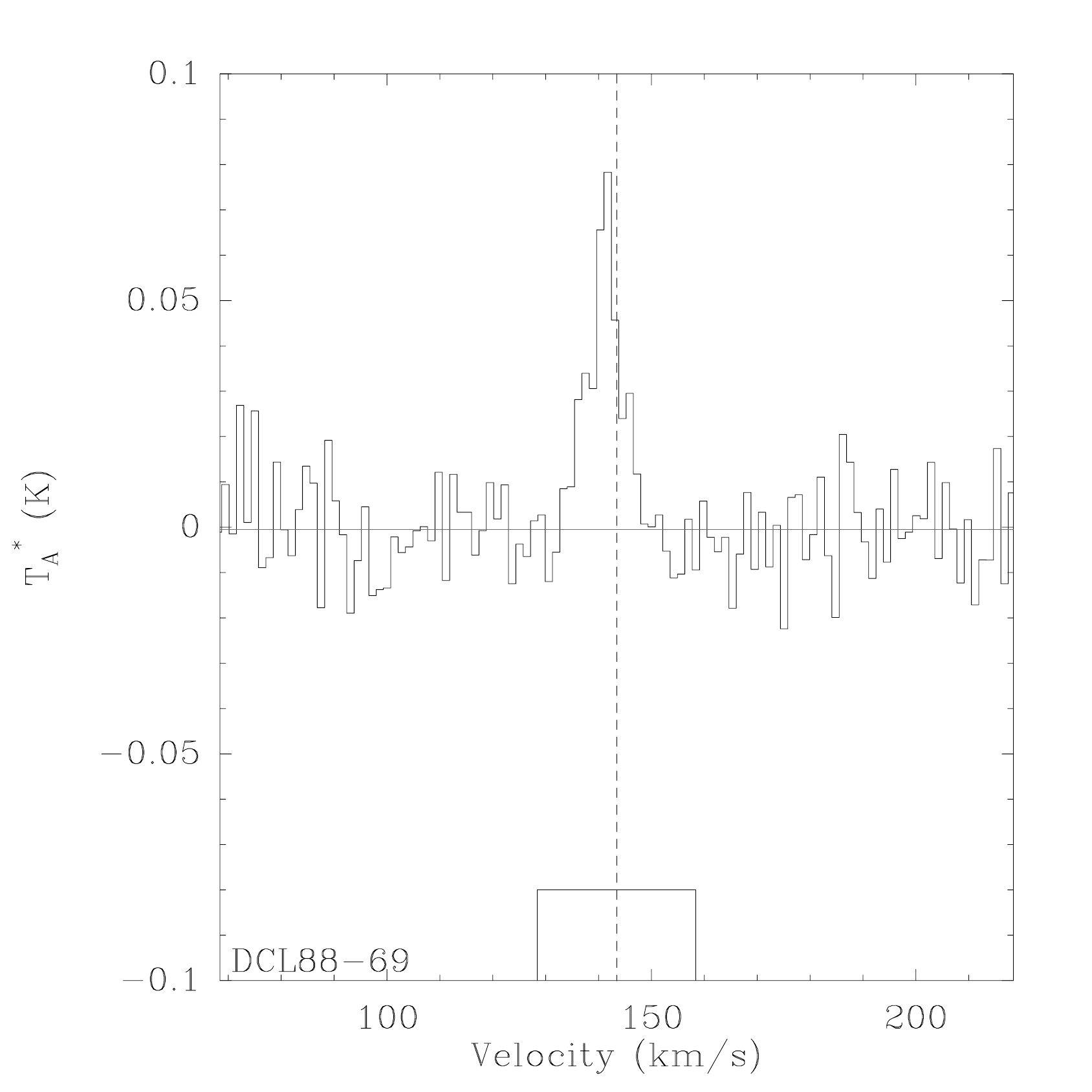}
\end{minipage}

\noindent\textbf{Figure~\ref{fig:stamps} -- continued.}
\end{figure*}

\begin{figure*}
%\ContinuedFloat

\begin{minipage}{0.24\linewidth}
\includegraphics[width=\linewidth]{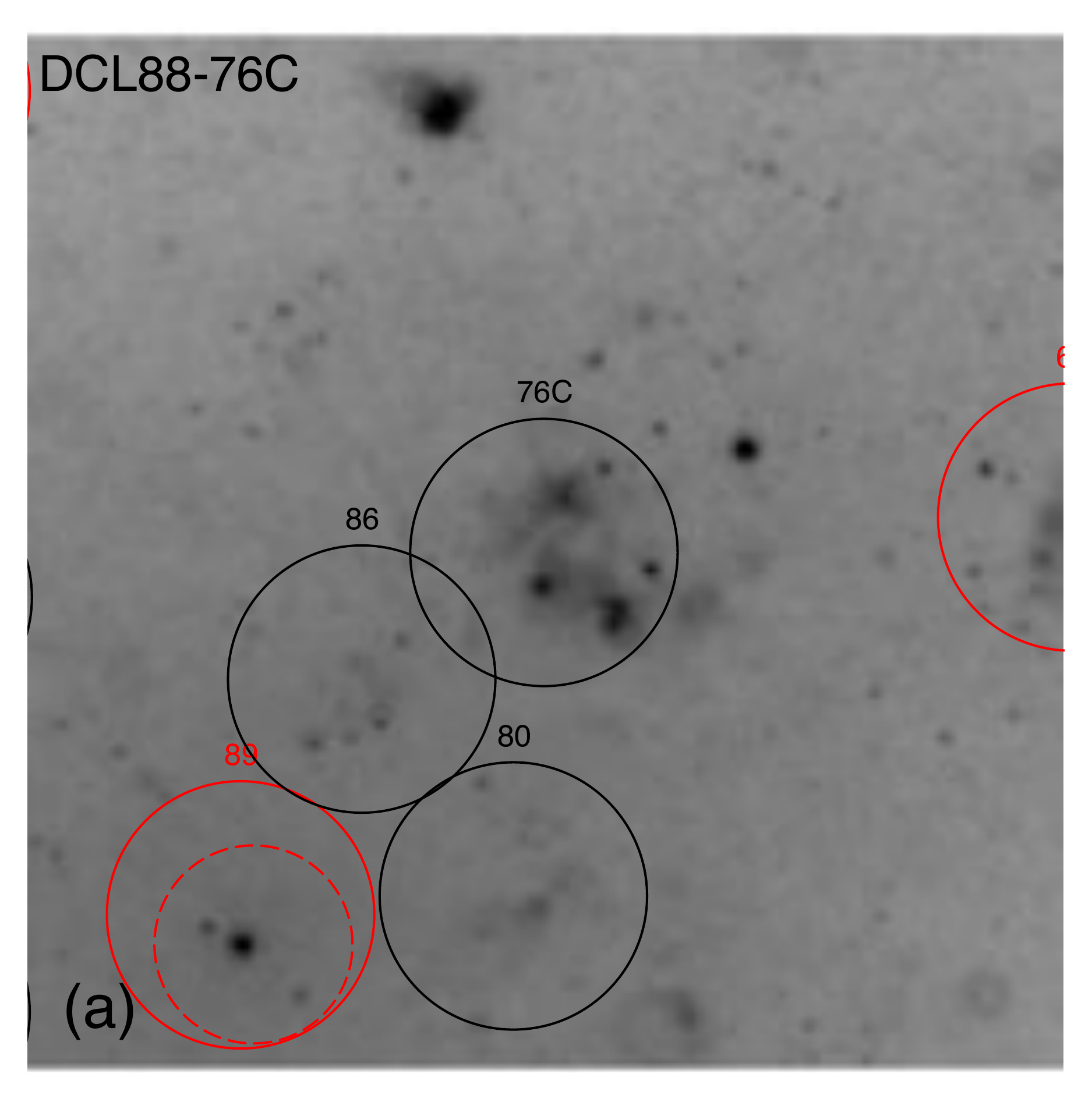}
\end{minipage}
\begin{minipage}{0.24\linewidth}
\includegraphics[width=\linewidth]{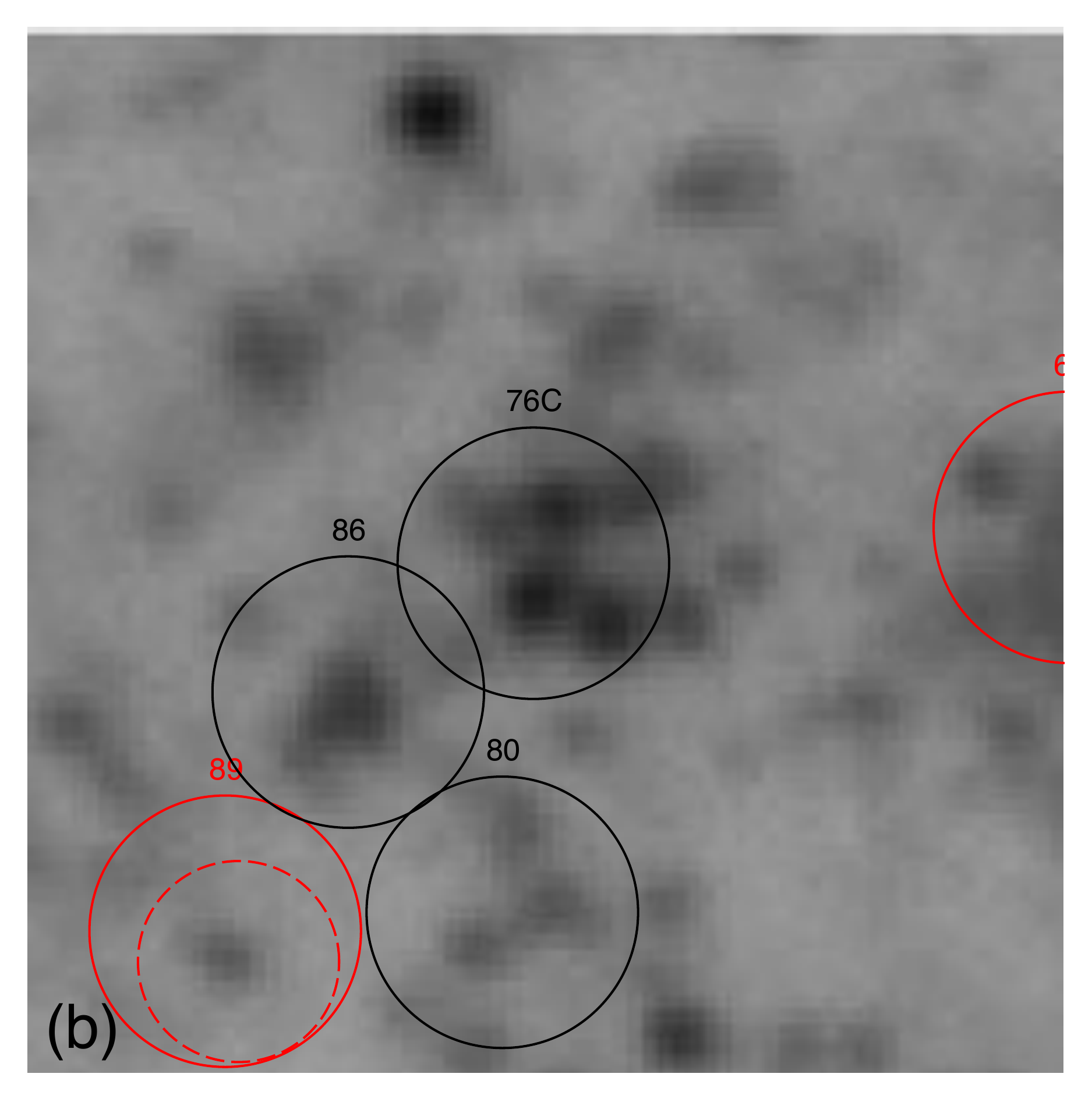}
\end{minipage}
\begin{minipage}{0.24\linewidth}
\includegraphics[width=\linewidth]{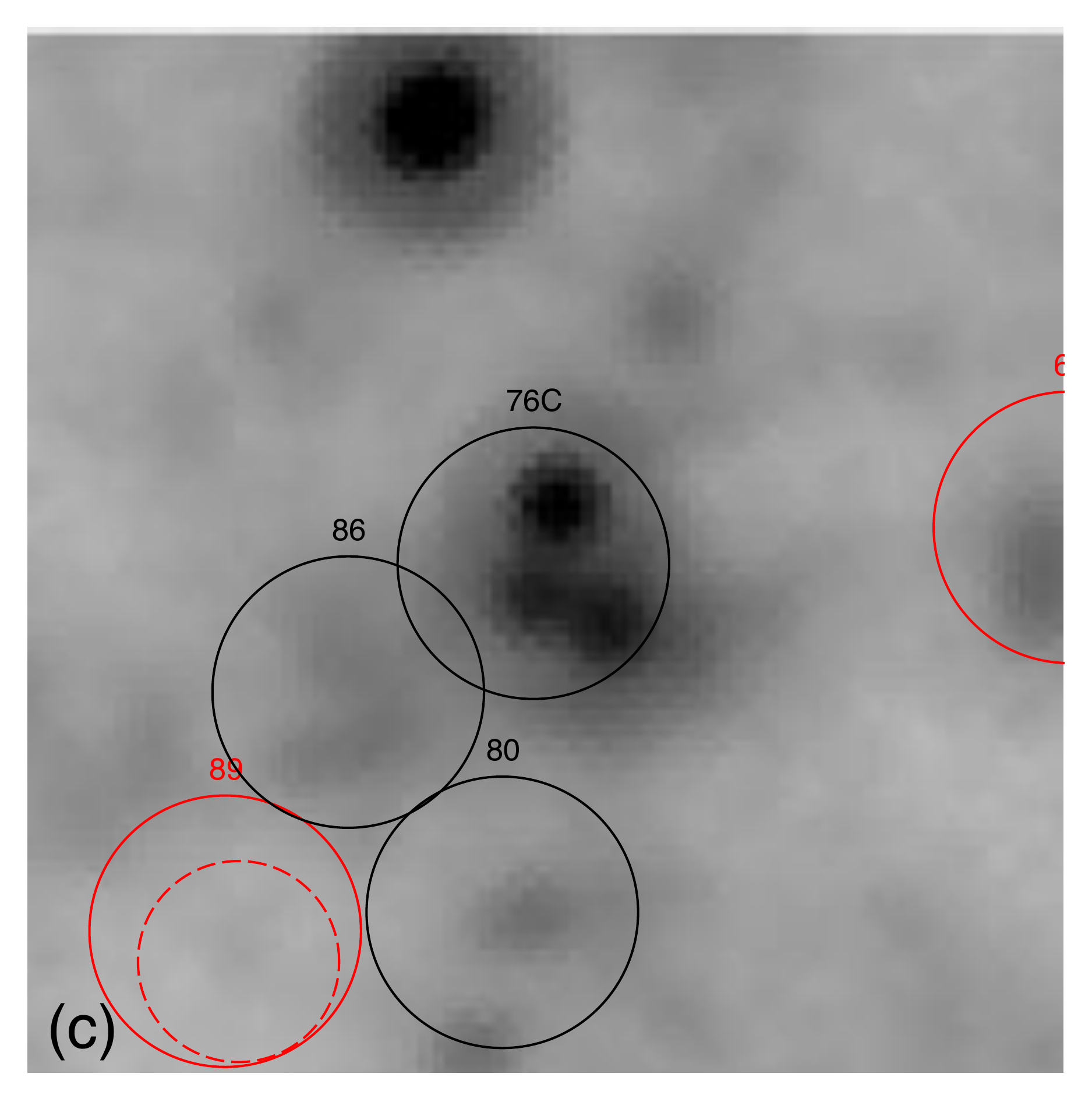}
\end{minipage}
\begin{minipage}{0.24\linewidth}
\includegraphics[width=\linewidth]{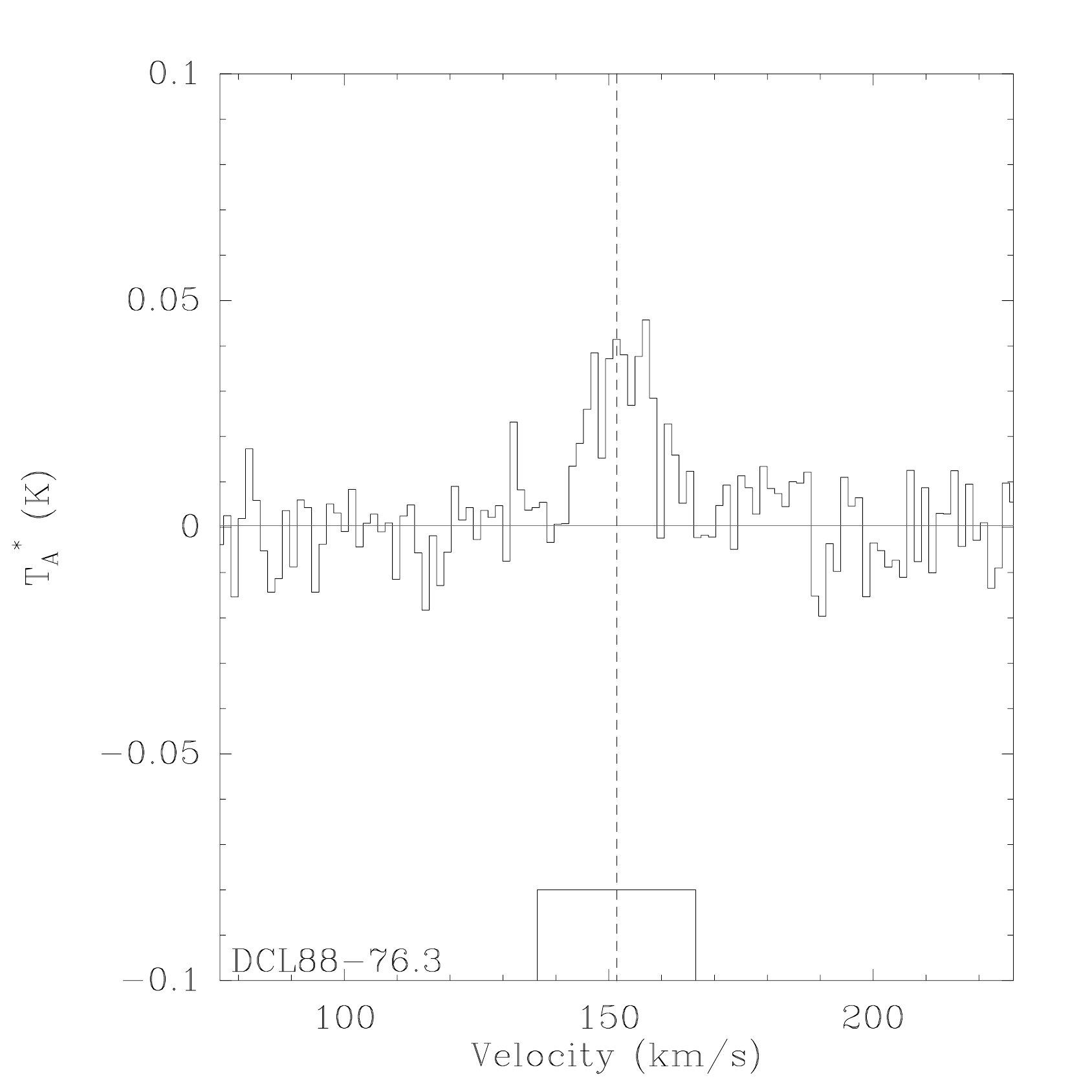}
\end{minipage}

\begin{minipage}{0.24\linewidth}
\includegraphics[width=\linewidth]{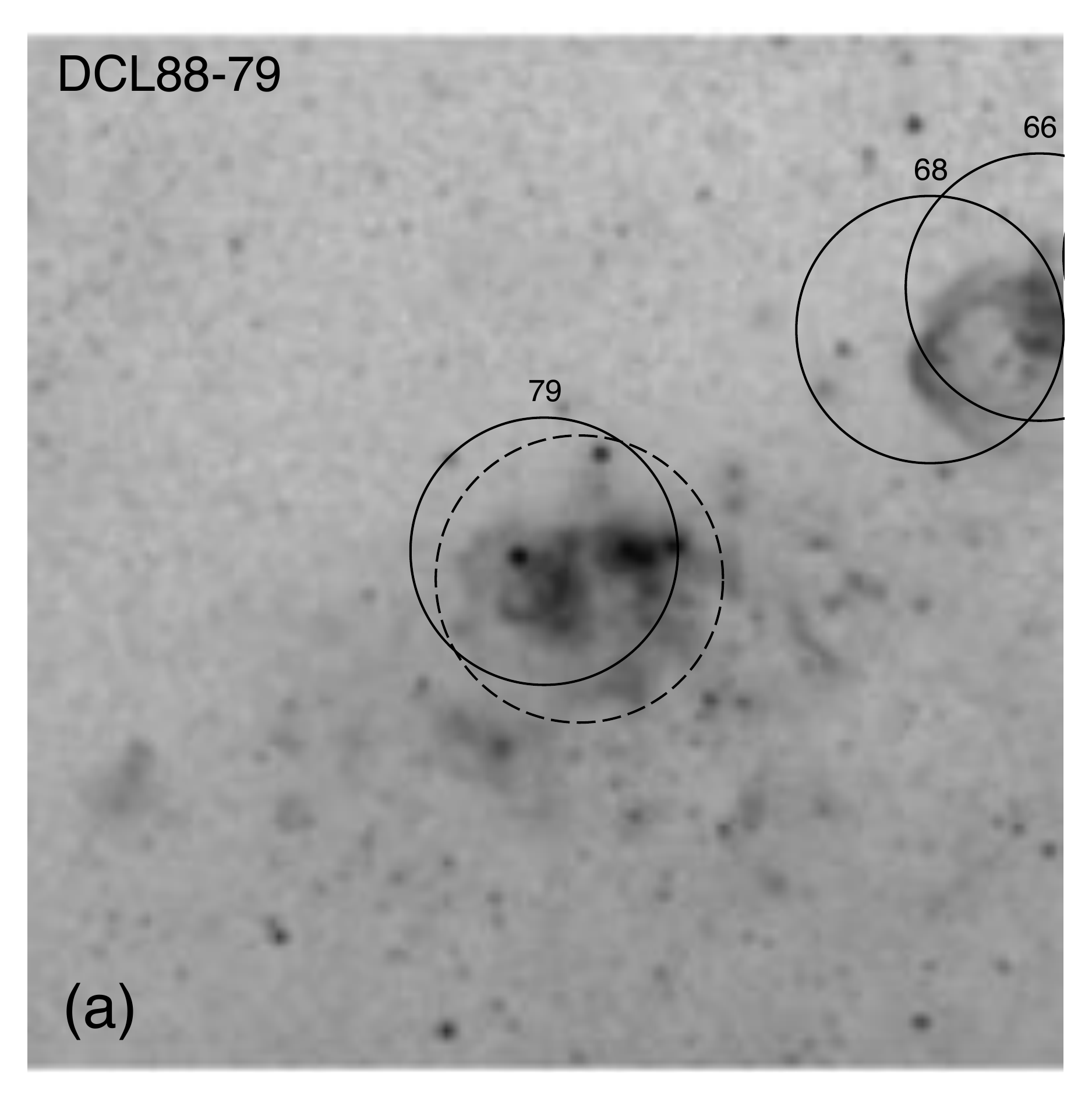}
\end{minipage}
\begin{minipage}{0.24\linewidth}
\includegraphics[width=\linewidth]{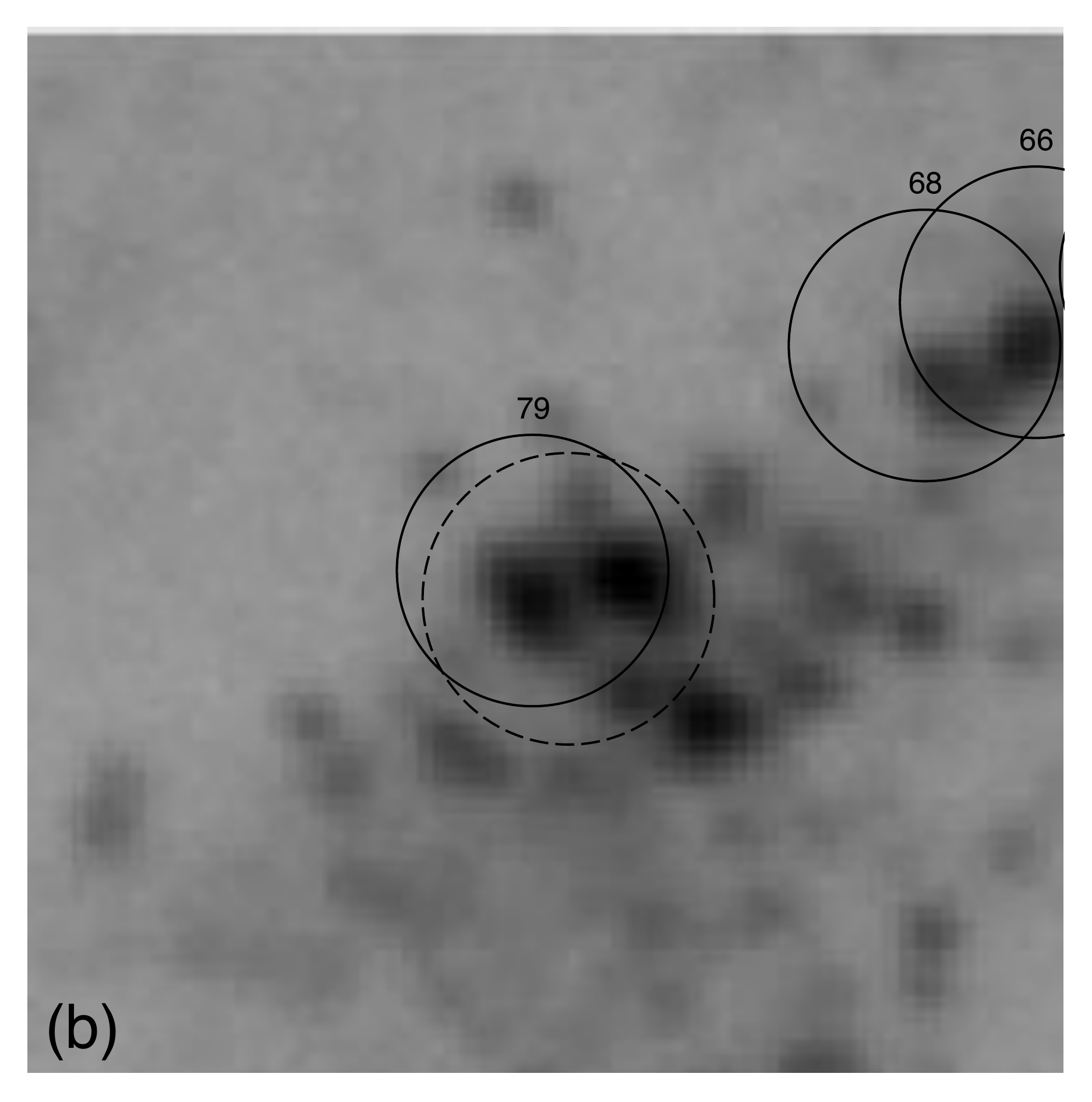}
\end{minipage}
\begin{minipage}{0.24\linewidth}
\includegraphics[width=\linewidth]{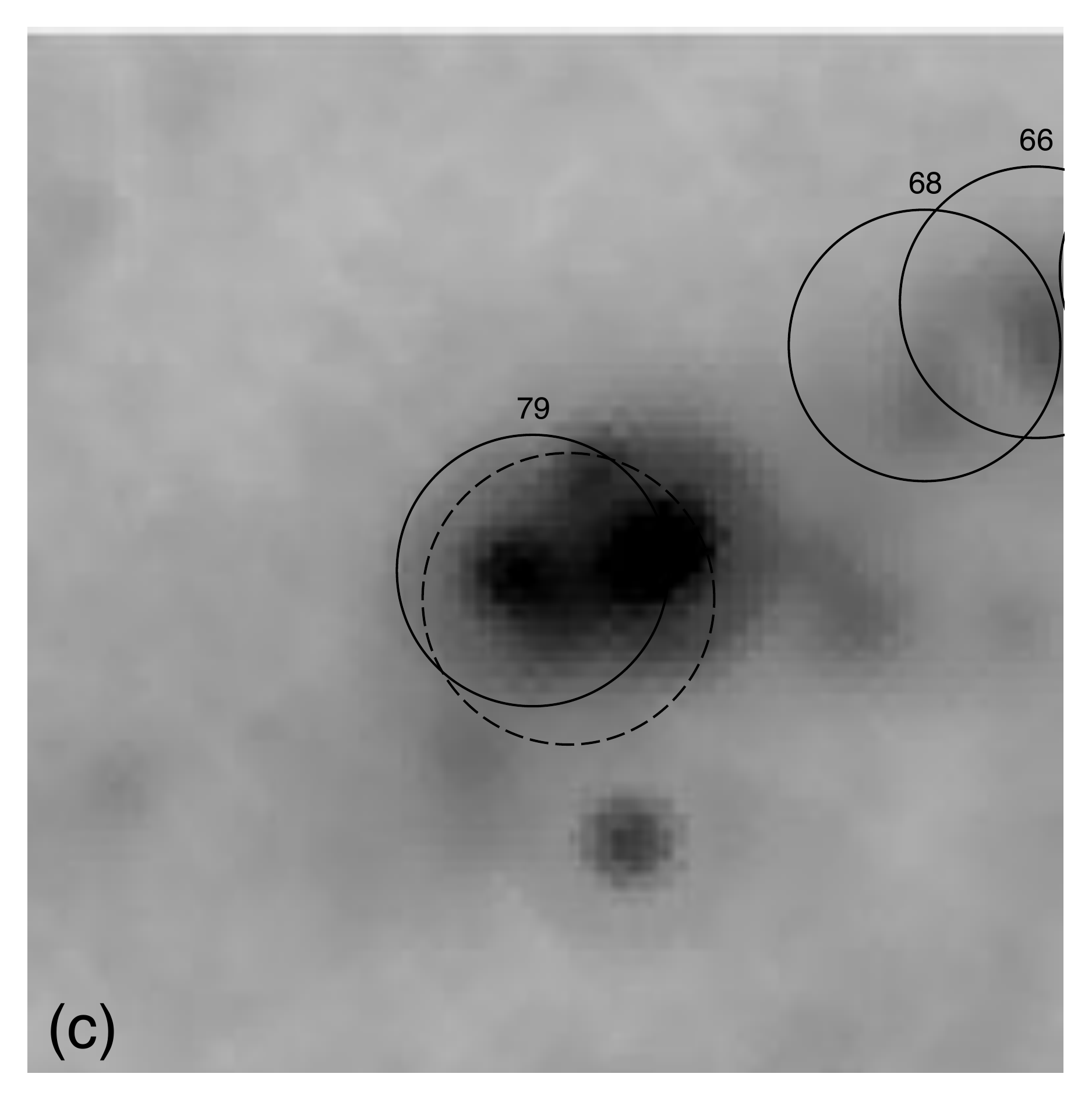}
\end{minipage}
\begin{minipage}{0.24\linewidth}
\includegraphics[width=\linewidth]{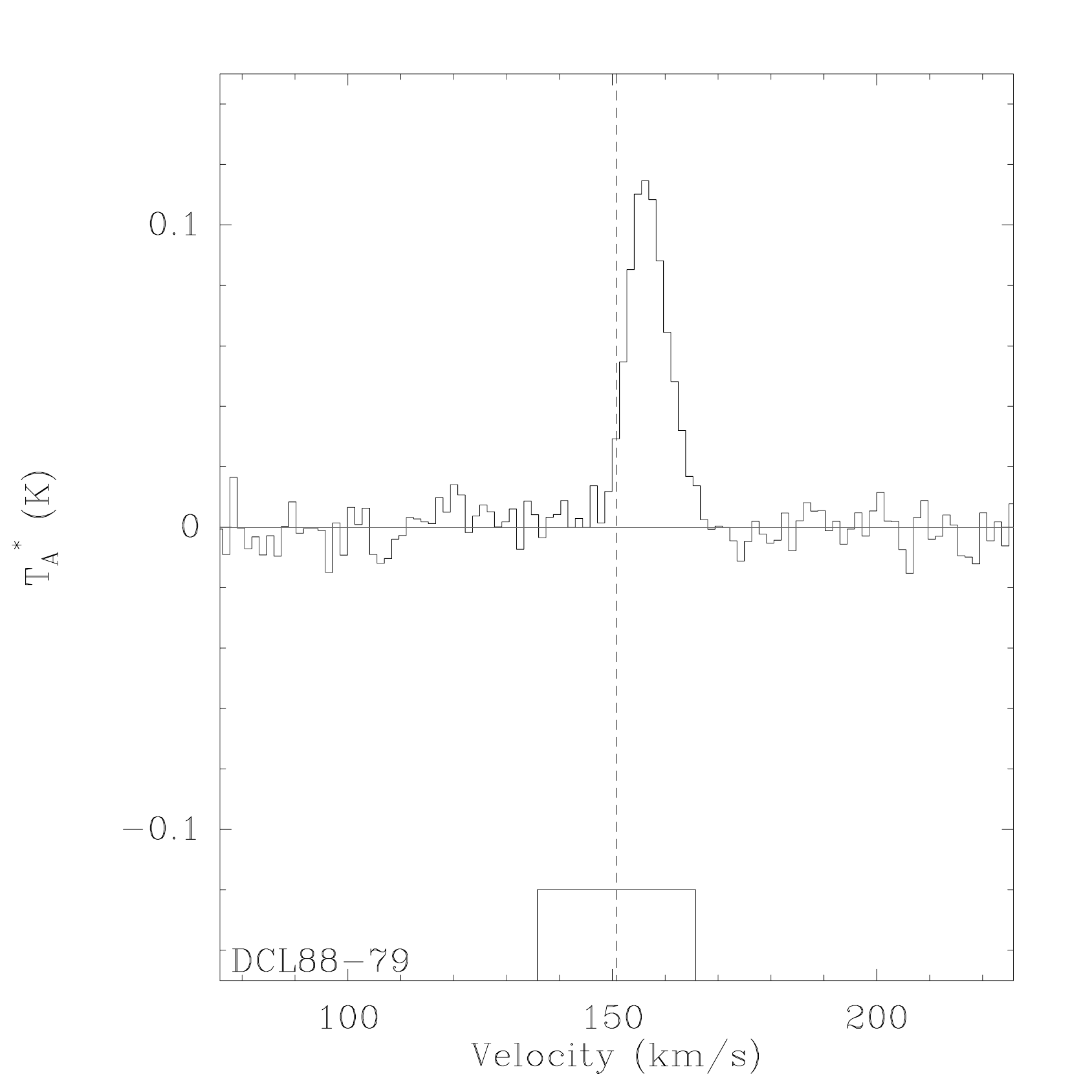}
\end{minipage}

\begin{minipage}{0.24\linewidth}
\includegraphics[width=\linewidth]{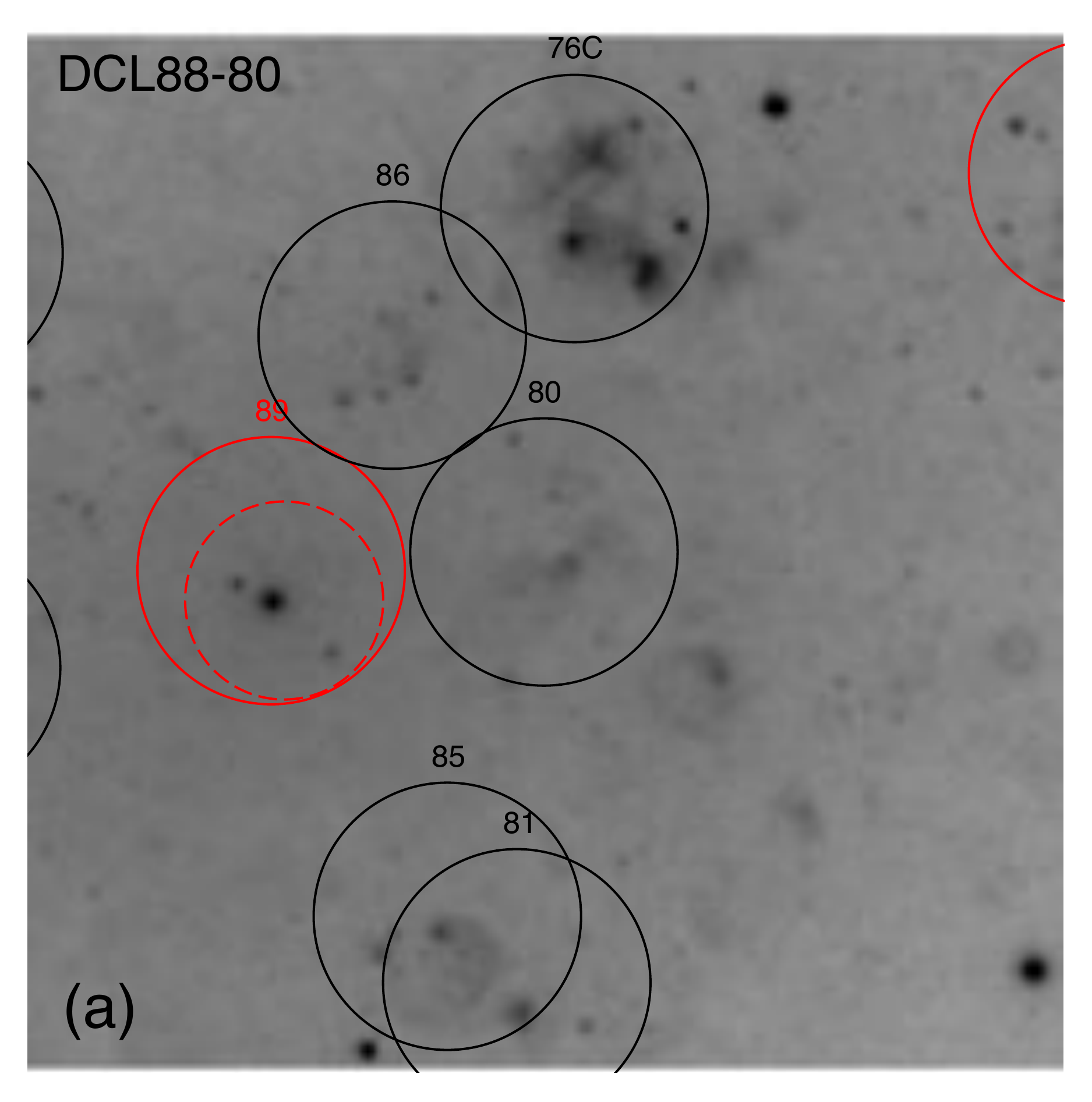}
\end{minipage}
\begin{minipage}{0.24\linewidth}
\includegraphics[width=\linewidth]{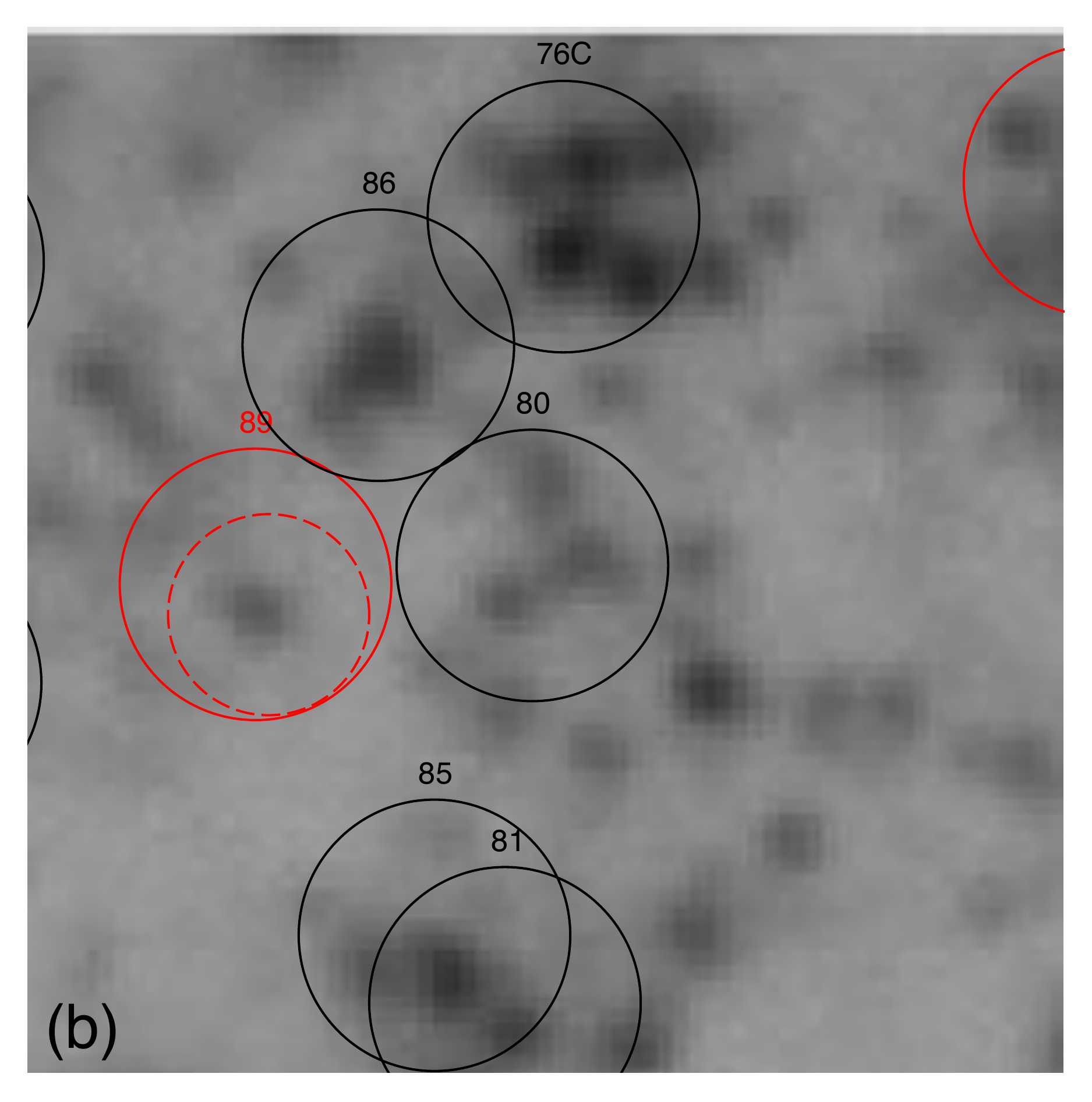}
\end{minipage}
\begin{minipage}{0.24\linewidth}
\includegraphics[width=\linewidth]{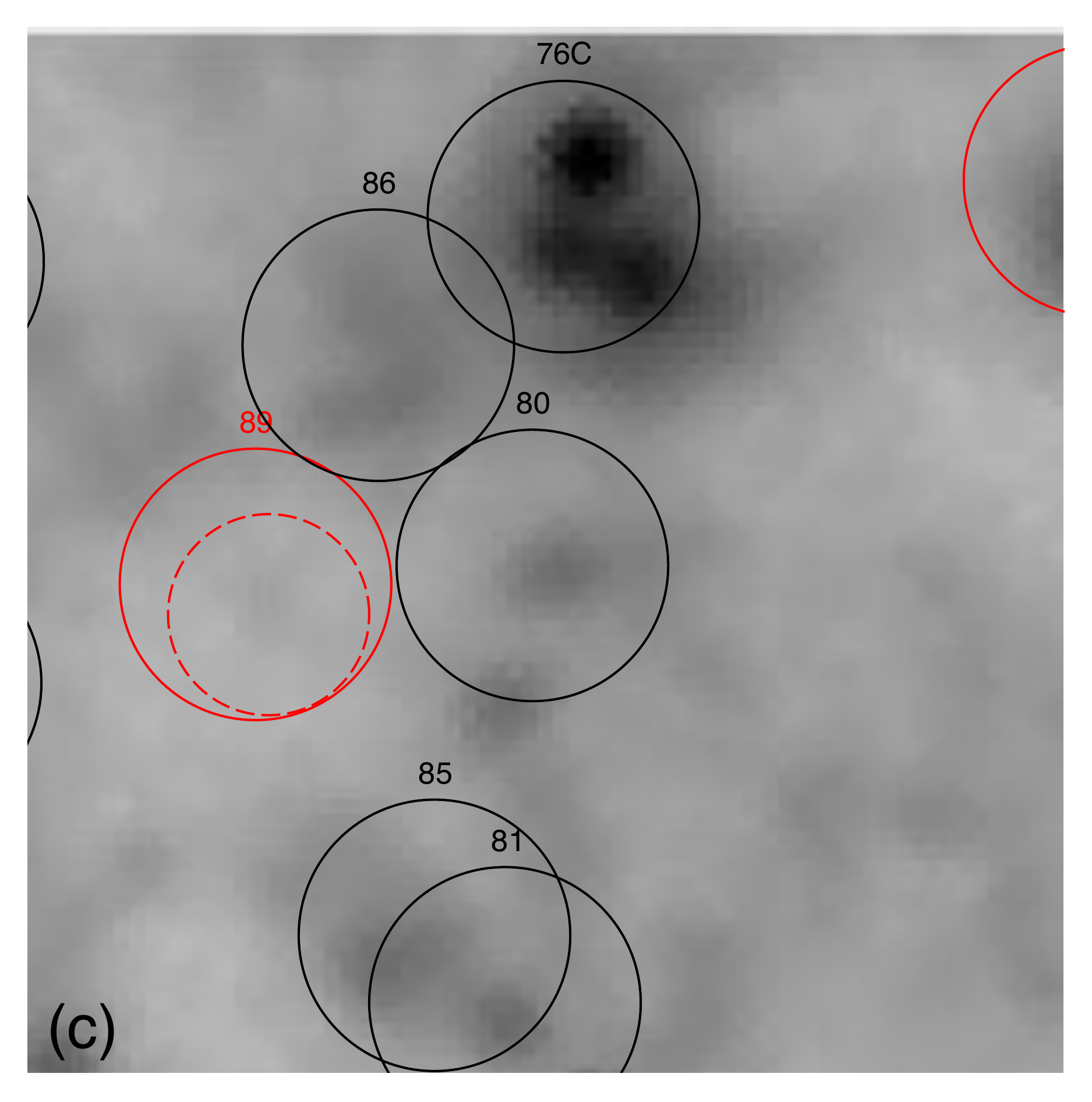}
\end{minipage}
\begin{minipage}{0.24\linewidth}
\includegraphics[width=\linewidth]{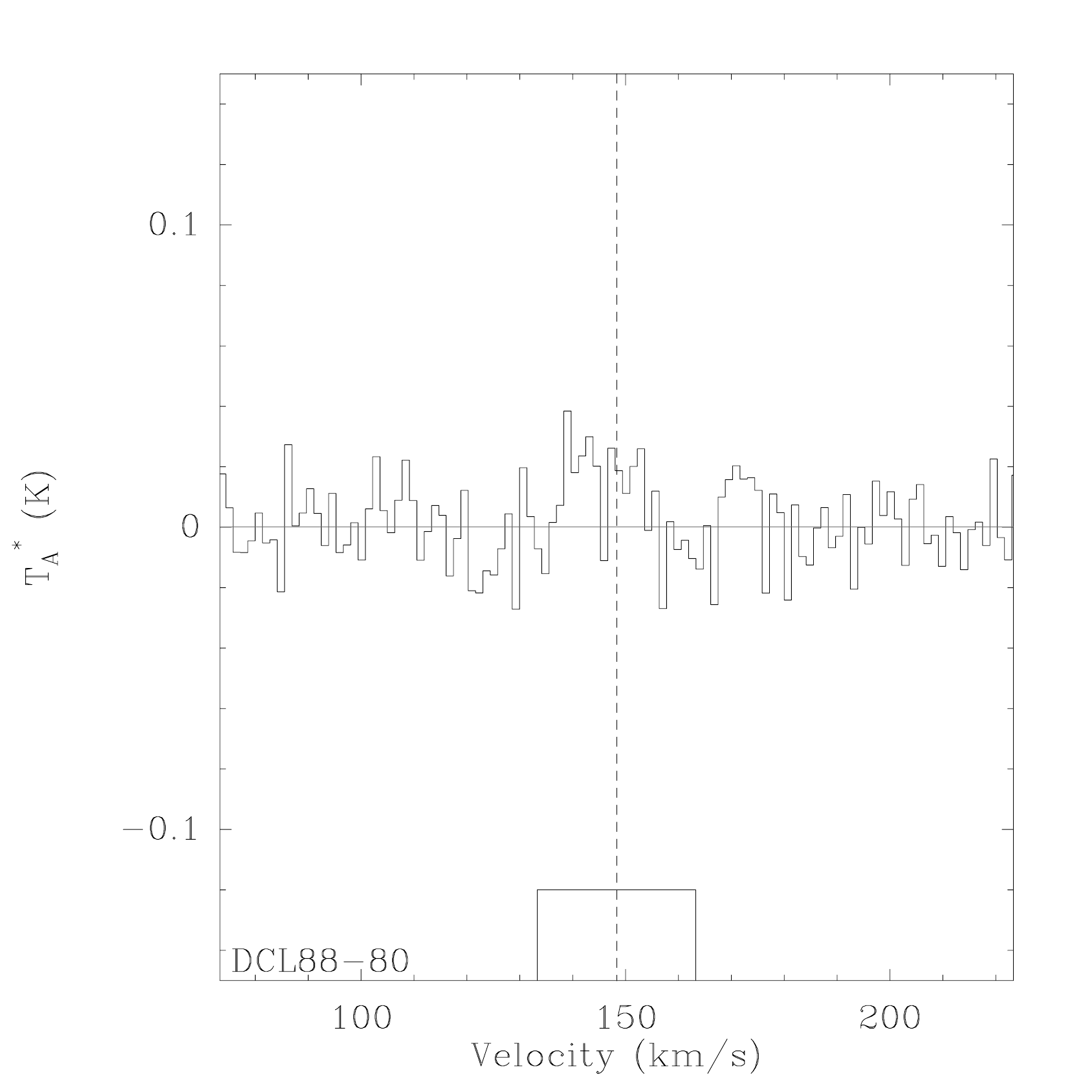}
\end{minipage}

\begin{minipage}{0.24\linewidth}
\includegraphics[width=\linewidth]{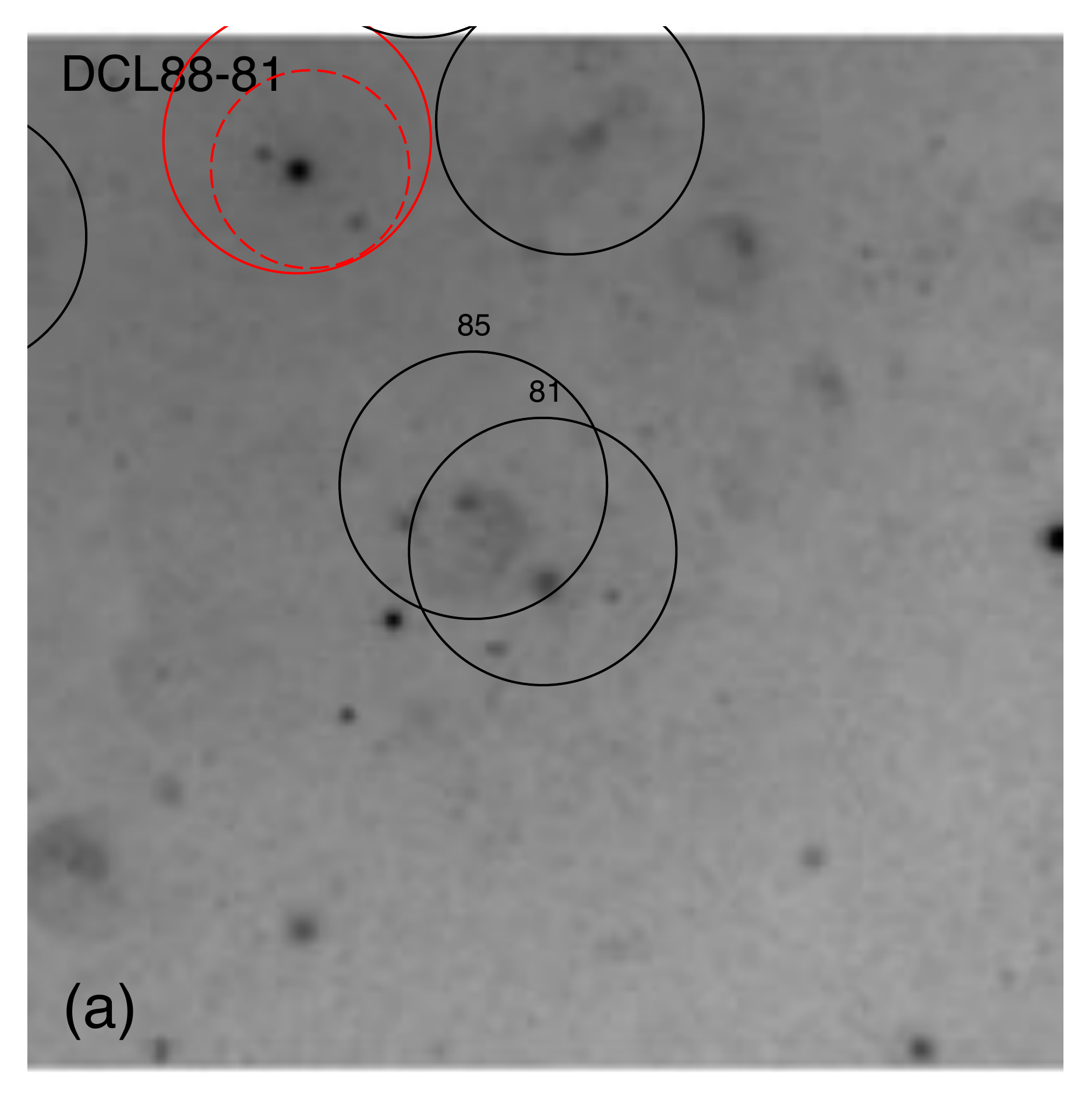}
\end{minipage}
\begin{minipage}{0.24\linewidth}
\includegraphics[width=\linewidth]{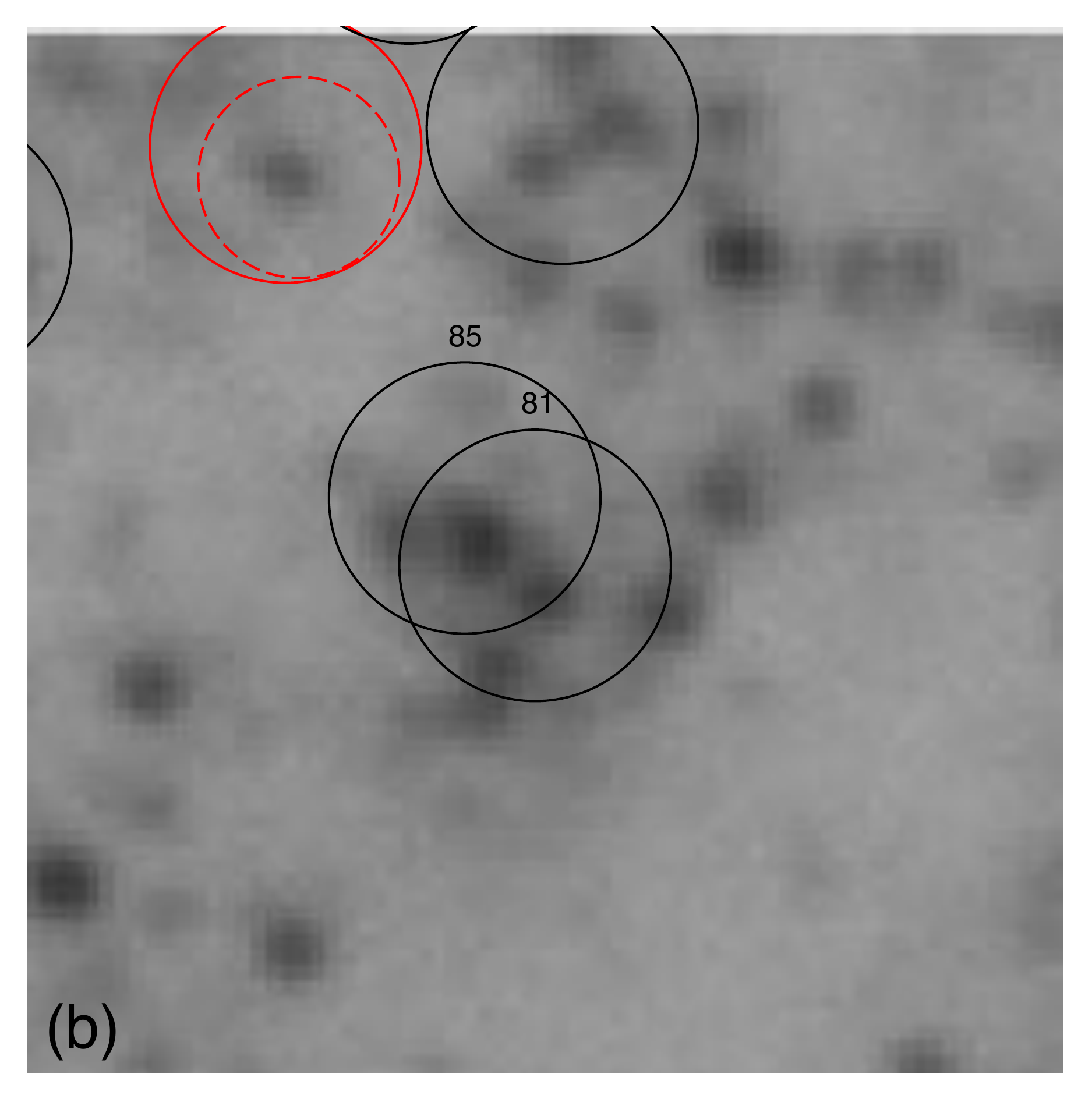}
\end{minipage}
\begin{minipage}{0.24\linewidth}
\includegraphics[width=\linewidth]{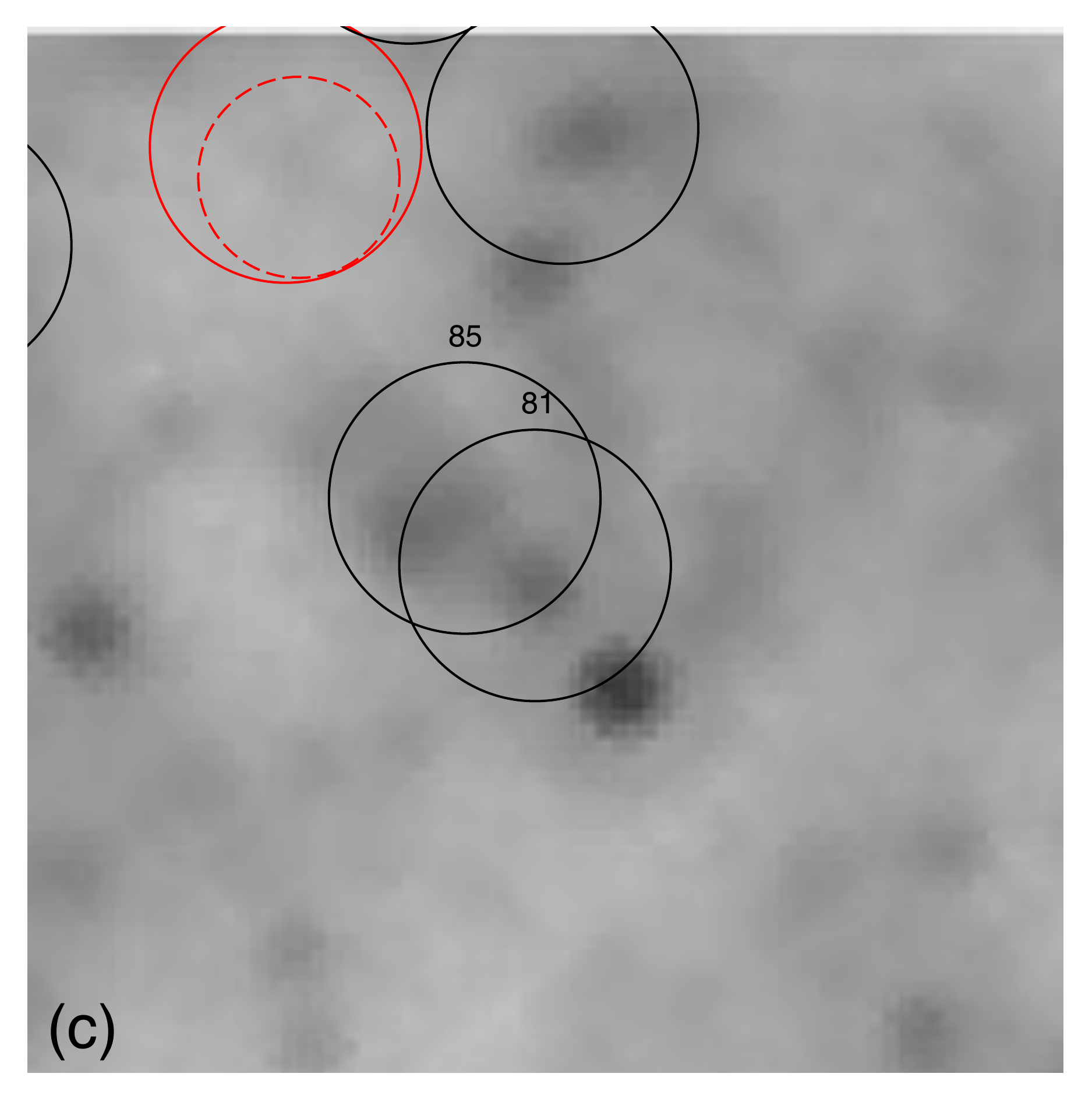}
\end{minipage}
\begin{minipage}{0.24\linewidth}
\includegraphics[width=\linewidth]{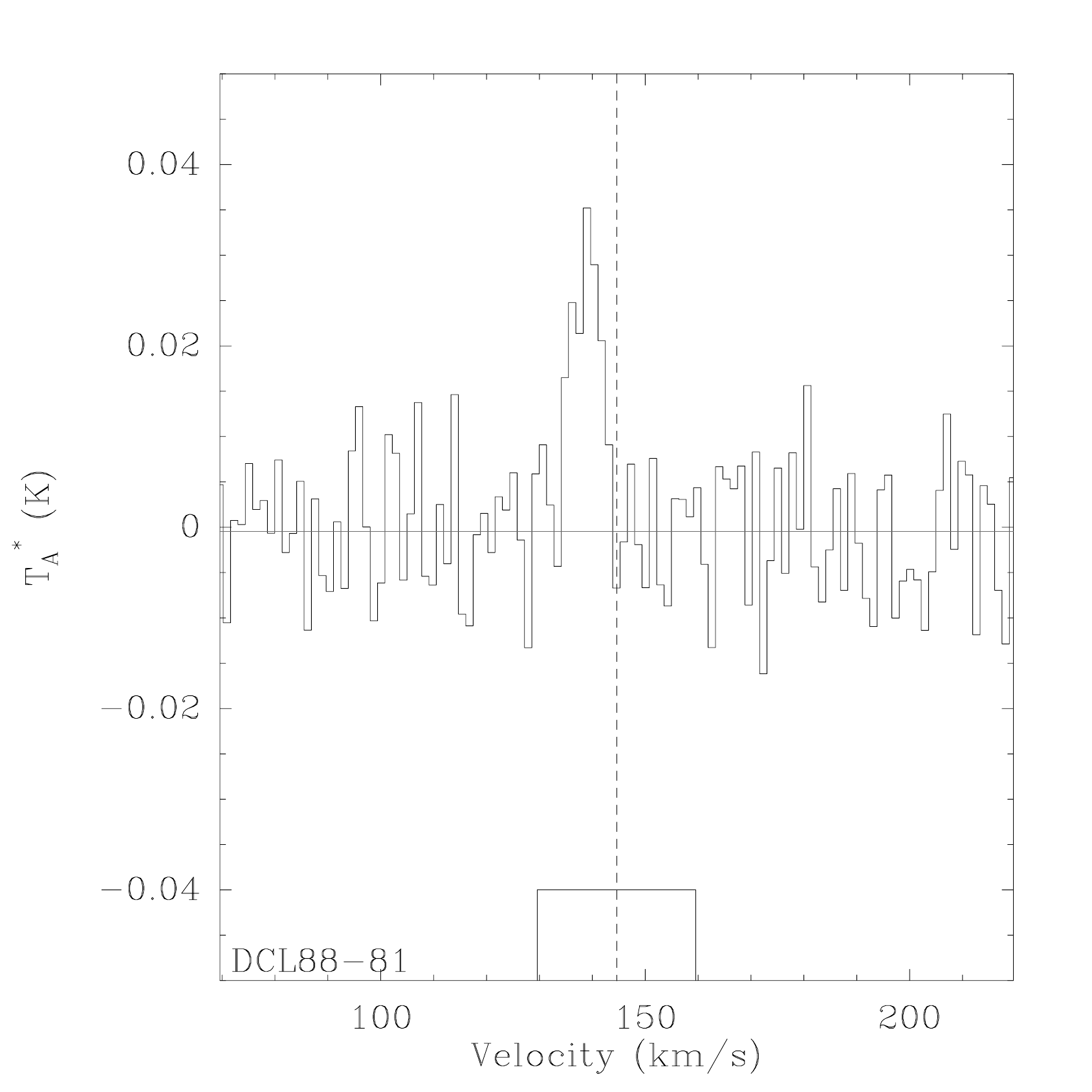}
\end{minipage}

\begin{minipage}{0.24\linewidth}
\includegraphics[width=\linewidth]{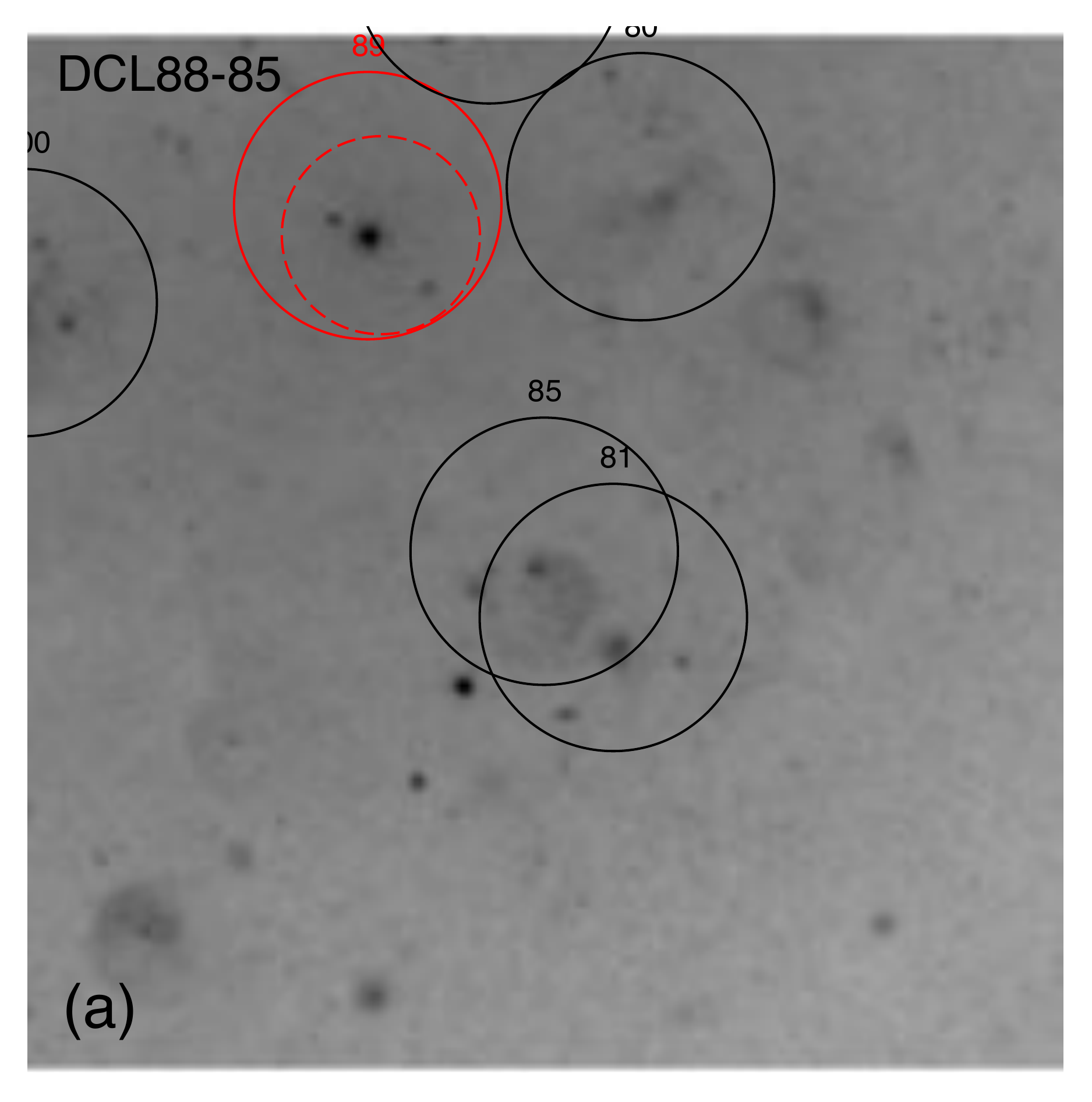}
\end{minipage}
\begin{minipage}{0.24\linewidth}
\includegraphics[width=\linewidth]{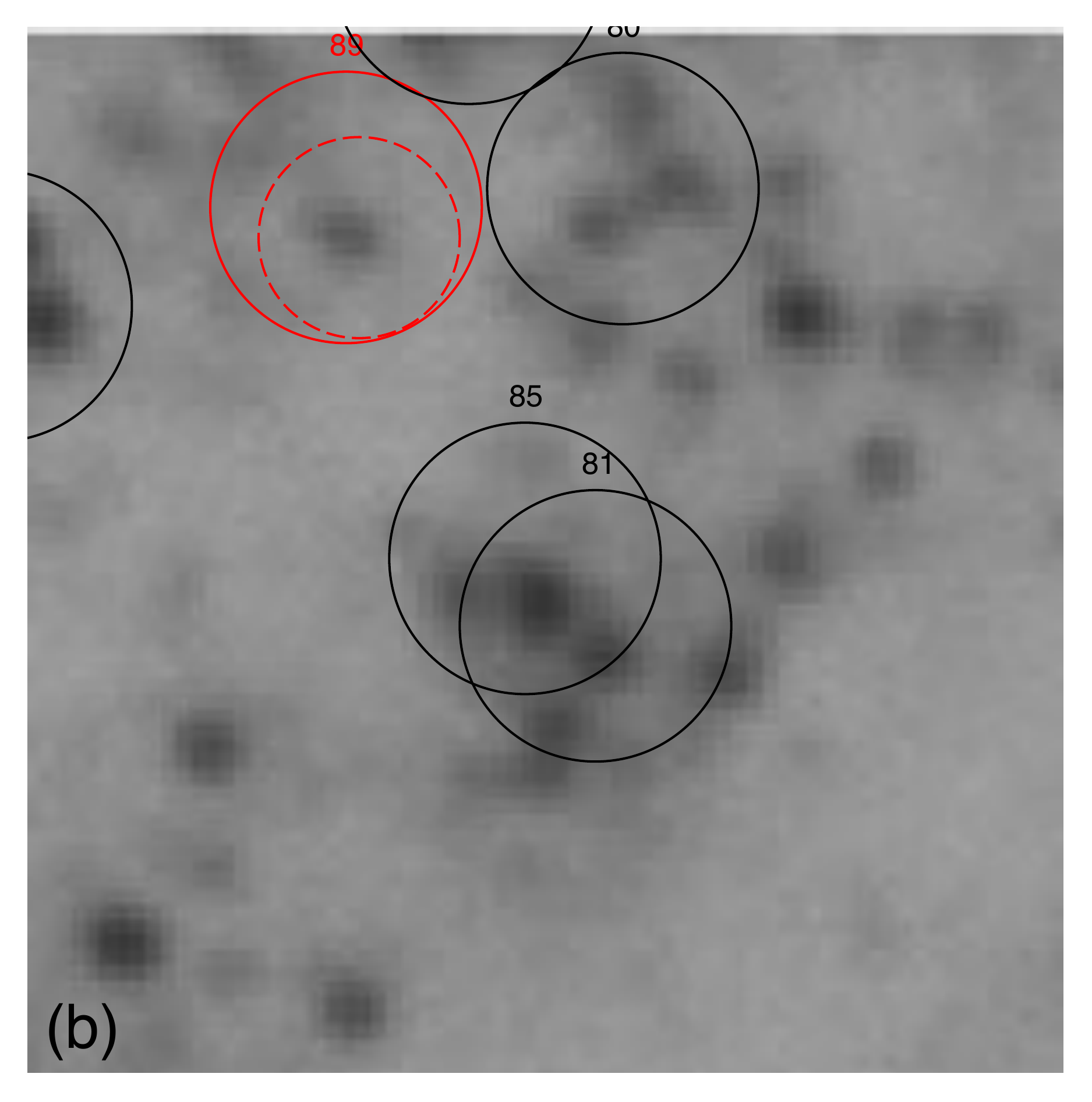}
\end{minipage}
\begin{minipage}{0.24\linewidth}
\includegraphics[width=\linewidth]{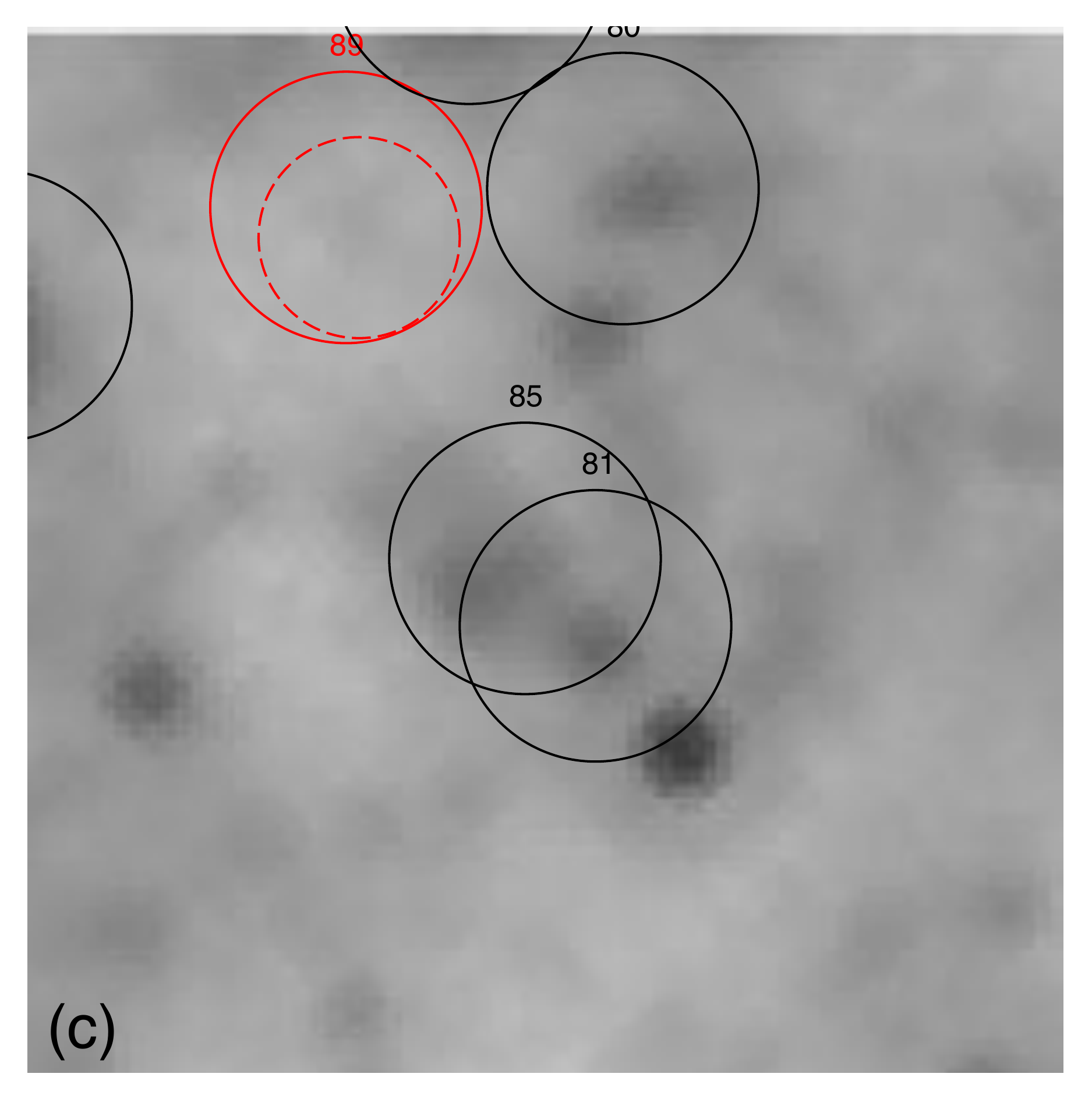}
\end{minipage}
\begin{minipage}{0.24\linewidth}
\includegraphics[width=\linewidth]{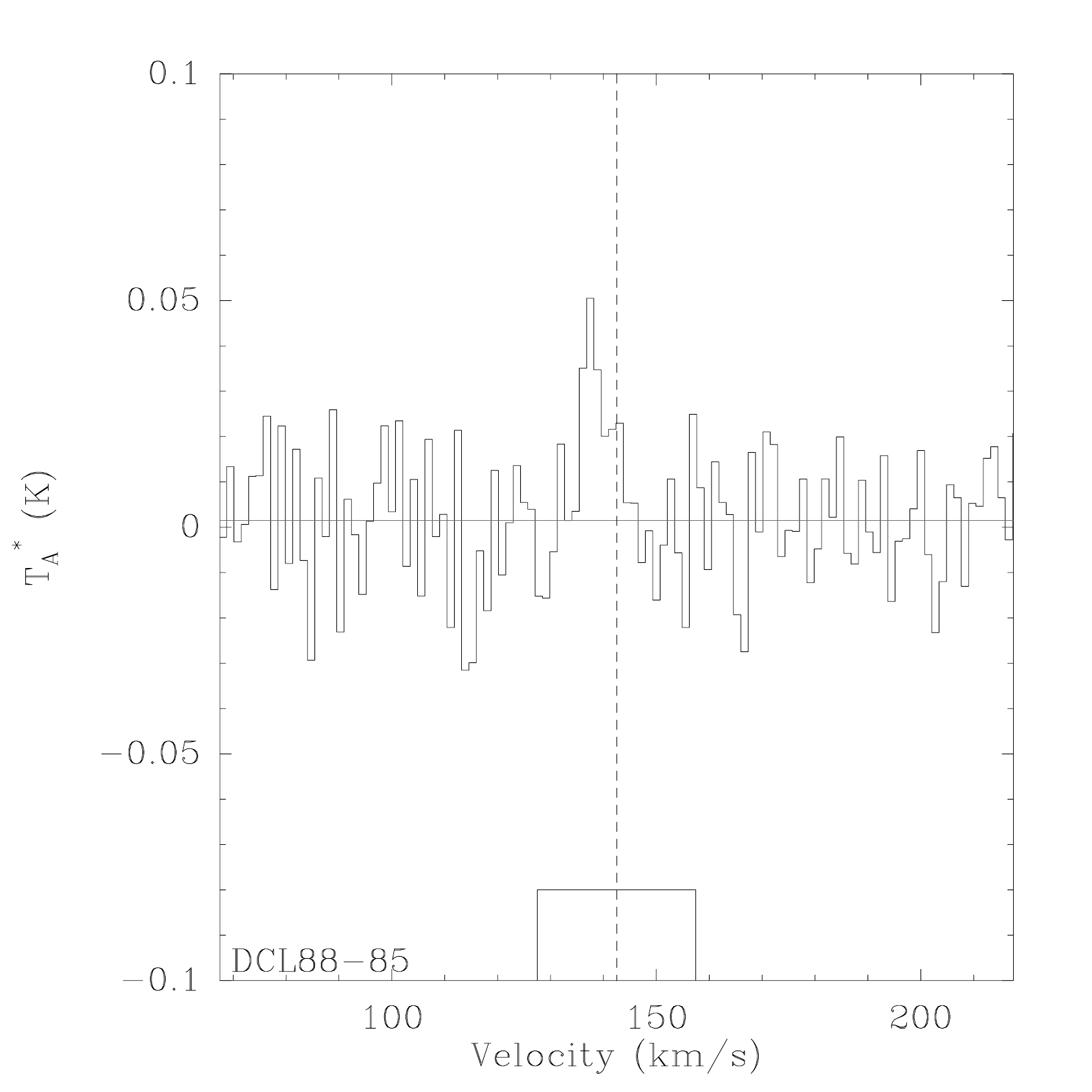}
\end{minipage}

\noindent\textbf{Figure~\ref{fig:stamps} -- continued.}
\end{figure*}

\begin{figure*}
%\ContinuedFloat

\begin{minipage}{0.24\linewidth}
\includegraphics[width=\linewidth]{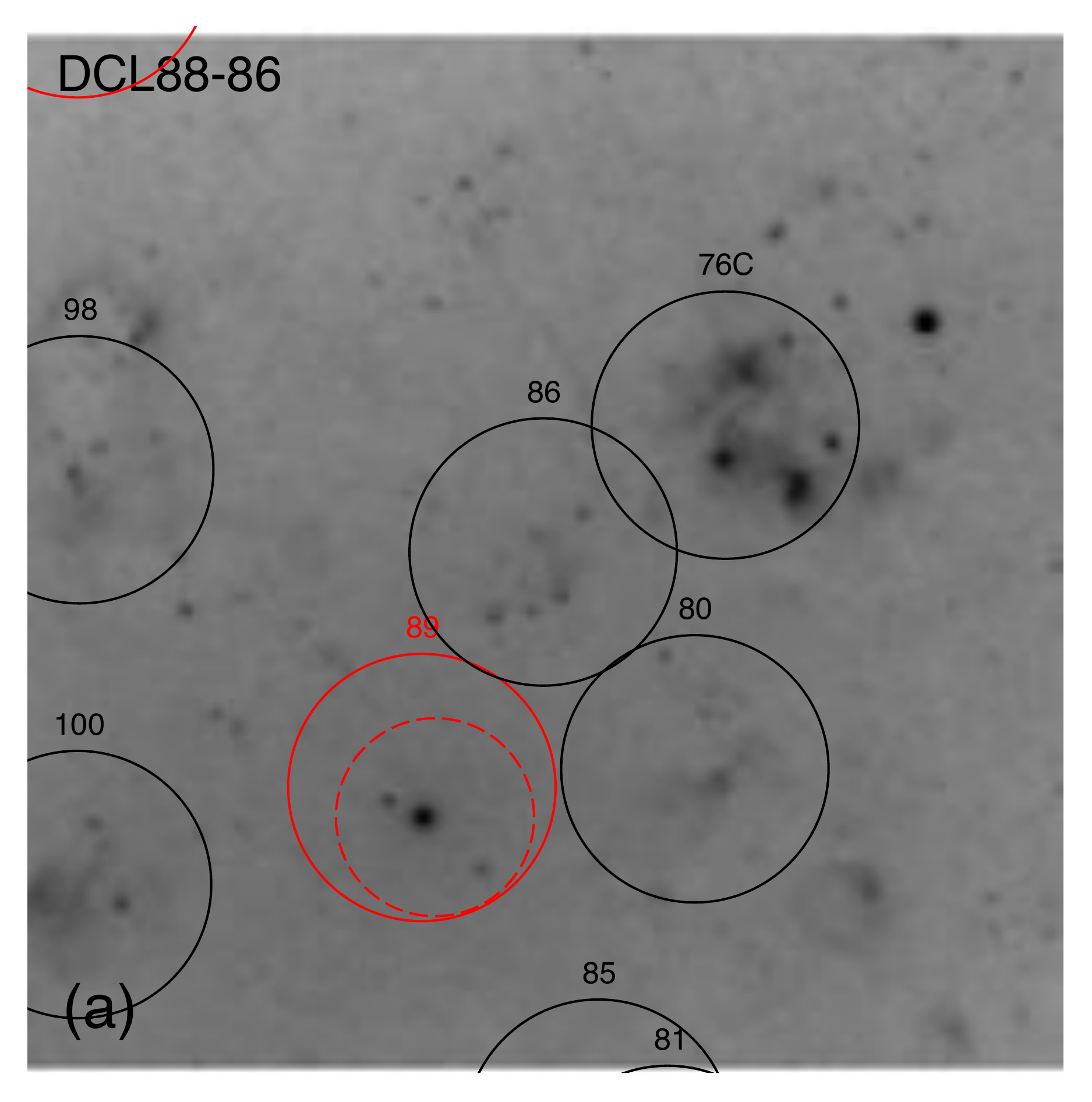}
\end{minipage}
\begin{minipage}{0.24\linewidth}
\includegraphics[width=\linewidth]{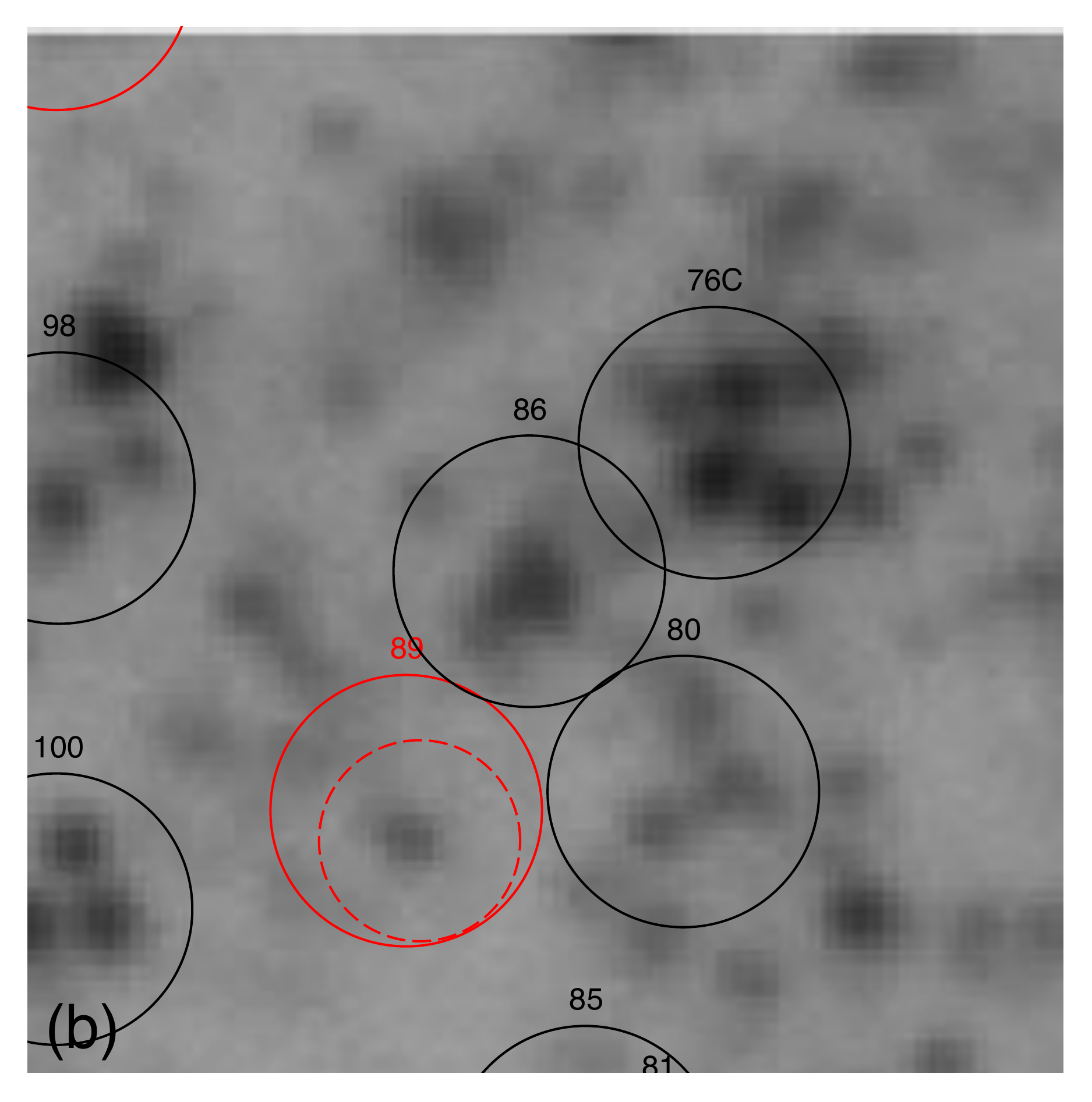}
\end{minipage}
\begin{minipage}{0.24\linewidth}
\includegraphics[width=\linewidth]{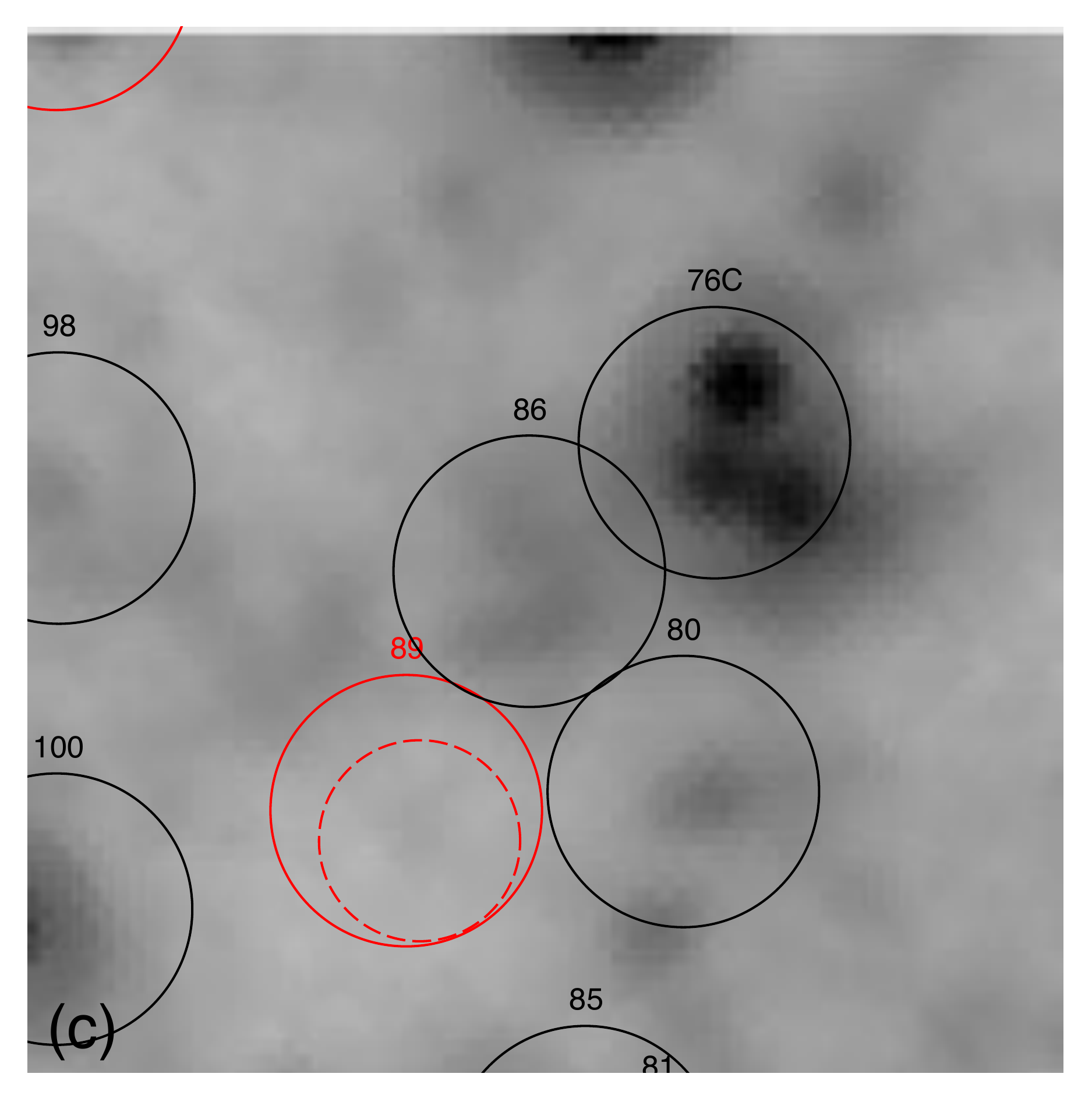}
\end{minipage}
\begin{minipage}{0.24\linewidth}
\includegraphics[width=\linewidth]{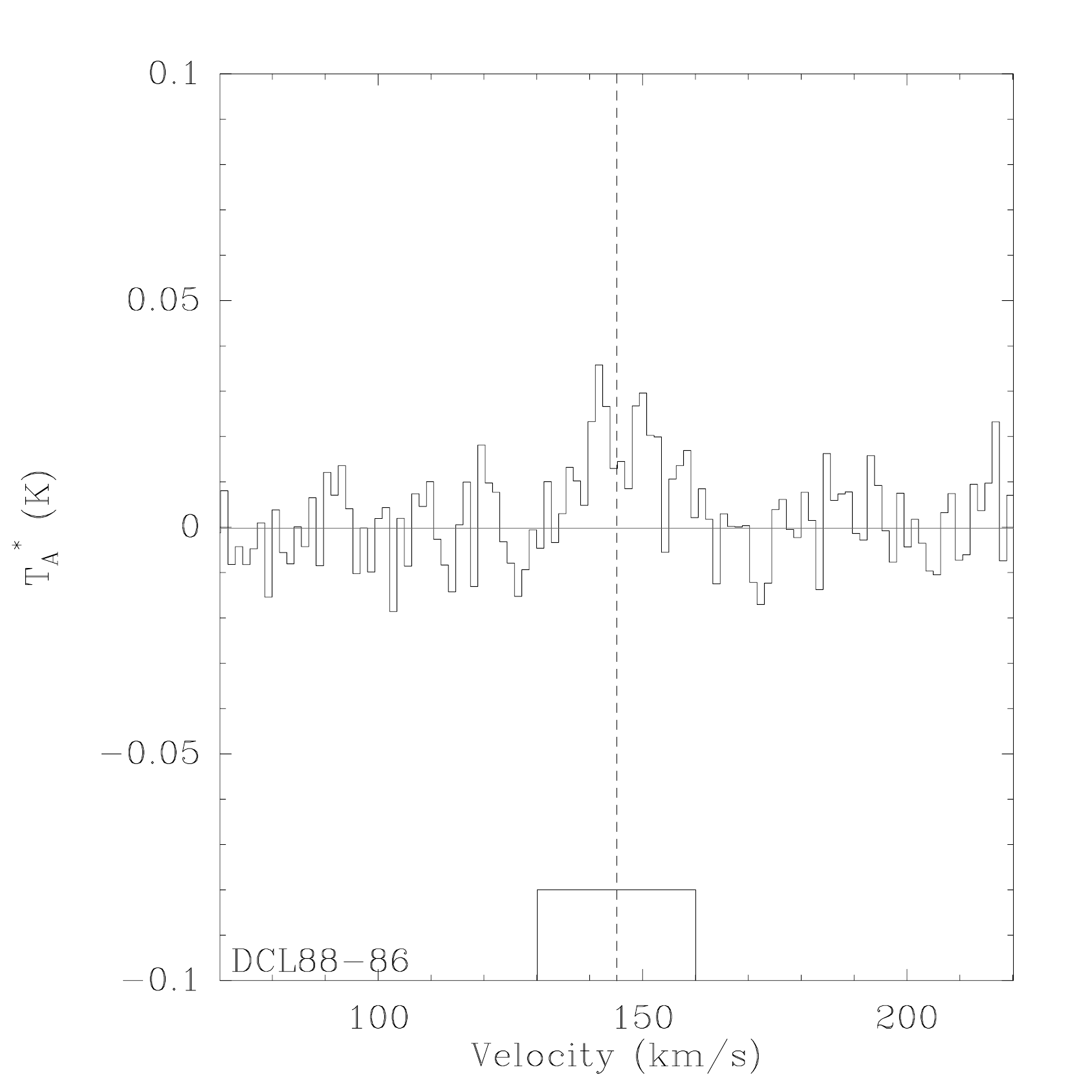}
\end{minipage}

\begin{minipage}{0.24\linewidth}
\includegraphics[width=\linewidth]{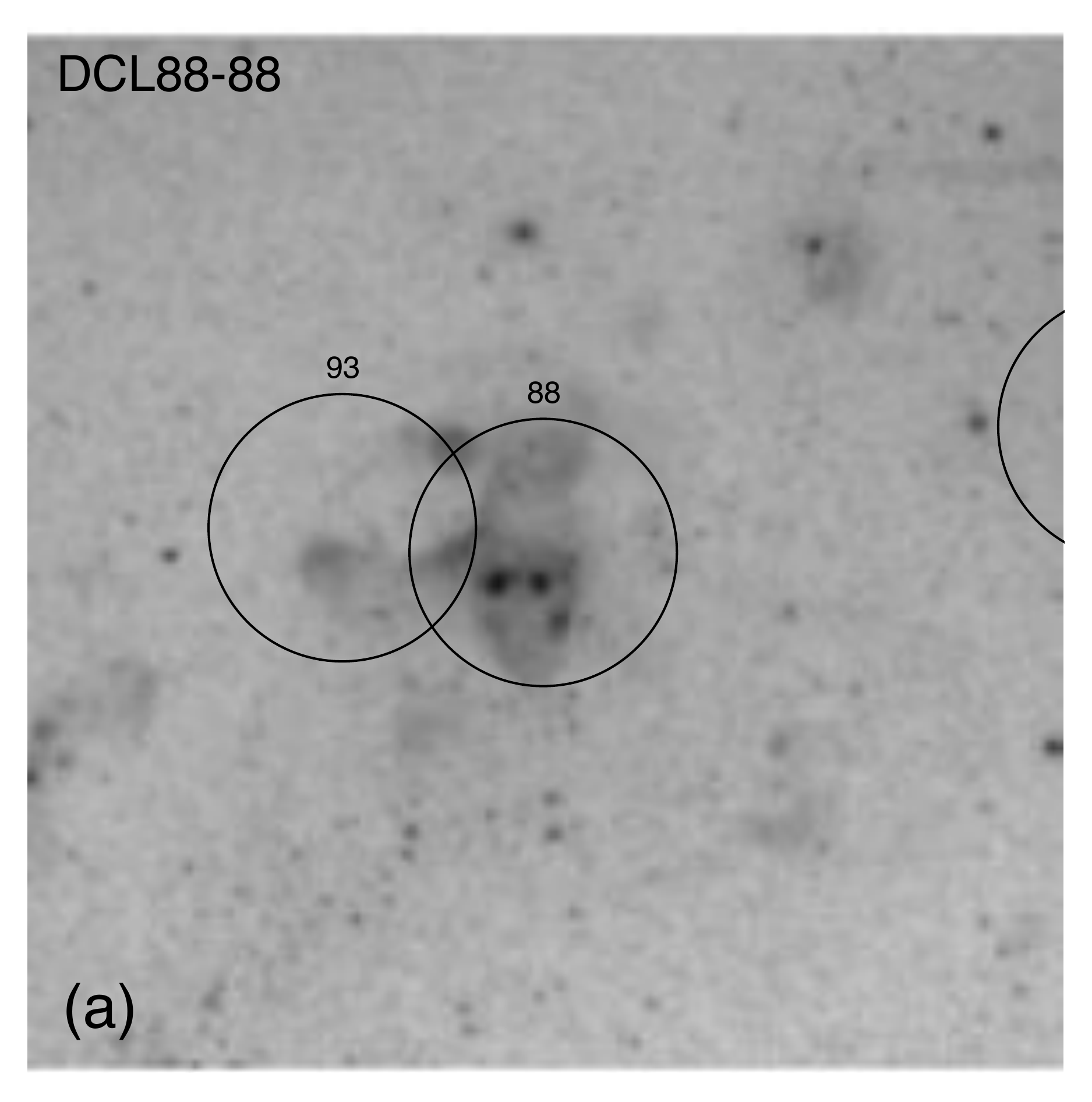}
\end{minipage}
\begin{minipage}{0.24\linewidth}
\includegraphics[width=\linewidth]{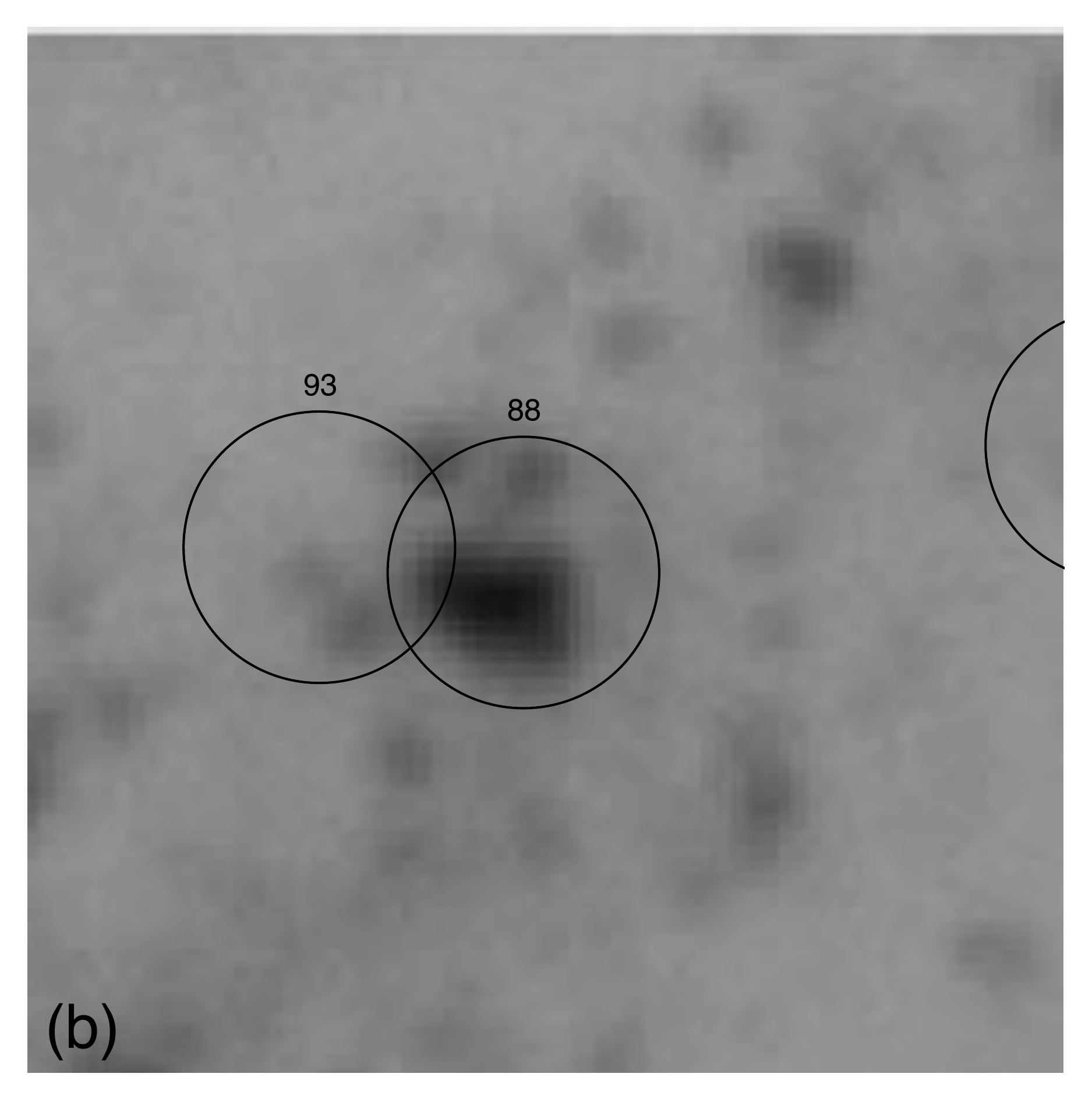}
\end{minipage}
\begin{minipage}{0.24\linewidth}
\includegraphics[width=\linewidth]{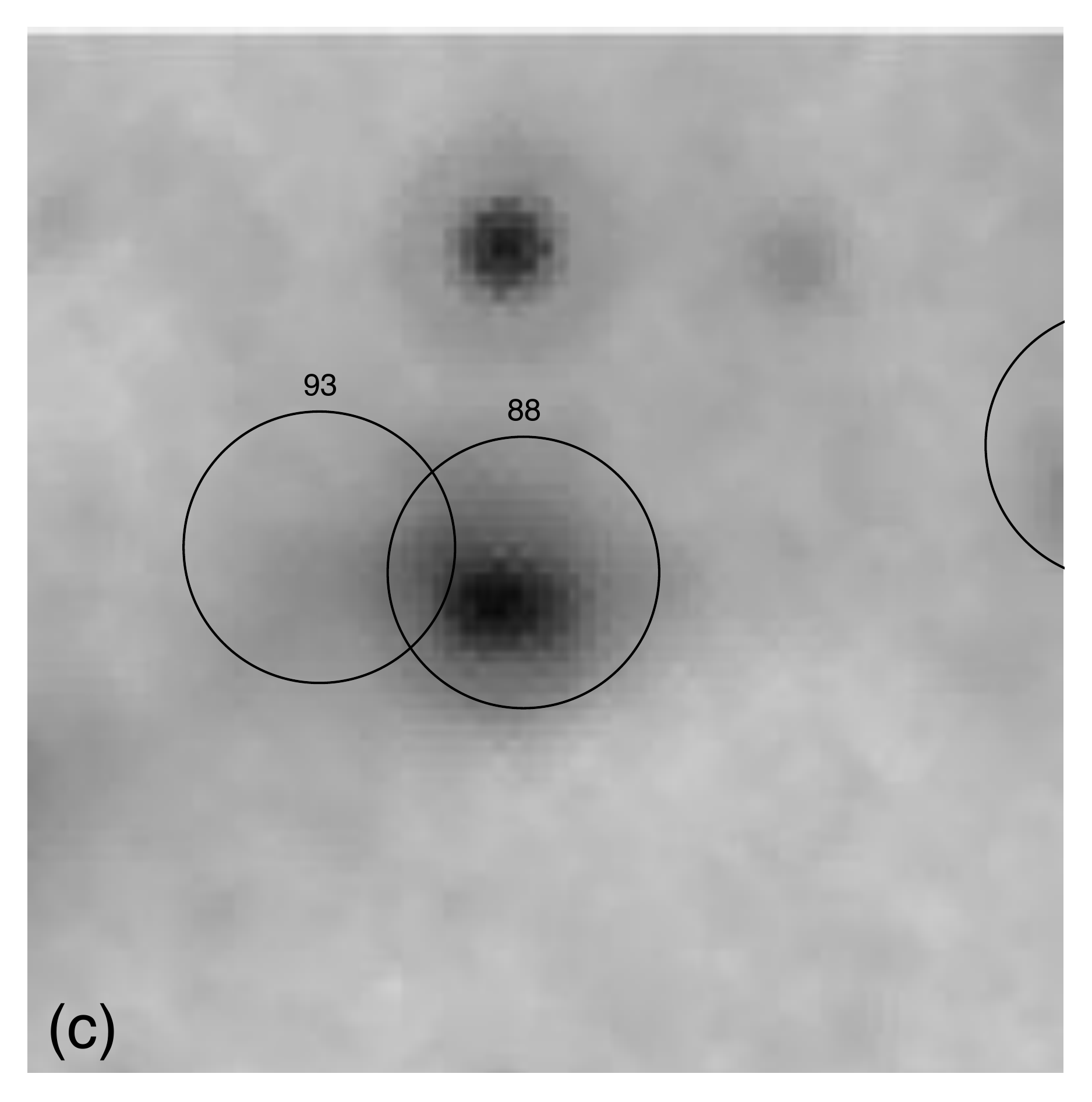}
\end{minipage}
\begin{minipage}{0.24\linewidth}
\includegraphics[width=\linewidth]{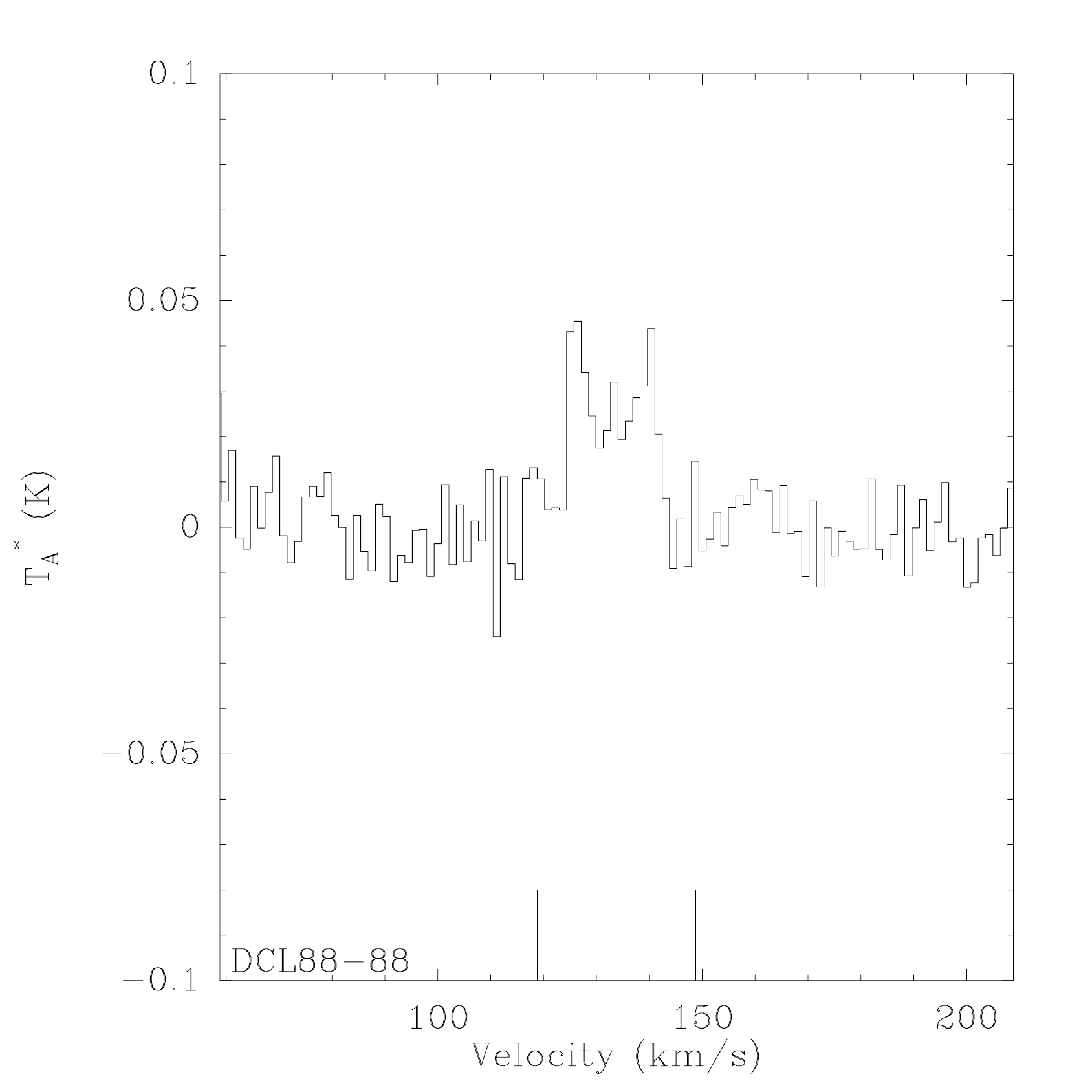}
\end{minipage}

\begin{minipage}{0.24\linewidth}
\includegraphics[width=\linewidth]{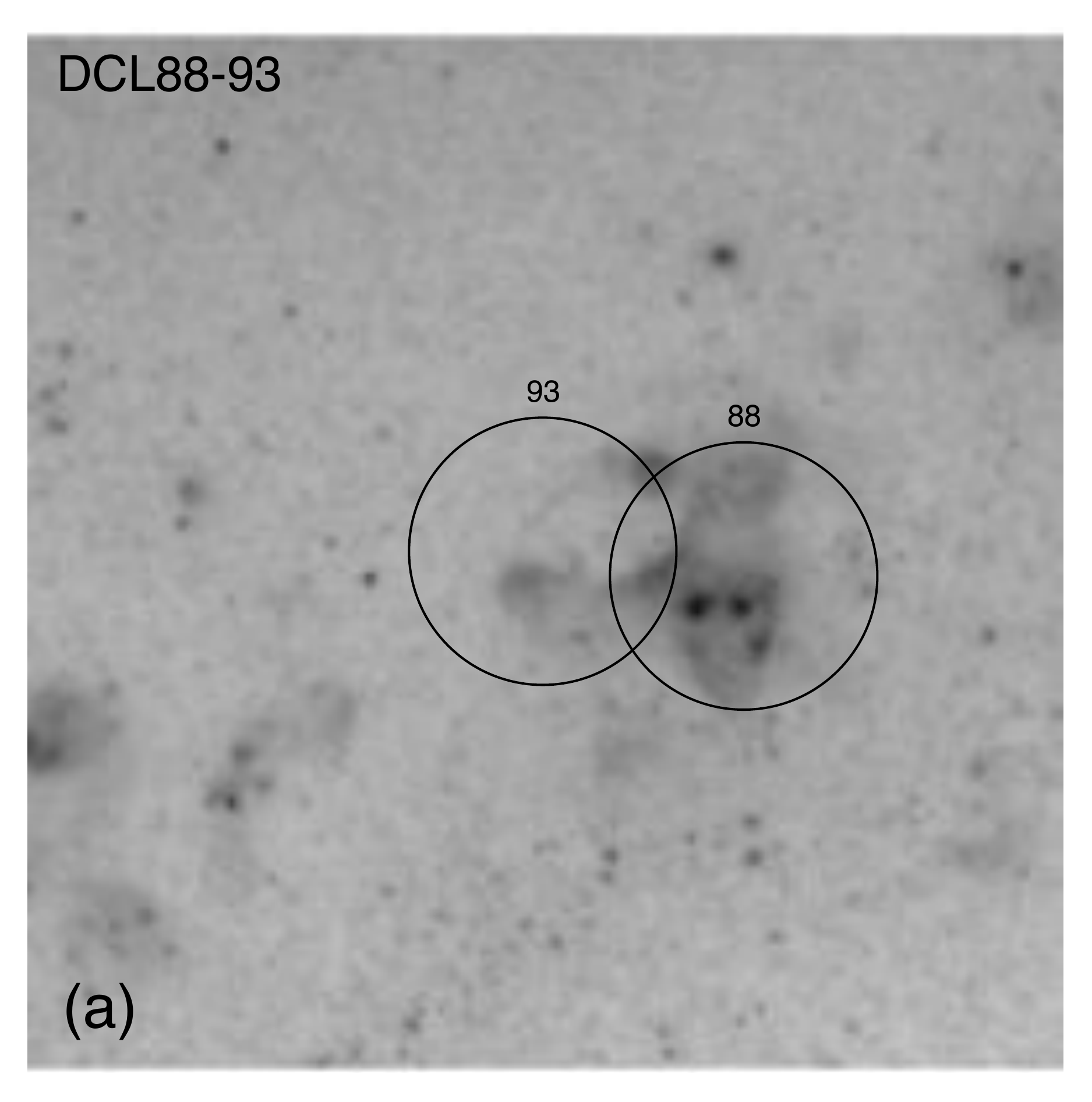}
\end{minipage}
\begin{minipage}{0.24\linewidth}
\includegraphics[width=\linewidth]{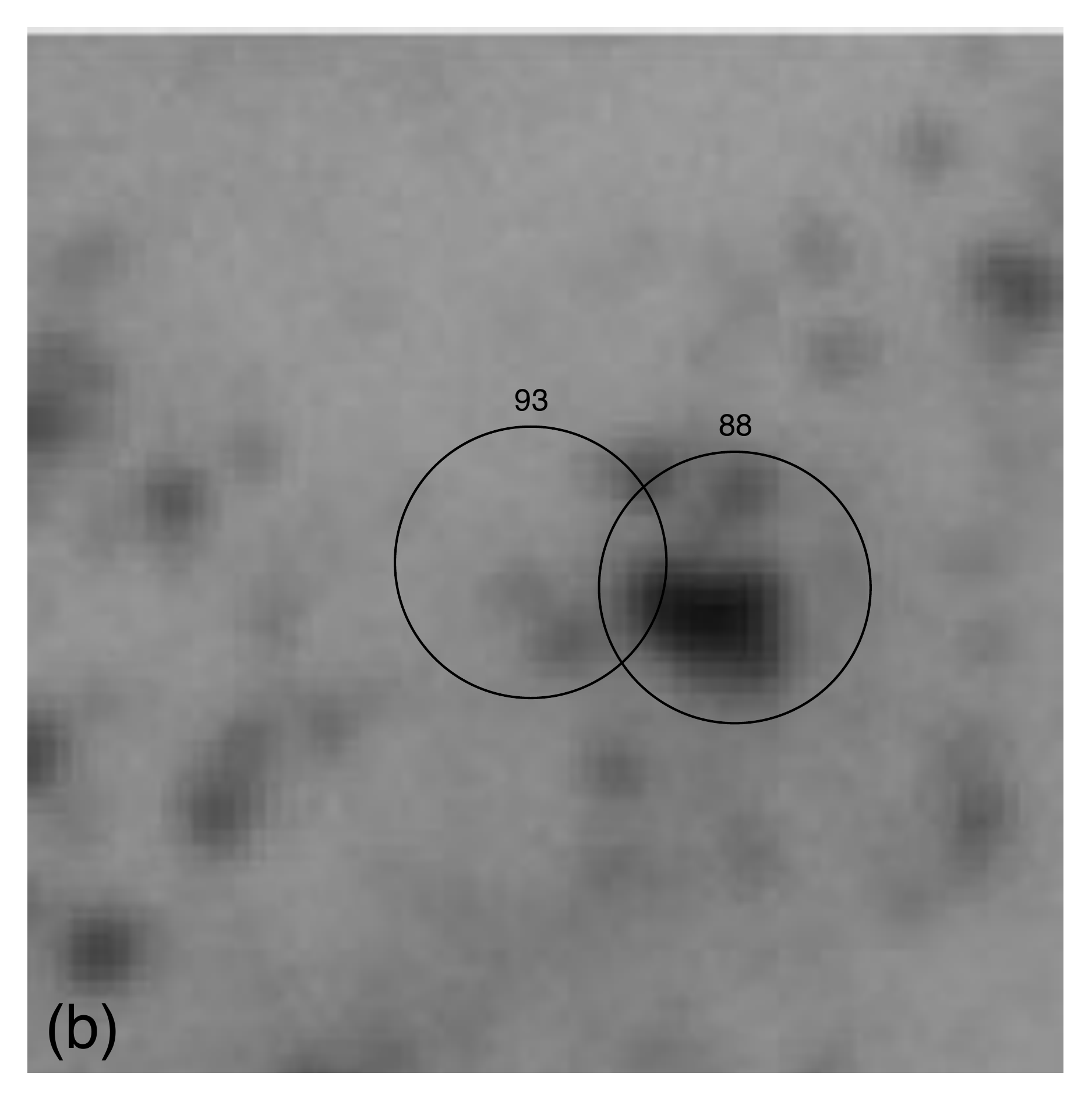}
\end{minipage}
\begin{minipage}{0.24\linewidth}
\includegraphics[width=\linewidth]{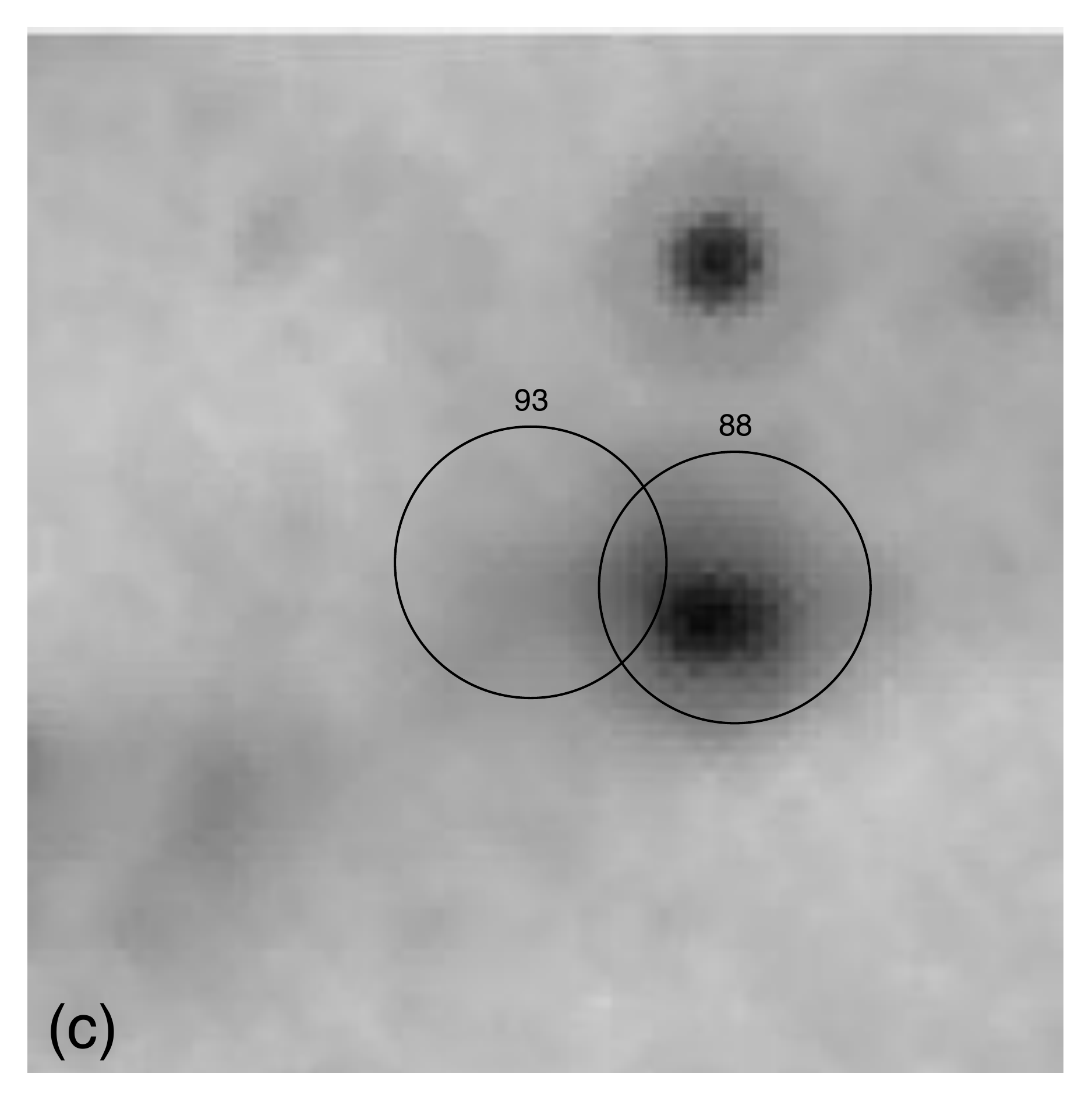}
\end{minipage}
\begin{minipage}{0.24\linewidth}
\includegraphics[width=\linewidth]{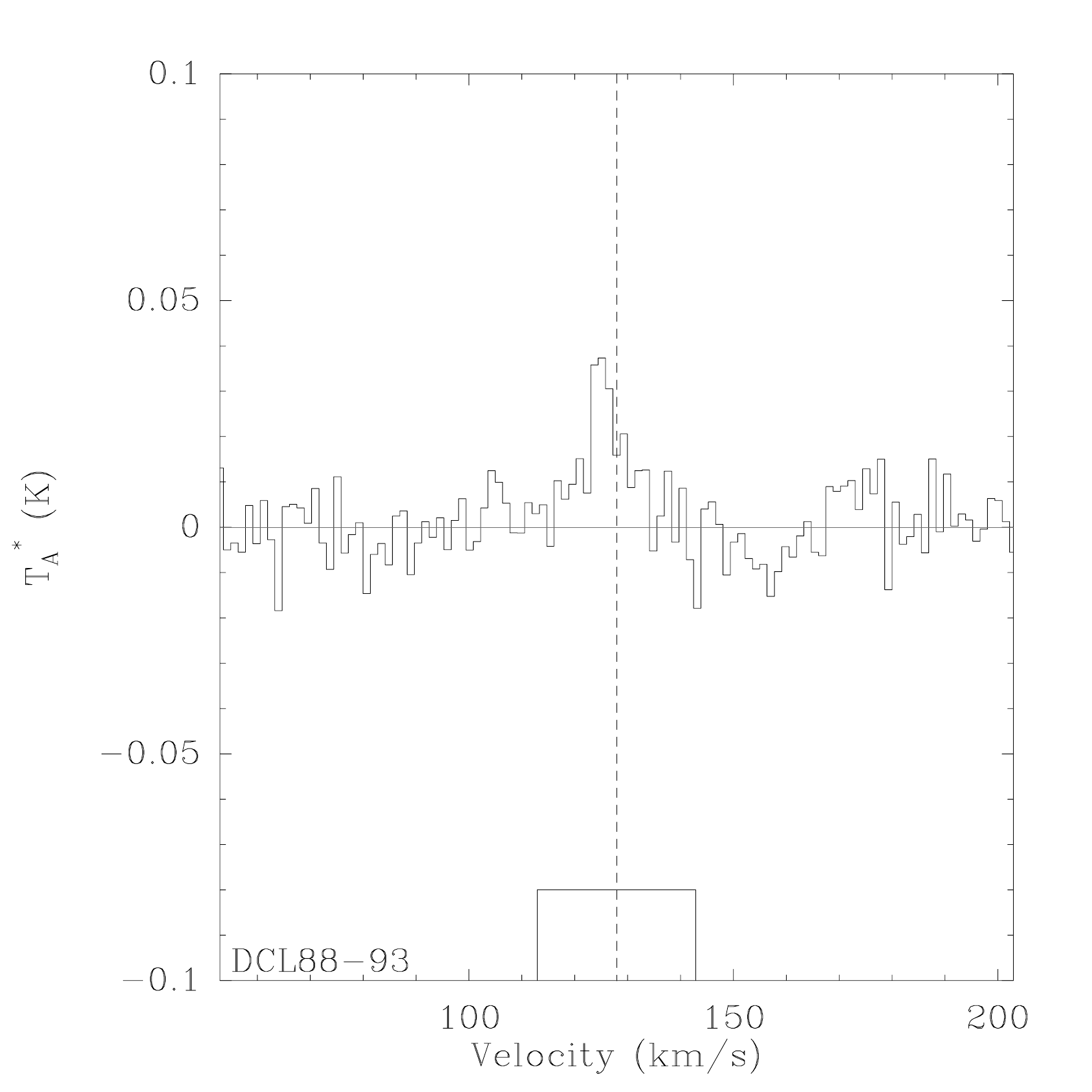}
\end{minipage}

\begin{minipage}{0.24\linewidth}
\includegraphics[width=\linewidth]{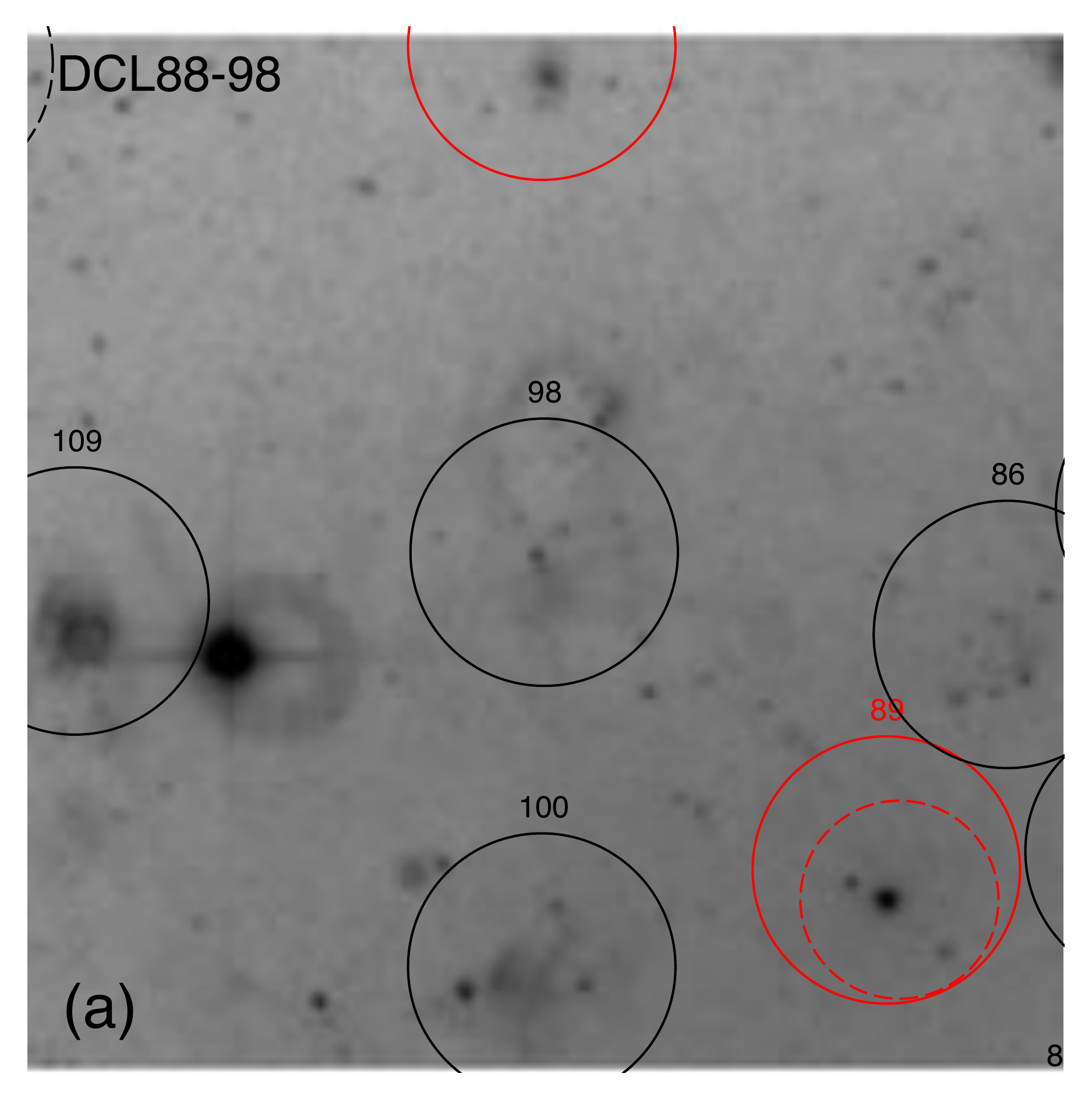}
\end{minipage}
\begin{minipage}{0.24\linewidth}
\includegraphics[width=\linewidth]{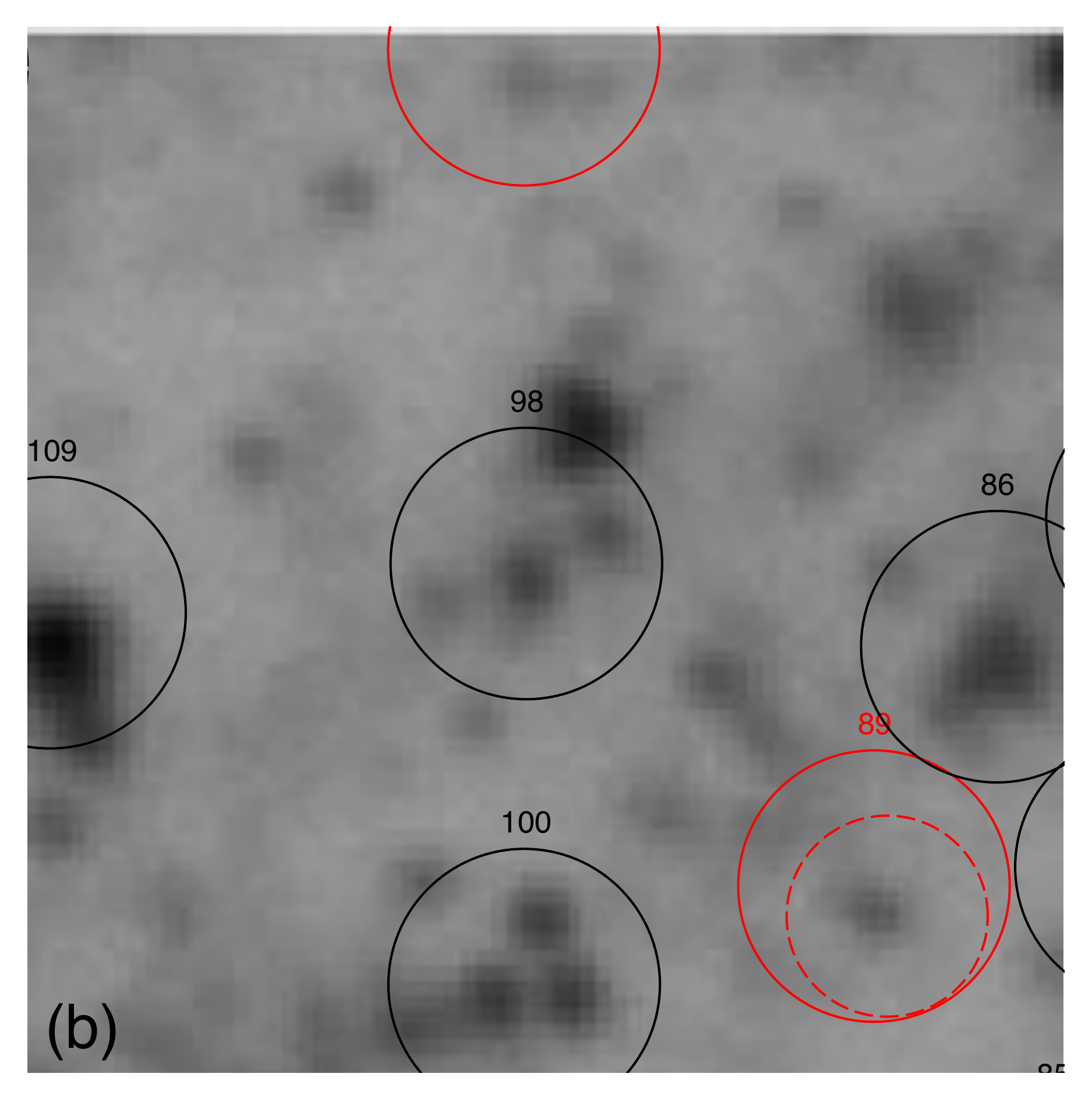}
\end{minipage}
\begin{minipage}{0.24\linewidth}
\includegraphics[width=\linewidth]{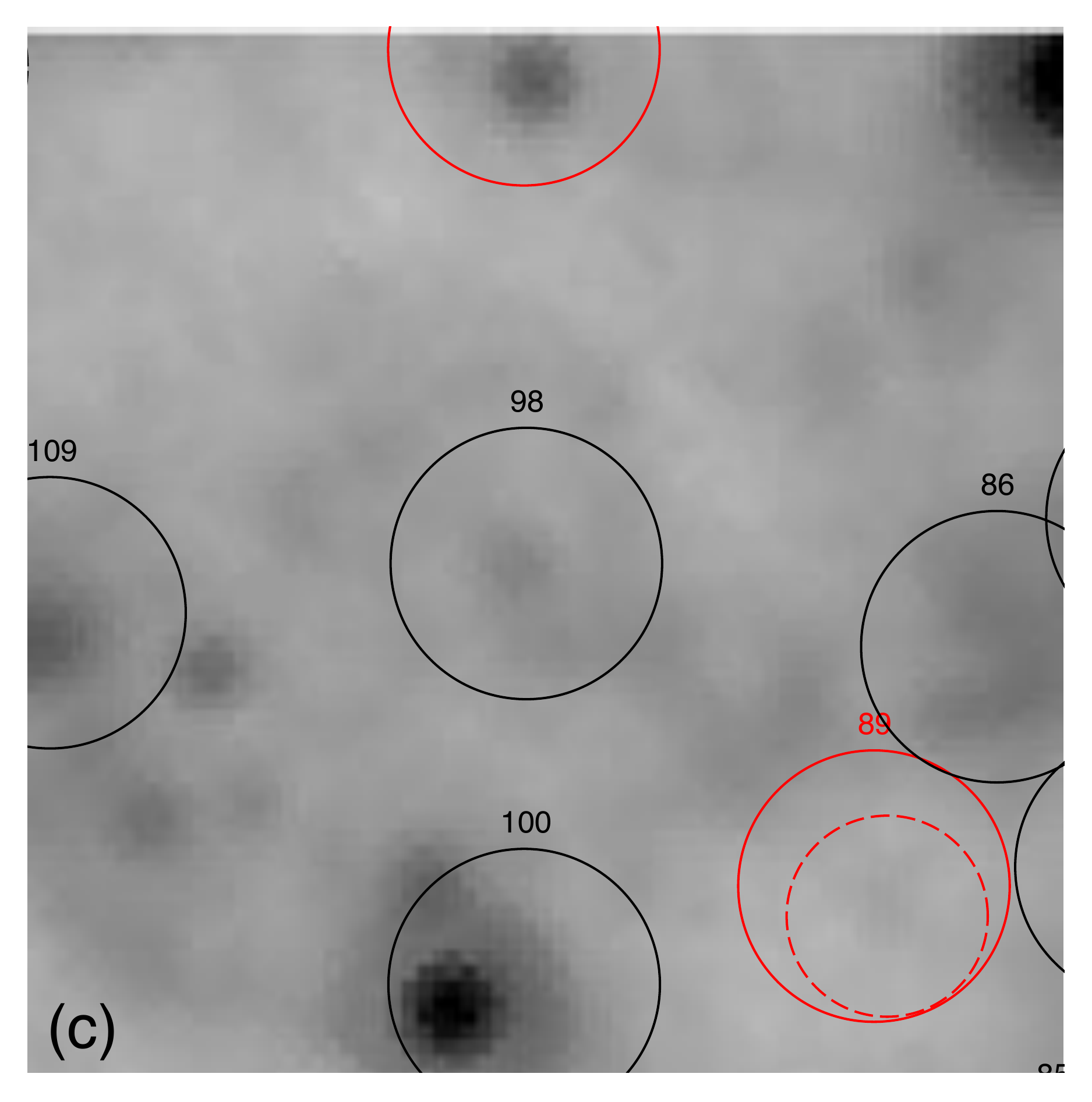}
\end{minipage}
\begin{minipage}{0.24\linewidth}
\includegraphics[width=\linewidth]{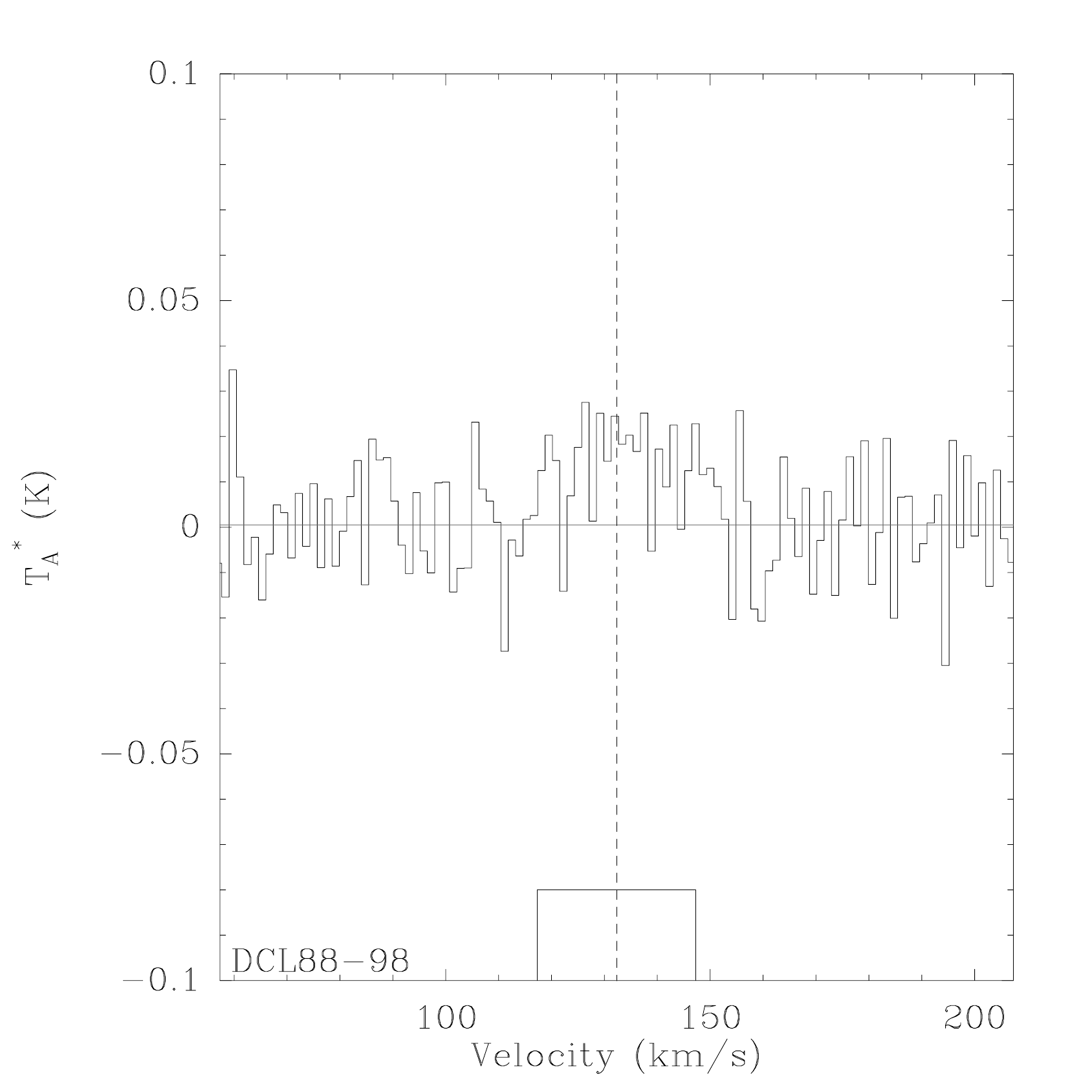}
\end{minipage}

\begin{minipage}{0.24\linewidth}
\includegraphics[width=\linewidth]{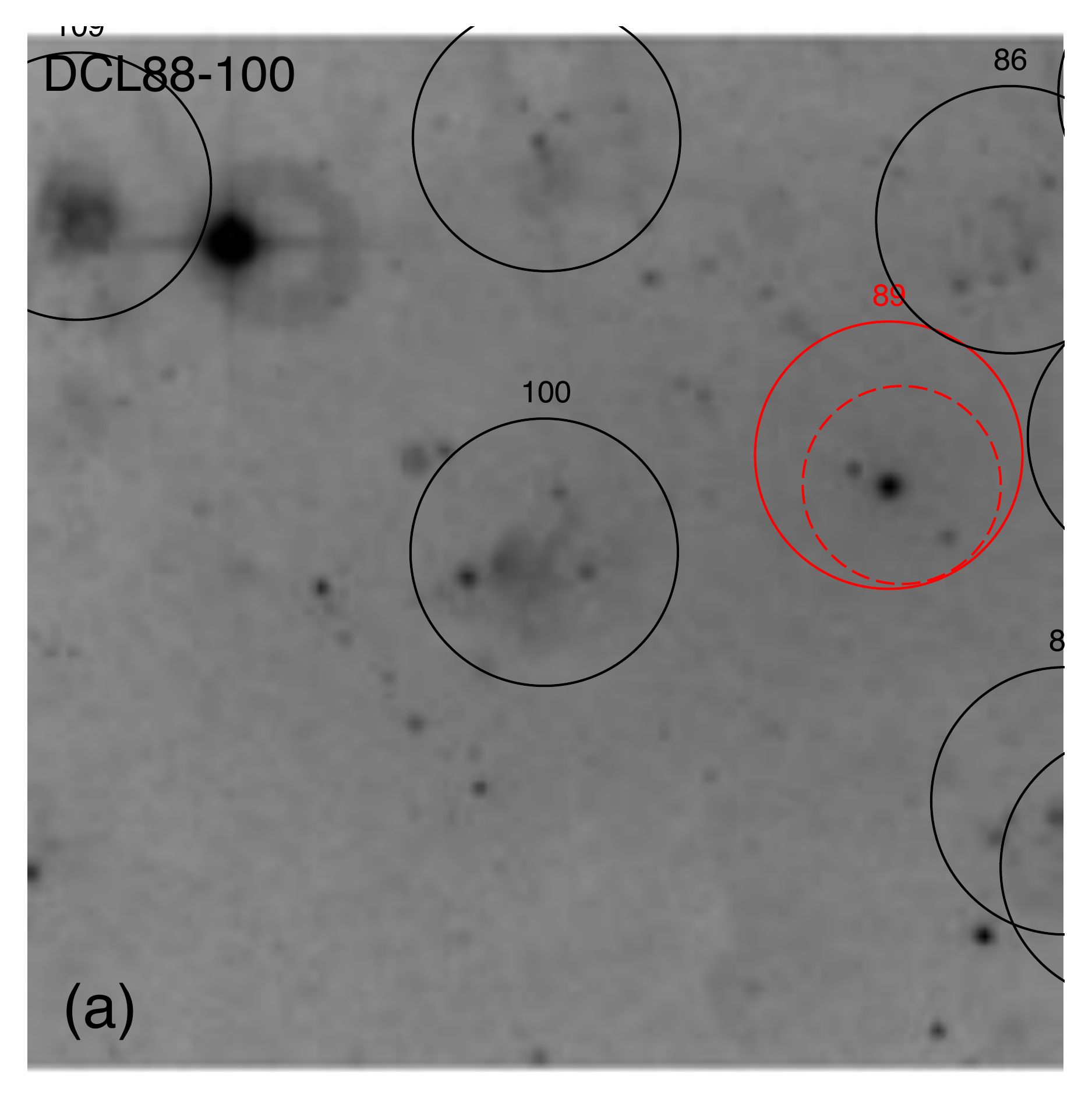}
\end{minipage}
\begin{minipage}{0.24\linewidth}
\includegraphics[width=\linewidth]{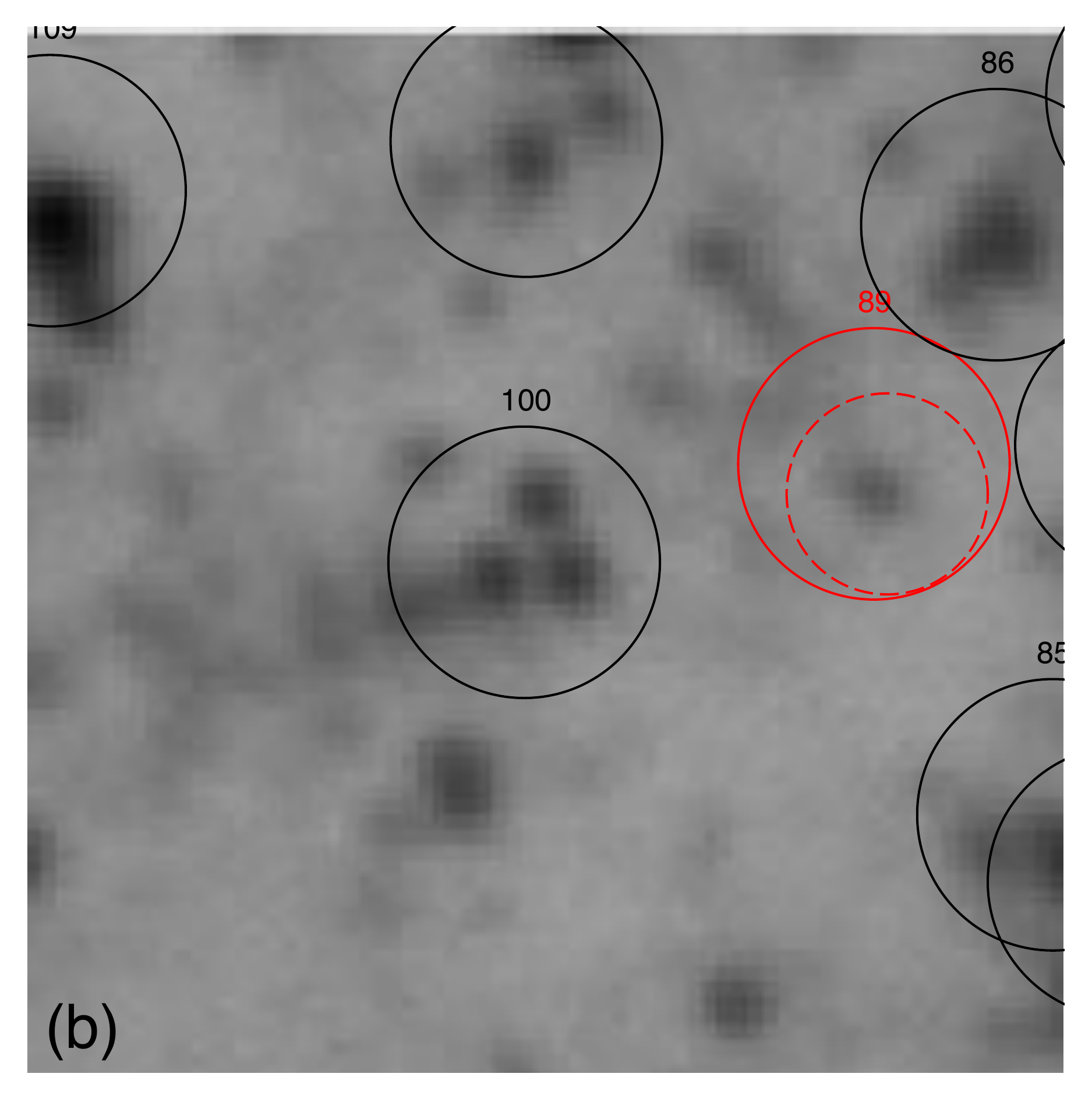}
\end{minipage}
\begin{minipage}{0.24\linewidth}
\includegraphics[width=\linewidth]{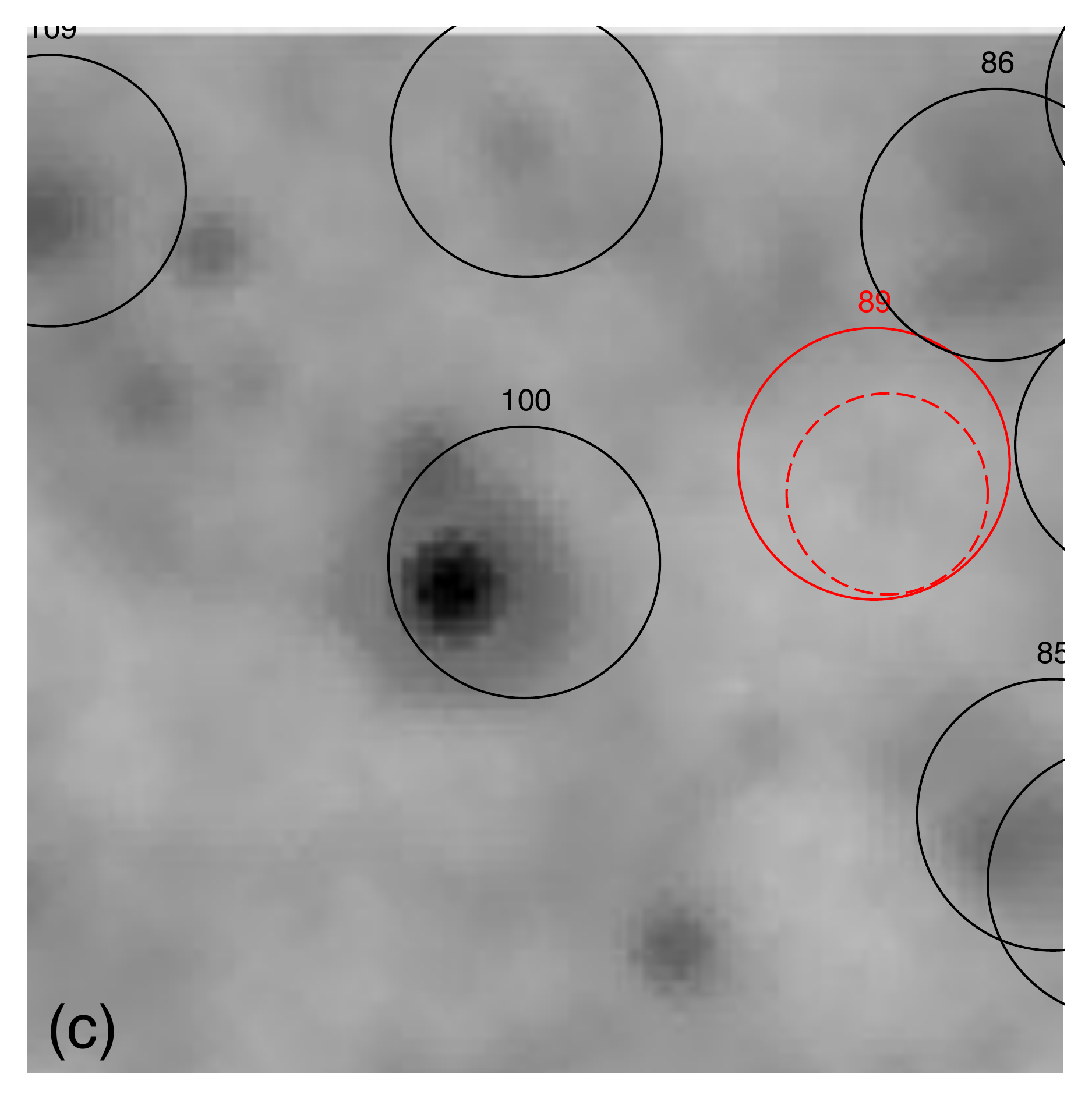}
\end{minipage}
\begin{minipage}{0.24\linewidth}
\includegraphics[width=\linewidth]{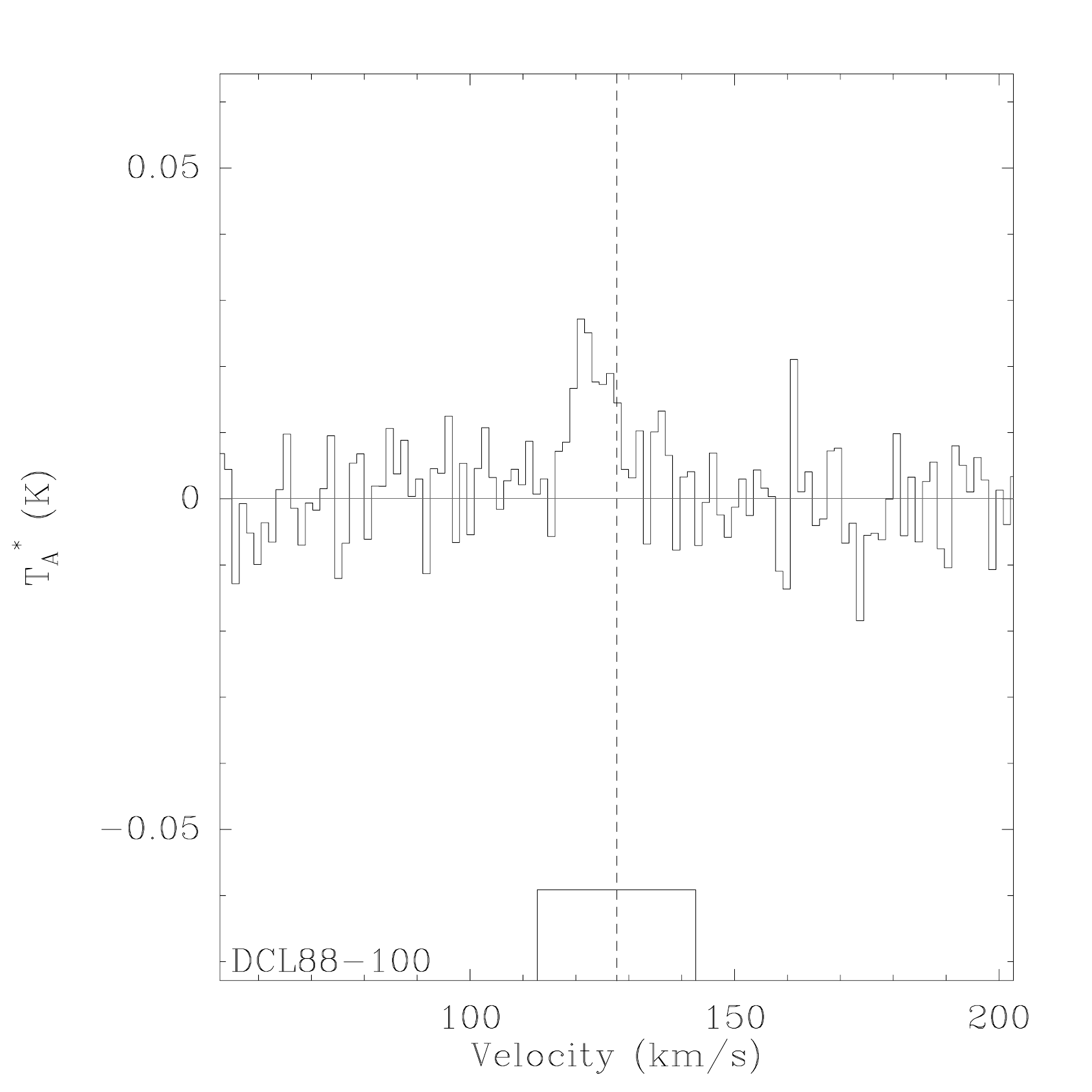}
\end{minipage}

\noindent\textbf{Figure~\ref{fig:stamps} -- continued.}

\end{figure*}

\begin{figure*}
%\ContinuedFloat

\begin{minipage}{0.24\linewidth}
\includegraphics[width=\linewidth]{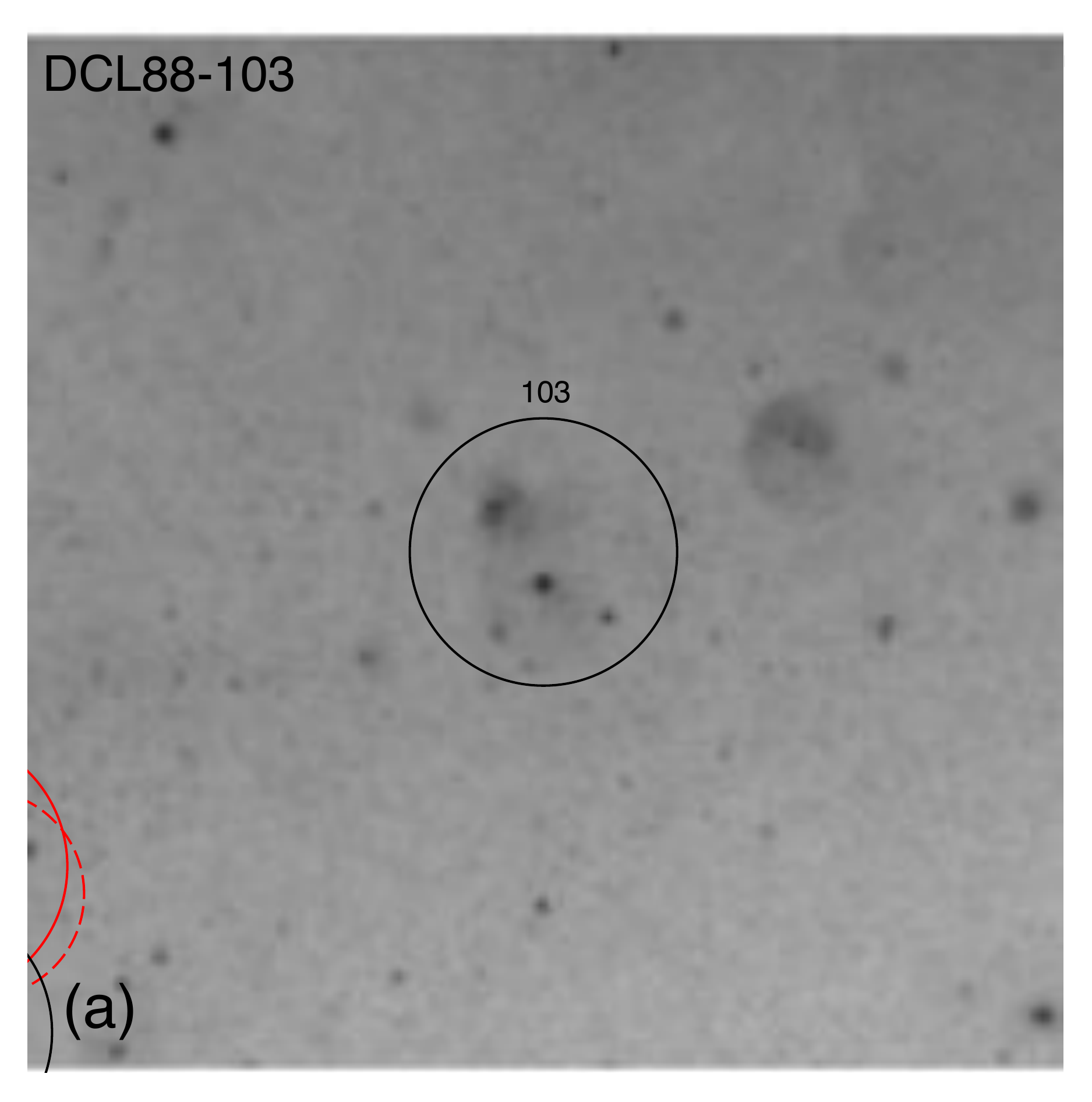}
\end{minipage}
\begin{minipage}{0.24\linewidth}
\includegraphics[width=\linewidth]{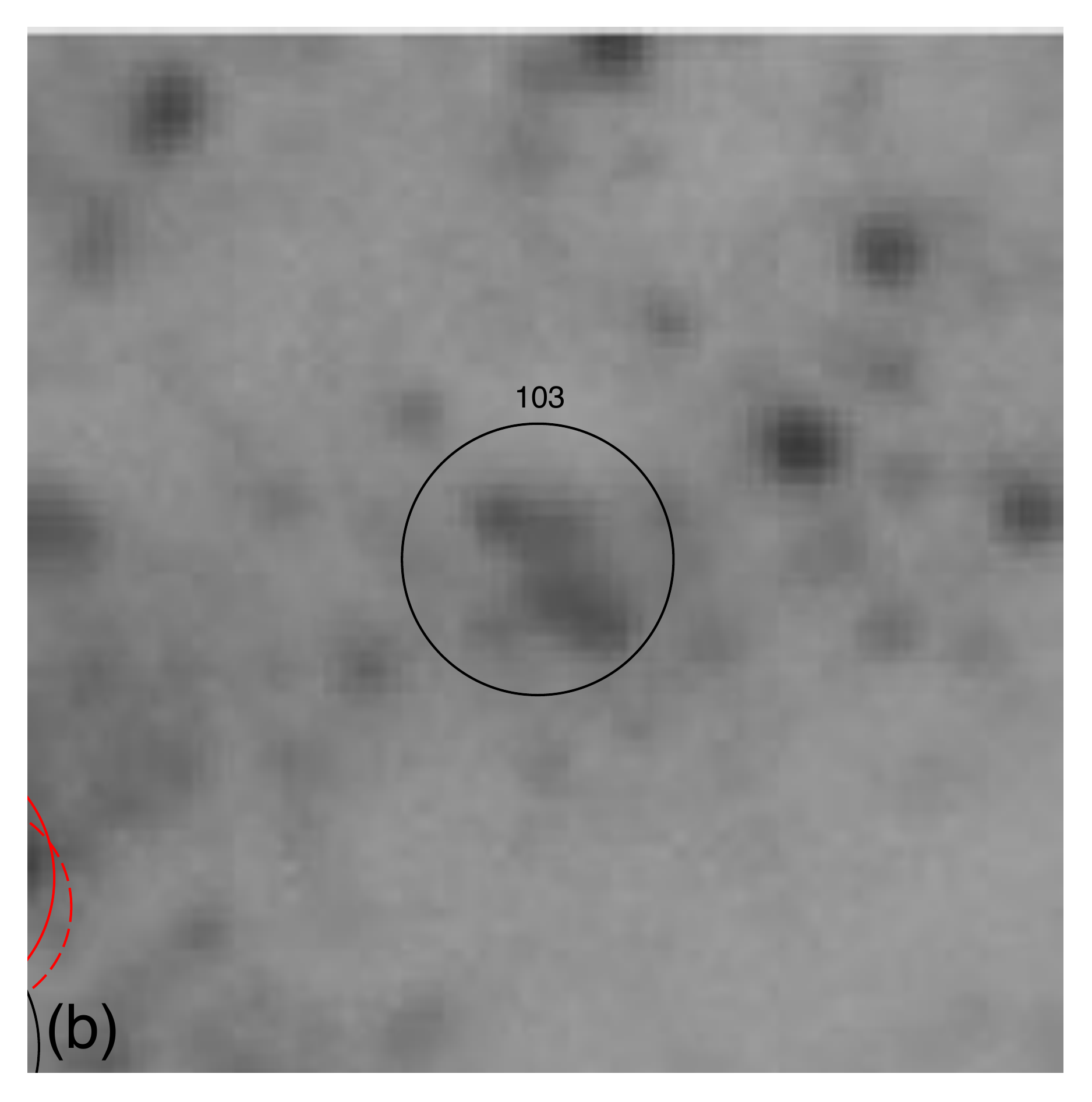}
\end{minipage}
\begin{minipage}{0.24\linewidth}
\includegraphics[width=\linewidth]{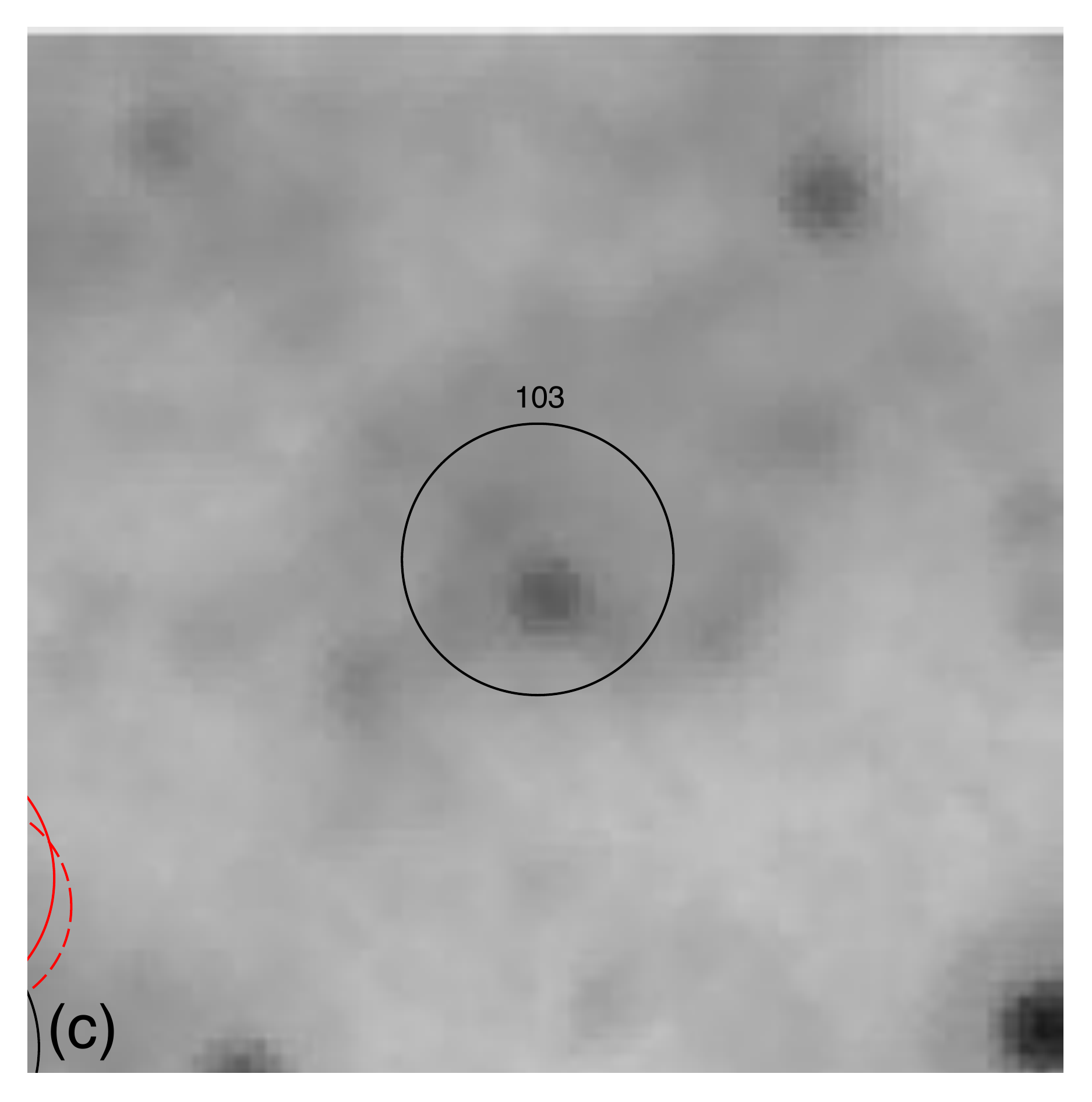}
\end{minipage}
\begin{minipage}{0.24\linewidth}
\includegraphics[width=\linewidth]{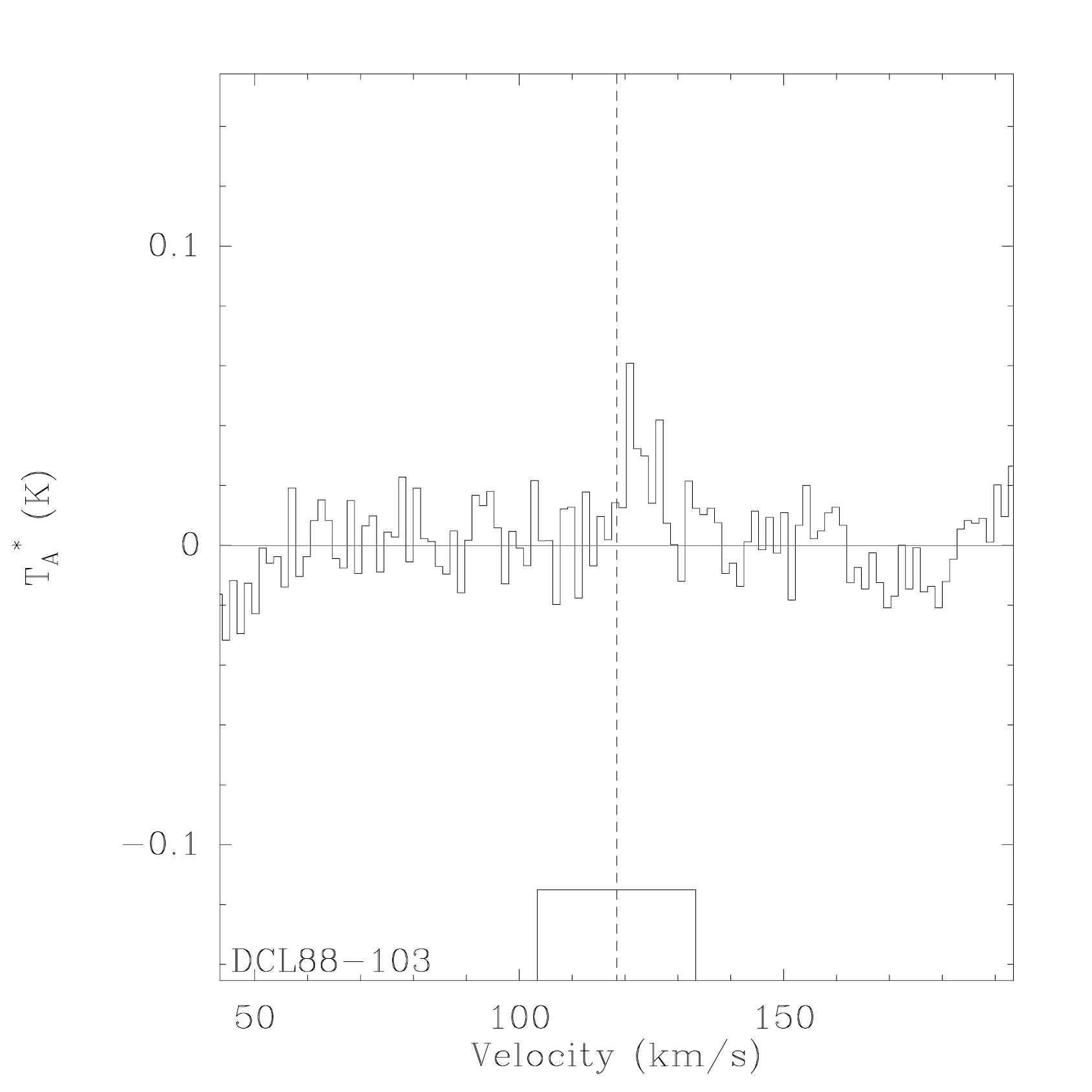}
\end{minipage}

\begin{minipage}{0.24\linewidth}
\includegraphics[width=\linewidth]{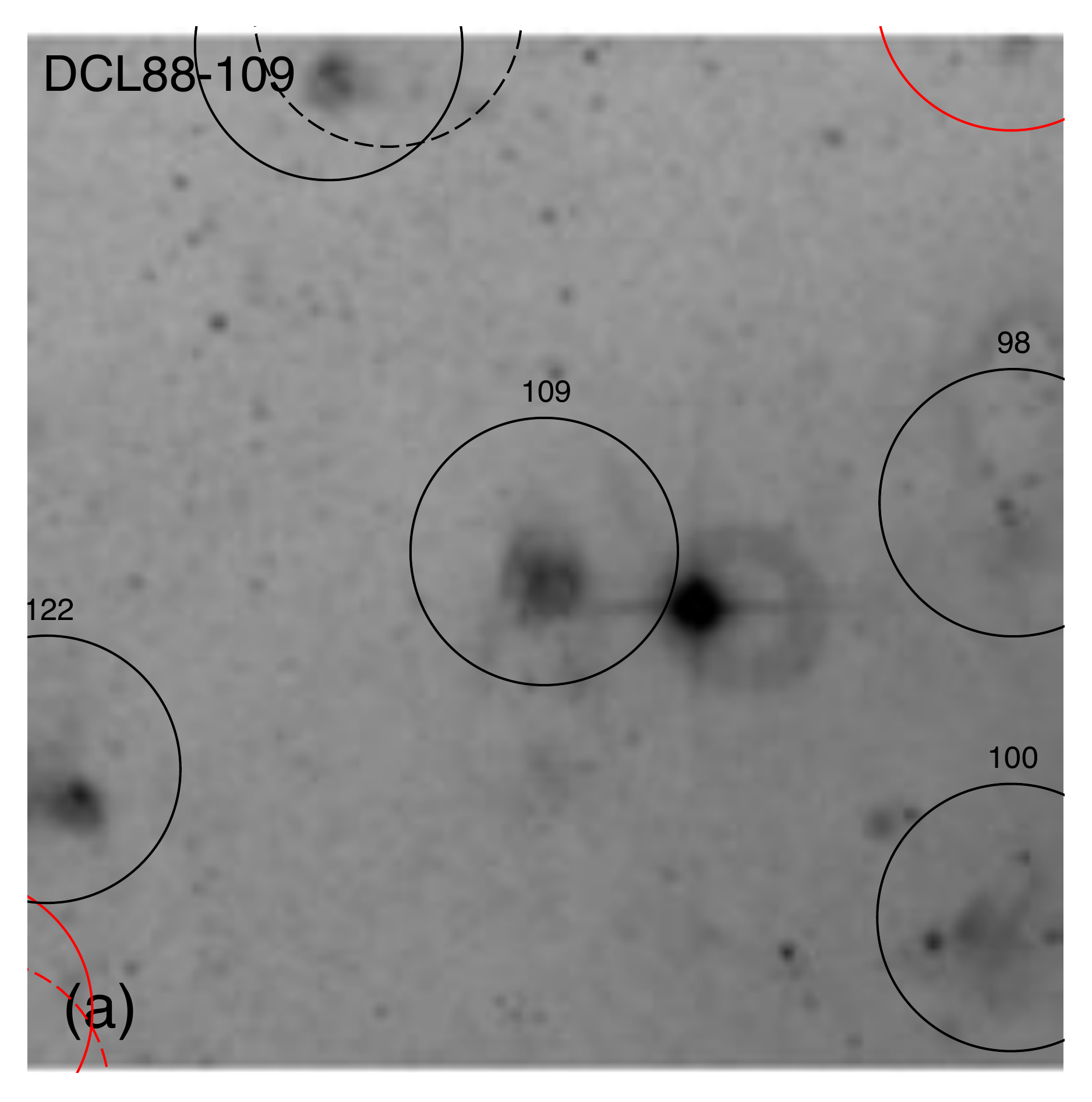}
\end{minipage}
\begin{minipage}{0.24\linewidth}
\includegraphics[width=\linewidth]{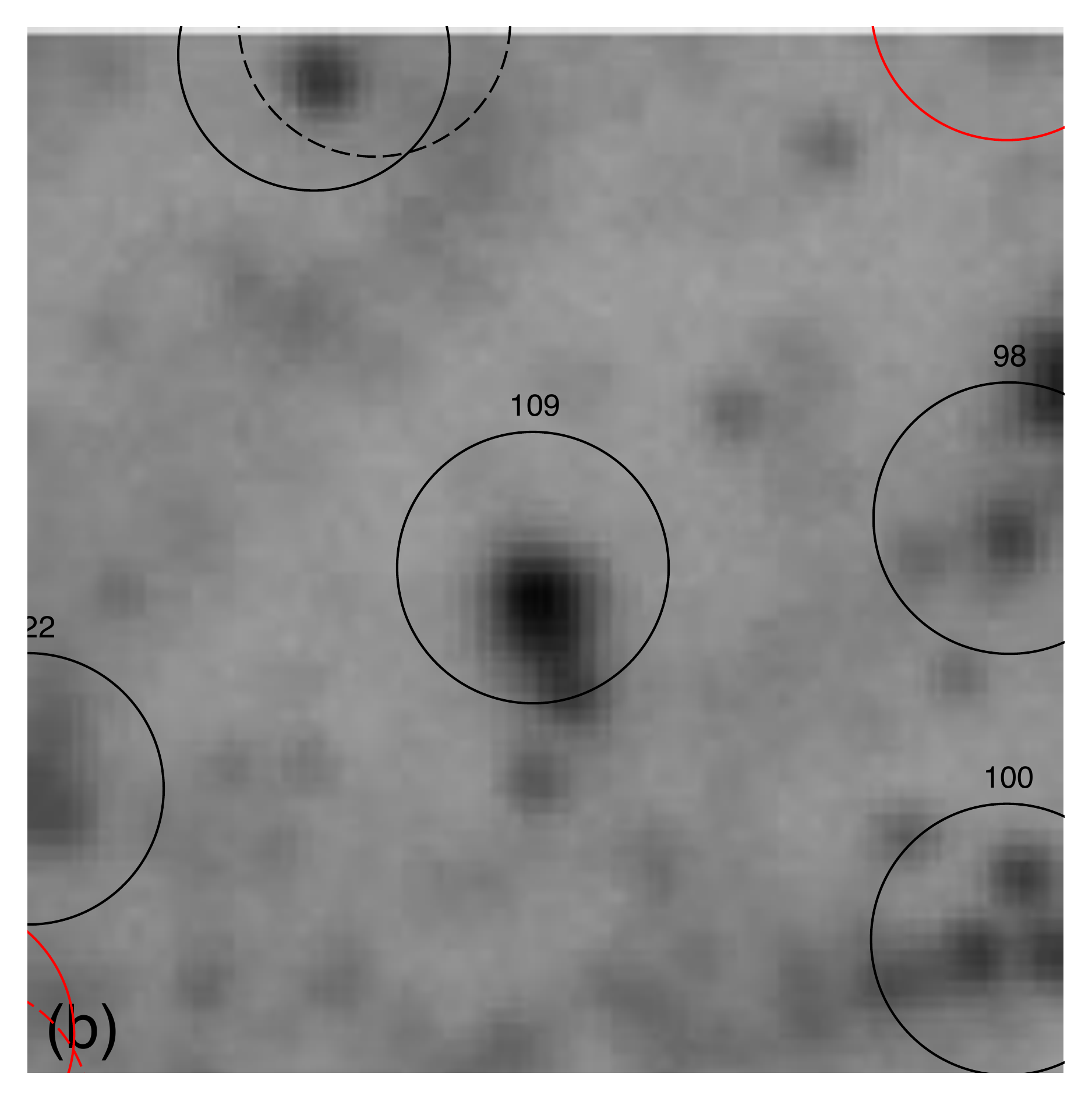}
\end{minipage}
\begin{minipage}{0.24\linewidth}
\includegraphics[width=\linewidth]{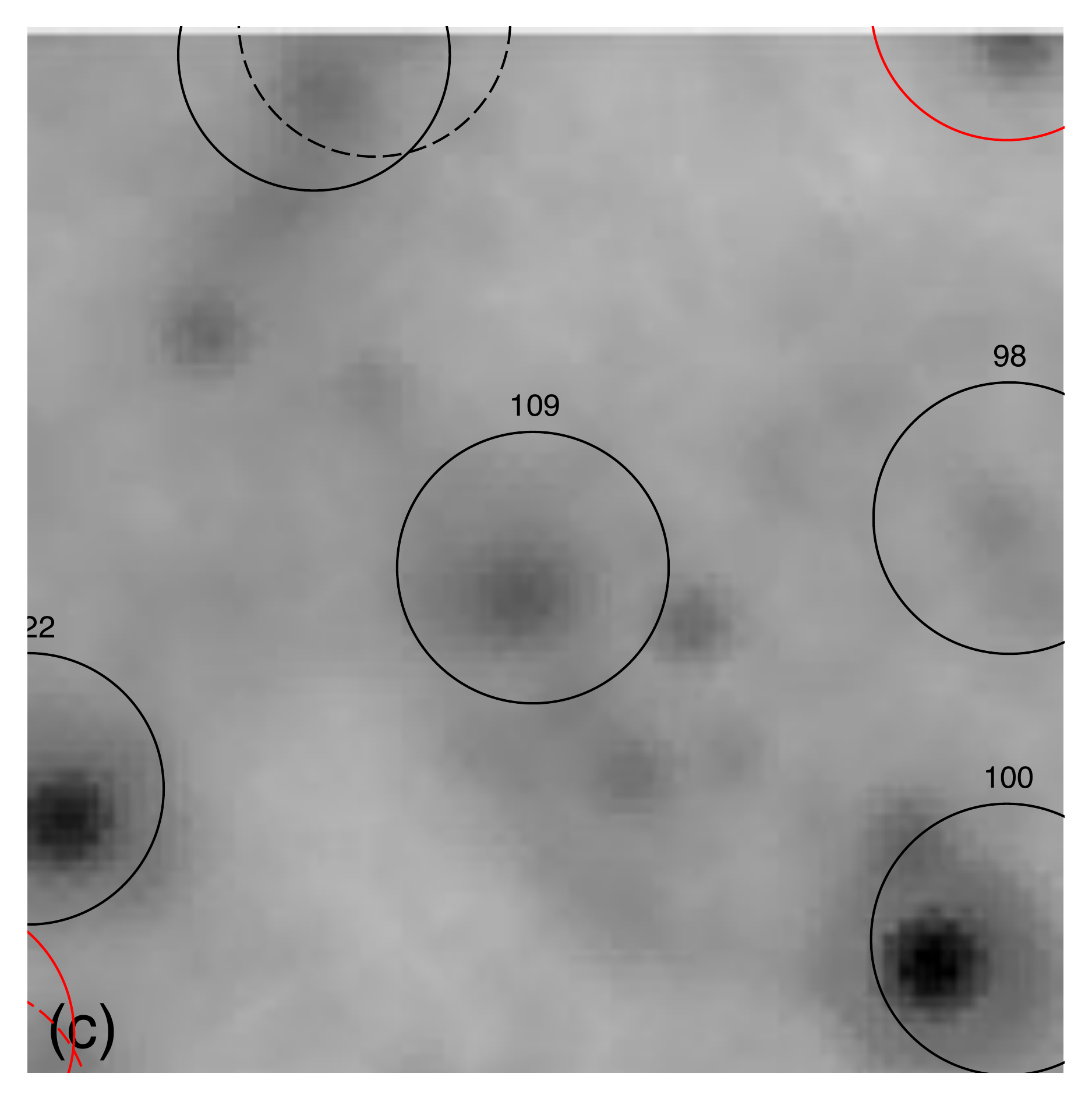}
\end{minipage}
\begin{minipage}{0.24\linewidth}
\includegraphics[width=\linewidth]{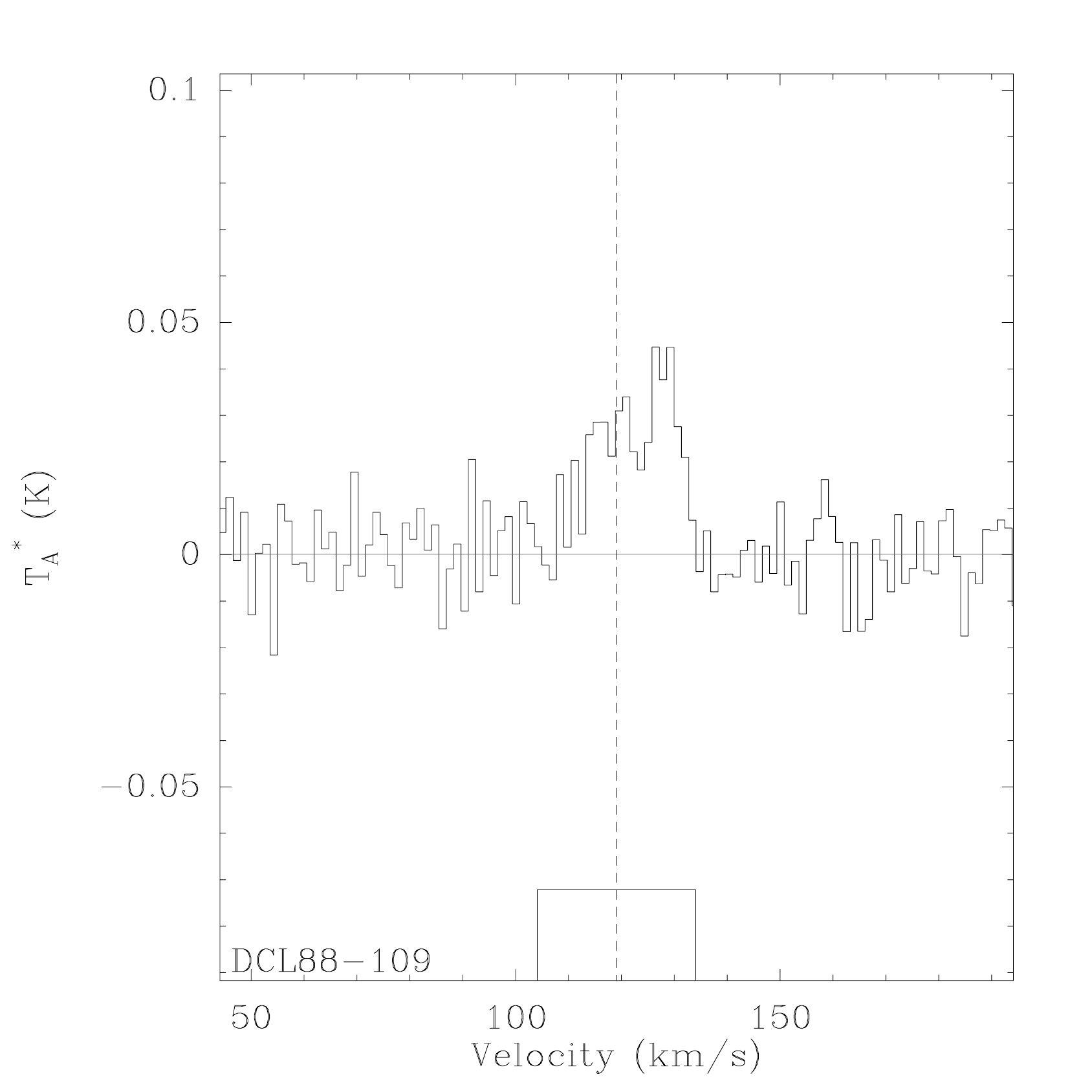}
\end{minipage}

\begin{minipage}{0.24\linewidth}
\includegraphics[width=\linewidth]{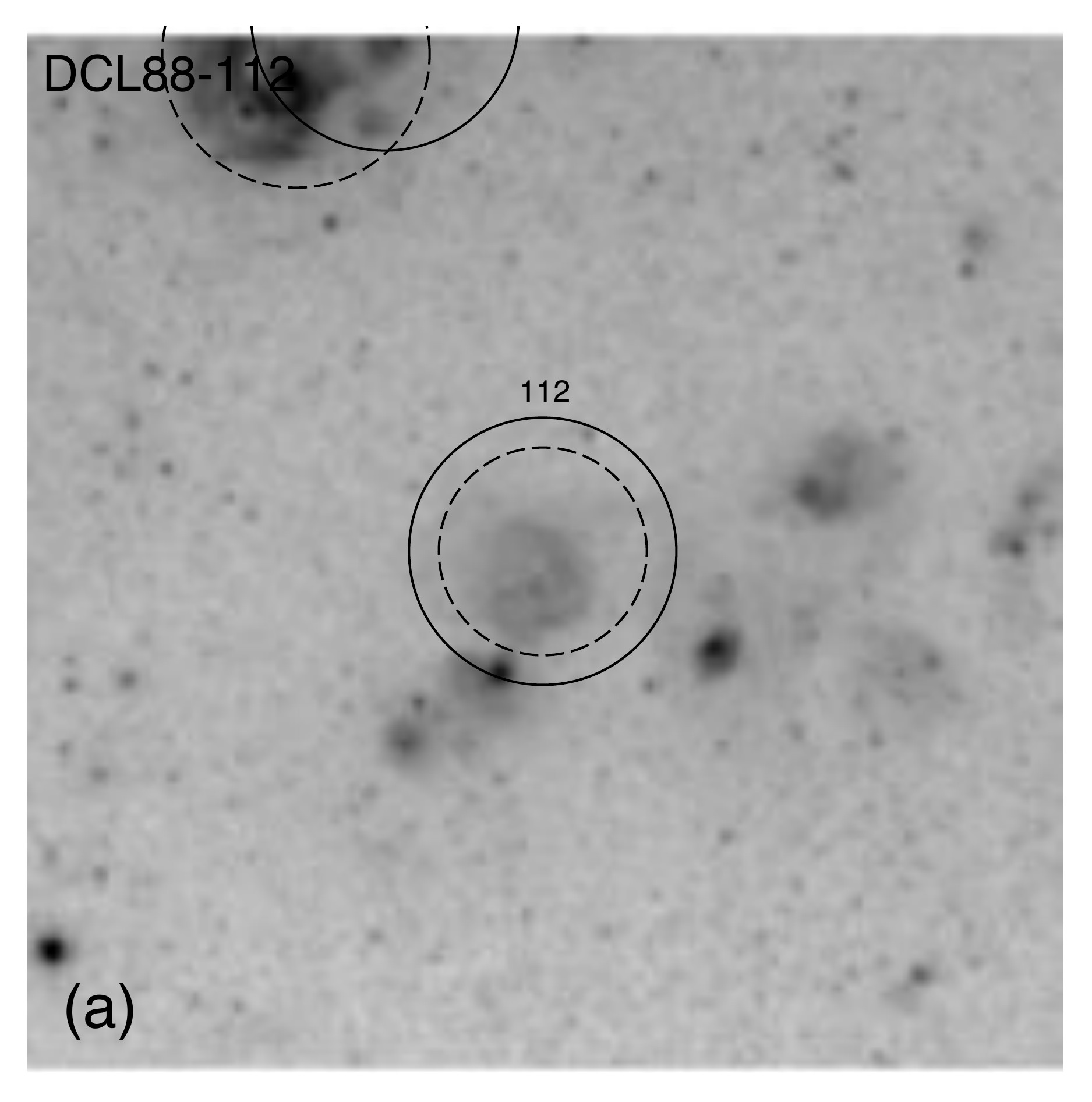}
\end{minipage}
\begin{minipage}{0.24\linewidth}
\includegraphics[width=\linewidth]{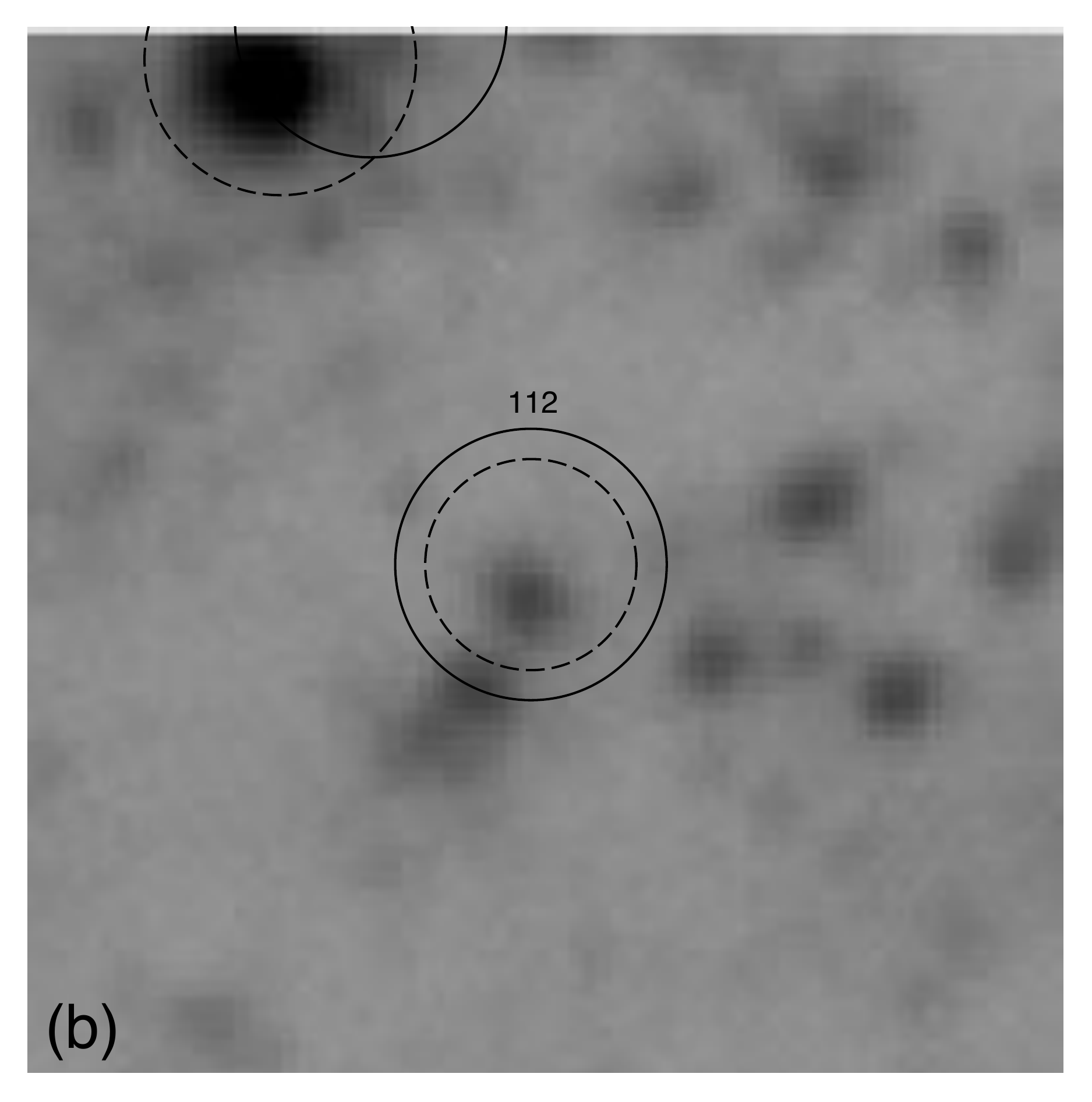}
\end{minipage}
\begin{minipage}{0.24\linewidth}
\includegraphics[width=\linewidth]{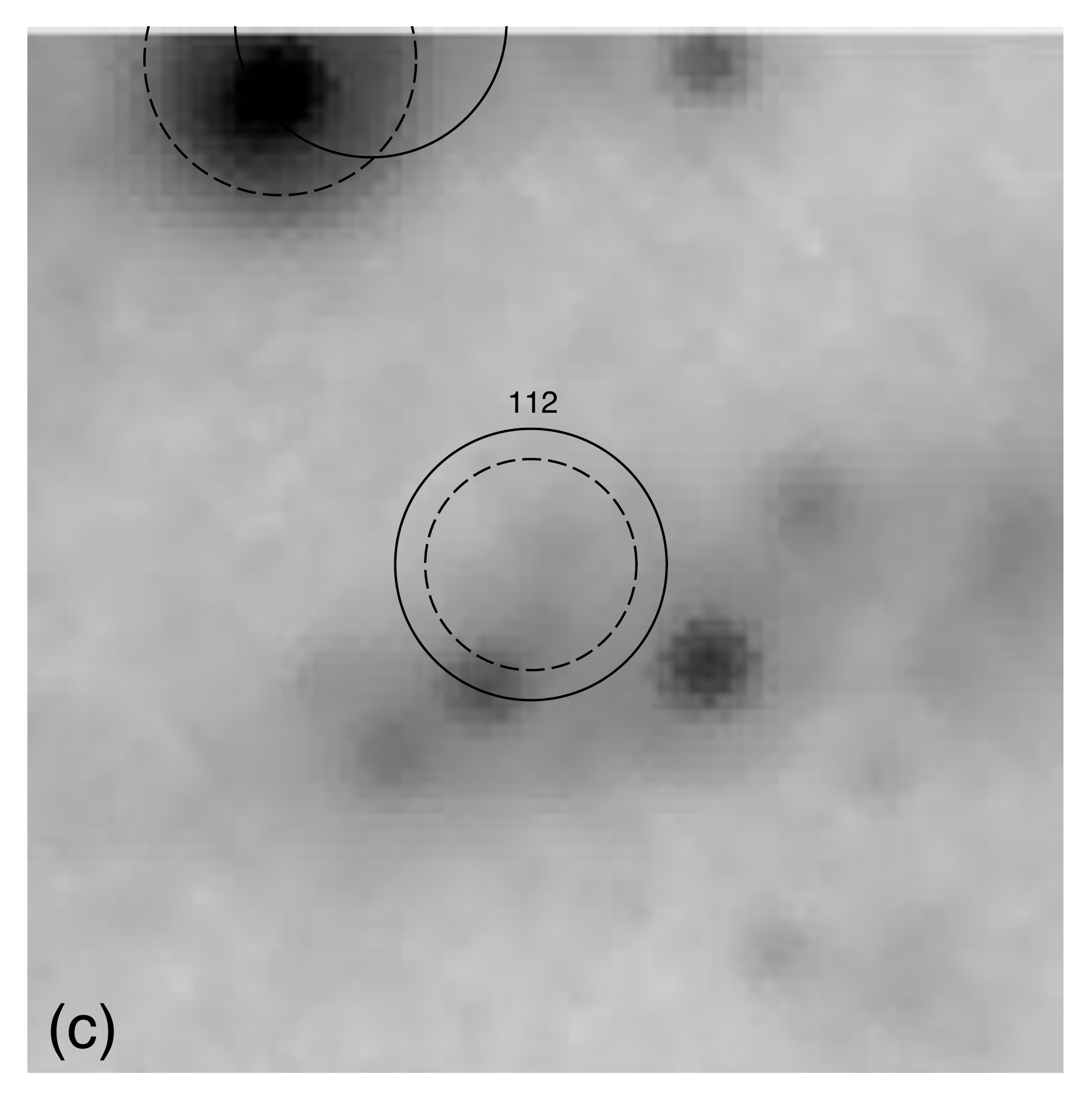}
\end{minipage}
\begin{minipage}{0.24\linewidth}
\includegraphics[width=\linewidth]{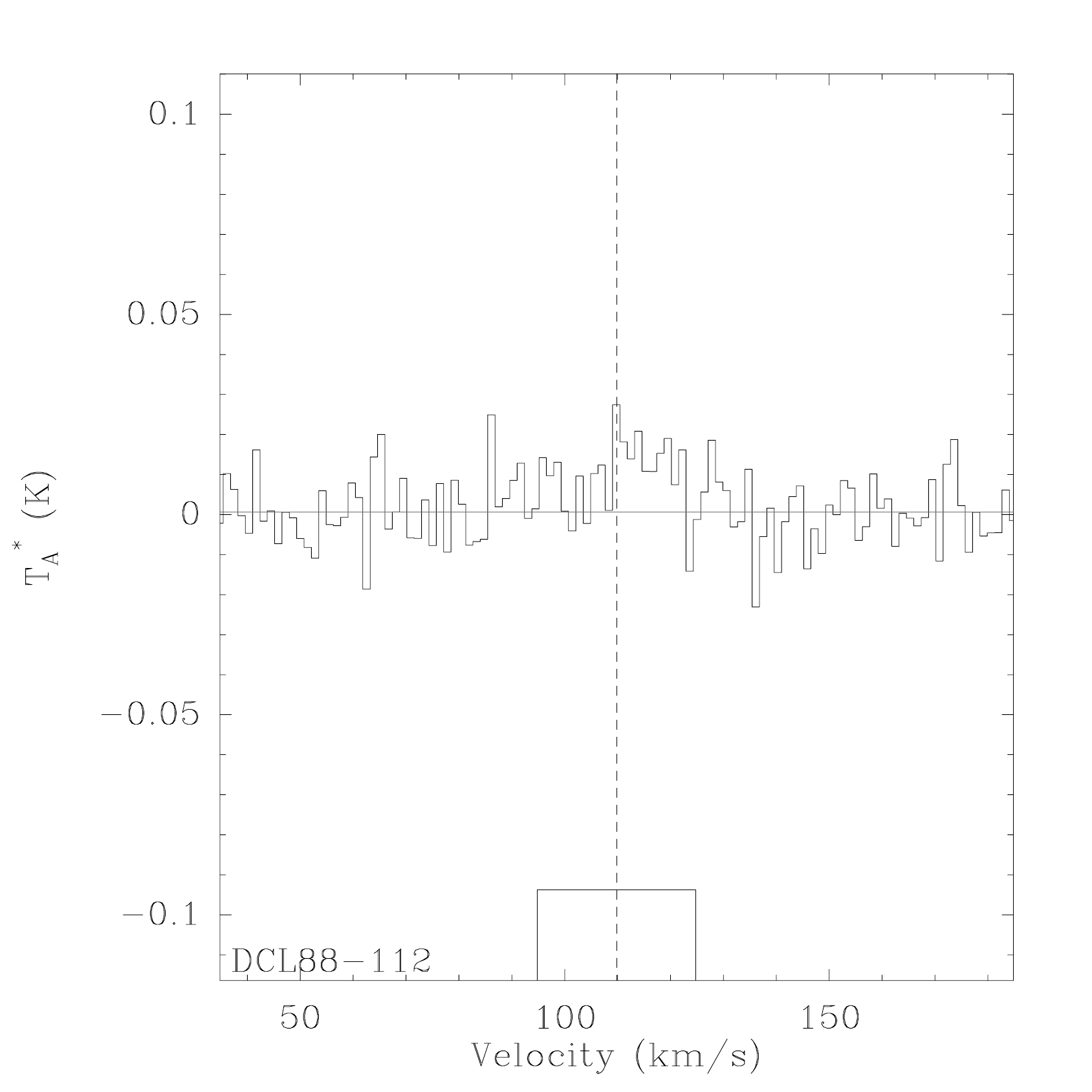}
\end{minipage}

\begin{minipage}{0.24\linewidth}
\includegraphics[width=\linewidth]{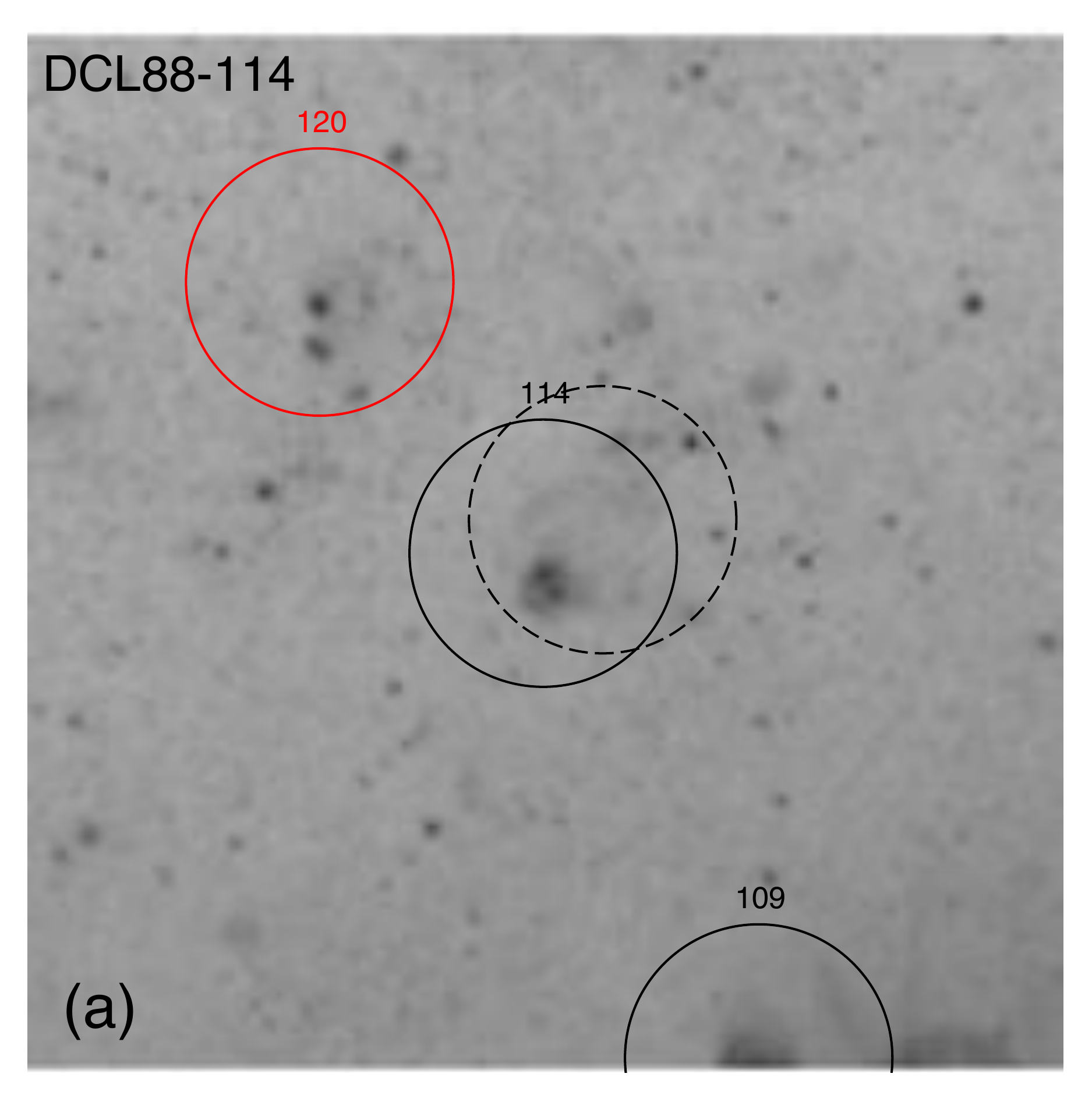}
\end{minipage}
\begin{minipage}{0.24\linewidth}
\includegraphics[width=\linewidth]{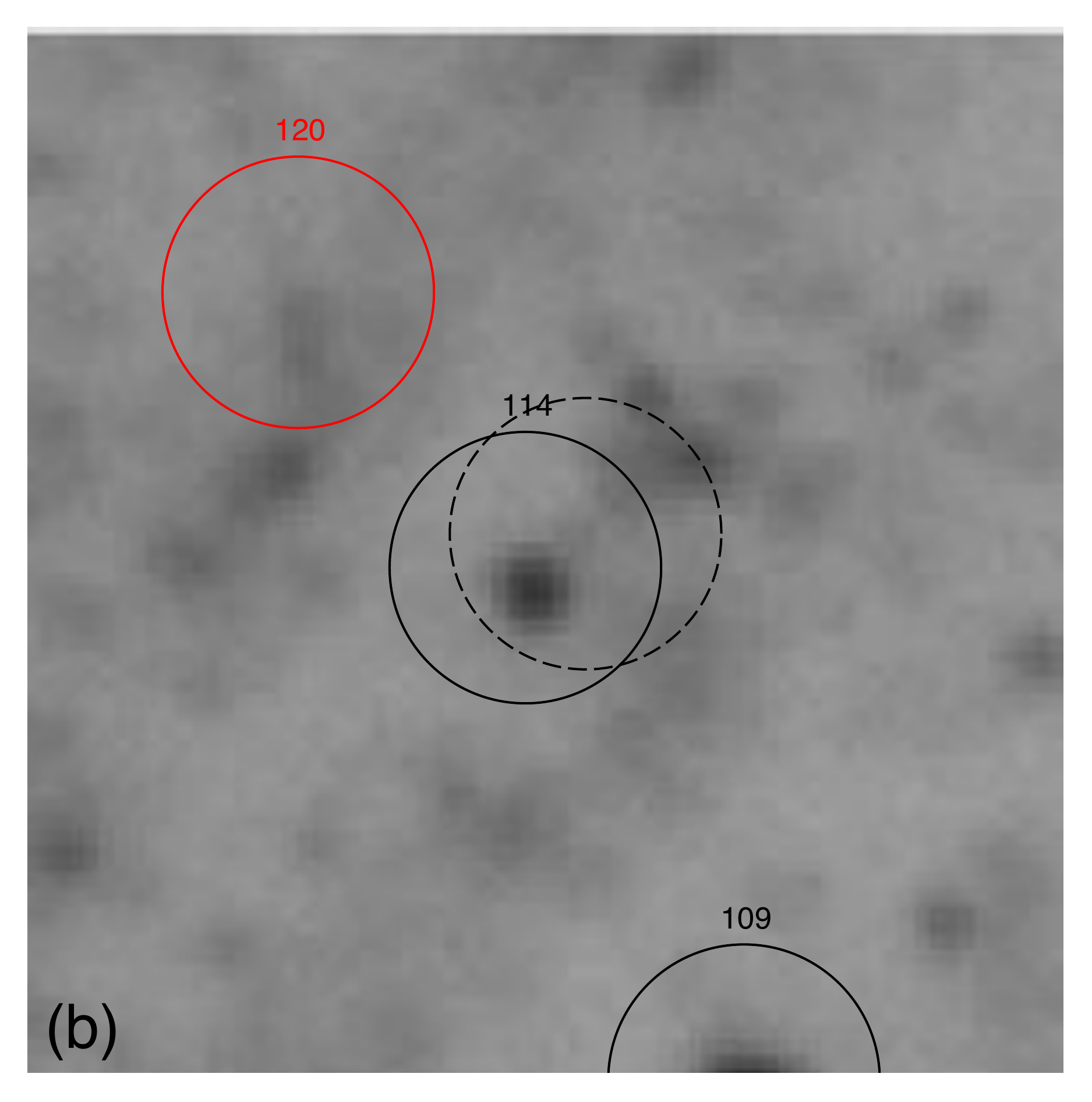}
\end{minipage}
\begin{minipage}{0.24\linewidth}
\includegraphics[width=\linewidth]{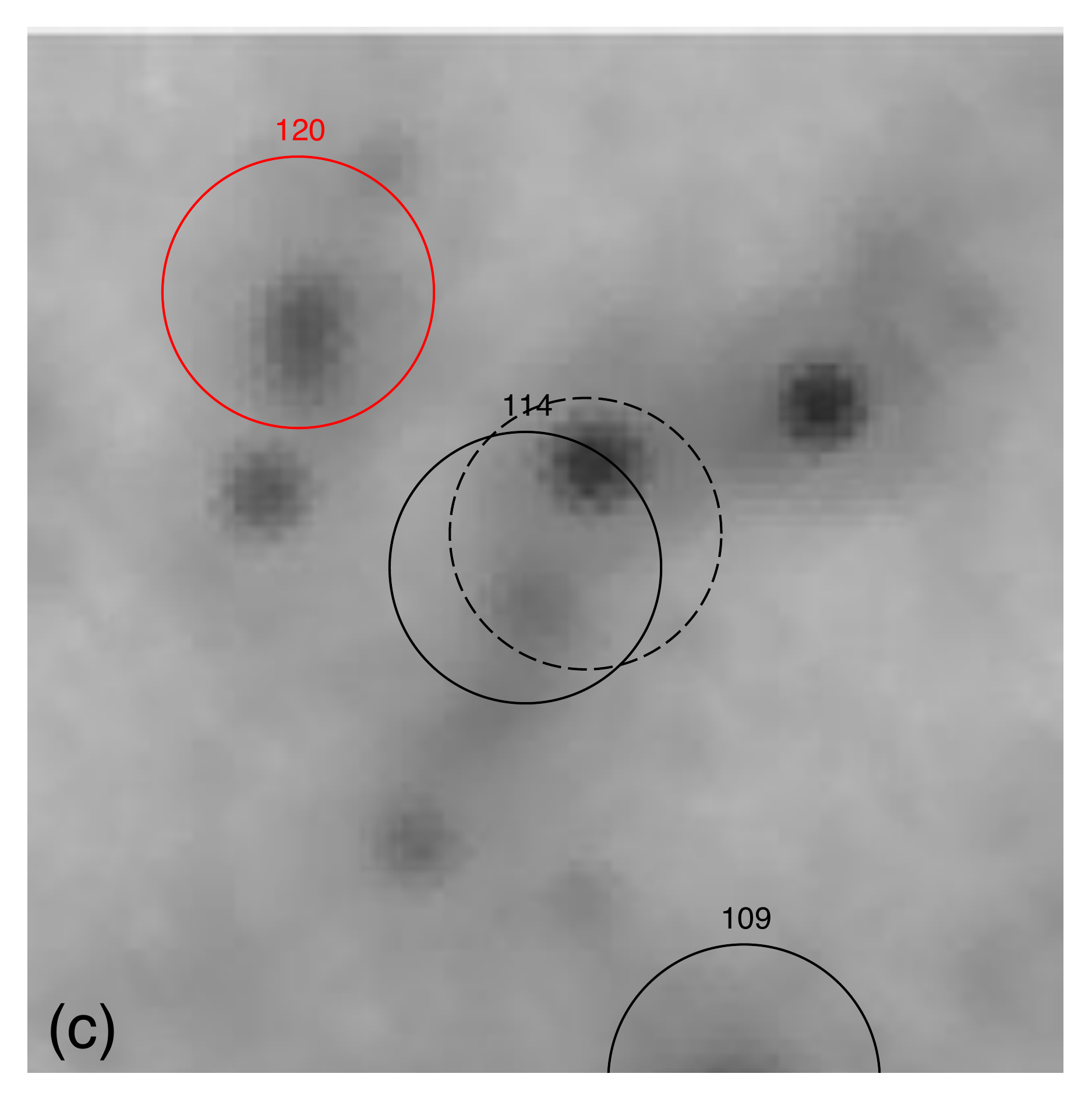}
\end{minipage}
\begin{minipage}{0.24\linewidth}
\includegraphics[width=\linewidth]{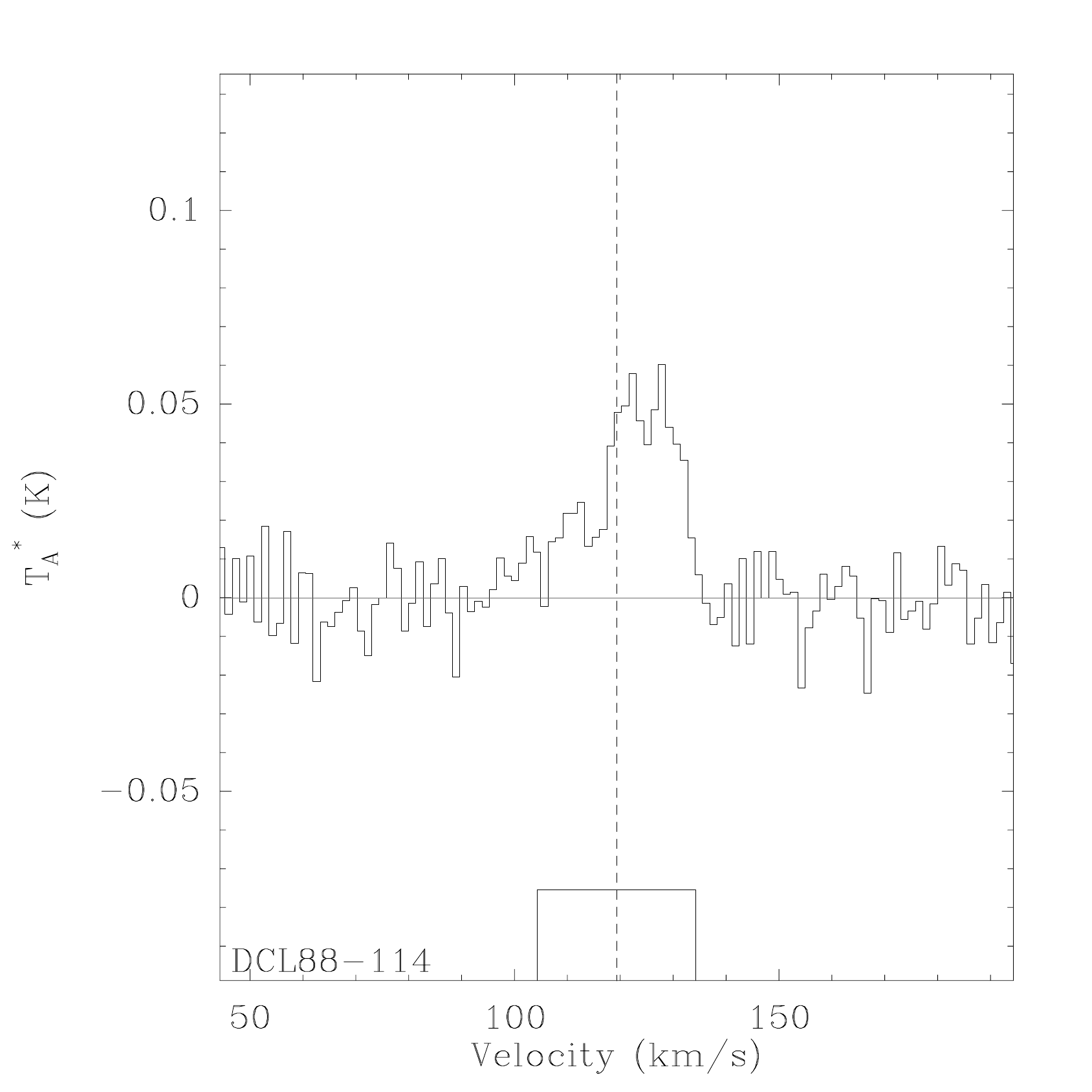}
\end{minipage}

\begin{minipage}{0.24\linewidth}
\includegraphics[width=\linewidth]{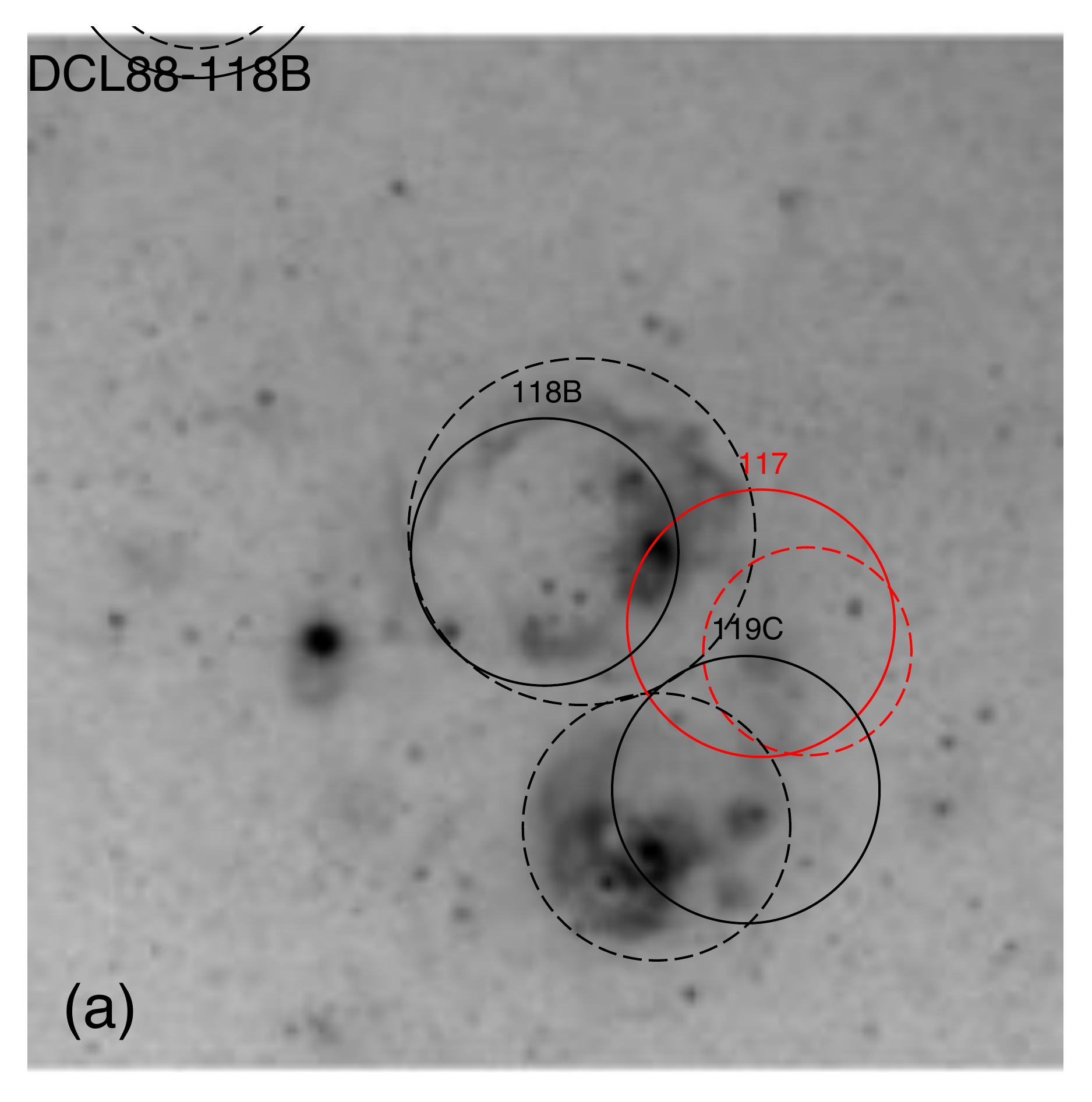}
\end{minipage}
\begin{minipage}{0.24\linewidth}
\includegraphics[width=\linewidth]{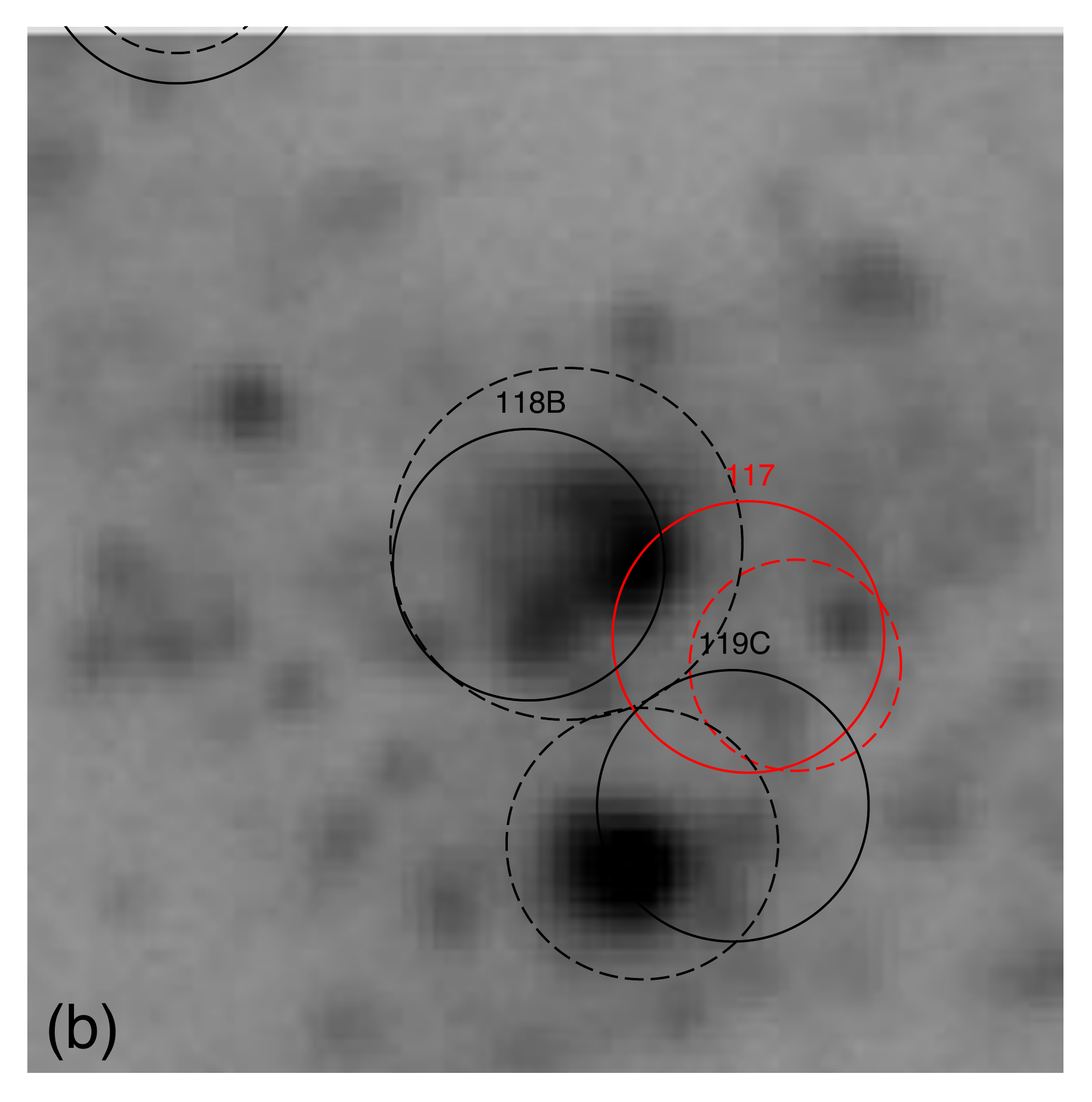}
\end{minipage}
\begin{minipage}{0.24\linewidth}
\includegraphics[width=\linewidth]{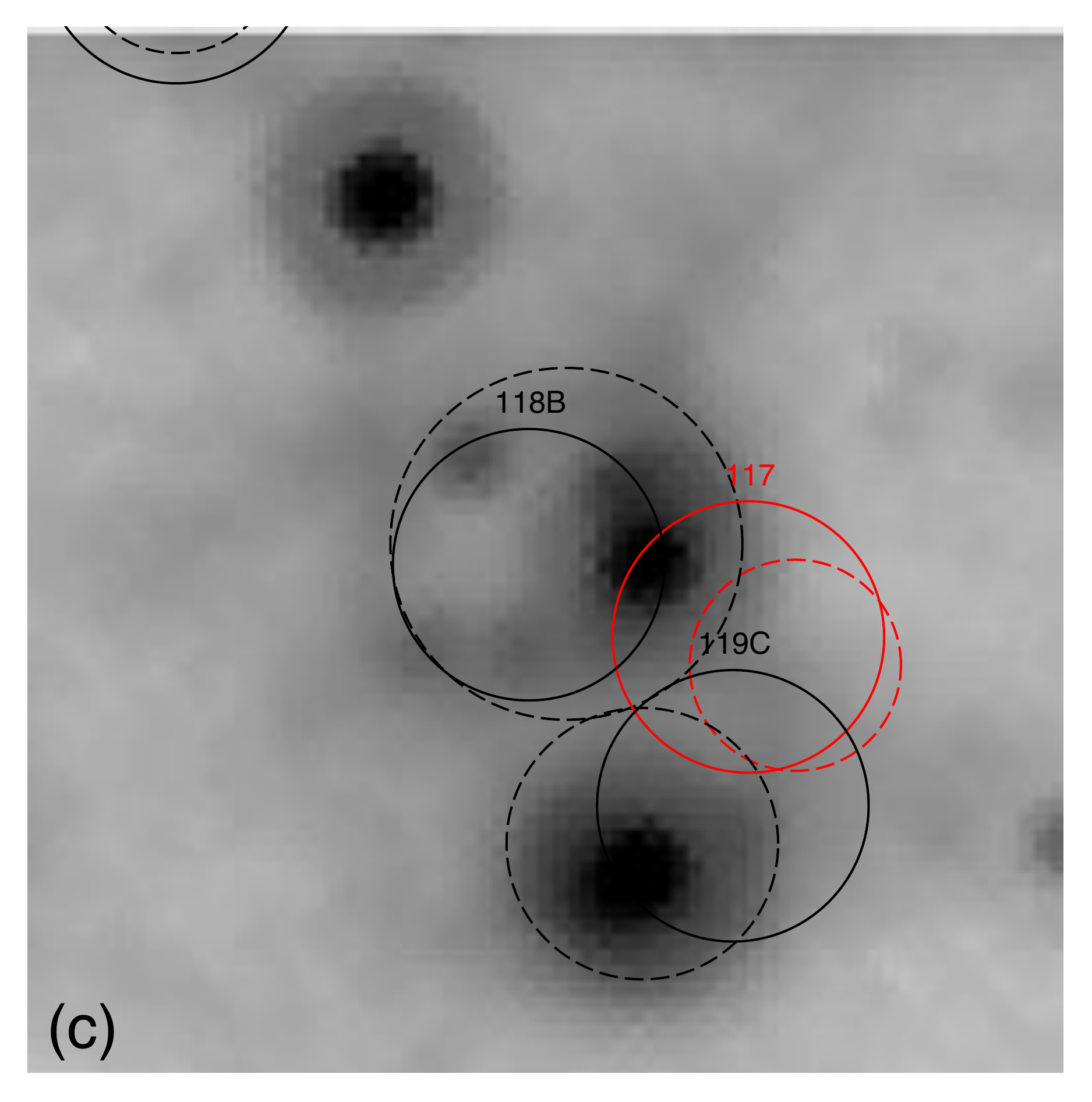}
\end{minipage}
\begin{minipage}{0.24\linewidth}
\includegraphics[width=\linewidth]{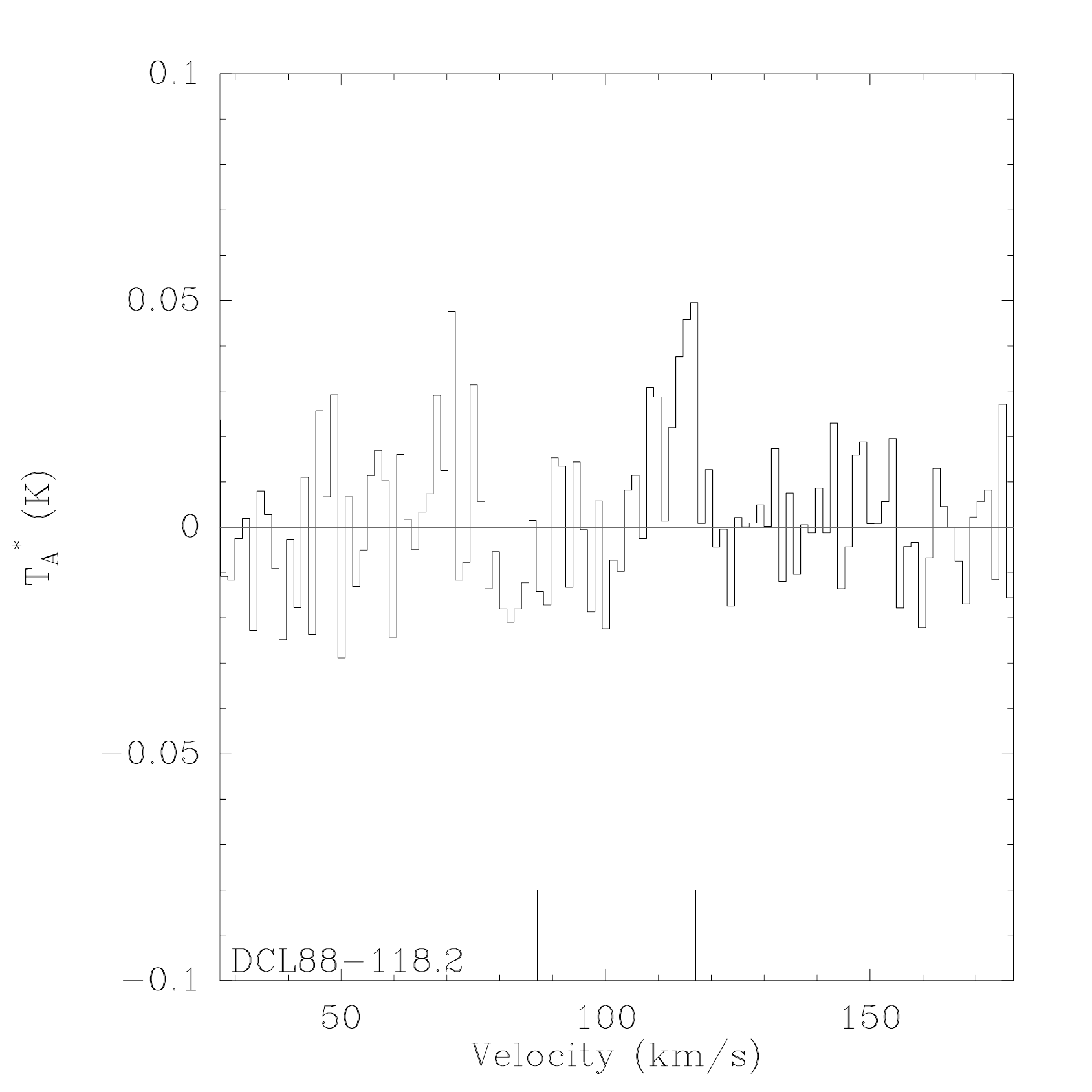}
\end{minipage}

\noindent\textbf{Figure~\ref{fig:stamps} -- continued.}

\end{figure*}

\begin{figure*}
%\ContinuedFloat

\begin{minipage}{0.24\linewidth}
\includegraphics[width=\linewidth]{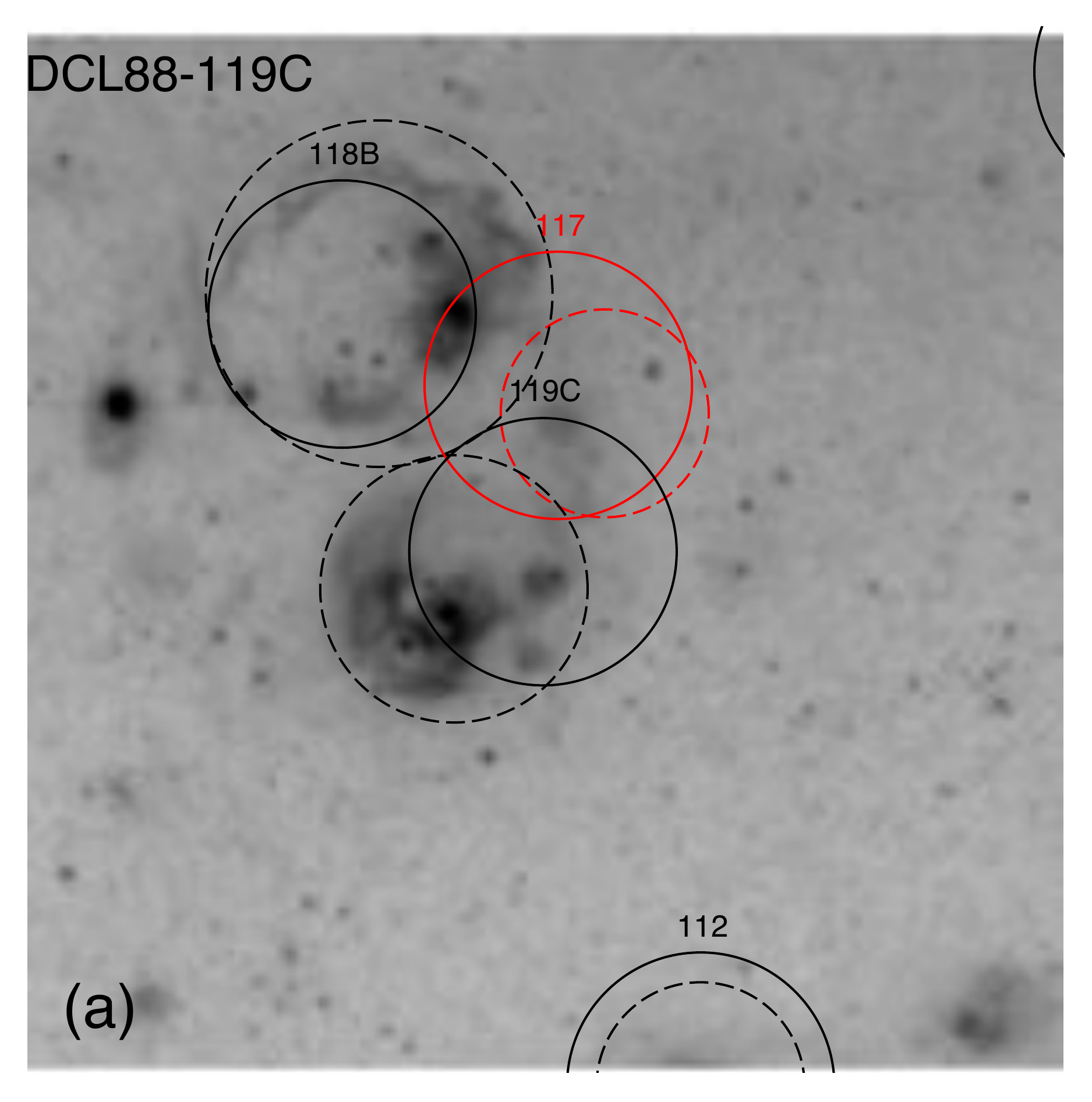}
\end{minipage}
\begin{minipage}{0.24\linewidth}
\includegraphics[width=\linewidth]{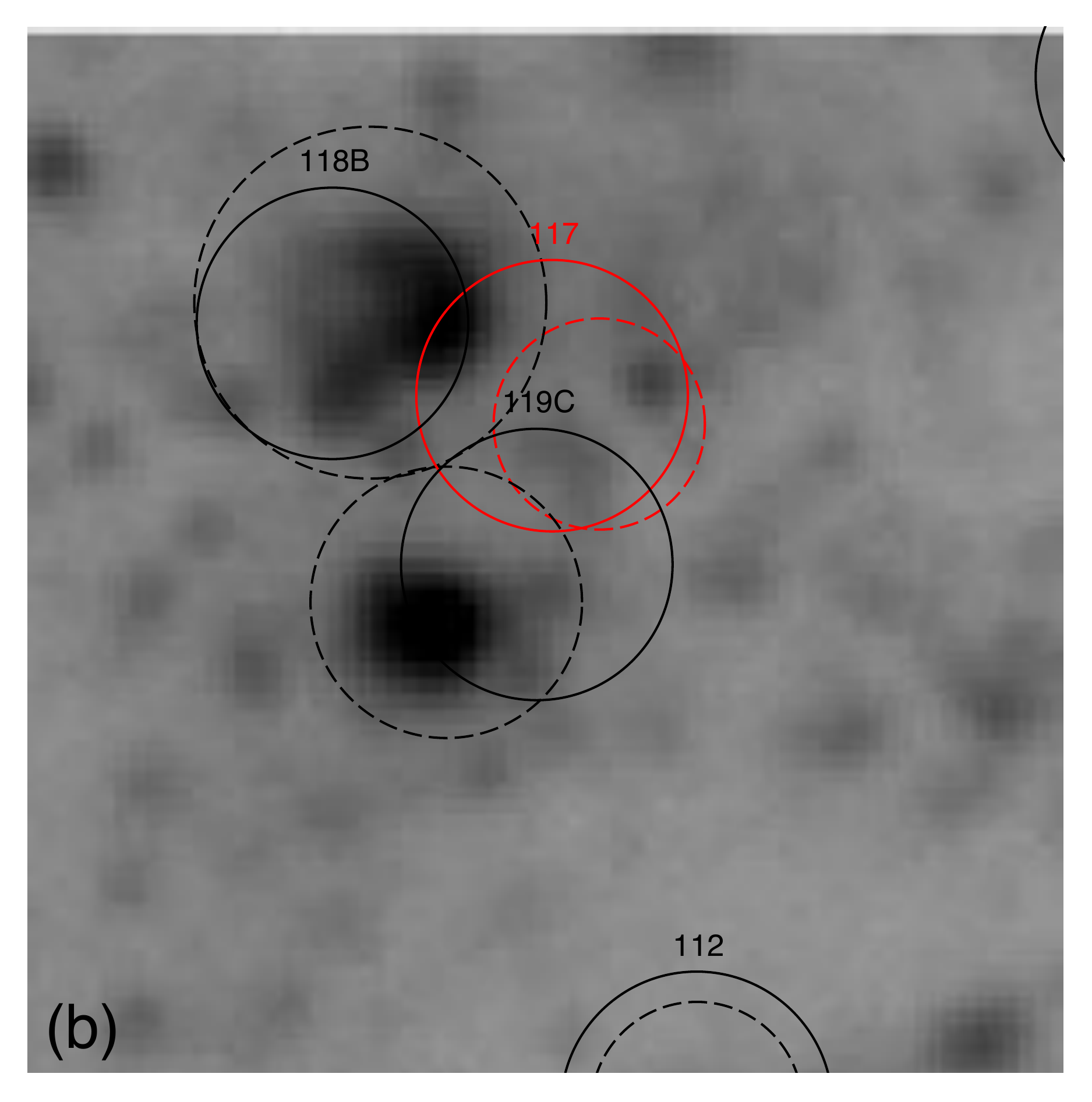}
\end{minipage}
\begin{minipage}{0.24\linewidth}
\includegraphics[width=\linewidth]{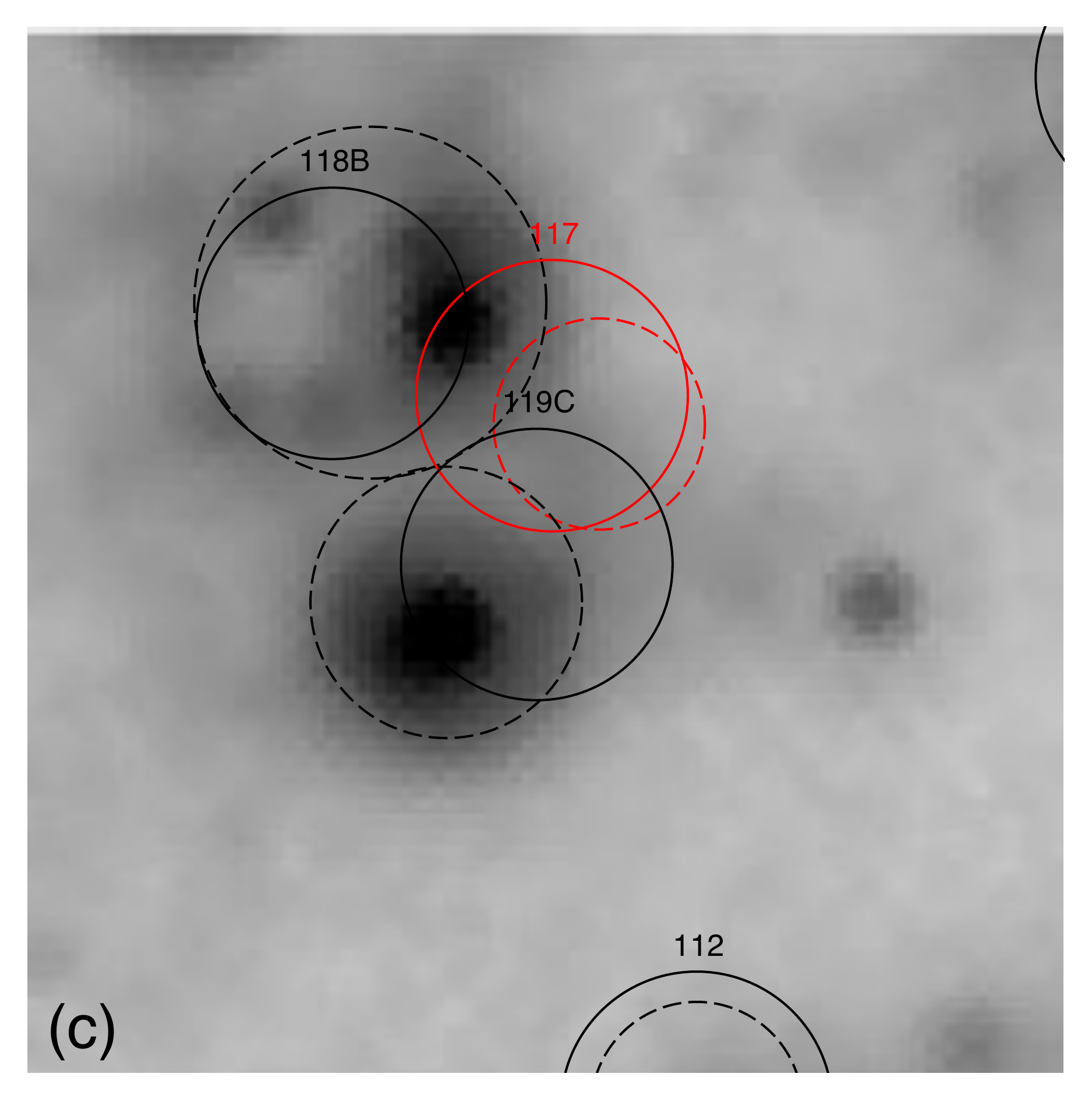}
\end{minipage}
\begin{minipage}{0.24\linewidth}
\includegraphics[width=\linewidth]{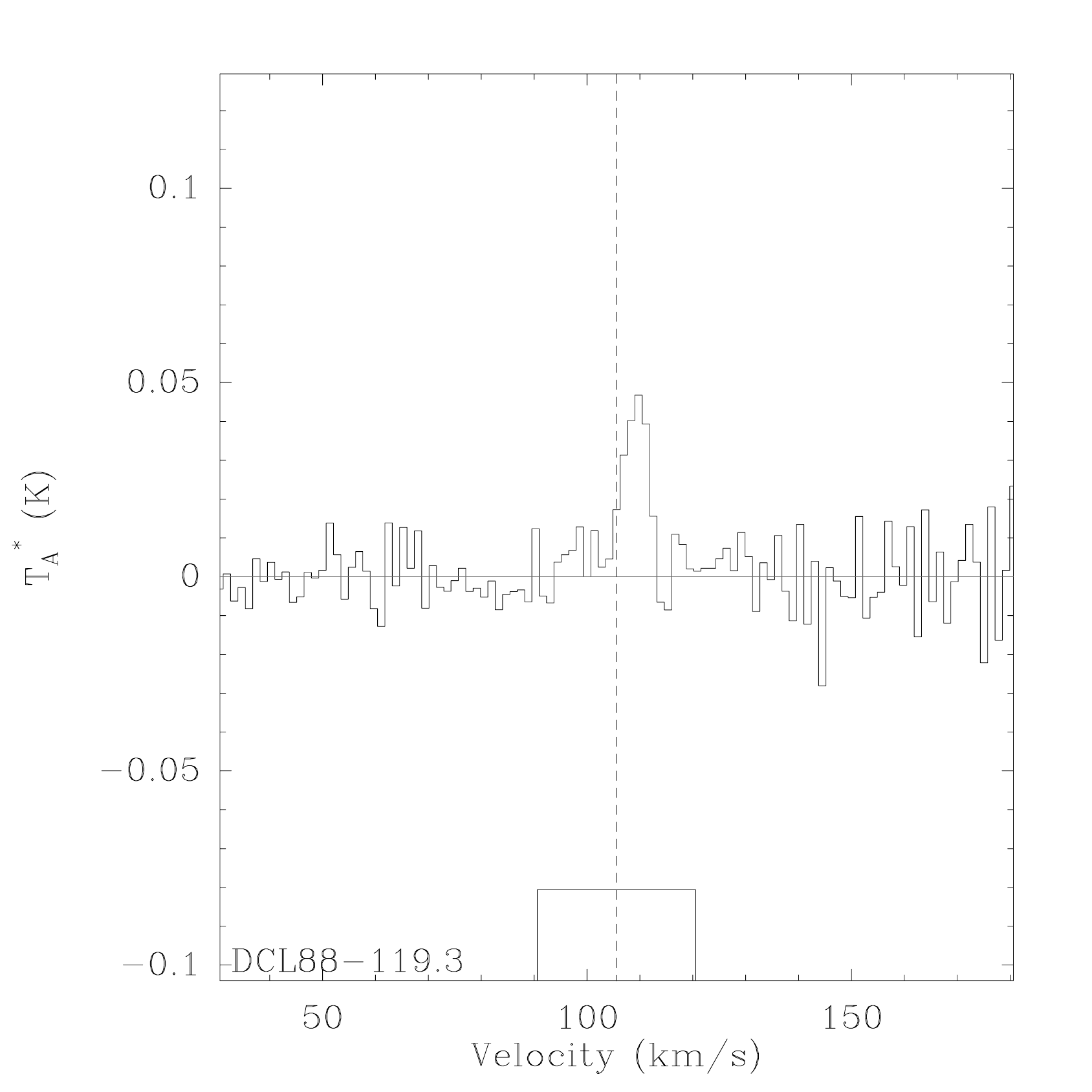}
\end{minipage}

\begin{minipage}{0.24\linewidth}
\includegraphics[width=\linewidth]{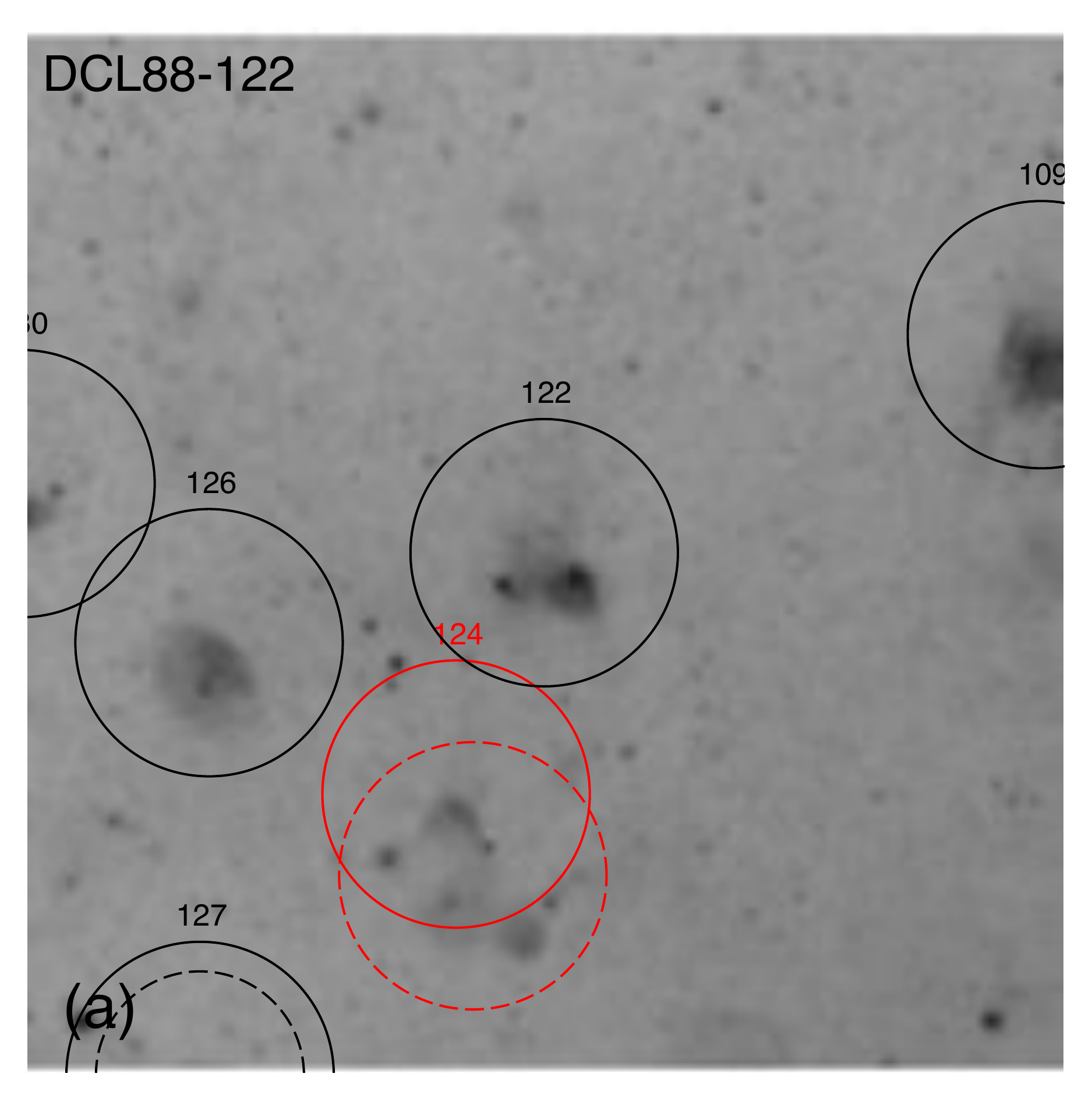}
\end{minipage}
\begin{minipage}{0.24\linewidth}
\includegraphics[width=\linewidth]{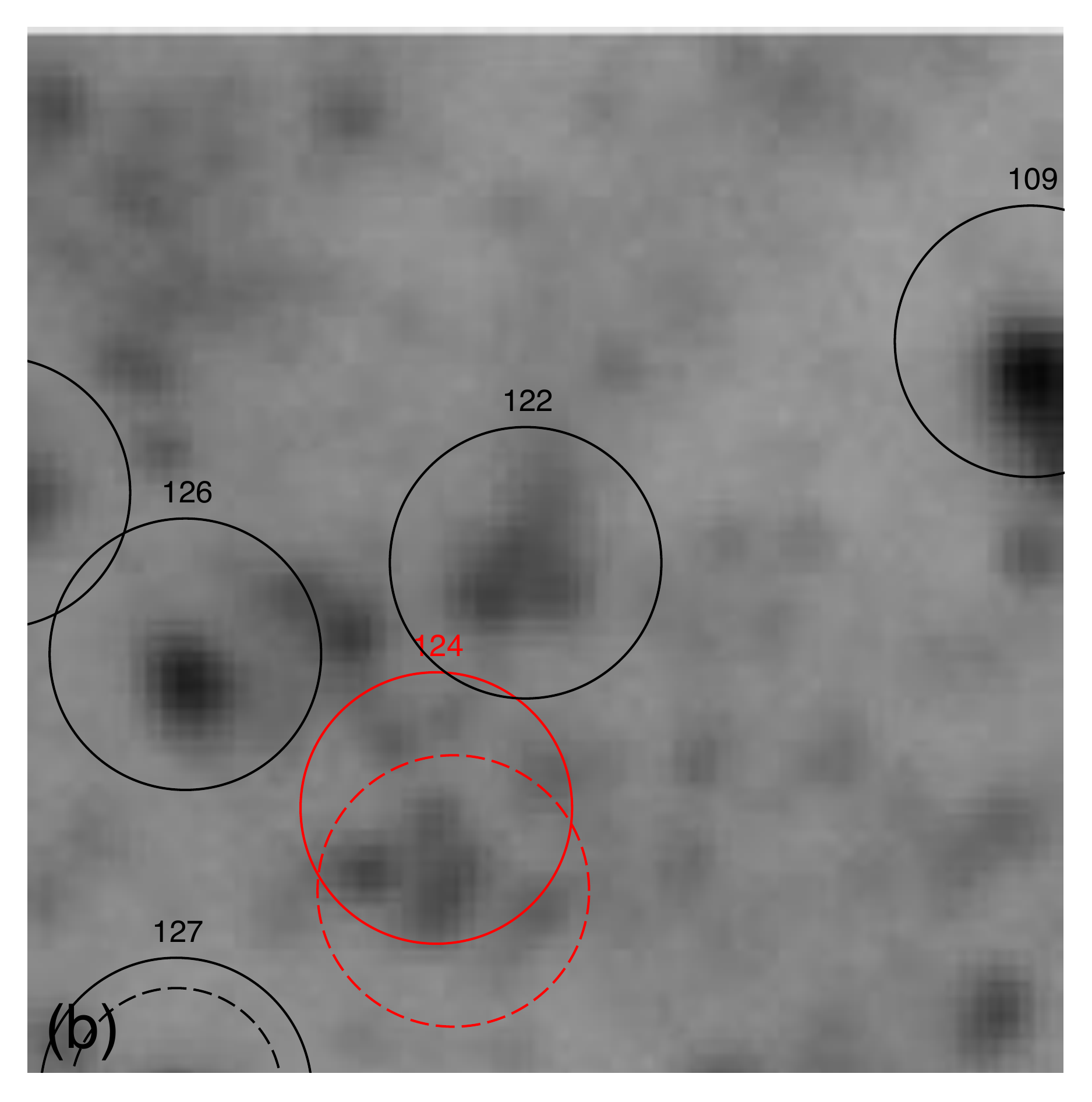}
\end{minipage}
\begin{minipage}{0.24\linewidth}
\includegraphics[width=\linewidth]{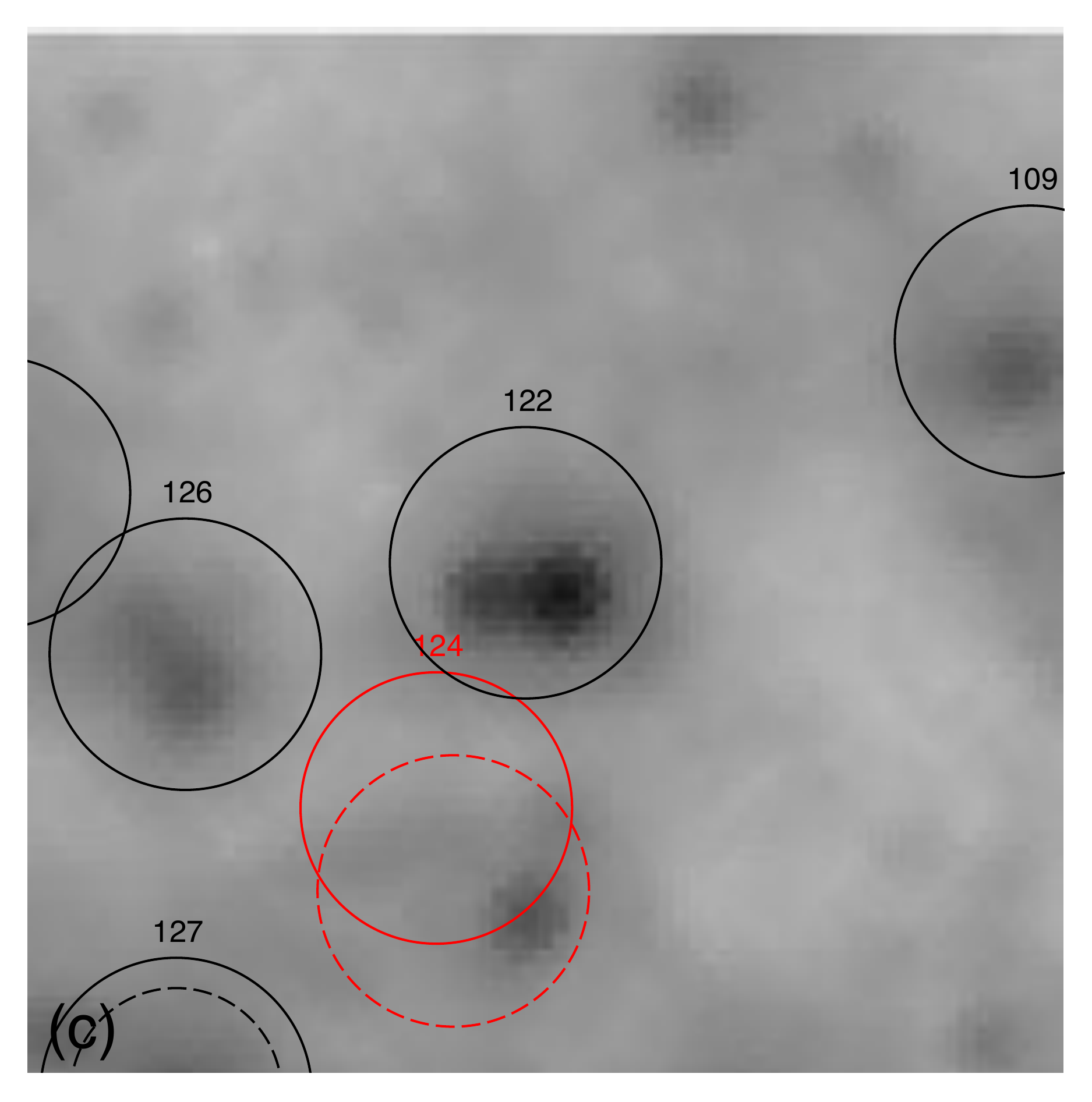}
\end{minipage}
\begin{minipage}{0.24\linewidth}
\includegraphics[width=\linewidth]{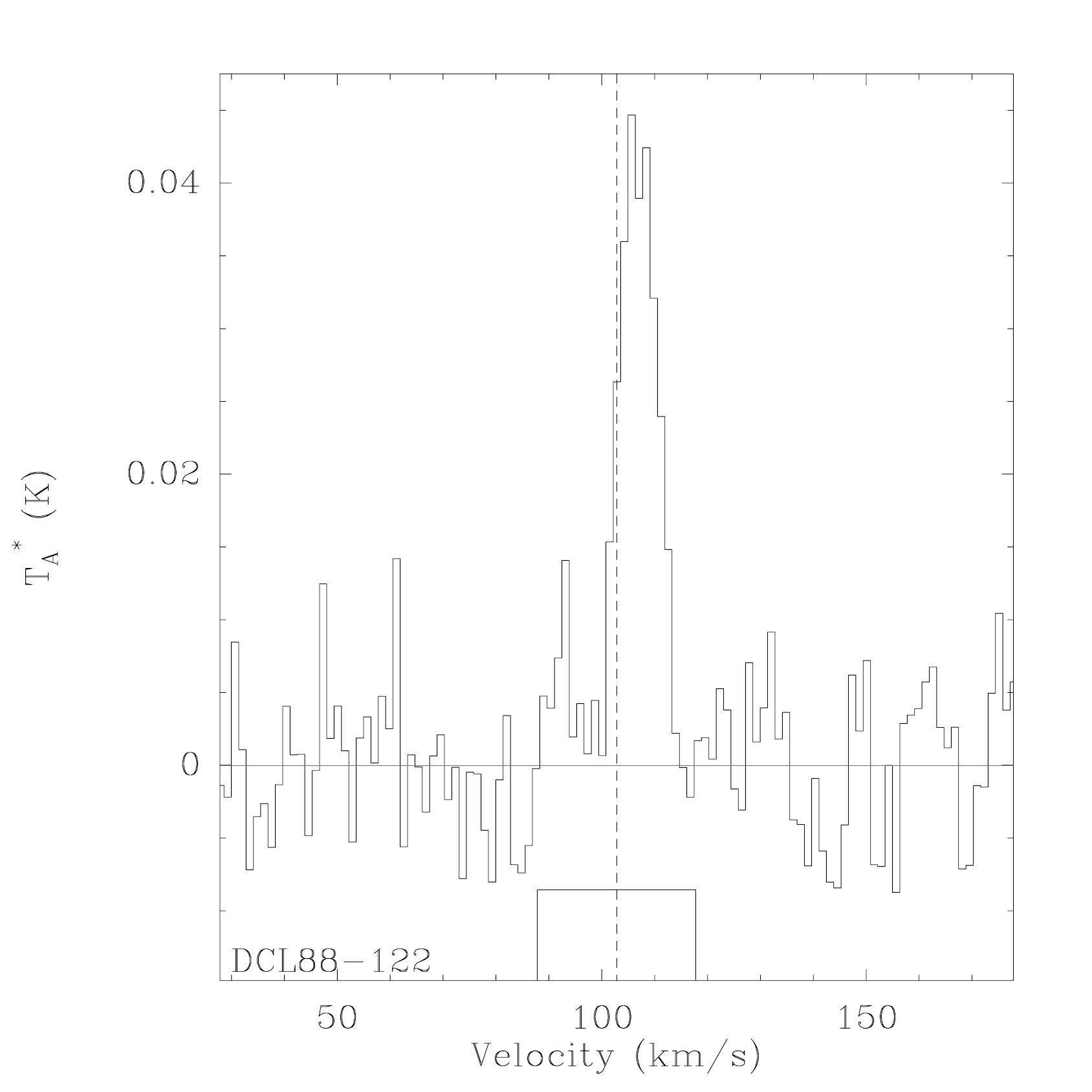}
\end{minipage}

\begin{minipage}{0.24\linewidth}
\includegraphics[width=\linewidth]{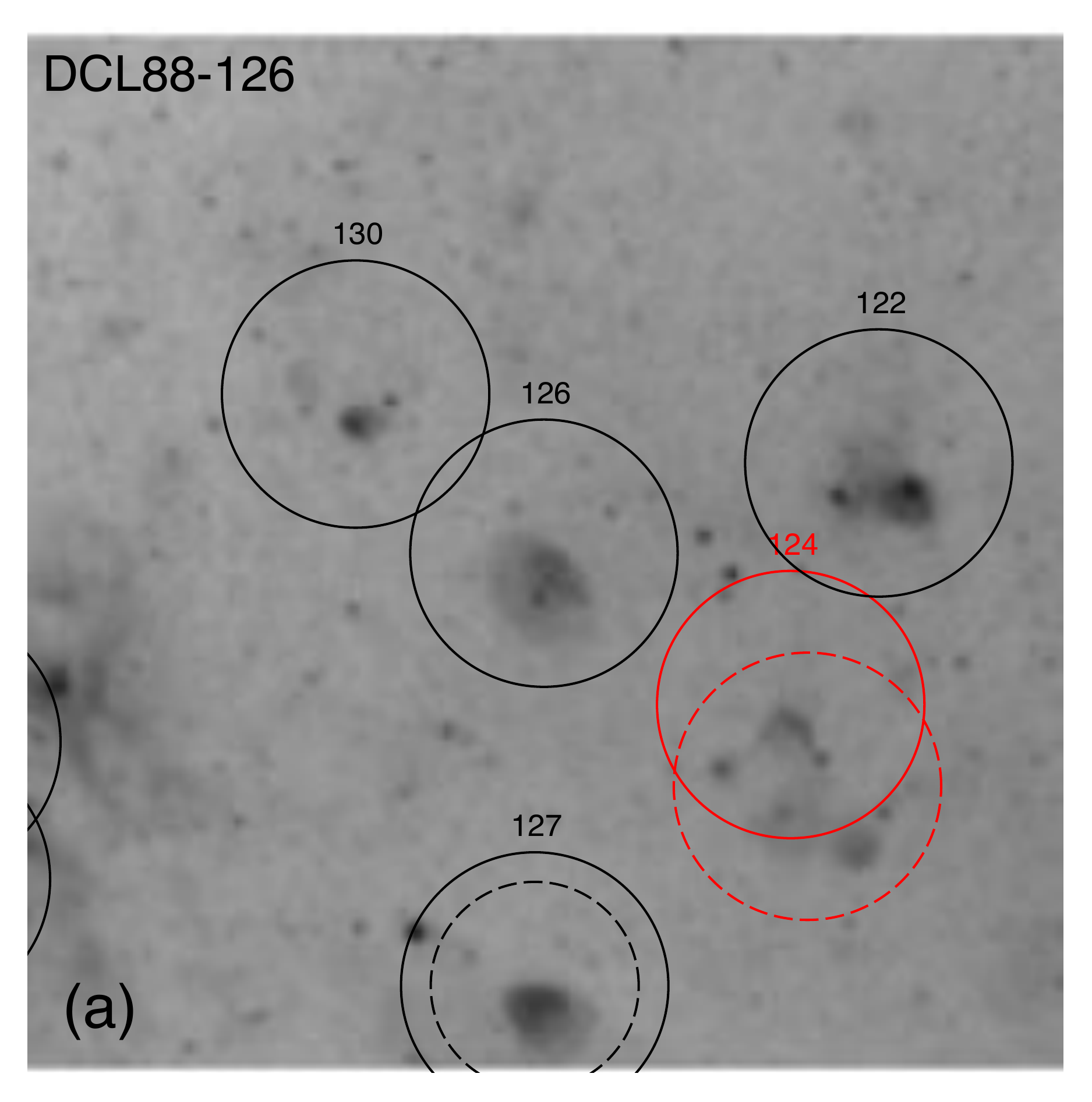}
\end{minipage}
\begin{minipage}{0.24\linewidth}
\includegraphics[width=\linewidth]{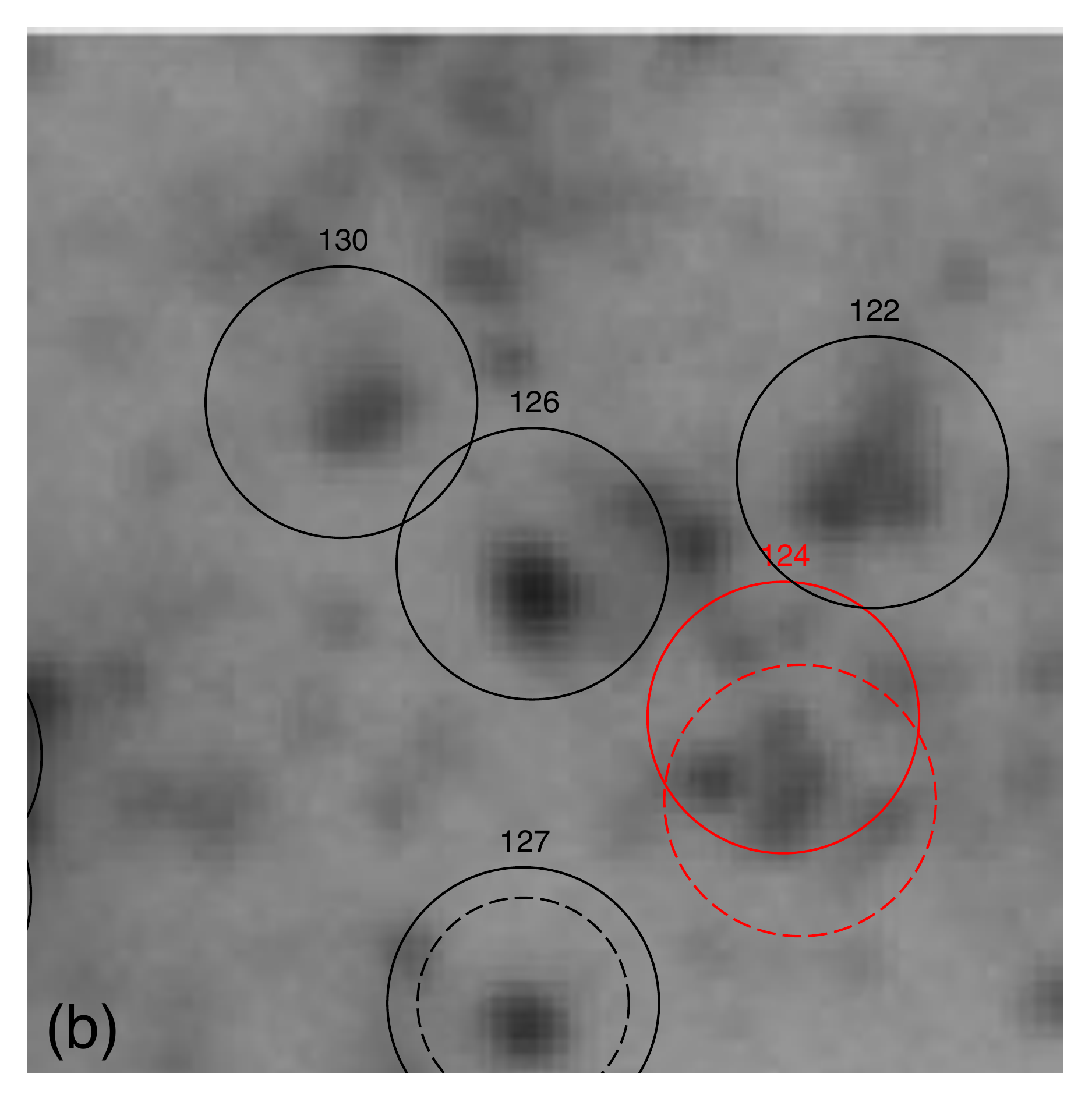}
\end{minipage}
\begin{minipage}{0.24\linewidth}
\includegraphics[width=\linewidth]{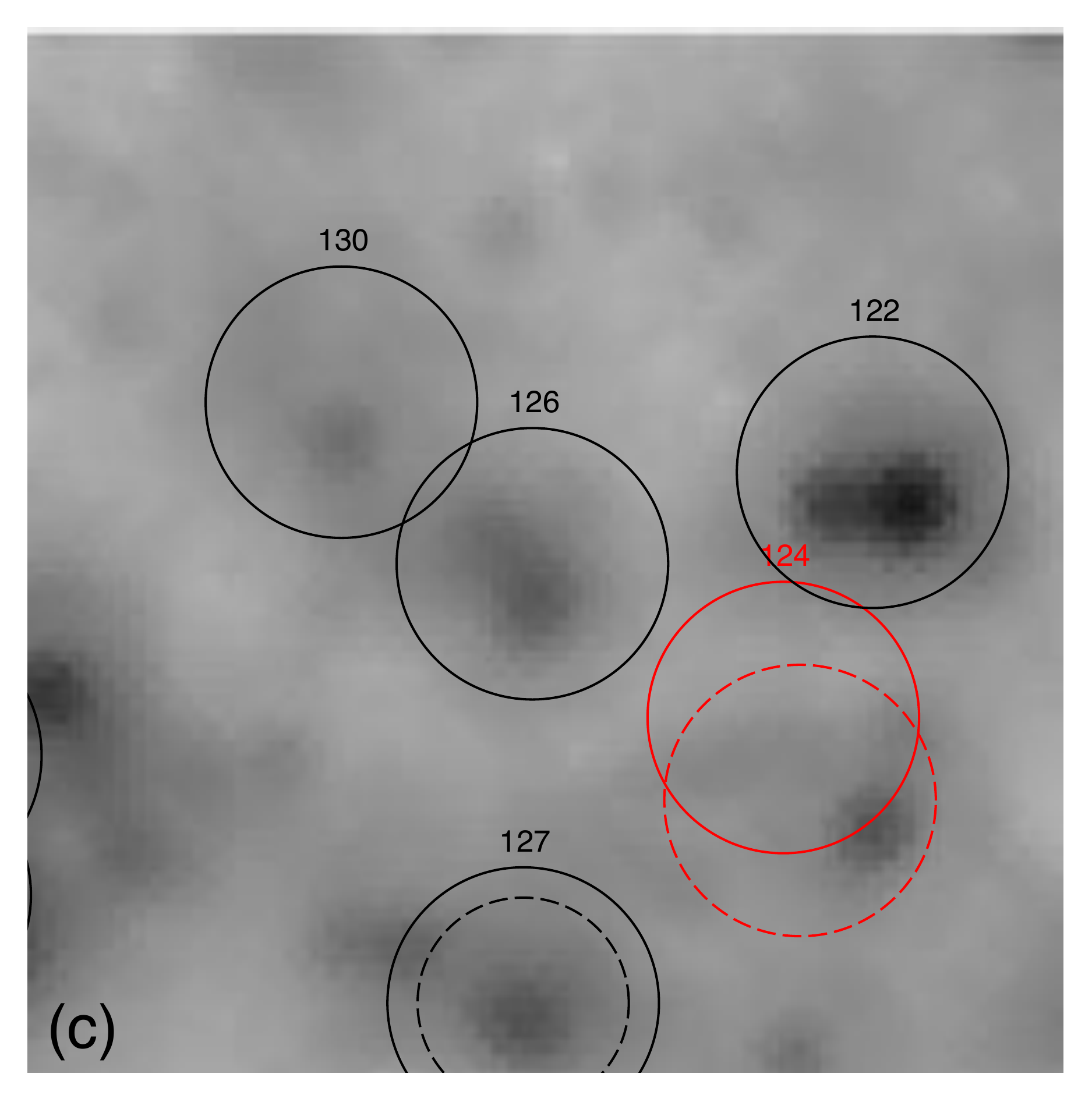}
\end{minipage}
\begin{minipage}{0.24\linewidth}
\includegraphics[width=\linewidth]{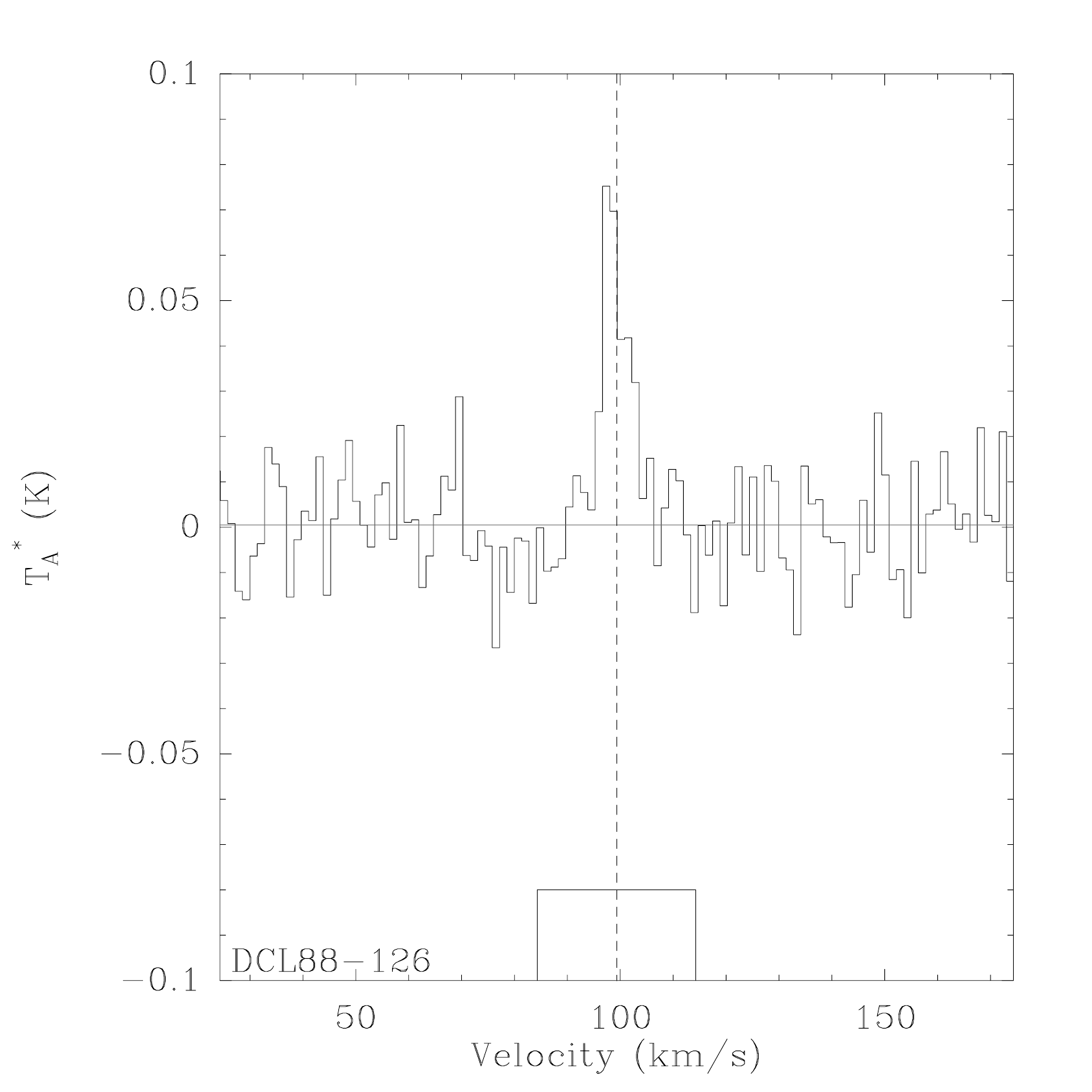}
\end{minipage}

\begin{minipage}{0.24\linewidth}
\includegraphics[width=\linewidth]{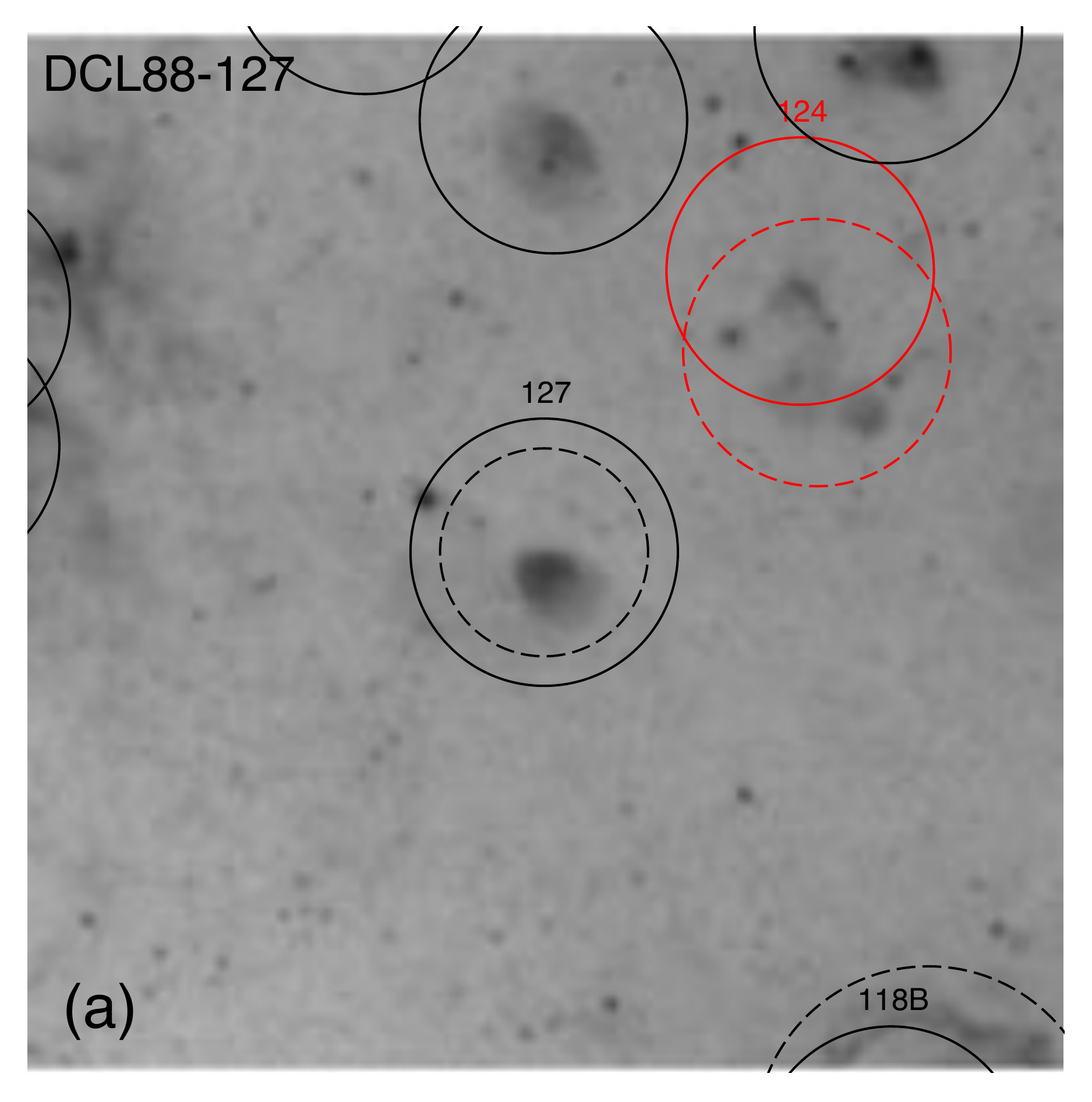}
\end{minipage}
\begin{minipage}{0.24\linewidth}
\includegraphics[width=\linewidth]{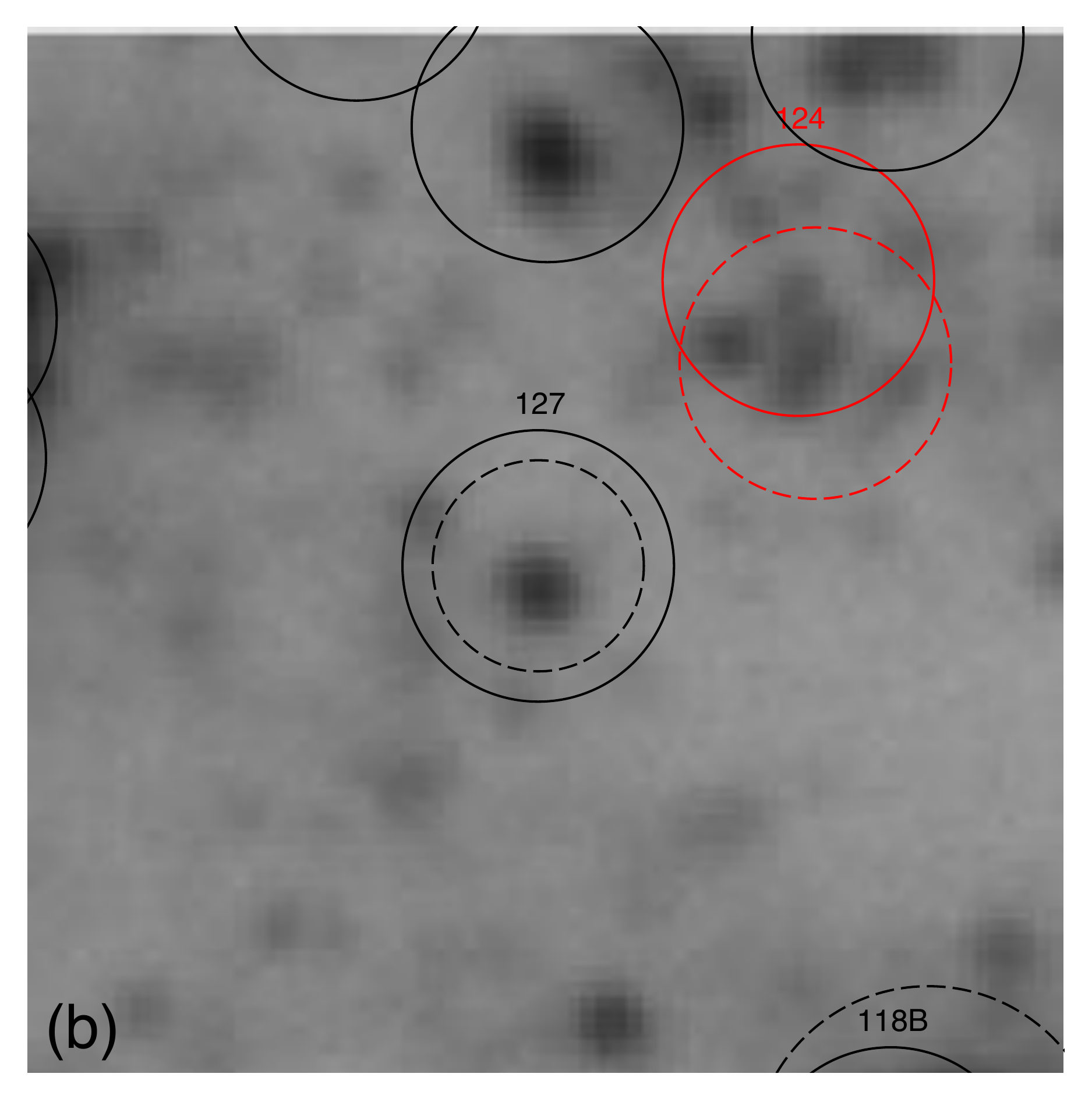}
\end{minipage}
\begin{minipage}{0.24\linewidth}
\includegraphics[width=\linewidth]{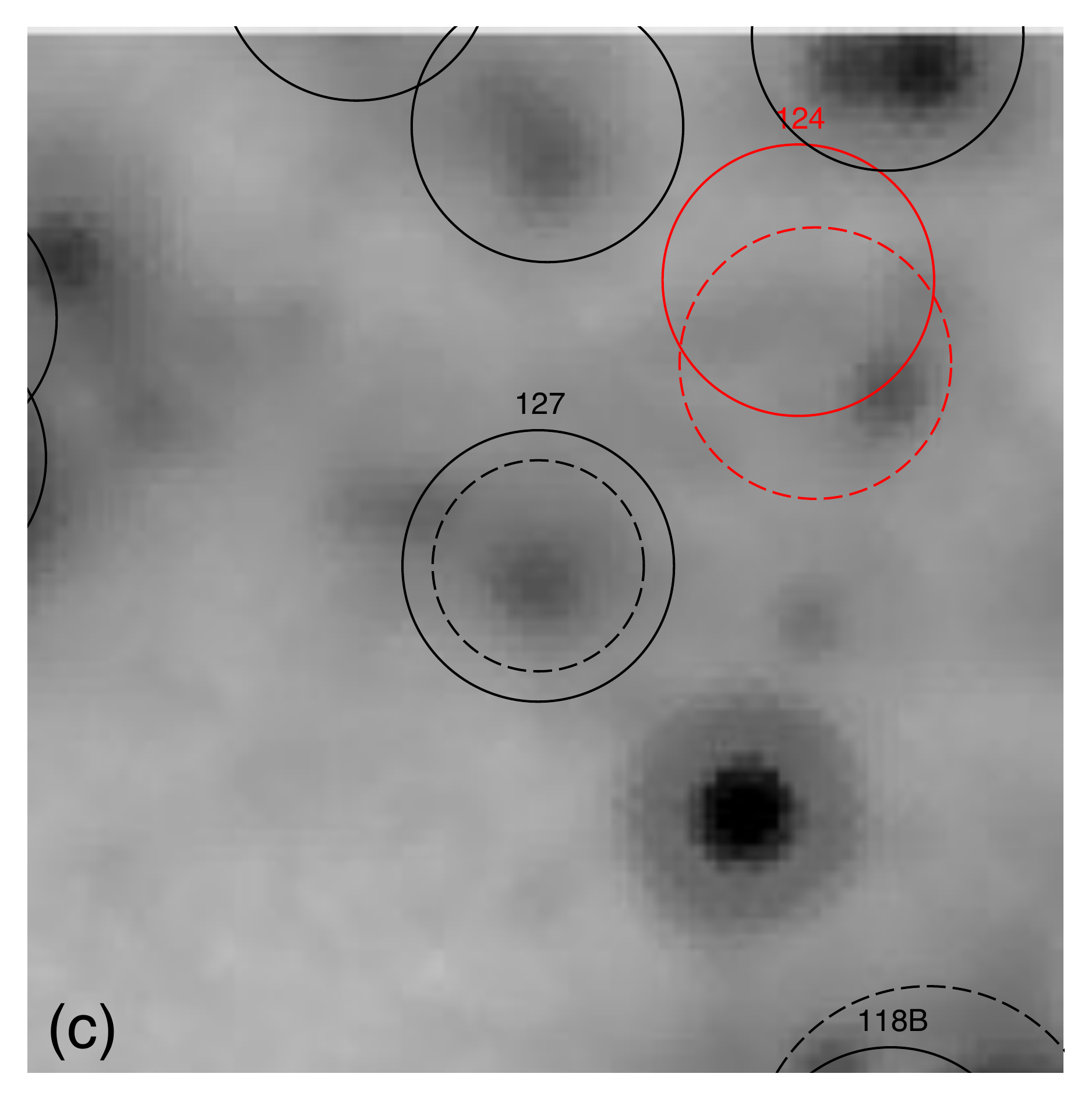}
\end{minipage}
\begin{minipage}{0.24\linewidth}
\includegraphics[width=\linewidth]{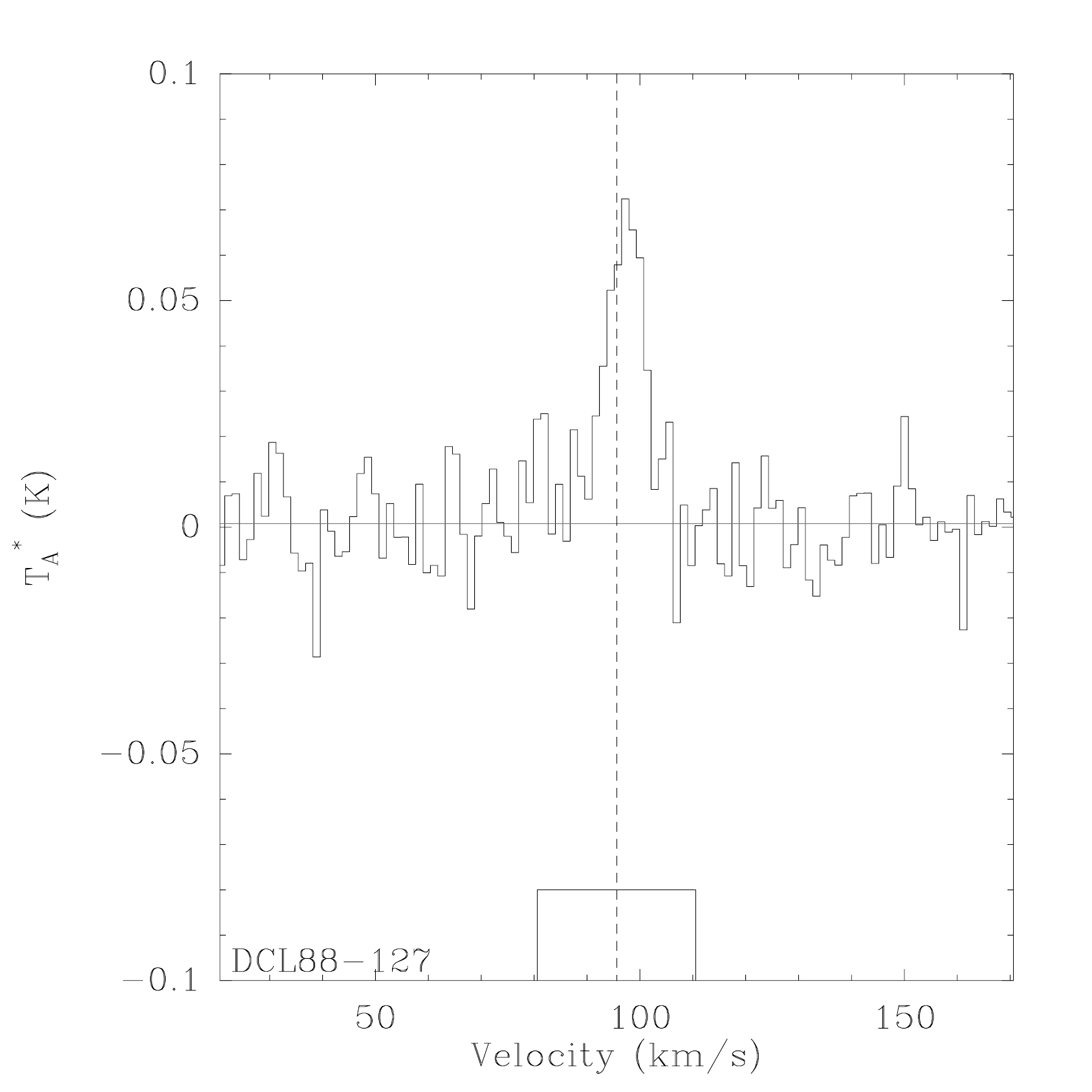}
\end{minipage}

\begin{minipage}{0.24\linewidth}
\includegraphics[width=\linewidth]{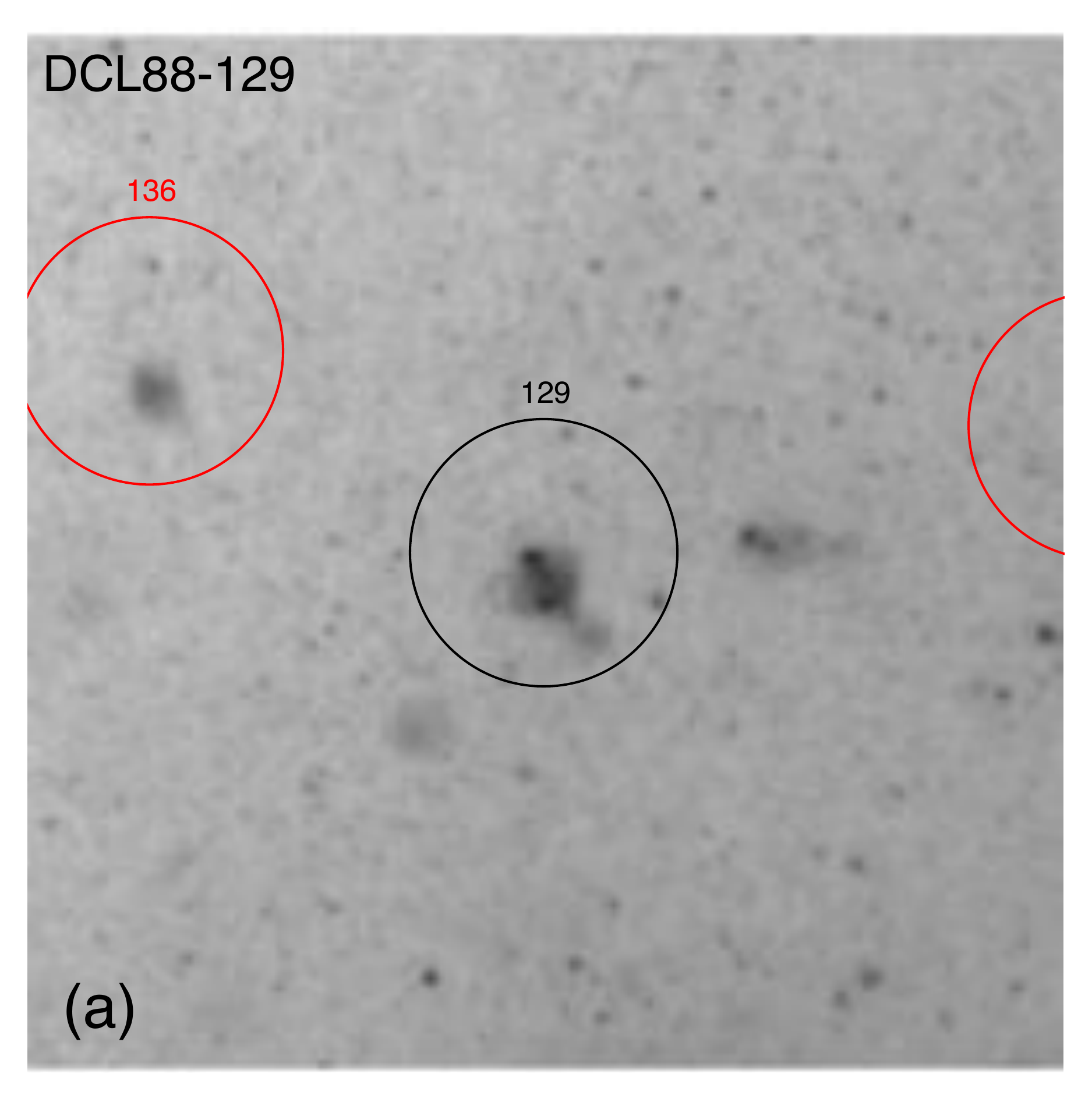}
\end{minipage}
\begin{minipage}{0.24\linewidth}
\includegraphics[width=\linewidth]{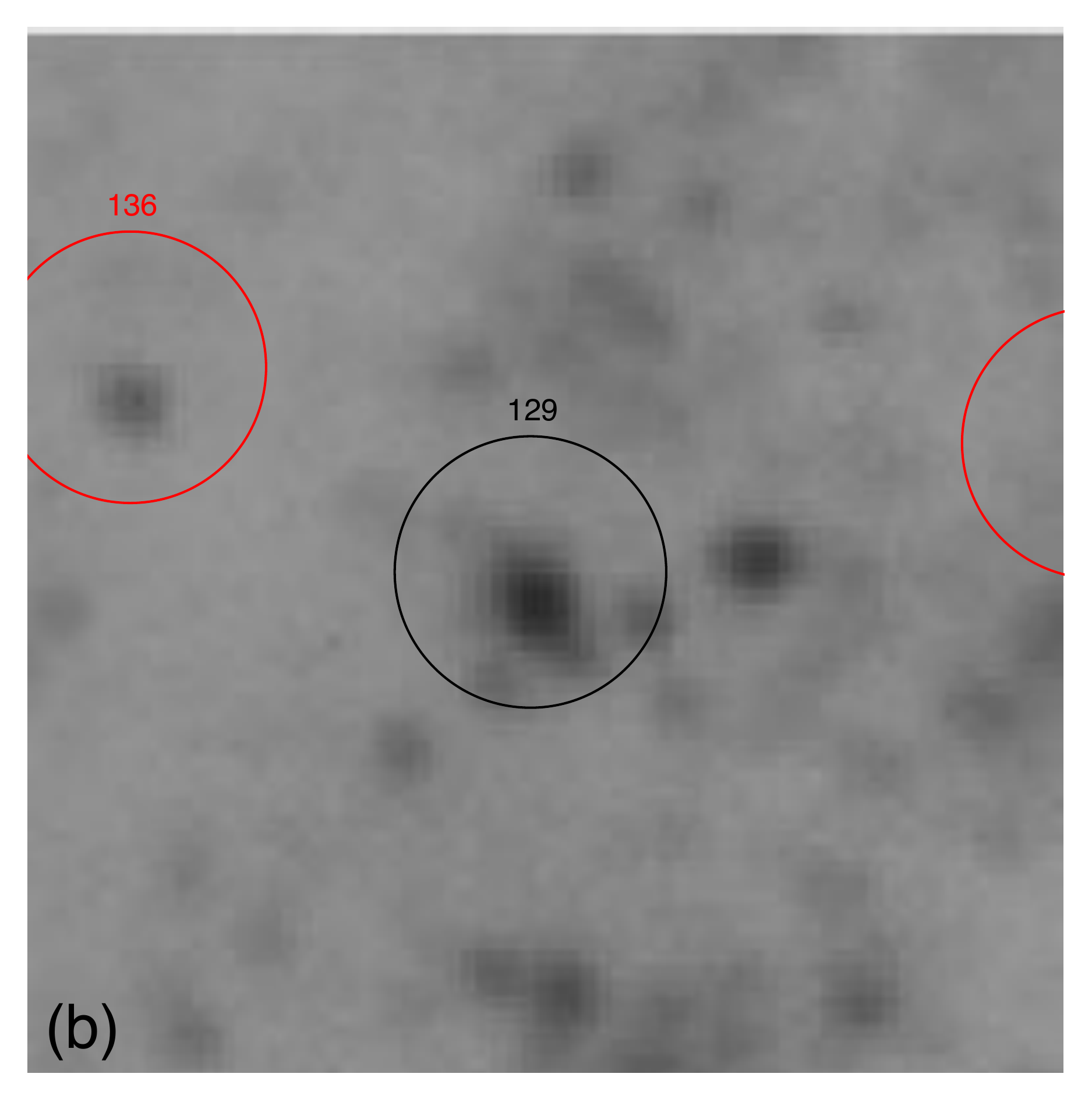}
\end{minipage}
\begin{minipage}{0.24\linewidth}
\includegraphics[width=\linewidth]{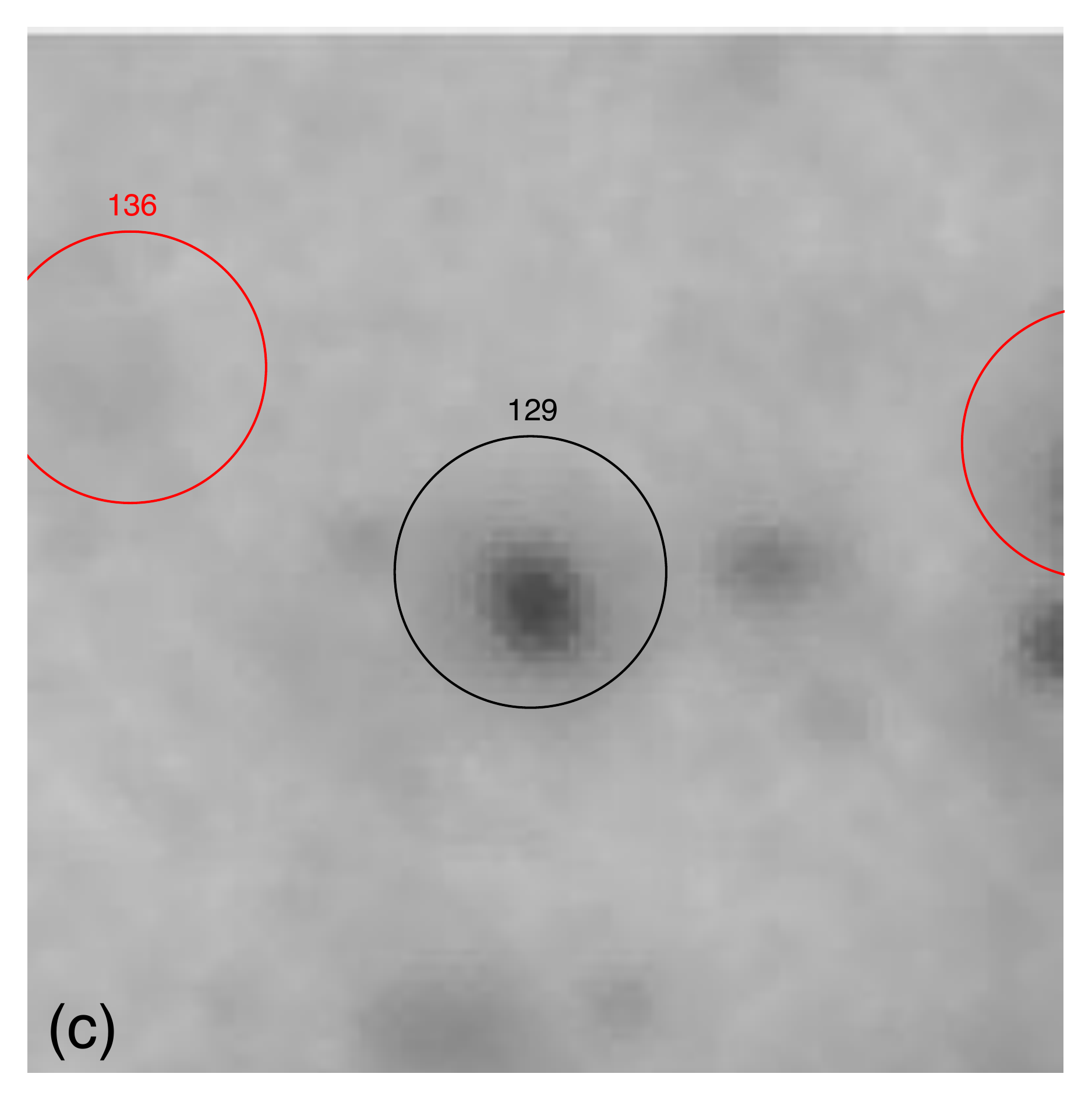}
\end{minipage}
\begin{minipage}{0.24\linewidth}
\includegraphics[width=\linewidth]{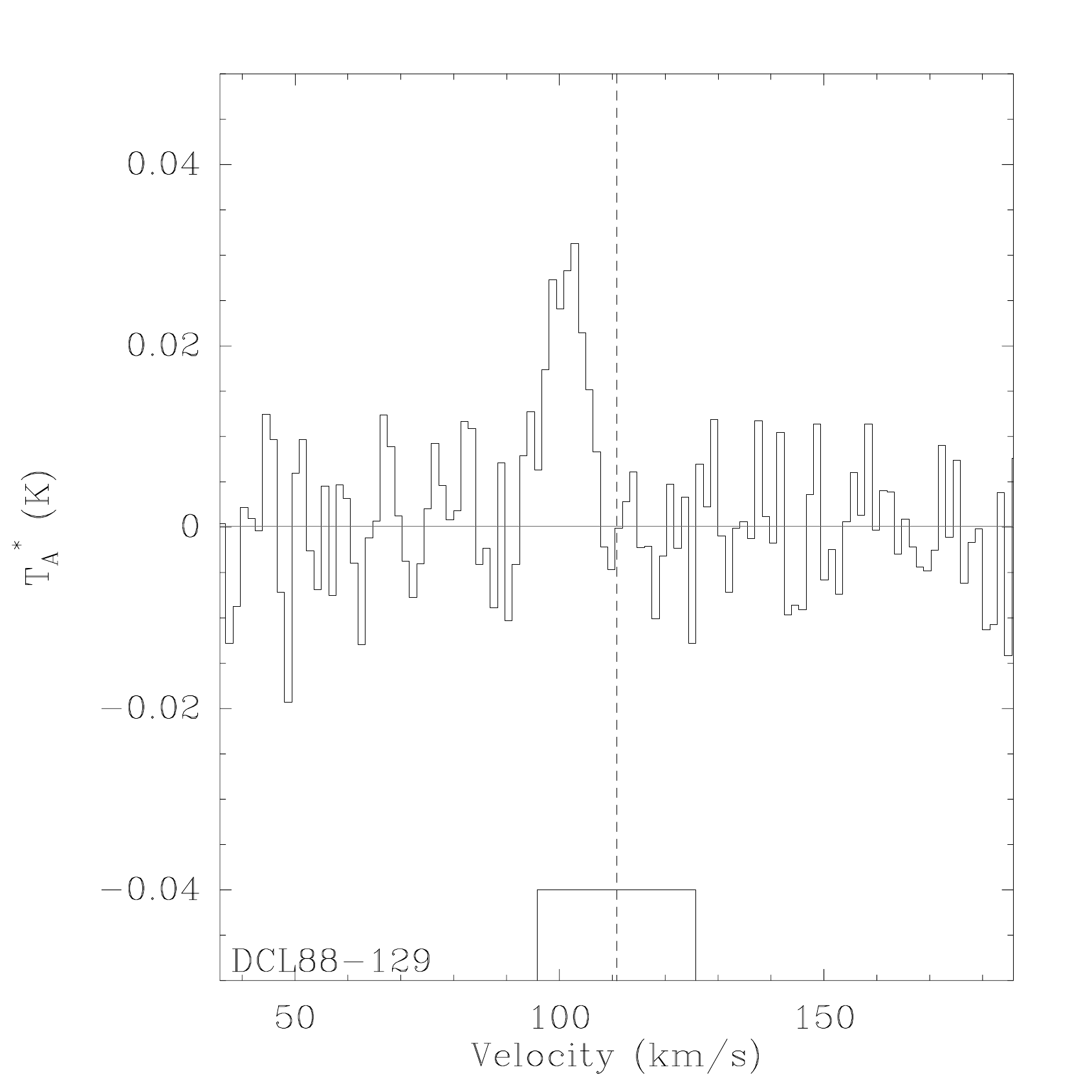}
\end{minipage}

\noindent\textbf{Figure~\ref{fig:stamps} -- continued.}

\end{figure*}

\begin{figure*}
%\ContinuedFloat

\begin{minipage}{0.24\linewidth}
\includegraphics[width=\linewidth]{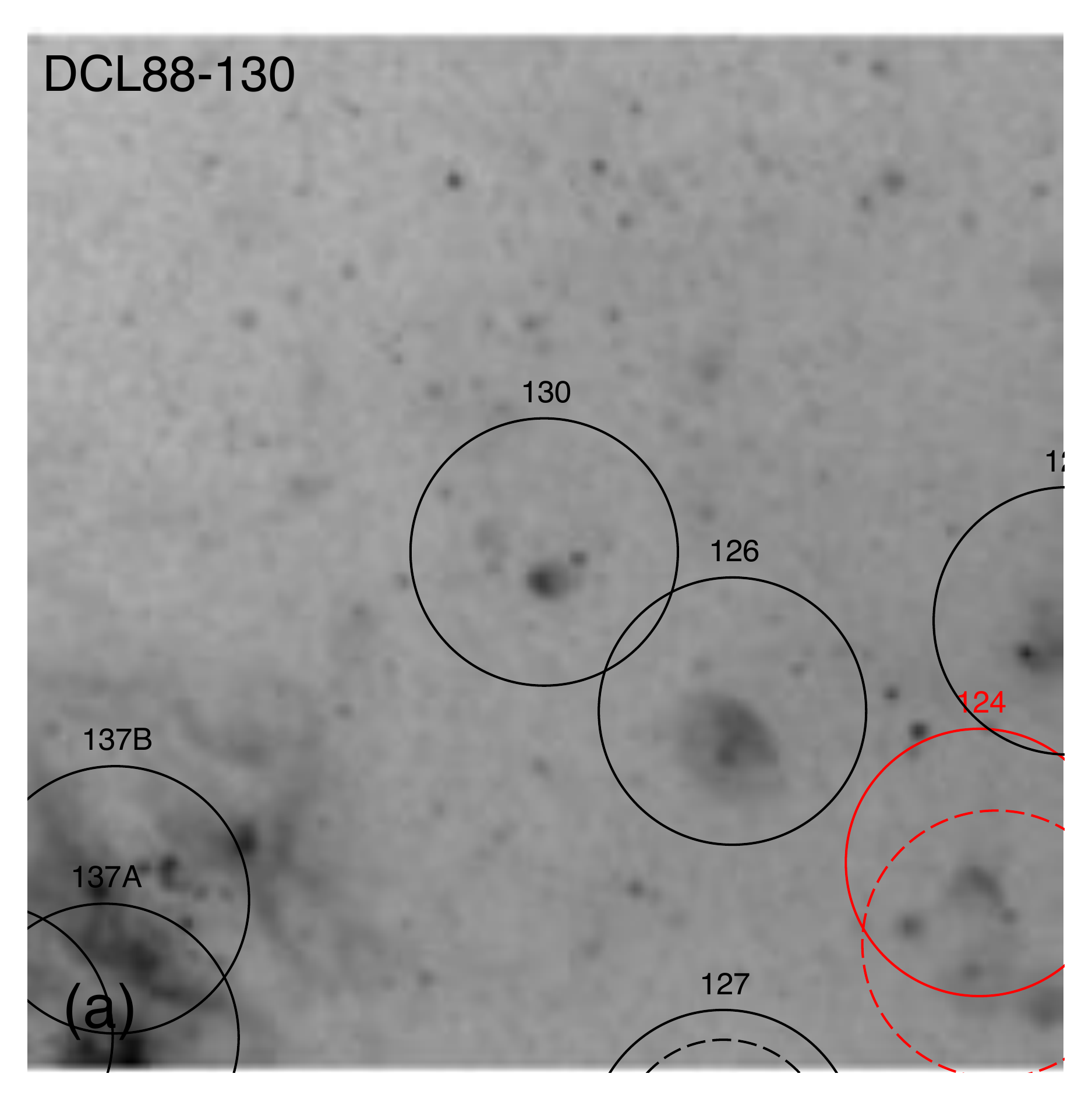}
\end{minipage}
\begin{minipage}{0.24\linewidth}
\includegraphics[width=\linewidth]{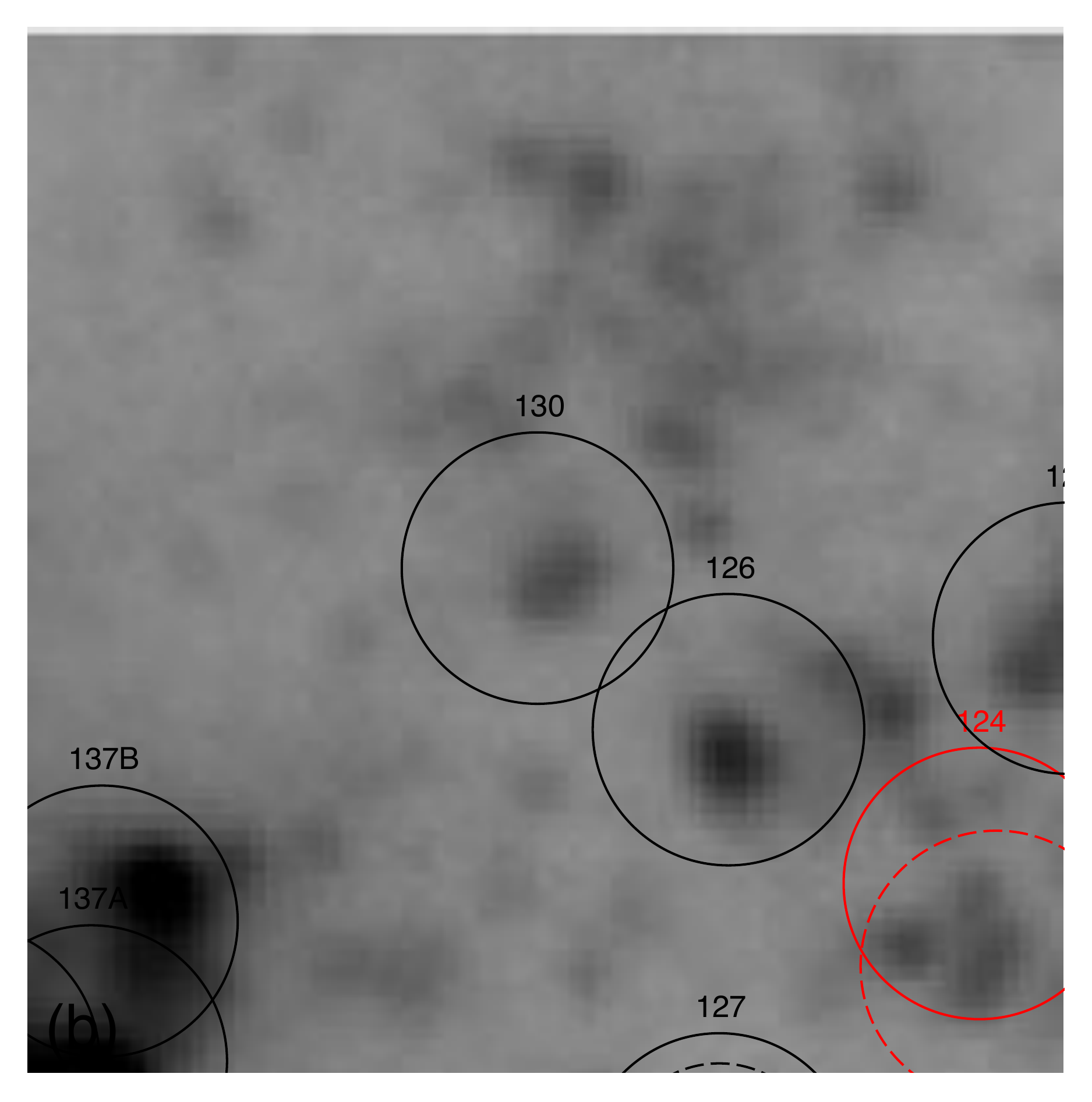}
\end{minipage}
\begin{minipage}{0.24\linewidth}
\includegraphics[width=\linewidth]{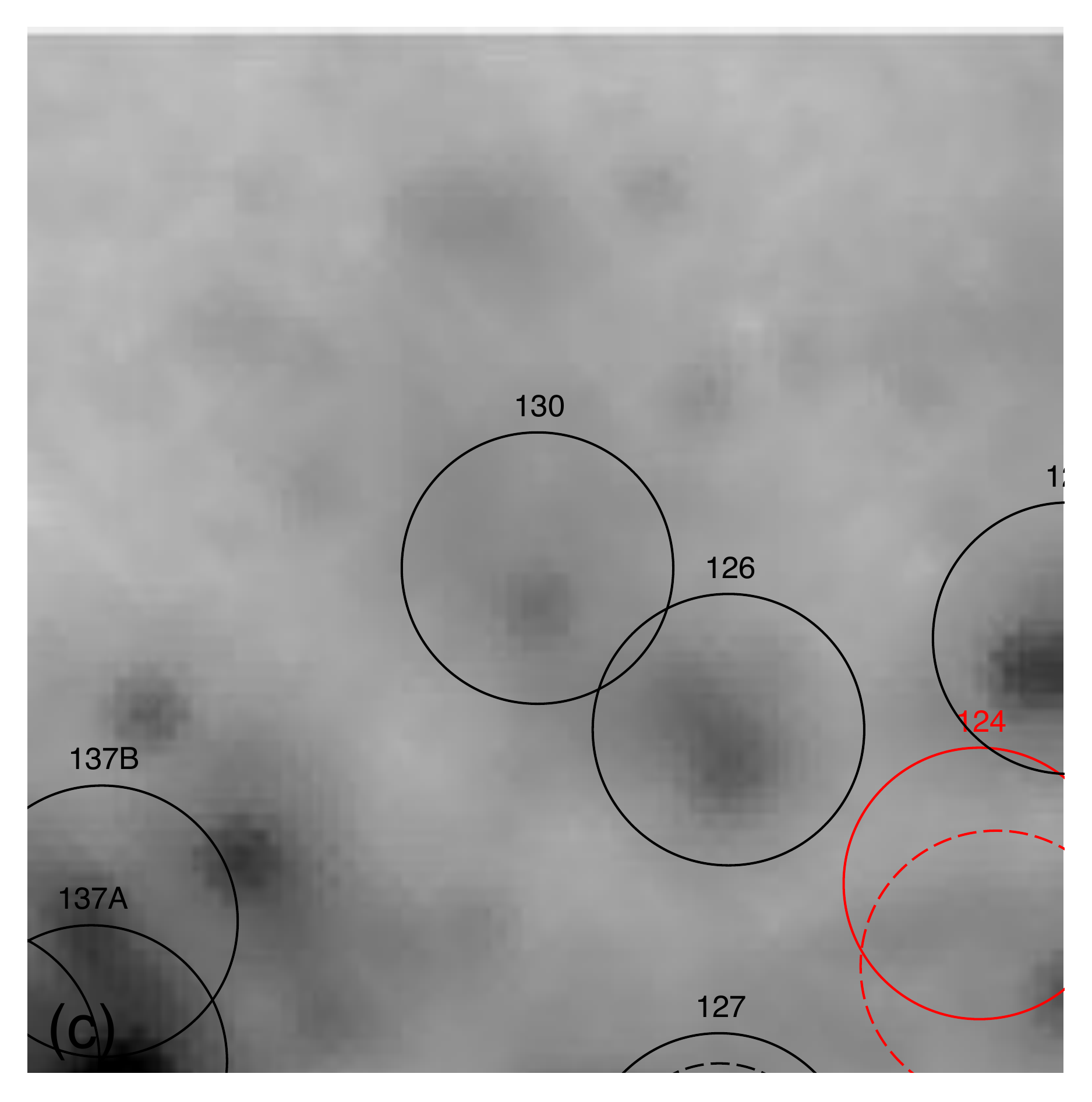}
\end{minipage}
\begin{minipage}{0.24\linewidth}
\includegraphics[width=\linewidth]{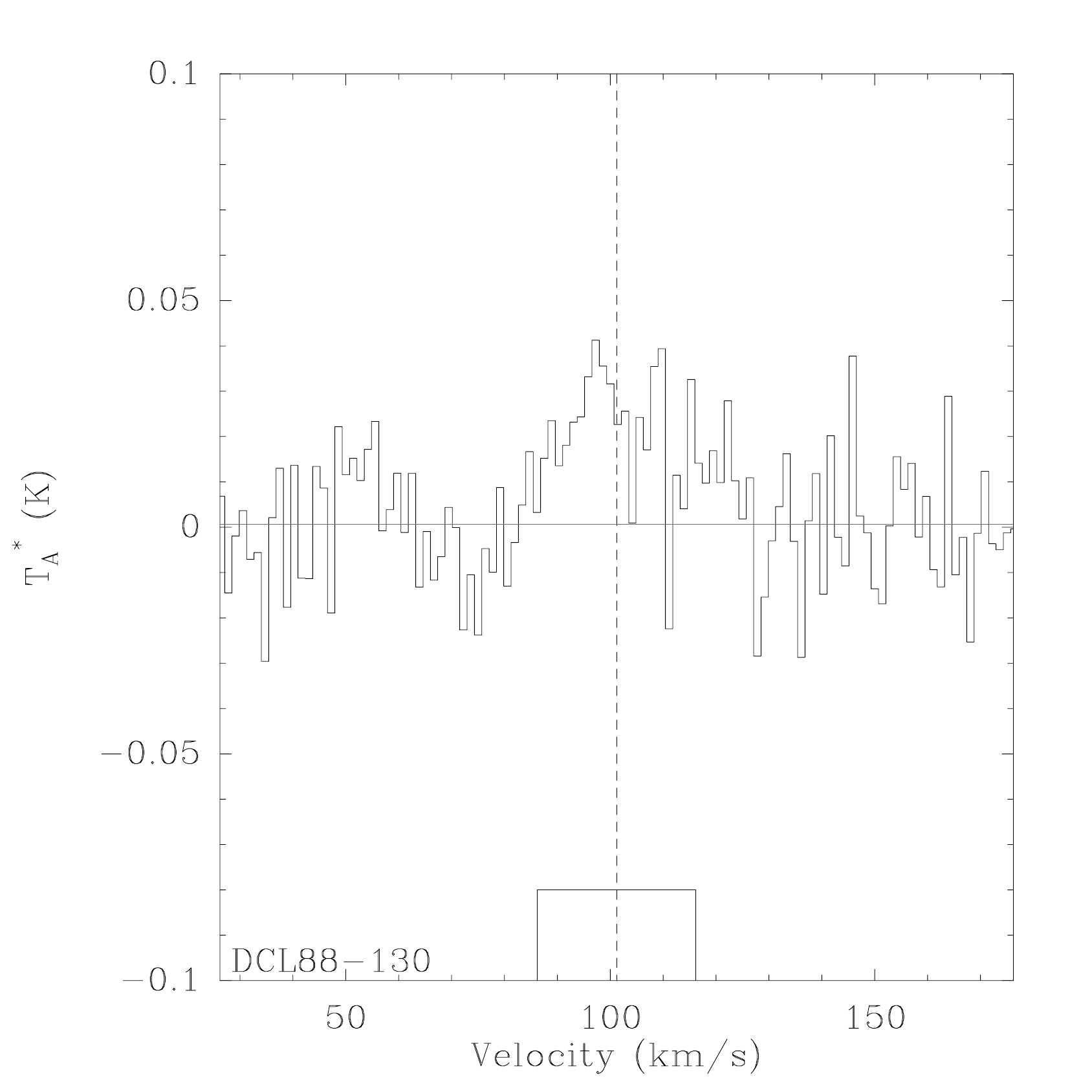}
\end{minipage}

\begin{minipage}{0.24\linewidth}
\includegraphics[width=\linewidth]{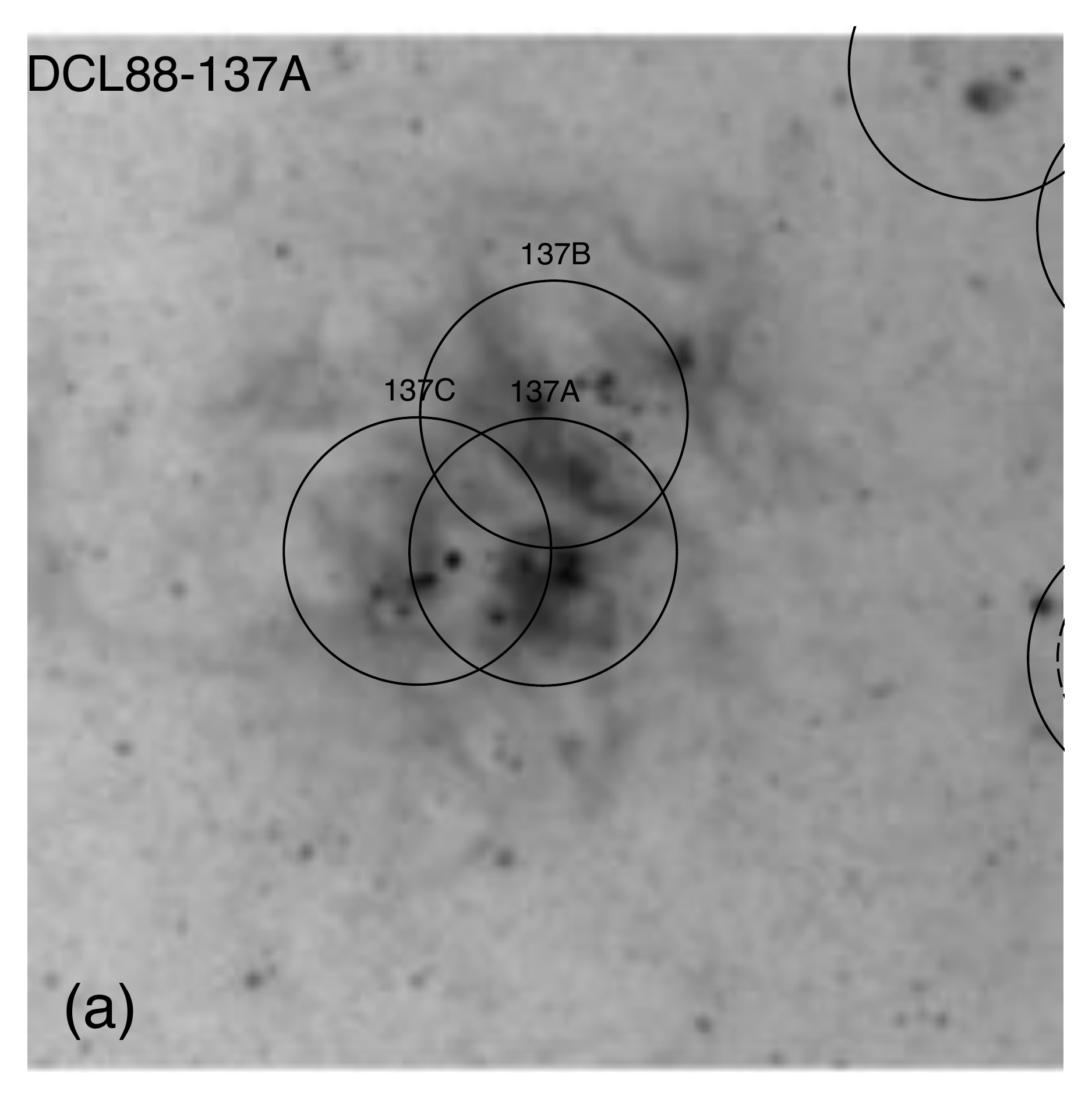}
\end{minipage}
\begin{minipage}{0.24\linewidth}
\includegraphics[width=\linewidth]{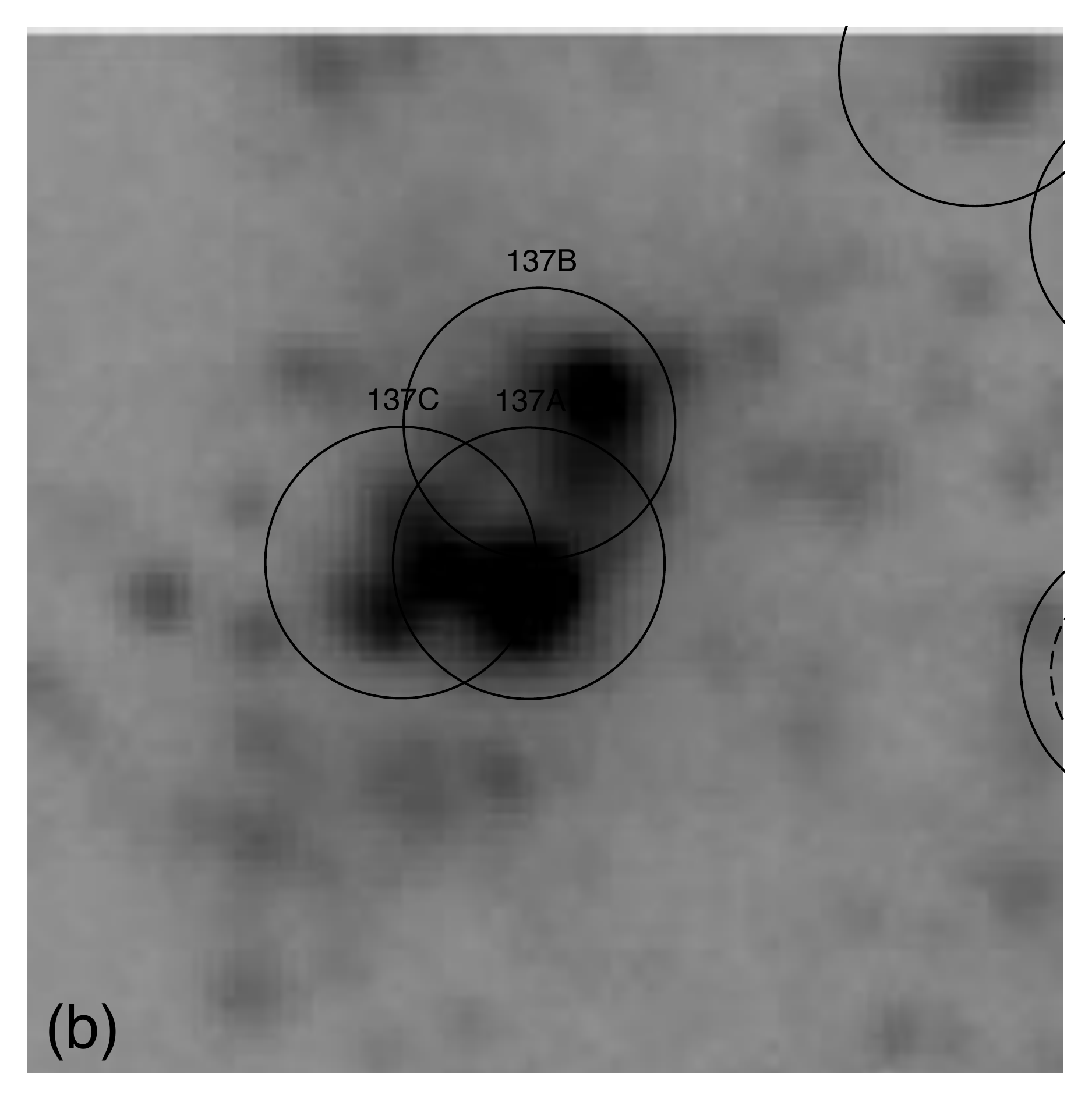}
\end{minipage}
\begin{minipage}{0.24\linewidth}
\includegraphics[width=\linewidth]{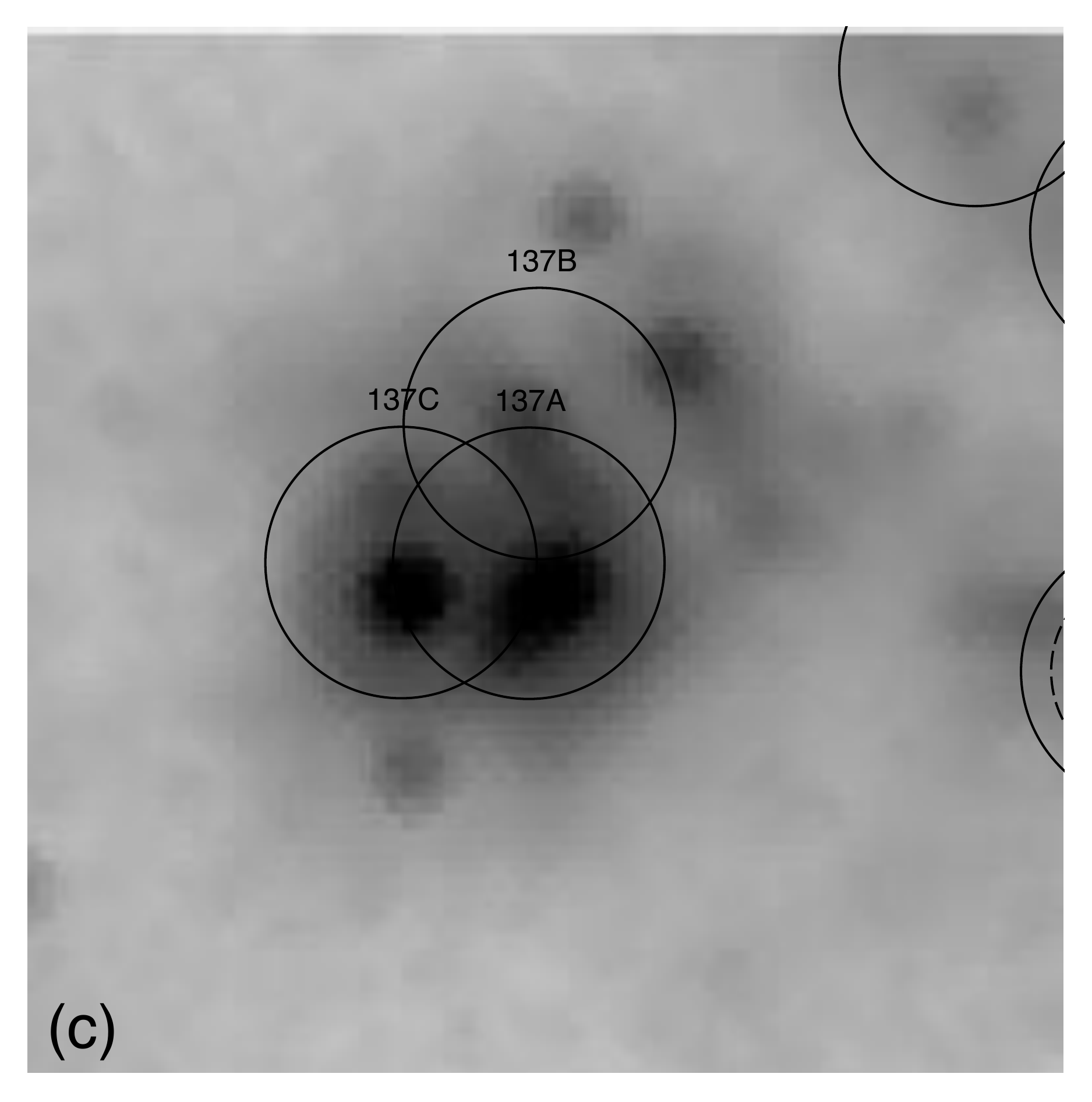}
\end{minipage}
\begin{minipage}{0.24\linewidth}
\includegraphics[width=\linewidth]{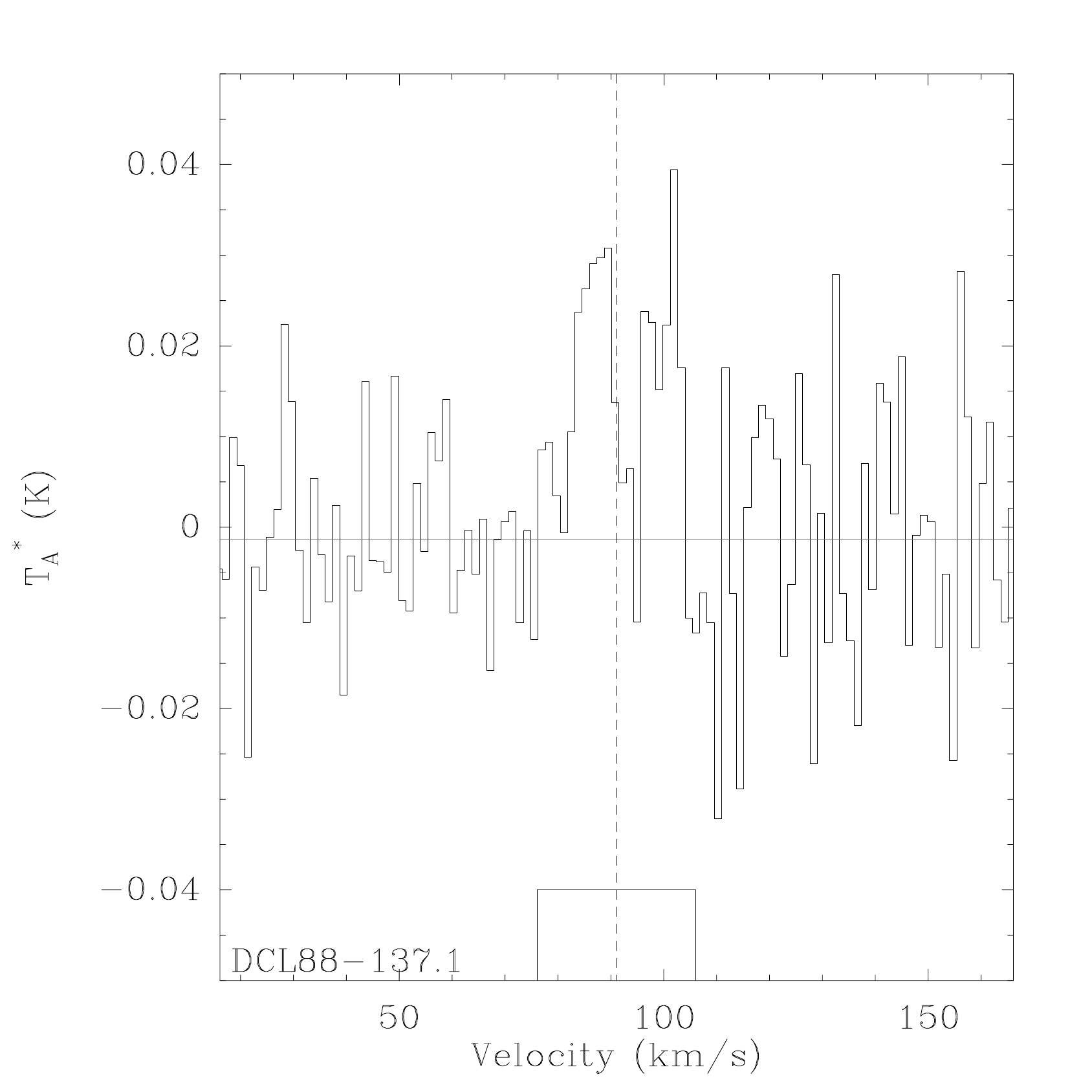}
\end{minipage}

\begin{minipage}{0.24\linewidth}
\includegraphics[width=\linewidth]{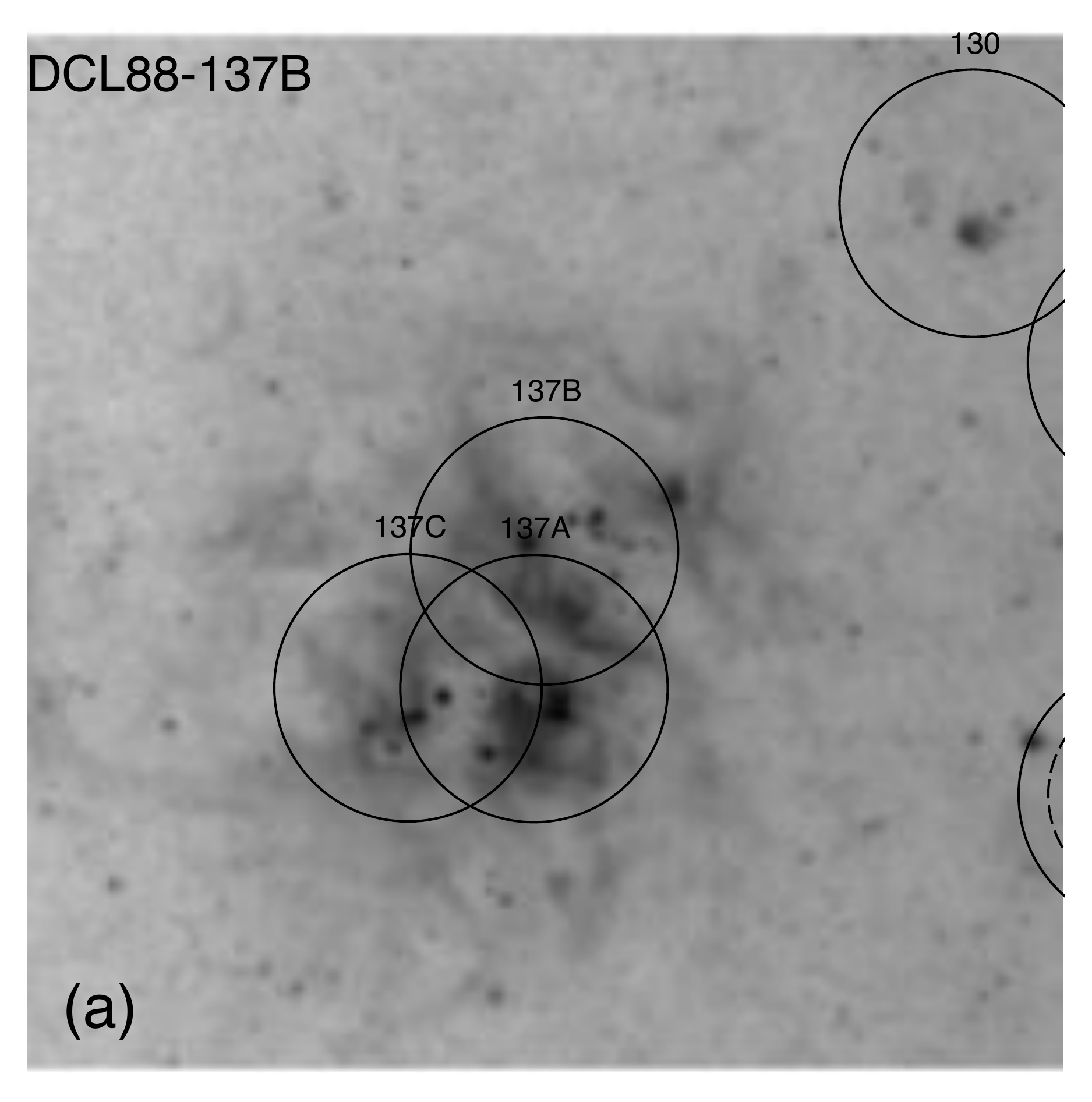}
\end{minipage}
\begin{minipage}{0.24\linewidth}
\includegraphics[width=\linewidth]{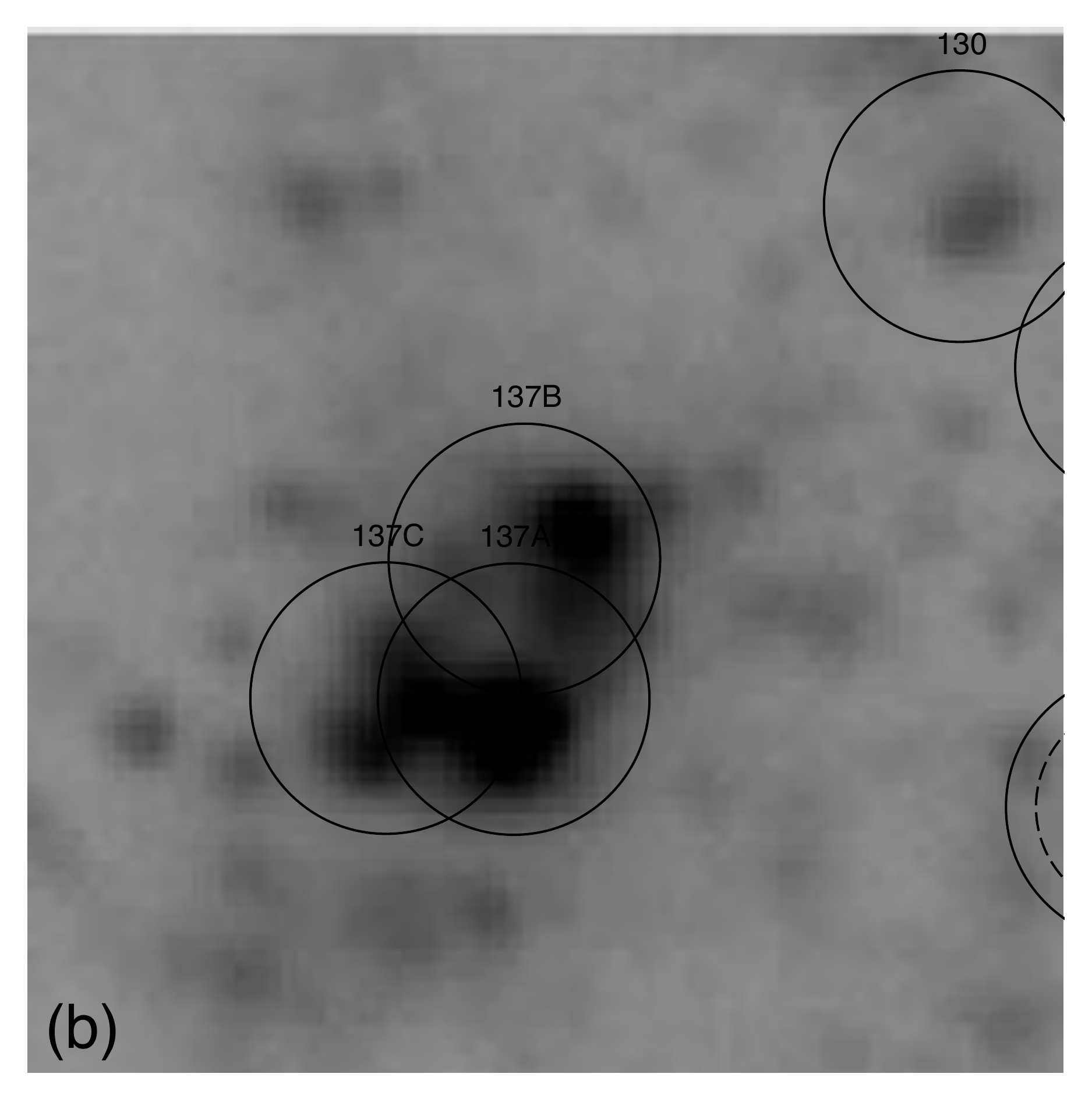}
\end{minipage}
\begin{minipage}{0.24\linewidth}
\includegraphics[width=\linewidth]{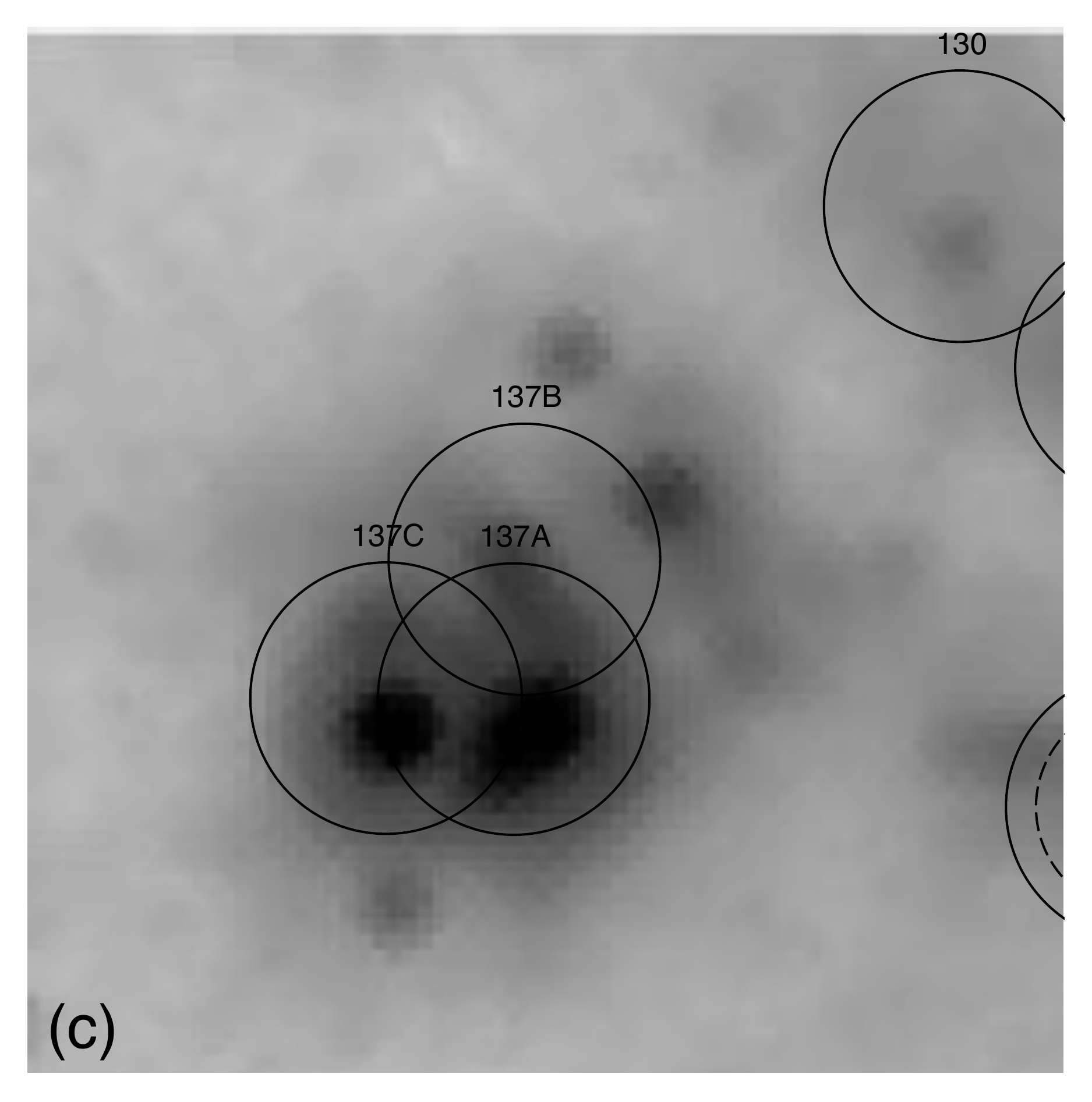}
\end{minipage}
\begin{minipage}{0.24\linewidth}
\includegraphics[width=\linewidth]{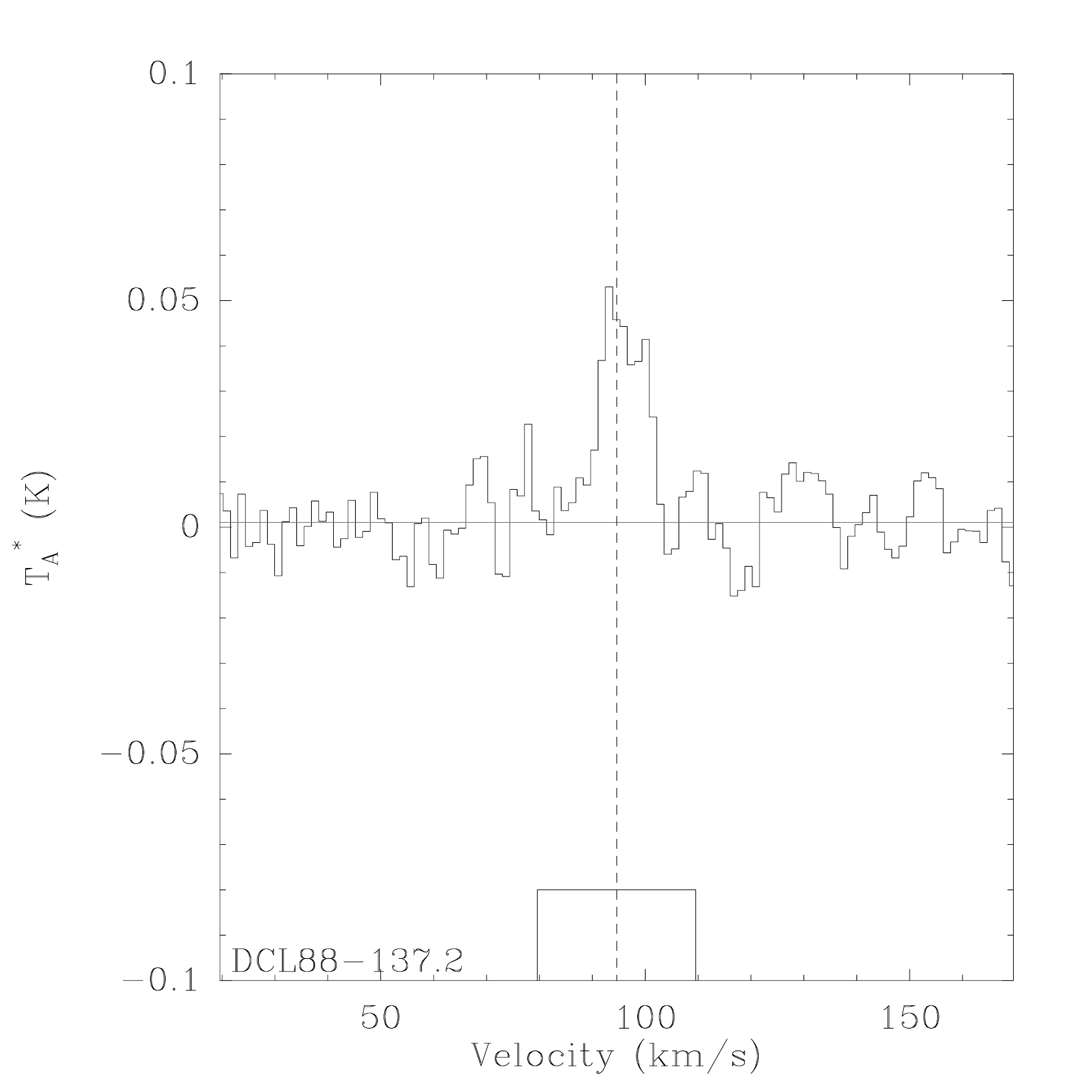}
\end{minipage}

\begin{minipage}{0.24\linewidth}
\includegraphics[width=\linewidth]{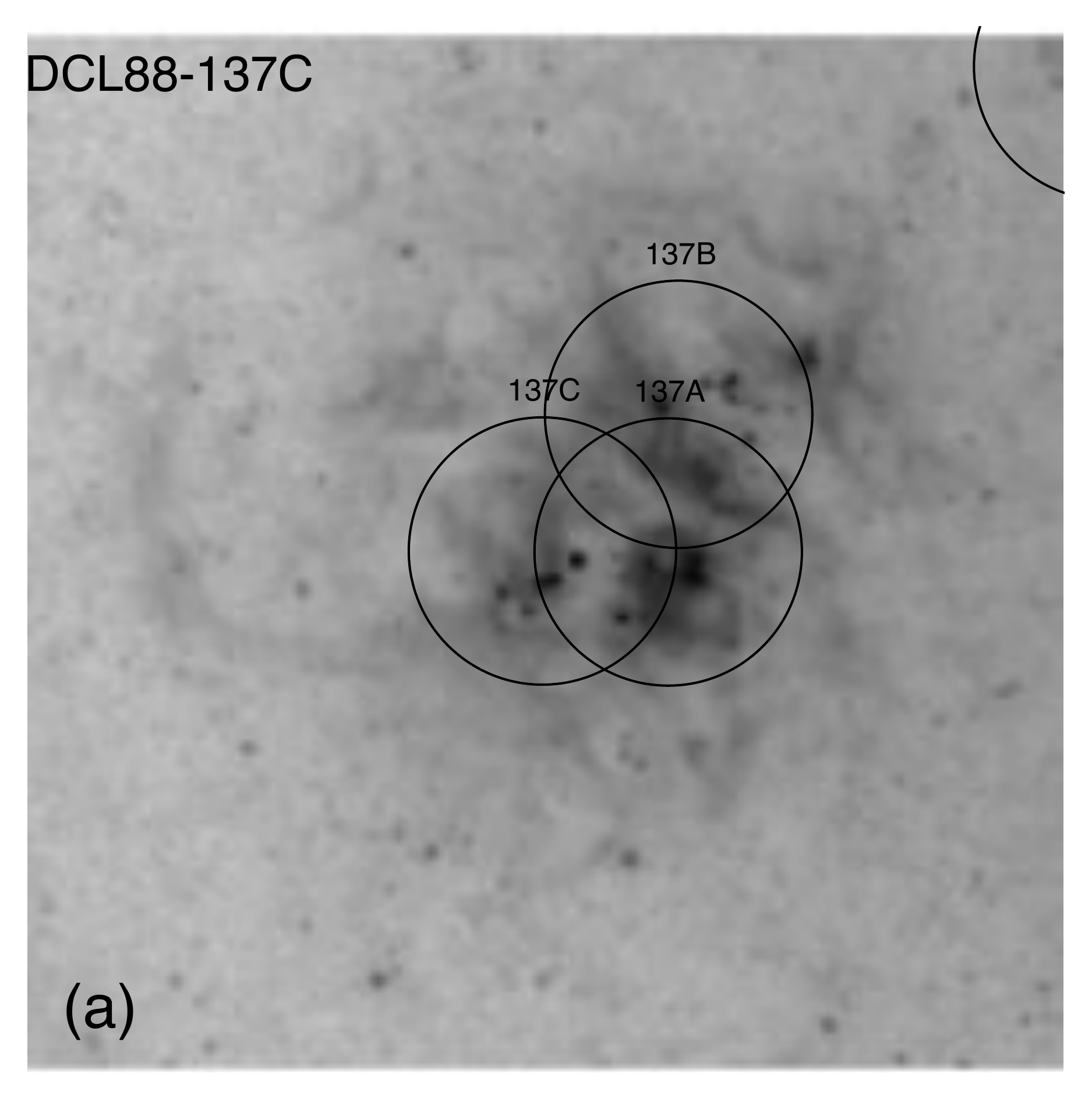}
\end{minipage}
\begin{minipage}{0.24\linewidth}
\includegraphics[width=\linewidth]{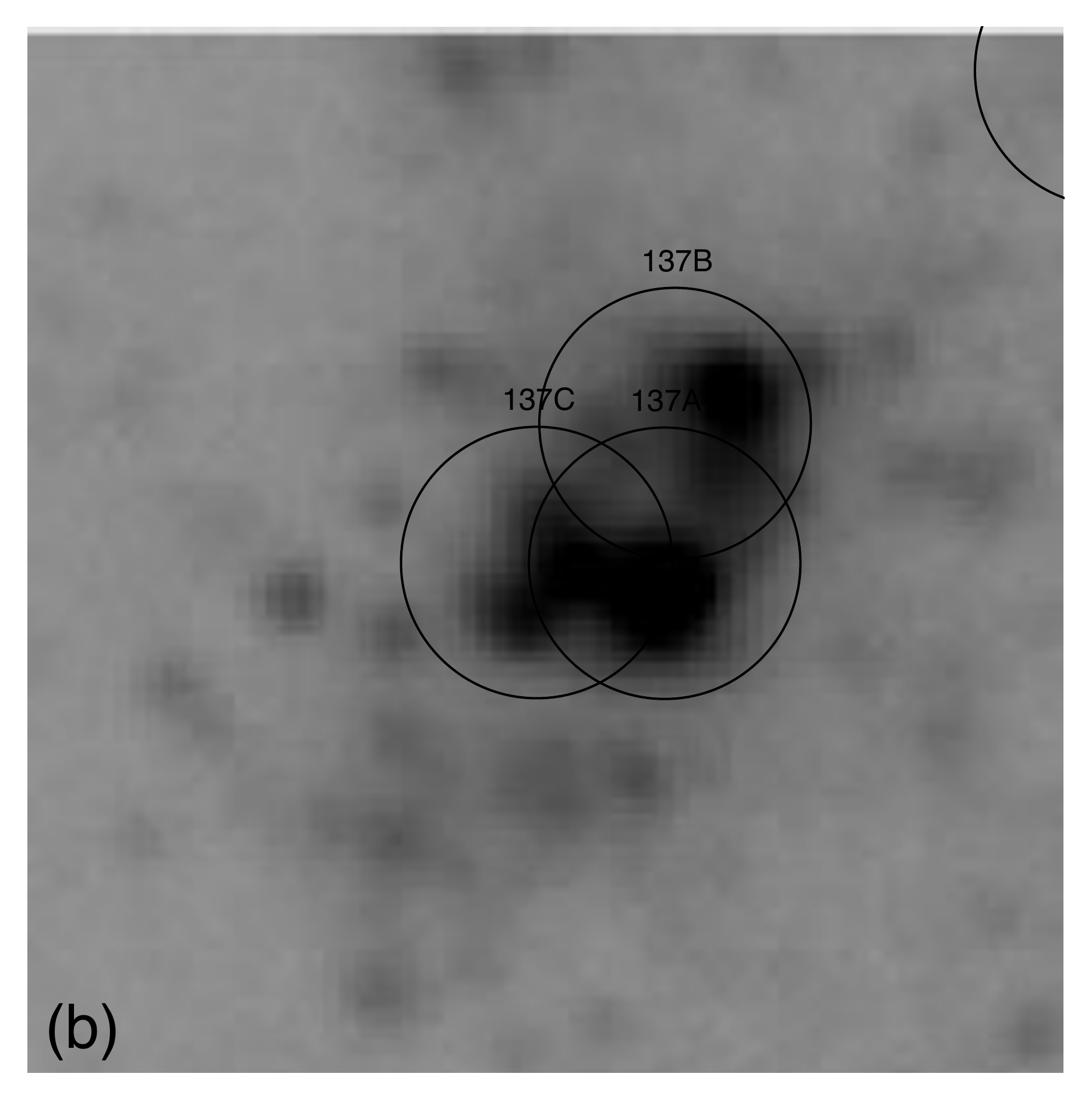}
\end{minipage}
\begin{minipage}{0.24\linewidth}
\includegraphics[width=\linewidth]{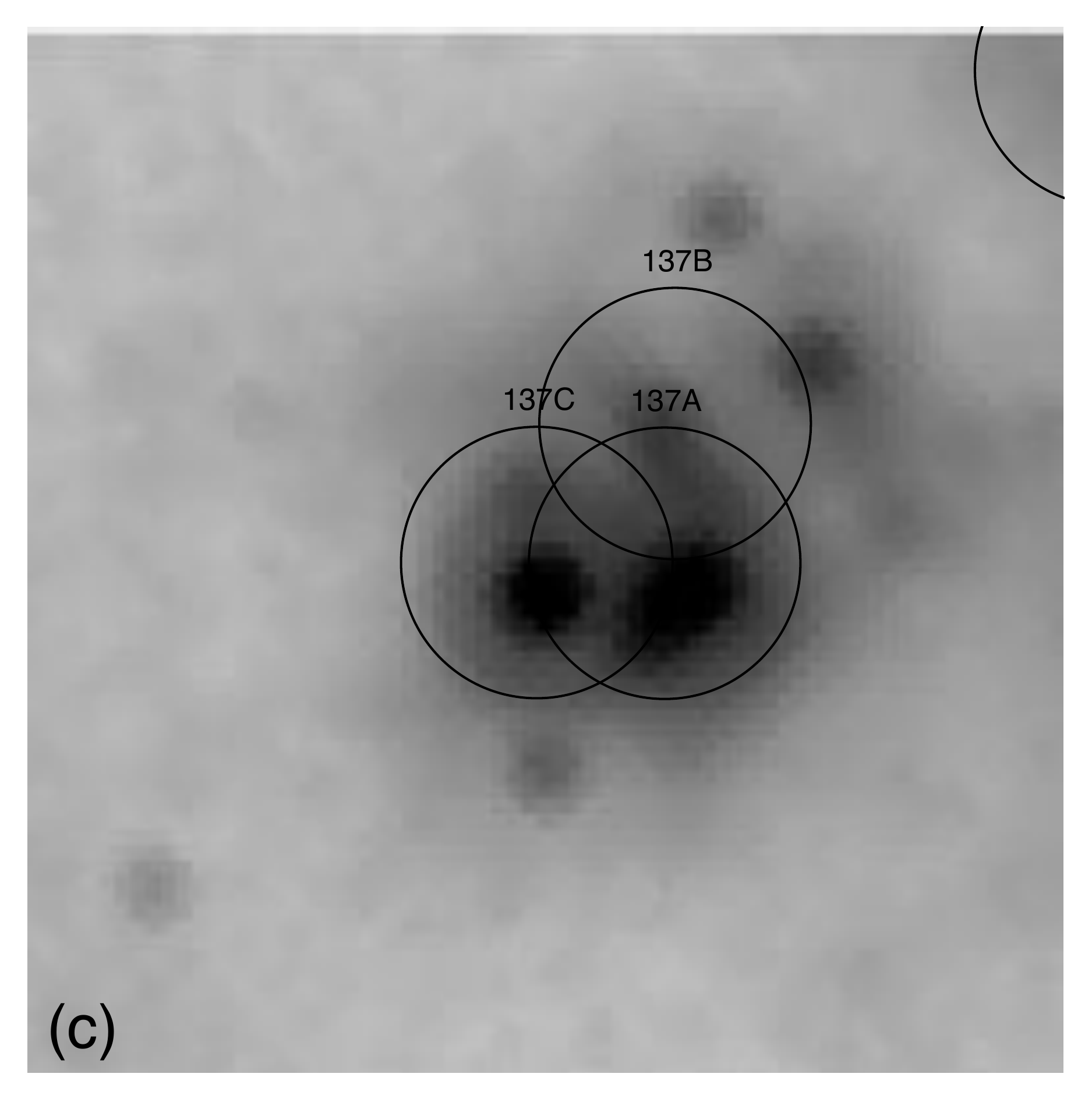}
\end{minipage}
\begin{minipage}{0.24\linewidth}
\includegraphics[width=\linewidth]{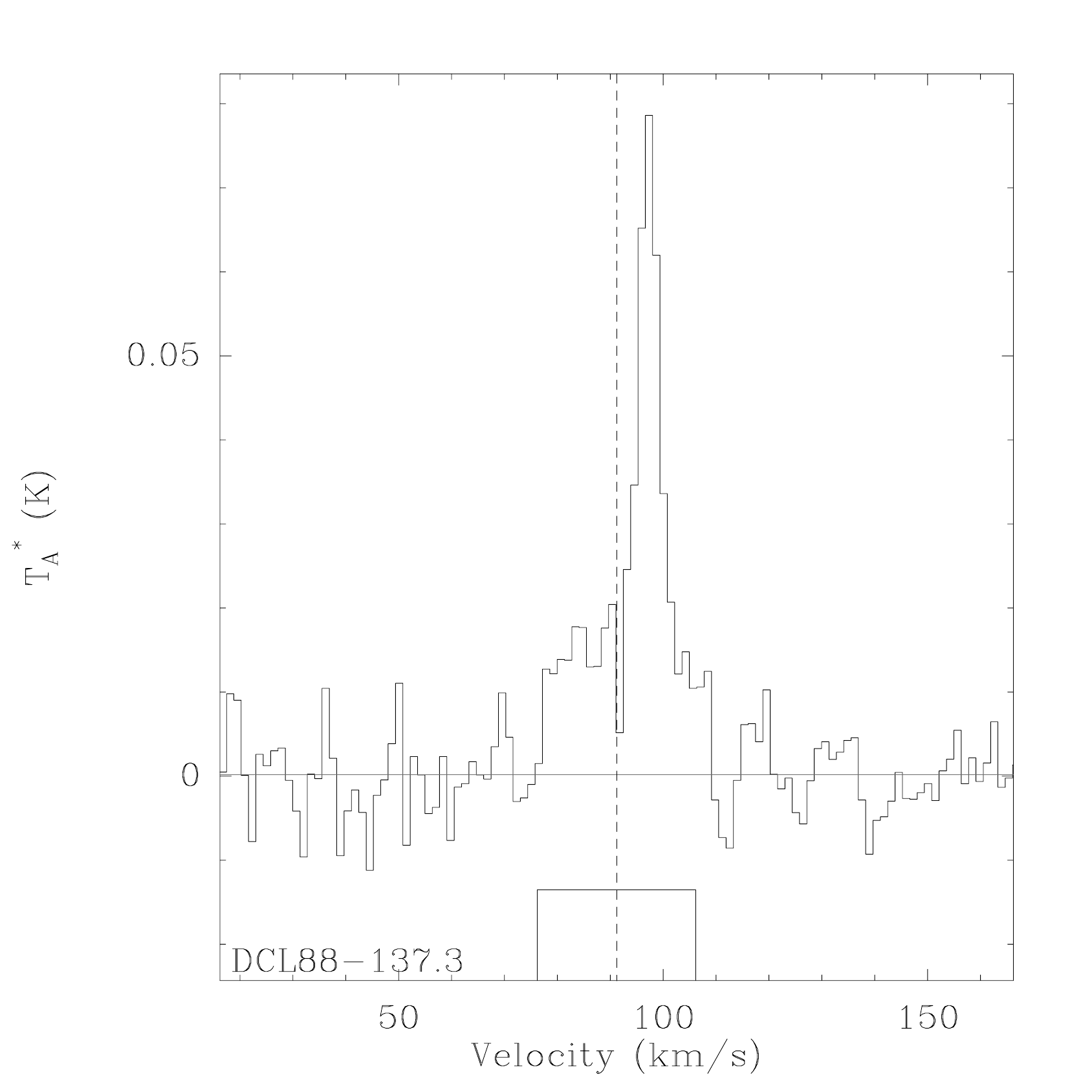}
\end{minipage}

\begin{minipage}{0.24\linewidth}
\includegraphics[width=\linewidth]{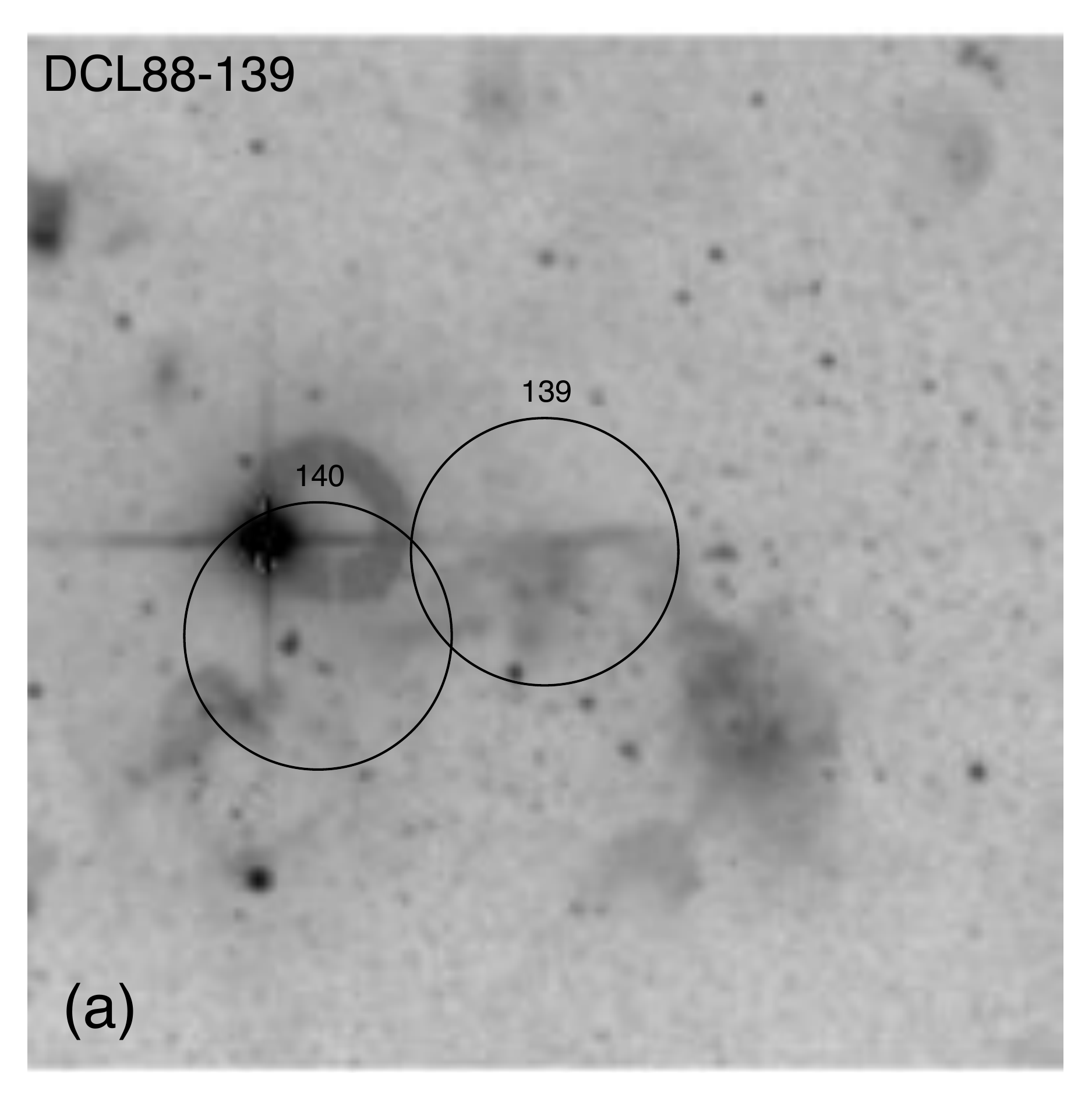}
\end{minipage}
\begin{minipage}{0.24\linewidth}
\includegraphics[width=\linewidth]{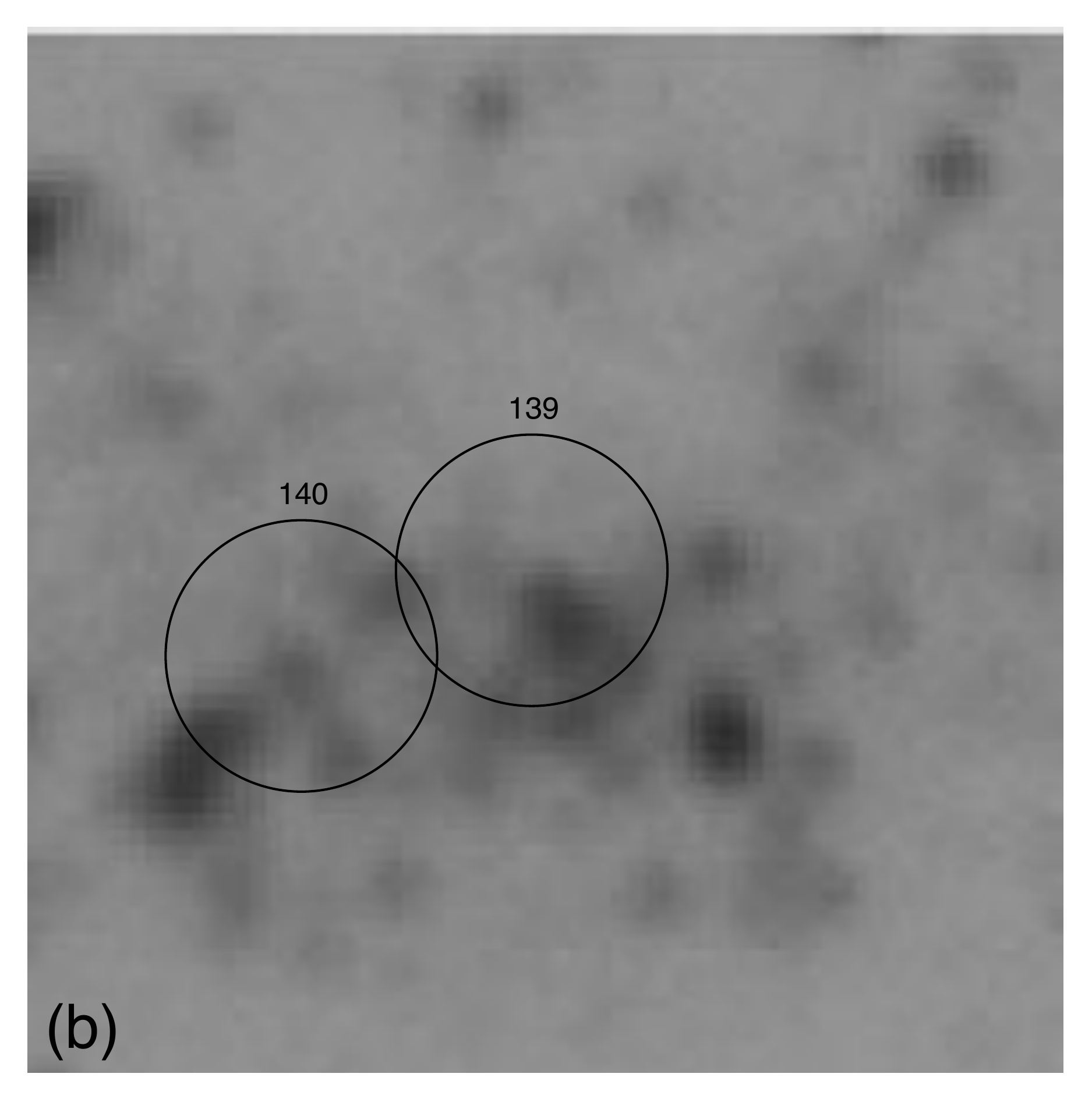}
\end{minipage}
\begin{minipage}{0.24\linewidth}
\includegraphics[width=\linewidth]{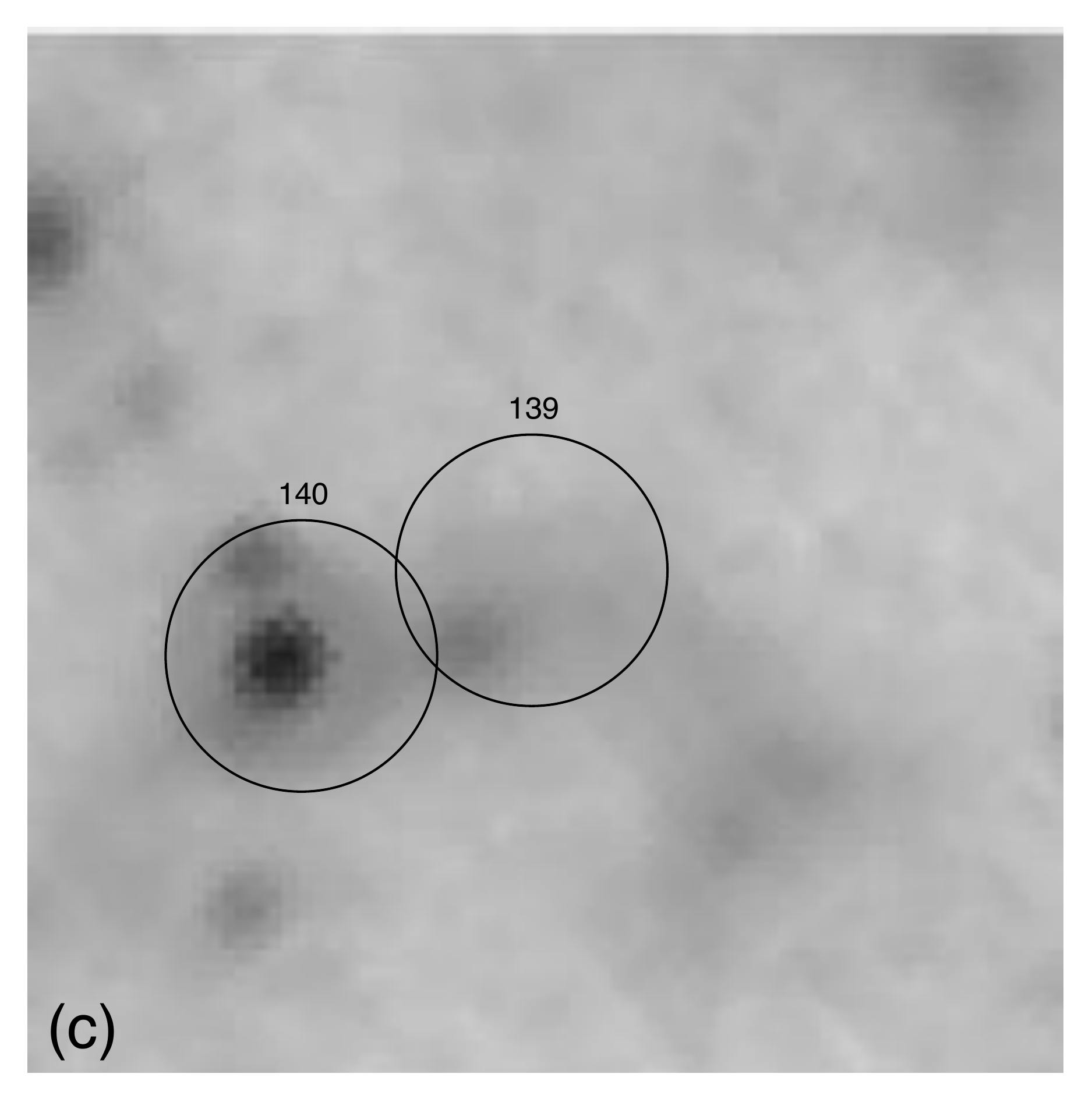}
\end{minipage}
\begin{minipage}{0.24\linewidth}
\includegraphics[width=\linewidth]{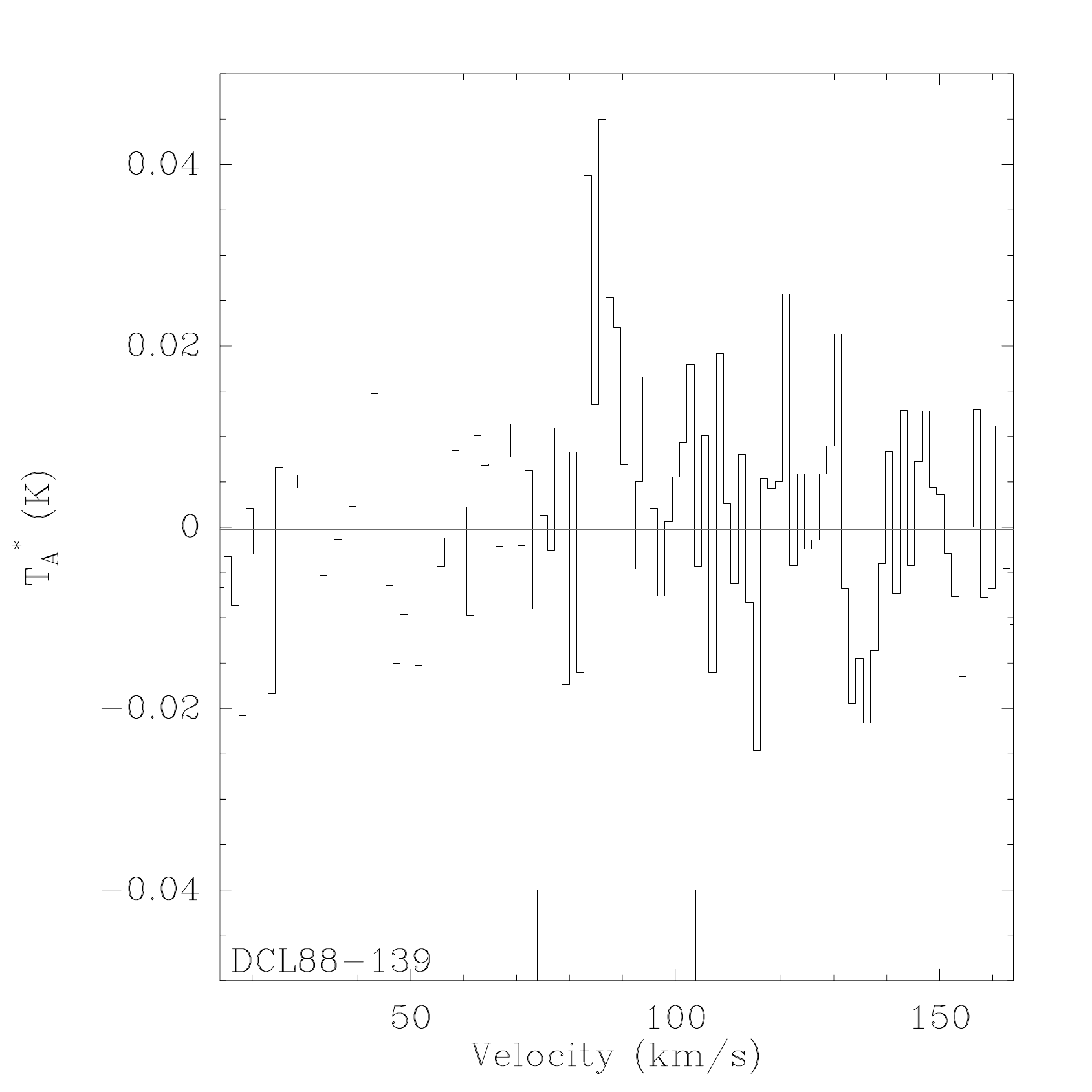}
\end{minipage}

\noindent\textbf{Figure~\ref{fig:stamps} -- continued.}

\end{figure*}

\begin{figure*}
%\ContinuedFloat

\begin{minipage}{0.24\linewidth}
\includegraphics[width=\linewidth]{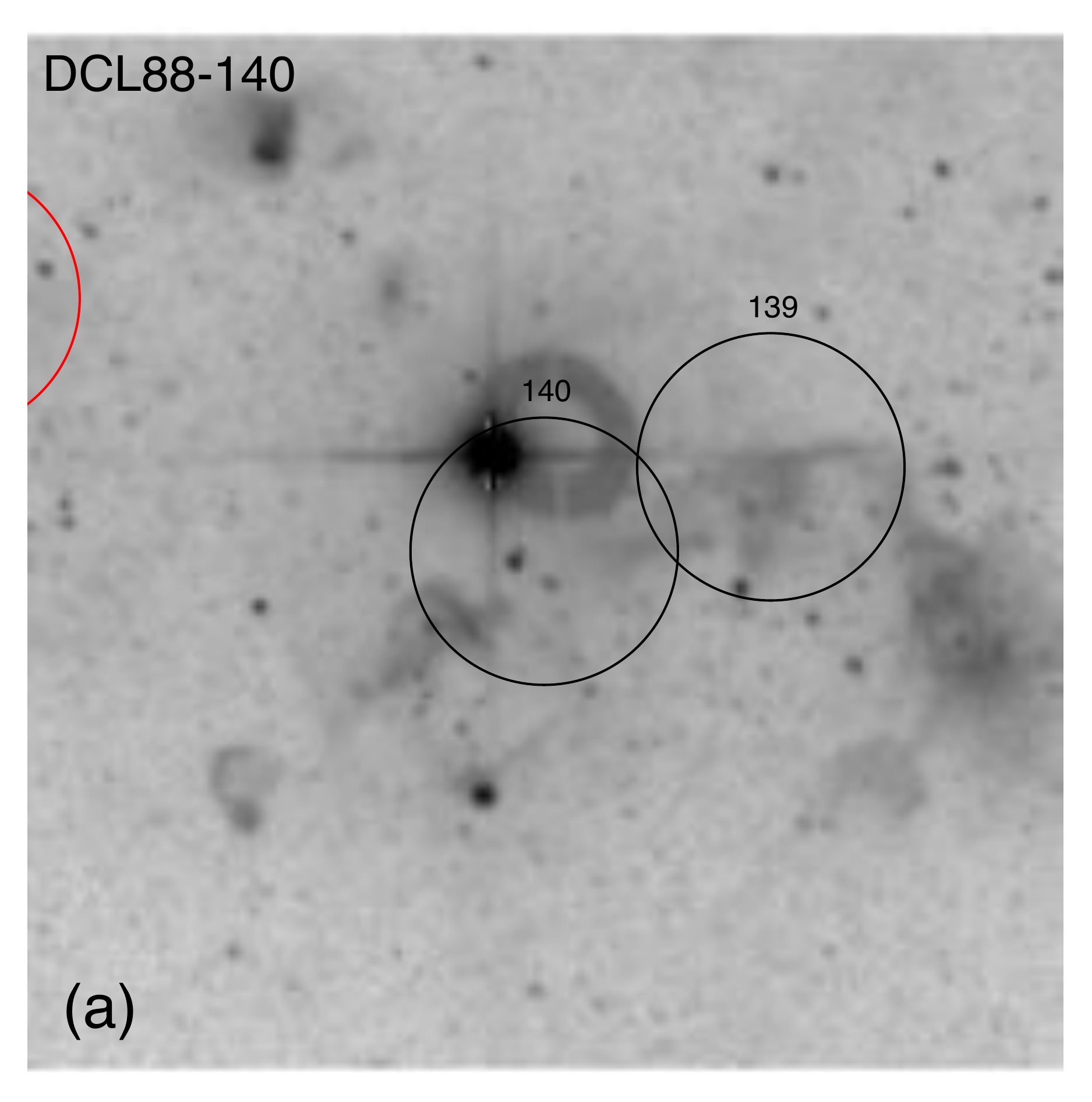}
\end{minipage}
\begin{minipage}{0.24\linewidth}
\includegraphics[width=\linewidth]{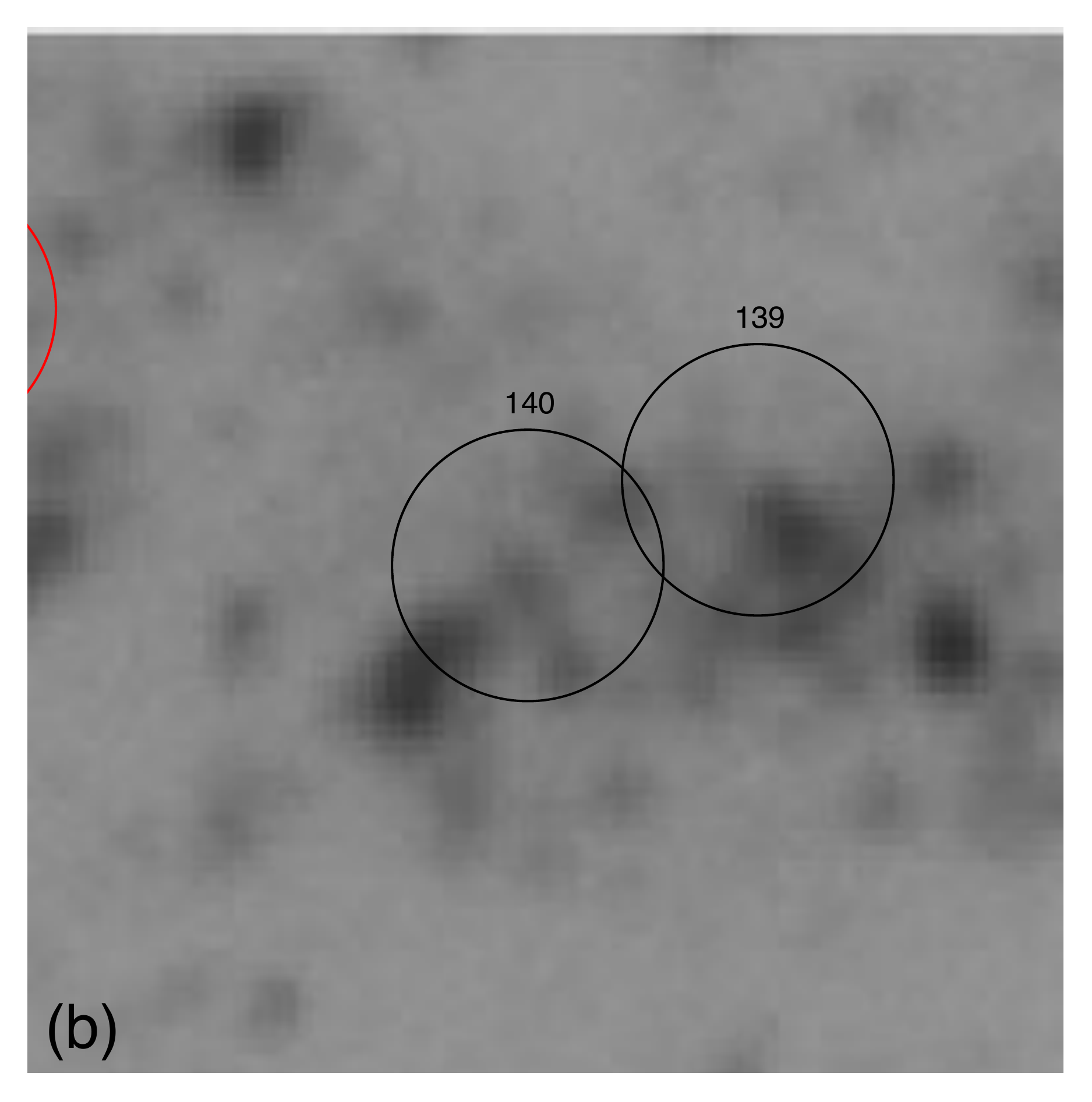}
\end{minipage}
\begin{minipage}{0.24\linewidth}
\includegraphics[width=\linewidth]{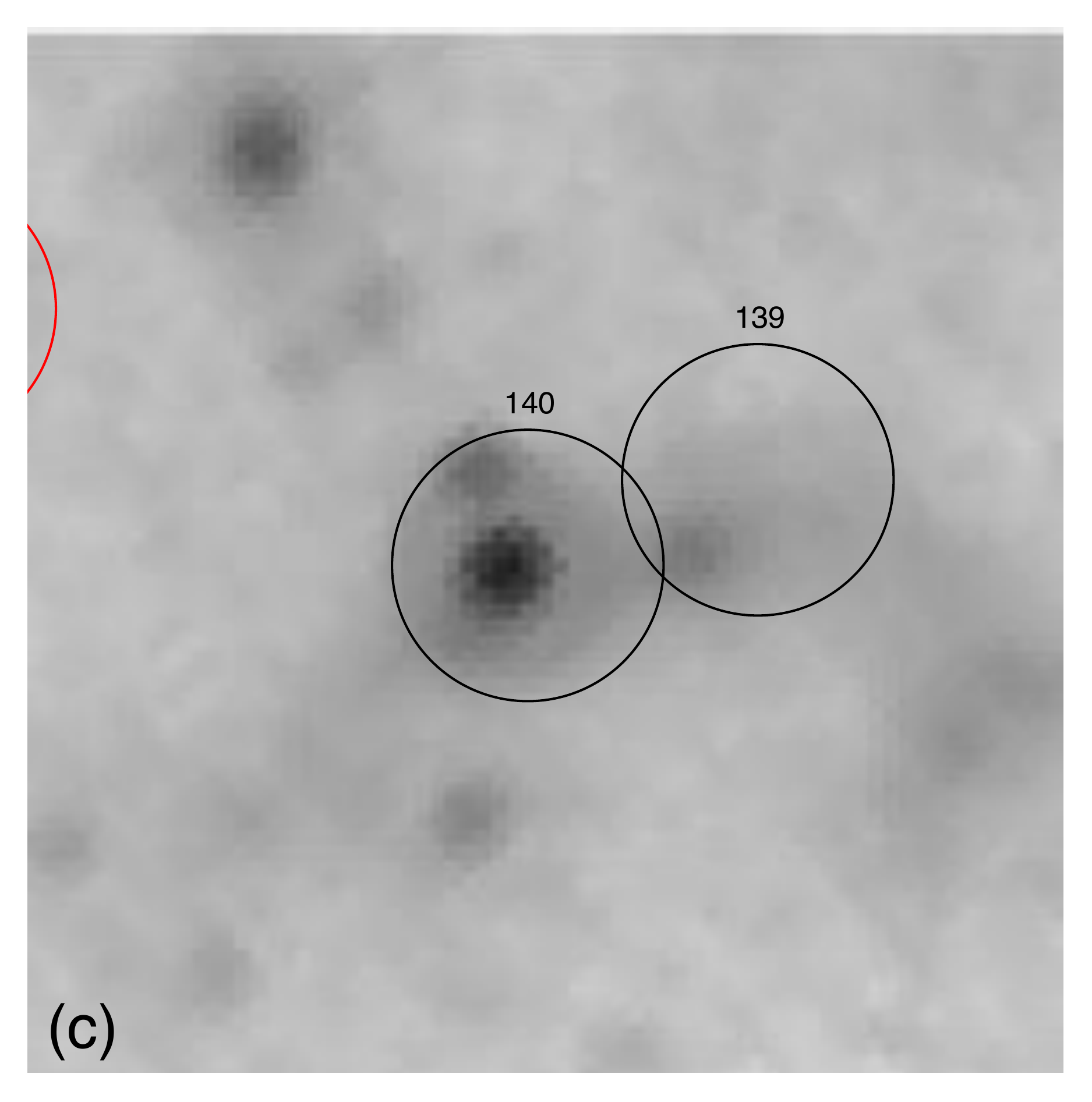}
\end{minipage}
\begin{minipage}{0.24\linewidth}
\includegraphics[width=\linewidth]{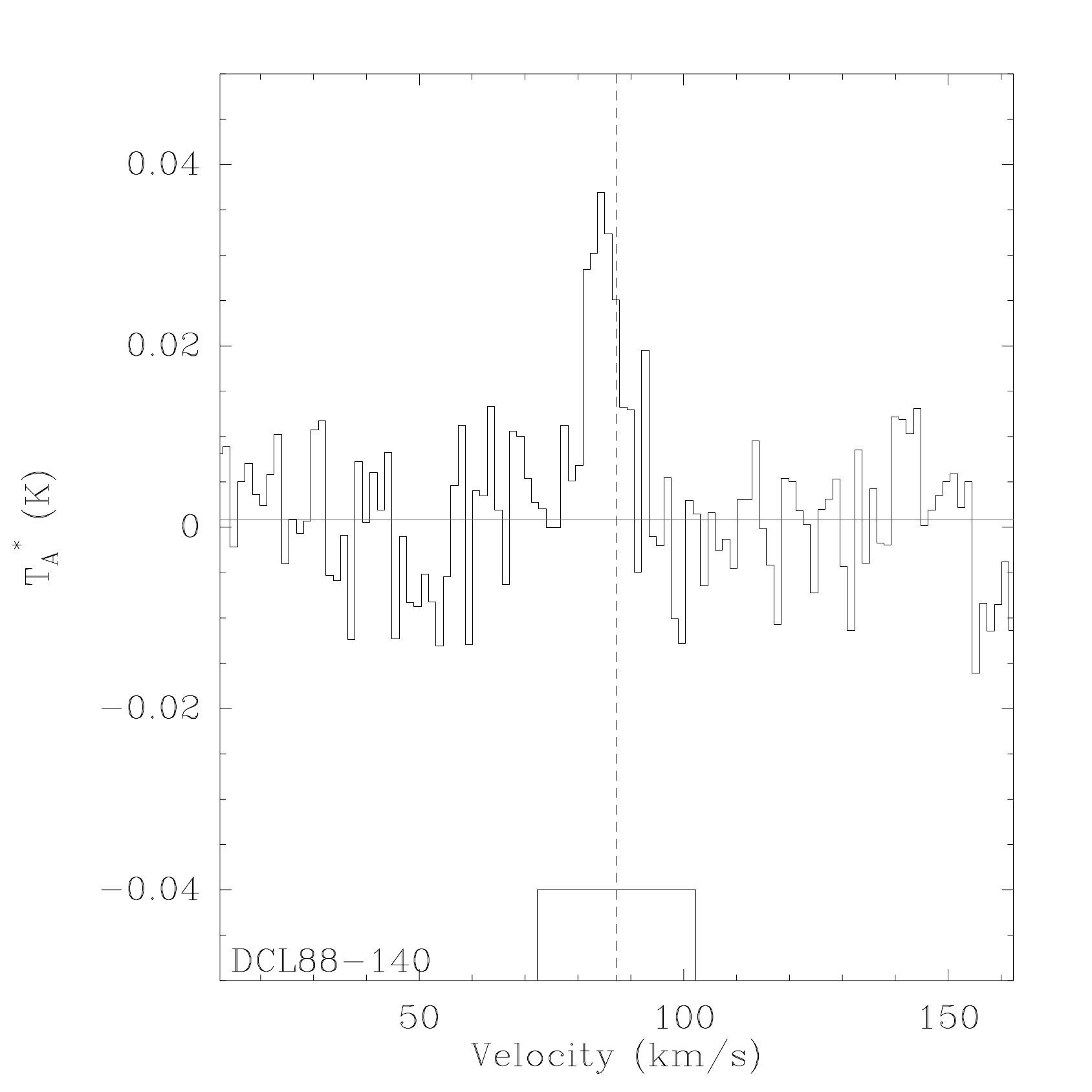}
\end{minipage}

\begin{minipage}{0.24\linewidth}
\includegraphics[width=\linewidth]{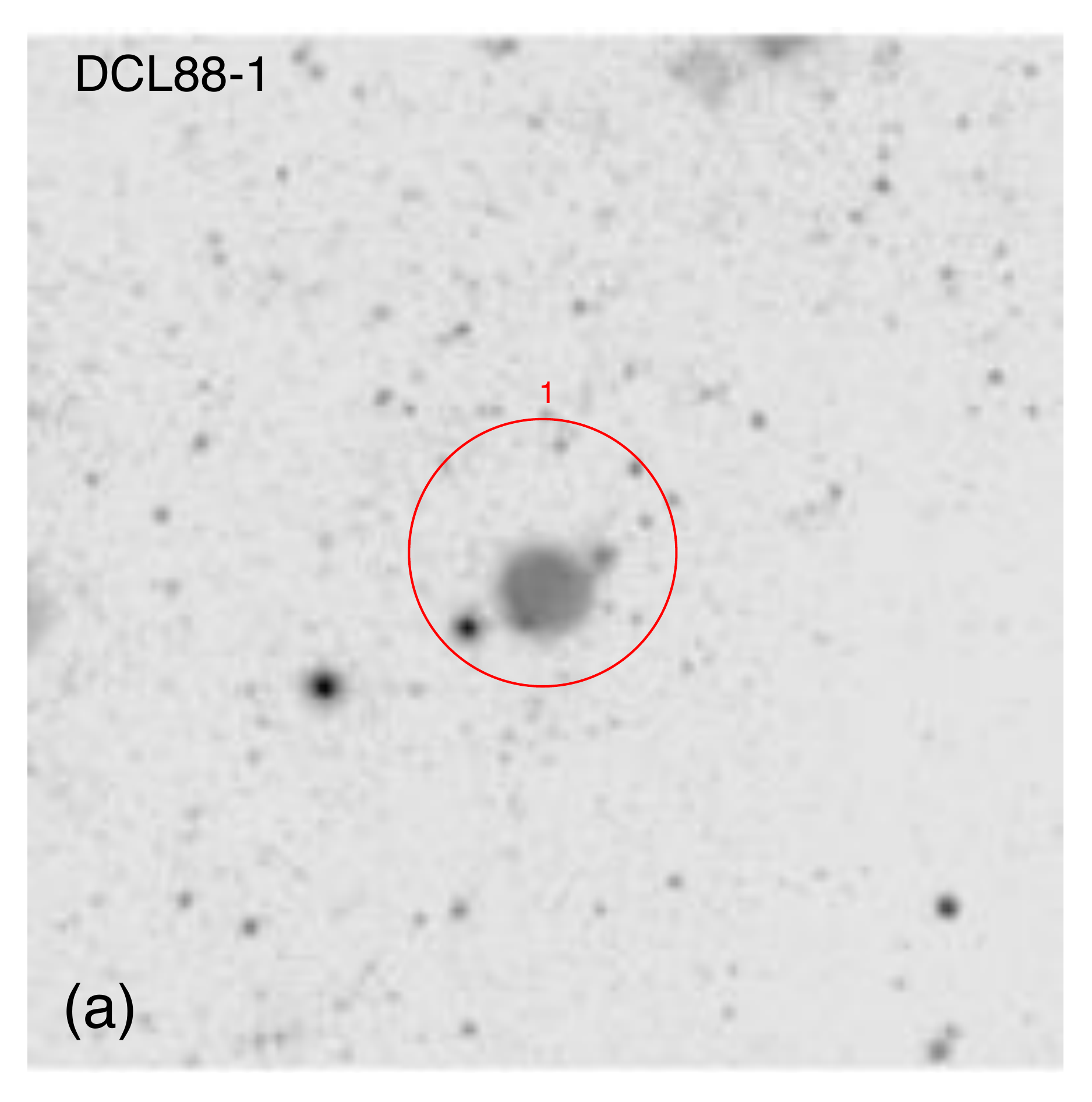}
\end{minipage}
\begin{minipage}{0.24\linewidth}
\includegraphics[width=\linewidth]{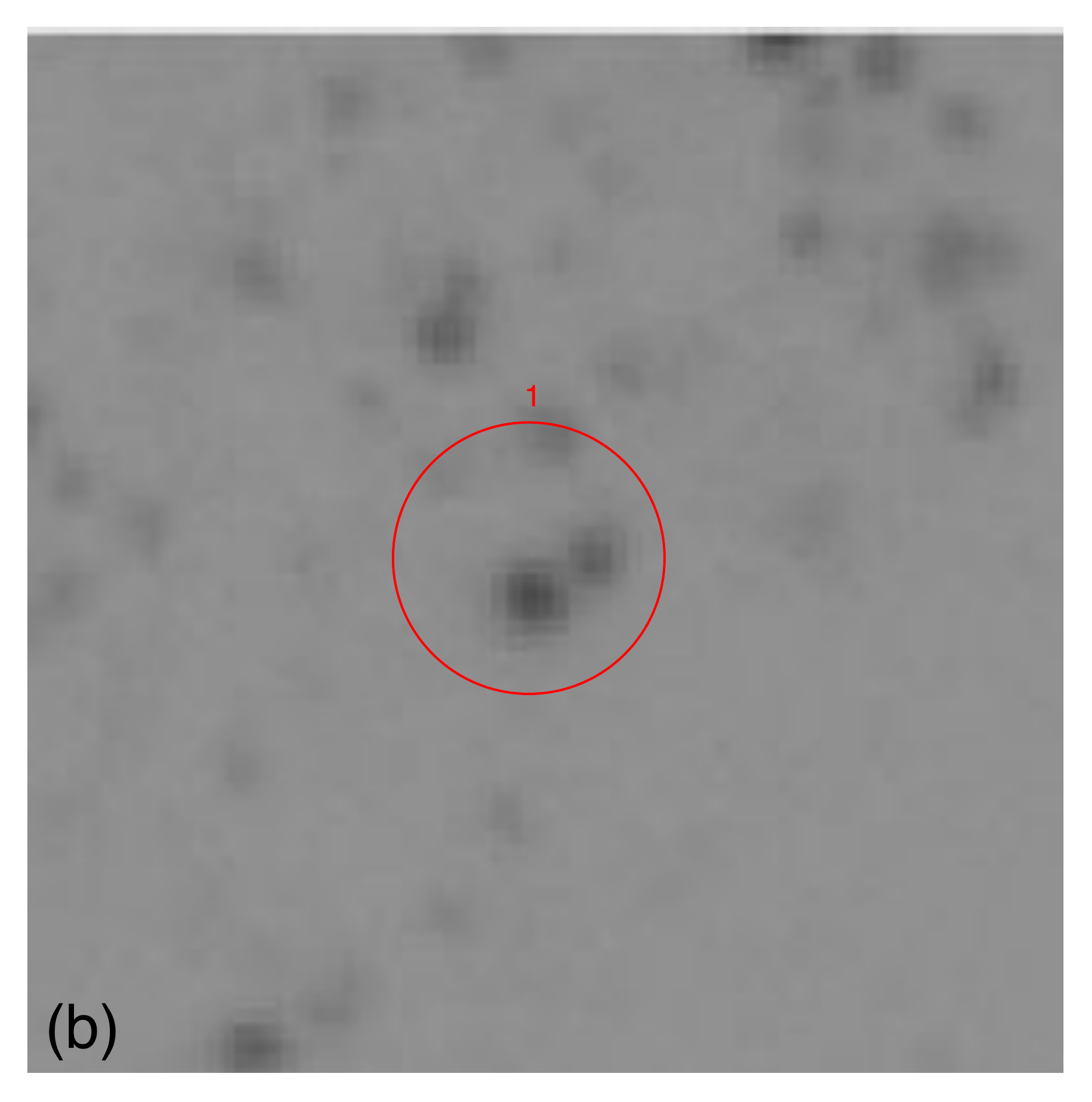}
\end{minipage}
\begin{minipage}{0.24\linewidth}
\includegraphics[width=\linewidth]{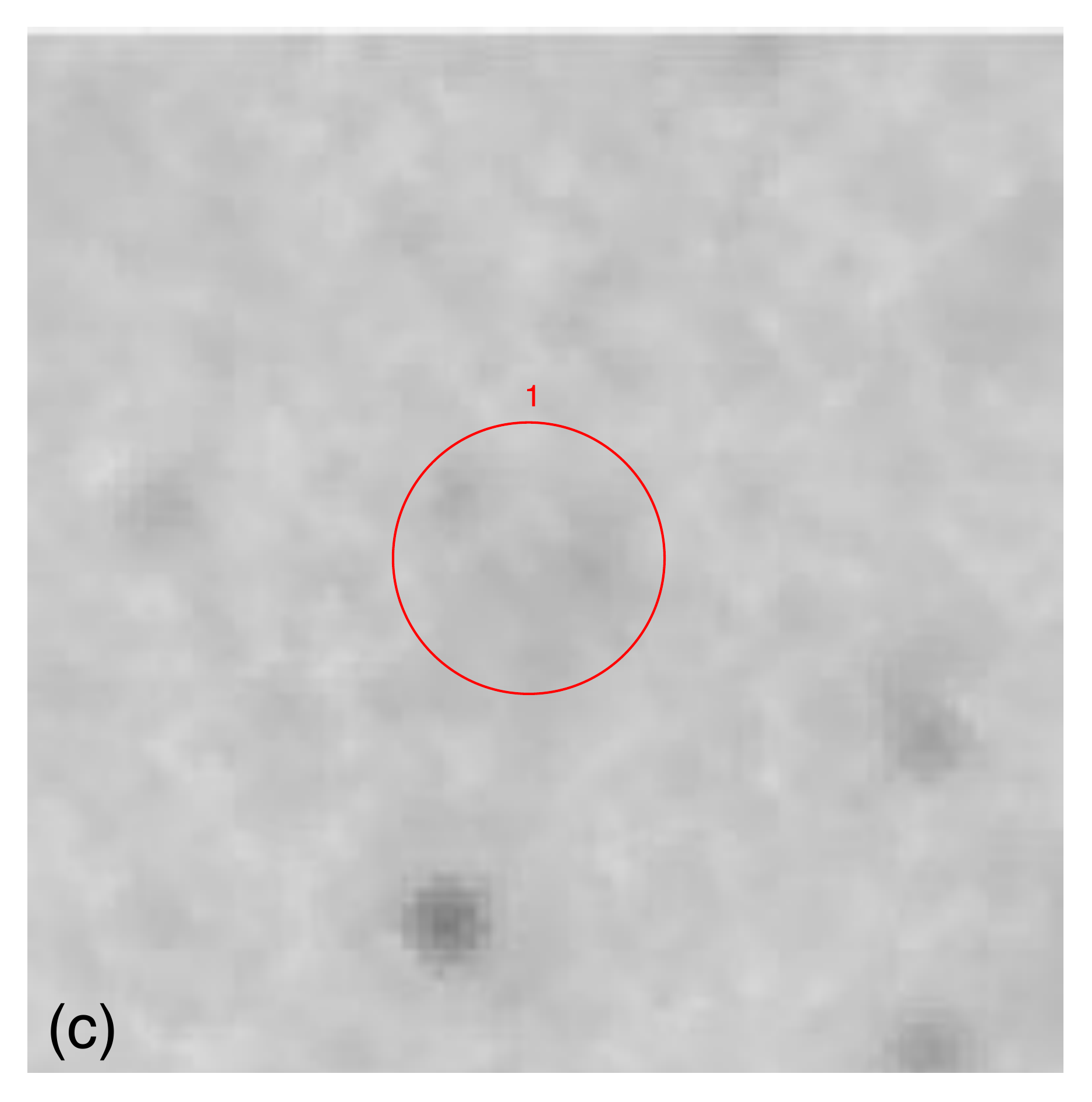}
\end{minipage}
\begin{minipage}{0.24\linewidth}
\includegraphics[width=\linewidth]{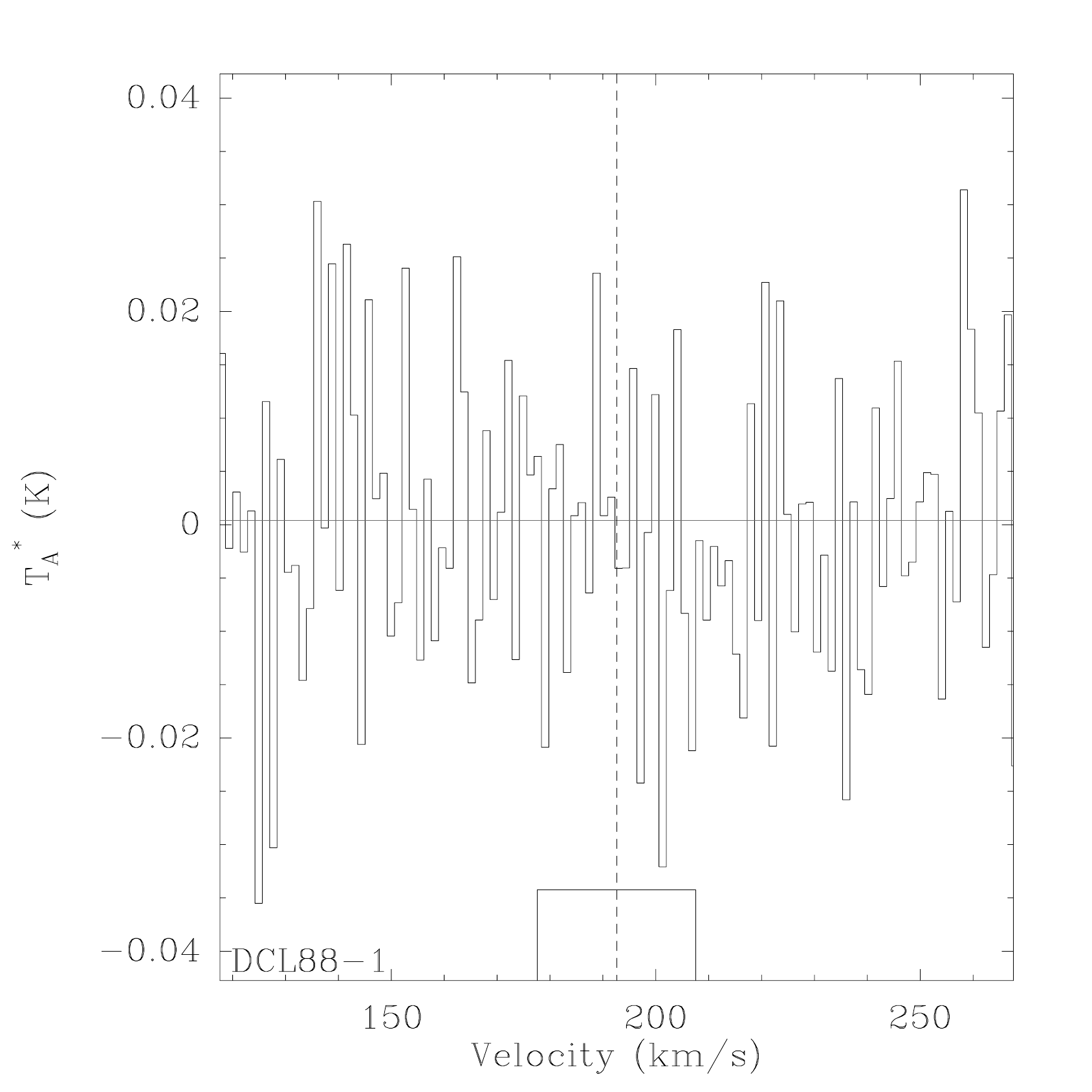}
\end{minipage}

\begin{minipage}{0.24\linewidth}
\includegraphics[width=\linewidth]{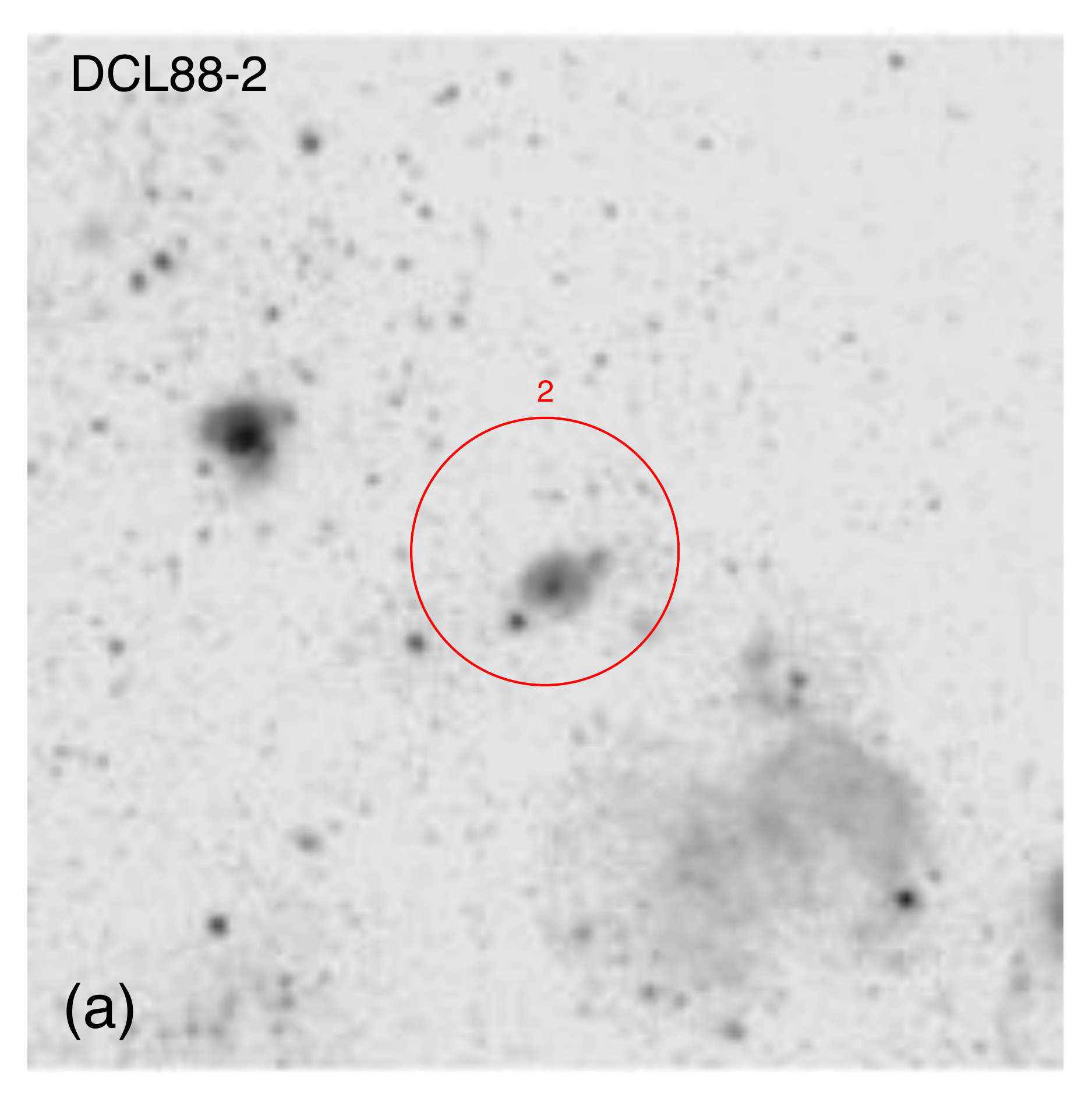}
\end{minipage}
\begin{minipage}{0.24\linewidth}
\includegraphics[width=\linewidth]{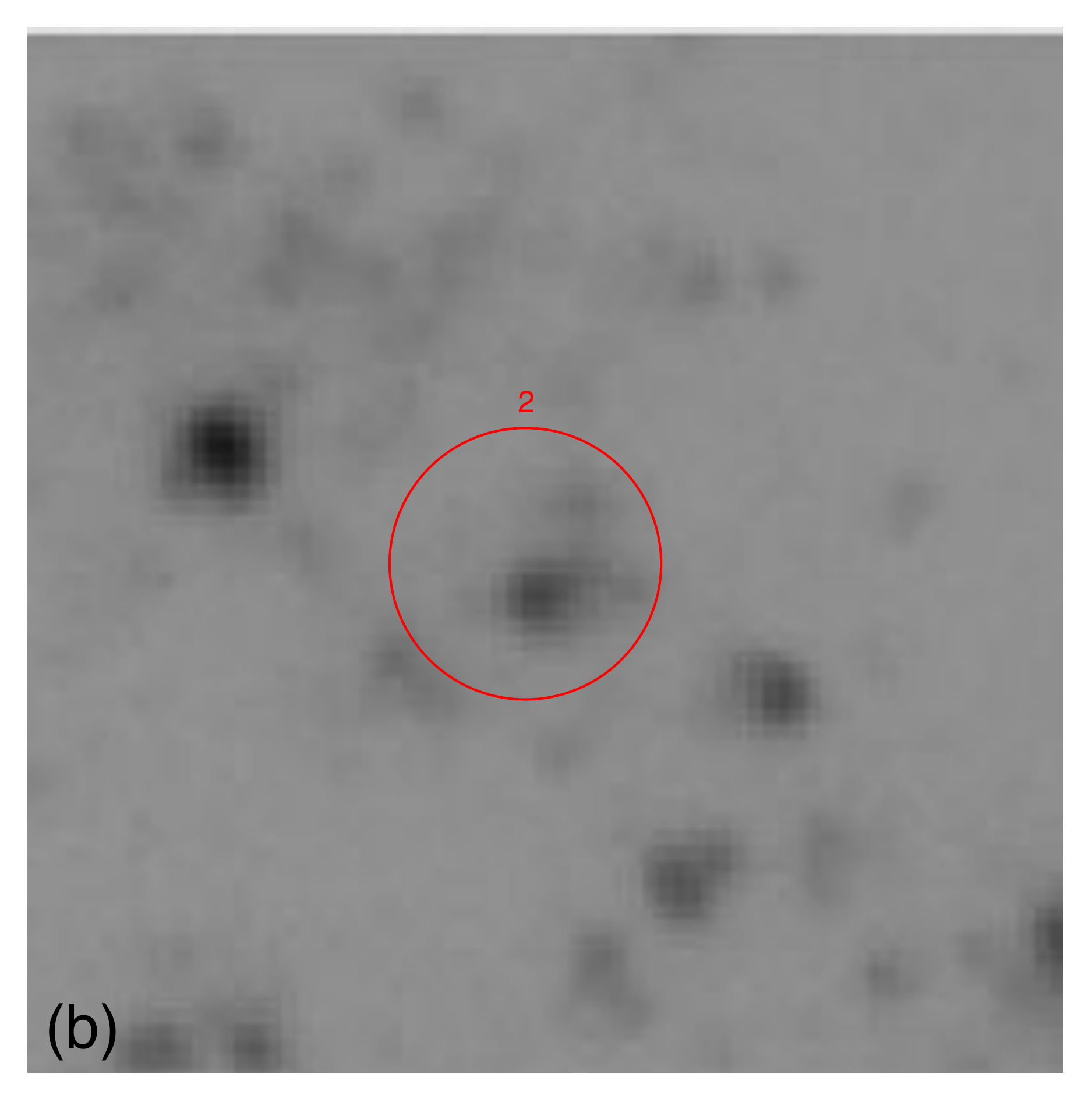}
\end{minipage}
\begin{minipage}{0.24\linewidth}
\includegraphics[width=\linewidth]{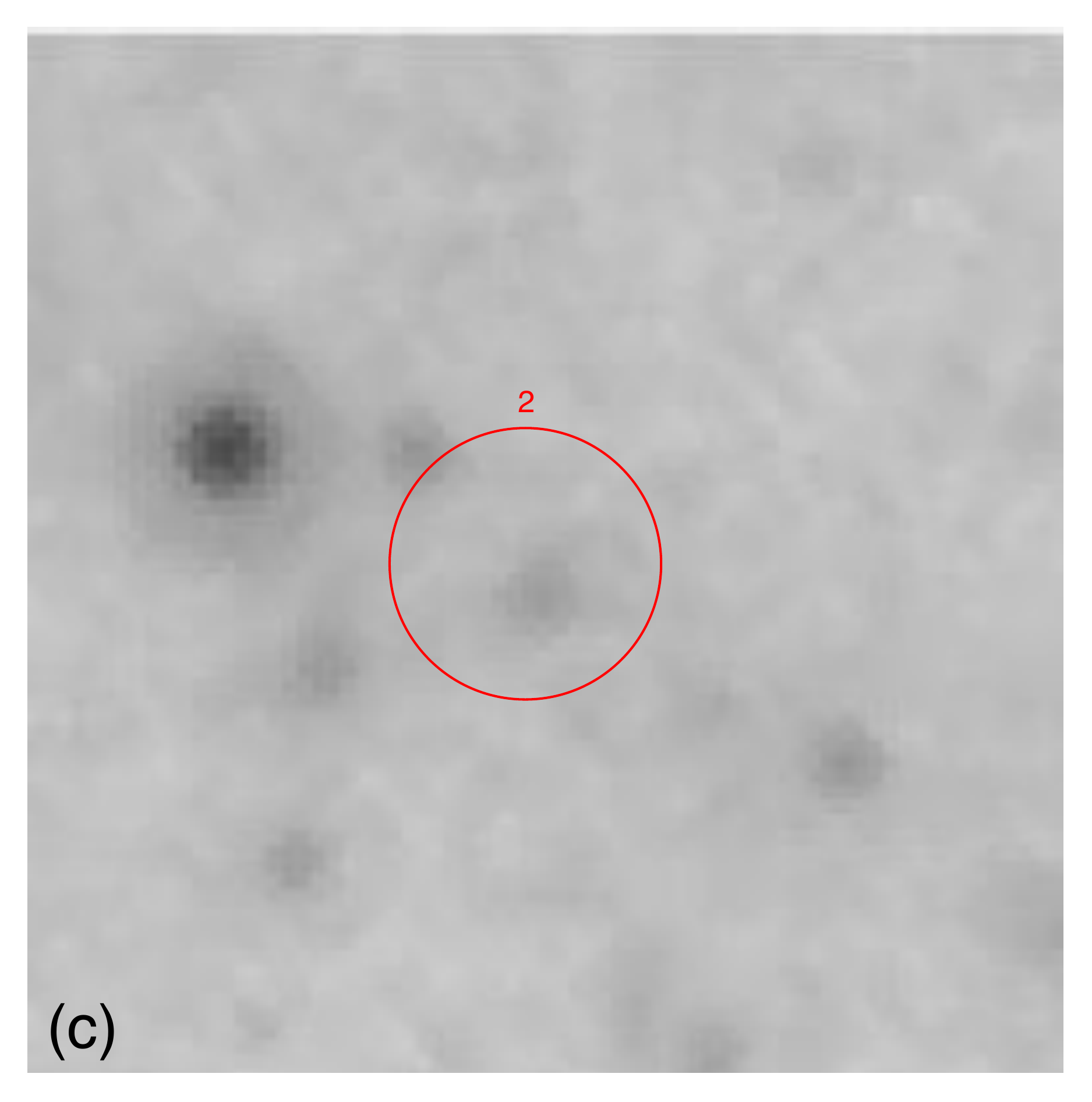}
\end{minipage}
\begin{minipage}{0.24\linewidth}
\includegraphics[width=\linewidth]{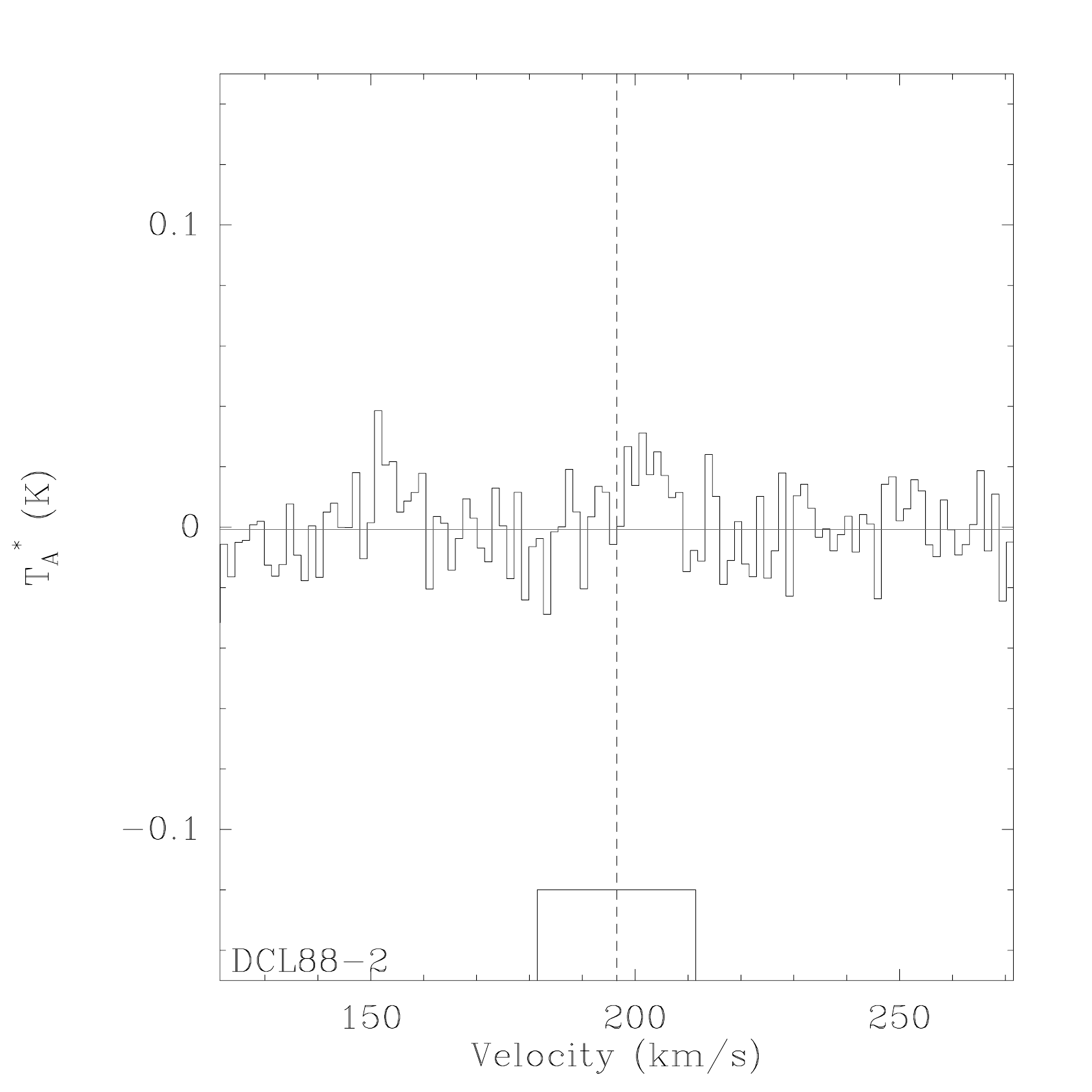}
\end{minipage}

\begin{minipage}{0.24\linewidth}
\includegraphics[width=\linewidth]{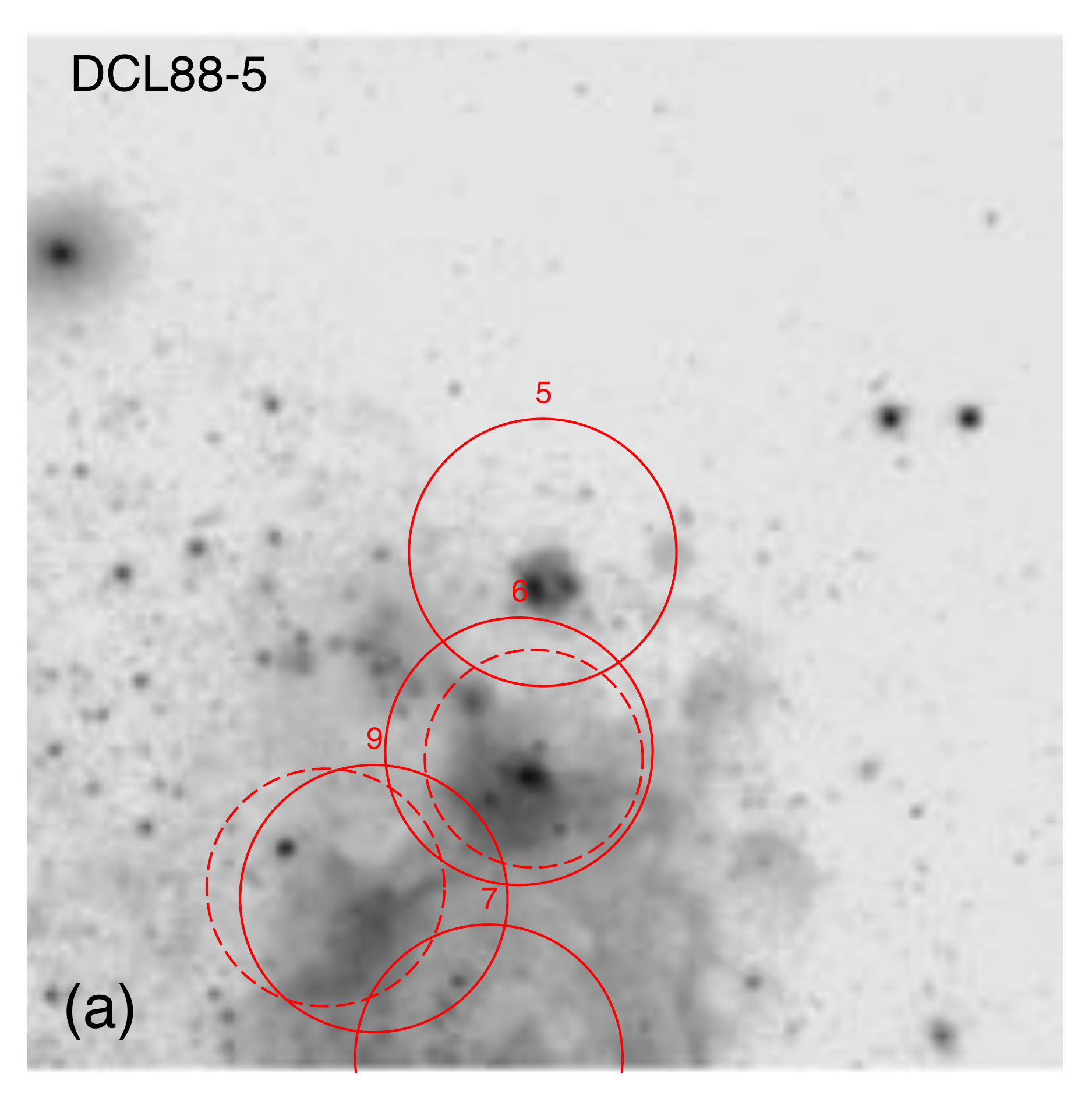}
\end{minipage}
\begin{minipage}{0.24\linewidth}
\includegraphics[width=\linewidth]{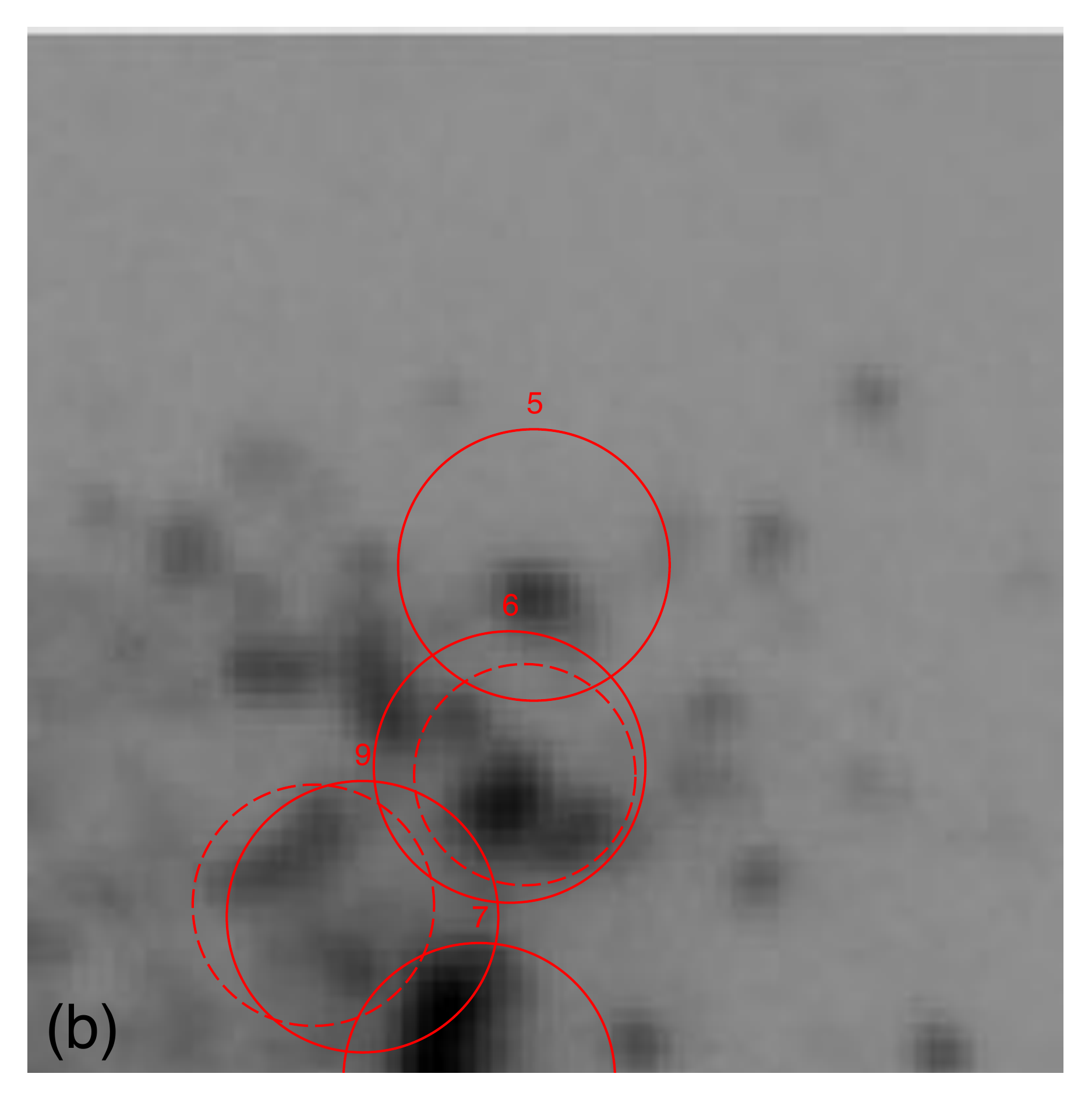}
\end{minipage}
\begin{minipage}{0.24\linewidth}
\includegraphics[width=\linewidth]{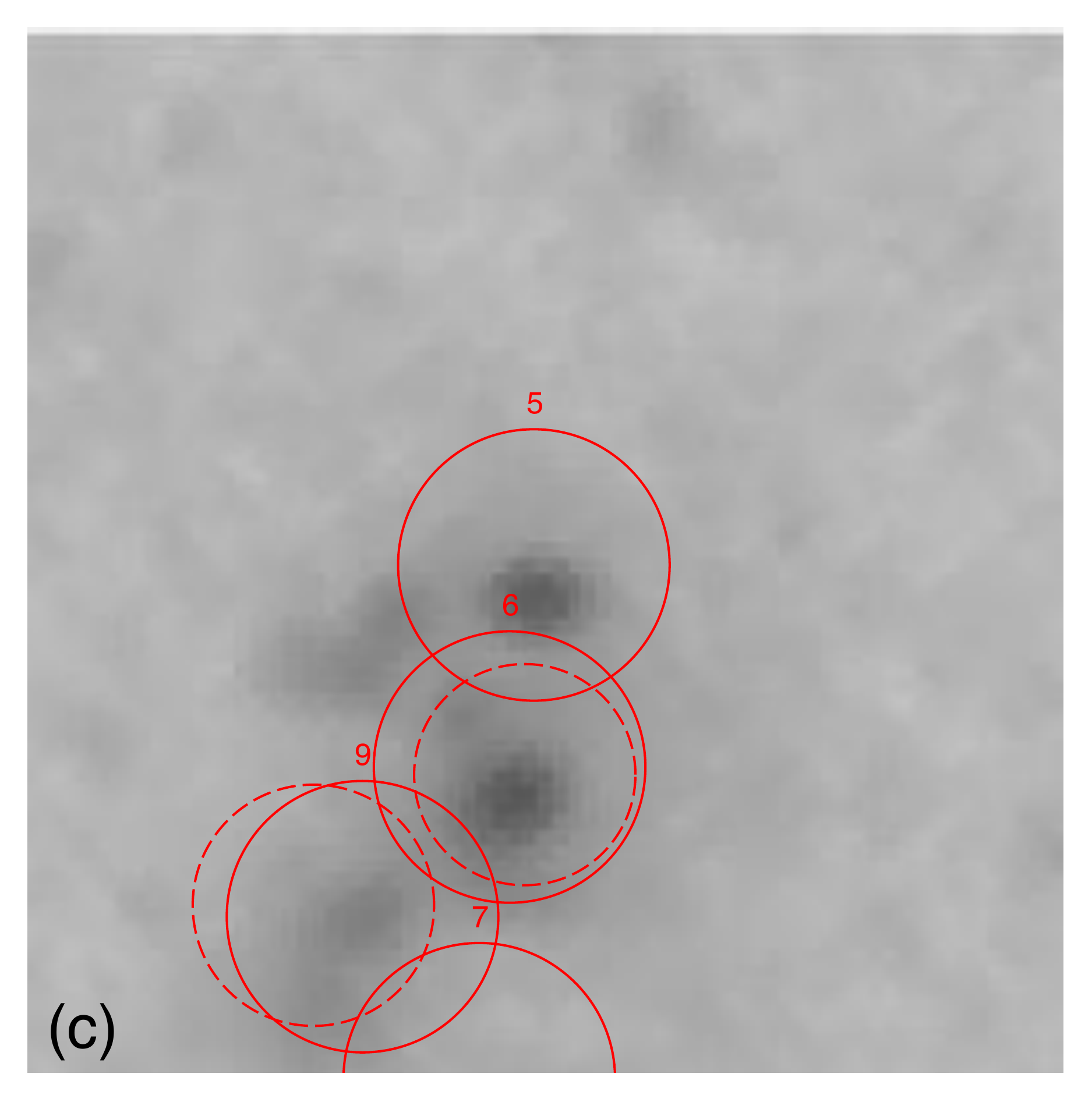}
\end{minipage}
\begin{minipage}{0.24\linewidth}
\includegraphics[width=\linewidth]{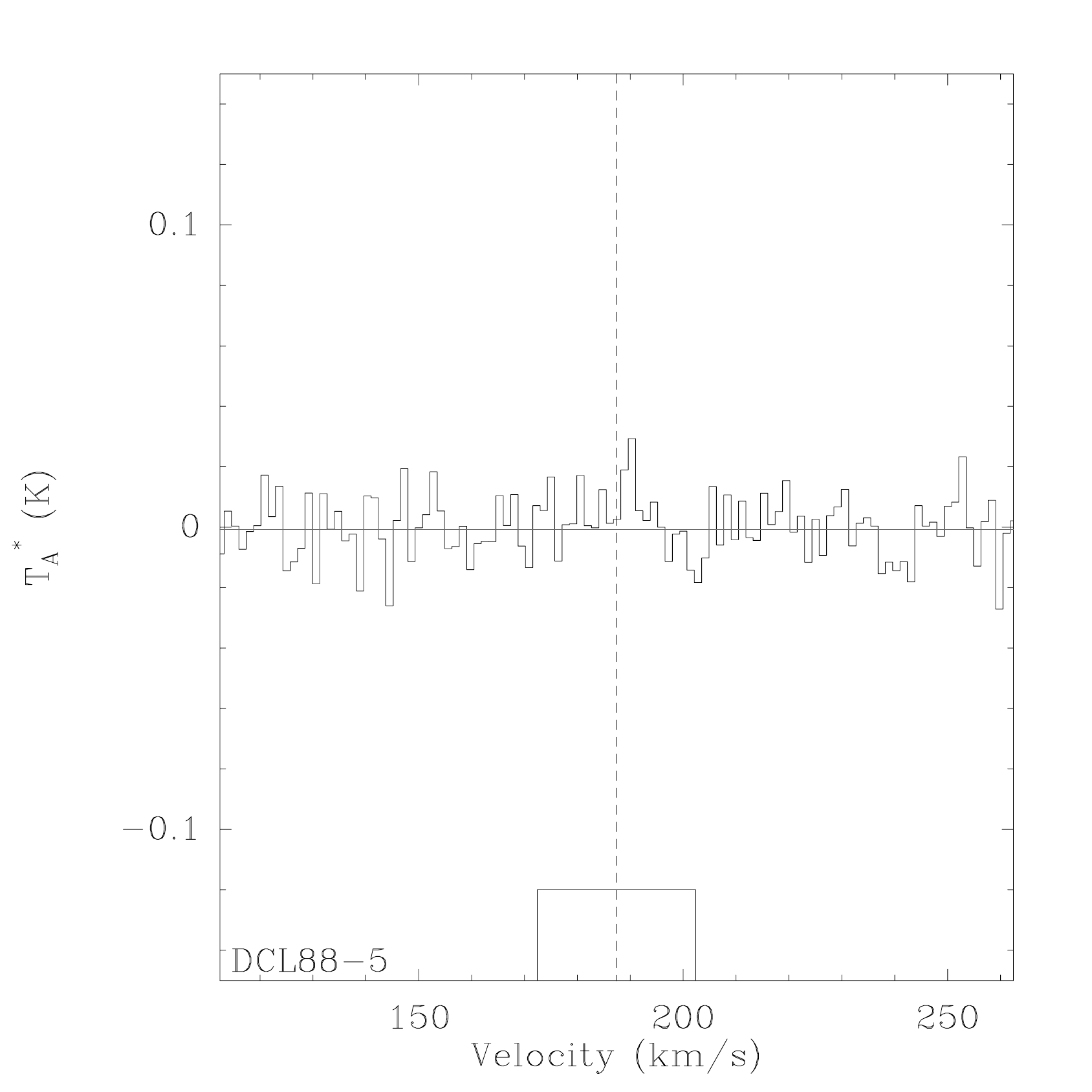}
\end{minipage}

\begin{minipage}{0.24\linewidth}
\includegraphics[width=\linewidth]{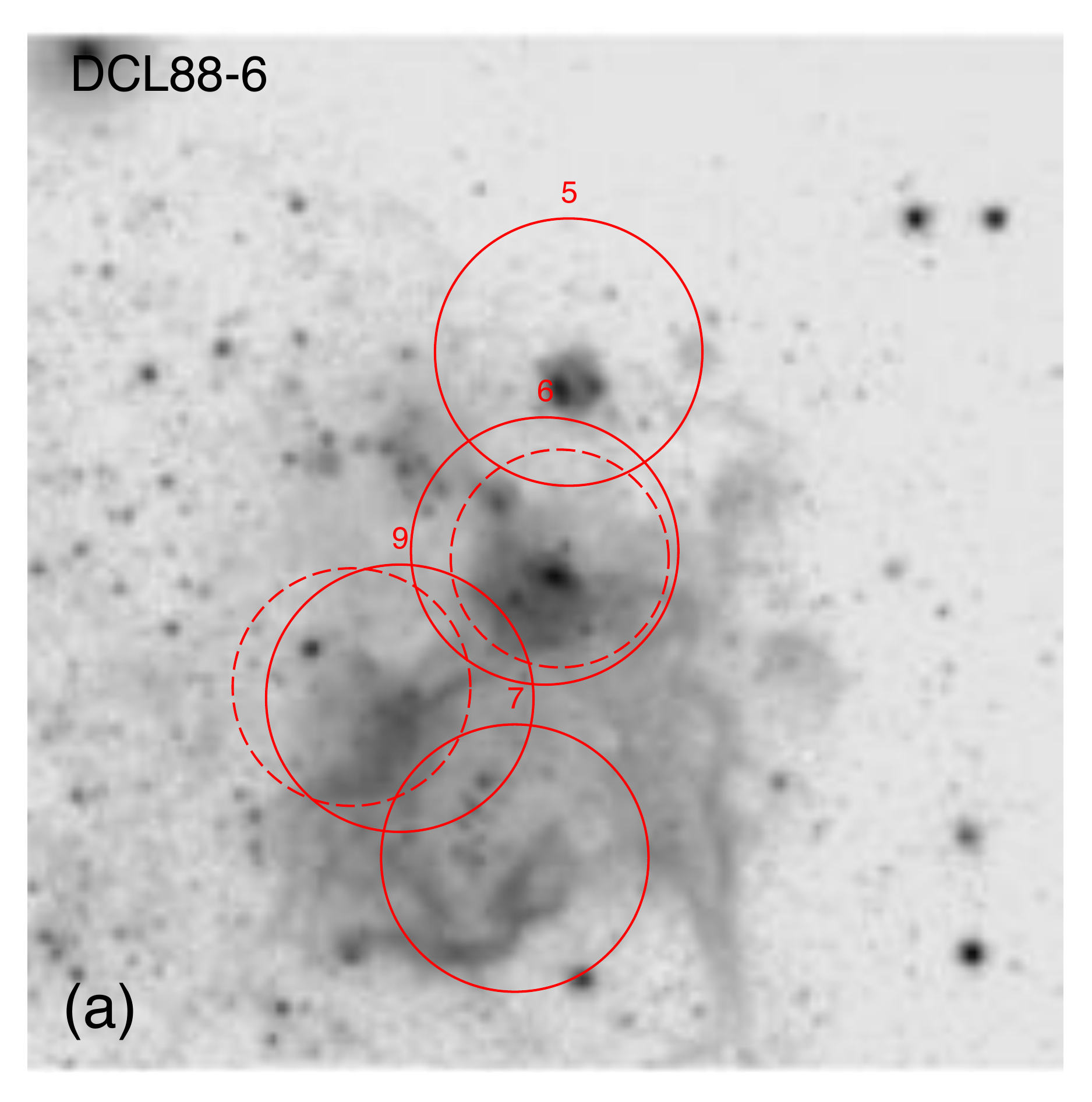}
\end{minipage}
\begin{minipage}{0.24\linewidth}
\includegraphics[width=\linewidth]{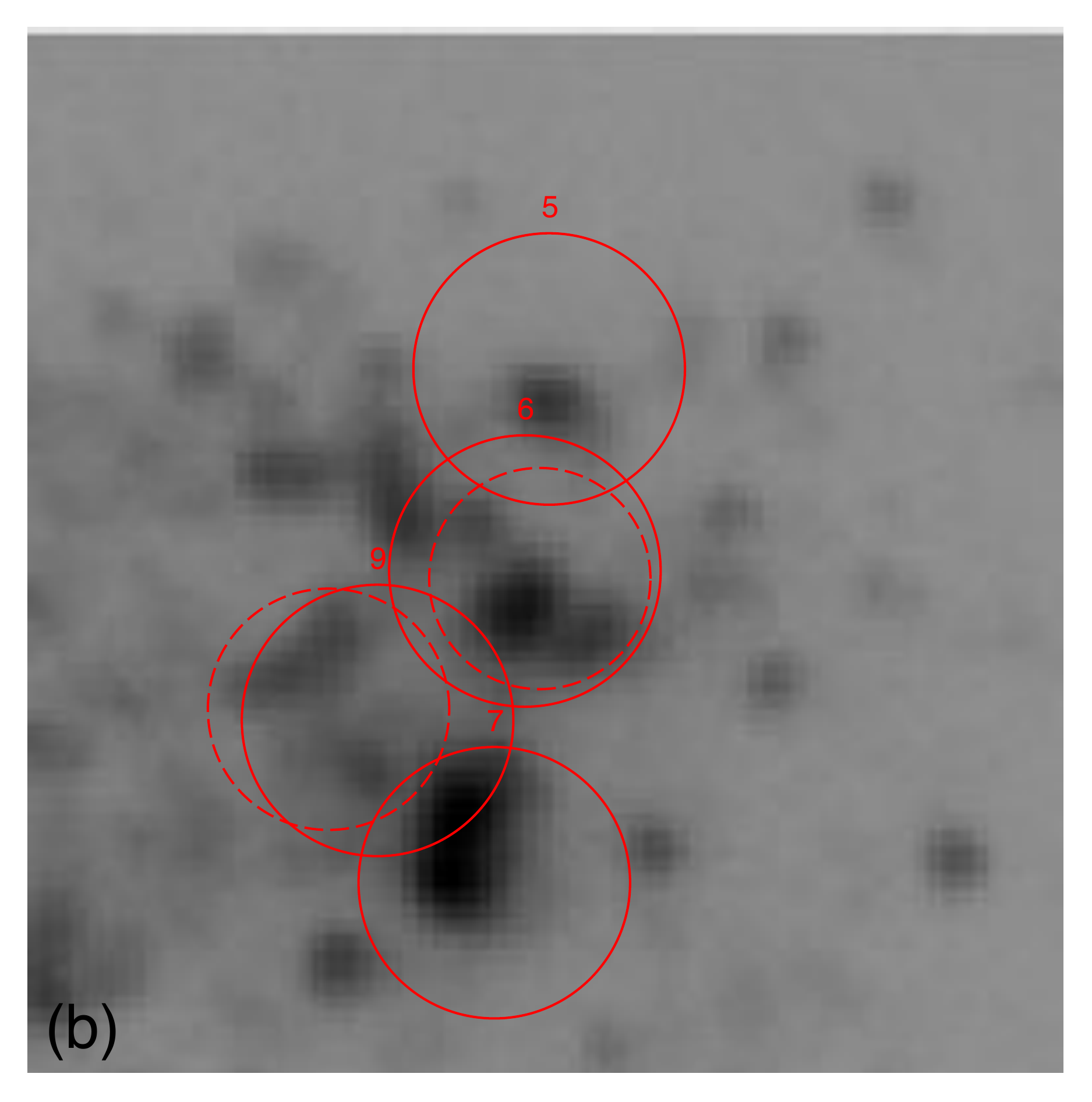}
\end{minipage}
\begin{minipage}{0.24\linewidth}
\includegraphics[width=\linewidth]{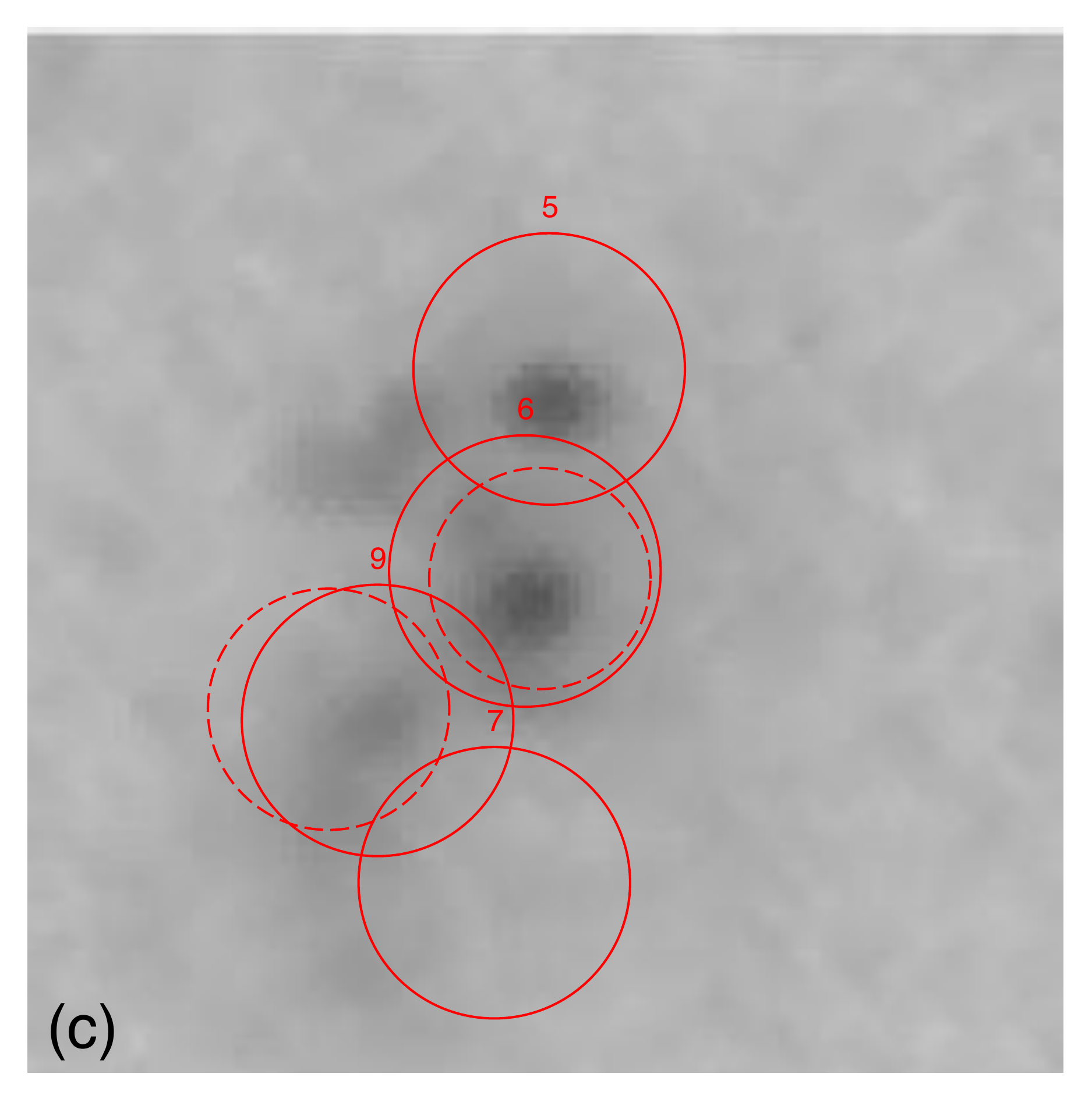}
\end{minipage}
\begin{minipage}{0.24\linewidth}
\includegraphics[width=\linewidth]{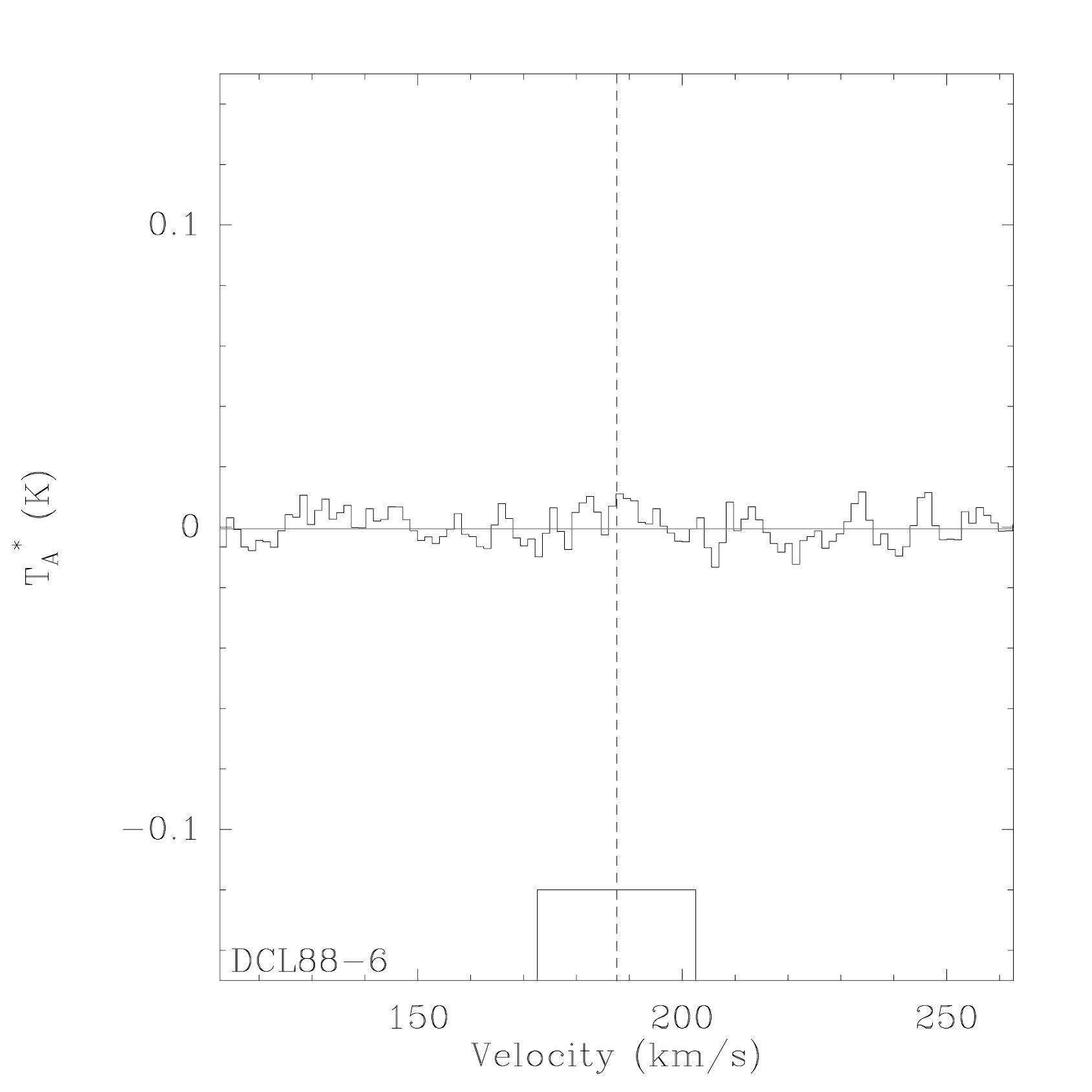}
\end{minipage}

\noindent\textbf{Figure~\ref{fig:stamps} -- continued.}

\end{figure*}

\begin{figure*}
%\ContinuedFloat

\begin{minipage}{0.24\linewidth}
\includegraphics[width=\linewidth]{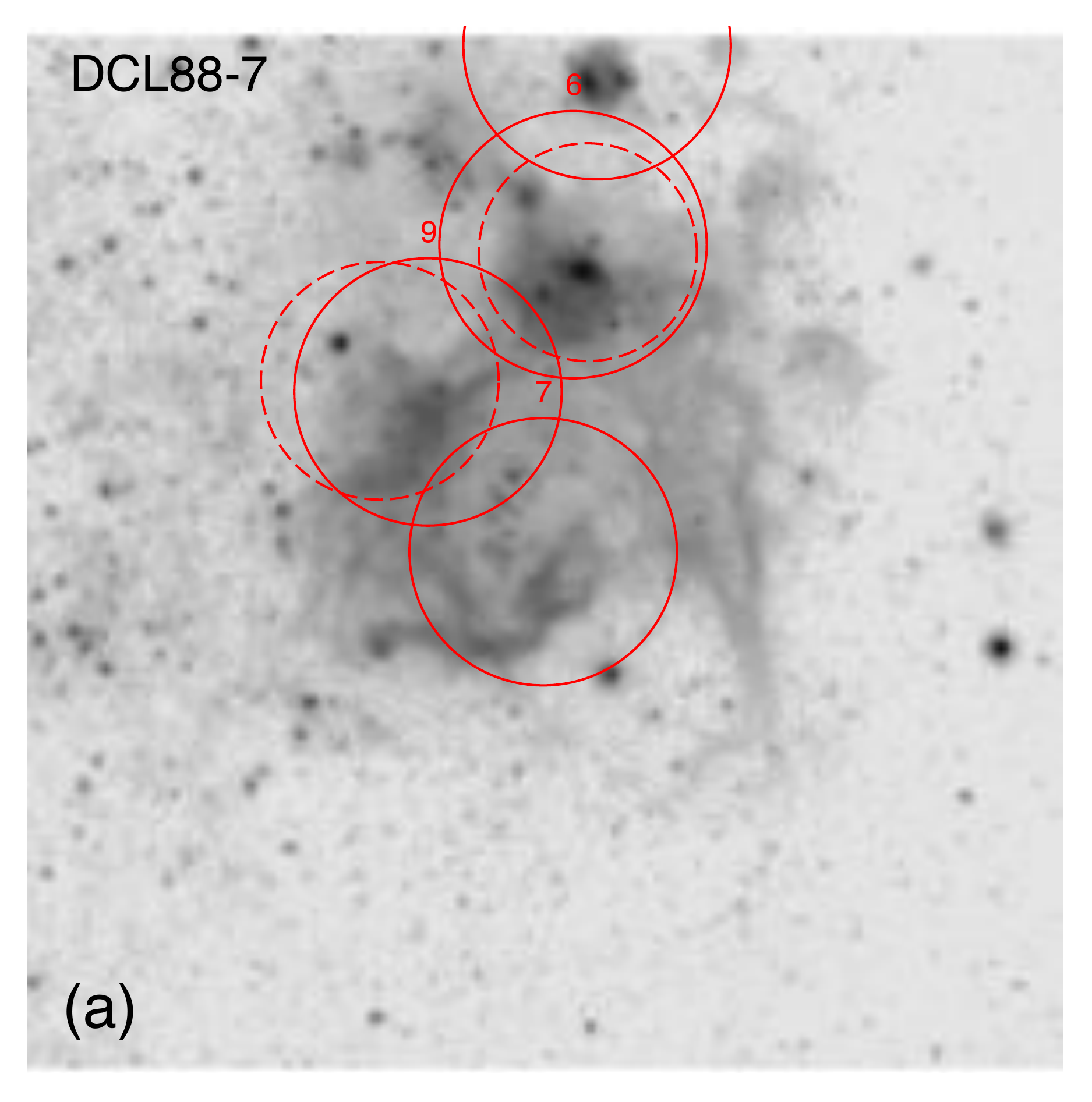}
\end{minipage}
\begin{minipage}{0.24\linewidth}
\includegraphics[width=\linewidth]{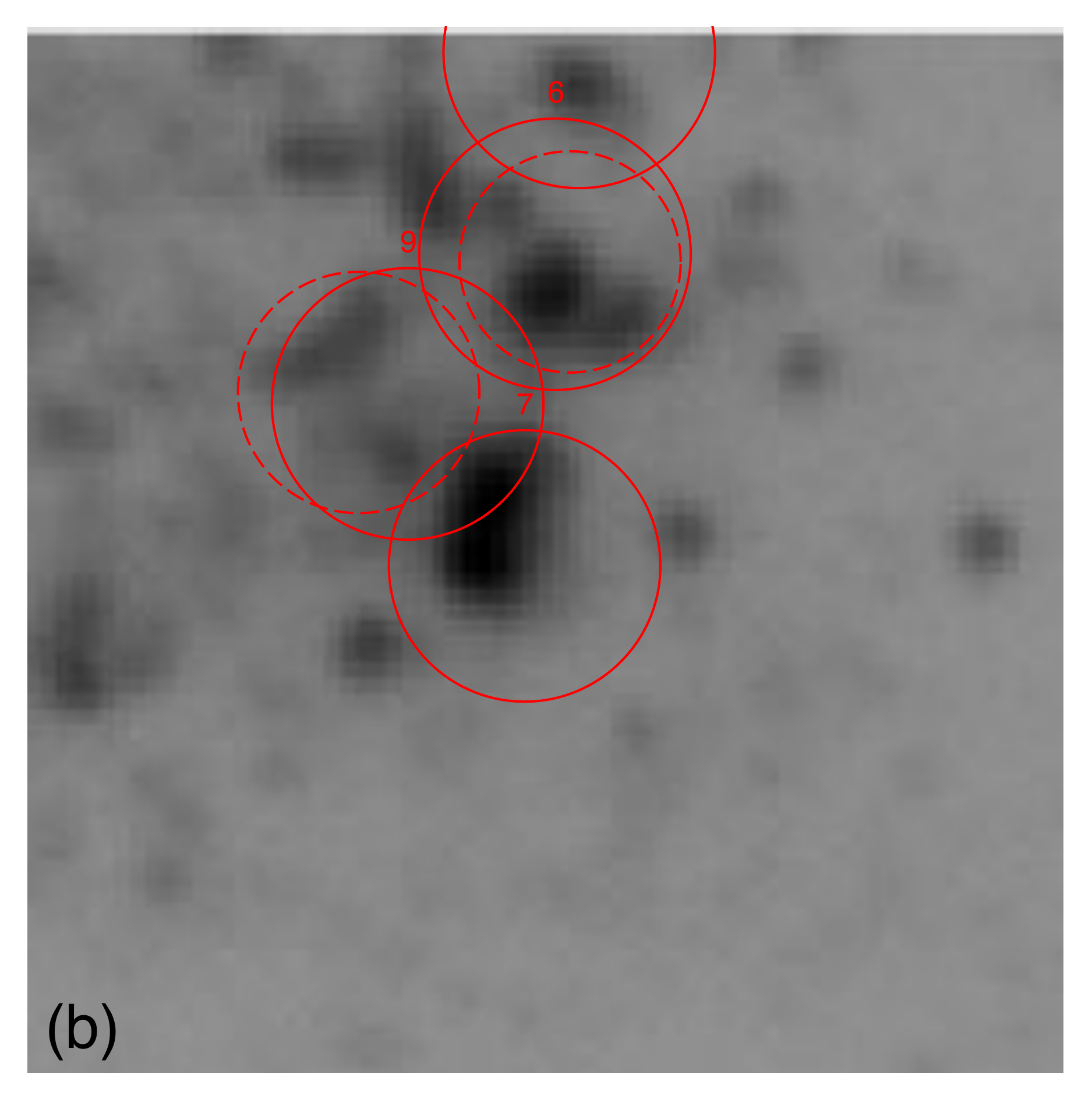}
\end{minipage}
\begin{minipage}{0.24\linewidth}
\includegraphics[width=\linewidth]{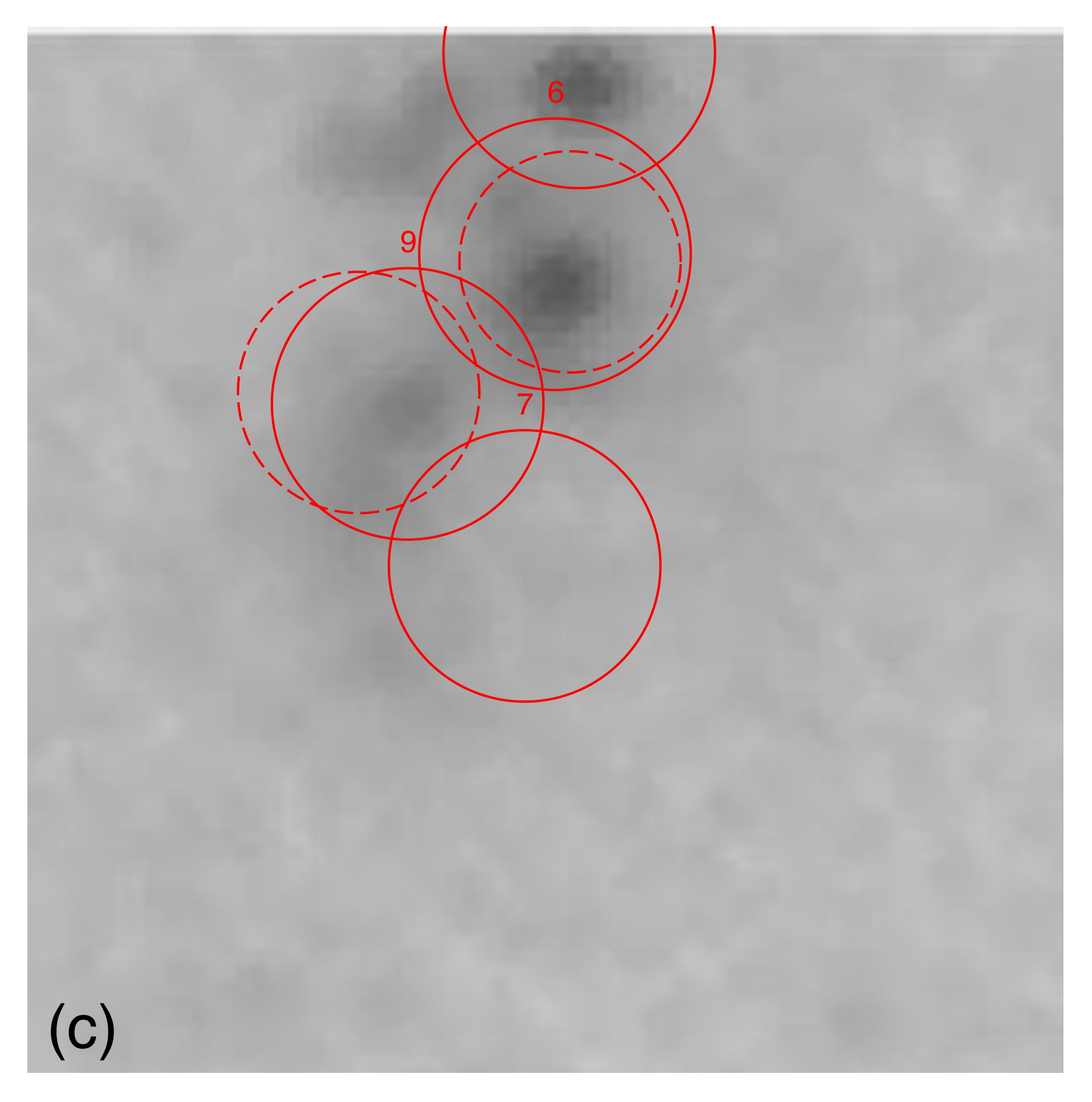}
\end{minipage}
\begin{minipage}{0.24\linewidth}
\includegraphics[width=\linewidth]{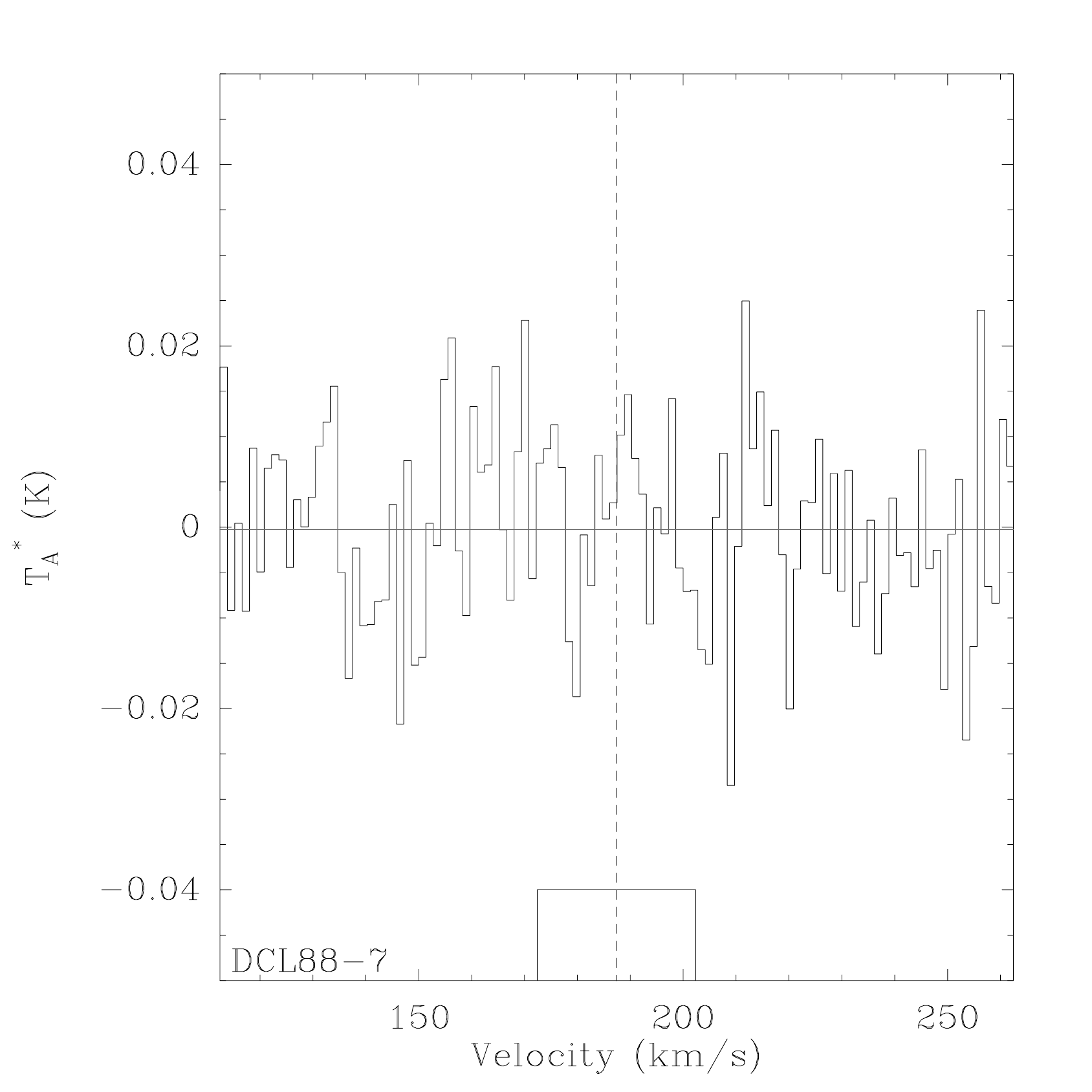}
\end{minipage}

\begin{minipage}{0.24\linewidth}
\includegraphics[width=\linewidth]{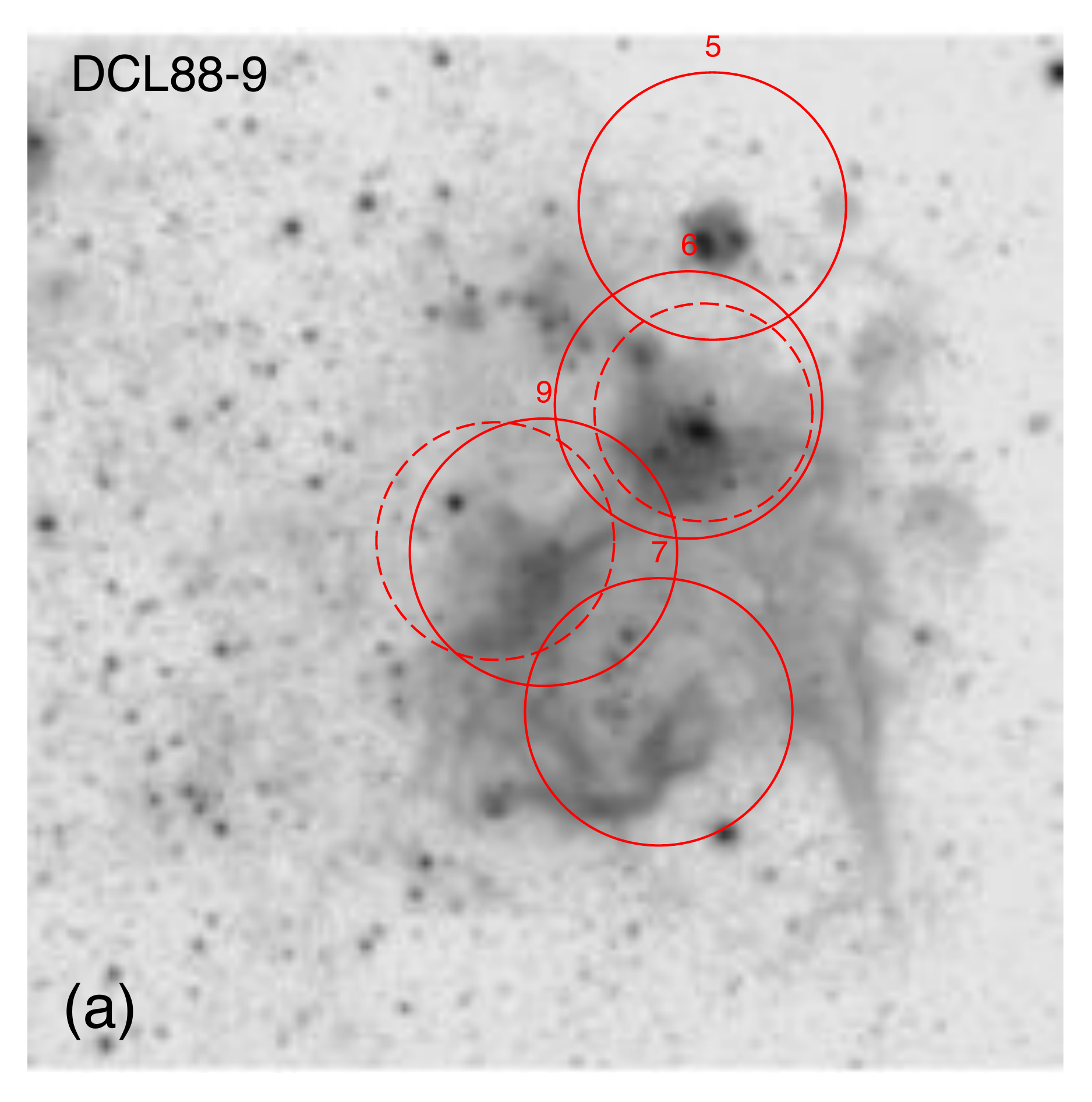}
\end{minipage}
\begin{minipage}{0.24\linewidth}
\includegraphics[width=\linewidth]{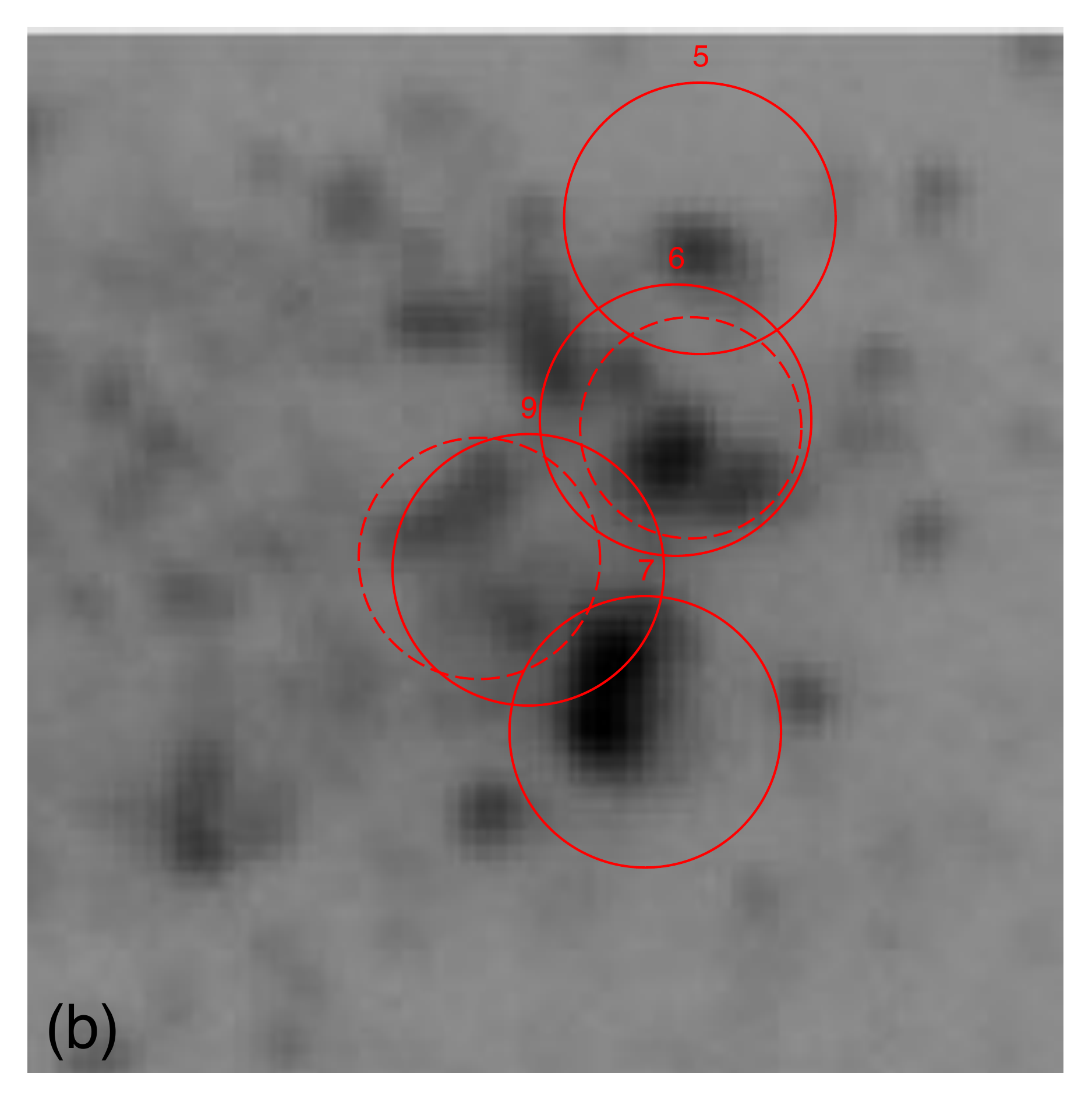}
\end{minipage}
\begin{minipage}{0.24\linewidth}
\includegraphics[width=\linewidth]{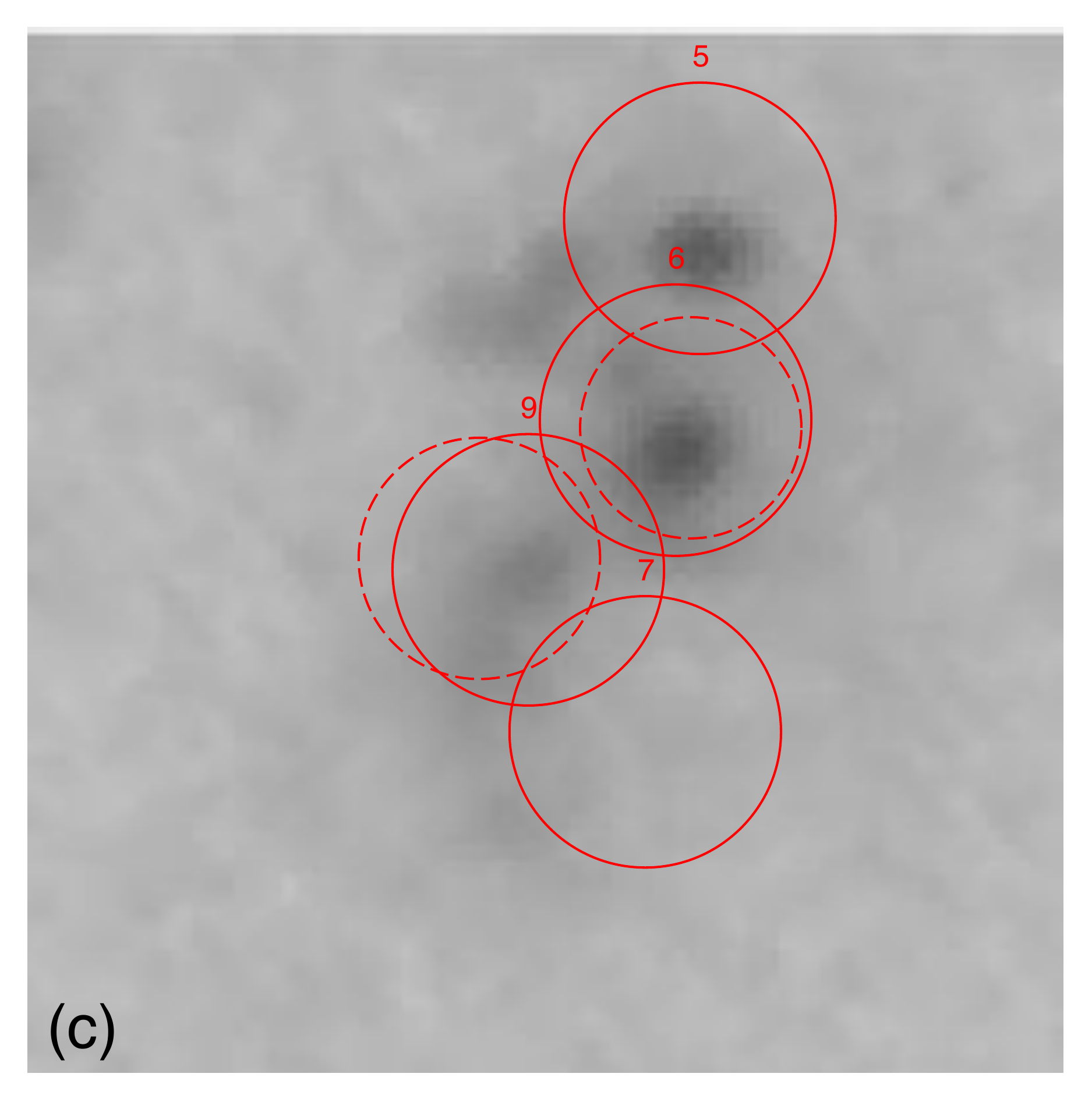}
\end{minipage}
\begin{minipage}{0.24\linewidth}
\includegraphics[width=\linewidth]{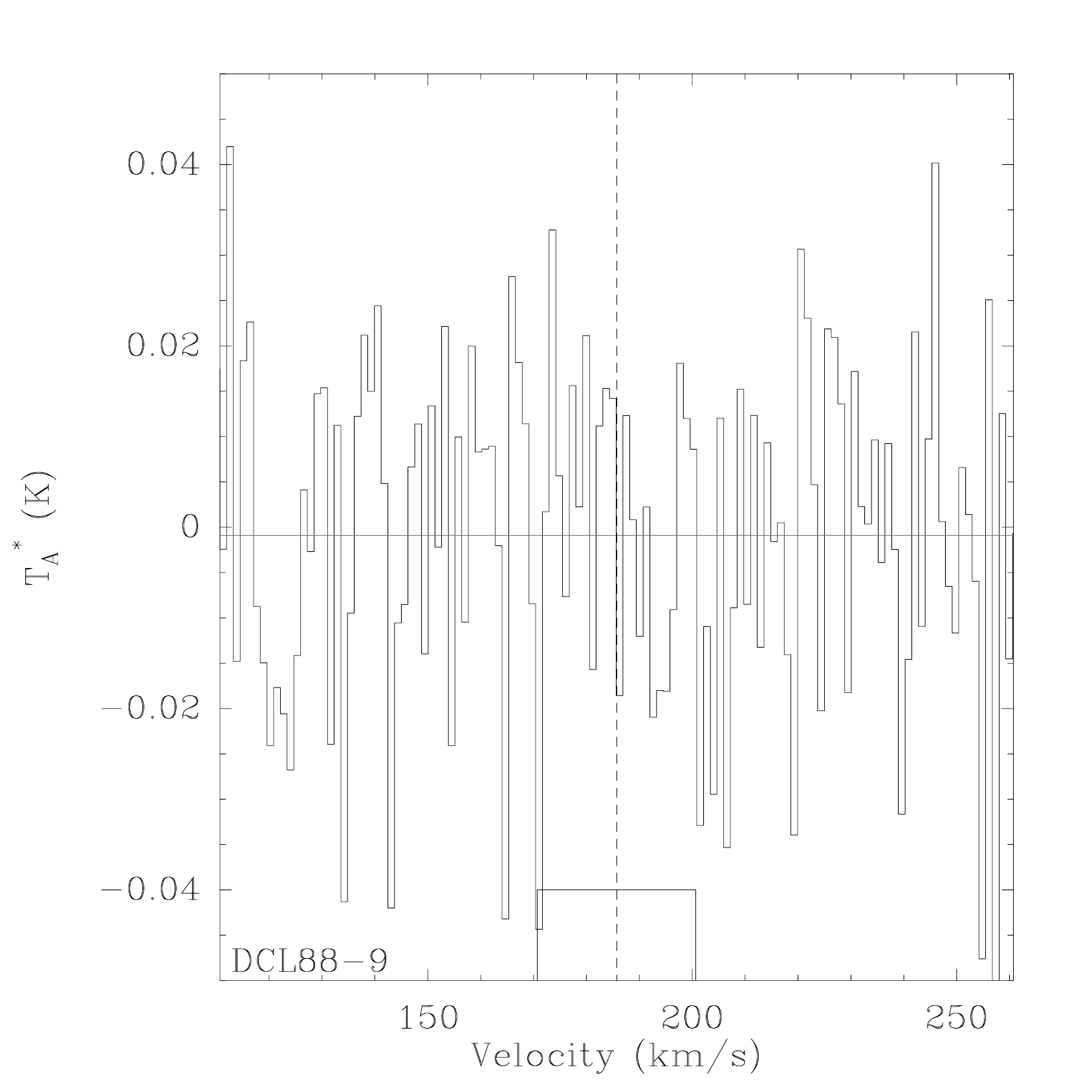}
\end{minipage}

\begin{minipage}{0.24\linewidth}
\includegraphics[width=\linewidth]{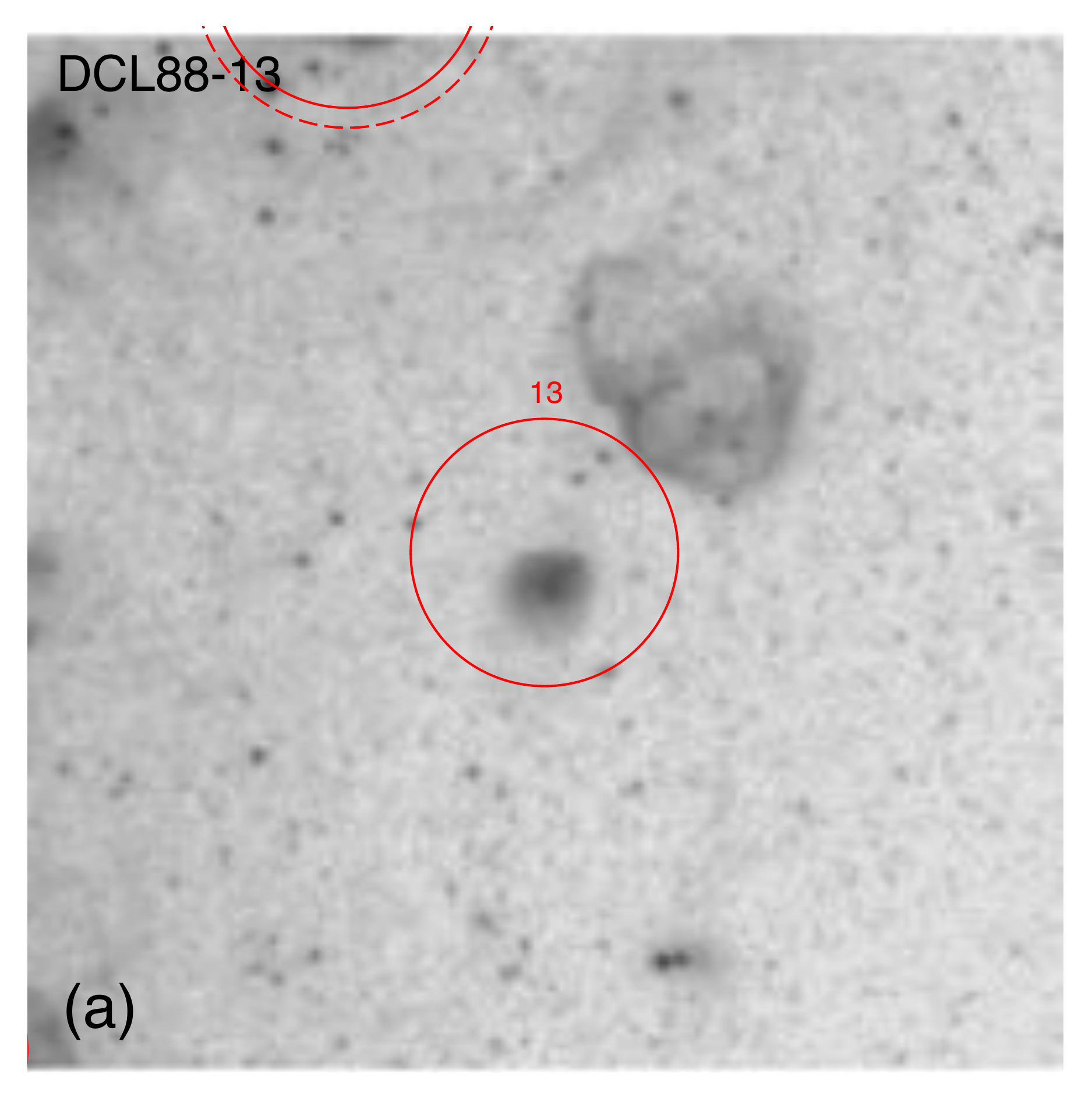}
\end{minipage}
\begin{minipage}{0.24\linewidth}
\includegraphics[width=\linewidth]{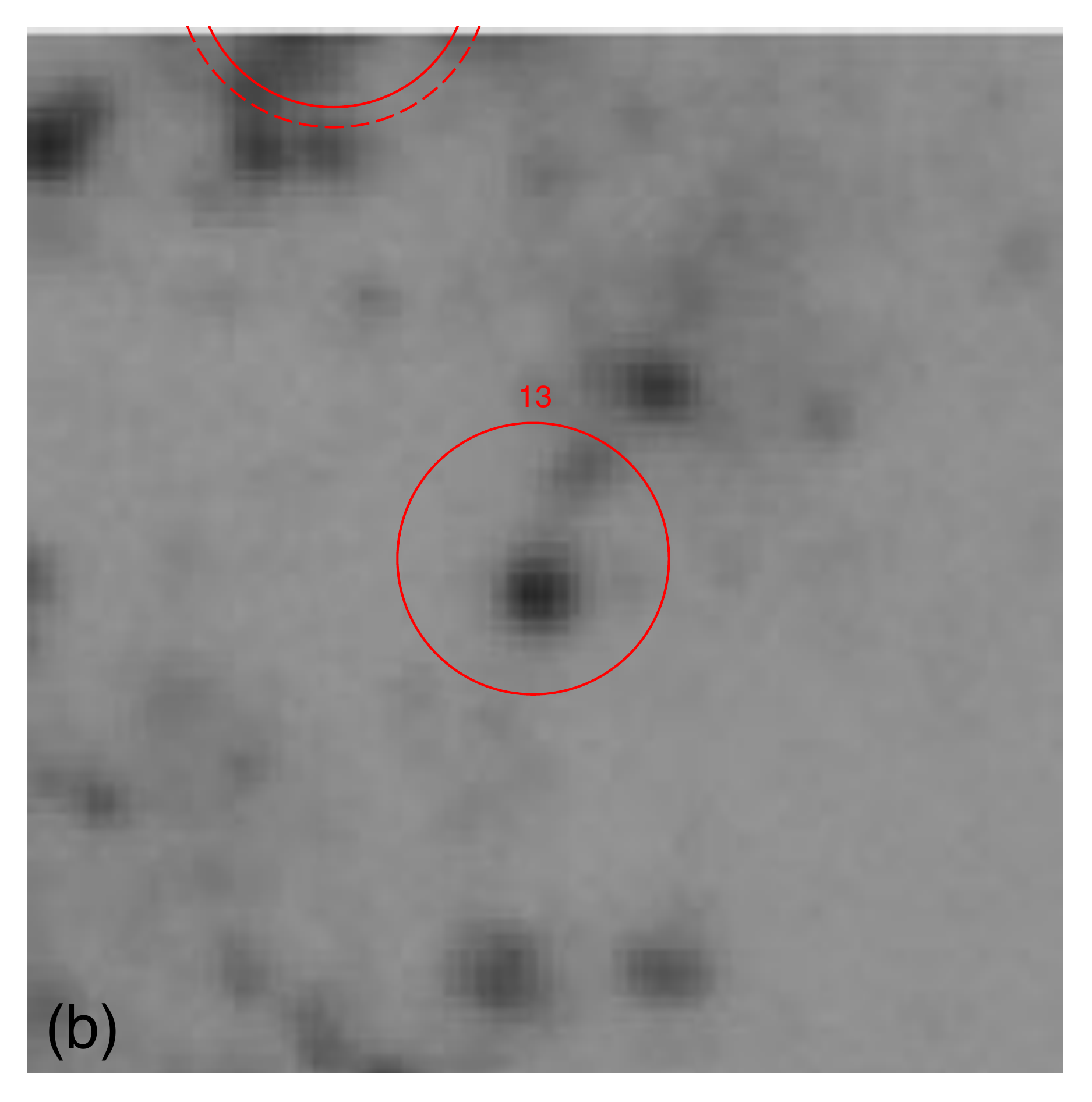}
\end{minipage}
\begin{minipage}{0.24\linewidth}
\includegraphics[width=\linewidth]{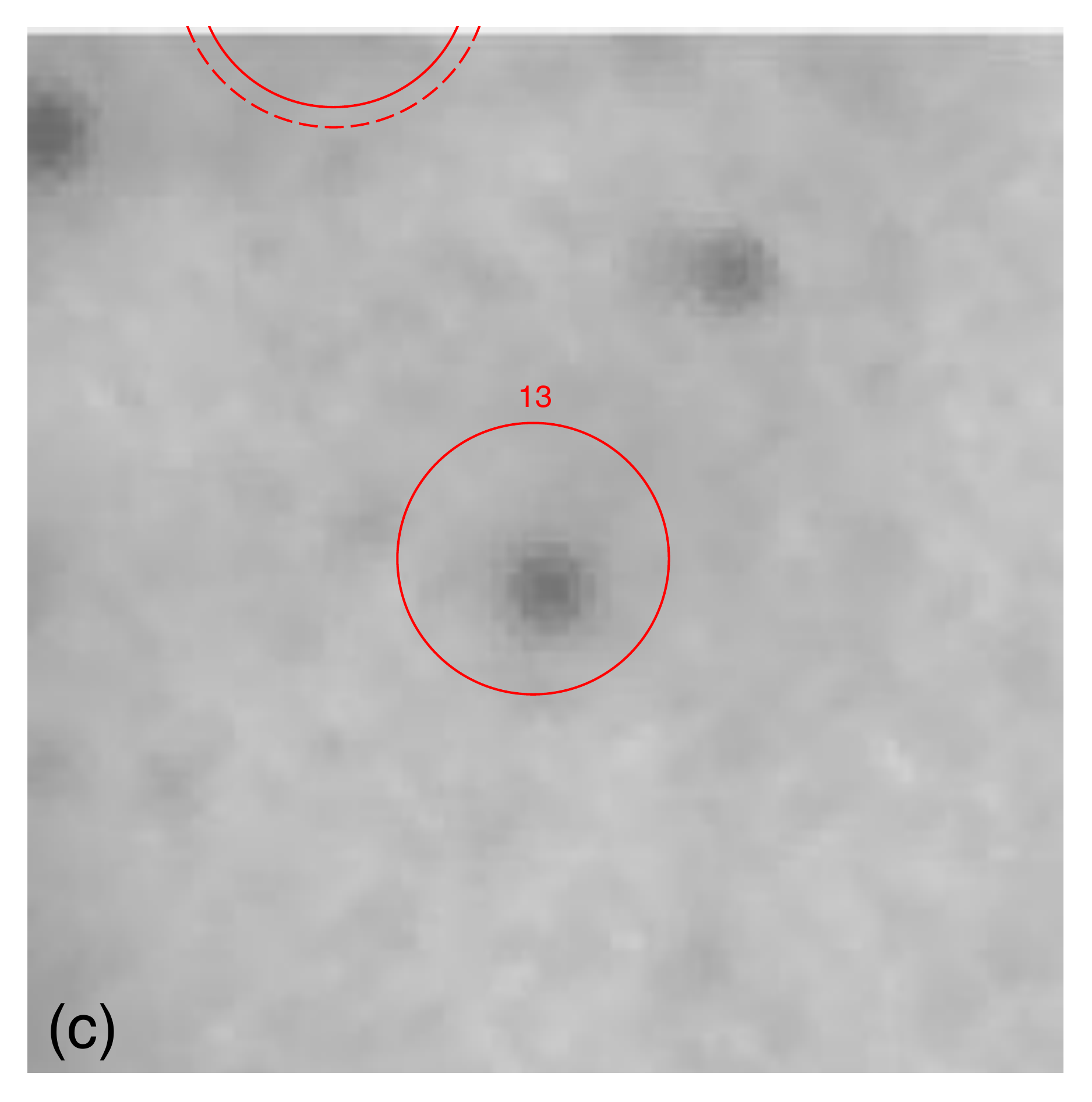}
\end{minipage}
\begin{minipage}{0.24\linewidth}
\includegraphics[width=\linewidth]{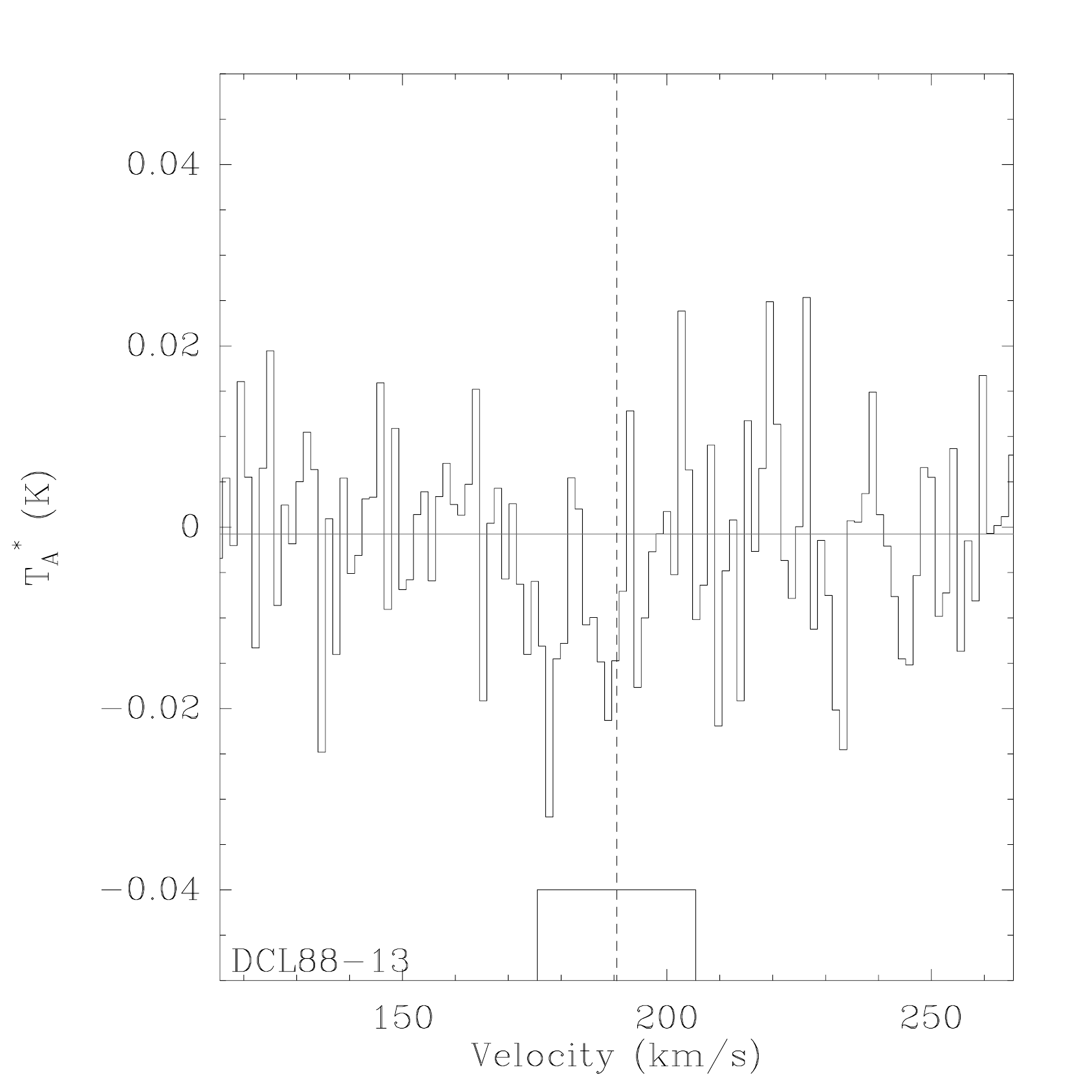}
\end{minipage}

\begin{minipage}{0.24\linewidth}
\includegraphics[width=\linewidth]{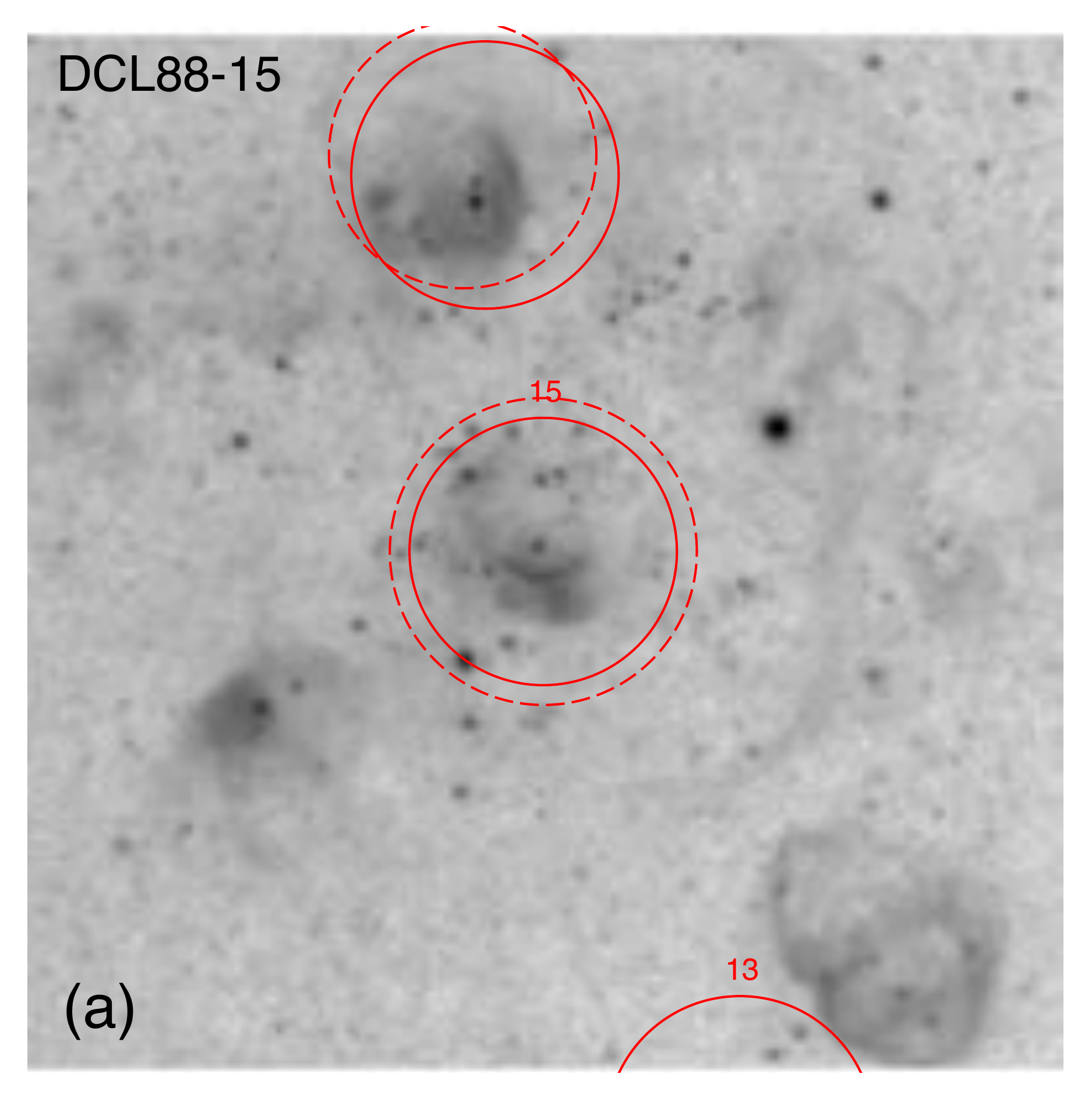}
\end{minipage}
\begin{minipage}{0.24\linewidth}
\includegraphics[width=\linewidth]{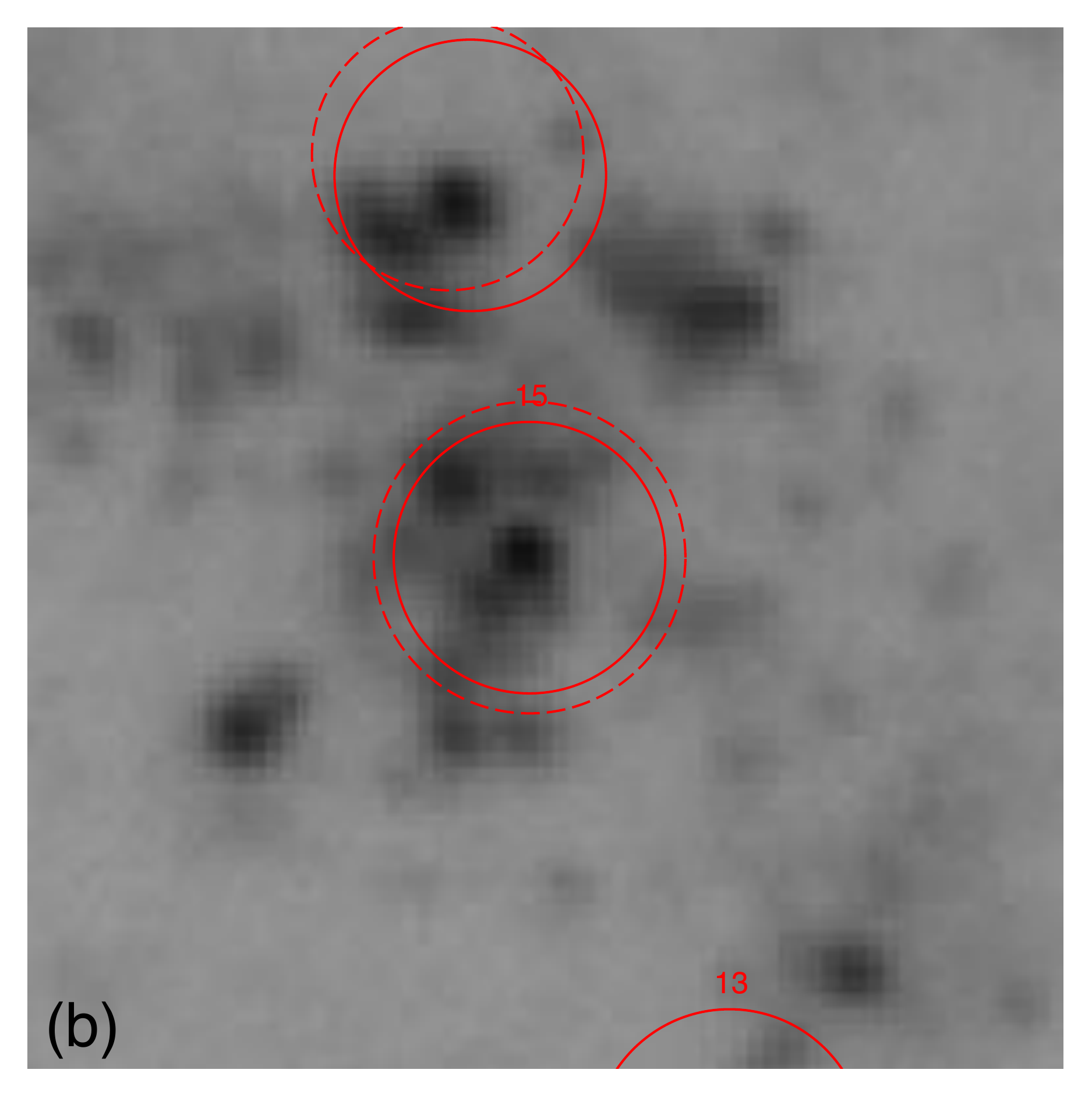}
\end{minipage}
\begin{minipage}{0.24\linewidth}
\includegraphics[width=\linewidth]{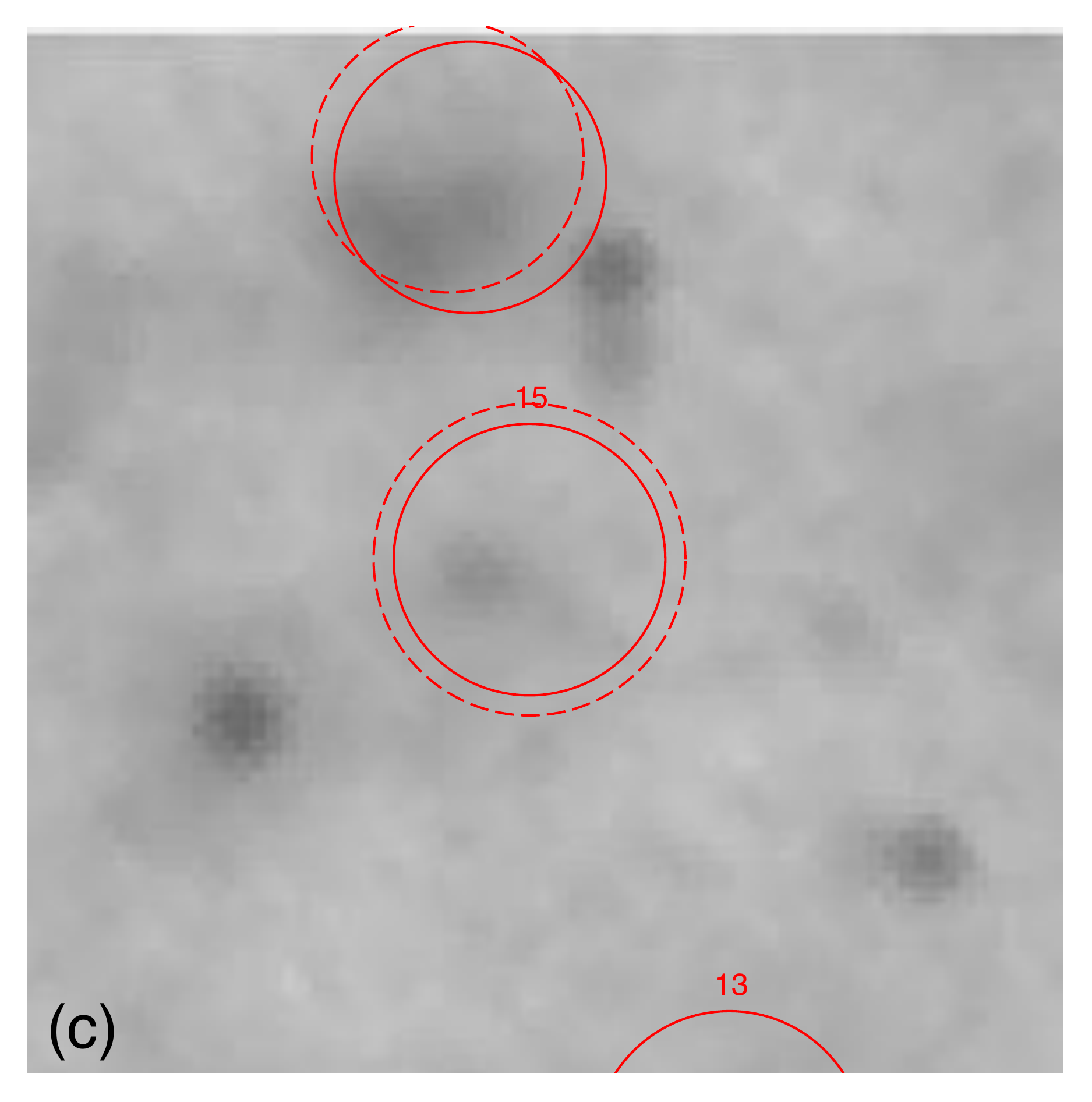}
\end{minipage}
\begin{minipage}{0.24\linewidth}
\includegraphics[width=\linewidth]{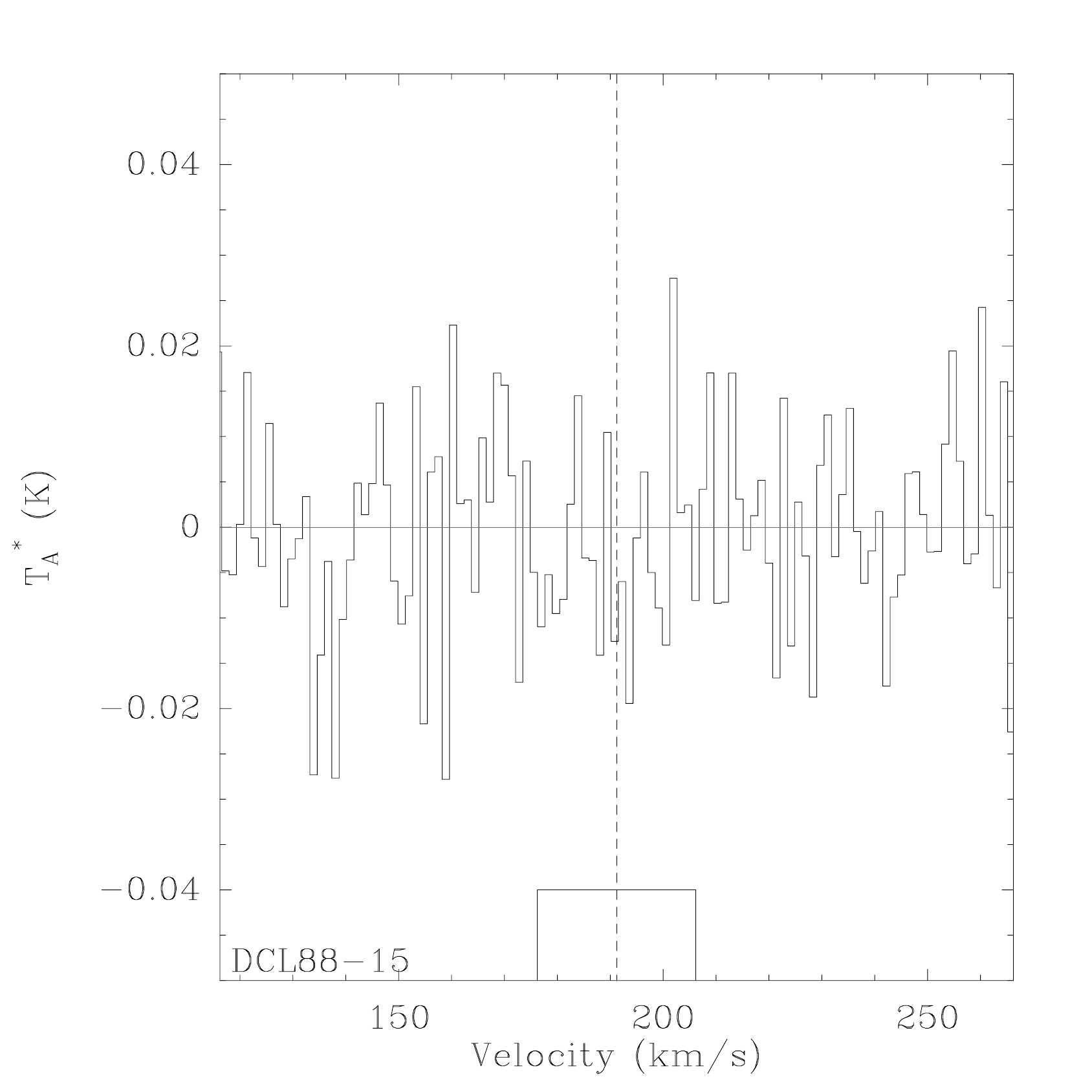}
\end{minipage}

\begin{minipage}{0.24\linewidth}
\includegraphics[width=\linewidth]{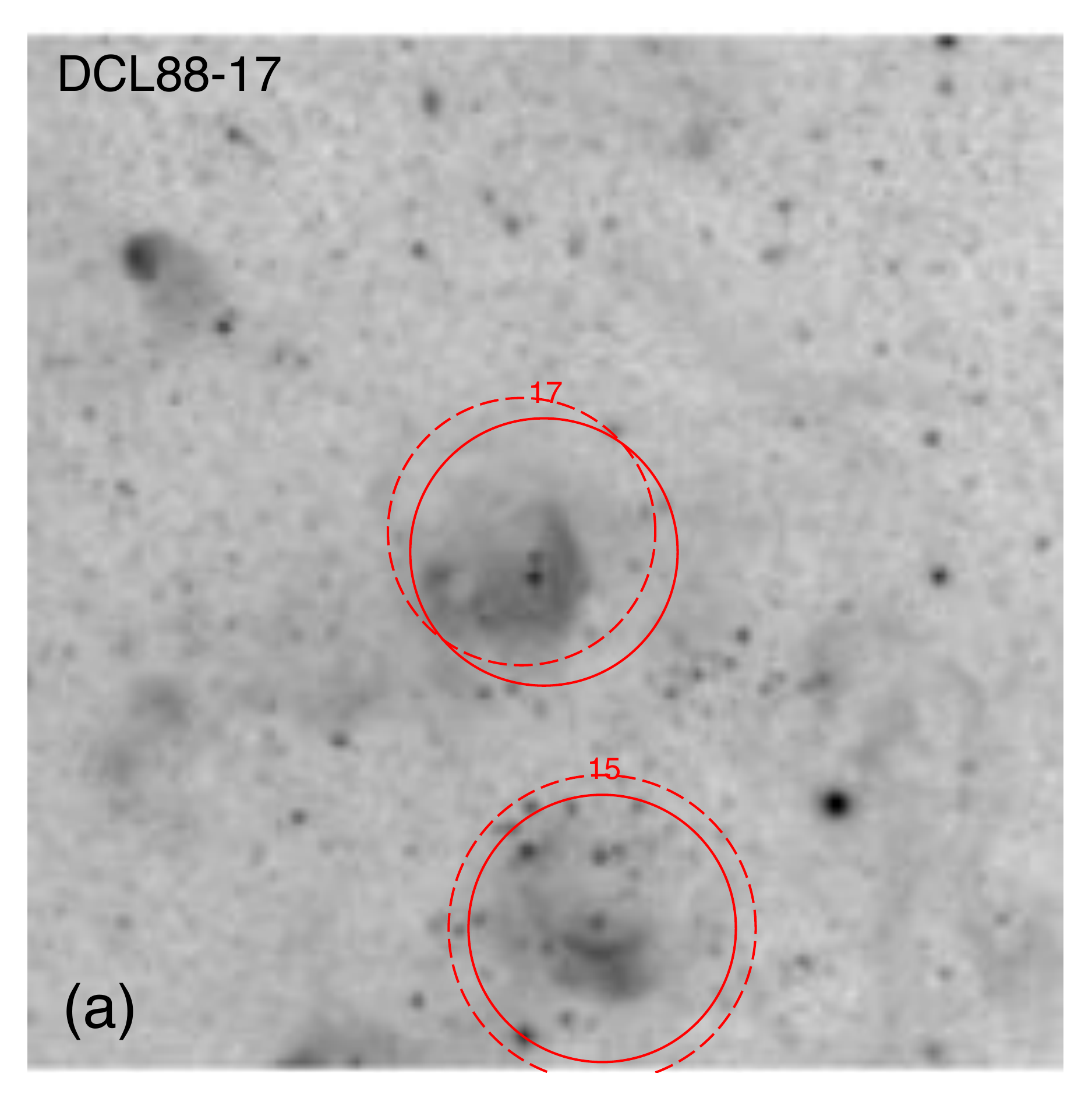}
\end{minipage}
\begin{minipage}{0.24\linewidth}
\includegraphics[width=\linewidth]{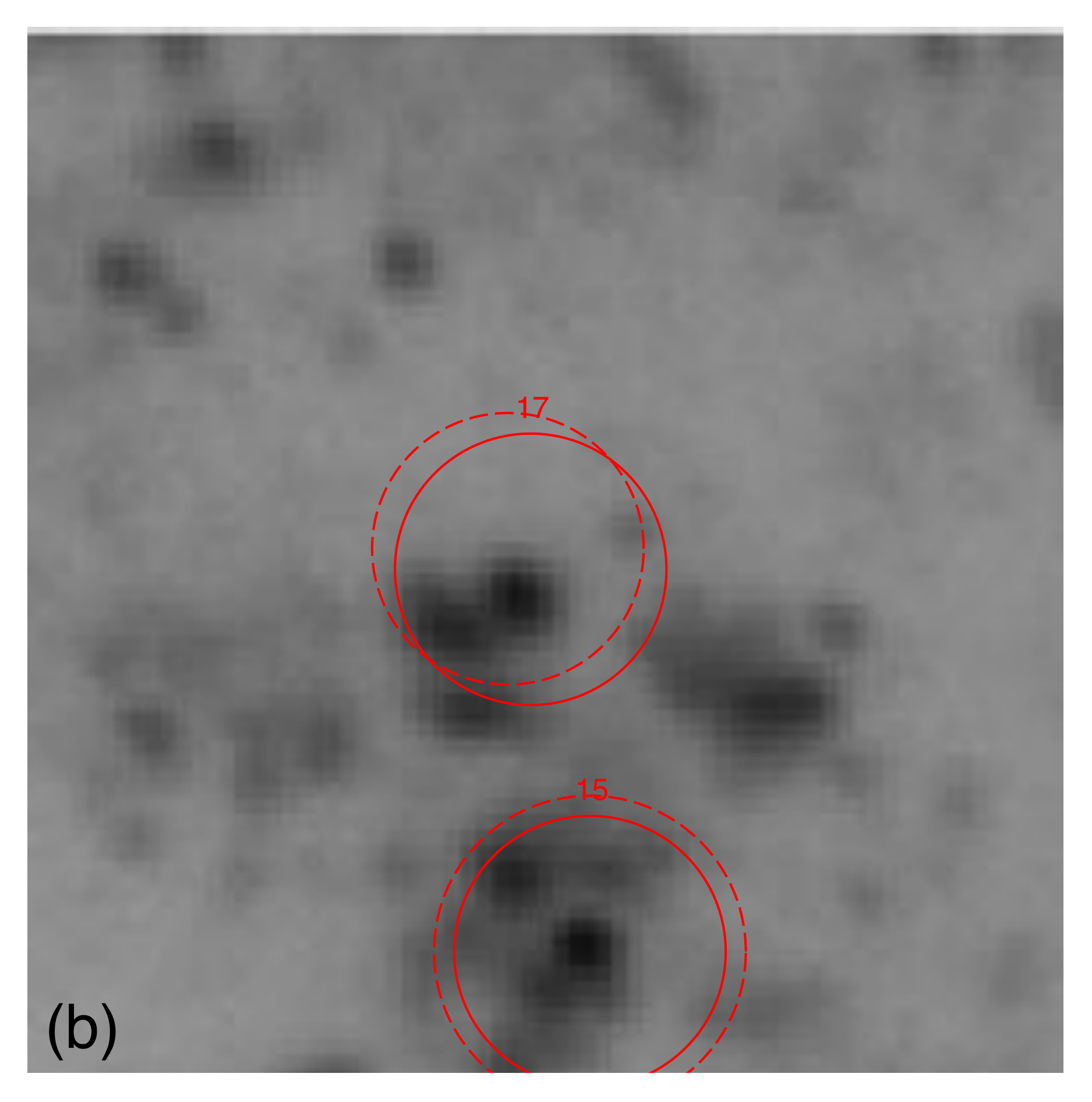}
\end{minipage}
\begin{minipage}{0.24\linewidth}
\includegraphics[width=\linewidth]{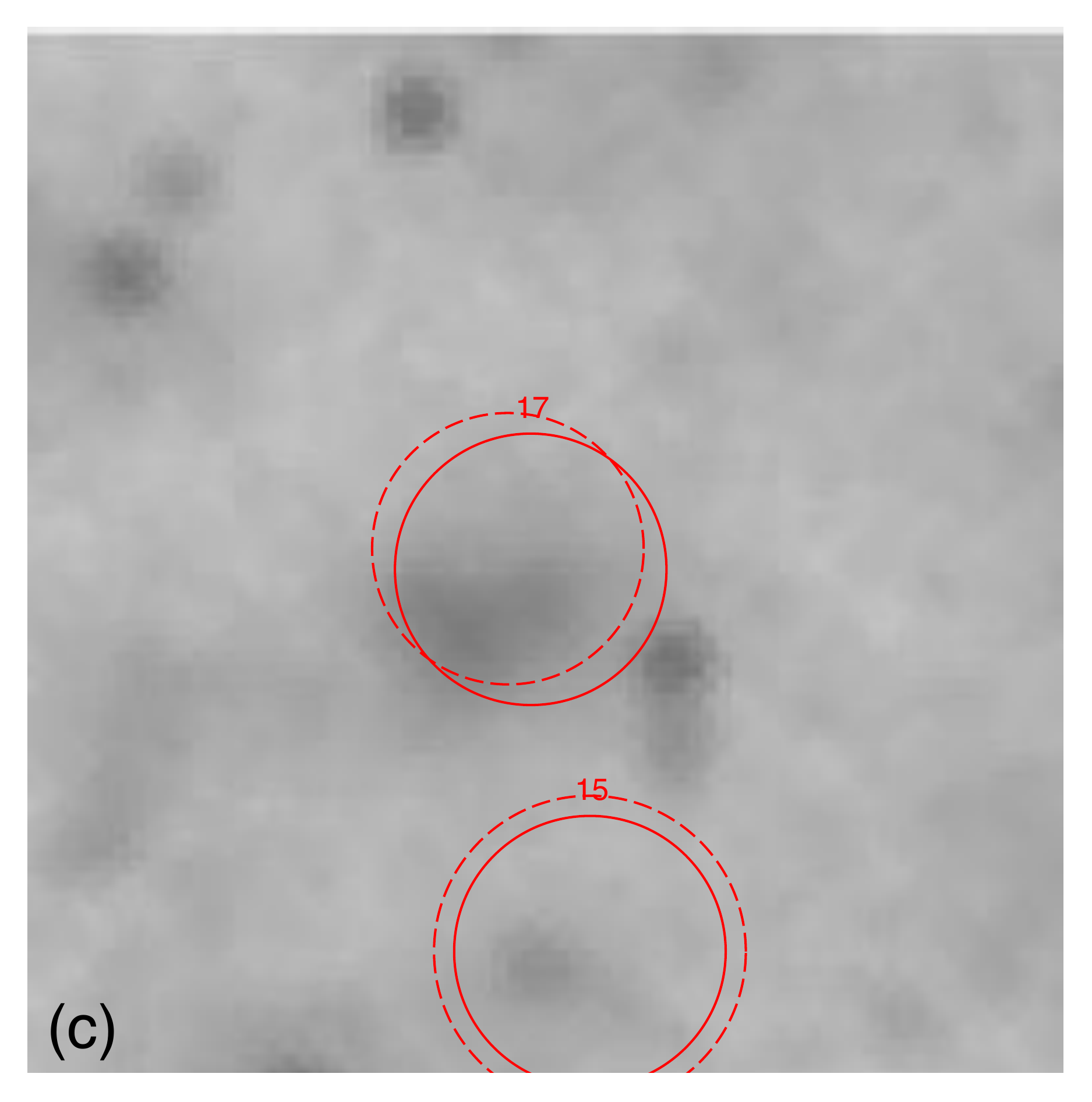}
\end{minipage}
\begin{minipage}{0.24\linewidth}
\includegraphics[width=\linewidth]{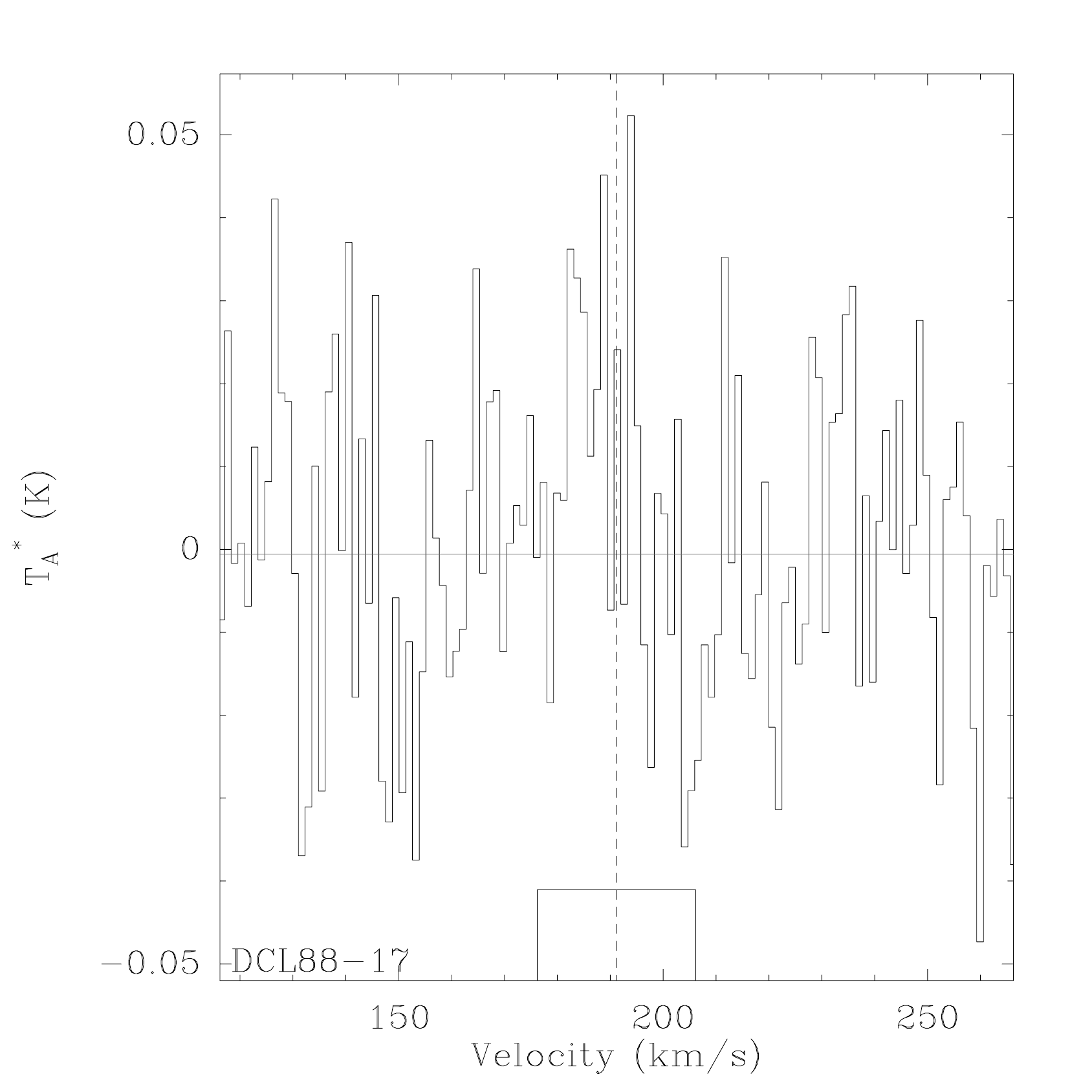}
\end{minipage}

\noindent\textbf{Figure~\ref{fig:stamps} -- continued.}

\end{figure*}

\begin{figure*}
%\ContinuedFloat

\begin{minipage}{0.24\linewidth}
\includegraphics[width=\linewidth]{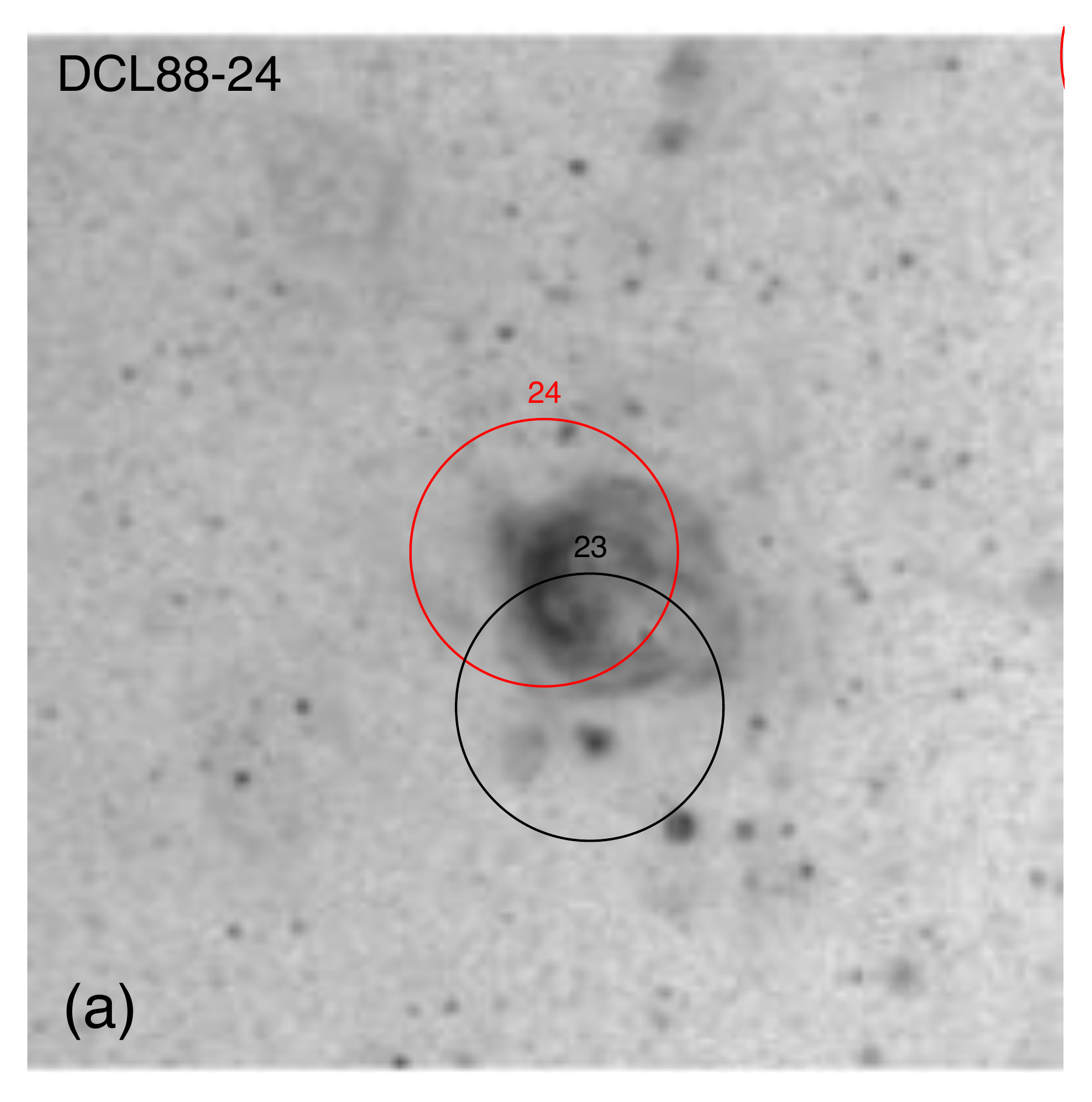}
\end{minipage}
\begin{minipage}{0.24\linewidth}
\includegraphics[width=\linewidth]{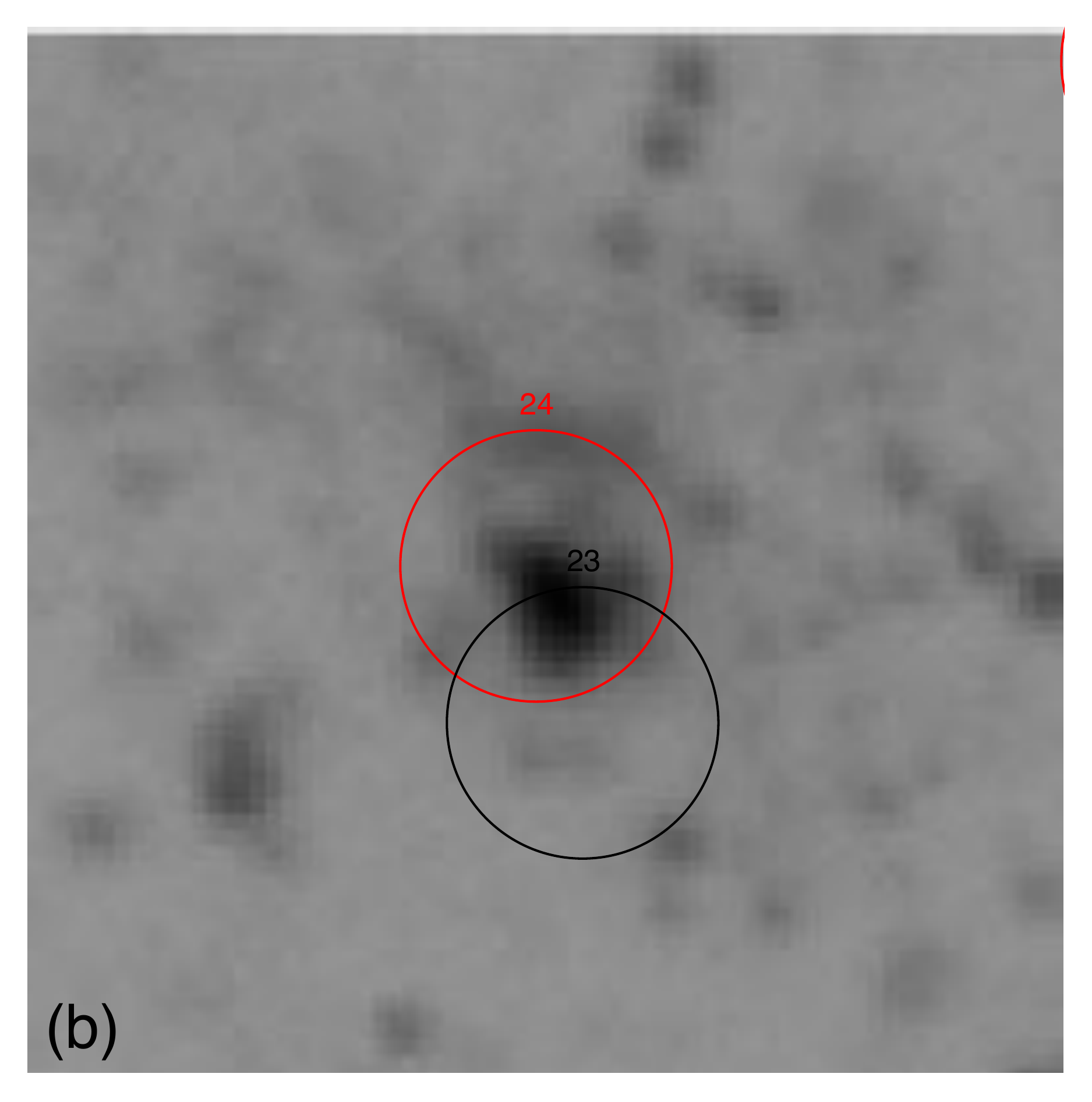}
\end{minipage}
\begin{minipage}{0.24\linewidth}
\includegraphics[width=\linewidth]{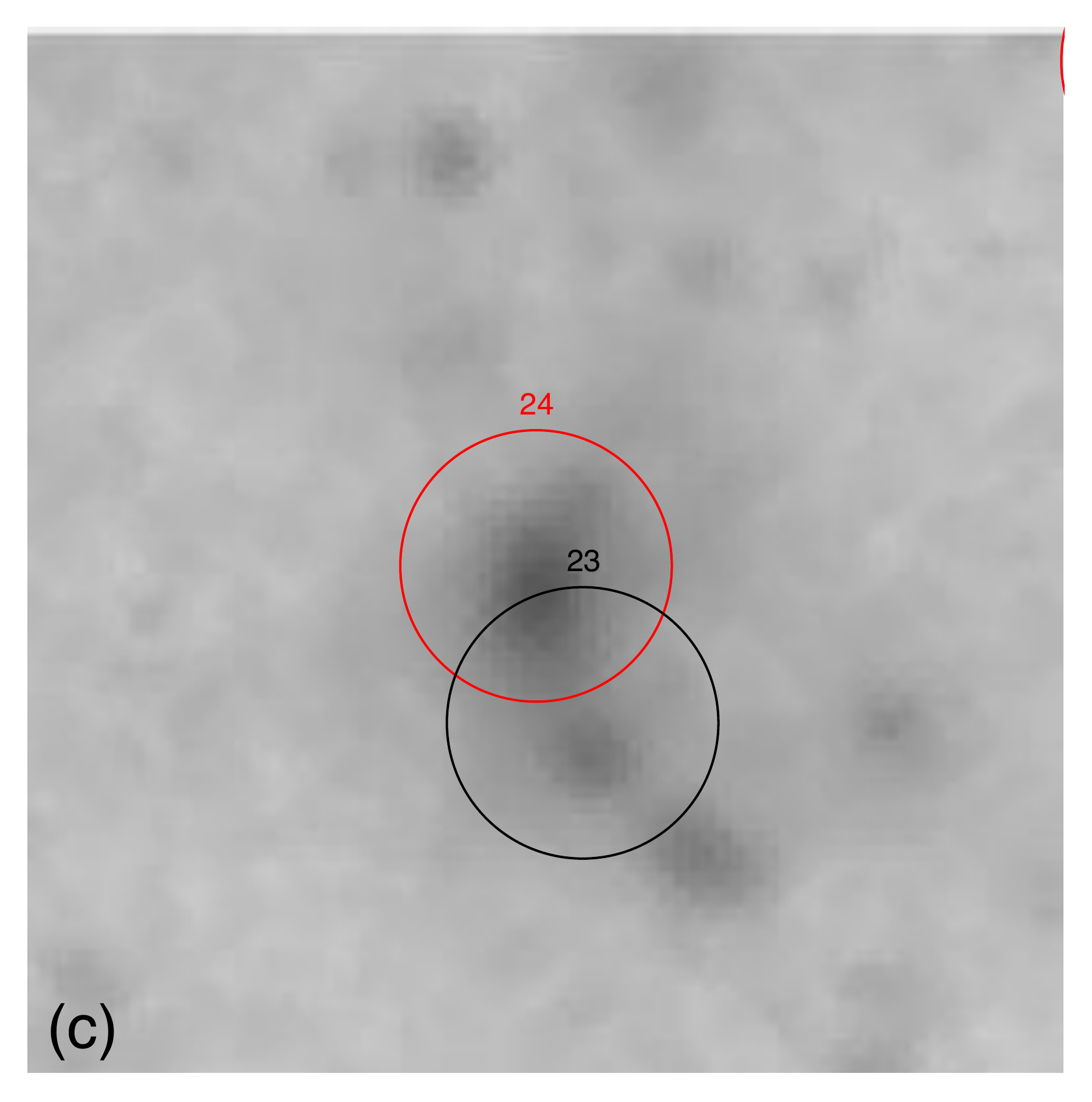}
\end{minipage}
\begin{minipage}{0.24\linewidth}
\includegraphics[width=\linewidth]{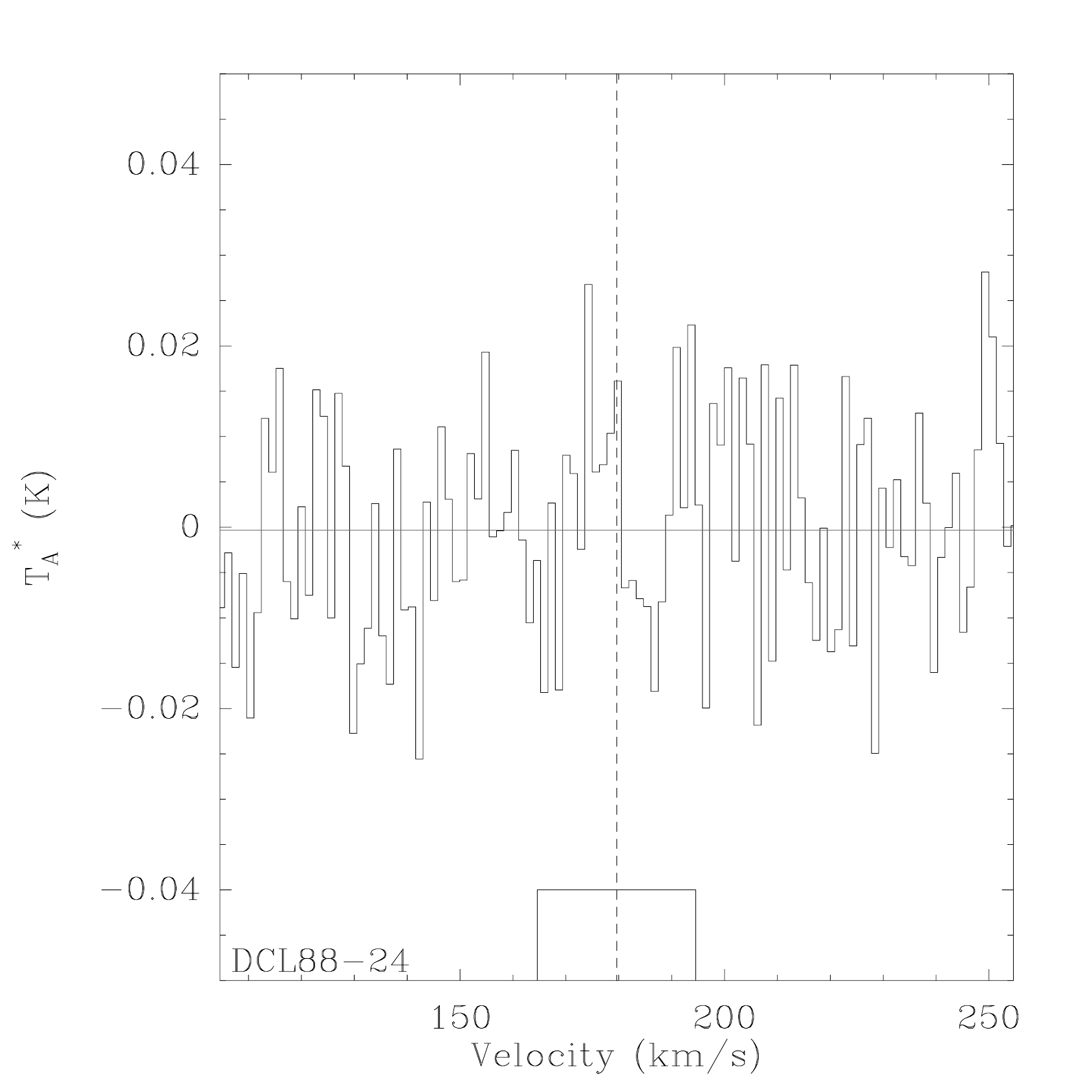}
\end{minipage}

\begin{minipage}{0.24\linewidth}
\includegraphics[width=\linewidth]{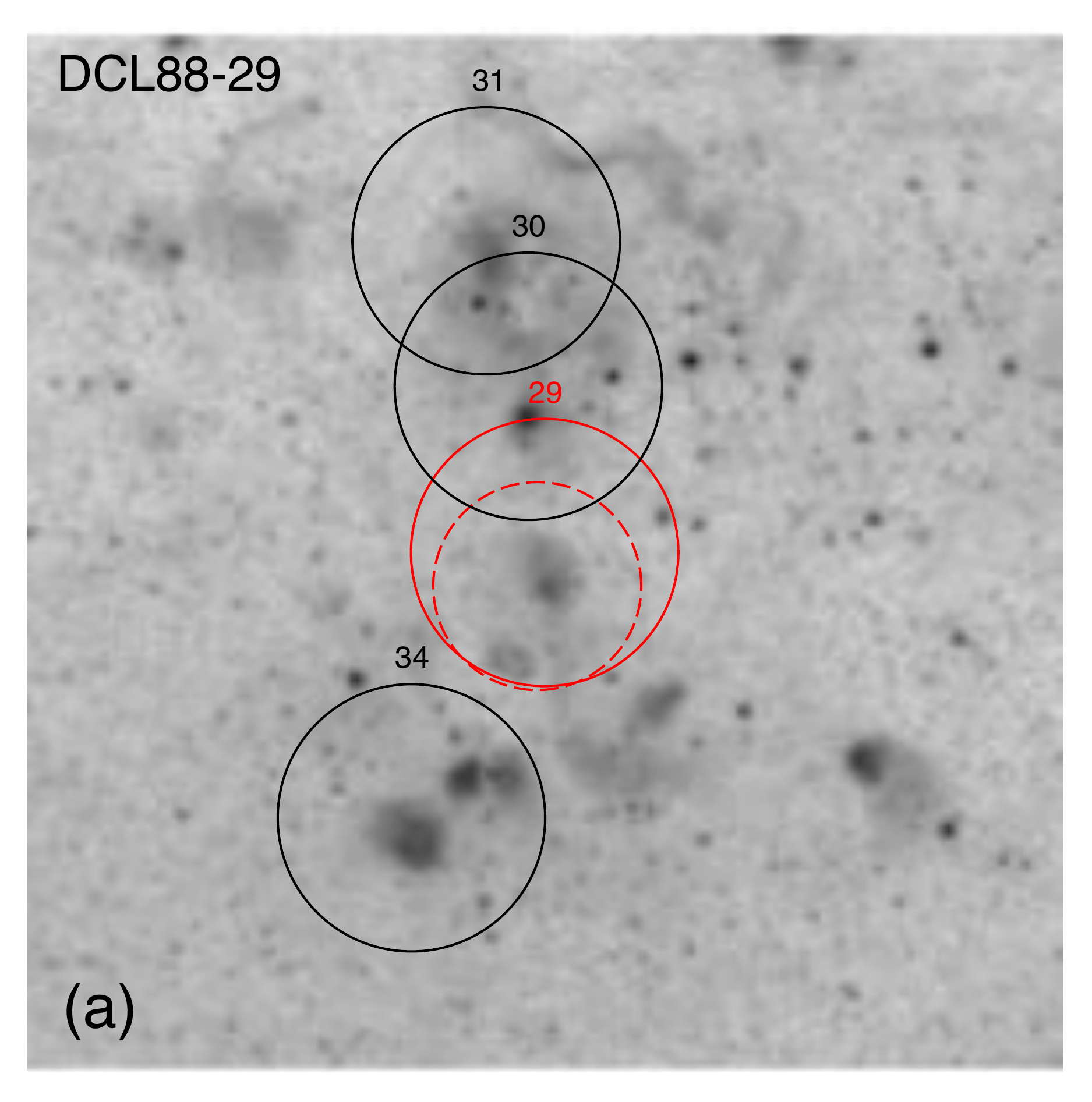}
\end{minipage}
\begin{minipage}{0.24\linewidth}
\includegraphics[width=\linewidth]{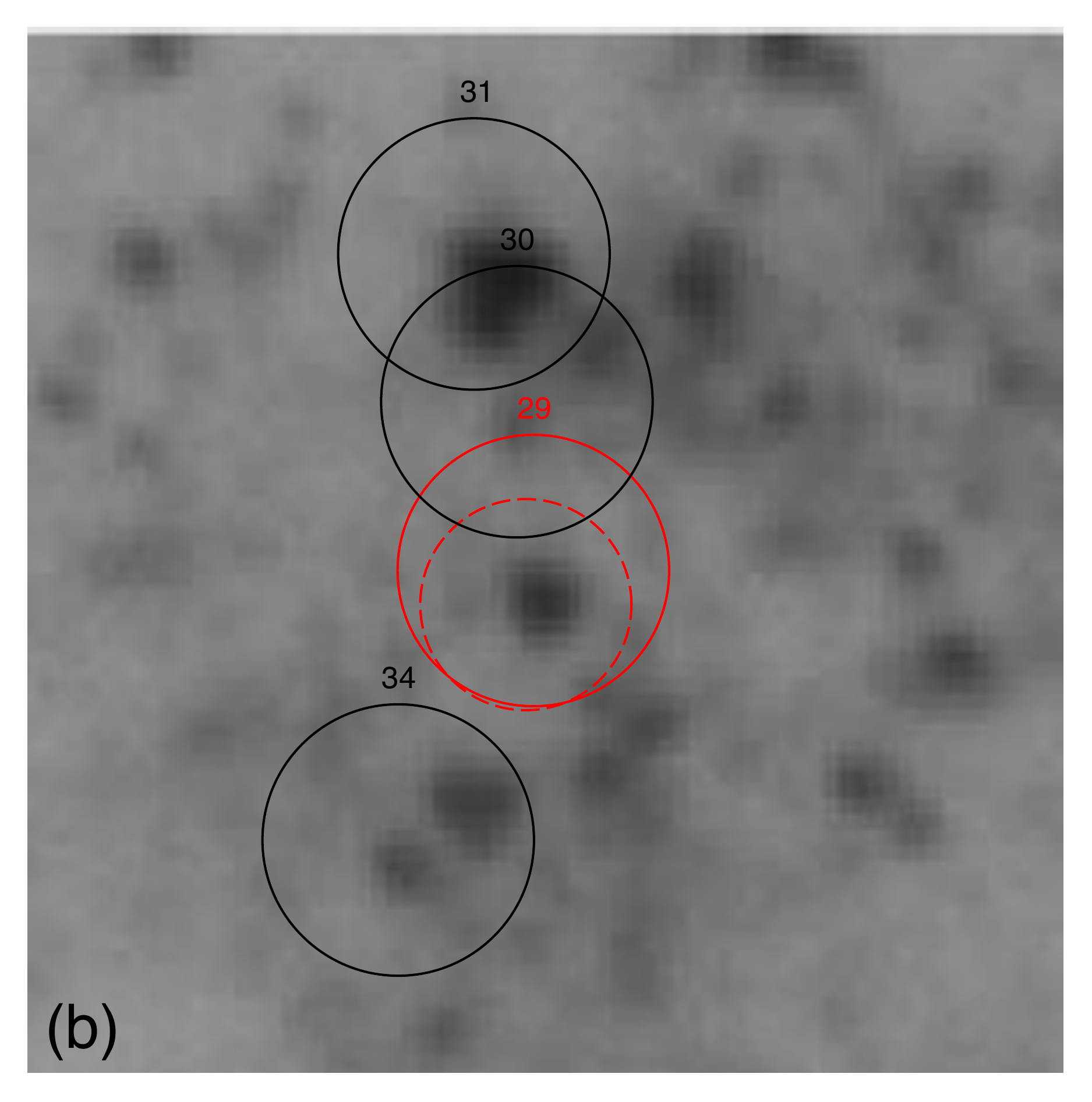}
\end{minipage}
\begin{minipage}{0.24\linewidth}
\includegraphics[width=\linewidth]{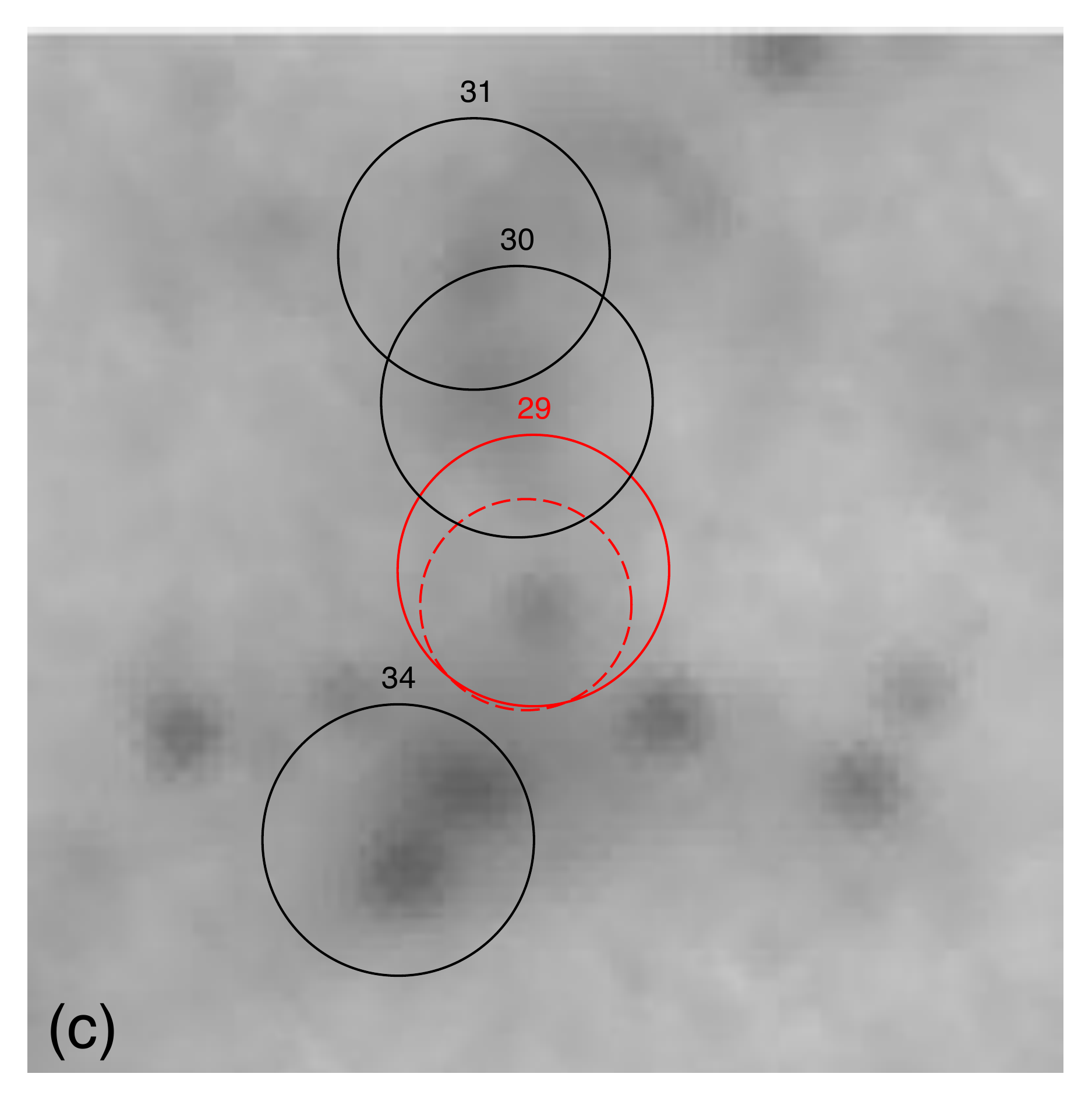}
\end{minipage}
\begin{minipage}{0.24\linewidth}
\includegraphics[width=\linewidth]{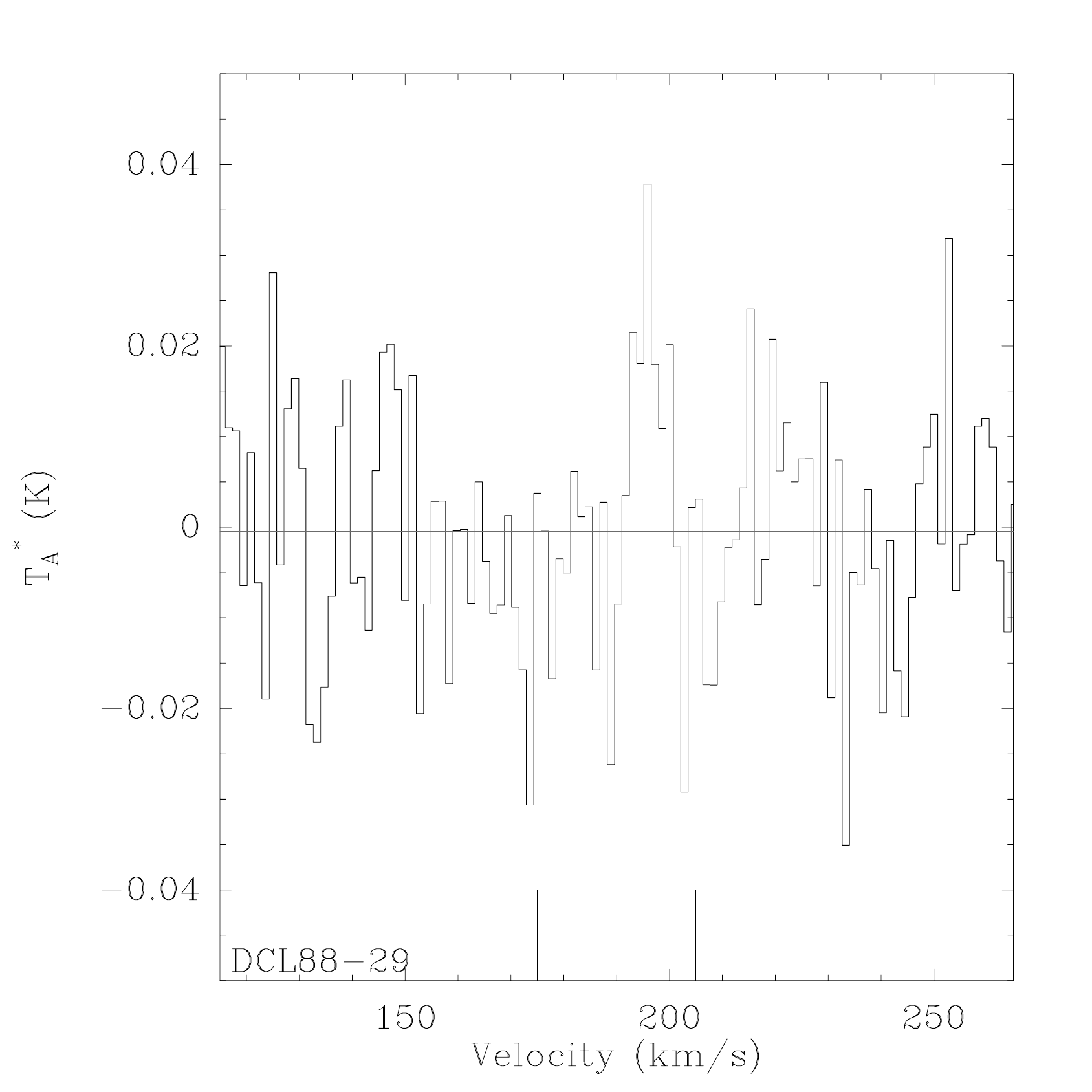}
\end{minipage}

\begin{minipage}{0.24\linewidth}
\includegraphics[width=\linewidth]{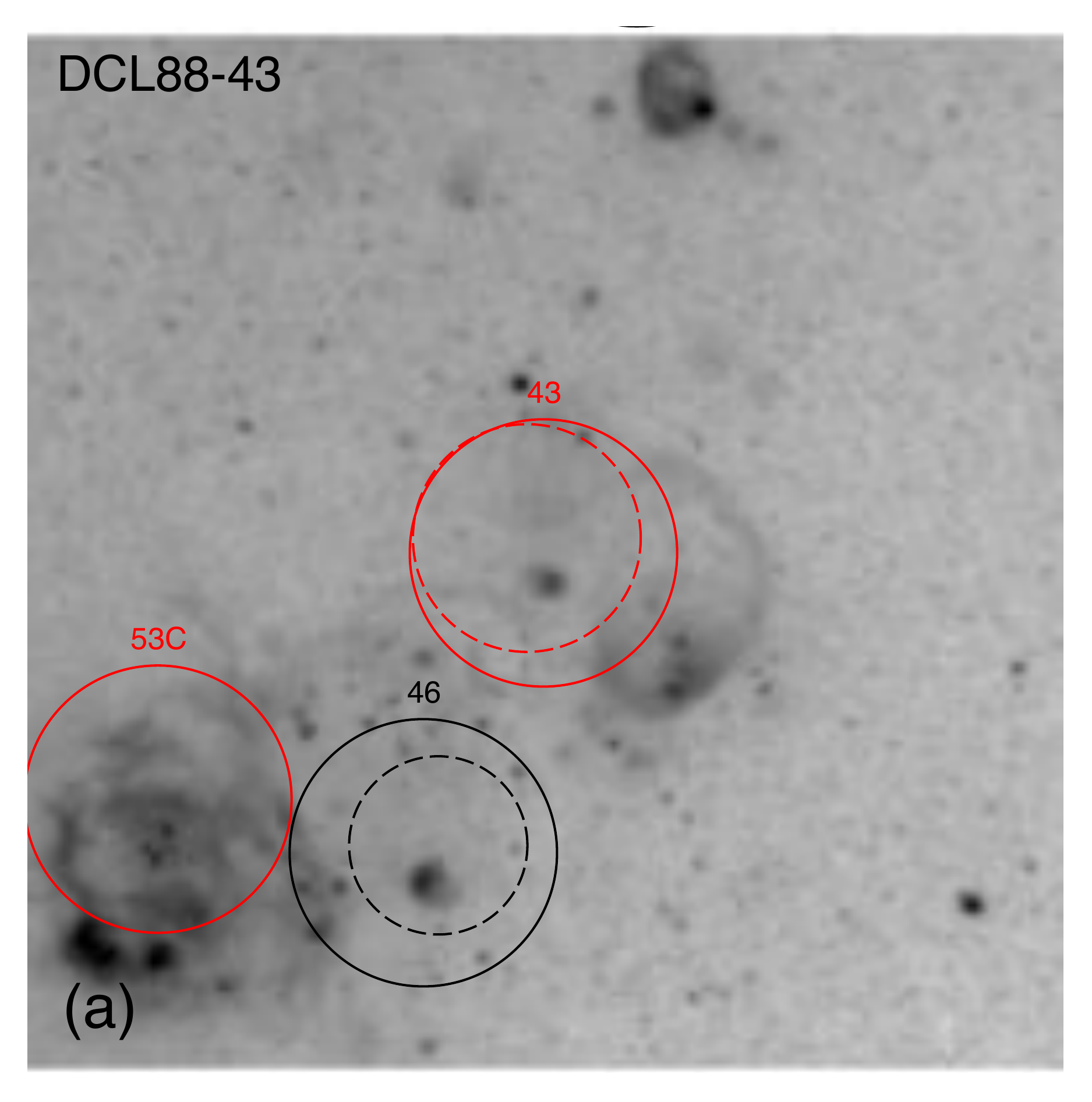}
\end{minipage}
\begin{minipage}{0.24\linewidth}
\includegraphics[width=\linewidth]{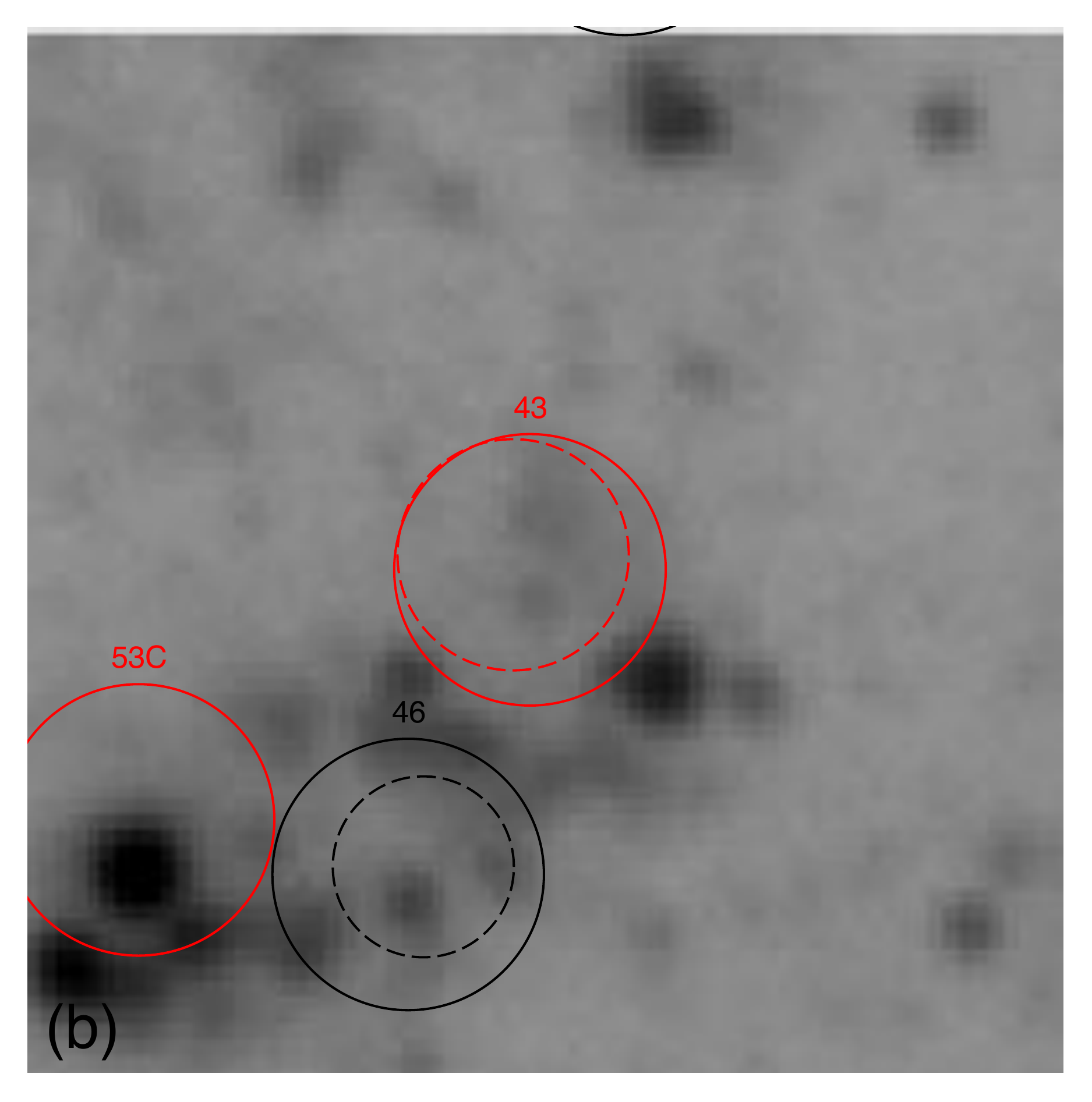}
\end{minipage}
\begin{minipage}{0.24\linewidth}
\includegraphics[width=\linewidth]{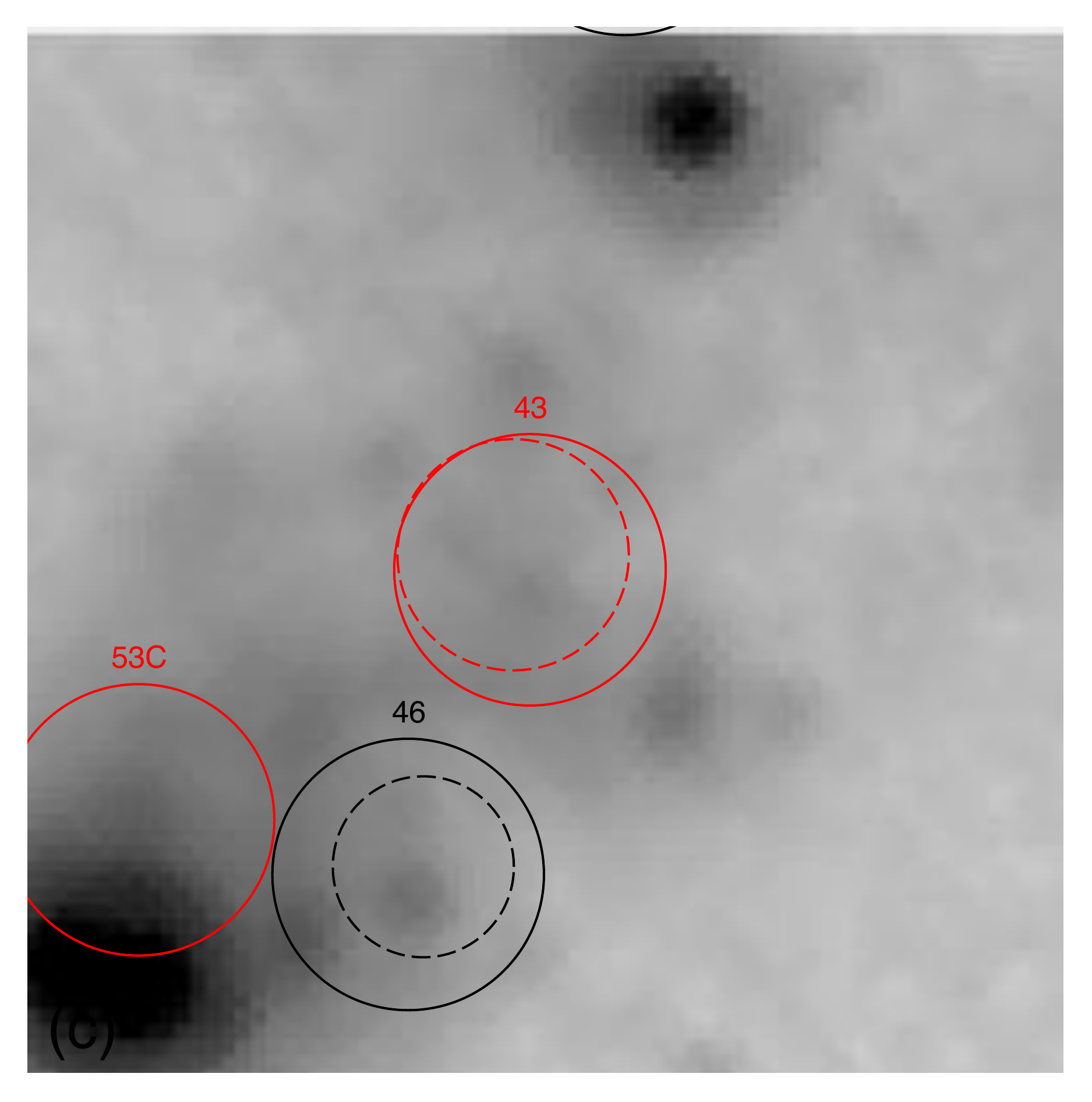}
\end{minipage}
\begin{minipage}{0.24\linewidth}
\includegraphics[width=\linewidth]{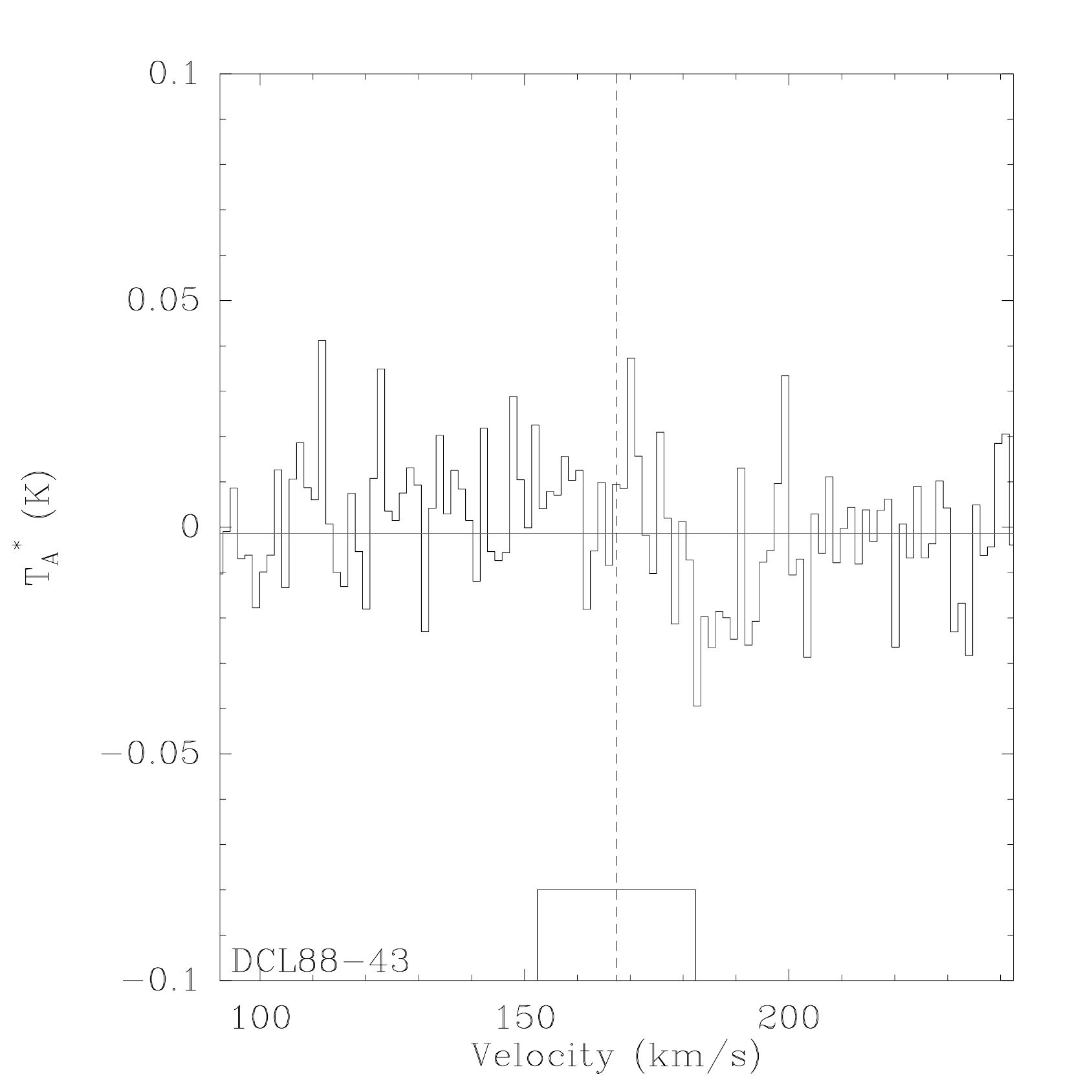}
\end{minipage}

\begin{minipage}{0.24\linewidth}
\includegraphics[width=\linewidth]{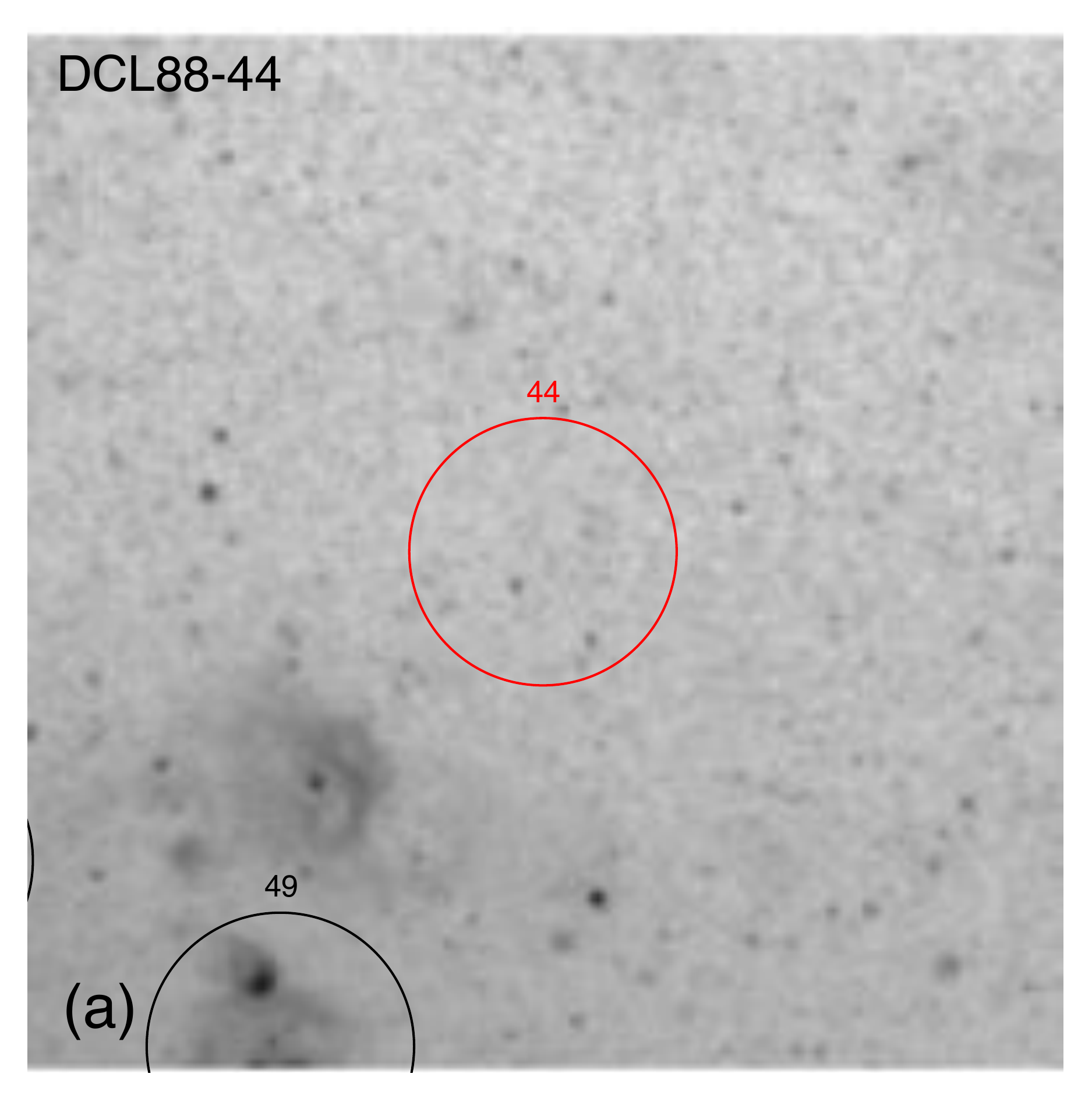}
\end{minipage}
\begin{minipage}{0.24\linewidth}
\includegraphics[width=\linewidth]{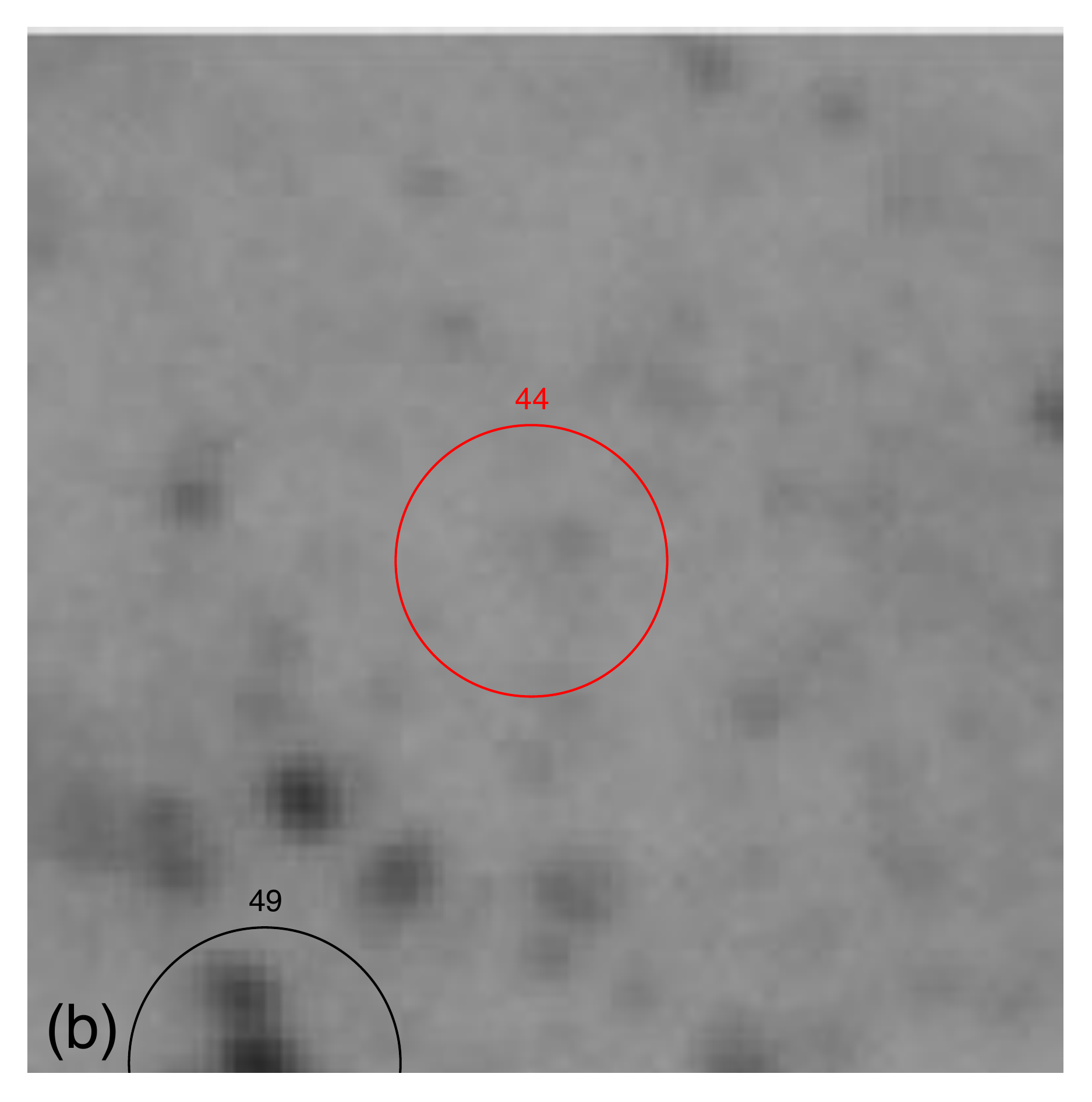}
\end{minipage}
\begin{minipage}{0.24\linewidth}
\includegraphics[width=\linewidth]{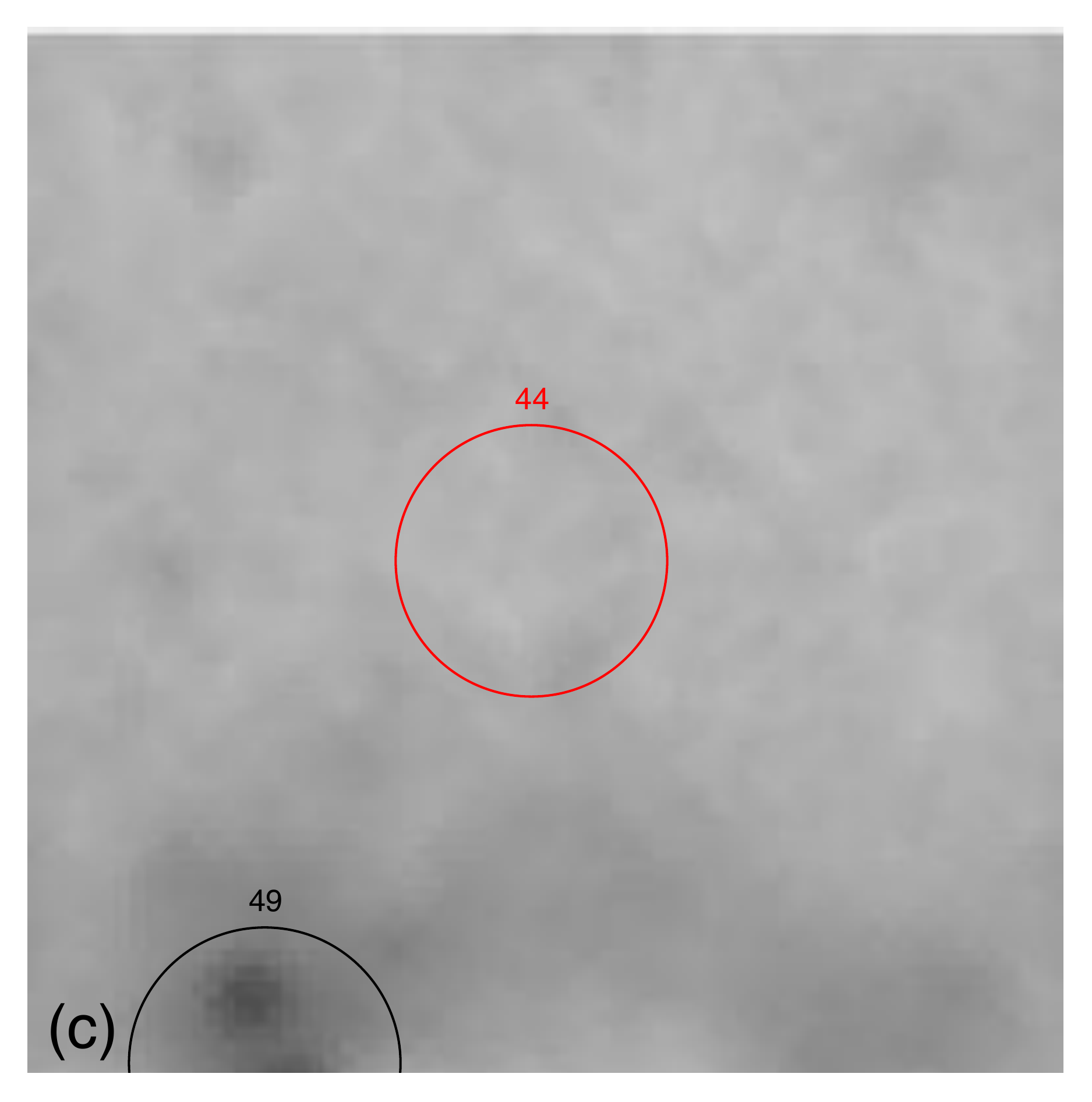}
\end{minipage}
\begin{minipage}{0.24\linewidth}
\includegraphics[width=\linewidth]{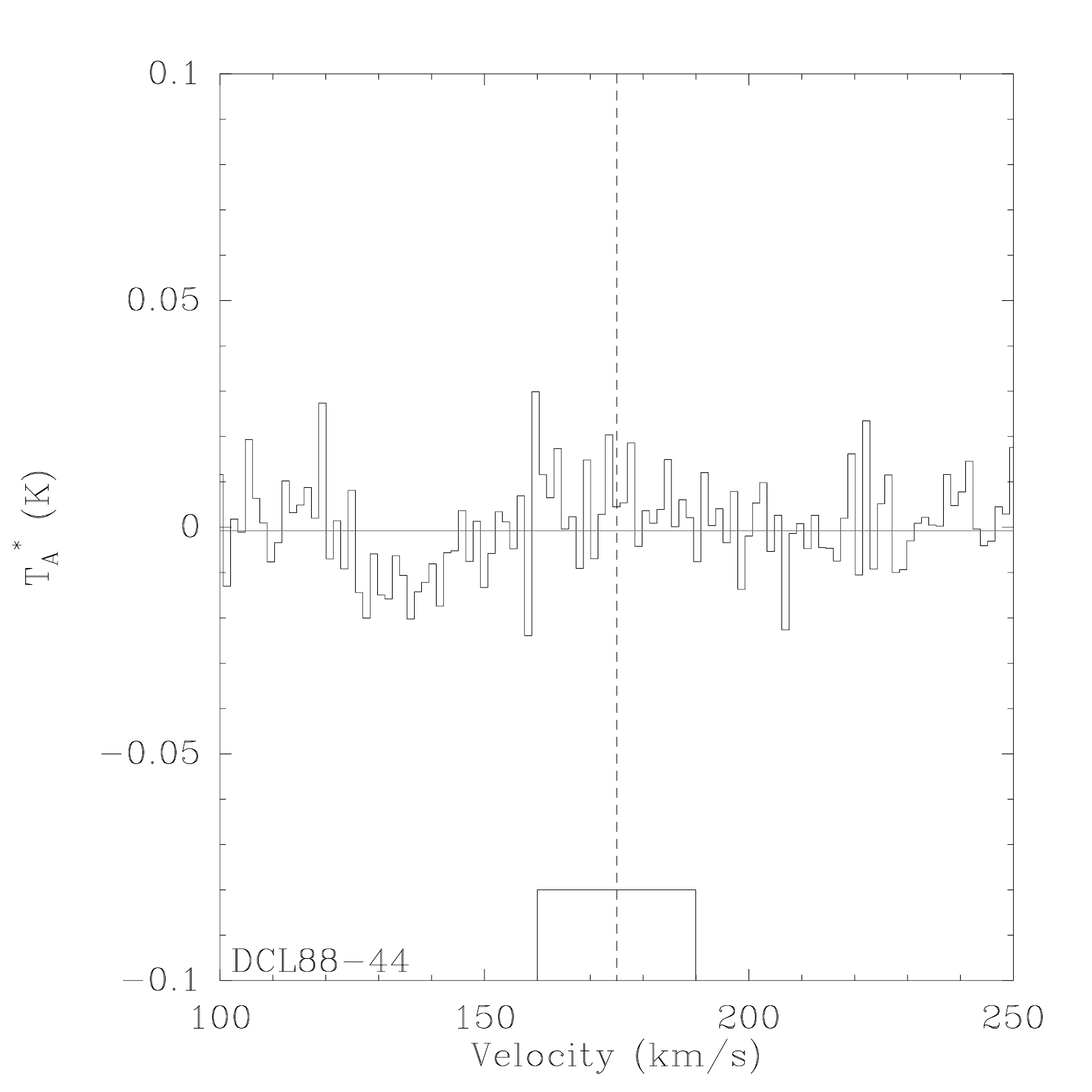}
\end{minipage}

\begin{minipage}{0.24\linewidth}
\includegraphics[width=\linewidth]{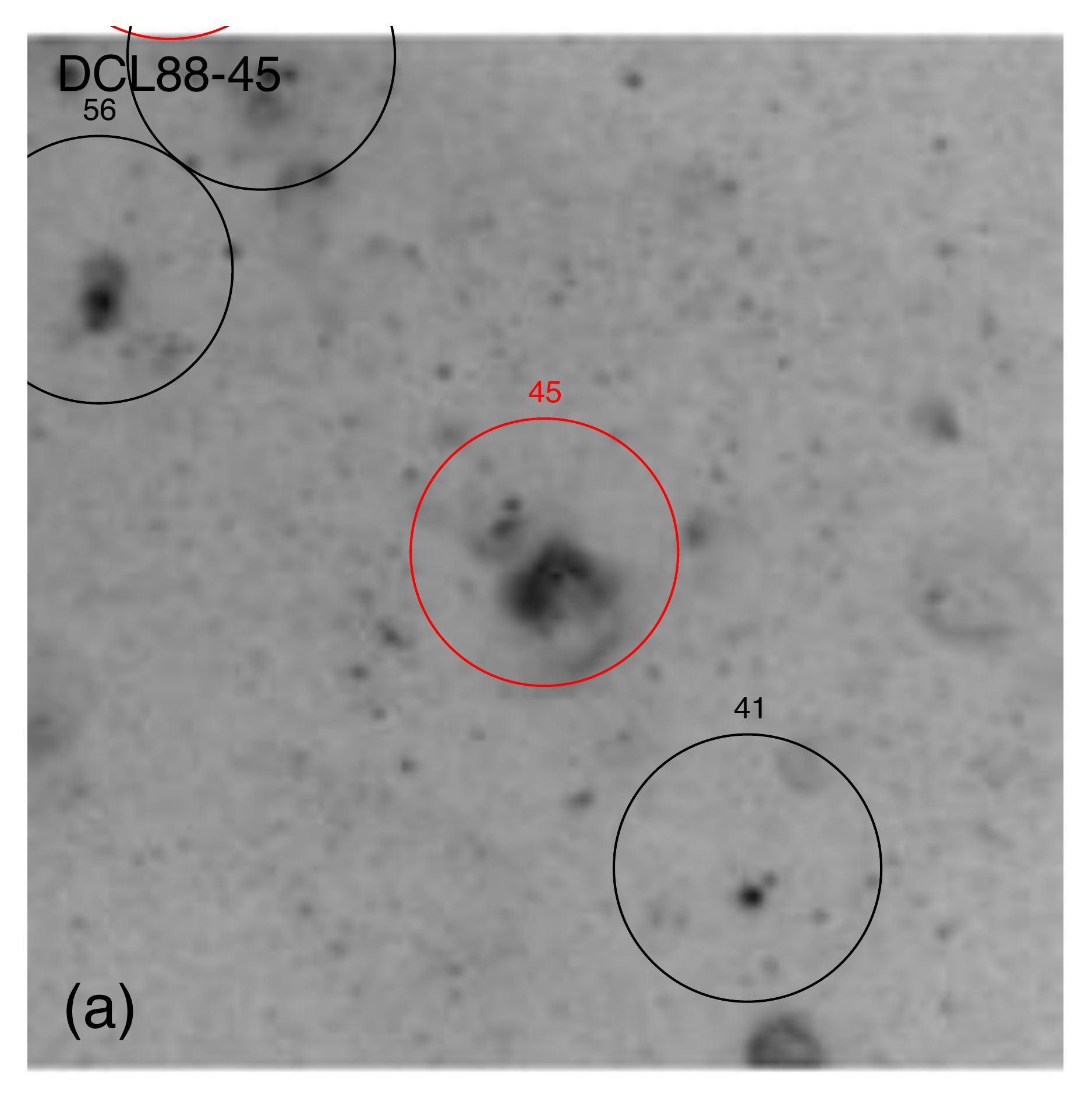}
\end{minipage}
\begin{minipage}{0.24\linewidth}
\includegraphics[width=\linewidth]{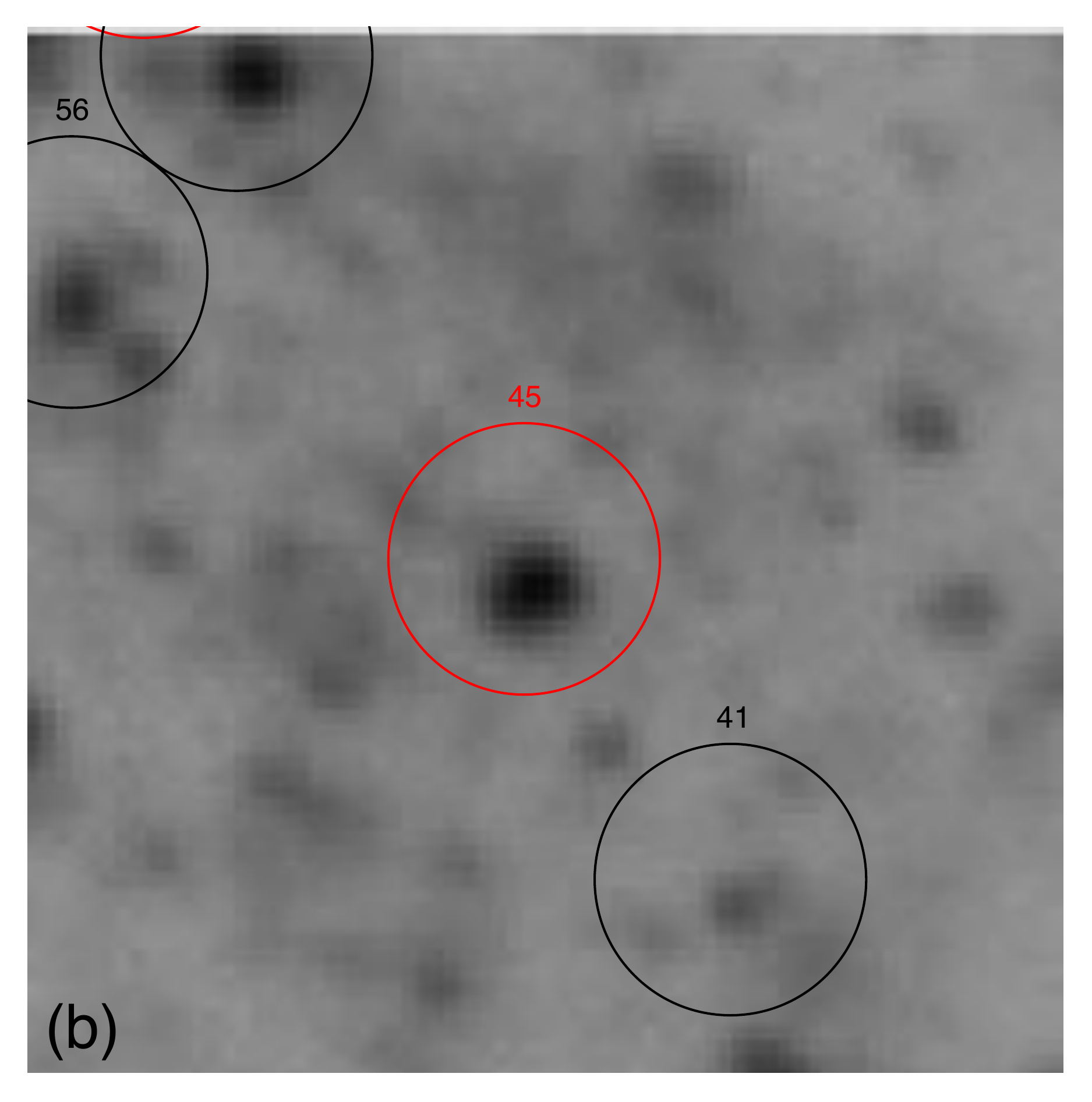}
\end{minipage}
\begin{minipage}{0.24\linewidth}
\includegraphics[width=\linewidth]{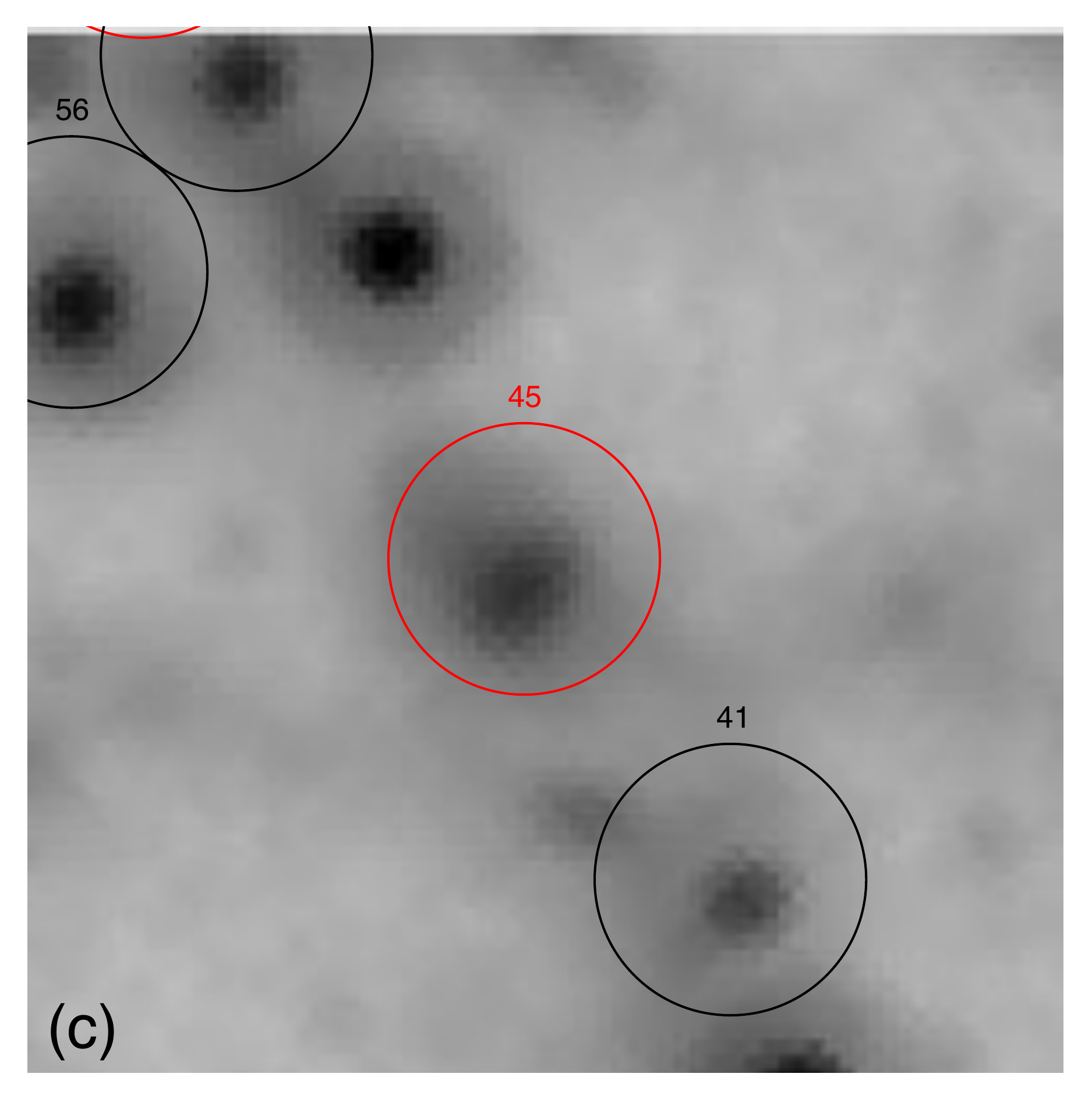}
\end{minipage}
\begin{minipage}{0.24\linewidth}
\includegraphics[width=\linewidth]{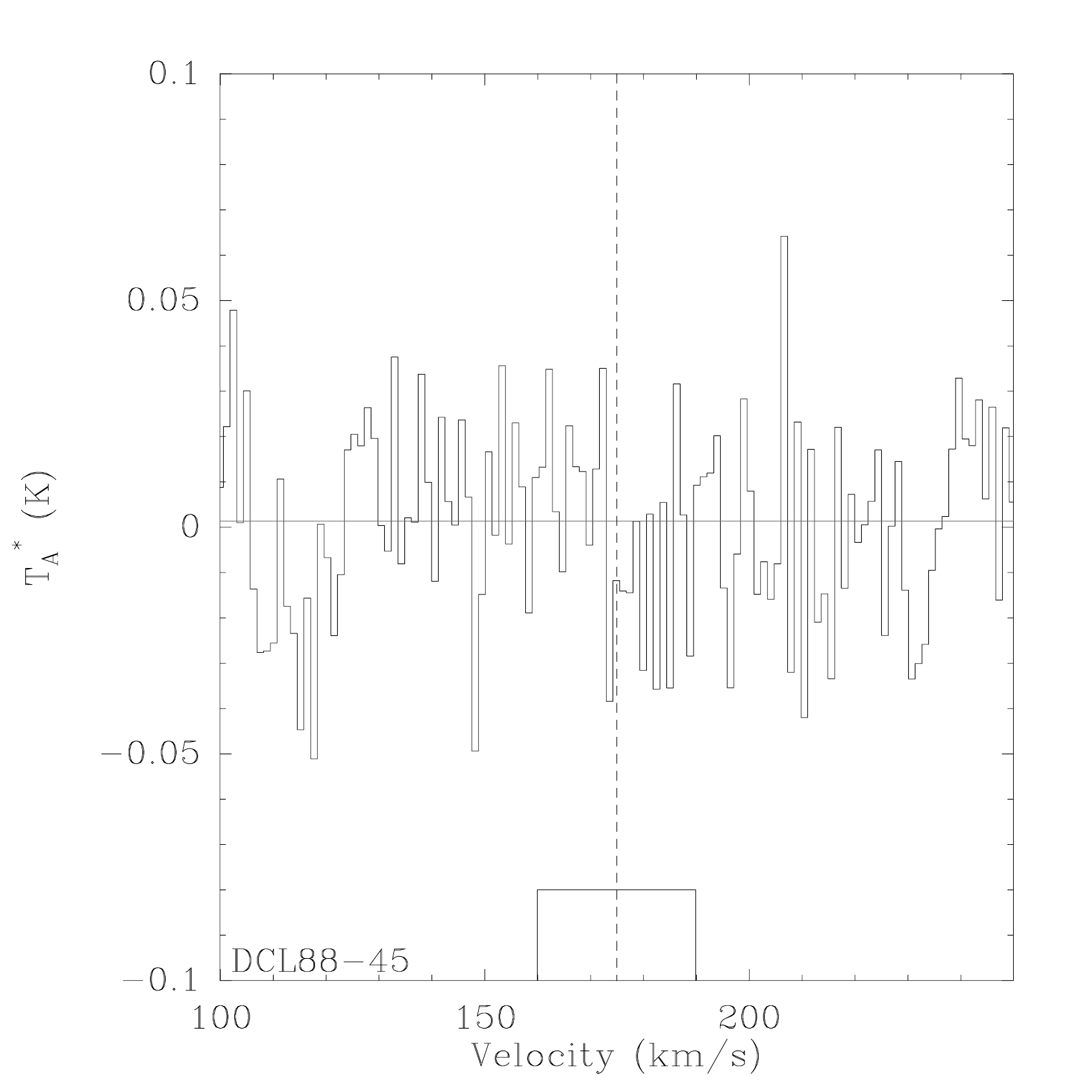}
\end{minipage}

\noindent\textbf{Figure~\ref{fig:stamps} -- continued.}

\end{figure*}

\begin{figure*}
%\ContinuedFloat

\begin{minipage}{0.24\linewidth}
\includegraphics[width=\linewidth]{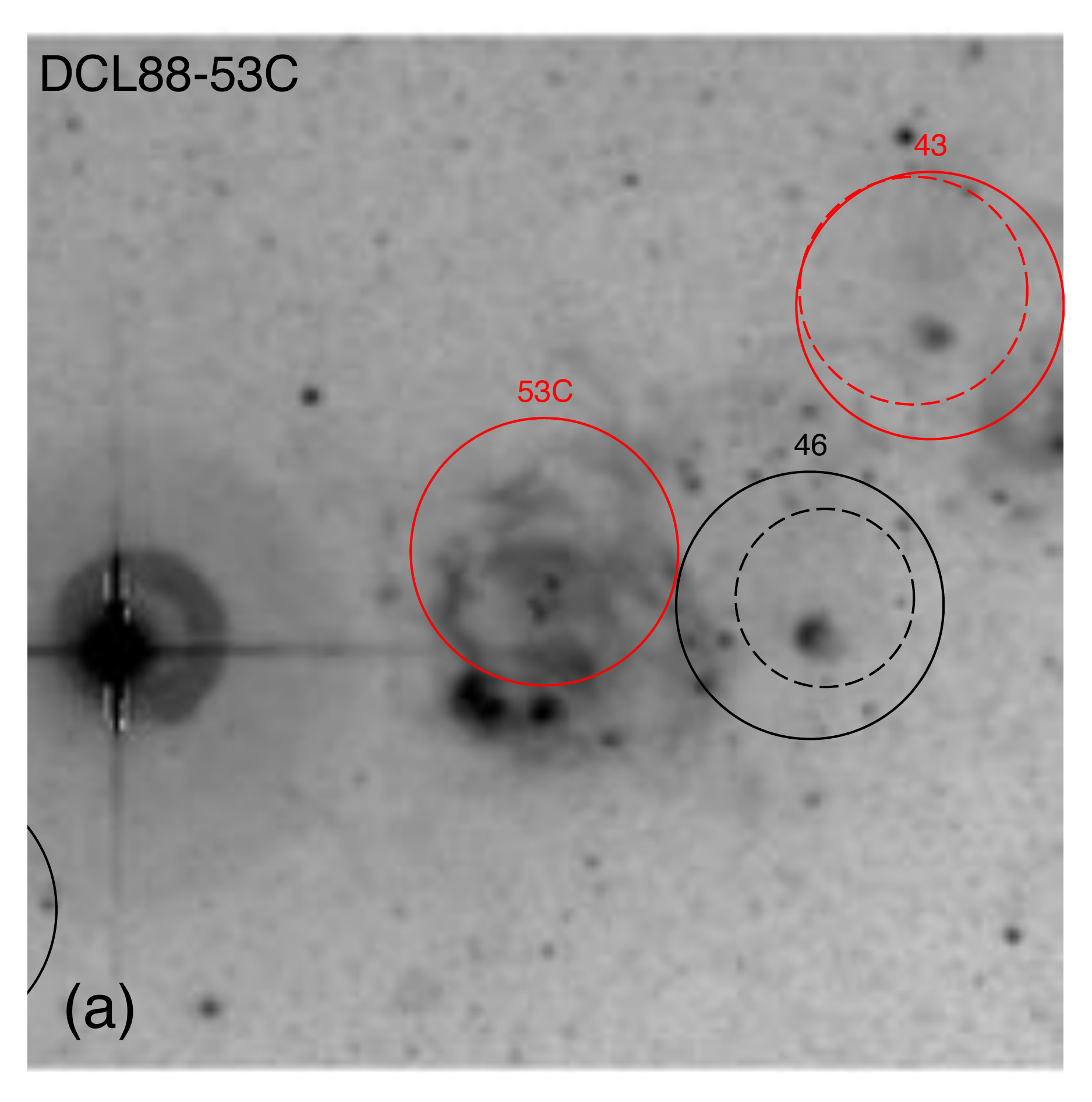}
\end{minipage}
\begin{minipage}{0.24\linewidth}
\includegraphics[width=\linewidth]{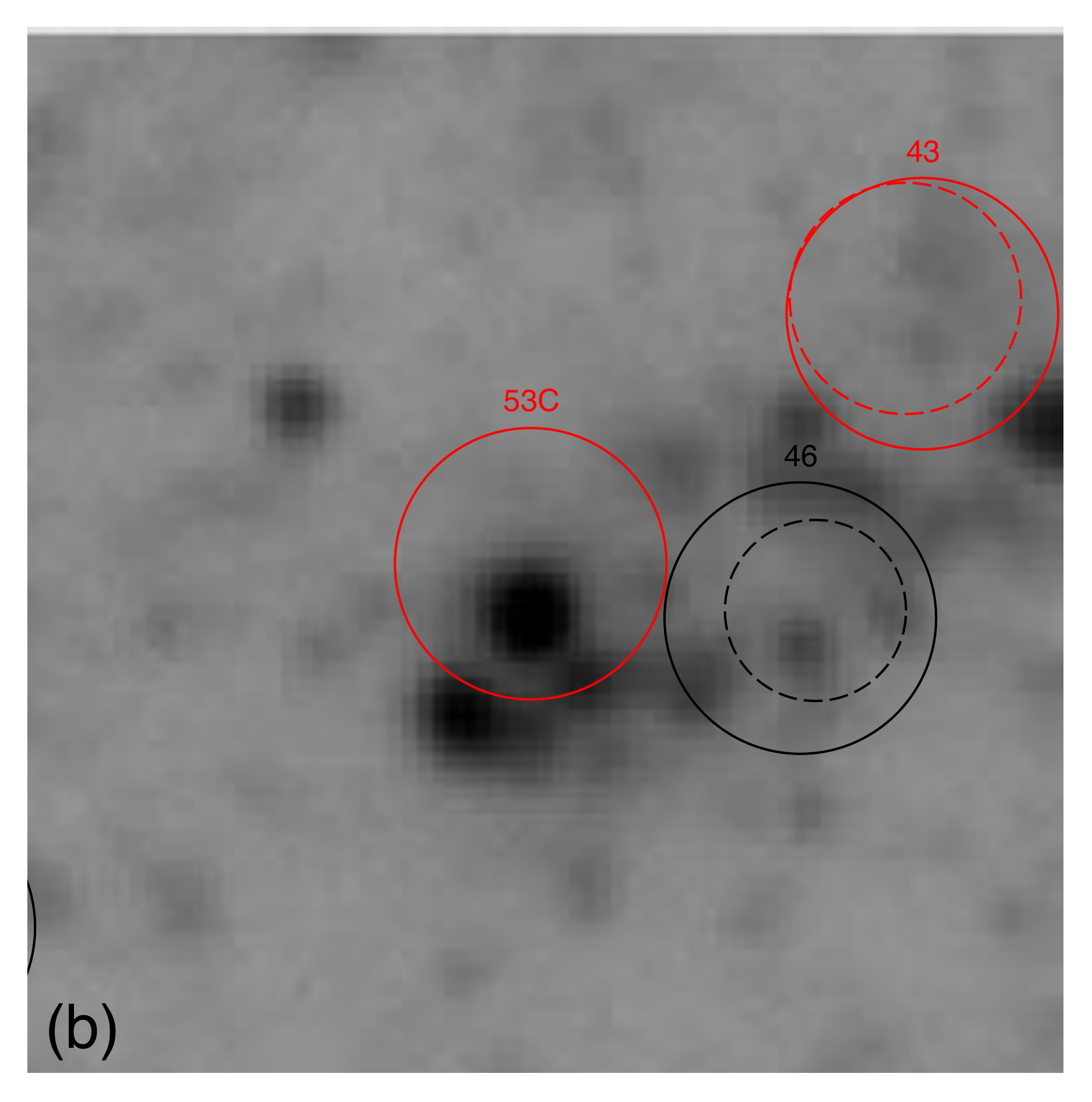}
\end{minipage}
\begin{minipage}{0.24\linewidth}
\includegraphics[width=\linewidth]{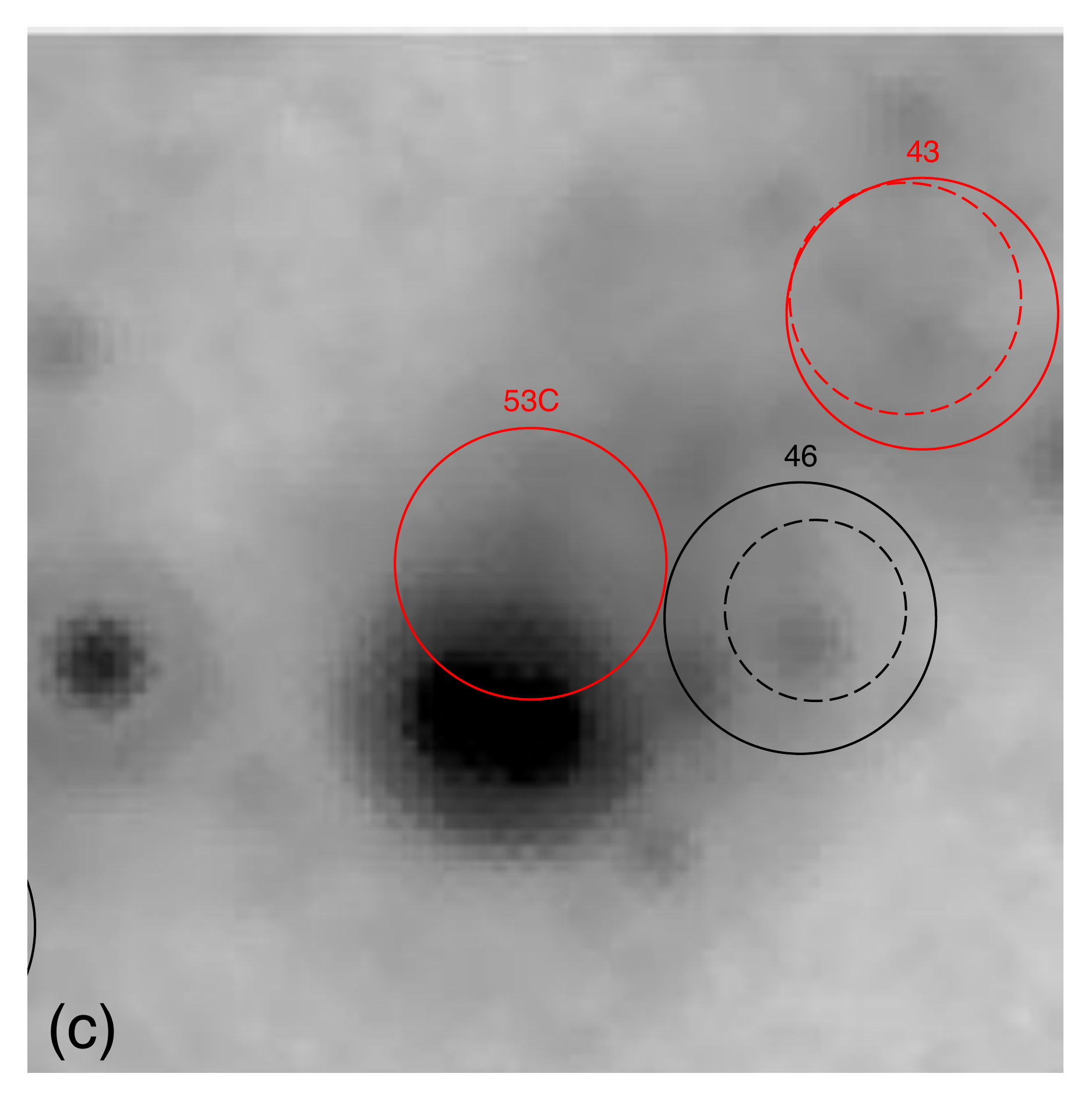}
\end{minipage}
\begin{minipage}{0.24\linewidth}
\includegraphics[width=\linewidth]{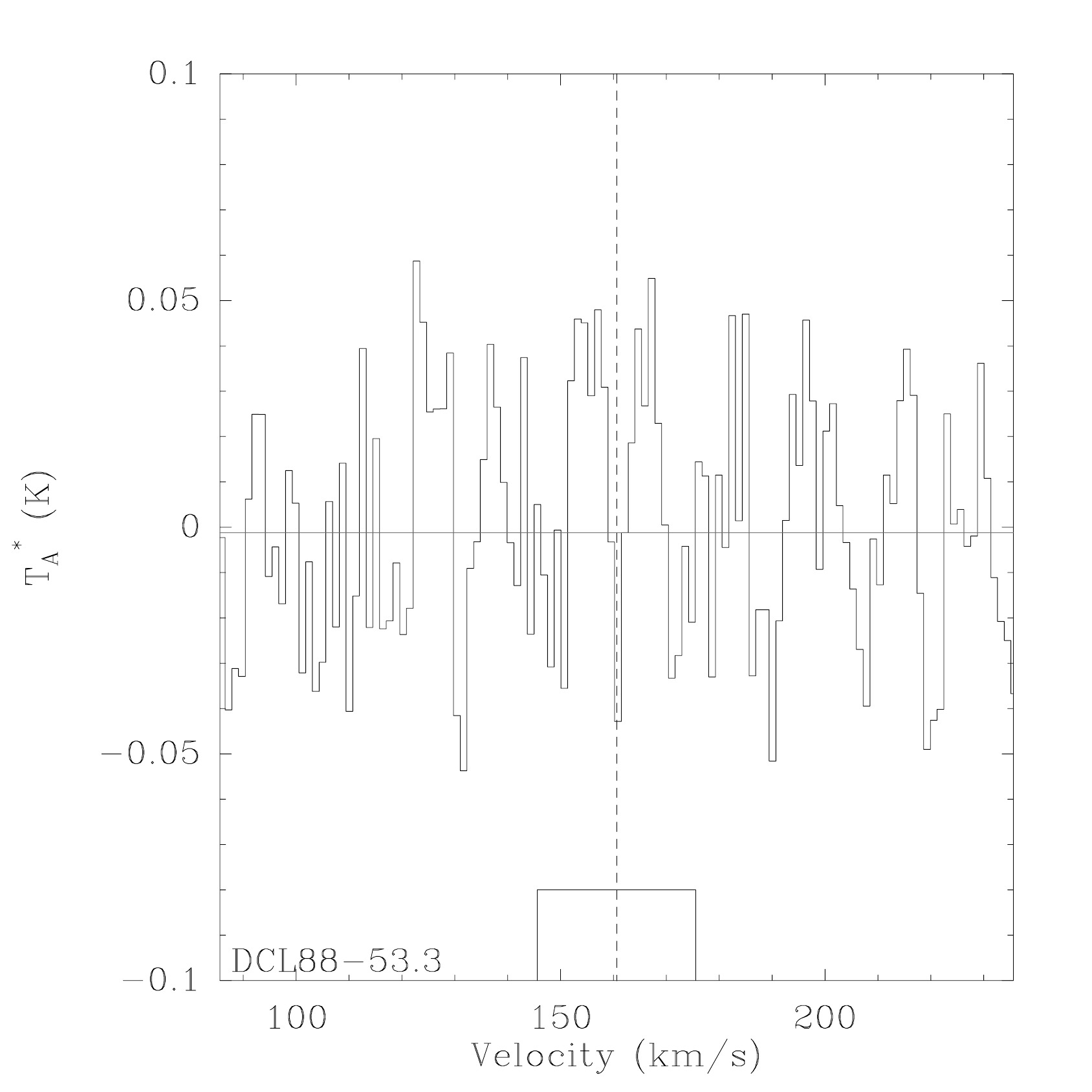}
\end{minipage}

\begin{minipage}{0.24\linewidth}
\includegraphics[width=\linewidth]{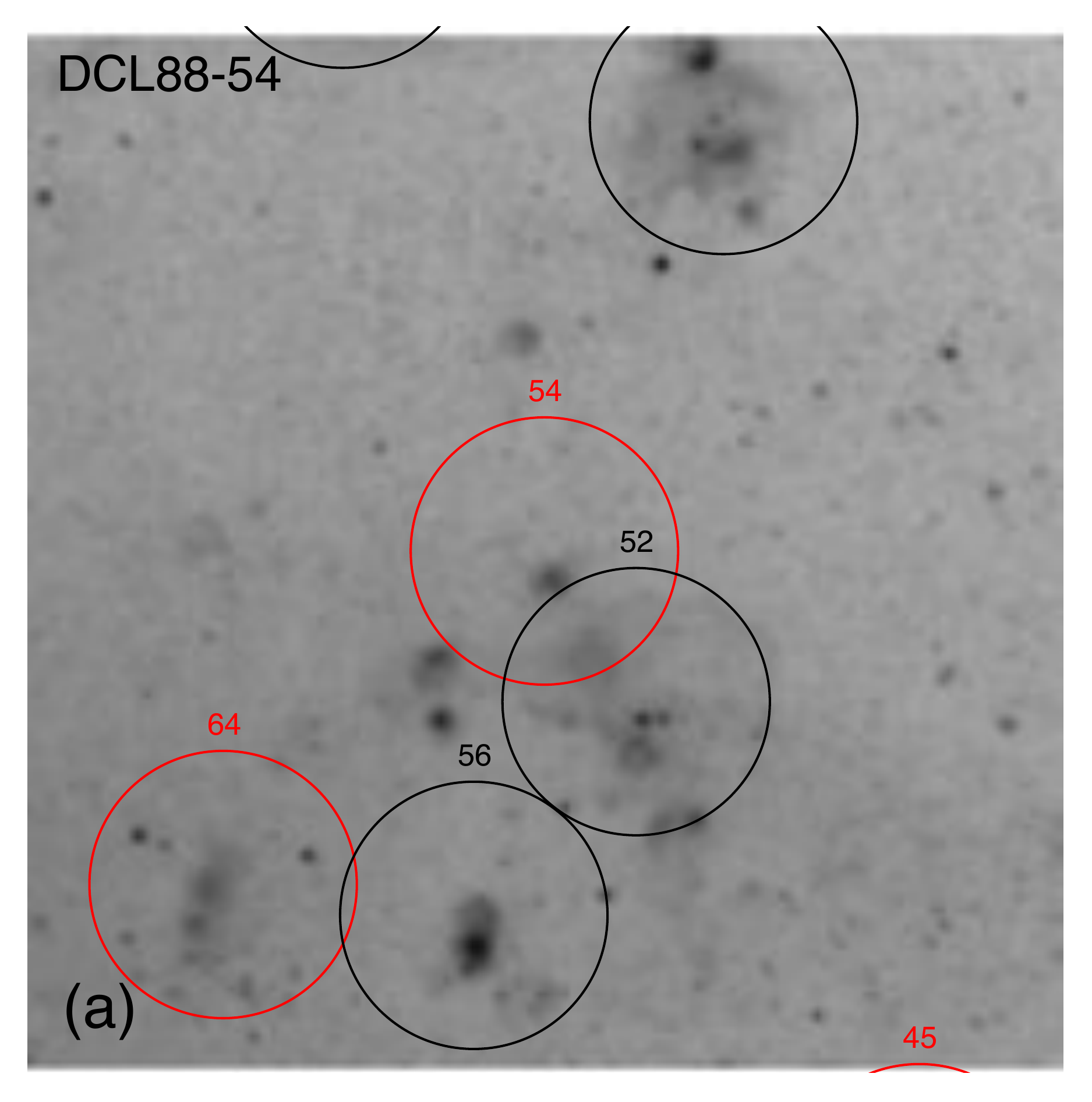}
\end{minipage}
\begin{minipage}{0.24\linewidth}
\includegraphics[width=\linewidth]{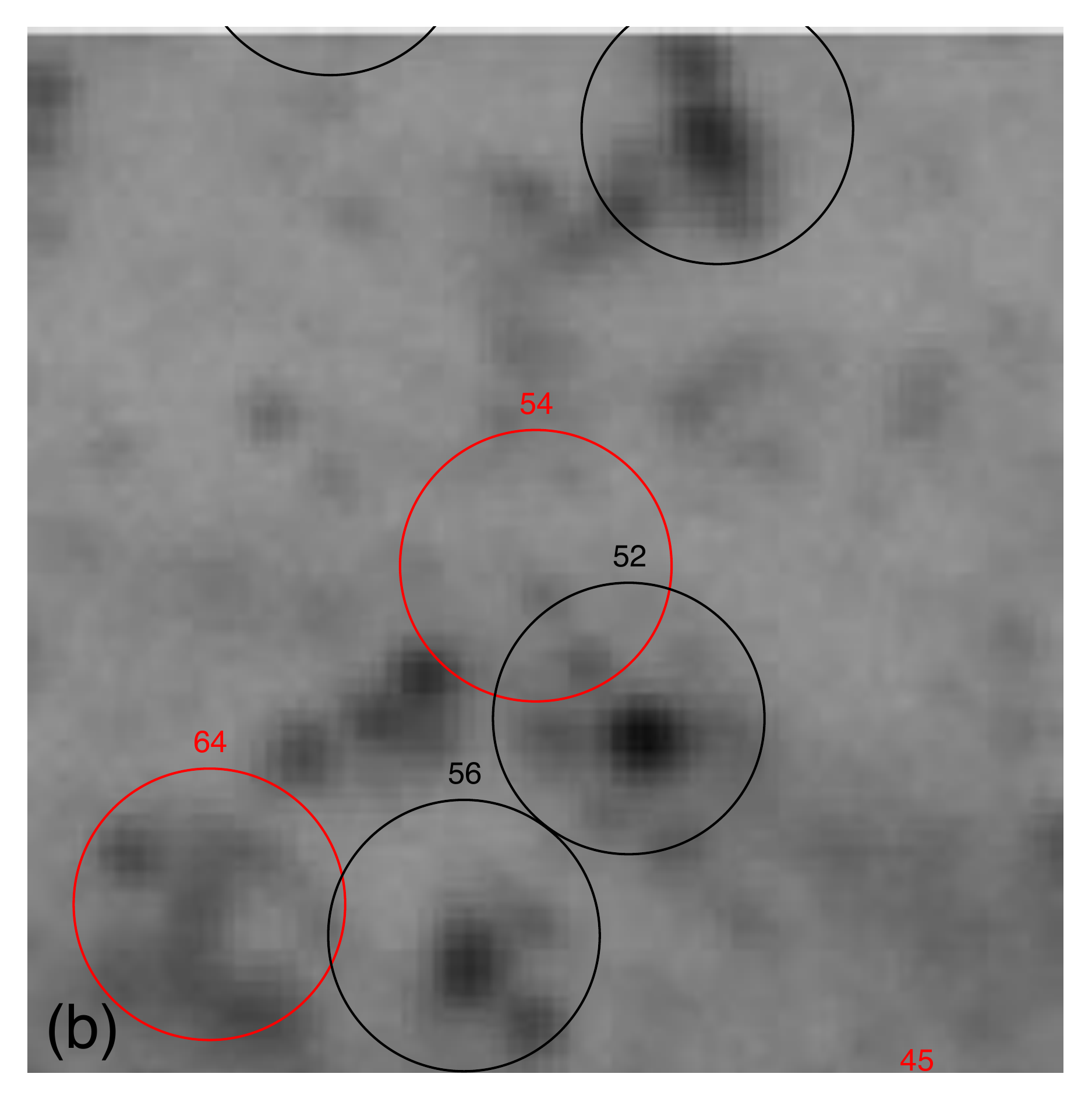}
\end{minipage}
\begin{minipage}{0.24\linewidth}
\includegraphics[width=\linewidth]{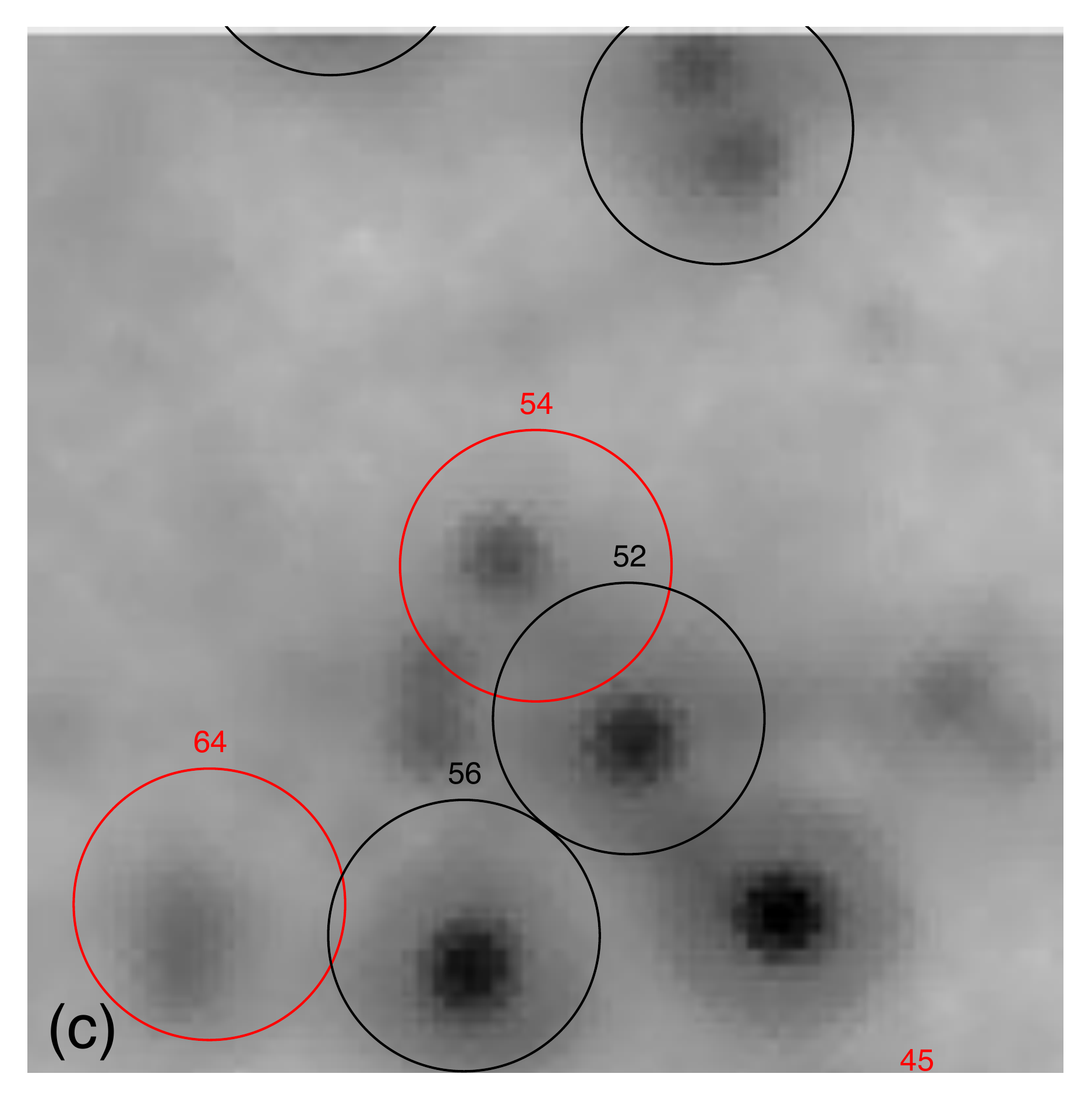}
\end{minipage}
\begin{minipage}{0.24\linewidth}
\includegraphics[width=\linewidth]{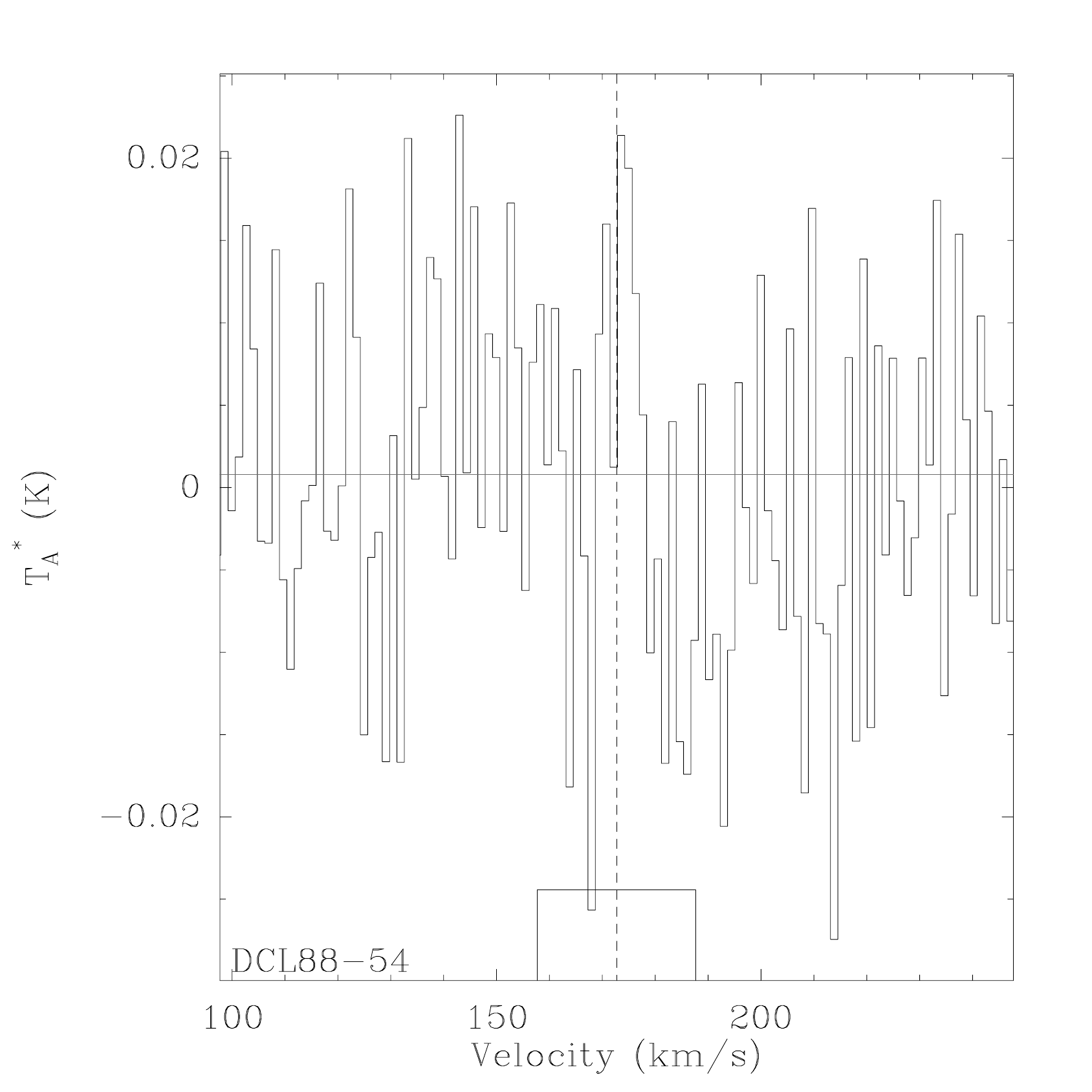}
\end{minipage}

\begin{minipage}{0.24\linewidth}
\includegraphics[width=\linewidth]{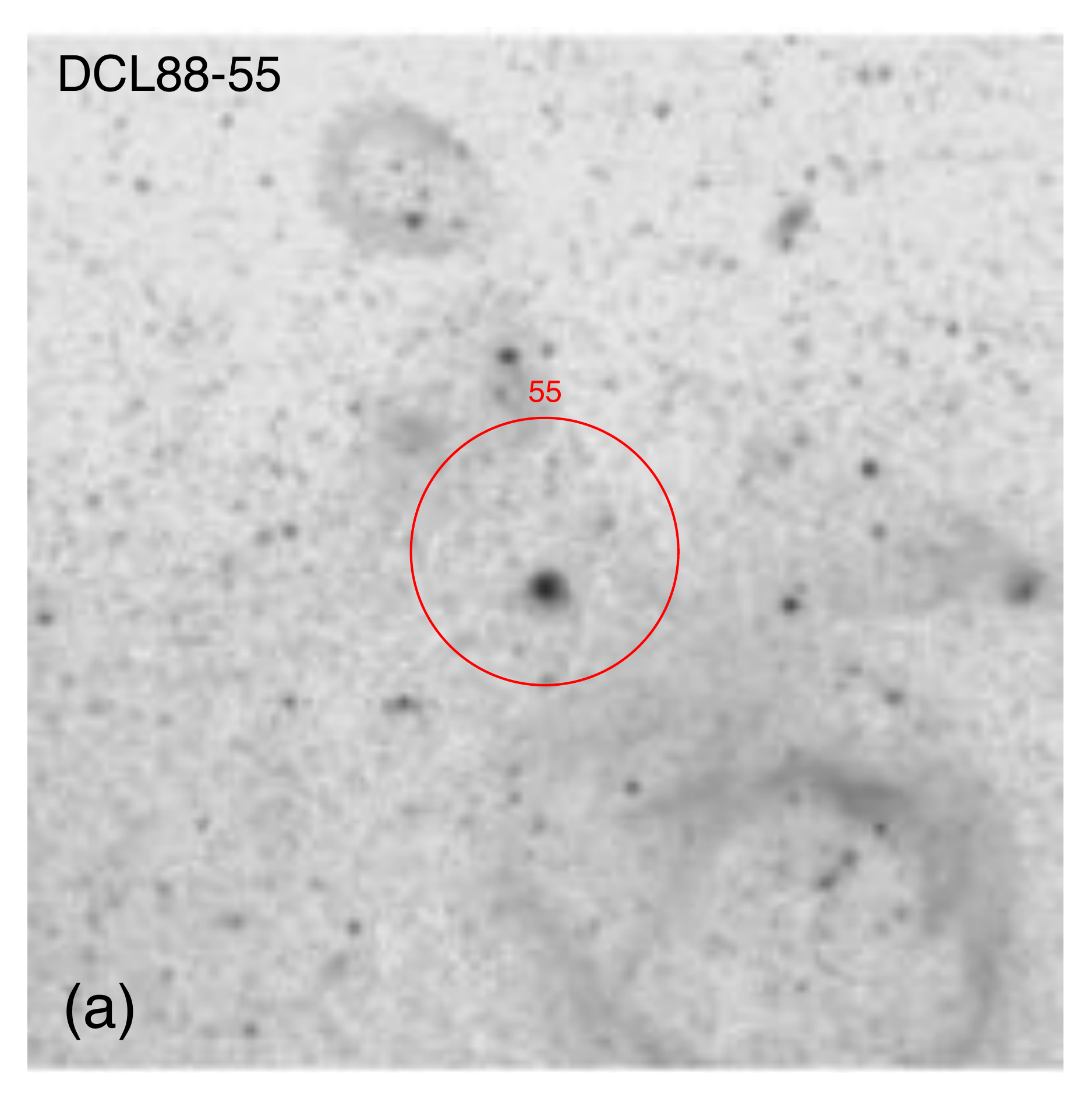}
\end{minipage}
\begin{minipage}{0.24\linewidth}
\includegraphics[width=\linewidth]{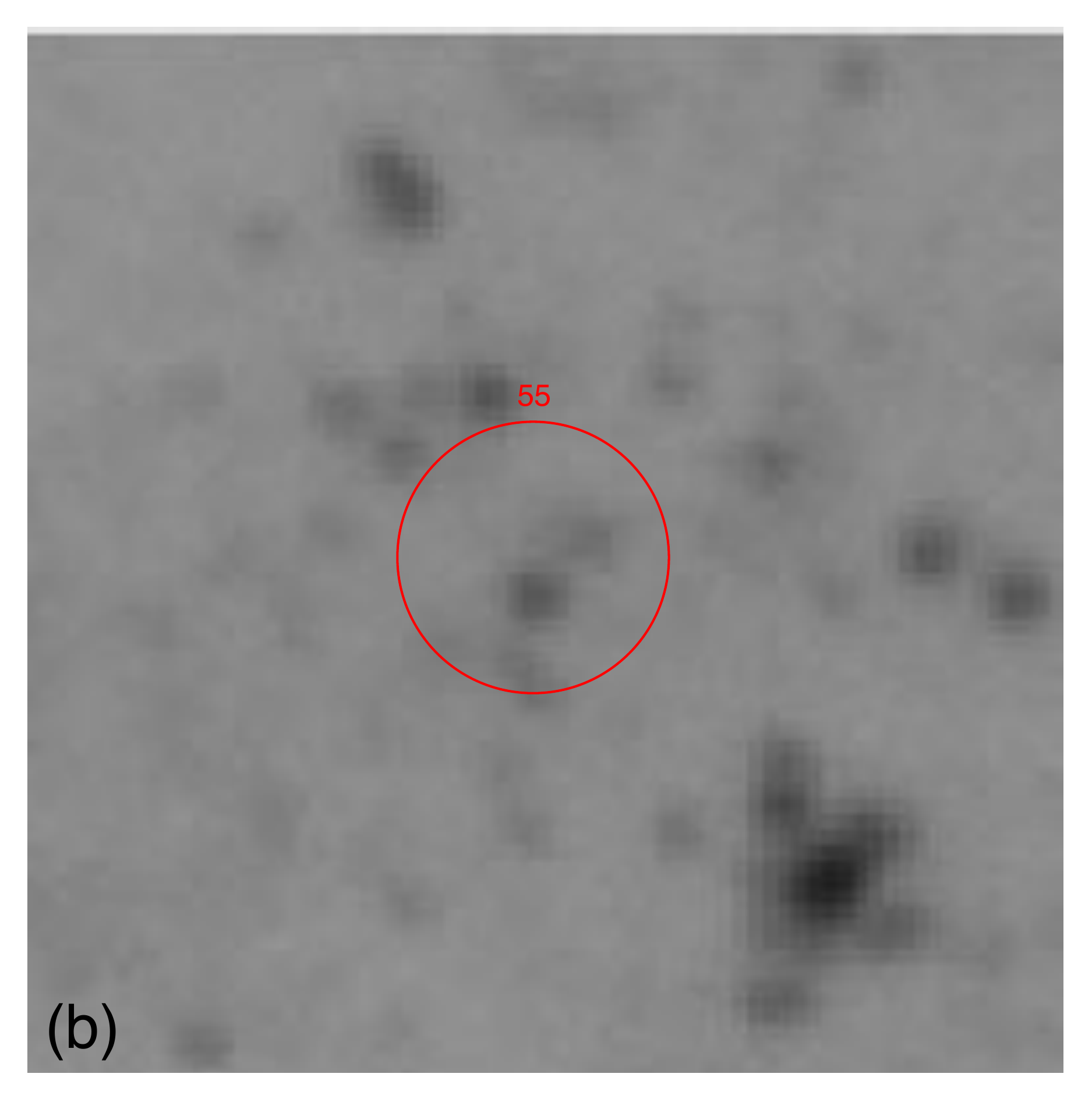}
\end{minipage}
\begin{minipage}{0.24\linewidth}
\includegraphics[width=\linewidth]{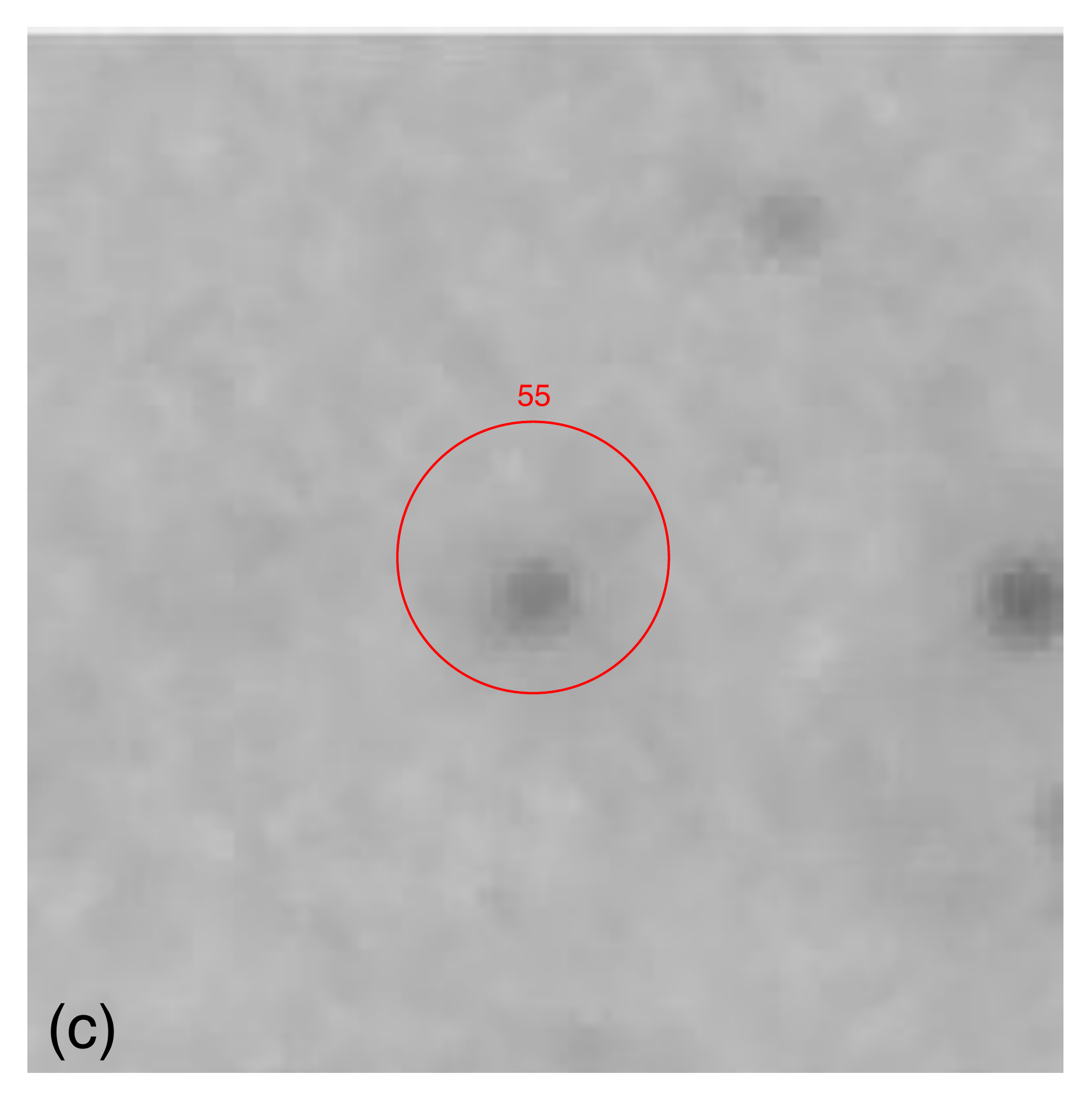}
\end{minipage}
\begin{minipage}{0.24\linewidth}
\includegraphics[width=\linewidth]{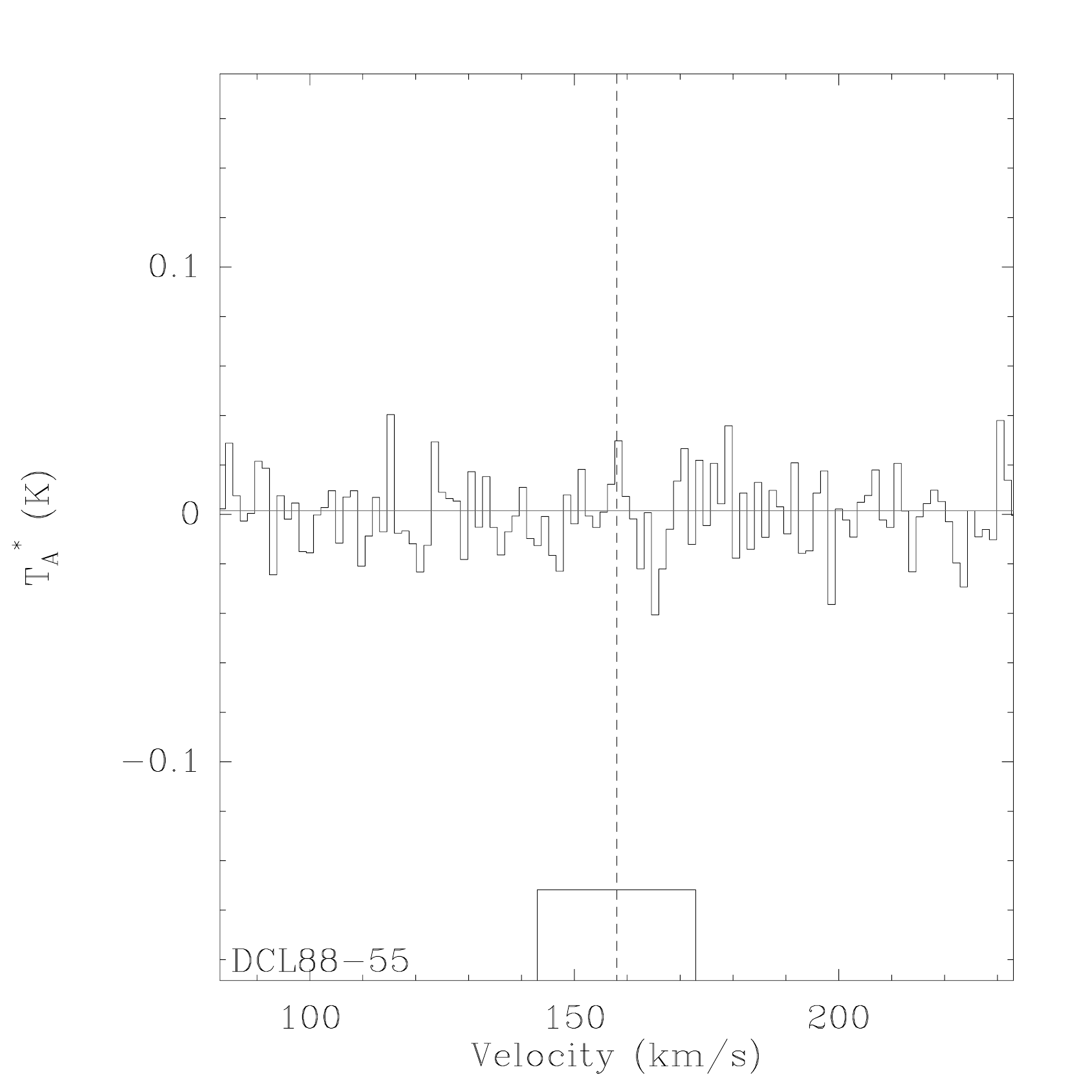}
\end{minipage}

\begin{minipage}{0.24\linewidth}
\includegraphics[width=\linewidth]{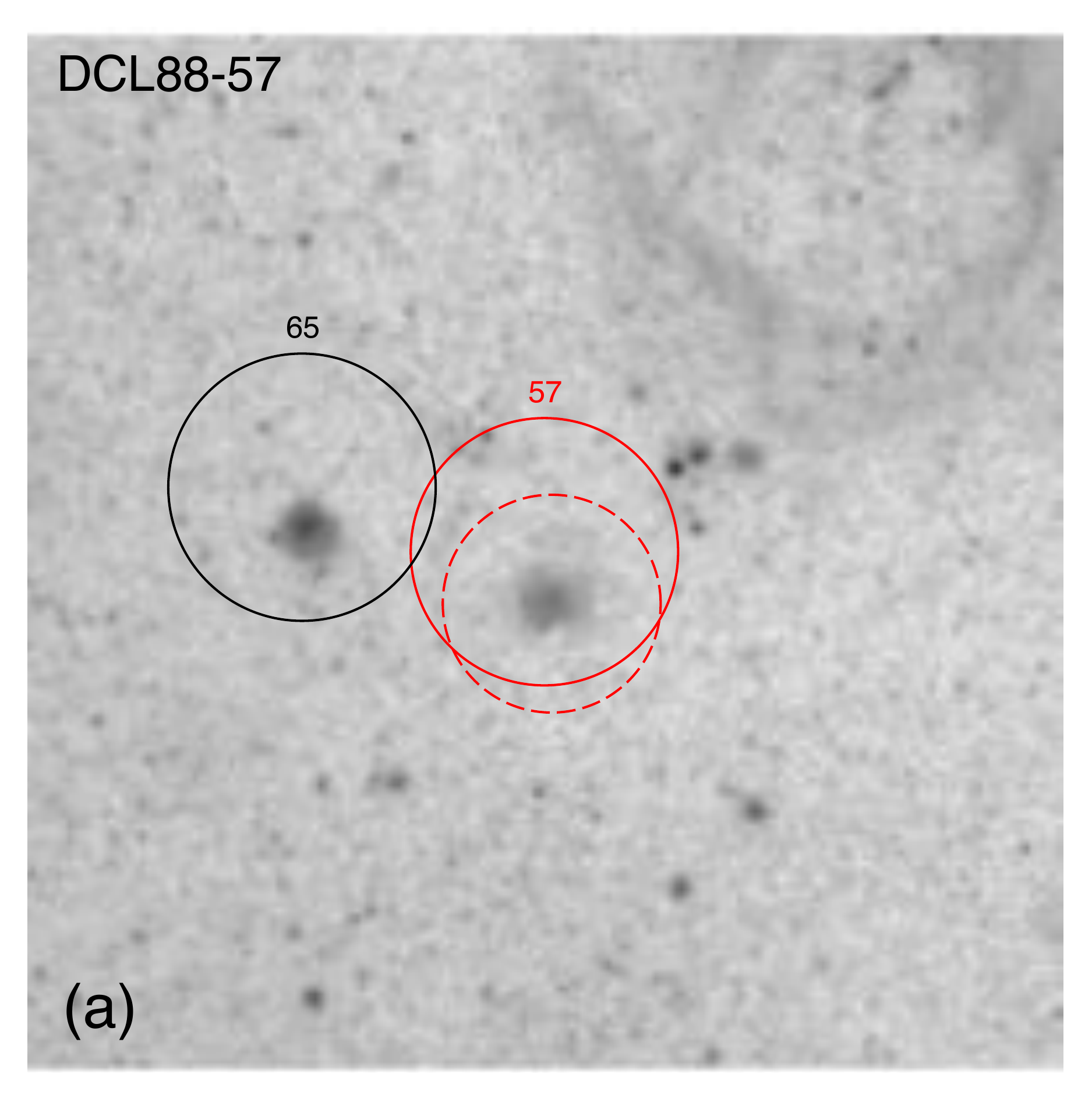}
\end{minipage}
\begin{minipage}{0.24\linewidth}
\includegraphics[width=\linewidth]{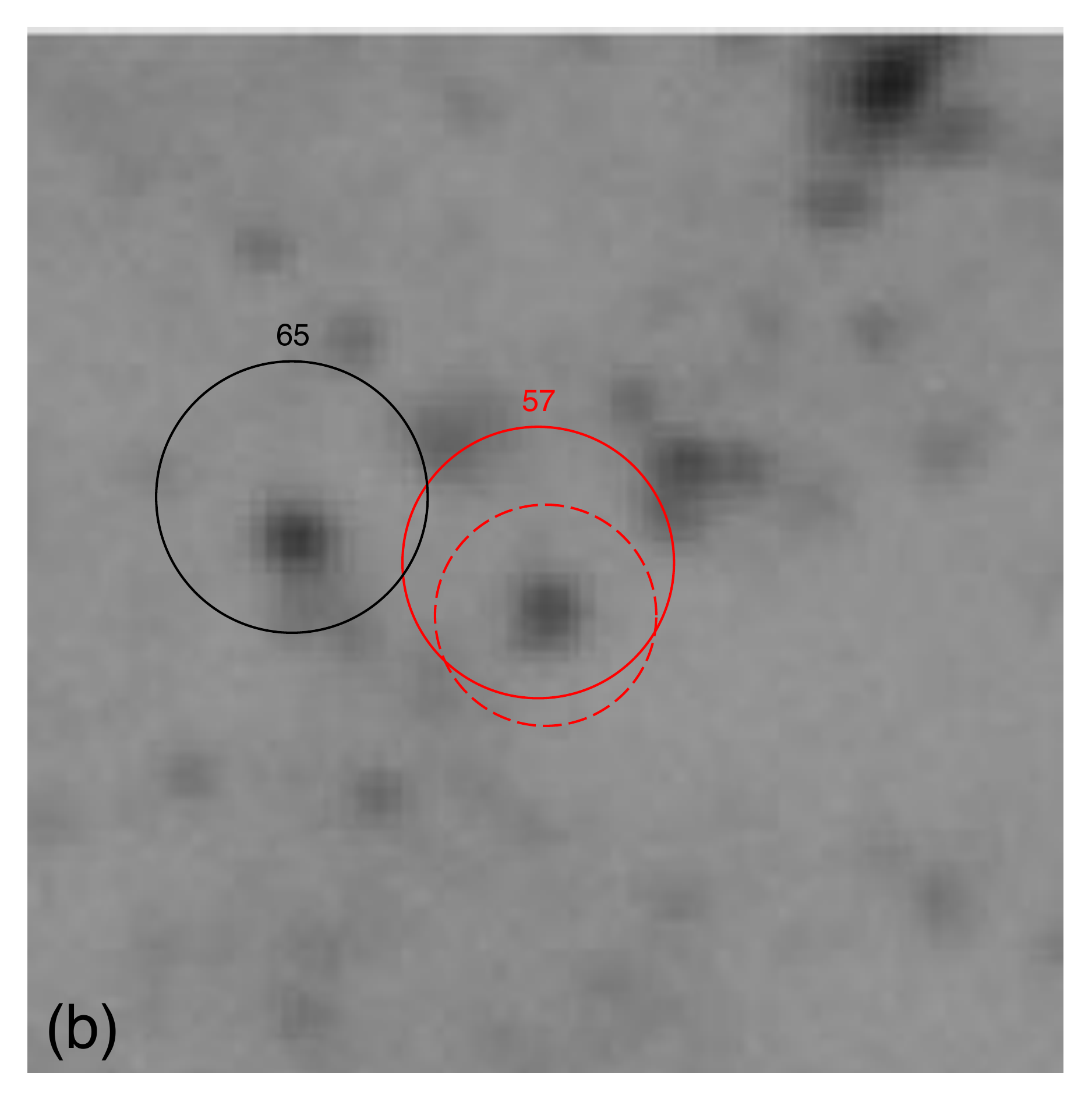}
\end{minipage}
\begin{minipage}{0.24\linewidth}
\includegraphics[width=\linewidth]{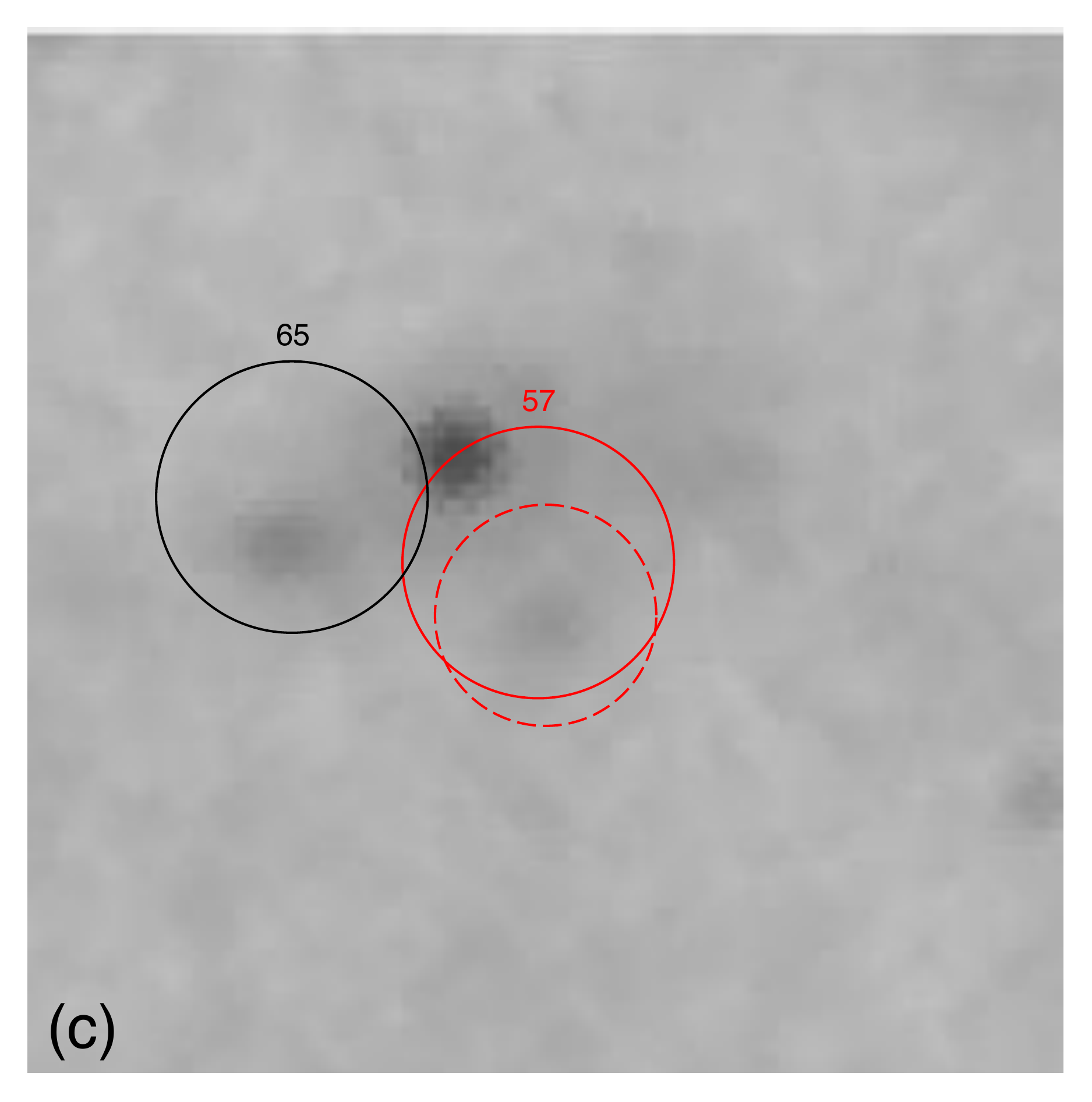}
\end{minipage}
\begin{minipage}{0.24\linewidth}
\includegraphics[width=\linewidth]{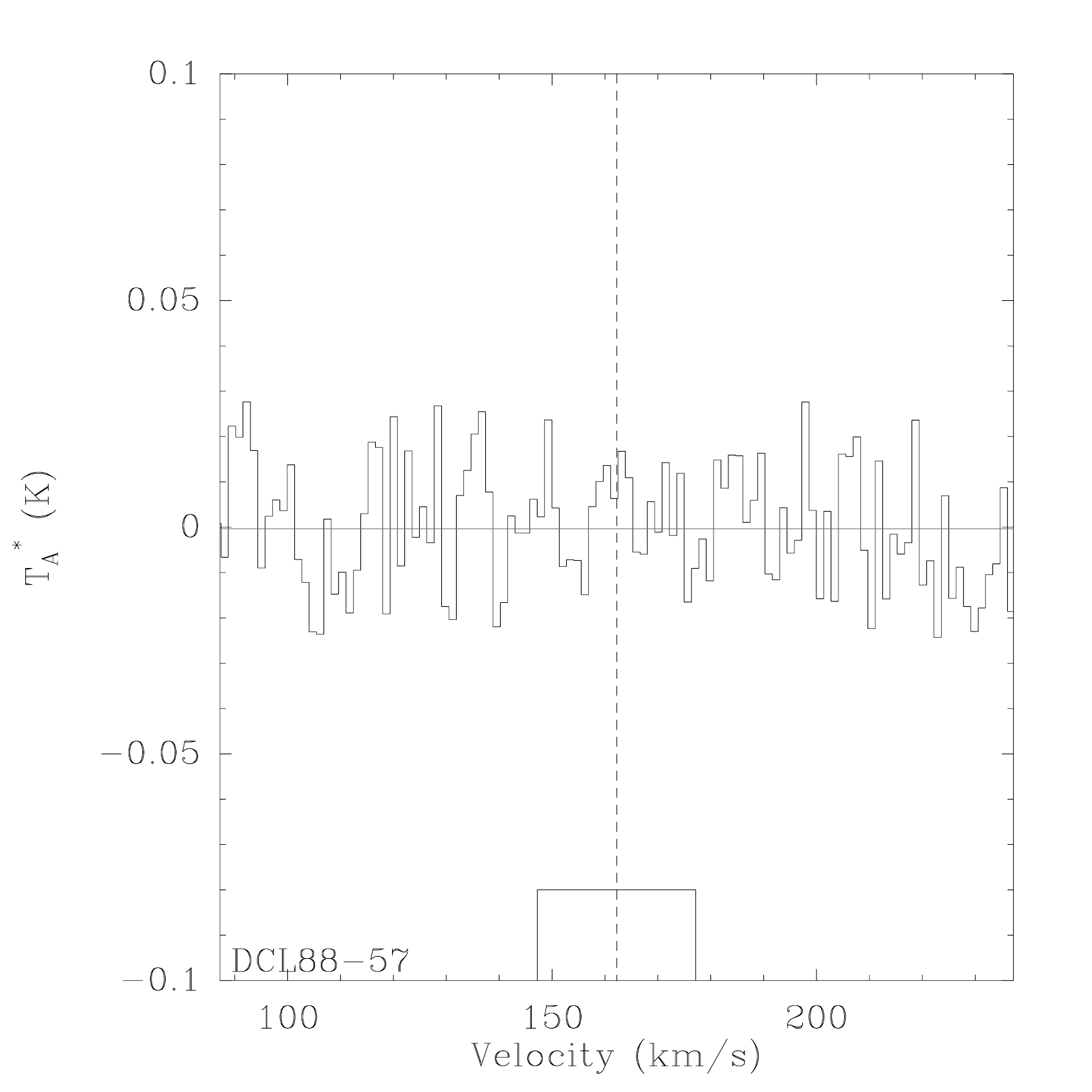}
\end{minipage}

\begin{minipage}{0.24\linewidth}
\includegraphics[width=\linewidth]{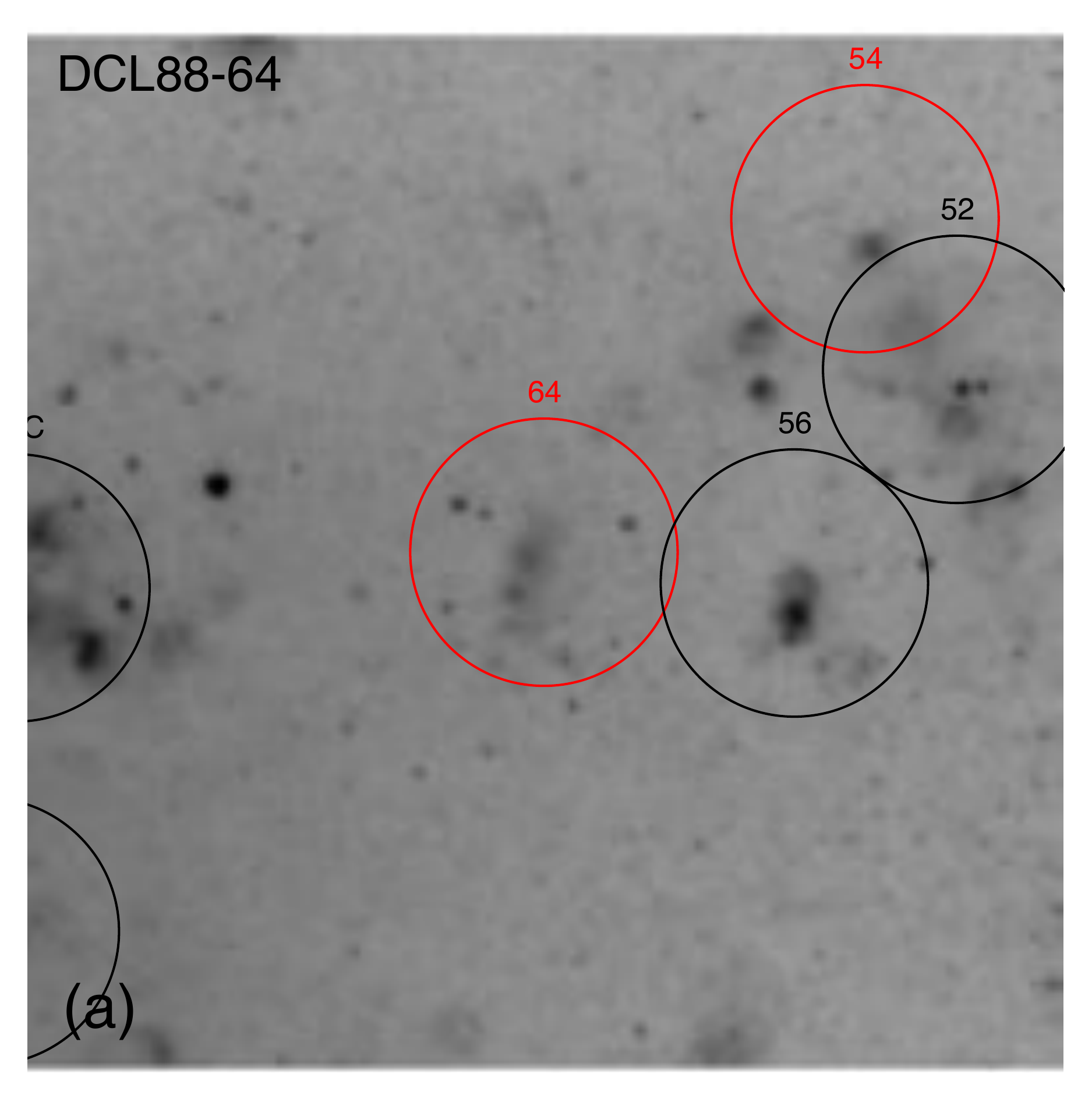}
\end{minipage}
\begin{minipage}{0.24\linewidth}
\includegraphics[width=\linewidth]{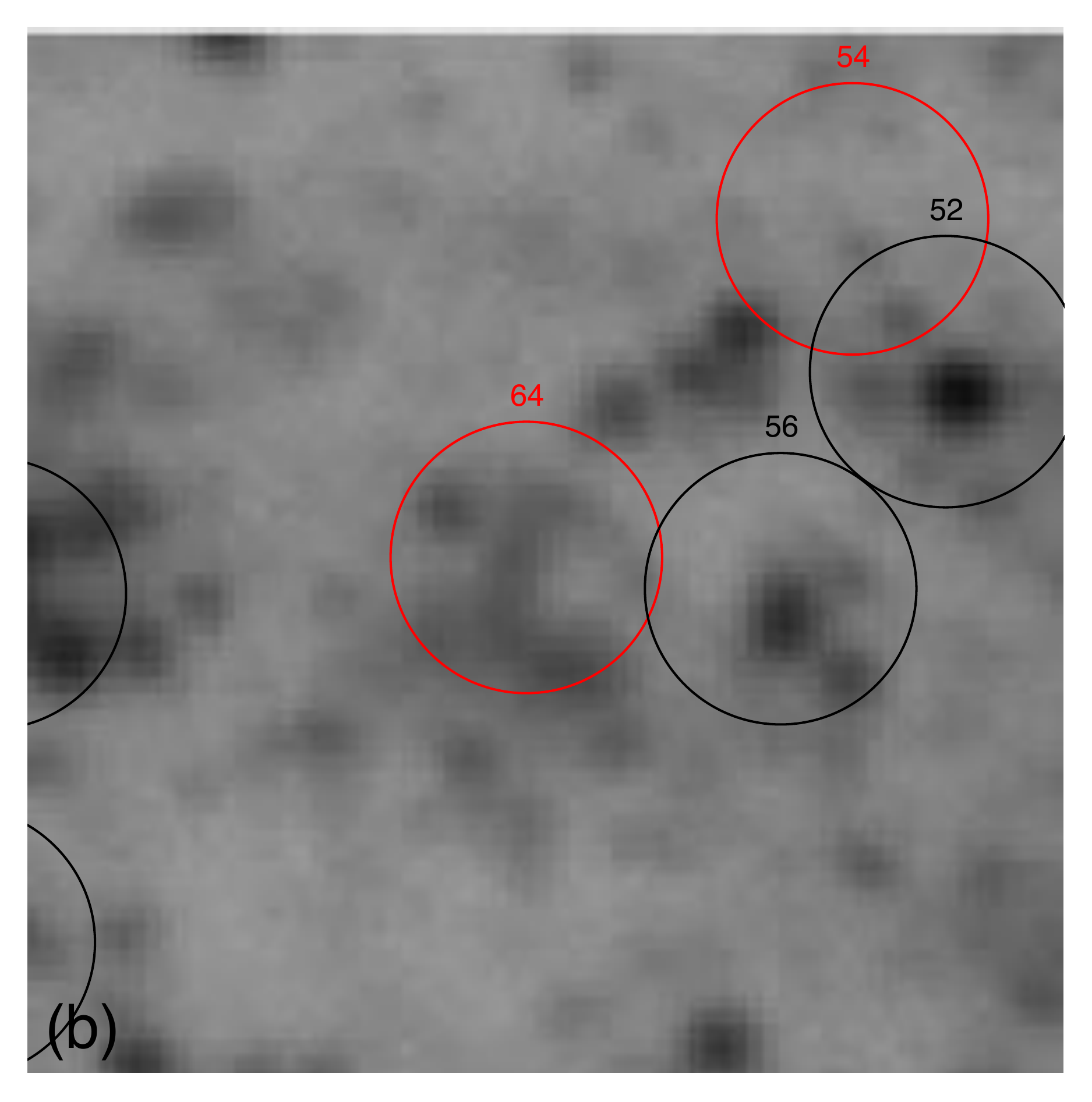}
\end{minipage}
\begin{minipage}{0.24\linewidth}
\includegraphics[width=\linewidth]{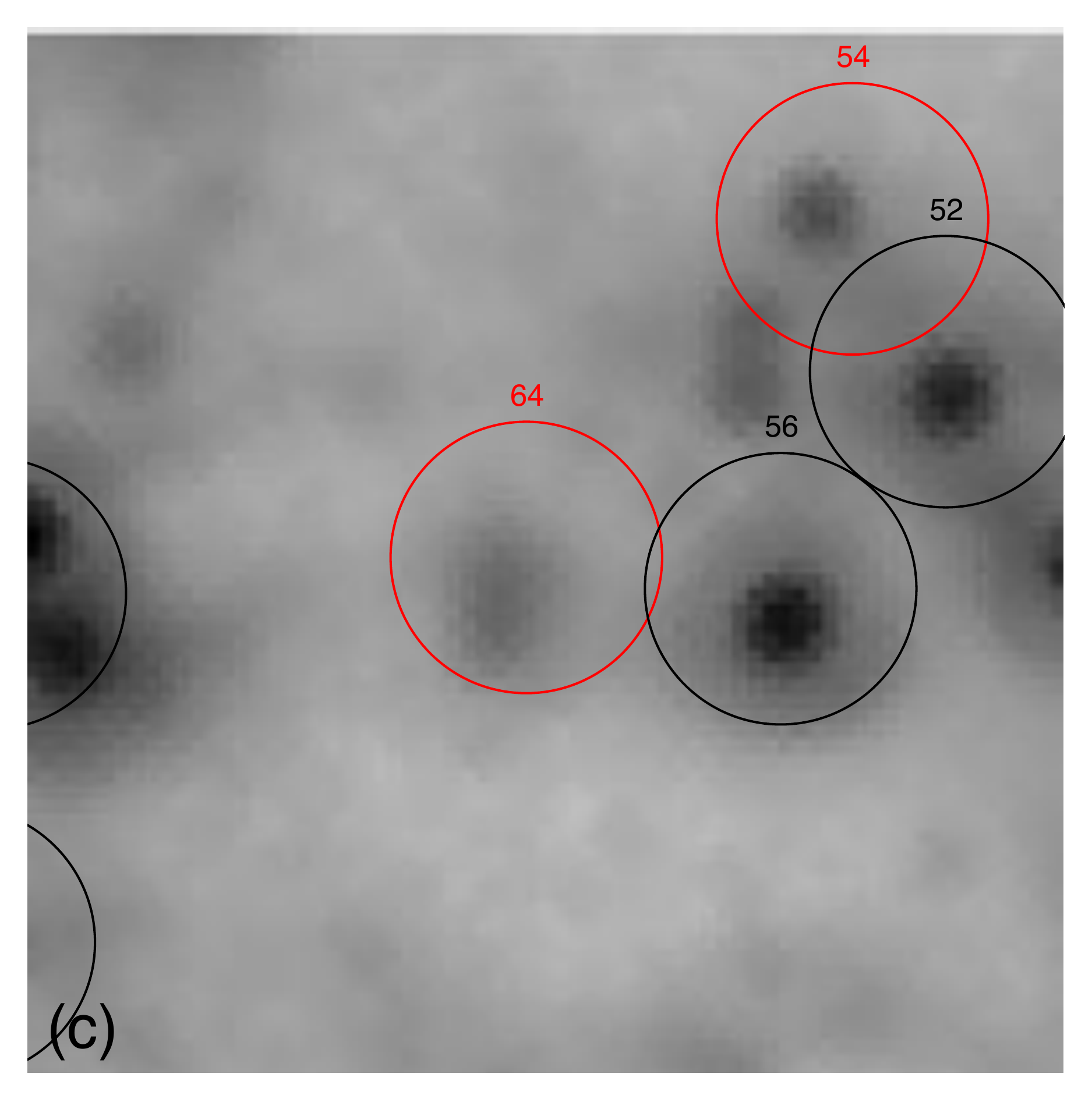}
\end{minipage}
\begin{minipage}{0.24\linewidth}
\includegraphics[width=\linewidth]{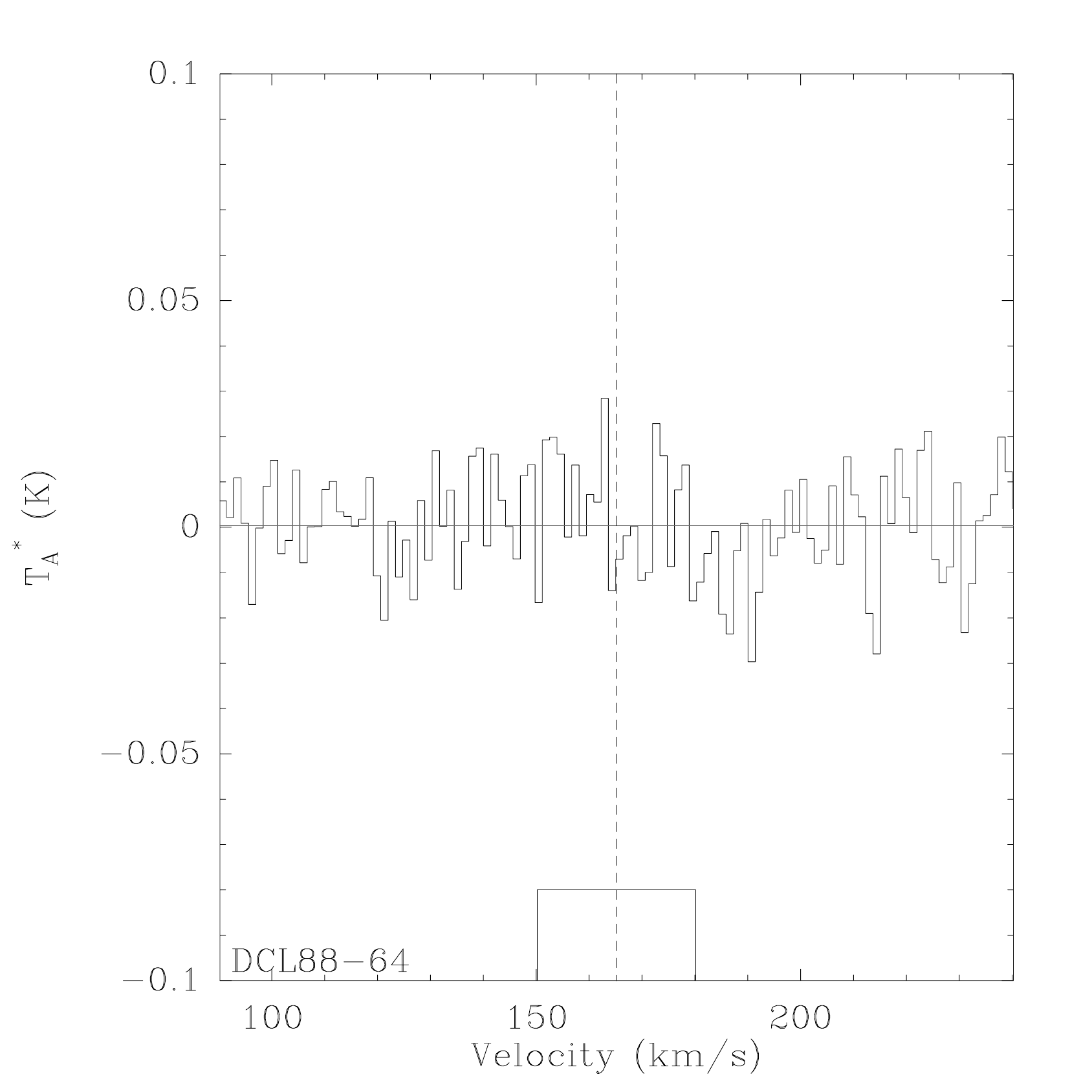}
\end{minipage}

\noindent\textbf{Figure~\ref{fig:stamps} -- continued.}

\end{figure*}

\begin{figure*}
%\ContinuedFloat

\begin{minipage}{0.24\linewidth}
\includegraphics[width=\linewidth]{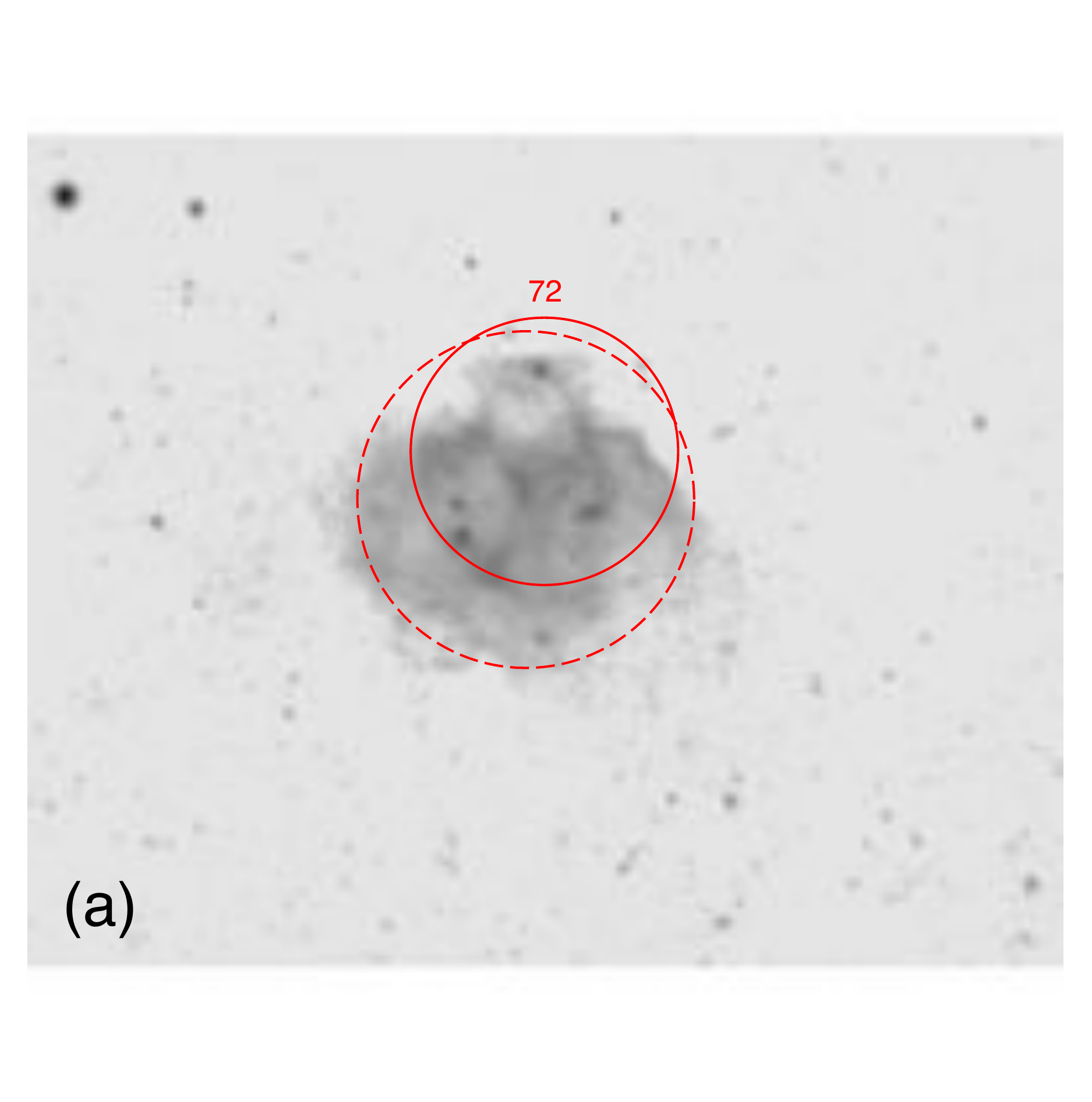}
\end{minipage}
\begin{minipage}{0.24\linewidth}
\includegraphics[width=\linewidth]{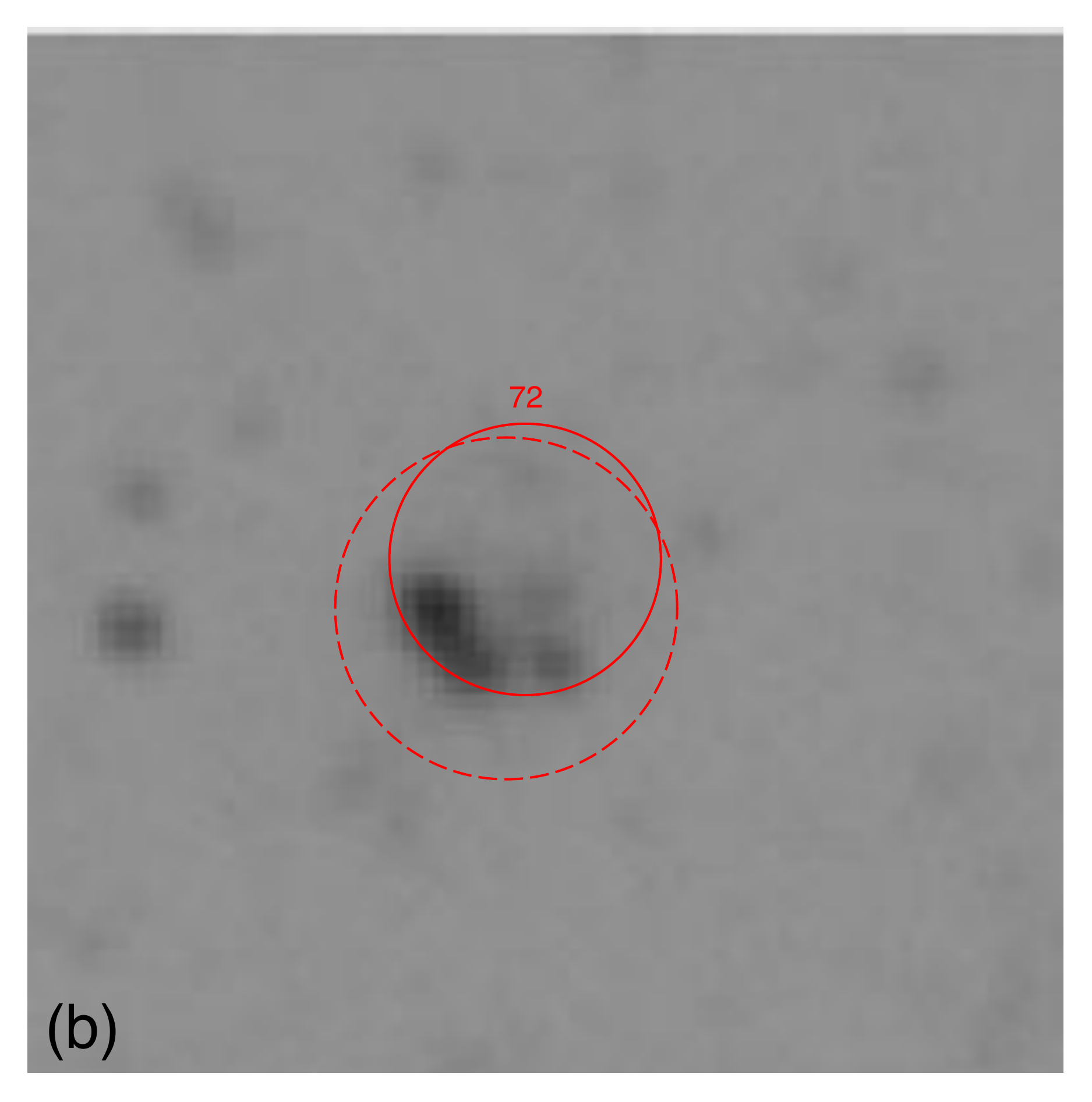}
\end{minipage}
\begin{minipage}{0.24\linewidth}
\includegraphics[width=\linewidth]{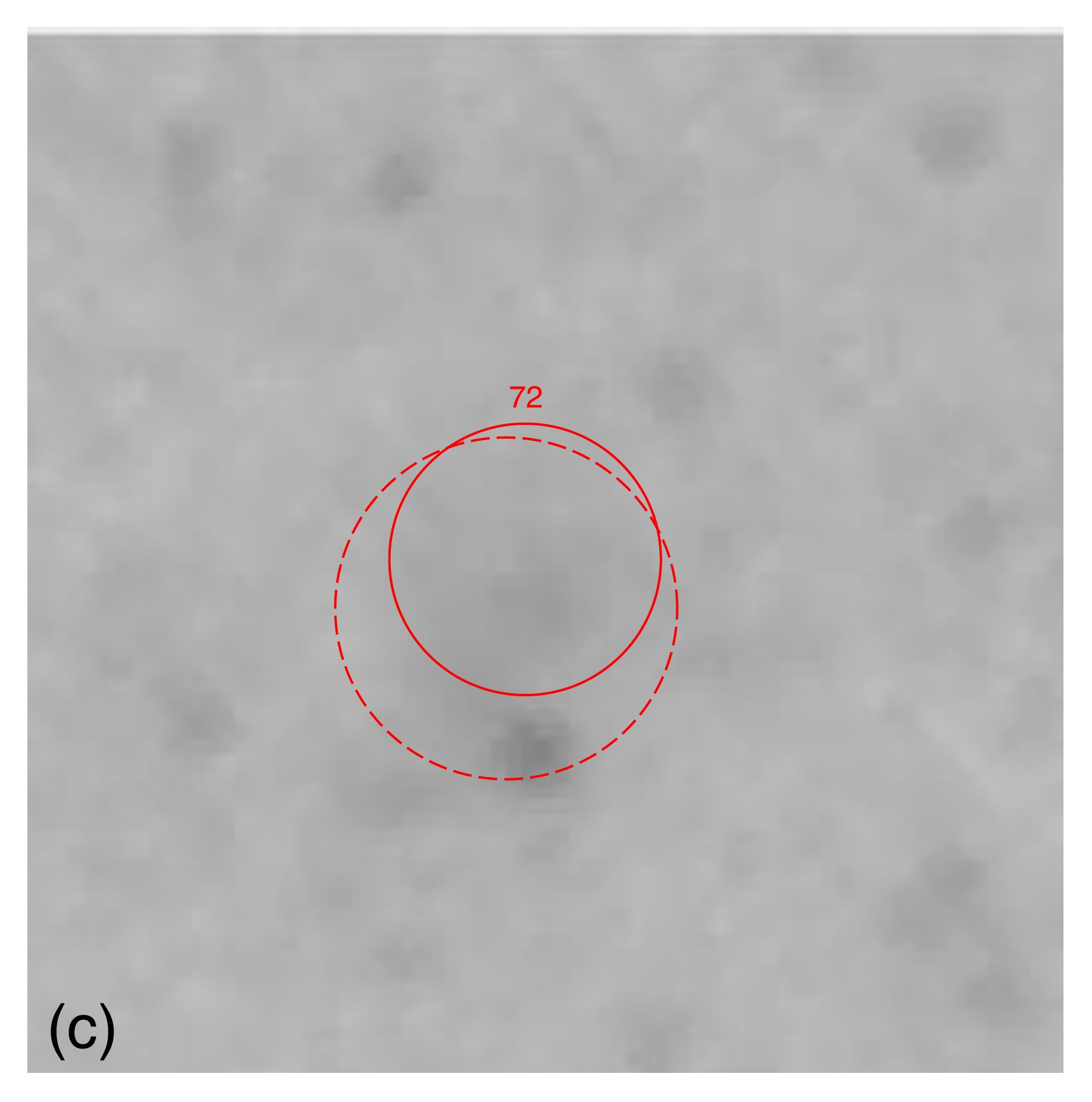}
\end{minipage}
\begin{minipage}{0.24\linewidth}
\includegraphics[width=\linewidth]{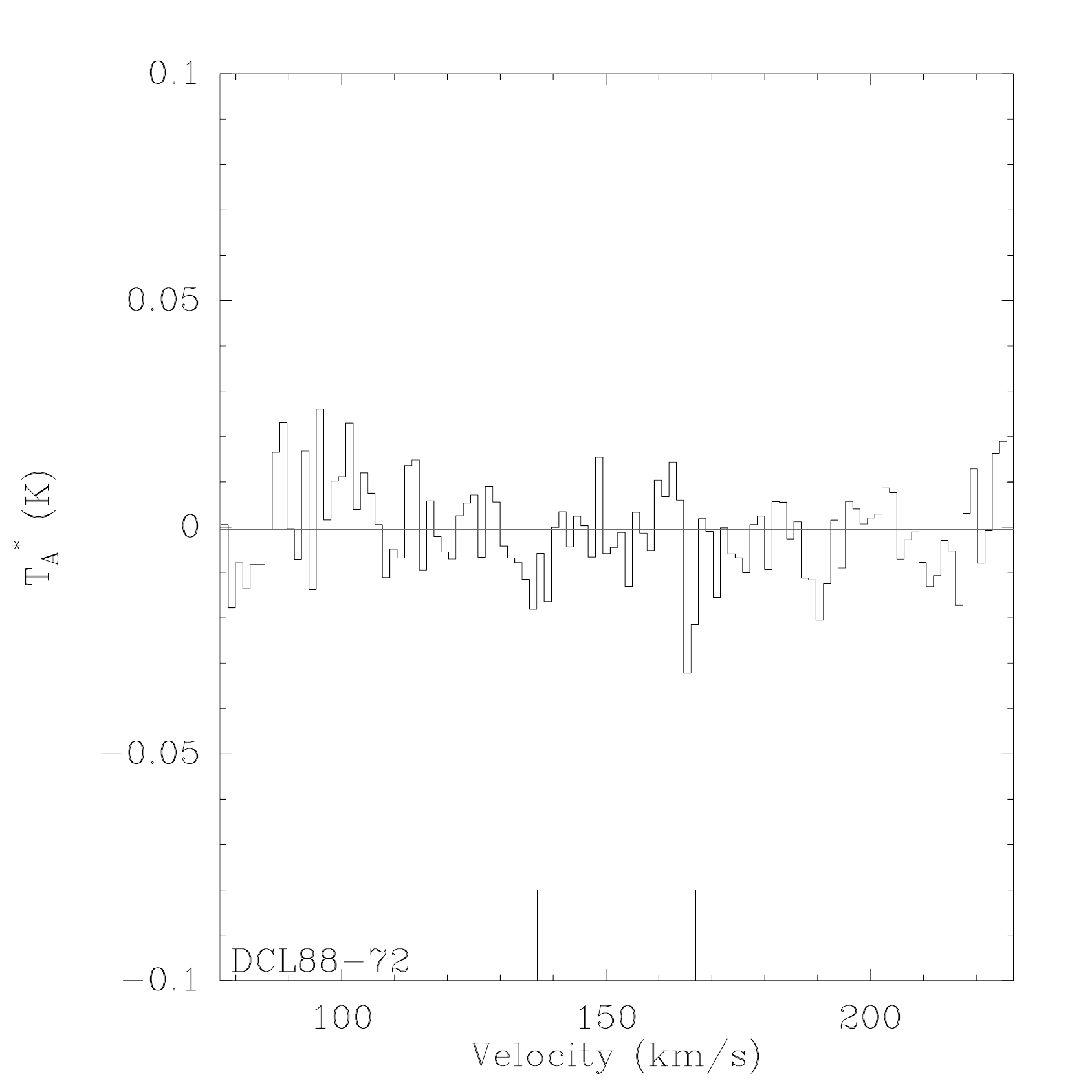}
\end{minipage}

\begin{minipage}{0.24\linewidth}
\includegraphics[width=\linewidth]{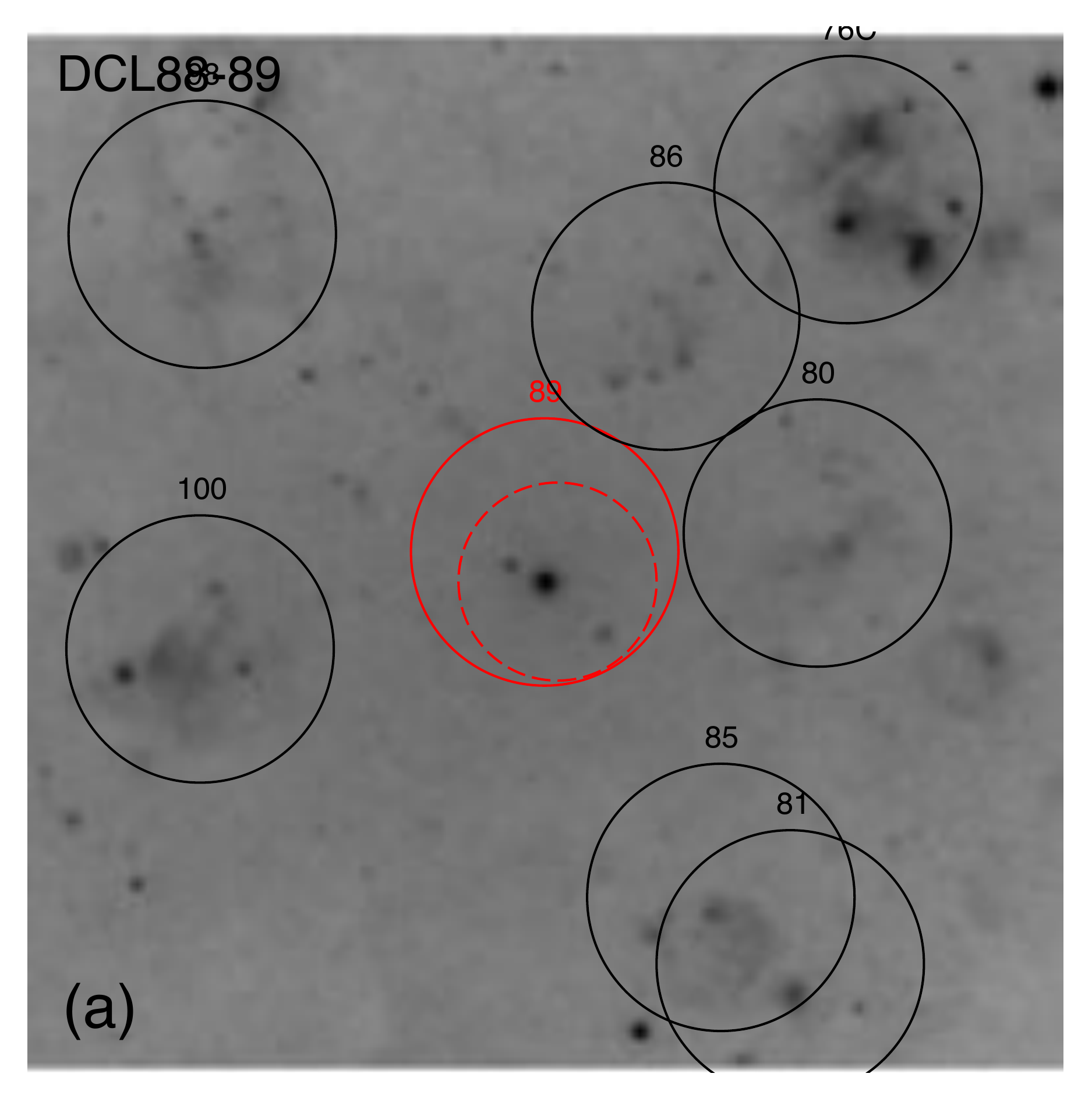}
\end{minipage}
\begin{minipage}{0.24\linewidth}
\includegraphics[width=\linewidth]{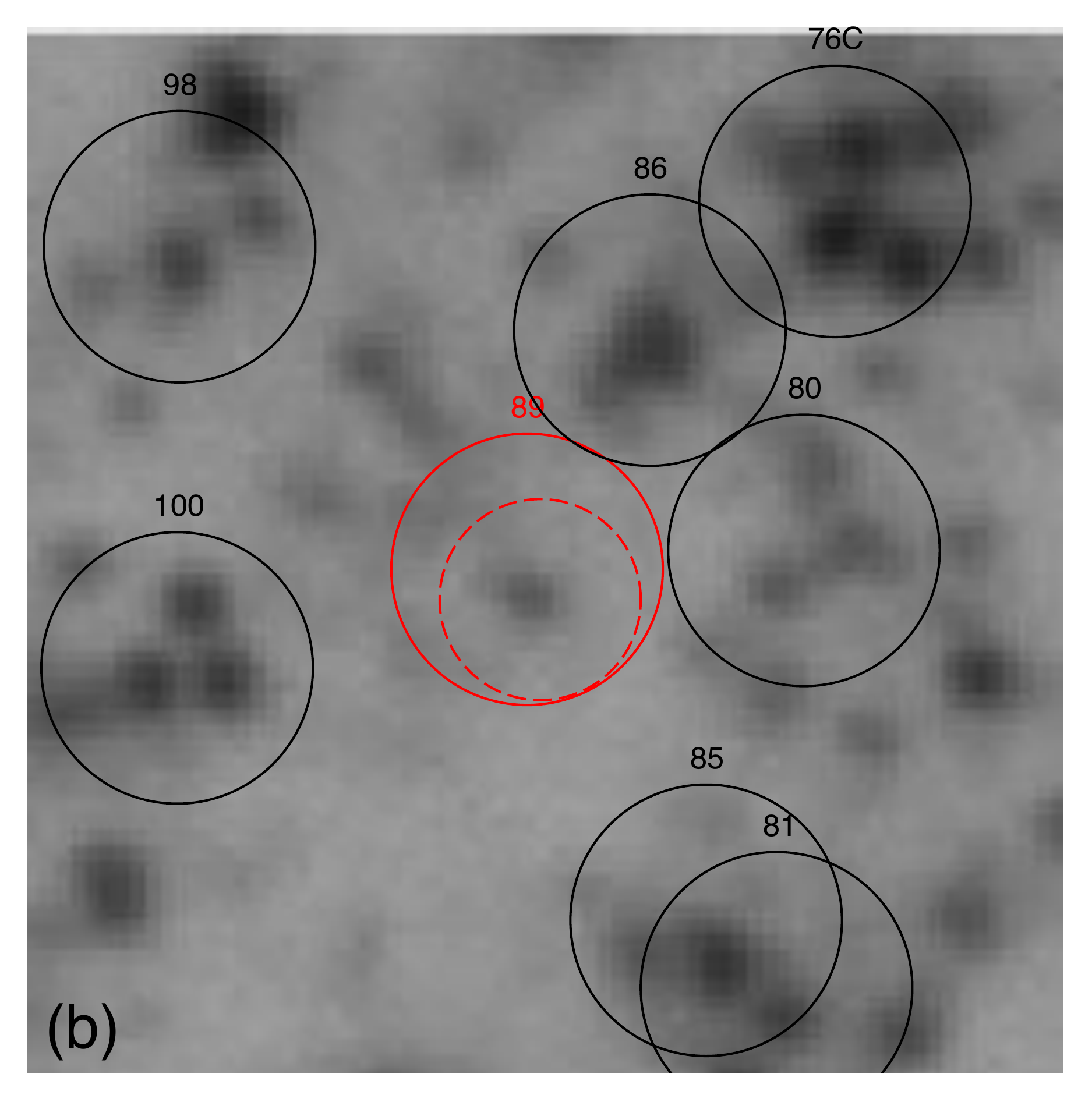}
\end{minipage}
\begin{minipage}{0.24\linewidth}
\includegraphics[width=\linewidth]{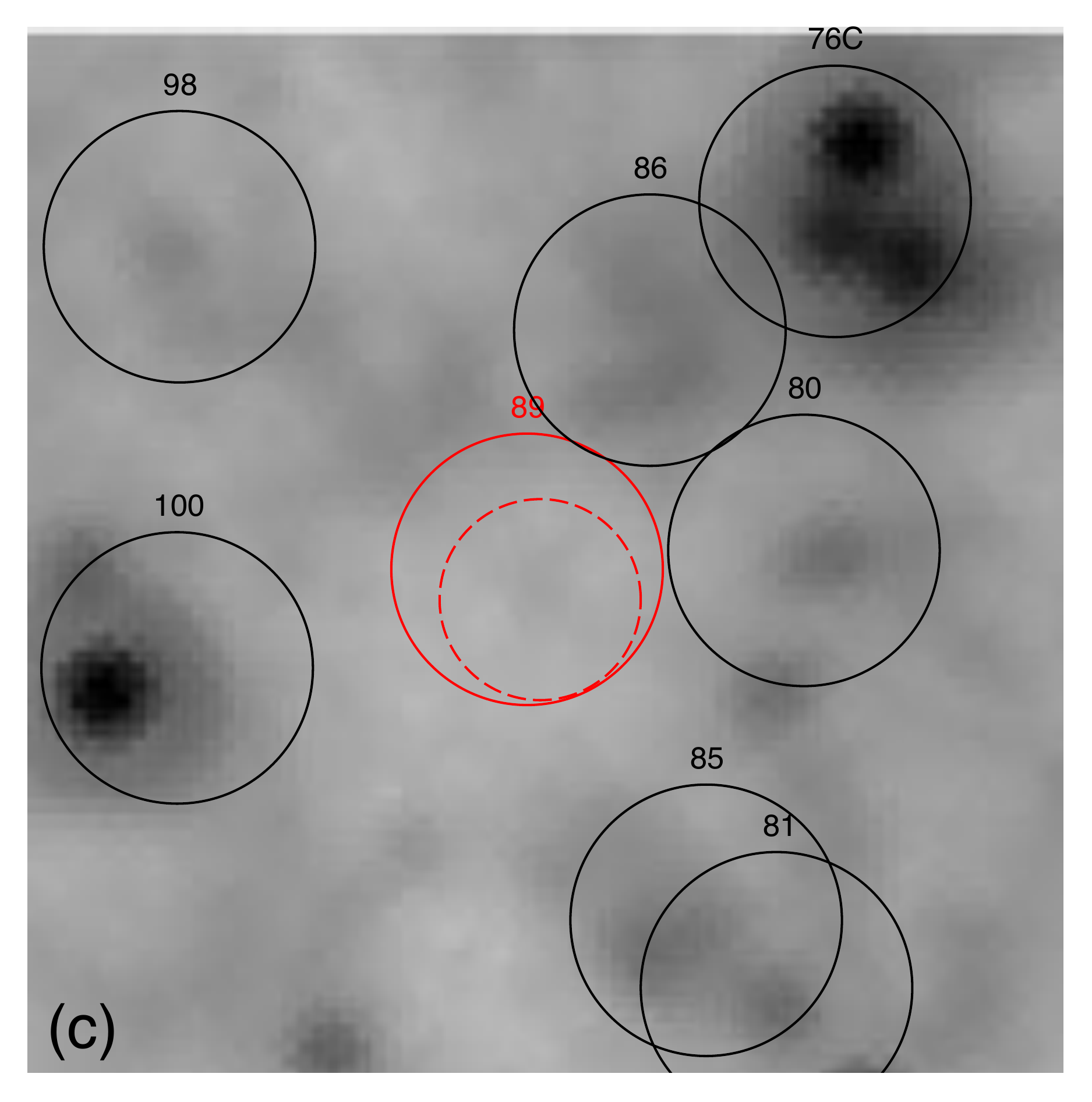}
\end{minipage}
\begin{minipage}{0.24\linewidth}
\includegraphics[width=\linewidth]{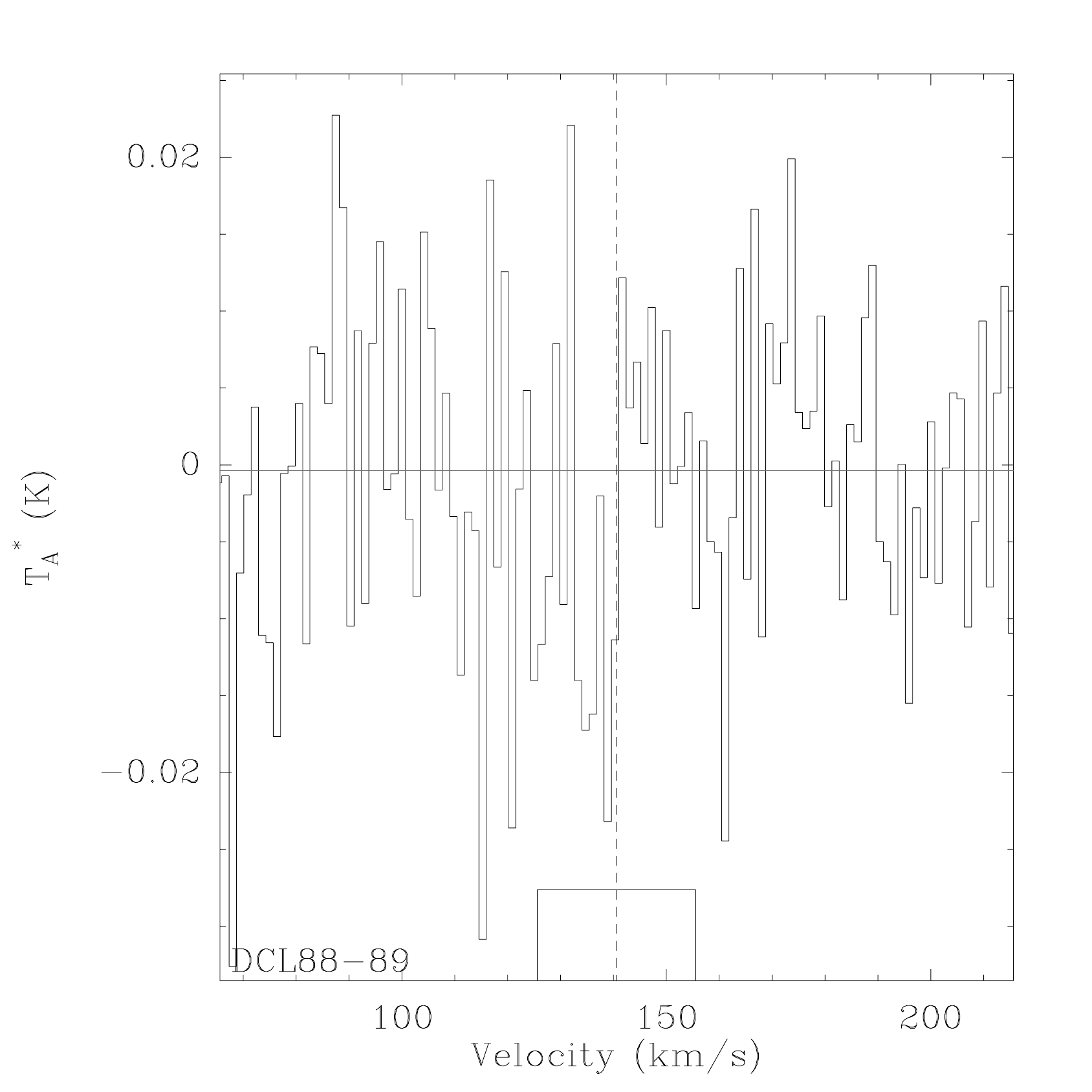}
\end{minipage}

\begin{minipage}{0.24\linewidth}
\includegraphics[width=\linewidth]{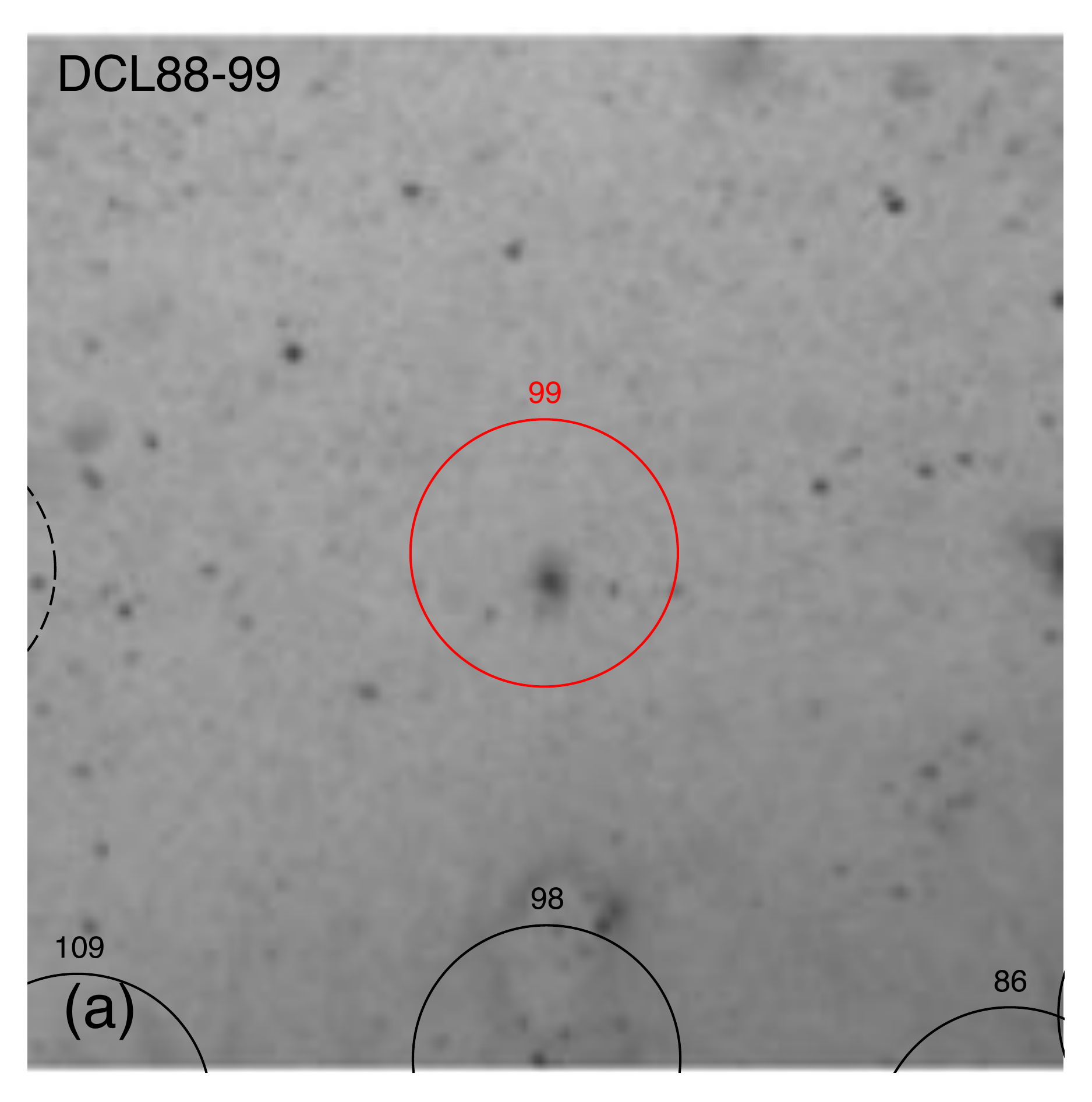}
\end{minipage}
\begin{minipage}{0.24\linewidth}
\includegraphics[width=\linewidth]{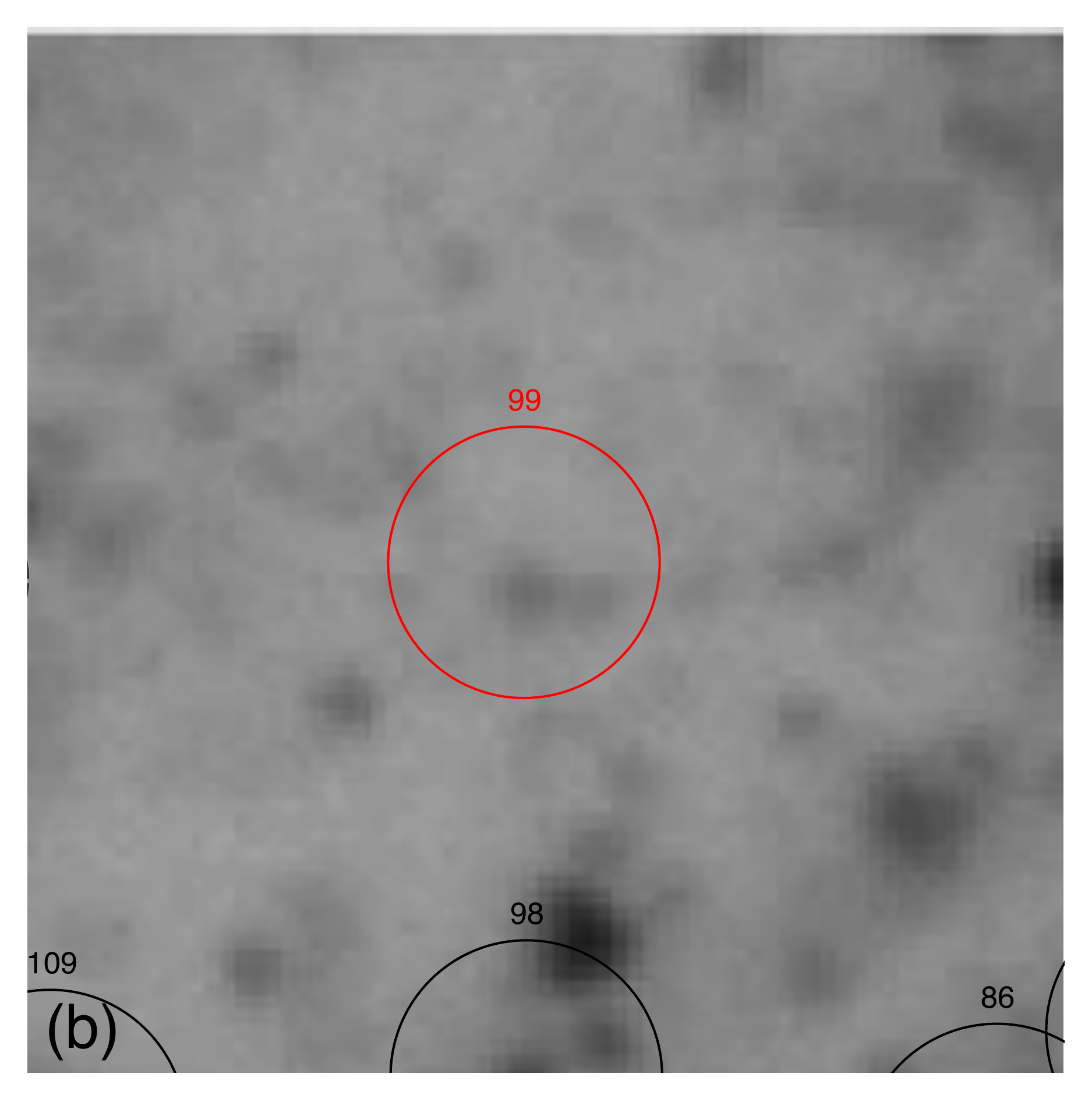}
\end{minipage}
\begin{minipage}{0.24\linewidth}
\includegraphics[width=\linewidth]{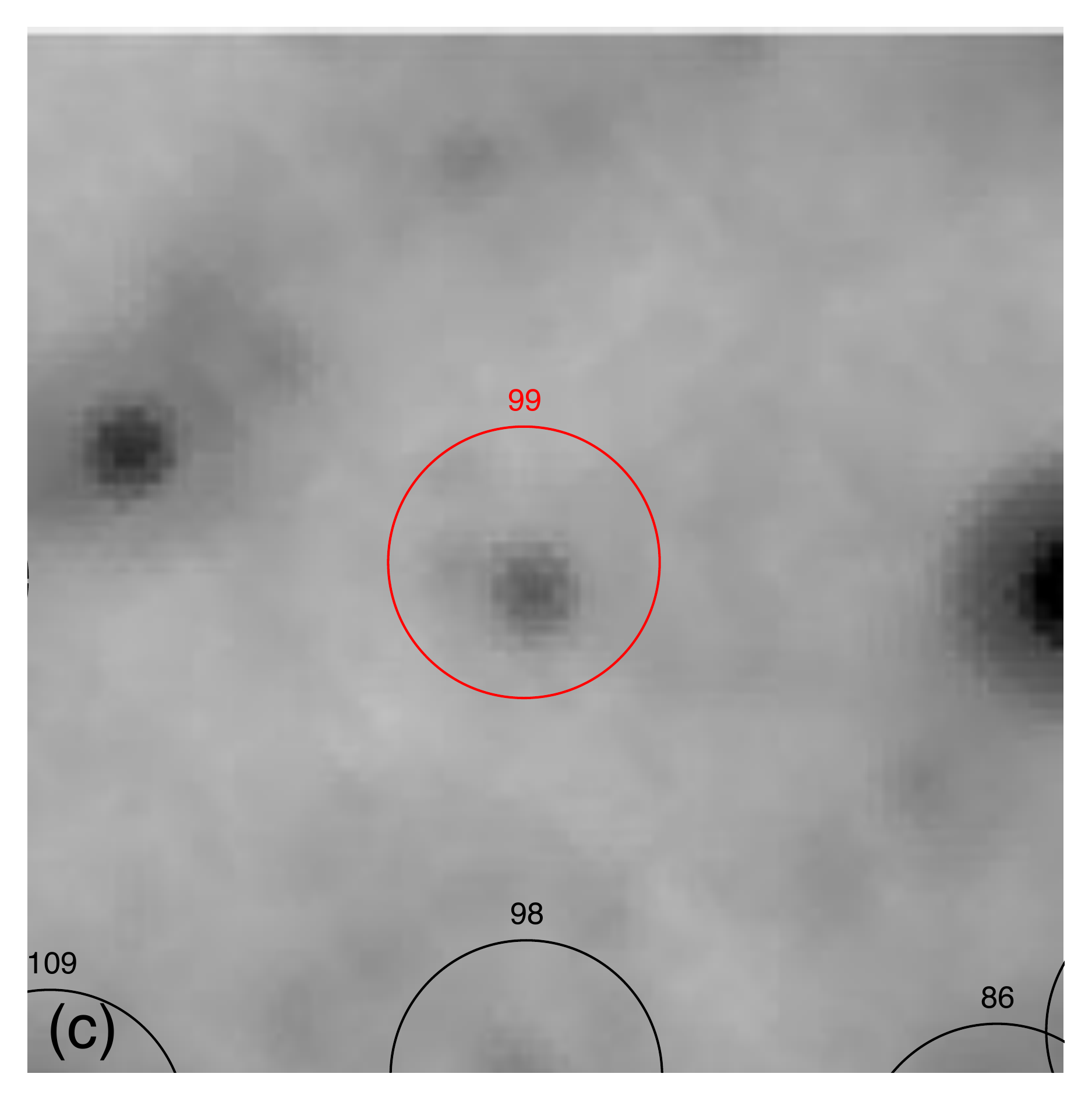}
\end{minipage}
\begin{minipage}{0.24\linewidth}
\includegraphics[width=\linewidth]{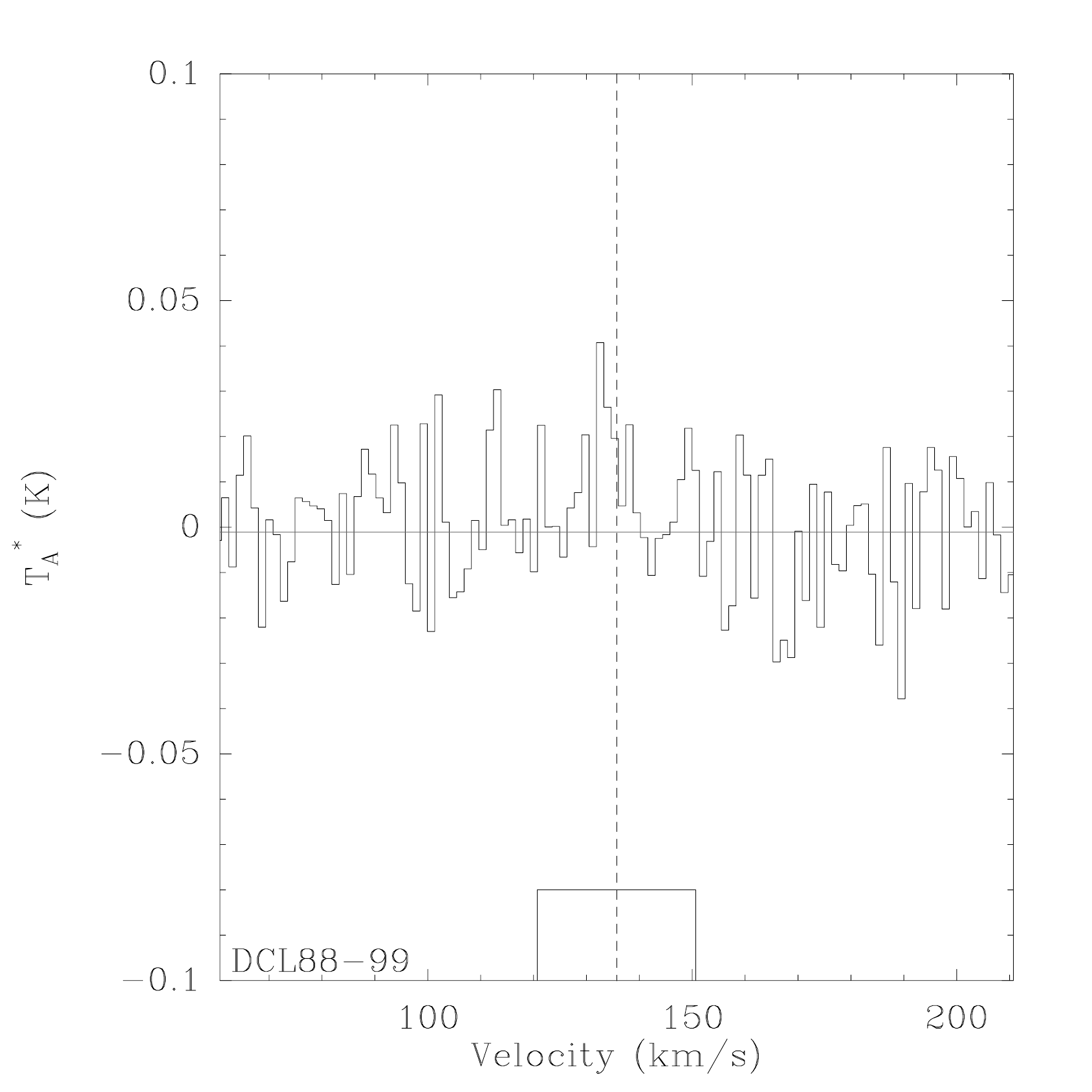}
\end{minipage}

\begin{minipage}{0.24\linewidth}
\includegraphics[width=\linewidth]{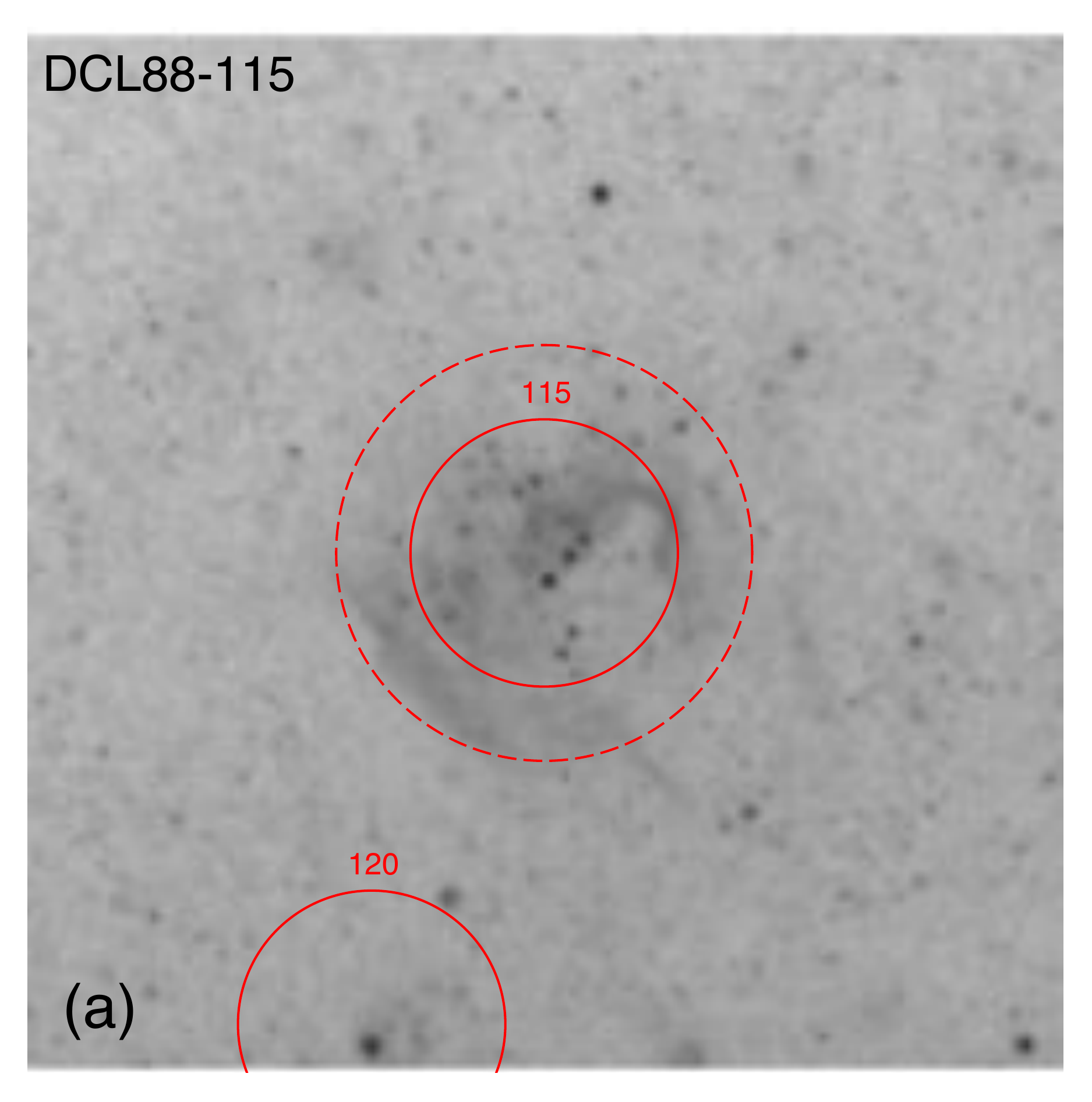}
\end{minipage}
\begin{minipage}{0.24\linewidth}
\includegraphics[width=\linewidth]{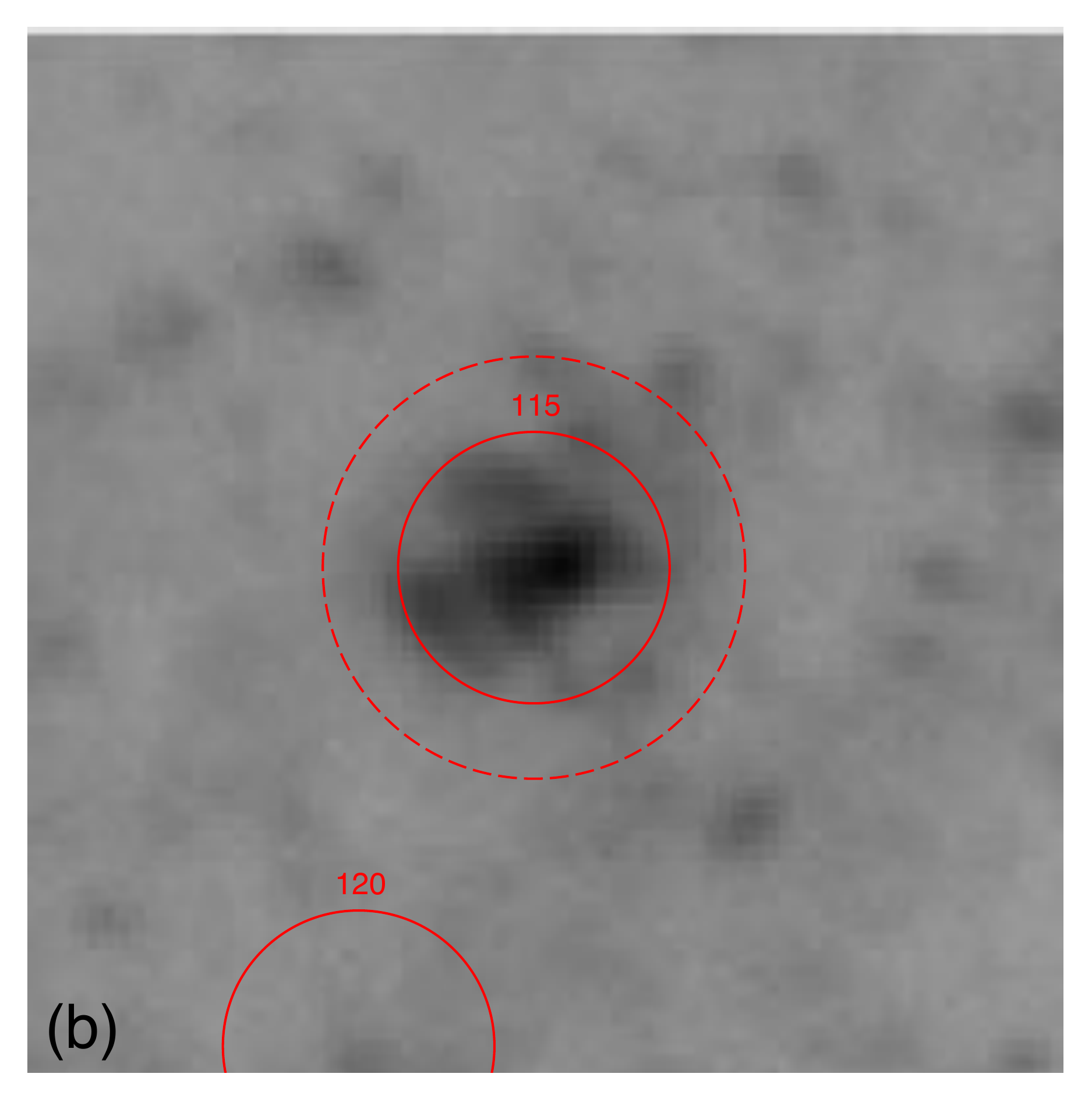}
\end{minipage}
\begin{minipage}{0.24\linewidth}
\includegraphics[width=\linewidth]{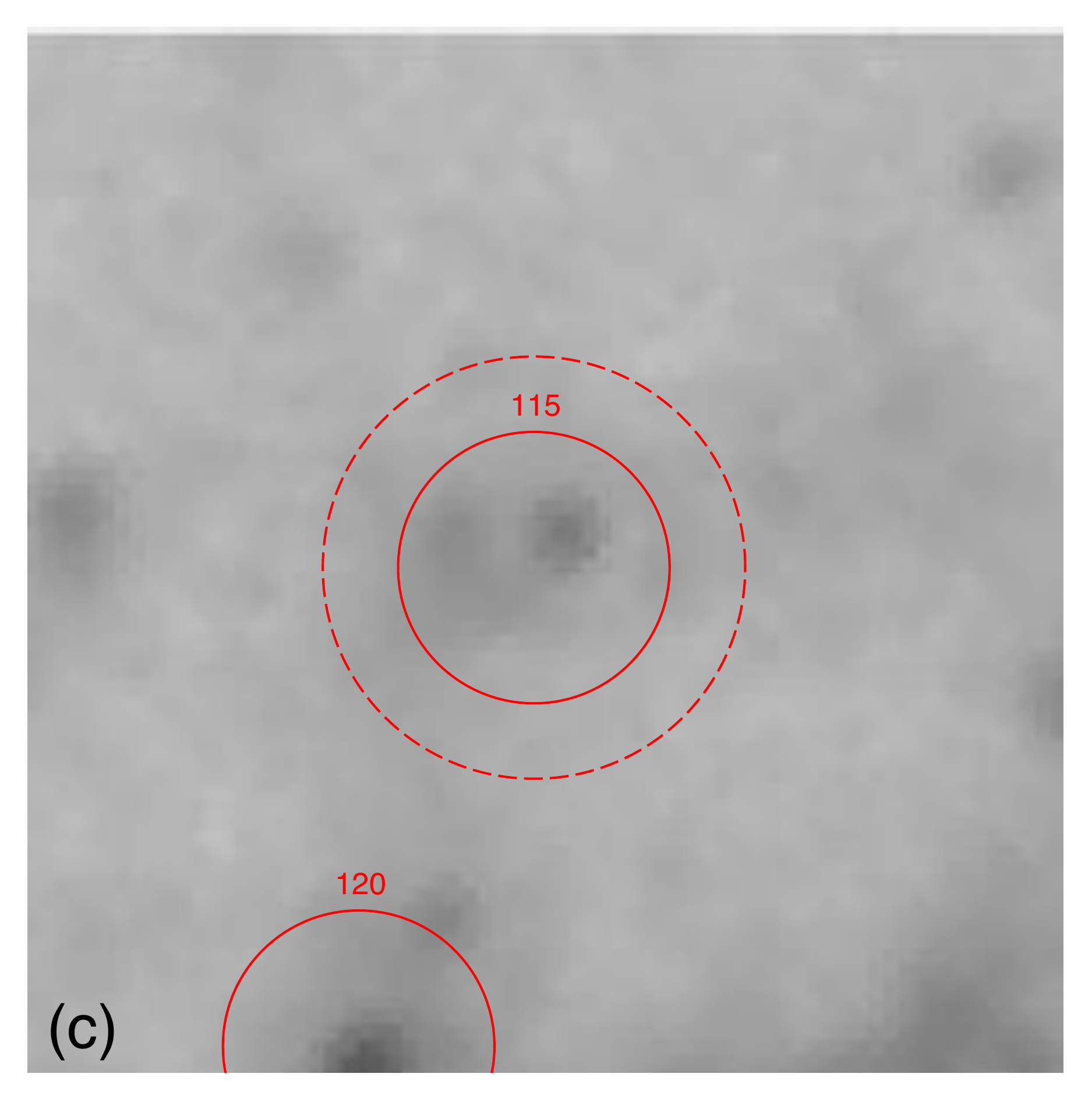}
\end{minipage}
\begin{minipage}{0.24\linewidth}
\includegraphics[width=\linewidth]{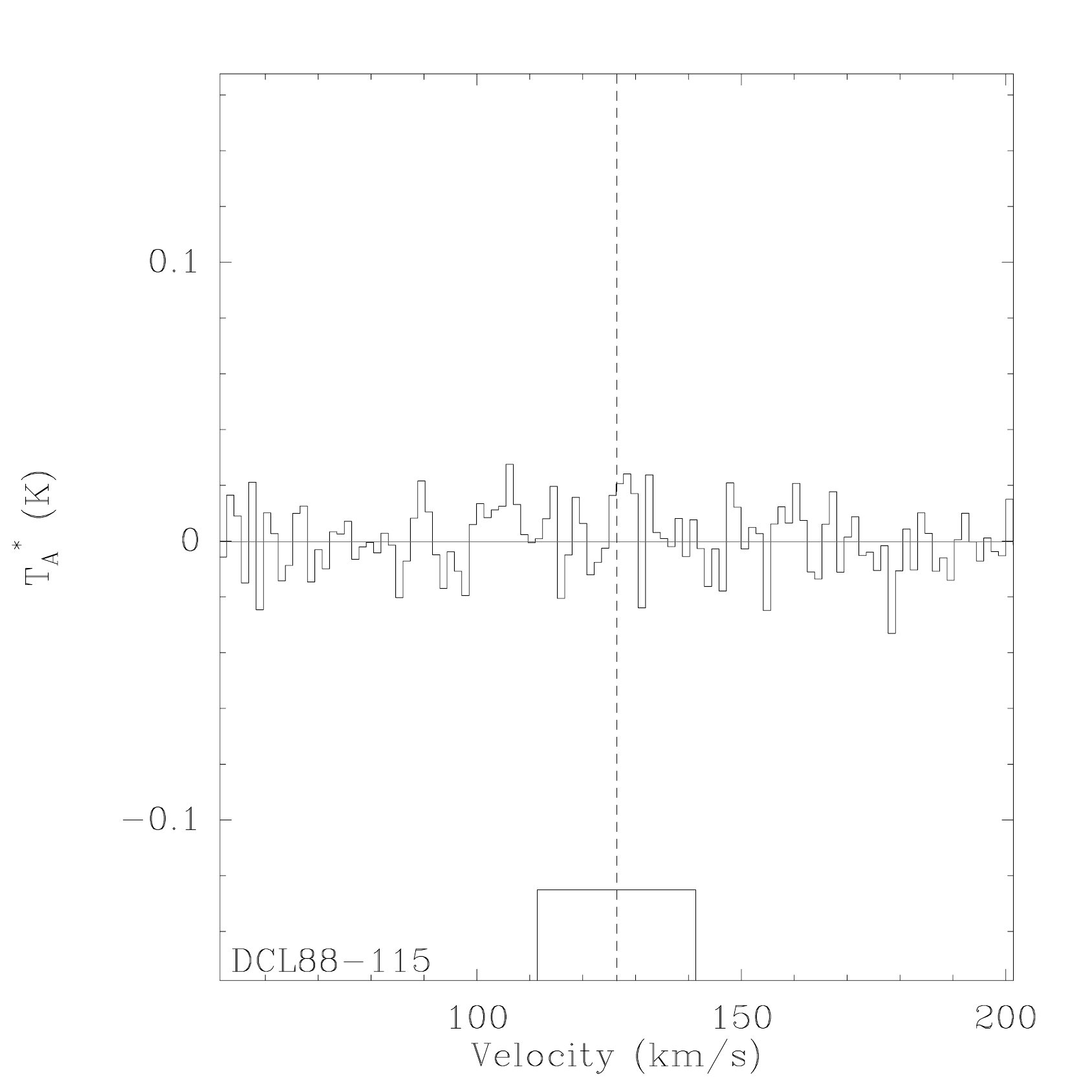}
\end{minipage}

\begin{minipage}{0.24\linewidth}
\includegraphics[width=\linewidth]{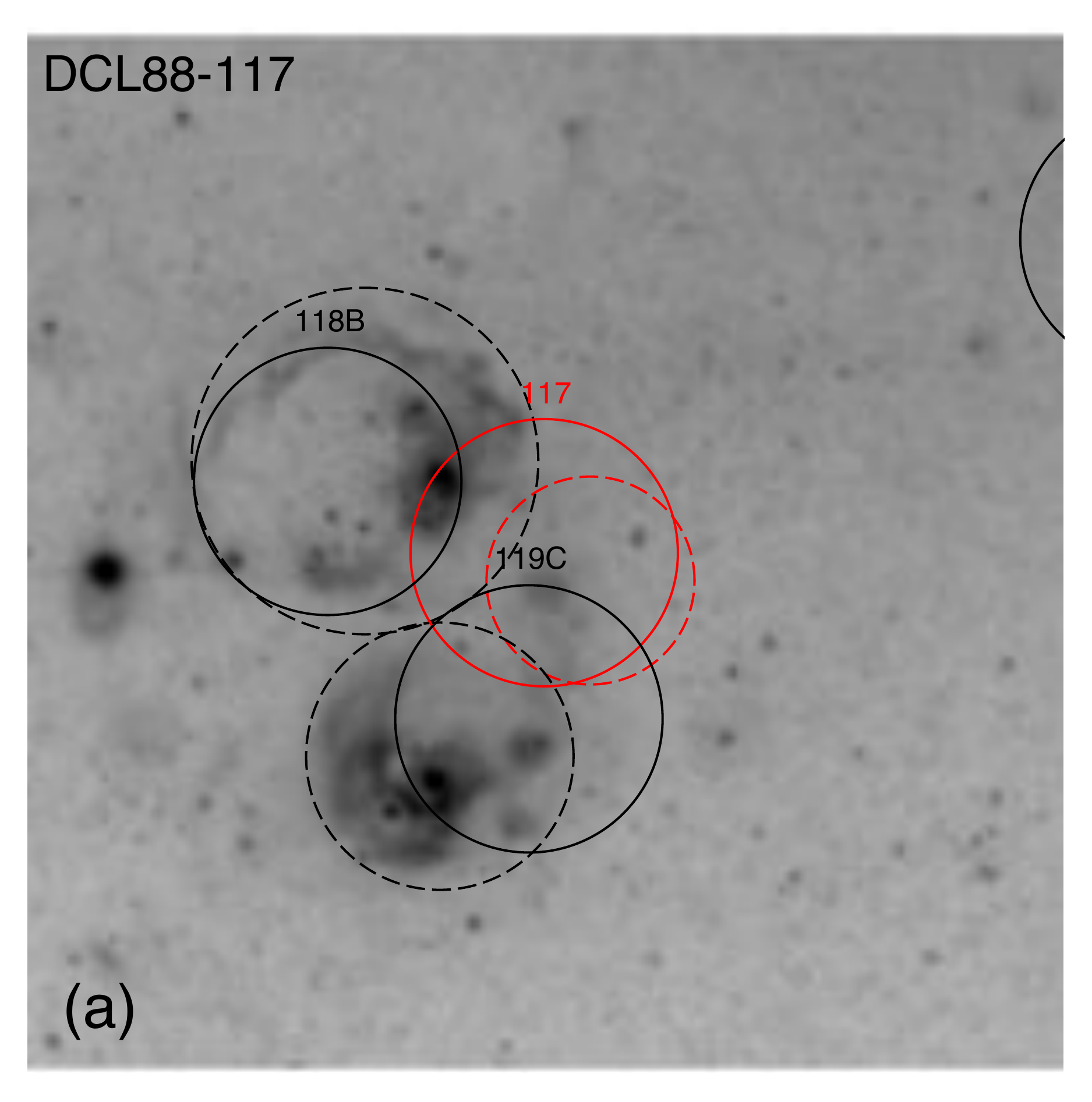}
\end{minipage}
\begin{minipage}{0.24\linewidth}
\includegraphics[width=\linewidth]{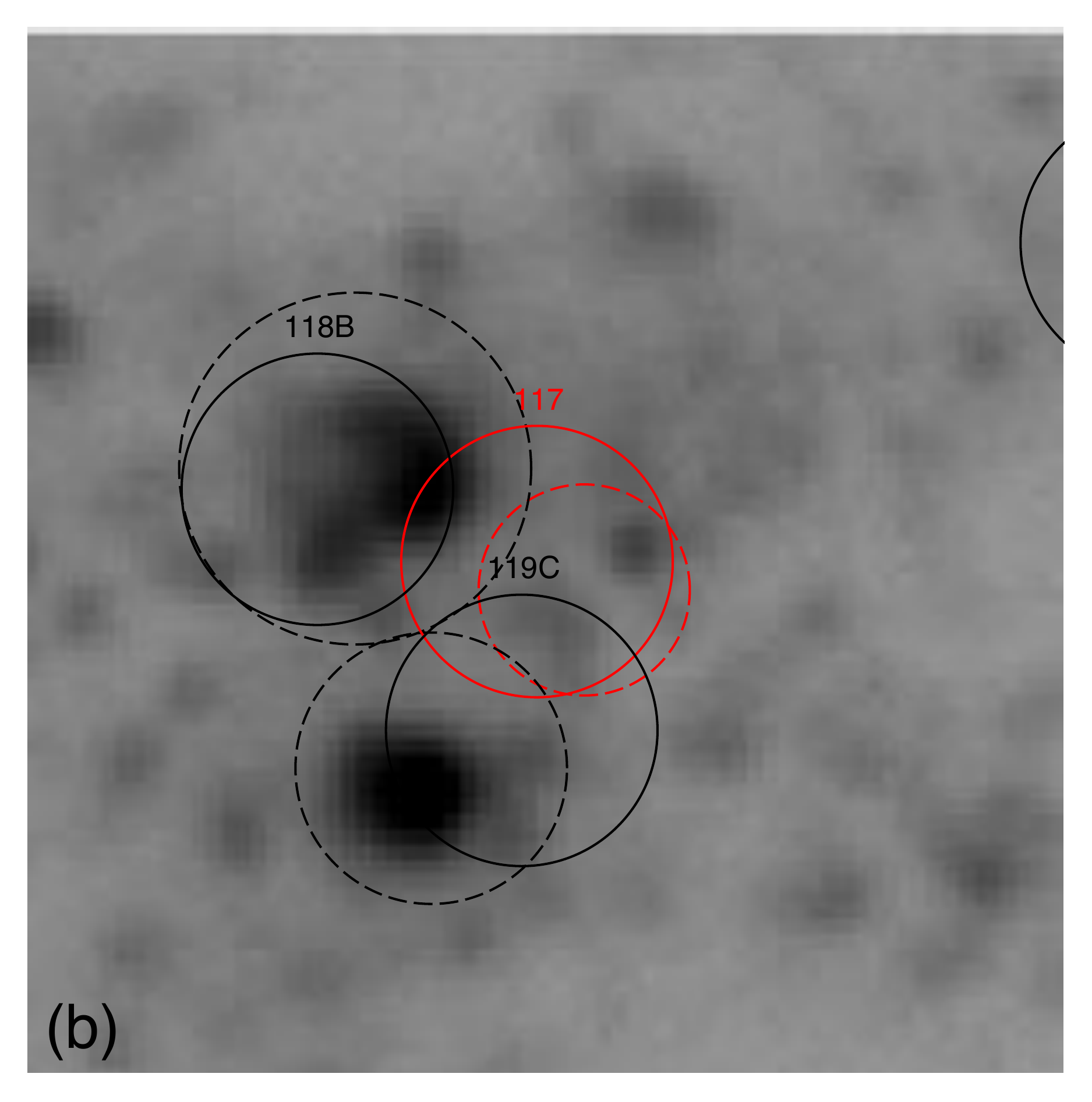}
\end{minipage}
\begin{minipage}{0.24\linewidth}
\includegraphics[width=\linewidth]{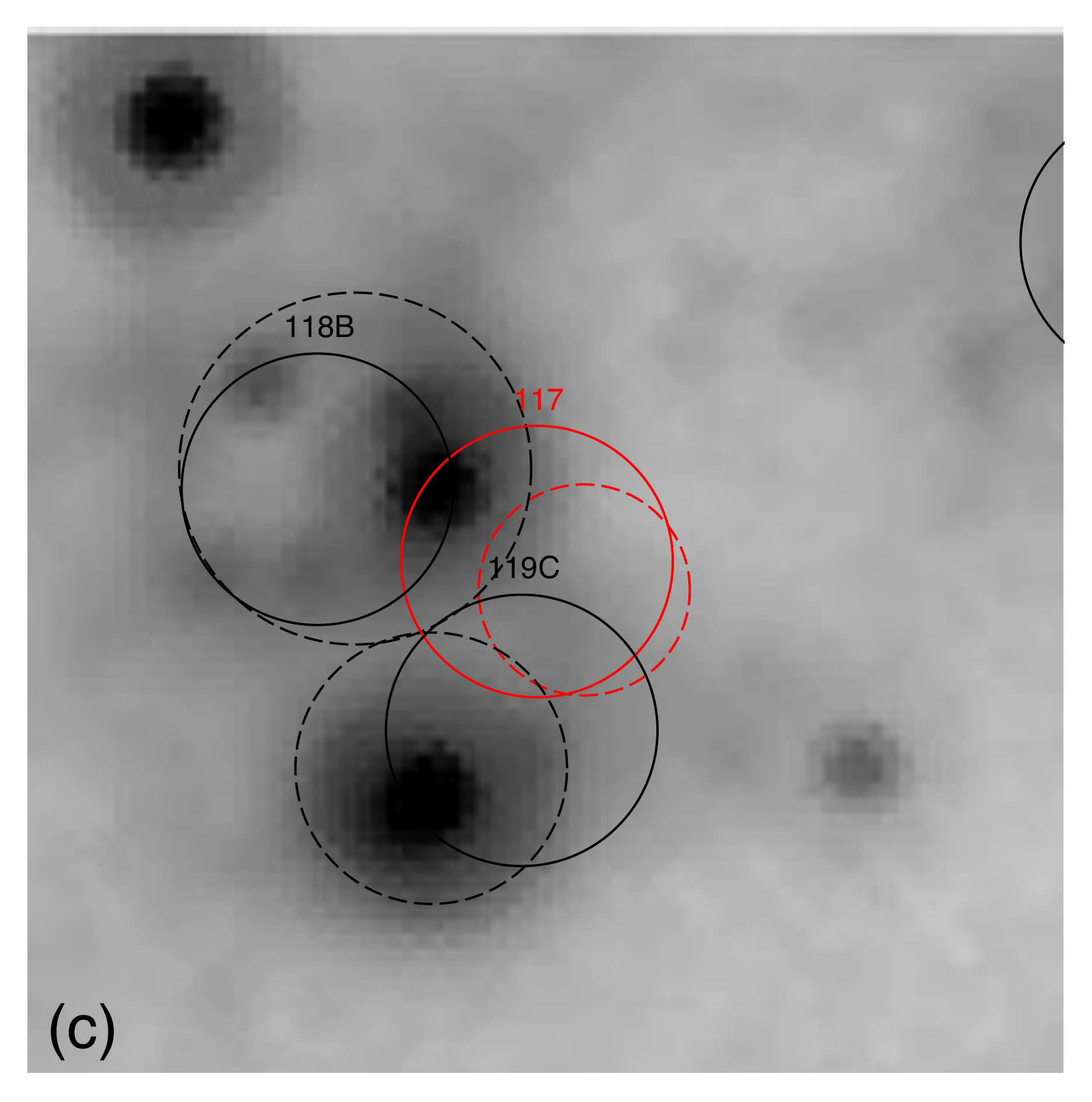}
\end{minipage}
\begin{minipage}{0.24\linewidth}
\includegraphics[width=\linewidth]{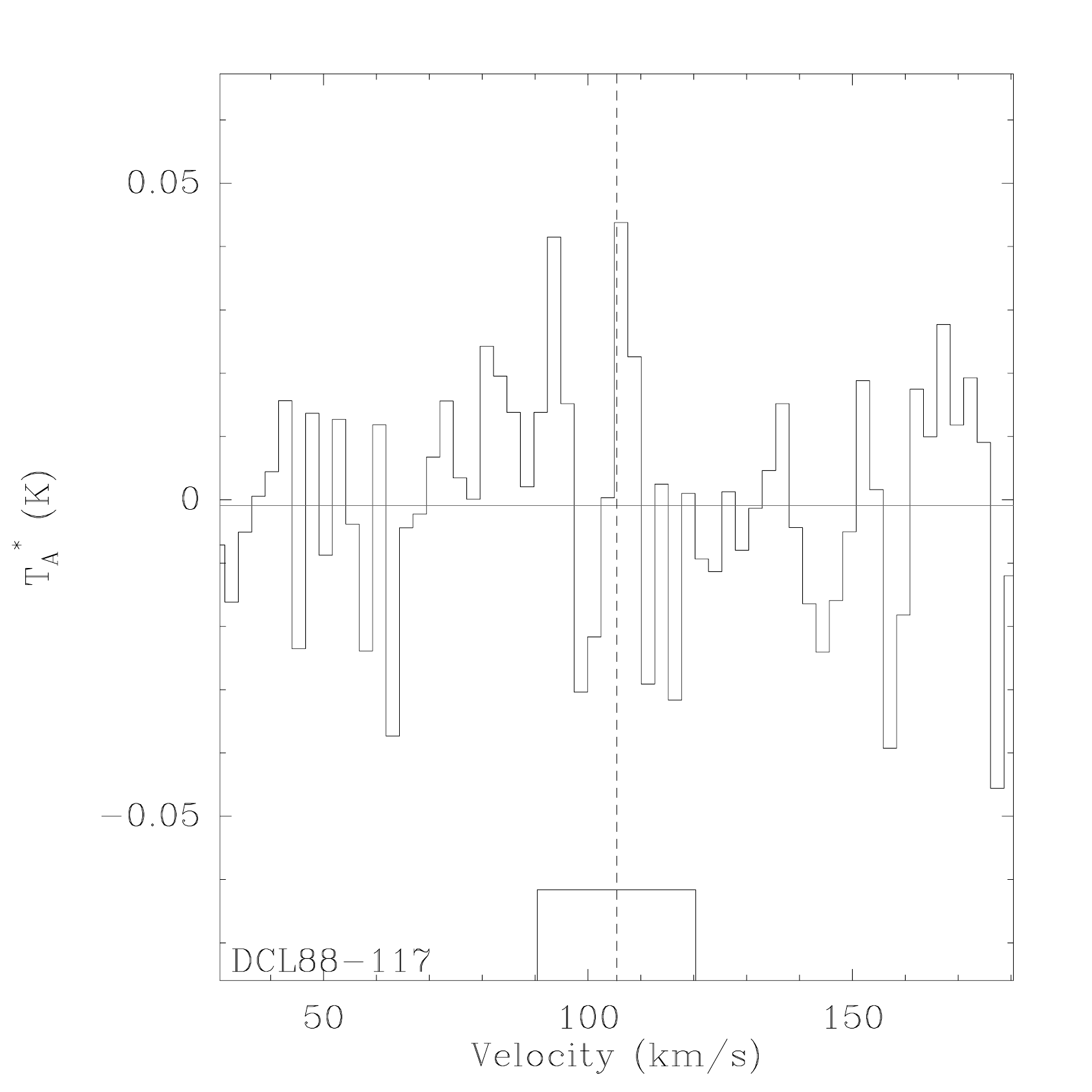}
\end{minipage}

\noindent\textbf{Figure~\ref{fig:stamps} -- continued.}

\end{figure*}

\begin{figure*}
%\ContinuedFloat

\begin{minipage}{0.24\linewidth}
\includegraphics[width=\linewidth]{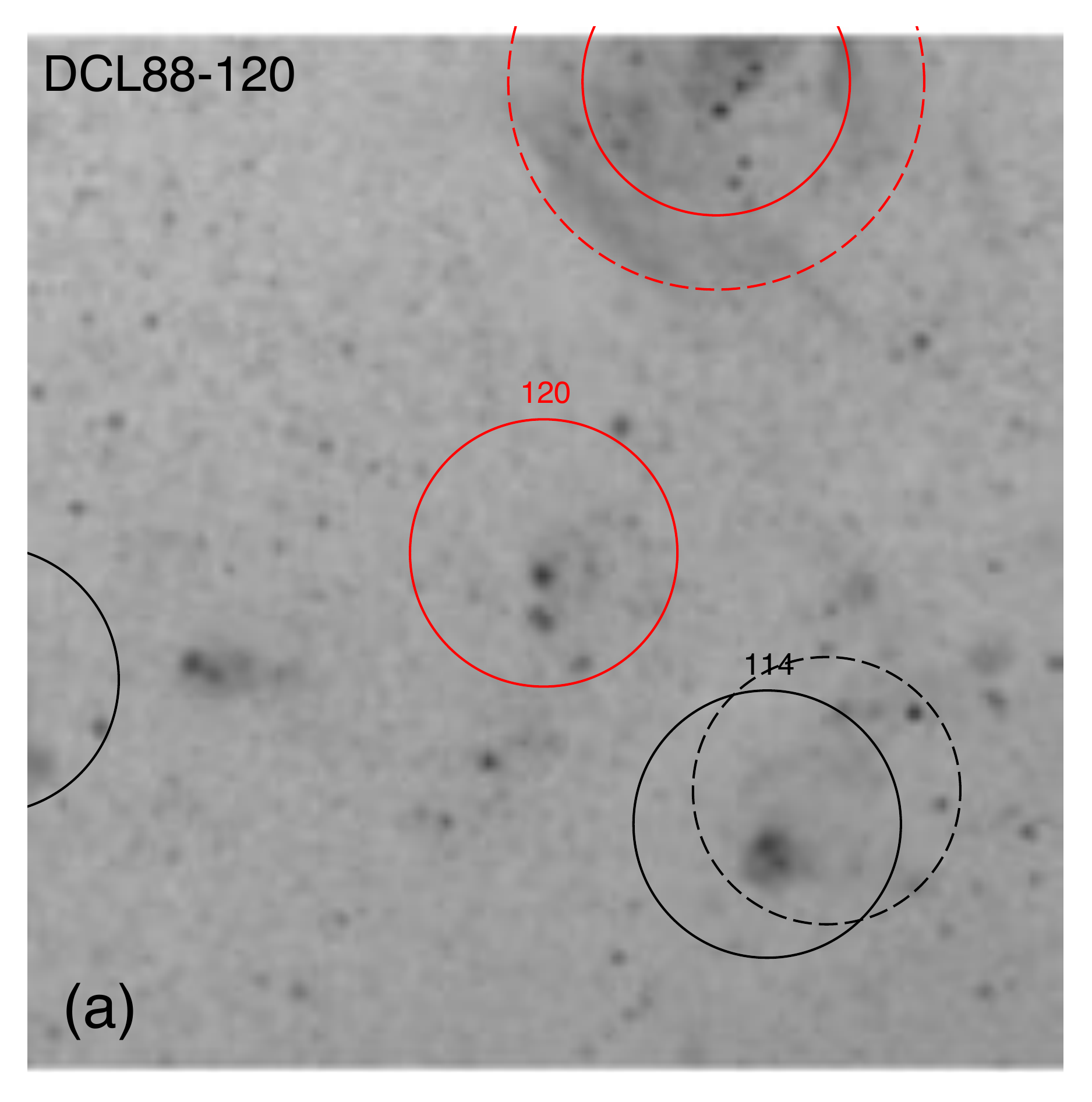}
\end{minipage}
\begin{minipage}{0.24\linewidth}
\includegraphics[width=\linewidth]{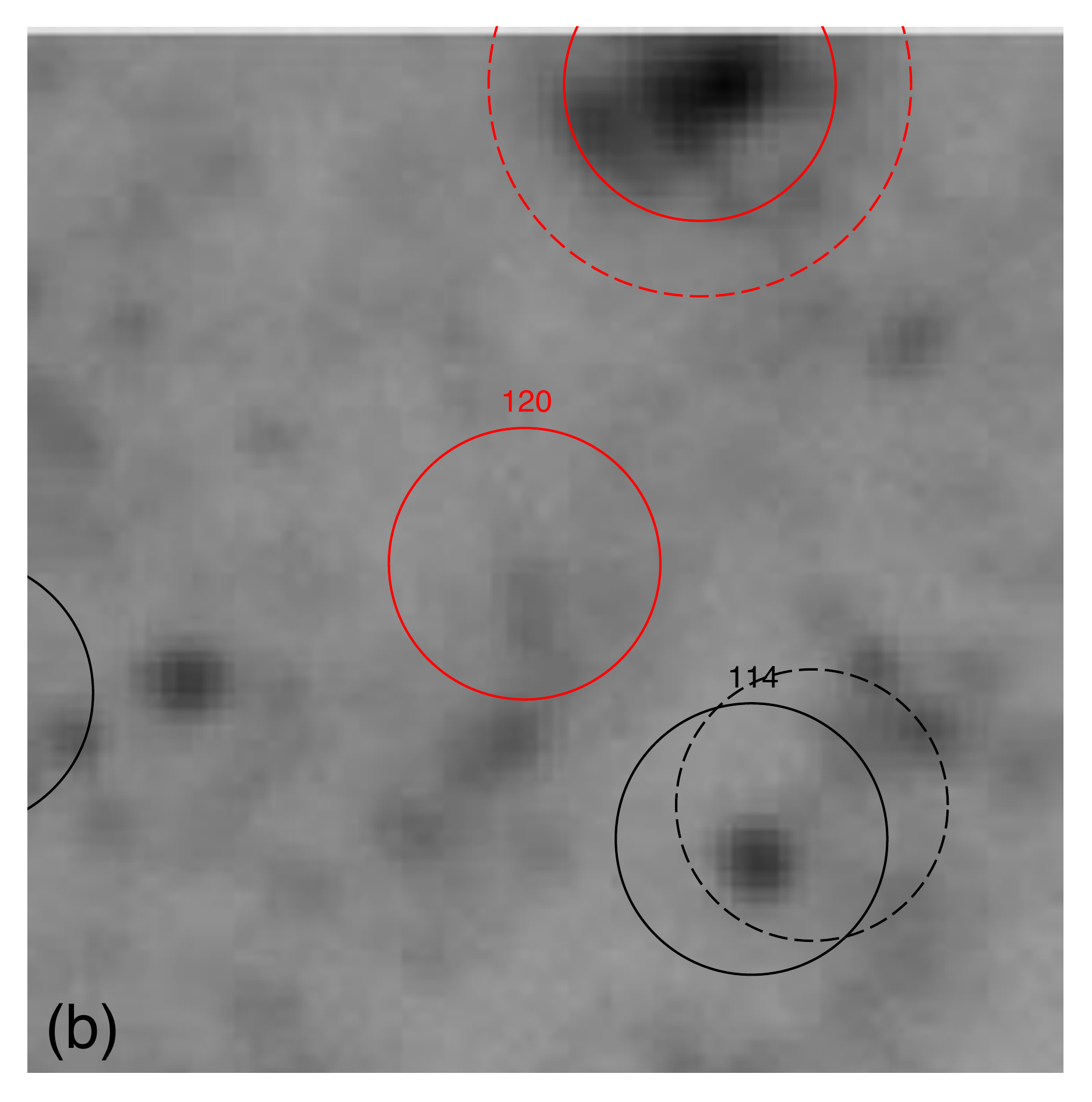}
\end{minipage}
\begin{minipage}{0.24\linewidth}
\includegraphics[width=\linewidth]{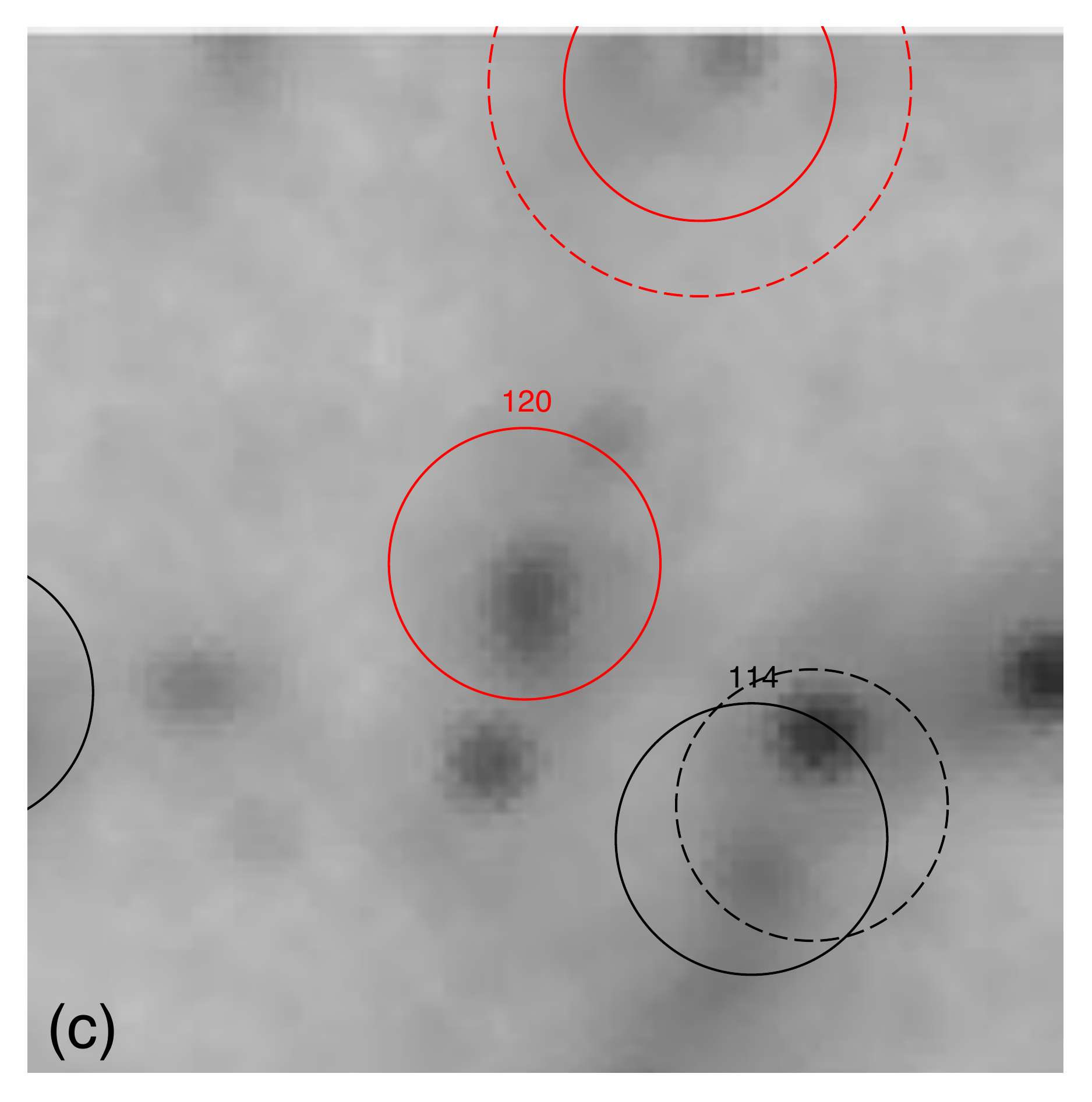}
\end{minipage}
\begin{minipage}{0.24\linewidth}
\includegraphics[width=\linewidth]{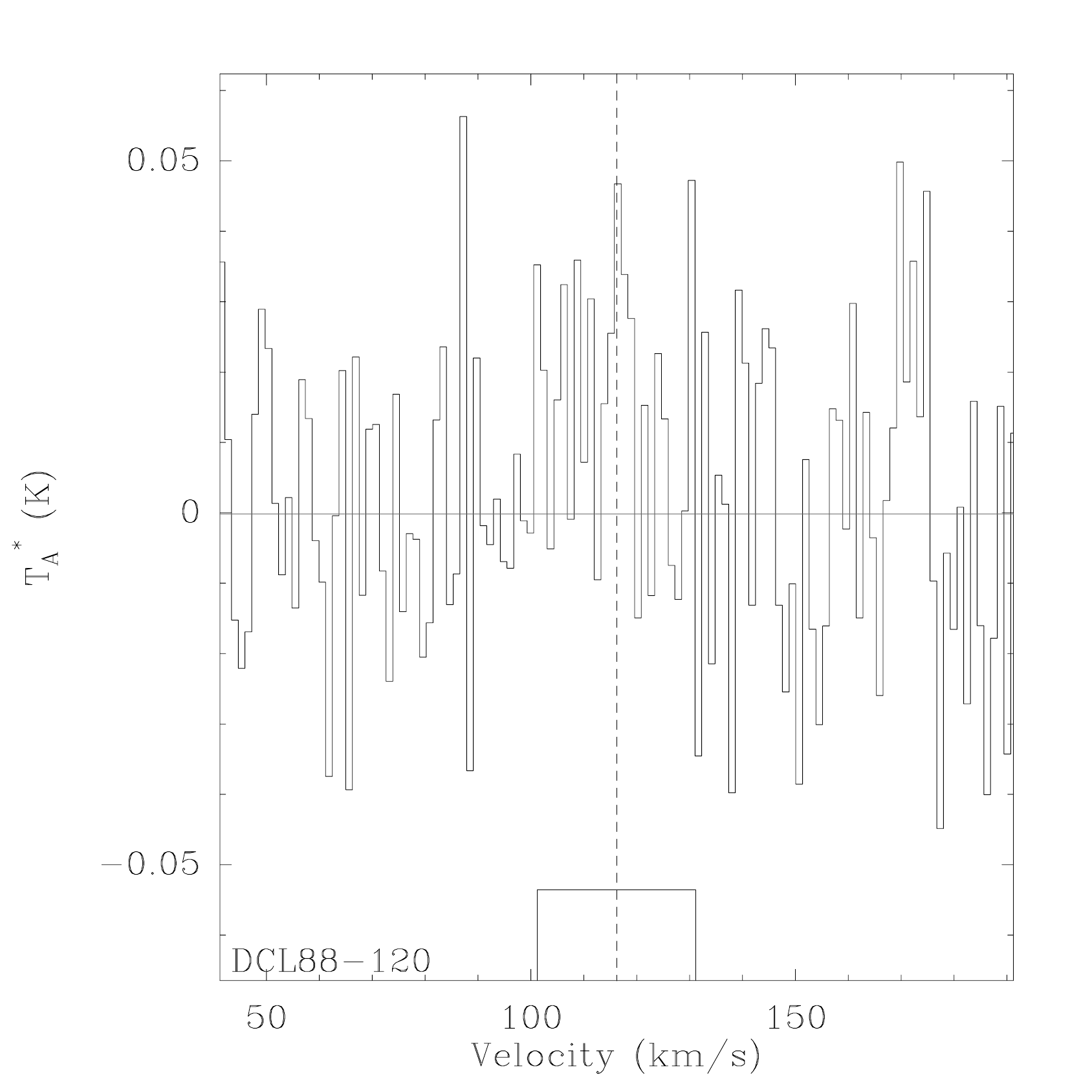}
\end{minipage}

\begin{minipage}{0.24\linewidth}
\includegraphics[width=\linewidth]{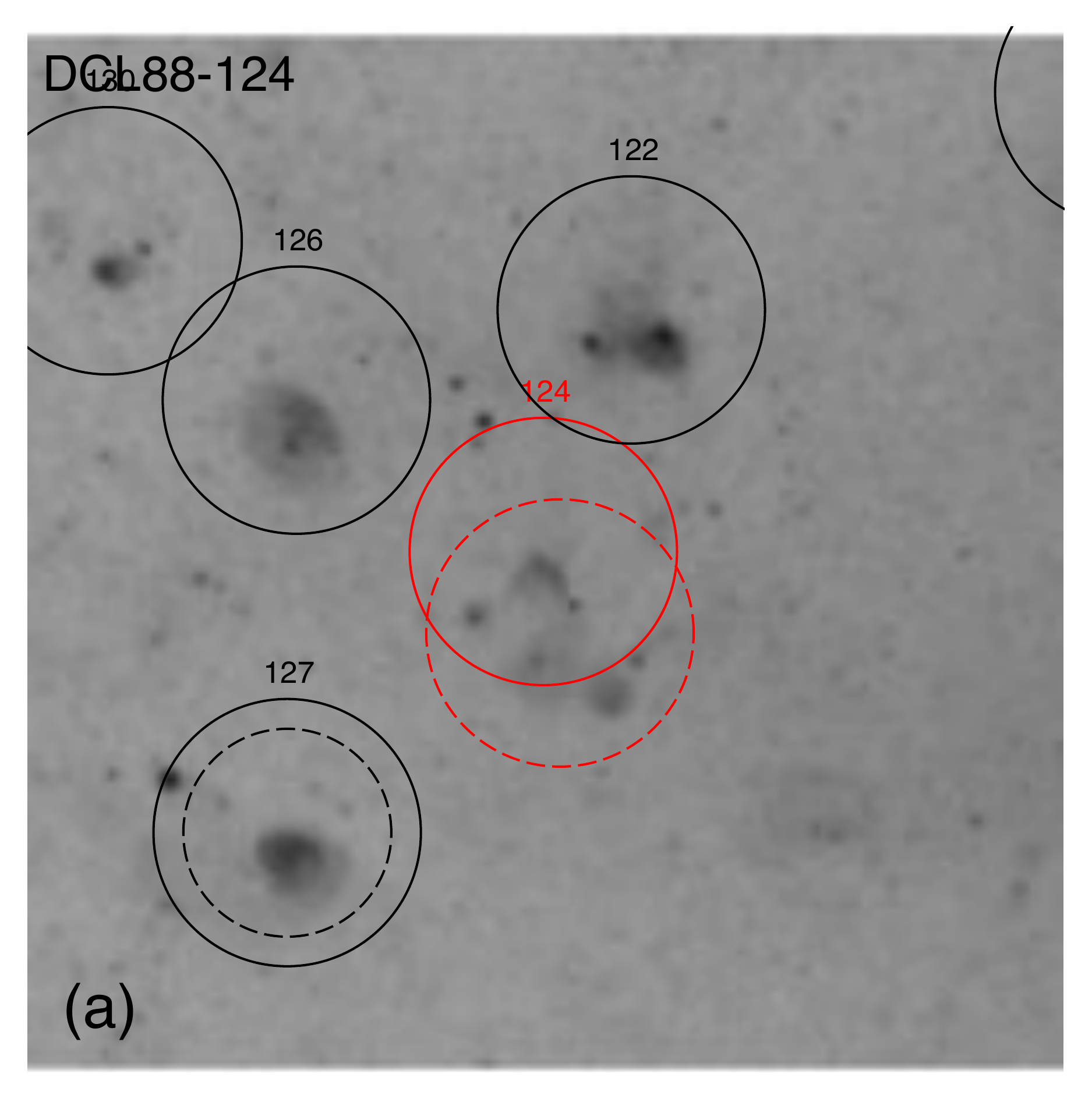}
\end{minipage}
\begin{minipage}{0.24\linewidth}
\includegraphics[width=\linewidth]{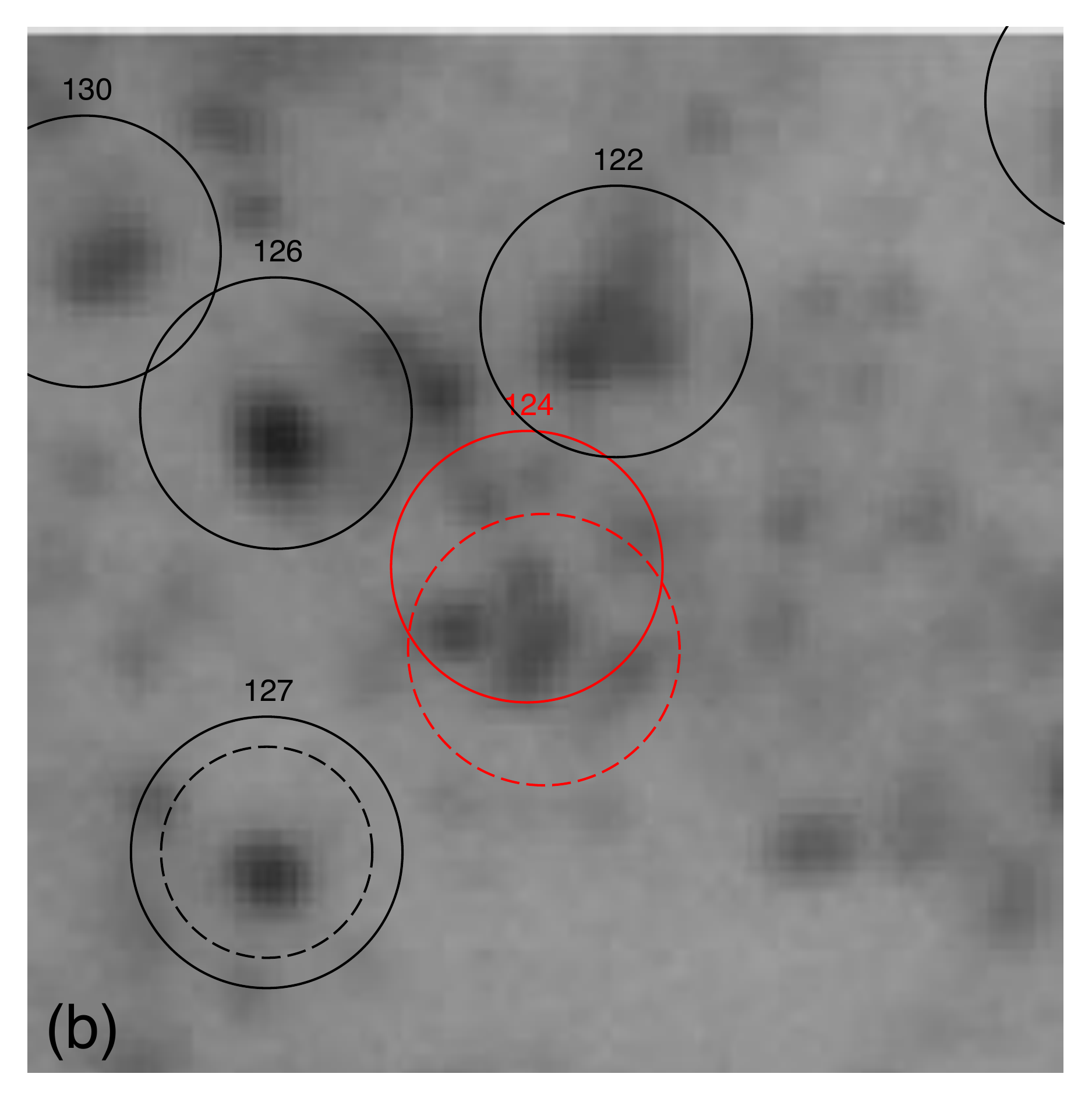}
\end{minipage}
\begin{minipage}{0.24\linewidth}
\includegraphics[width=\linewidth]{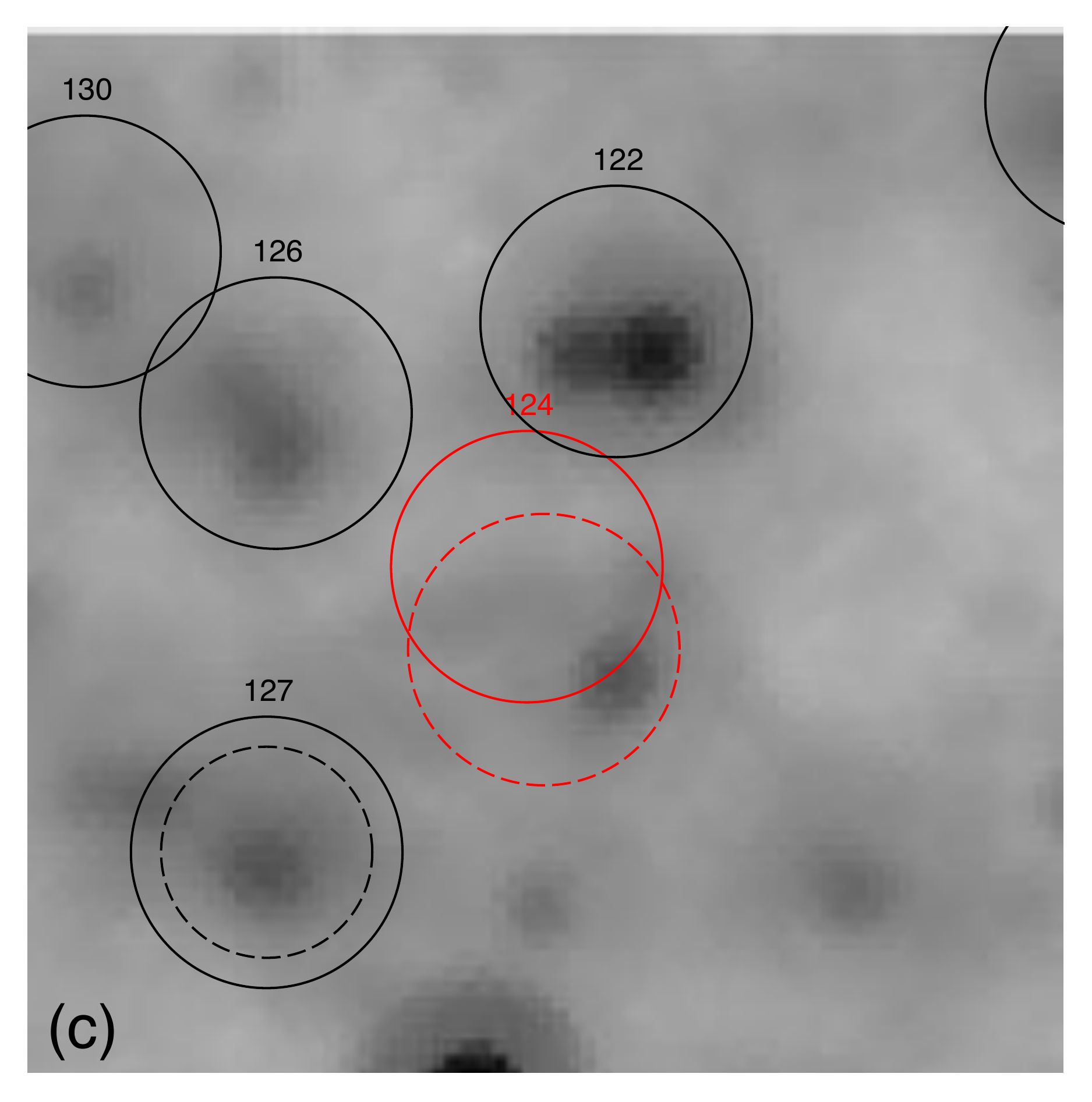}
\end{minipage}
\begin{minipage}{0.24\linewidth}
\includegraphics[width=\linewidth]{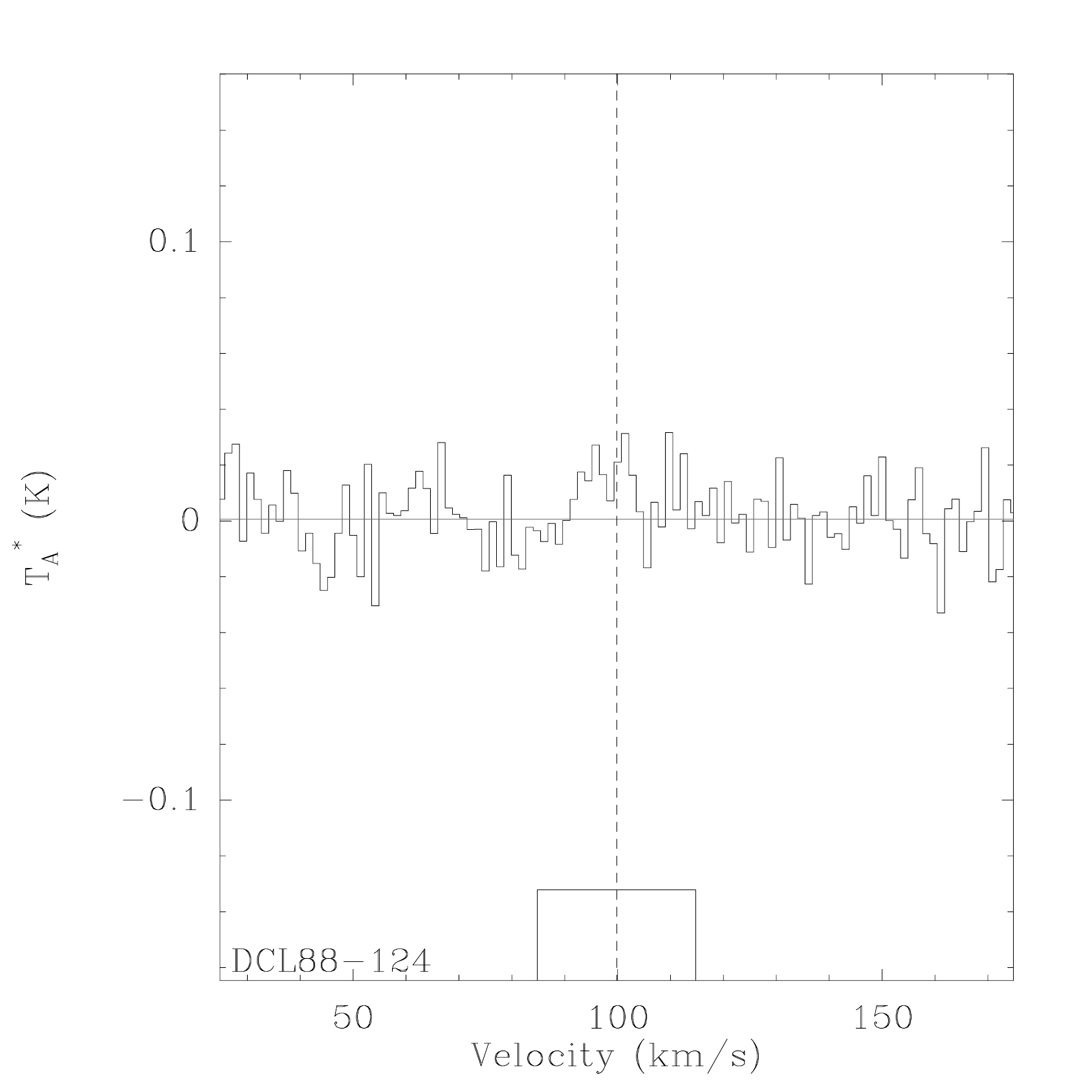}
\end{minipage}

\begin{minipage}{0.24\linewidth}
\includegraphics[width=\linewidth]{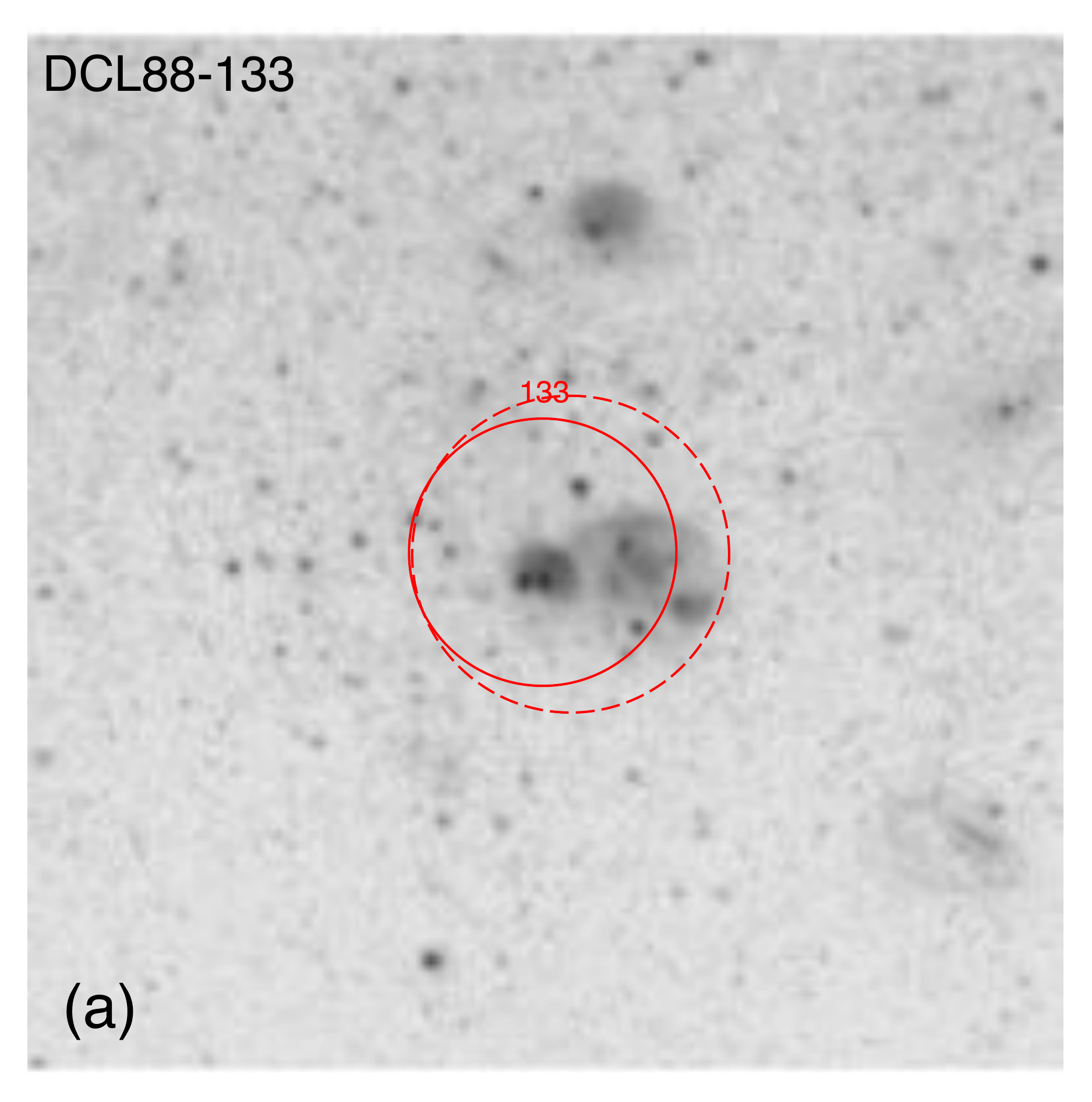}
\end{minipage}
\begin{minipage}{0.24\linewidth}
\includegraphics[width=\linewidth]{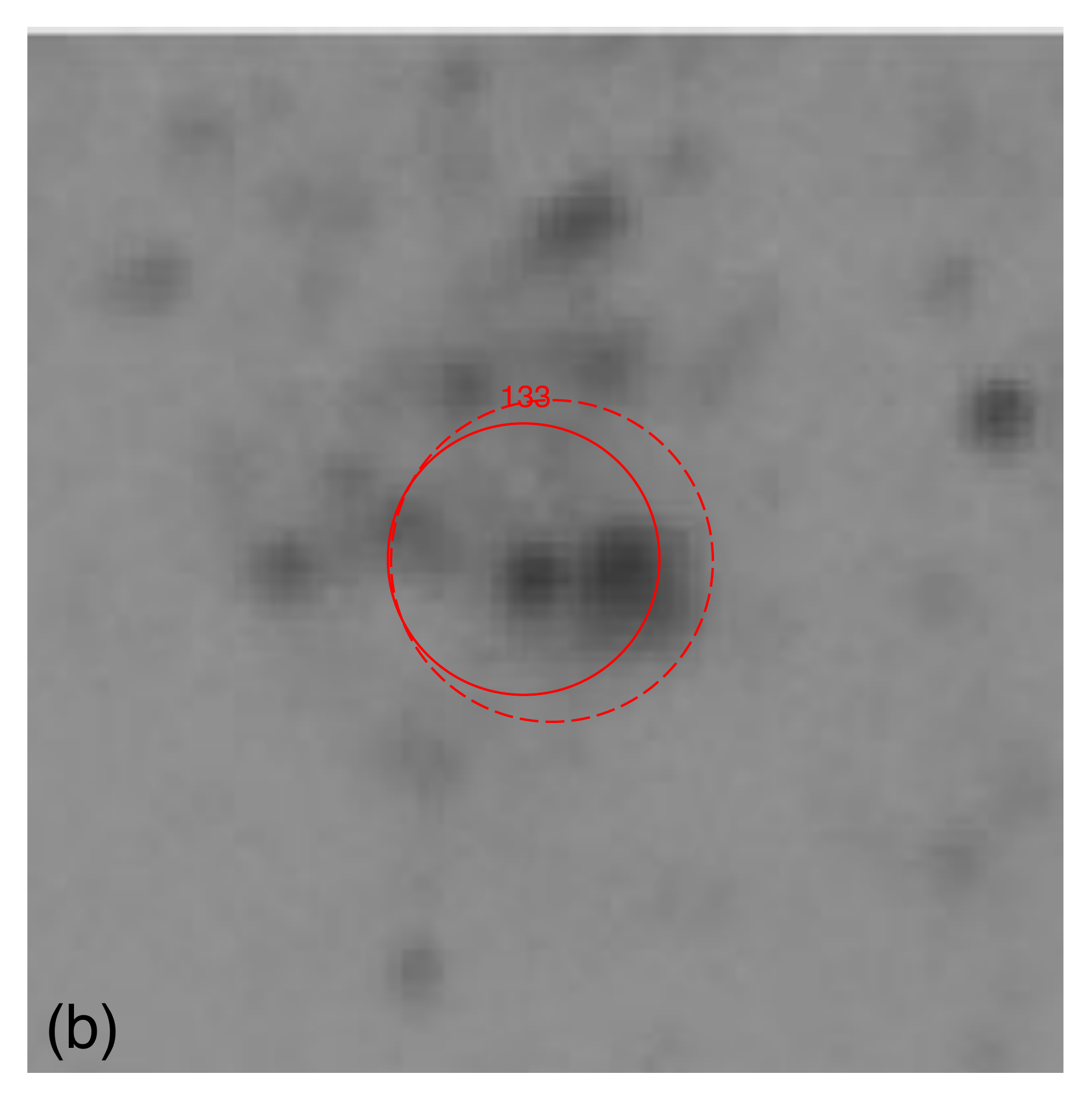}
\end{minipage}
\begin{minipage}{0.24\linewidth}
\includegraphics[width=\linewidth]{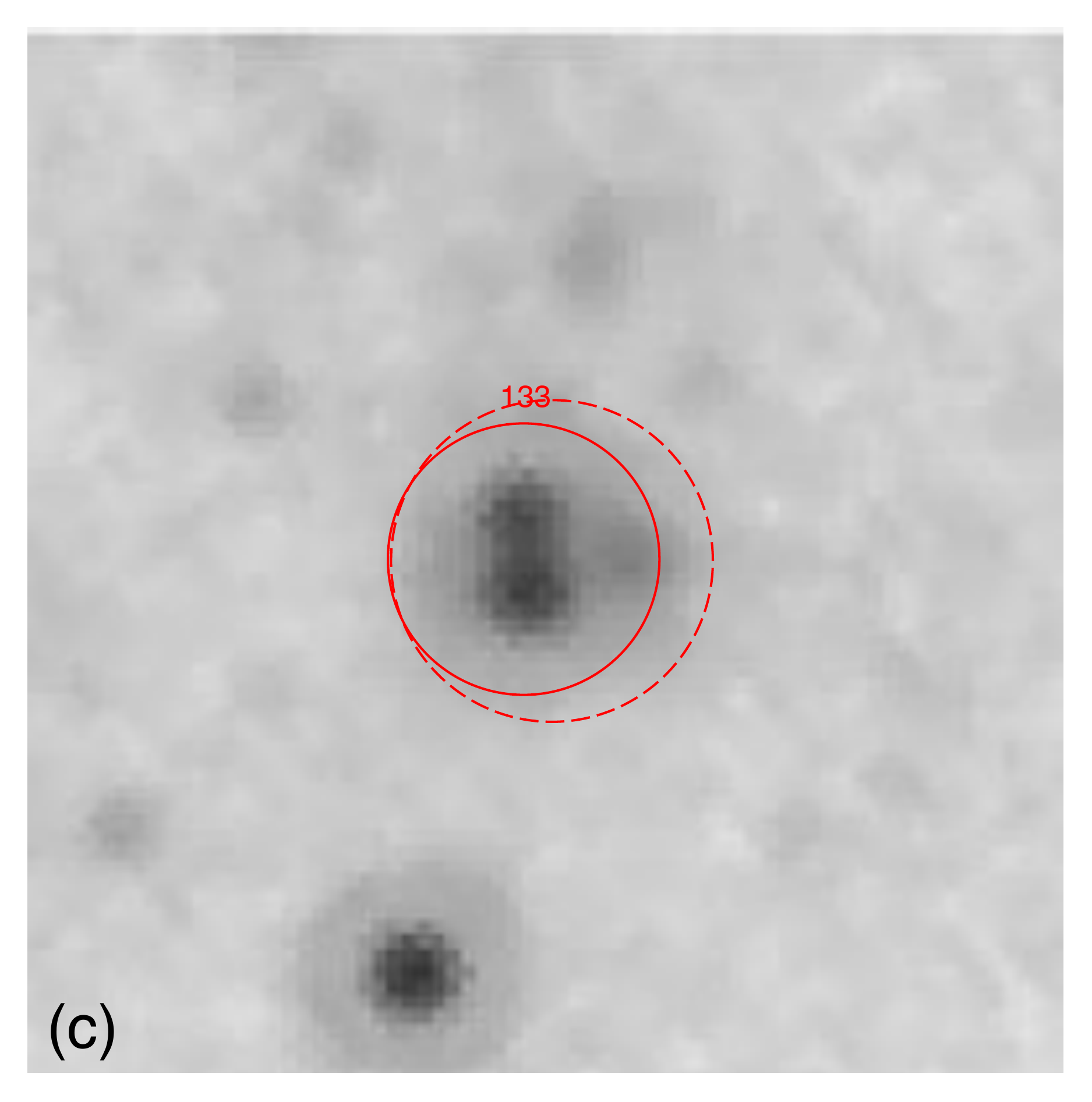}
\end{minipage}
\begin{minipage}{0.24\linewidth}
\includegraphics[width=\linewidth]{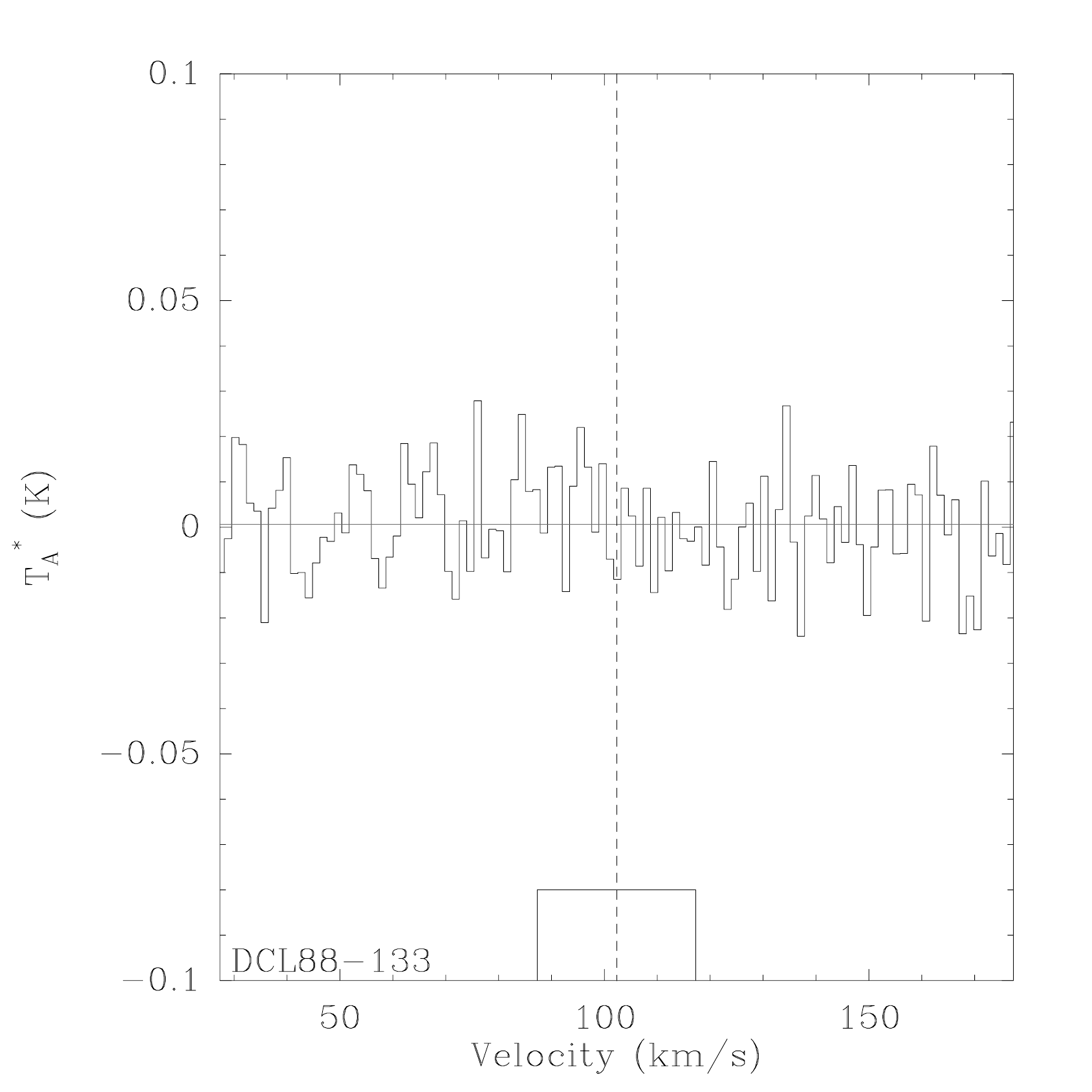}
\end{minipage}

\begin{minipage}{0.24\linewidth}
\includegraphics[width=\linewidth]{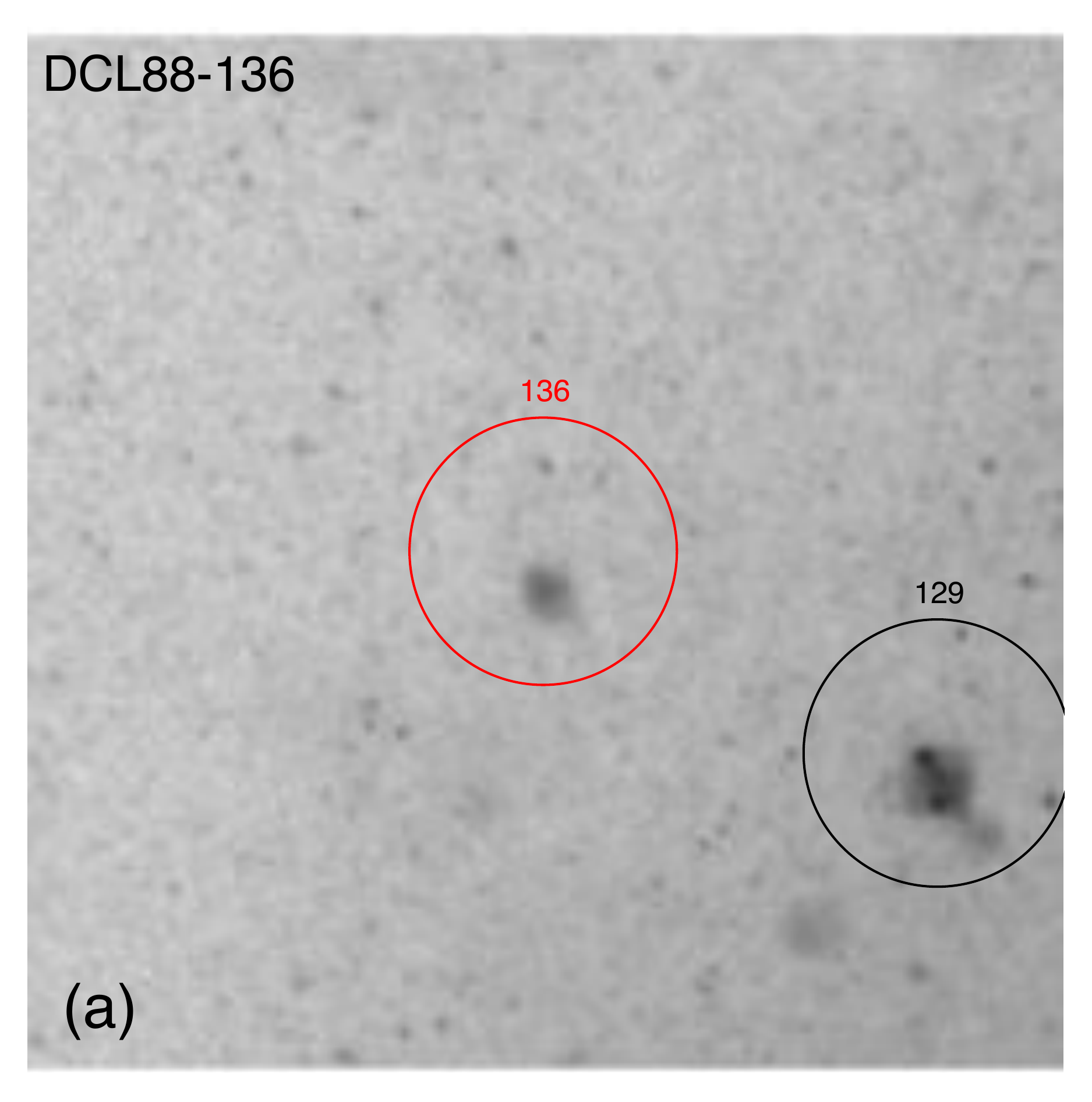}
\end{minipage}
\begin{minipage}{0.24\linewidth}
\includegraphics[width=\linewidth]{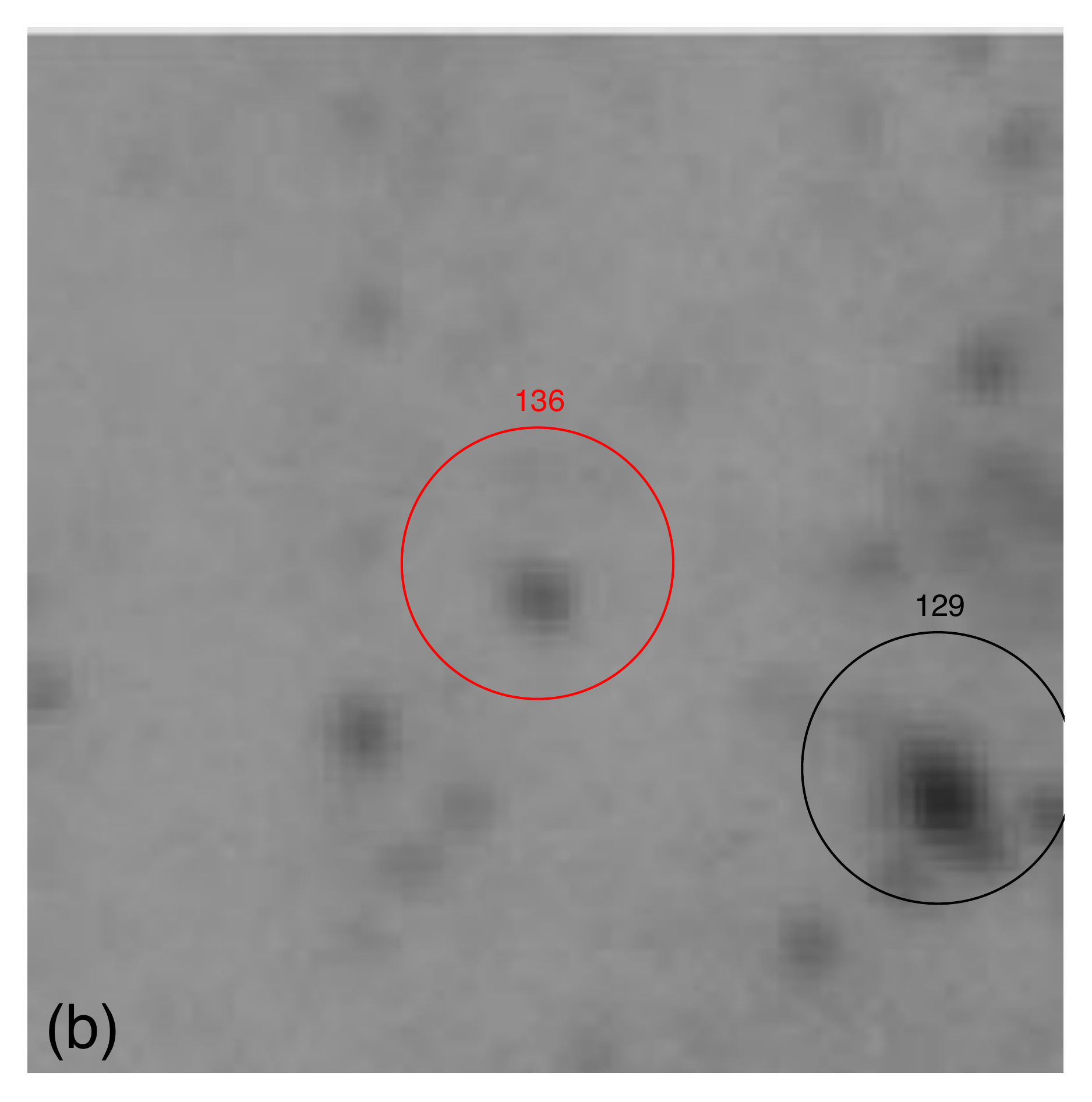}
\end{minipage}
\begin{minipage}{0.24\linewidth}
\includegraphics[width=\linewidth]{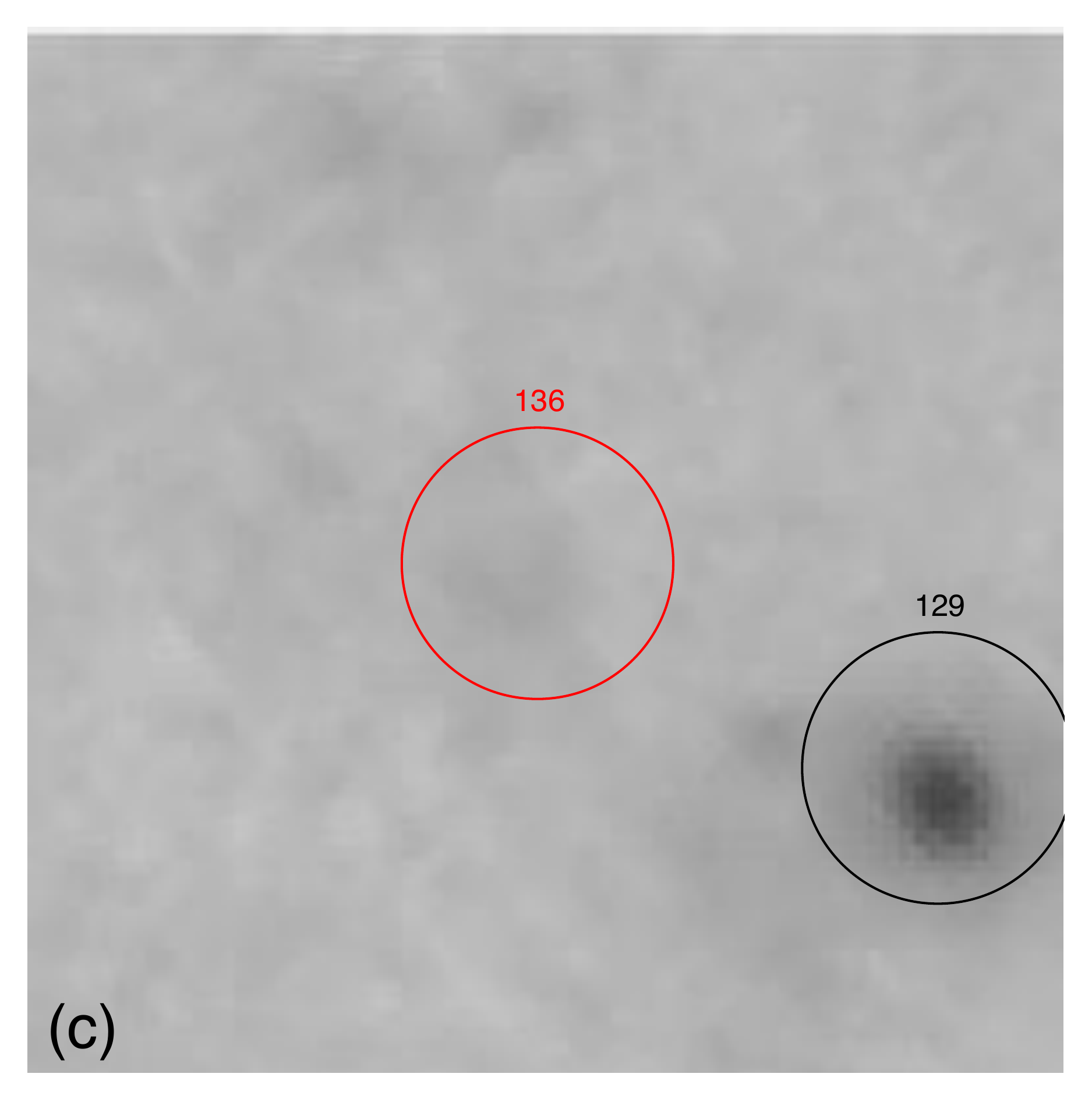}
\end{minipage}
\begin{minipage}{0.24\linewidth}
\includegraphics[width=\linewidth]{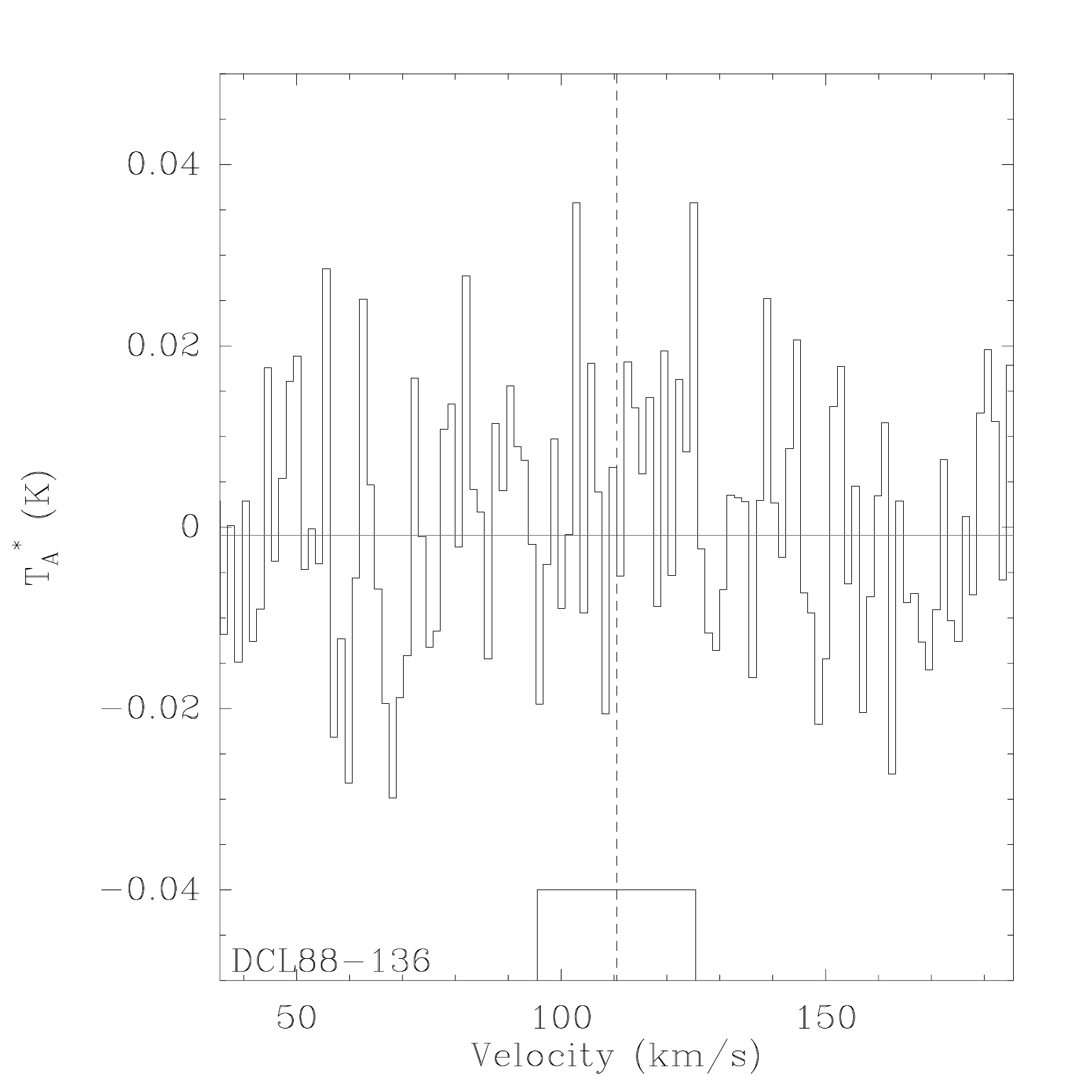}
\end{minipage}

\begin{minipage}{0.24\linewidth}
\includegraphics[width=\linewidth]{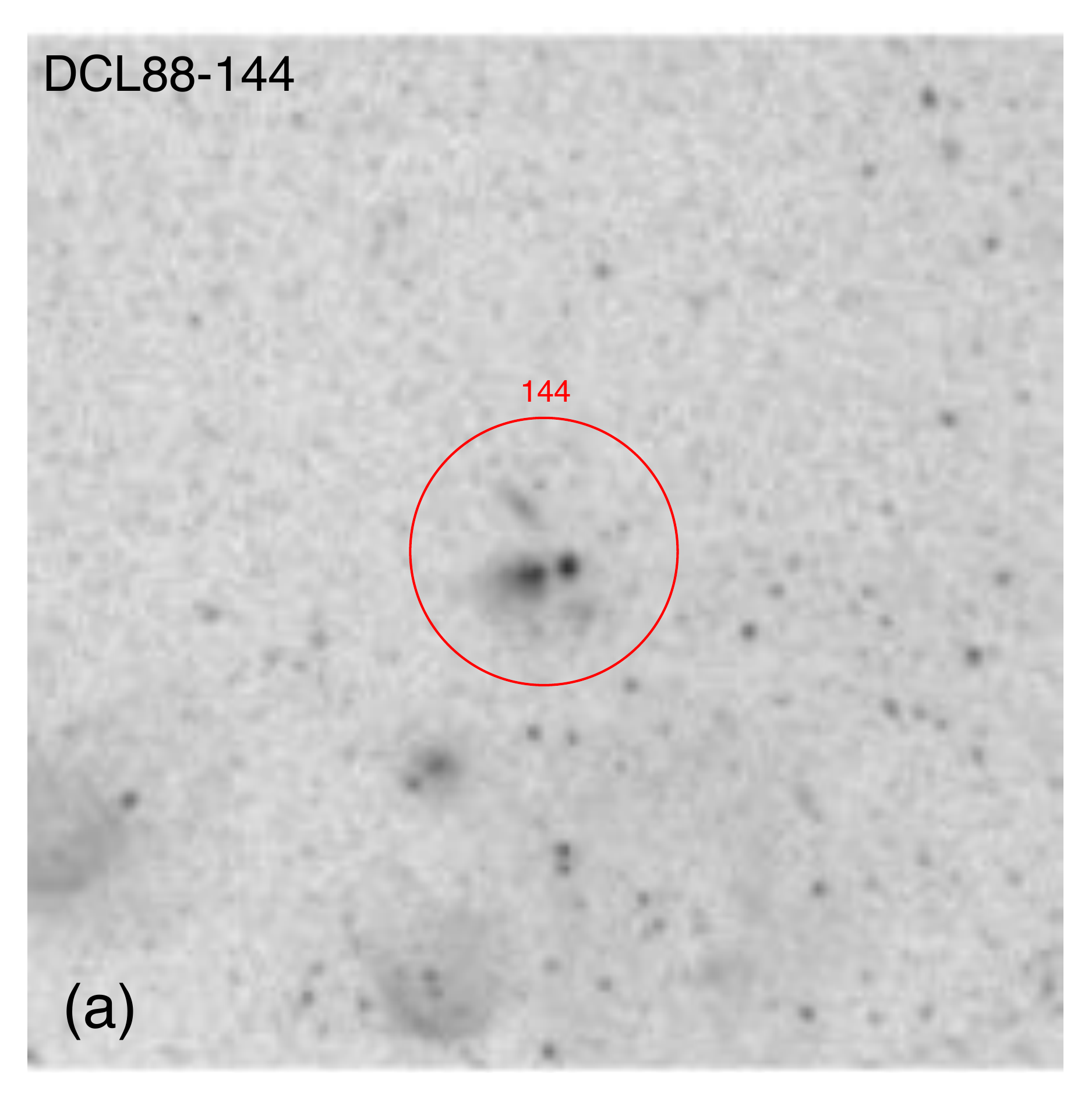}
\end{minipage}
\begin{minipage}{0.24\linewidth}
\includegraphics[width=\linewidth]{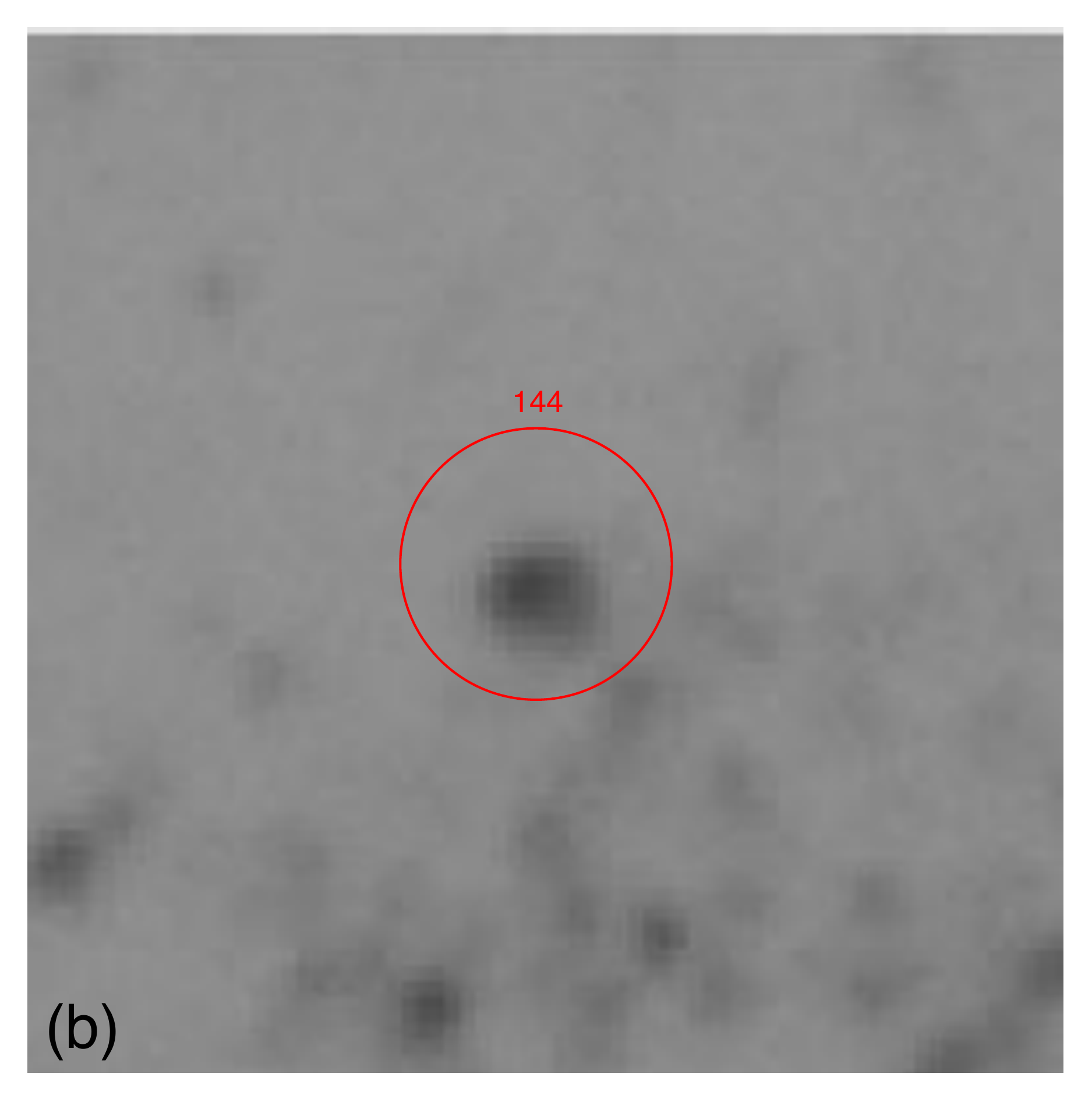}
\end{minipage}
\begin{minipage}{0.24\linewidth}
\includegraphics[width=\linewidth]{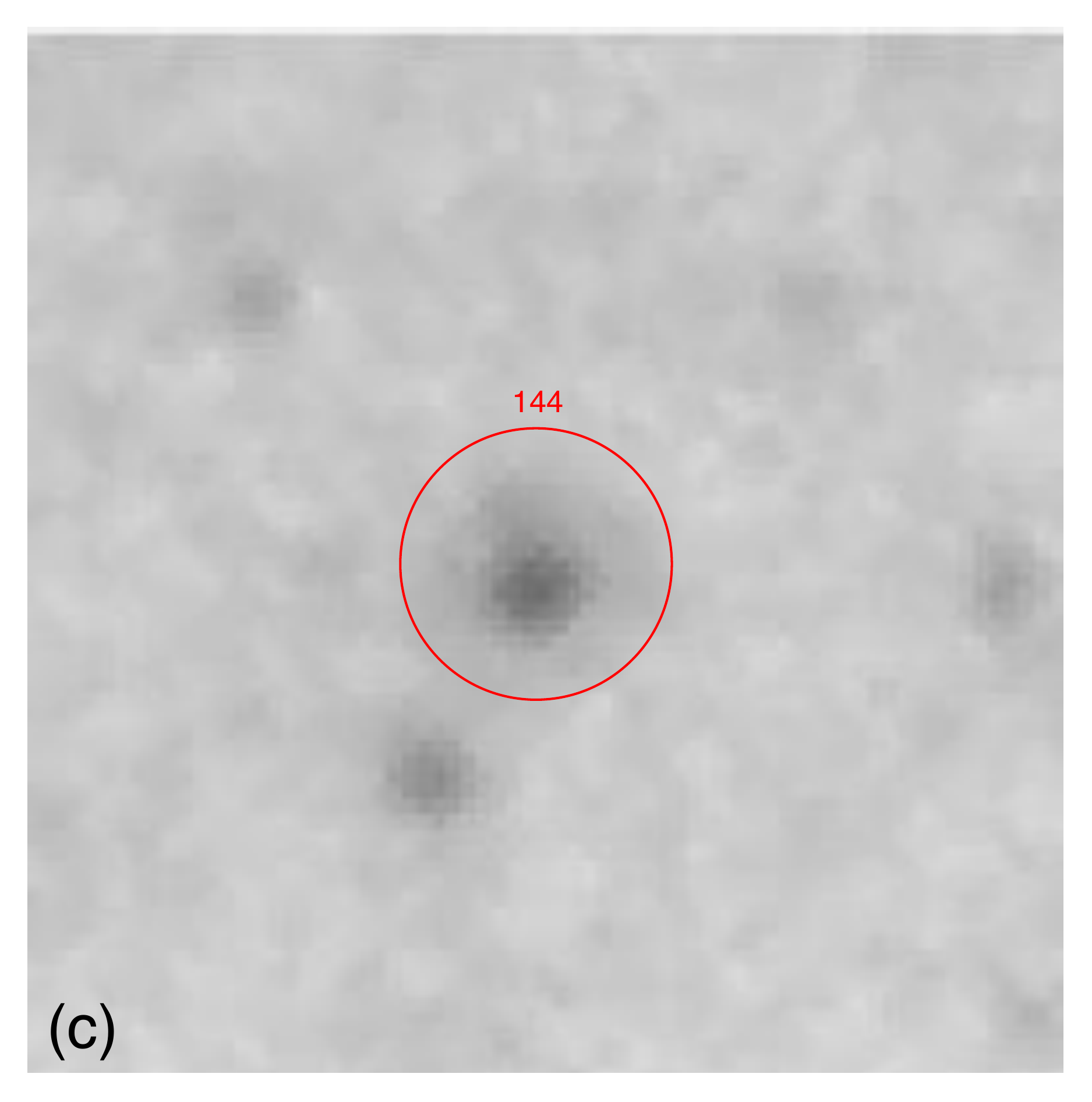}
\end{minipage}
\begin{minipage}{0.24\linewidth}
\includegraphics[width=\linewidth]{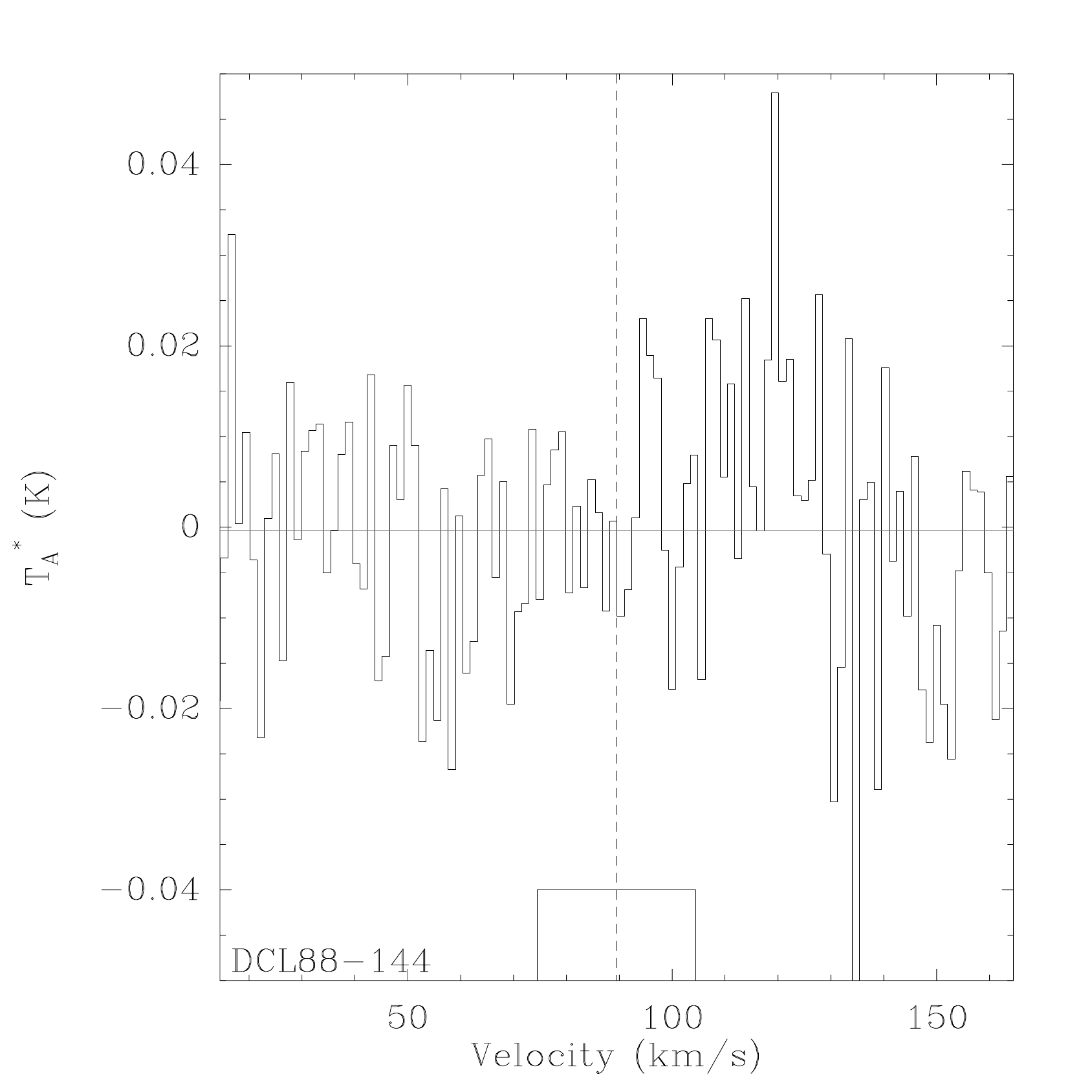}
\end{minipage}

\noindent\textbf{Figure~\ref{fig:stamps} -- continued.}

\end{figure*}

\begin{figure*}
%\ContinuedFloat

\begin{minipage}{0.24\linewidth}
\includegraphics[width=\linewidth]{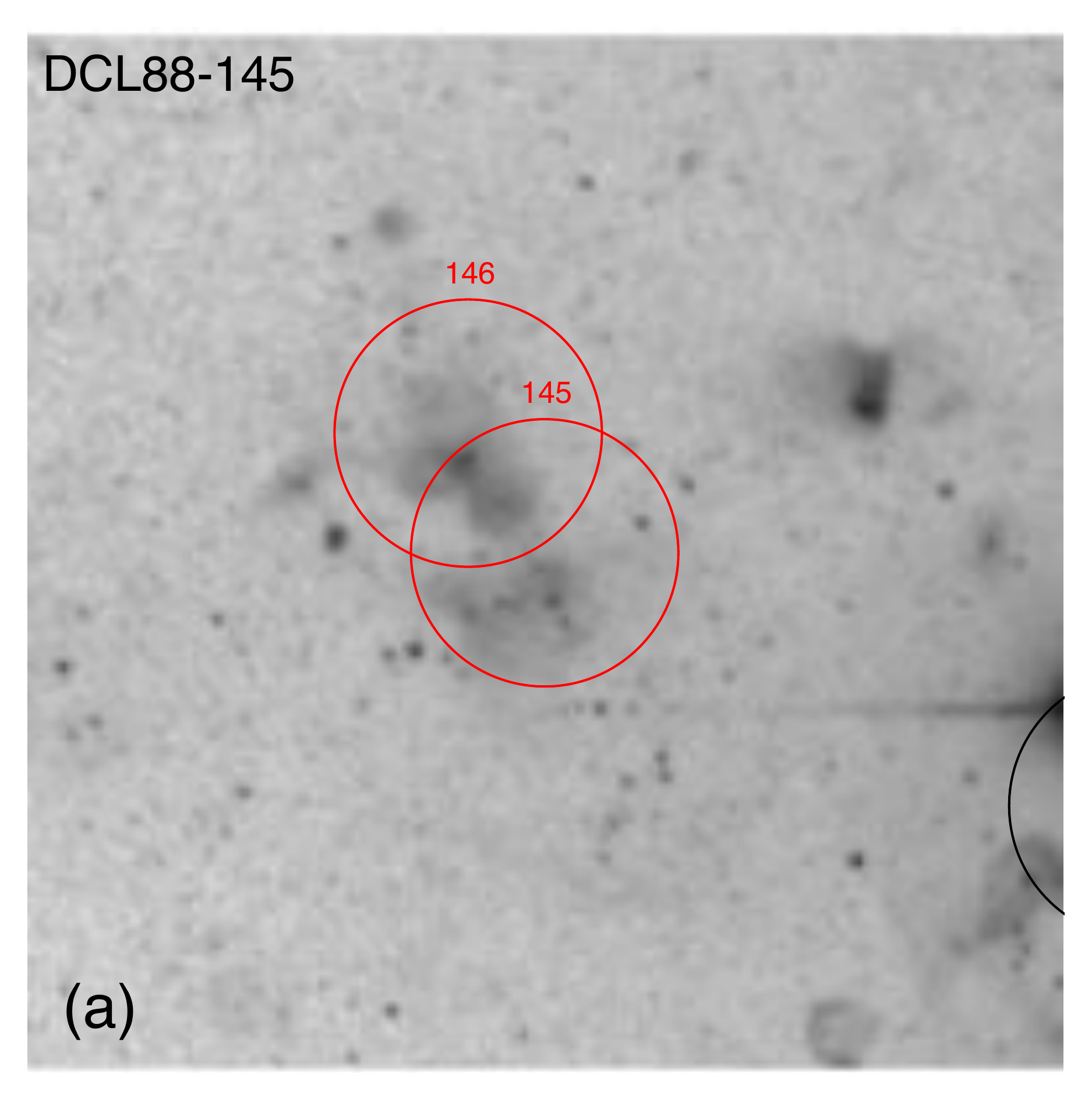}
\end{minipage}
\begin{minipage}{0.24\linewidth}
\includegraphics[width=\linewidth]{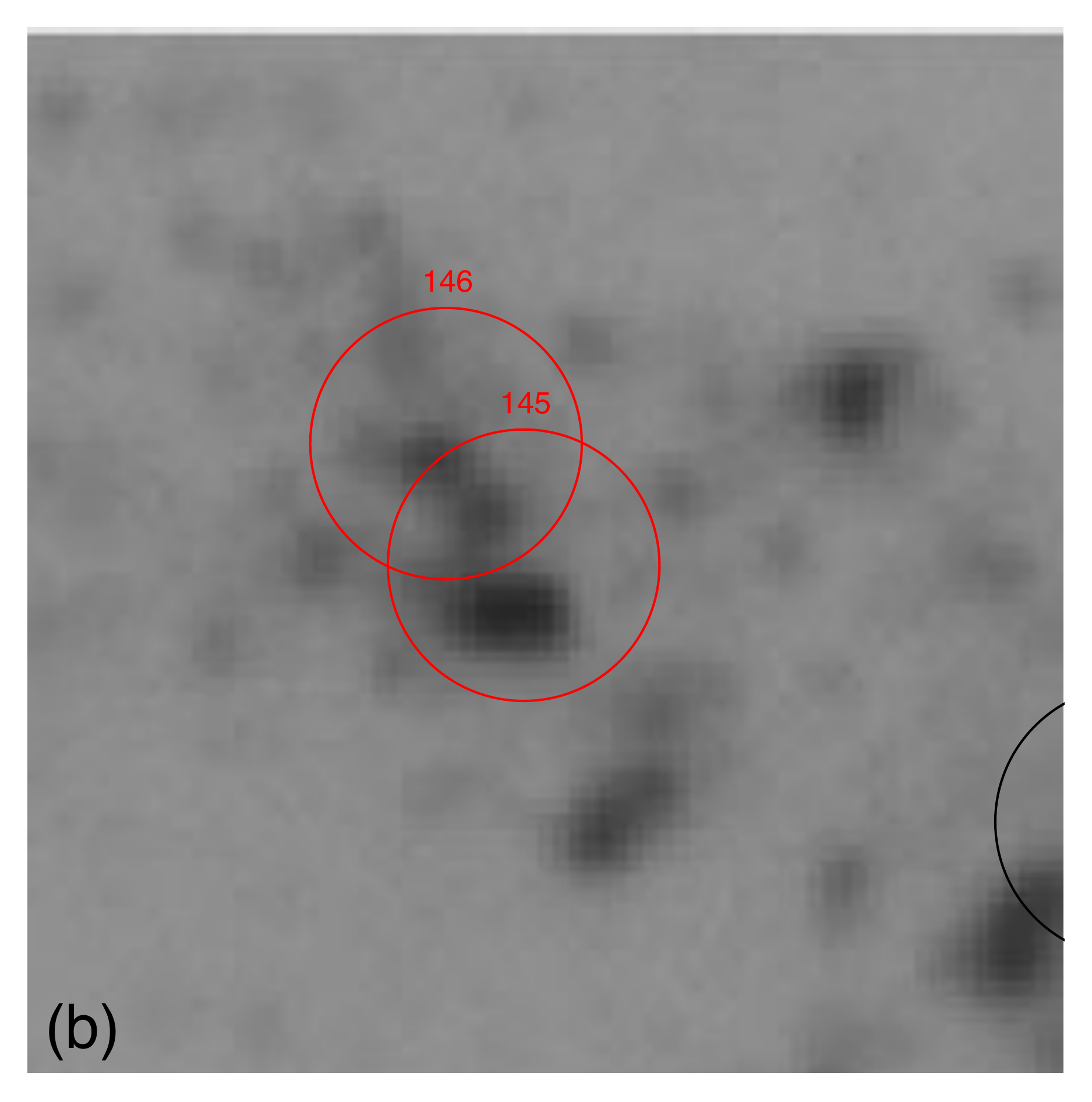}
\end{minipage}
\begin{minipage}{0.24\linewidth}
\includegraphics[width=\linewidth]{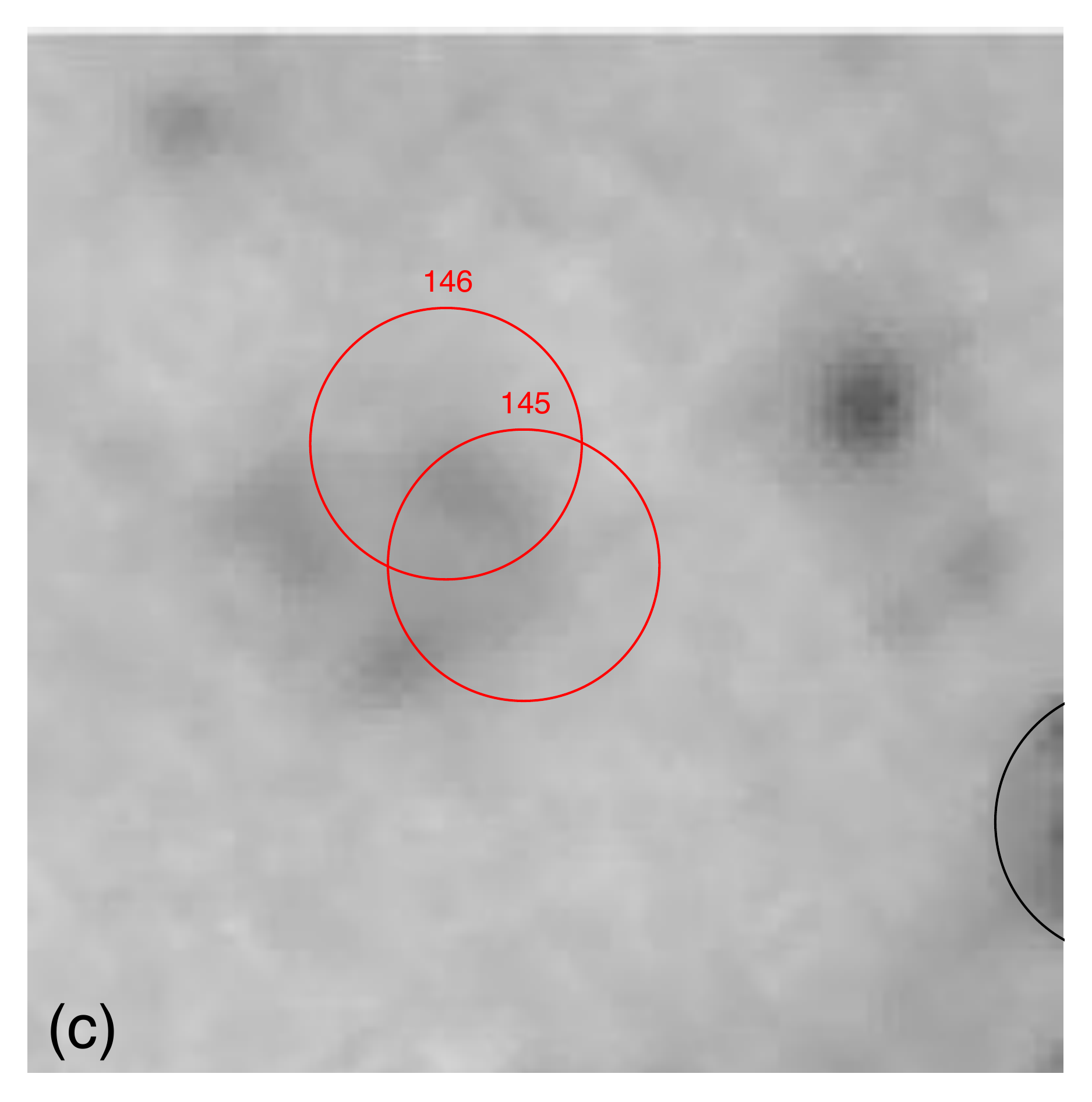}
\end{minipage}
\begin{minipage}{0.24\linewidth}
\includegraphics[width=\linewidth]{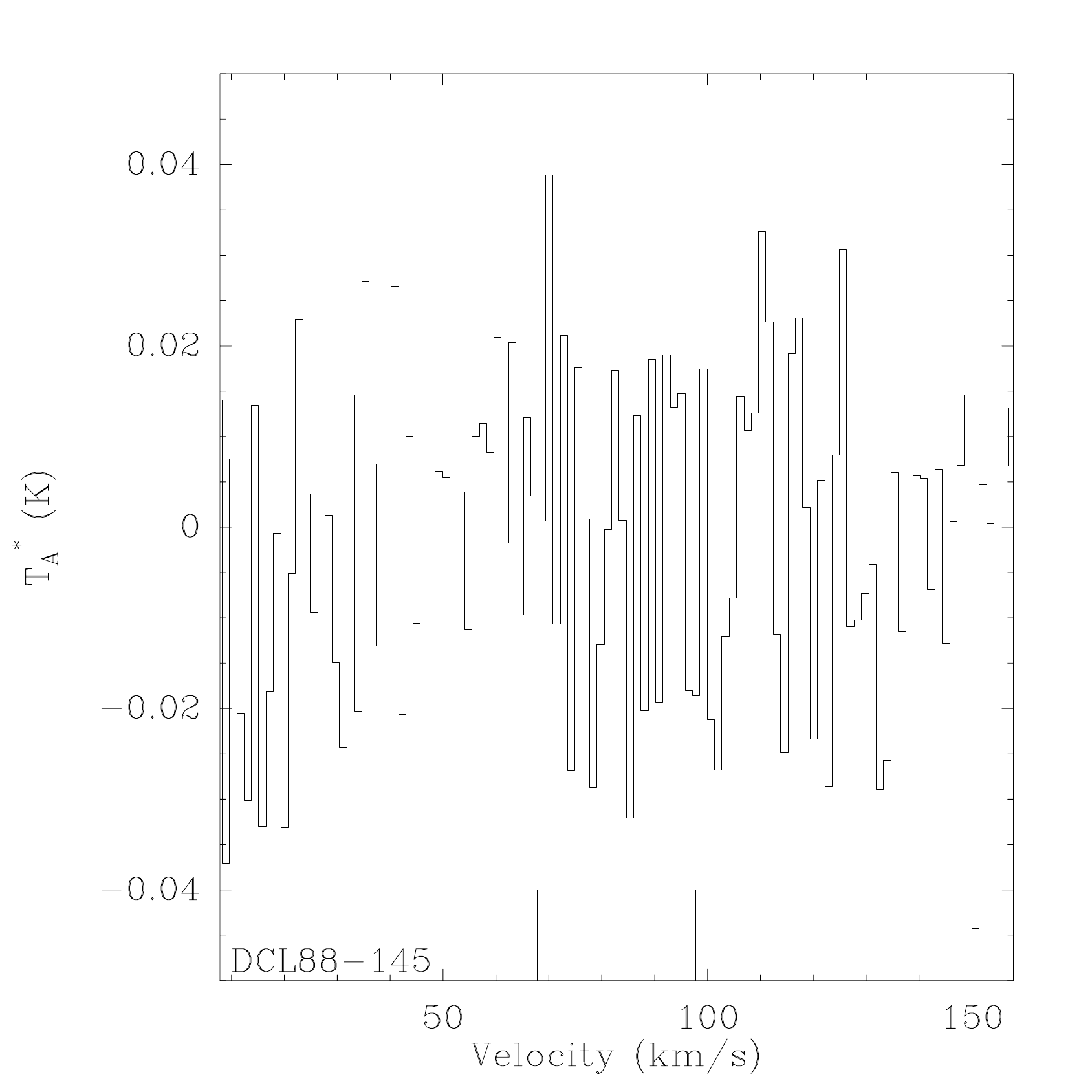}
\end{minipage}

\begin{minipage}{0.24\linewidth}
\includegraphics[width=\linewidth]{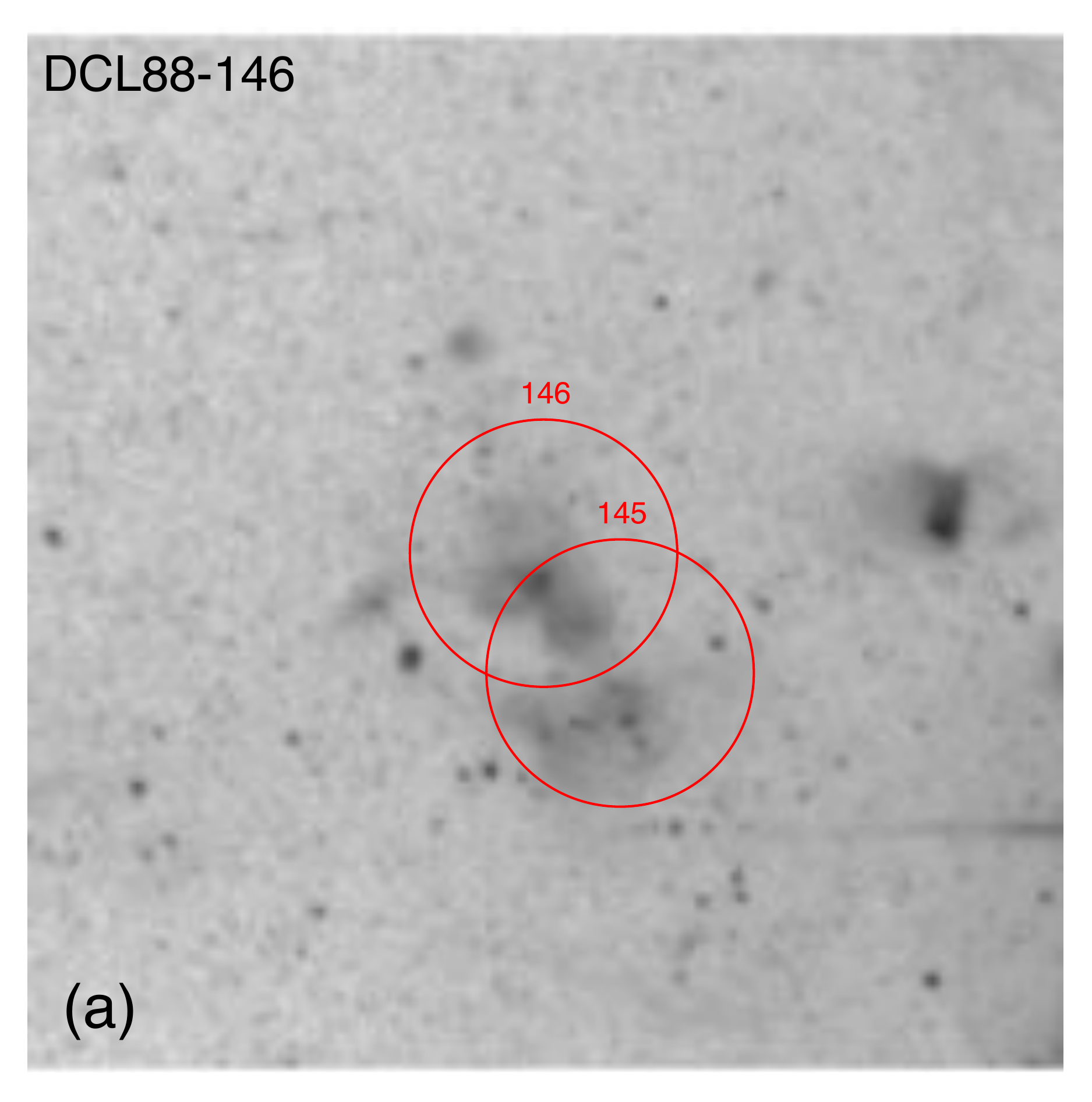}
\end{minipage}
\begin{minipage}{0.24\linewidth}
\includegraphics[width=\linewidth]{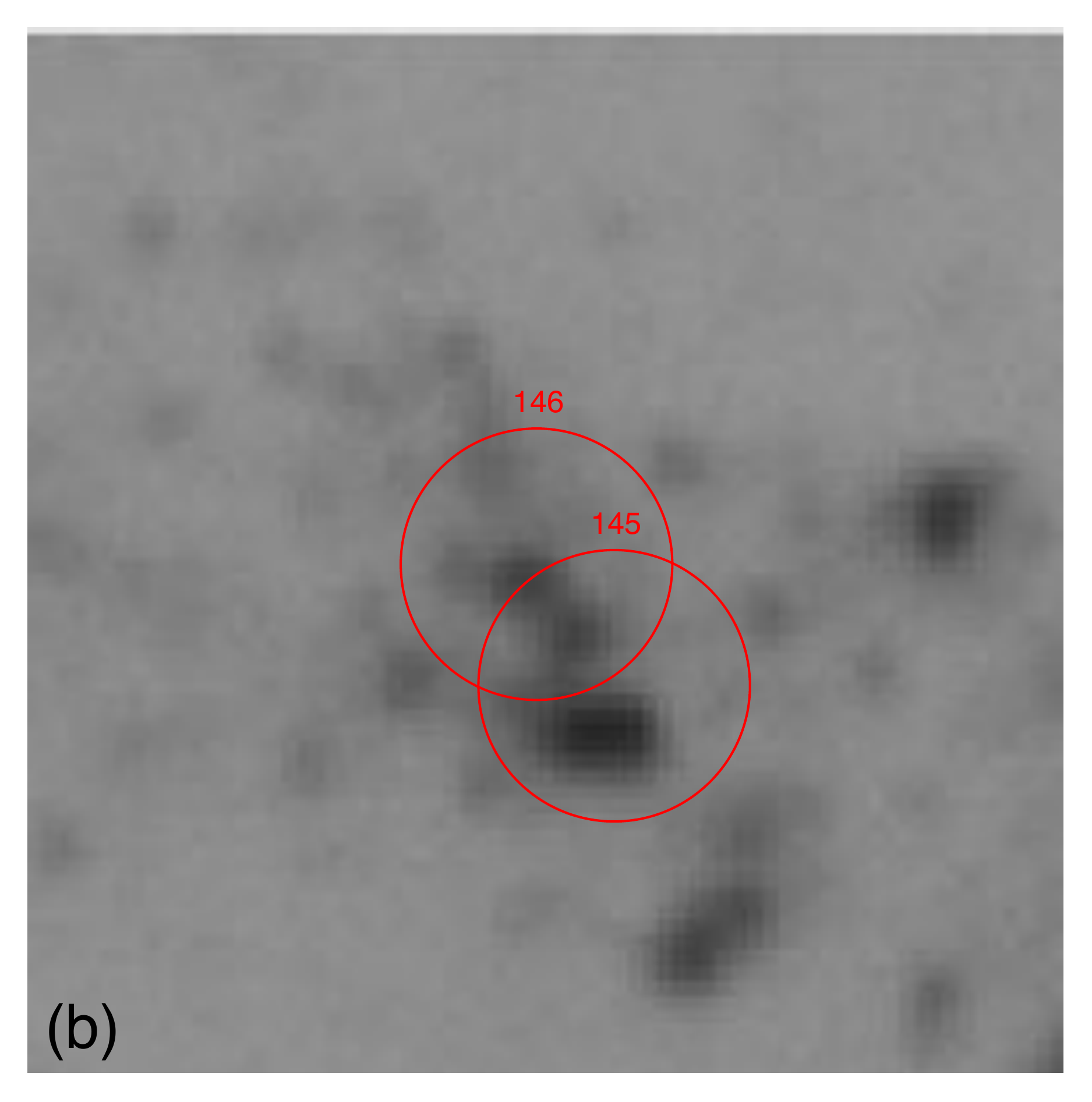}
\end{minipage}
\begin{minipage}{0.24\linewidth}
\includegraphics[width=\linewidth]{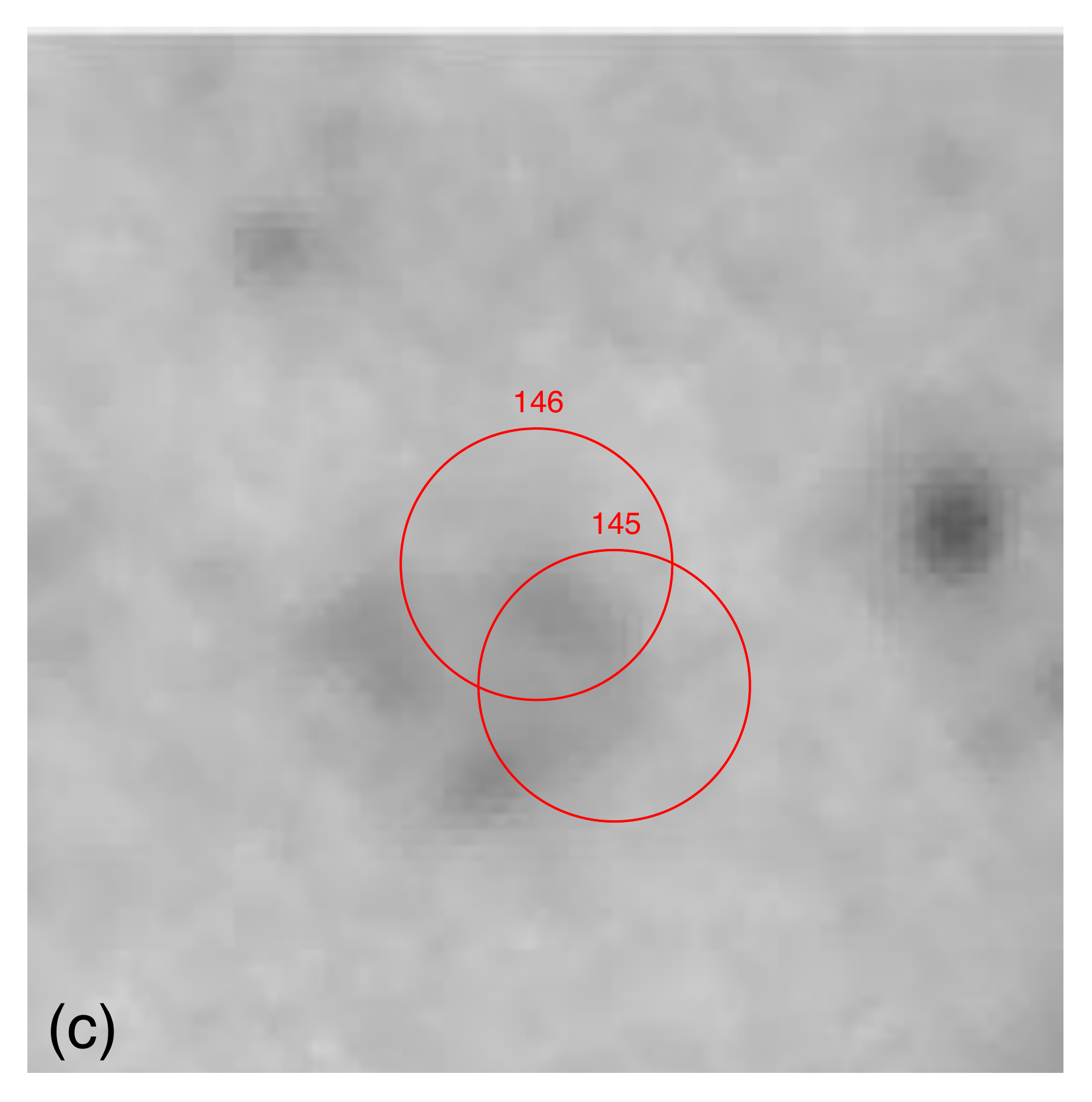}
\end{minipage}
\begin{minipage}{0.24\linewidth}
\includegraphics[width=\linewidth]{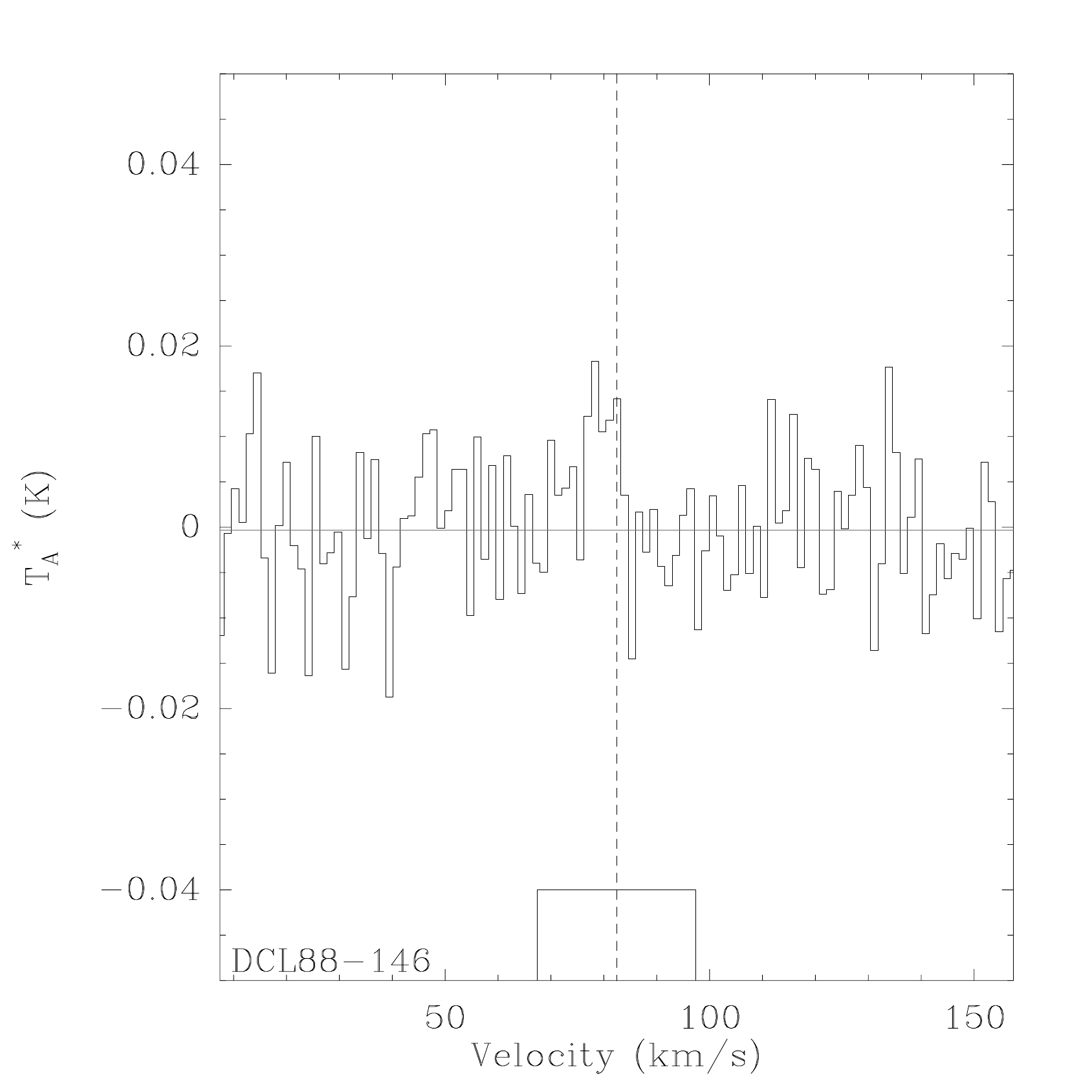}
\end{minipage}

\begin{minipage}{0.24\linewidth}
\includegraphics[width=\linewidth]{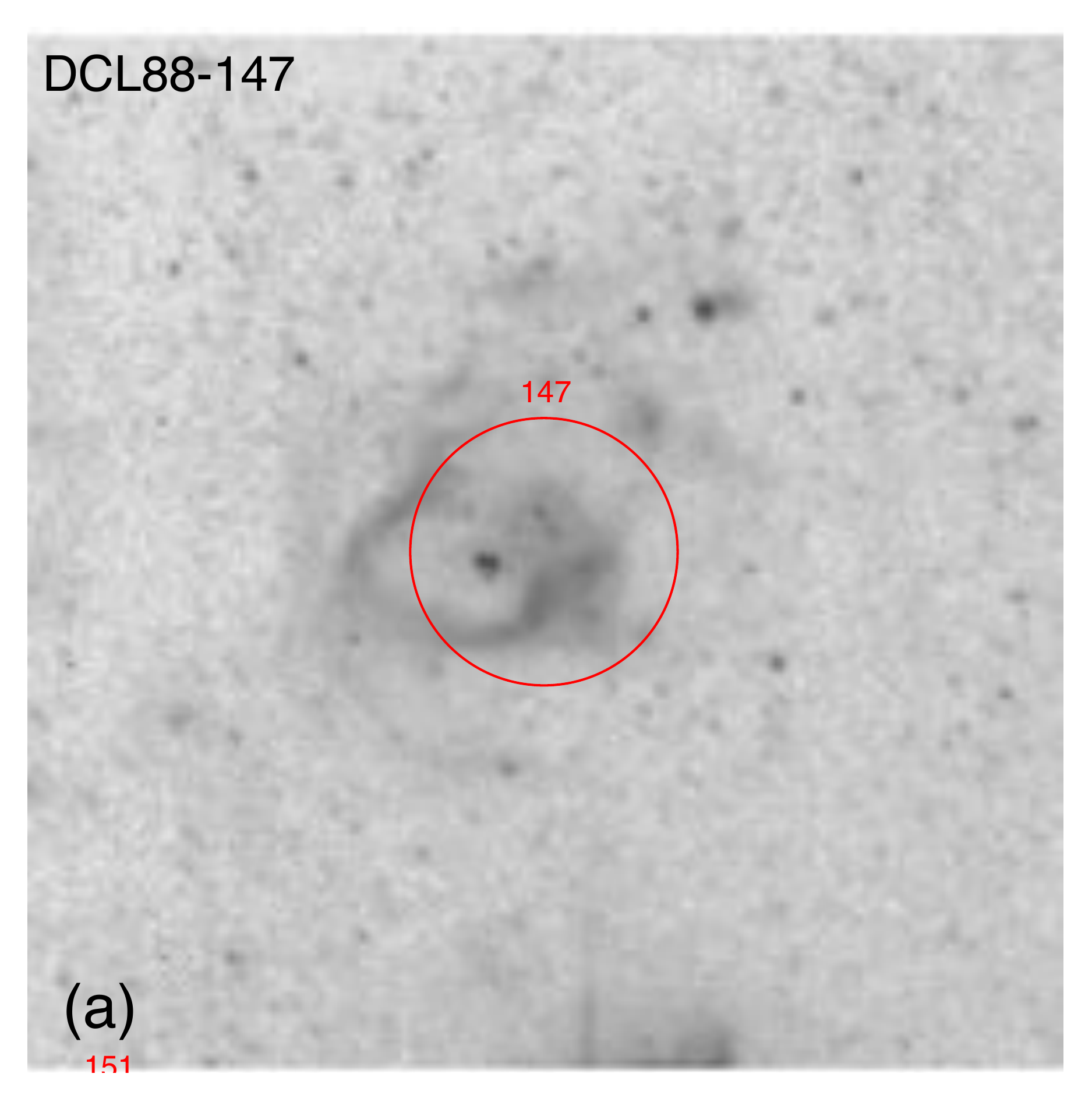}
\end{minipage}
\begin{minipage}{0.24\linewidth}
\includegraphics[width=\linewidth]{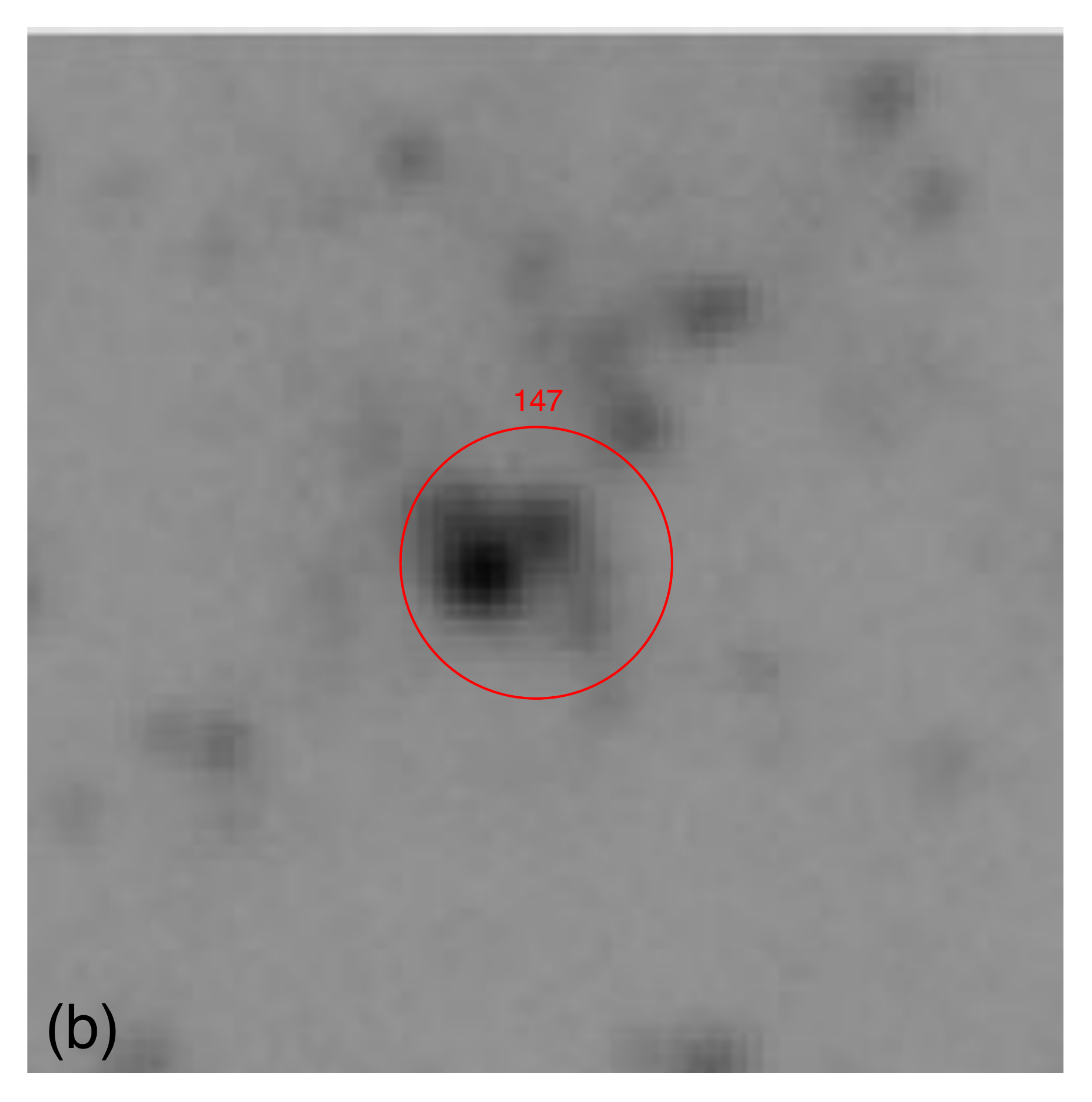}
\end{minipage}
\begin{minipage}{0.24\linewidth}
\includegraphics[width=\linewidth]{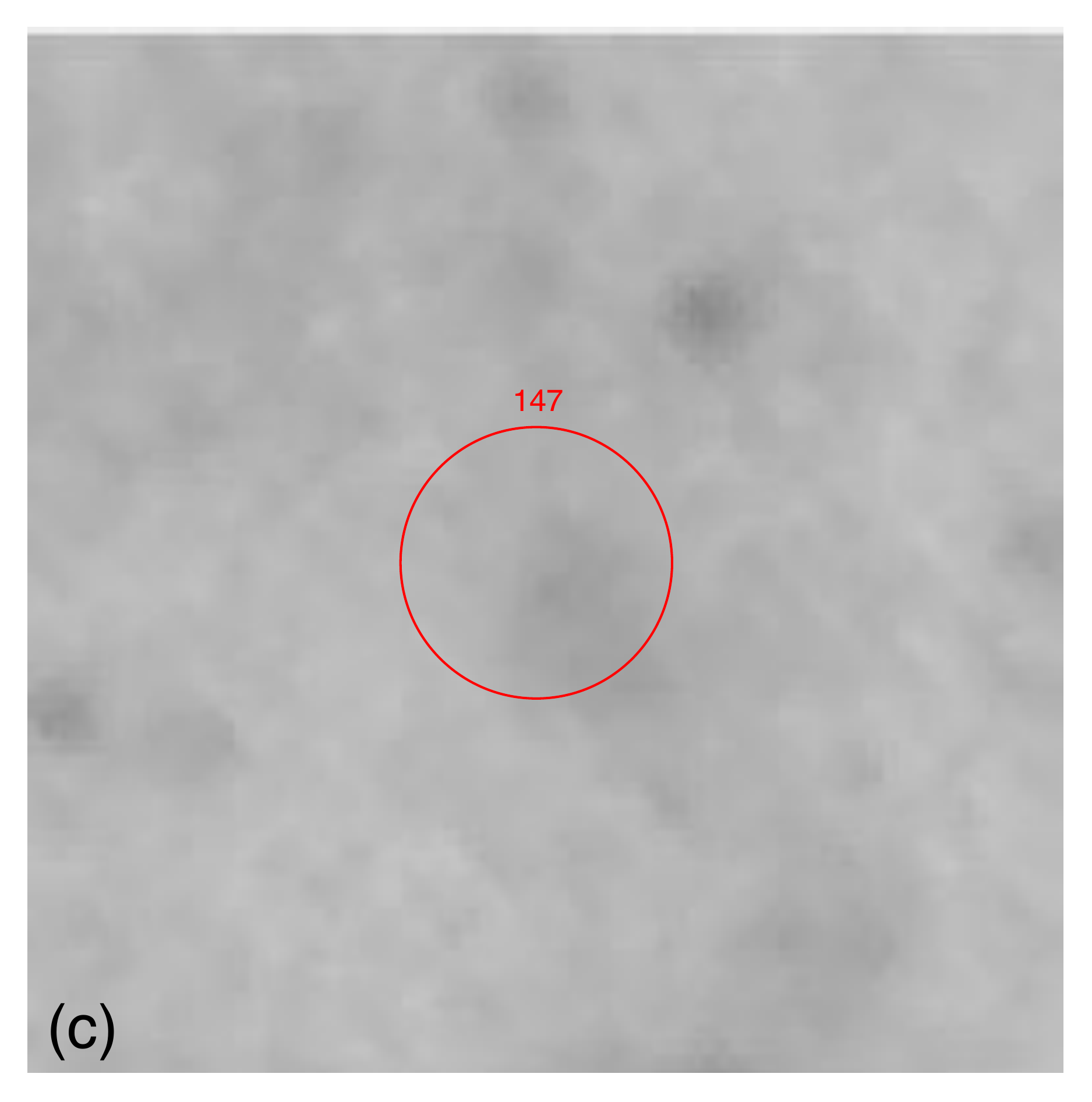}
\end{minipage}
\begin{minipage}{0.24\linewidth}
\includegraphics[width=\linewidth]{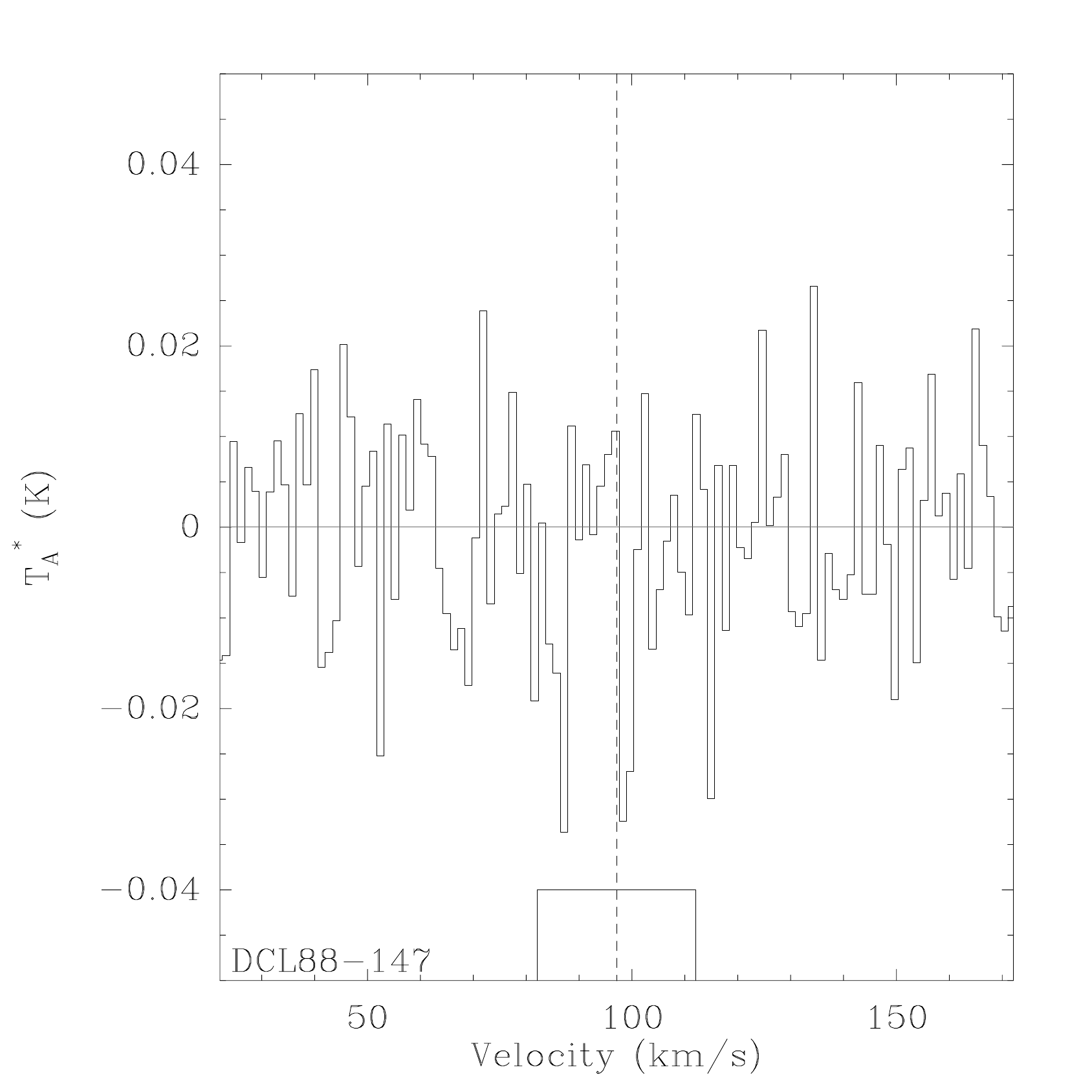}
\end{minipage}

\begin{minipage}{0.24\linewidth}
\includegraphics[width=\linewidth]{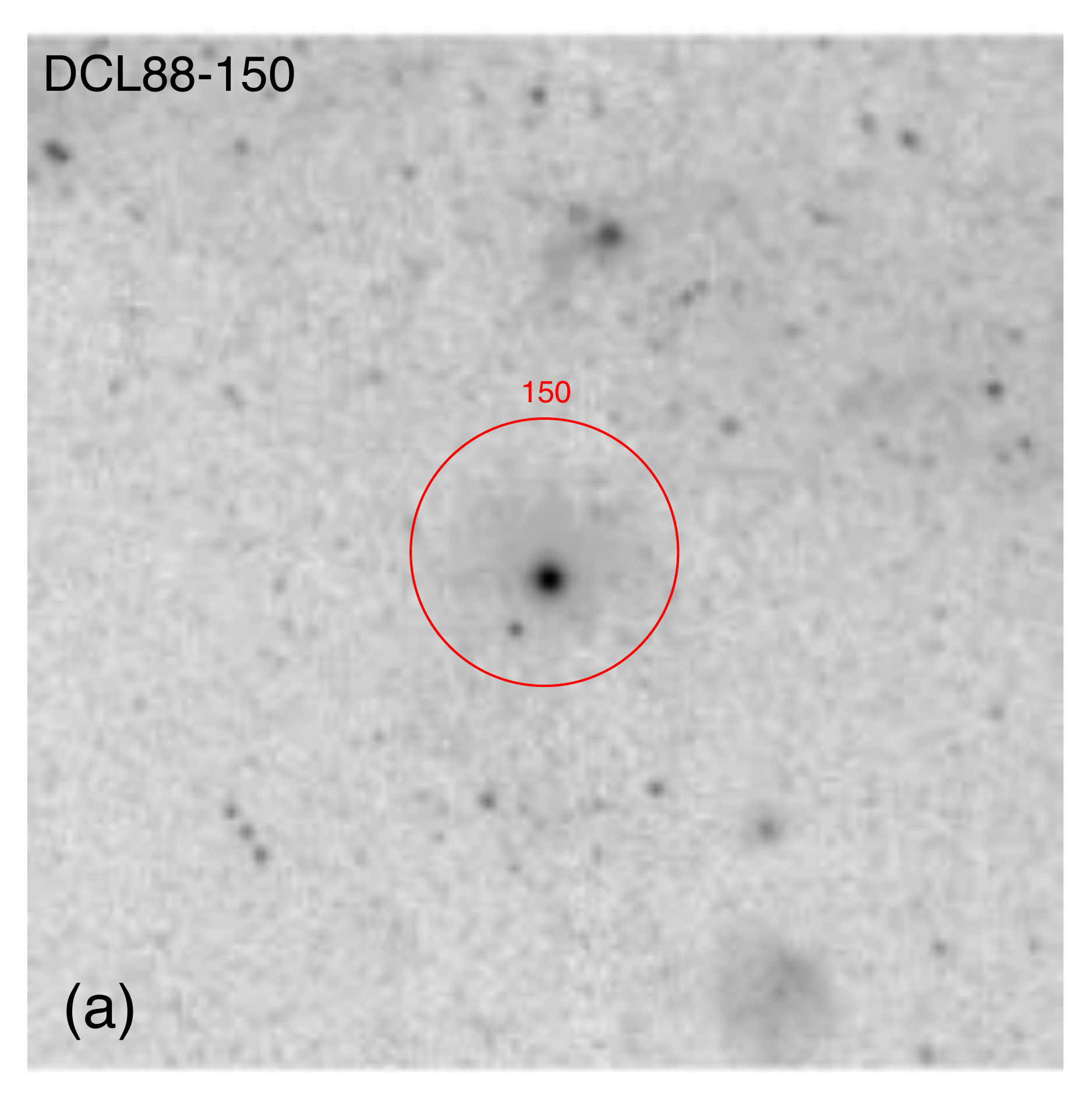}
\end{minipage}
\begin{minipage}{0.24\linewidth}
\includegraphics[width=\linewidth]{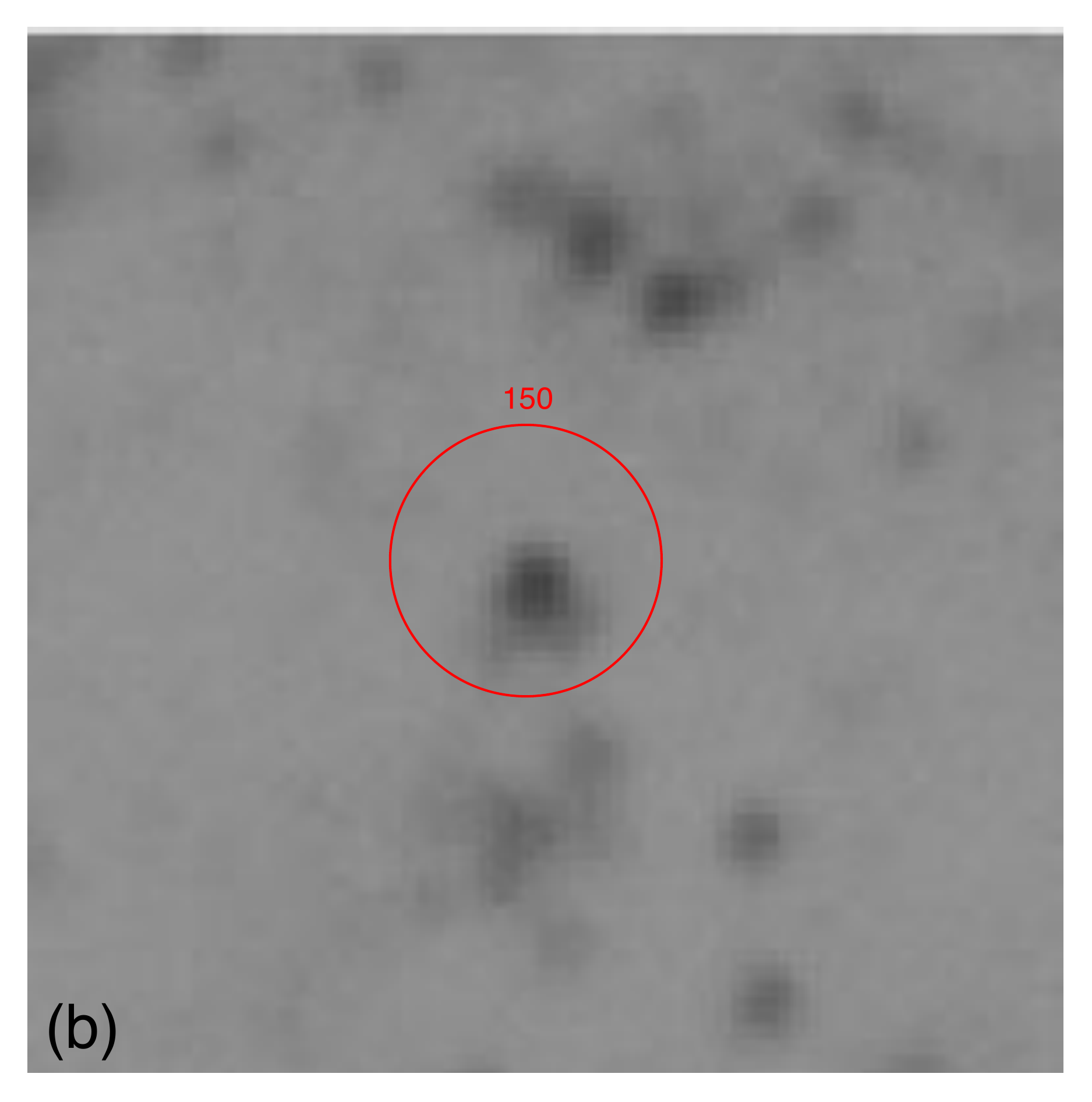}
\end{minipage}
\begin{minipage}{0.24\linewidth}
\includegraphics[width=\linewidth]{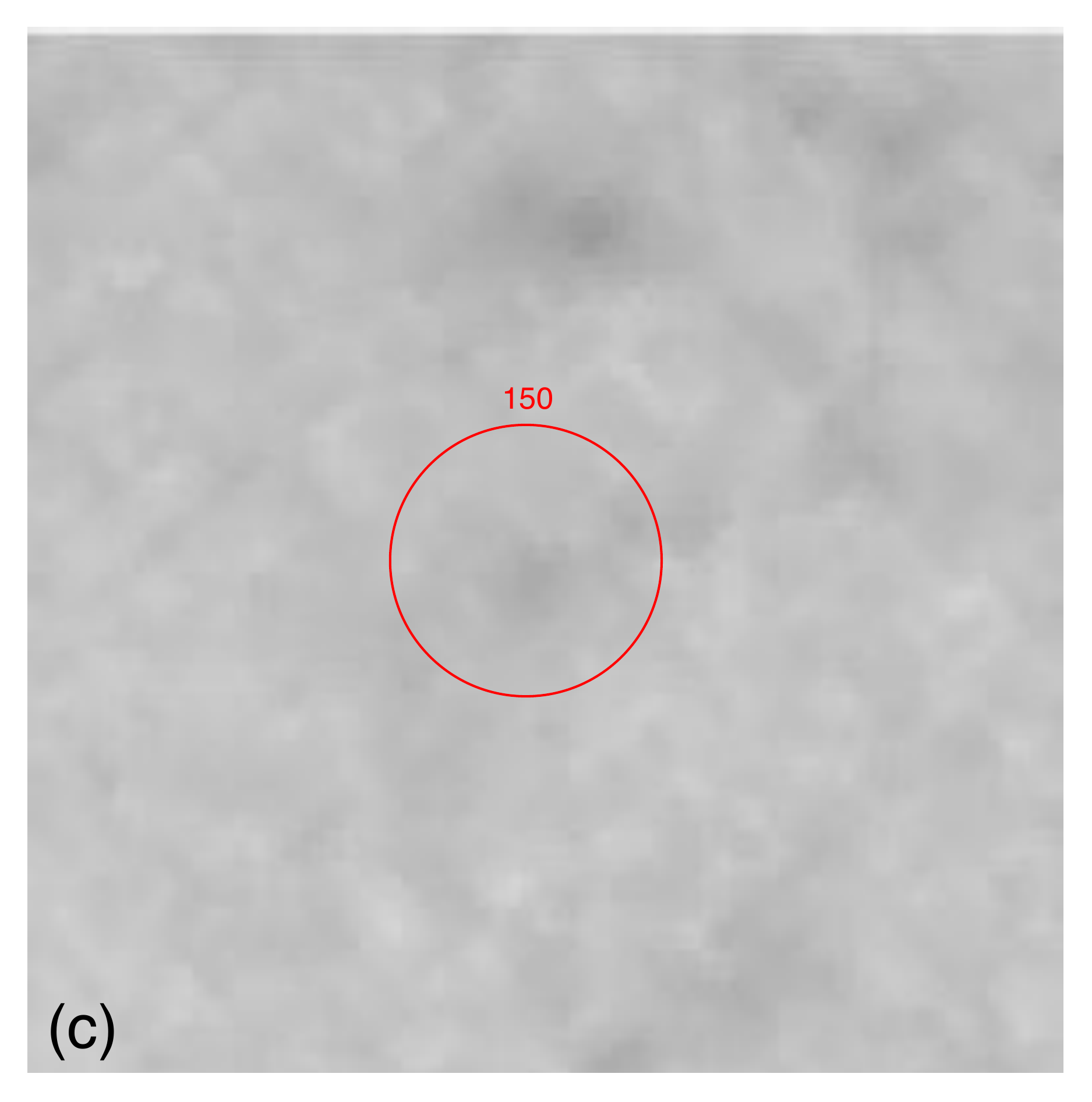}
\end{minipage}
\begin{minipage}{0.24\linewidth}
\includegraphics[width=\linewidth]{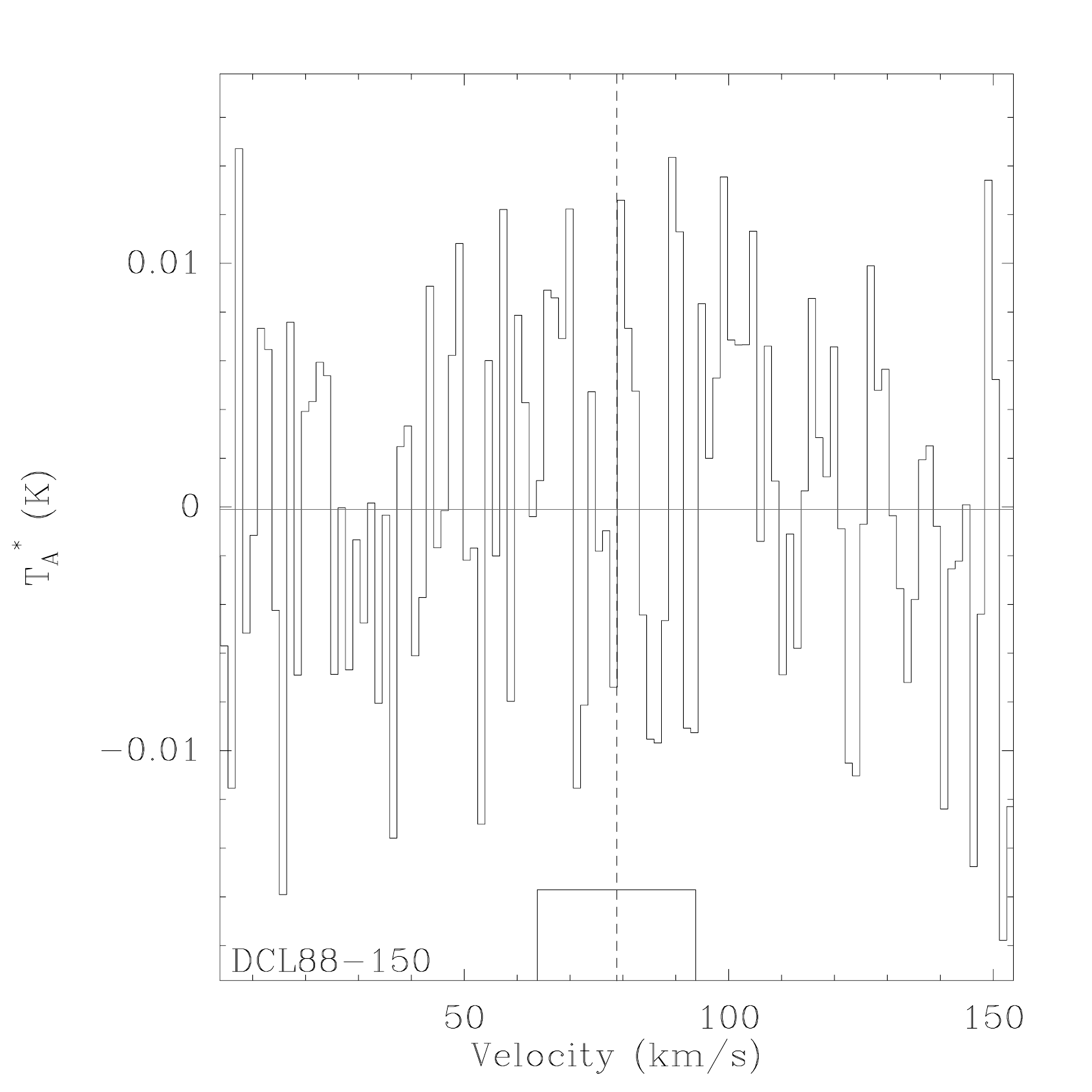}
\end{minipage}

\begin{minipage}{0.24\linewidth}
\includegraphics[width=\linewidth]{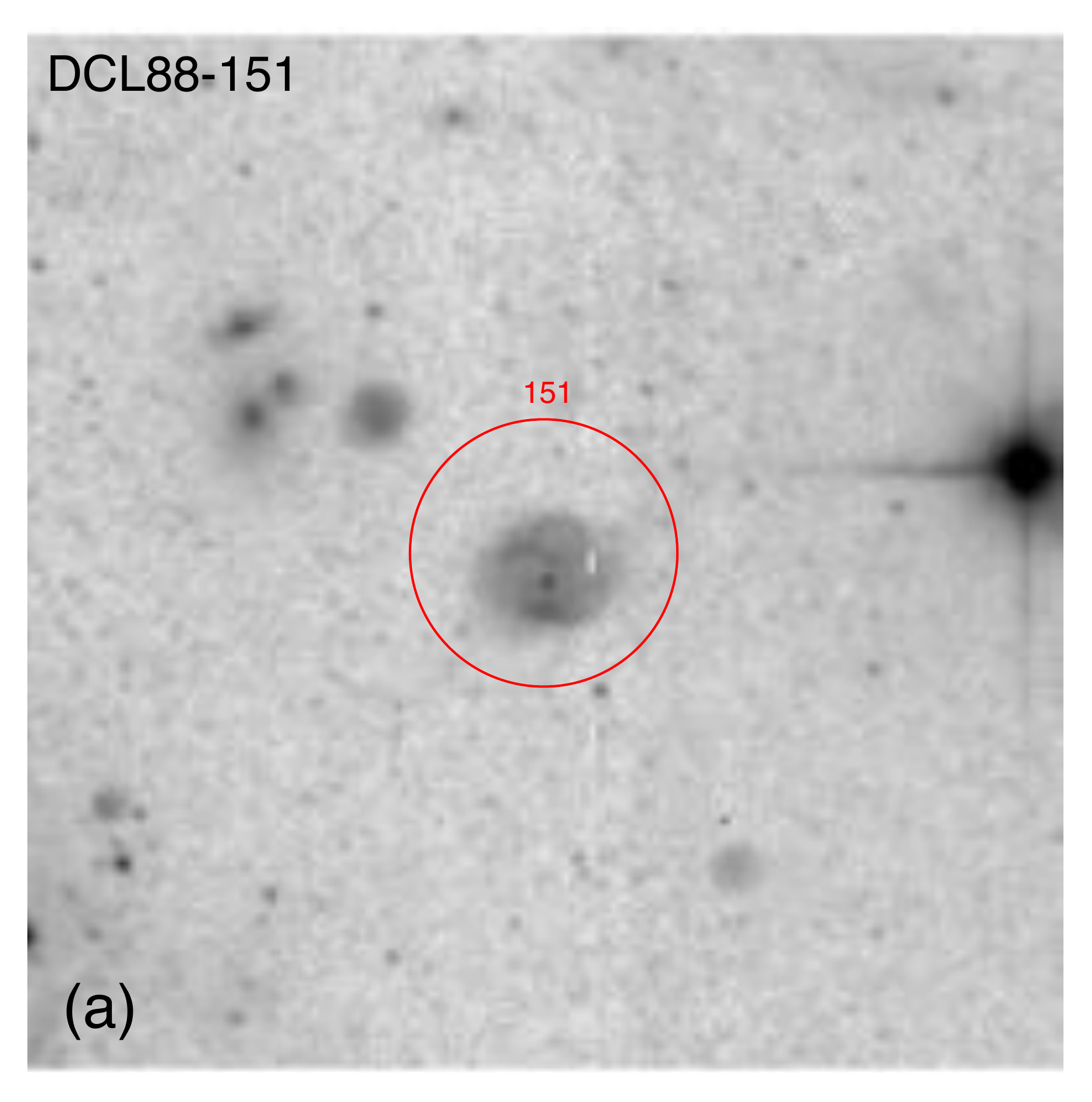}
\end{minipage}
\begin{minipage}{0.24\linewidth}
\includegraphics[width=\linewidth]{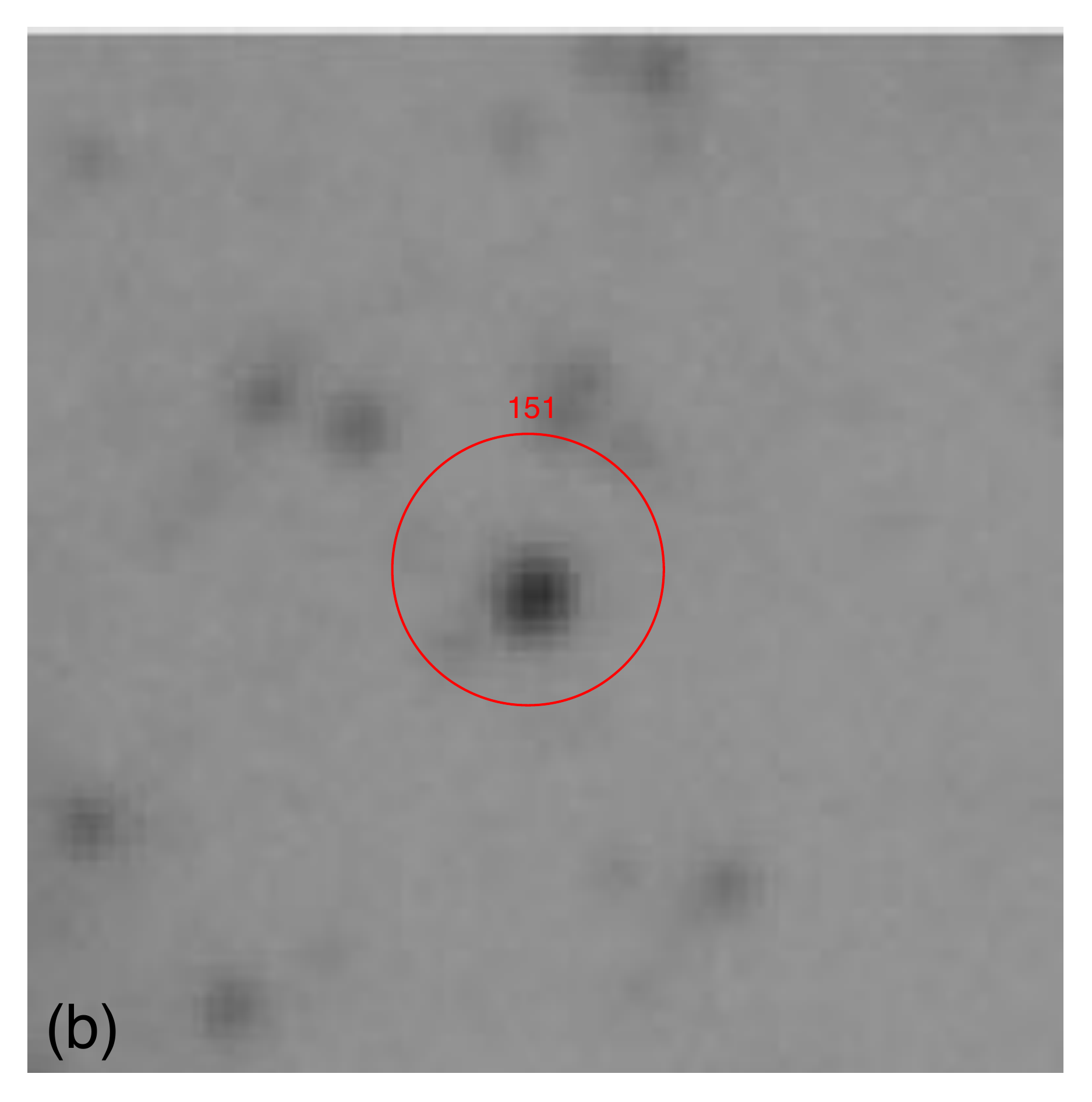}
\end{minipage}
\begin{minipage}{0.24\linewidth}
\includegraphics[width=\linewidth]{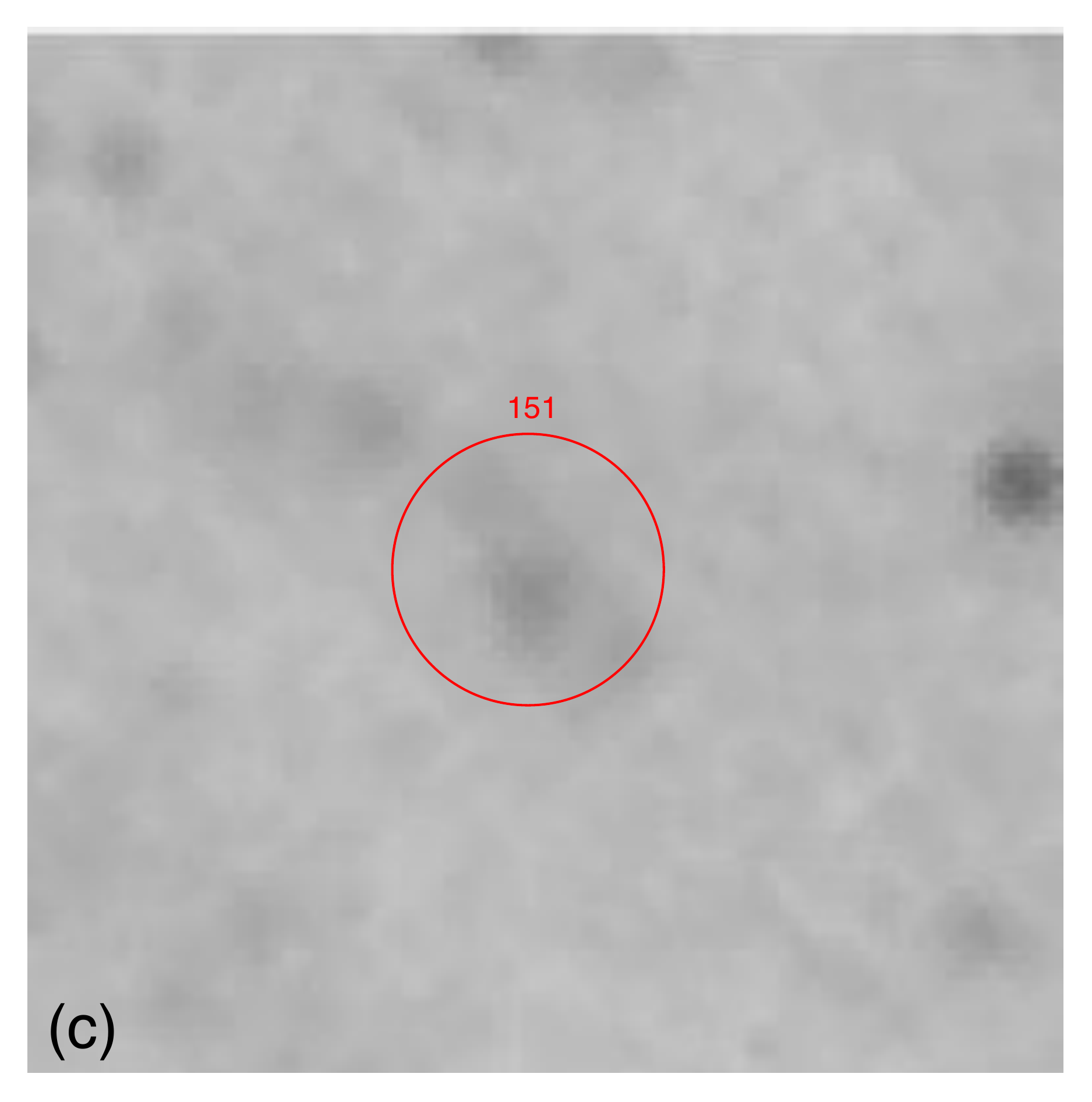}
\end{minipage}
\begin{minipage}{0.24\linewidth}
\includegraphics[width=\linewidth]{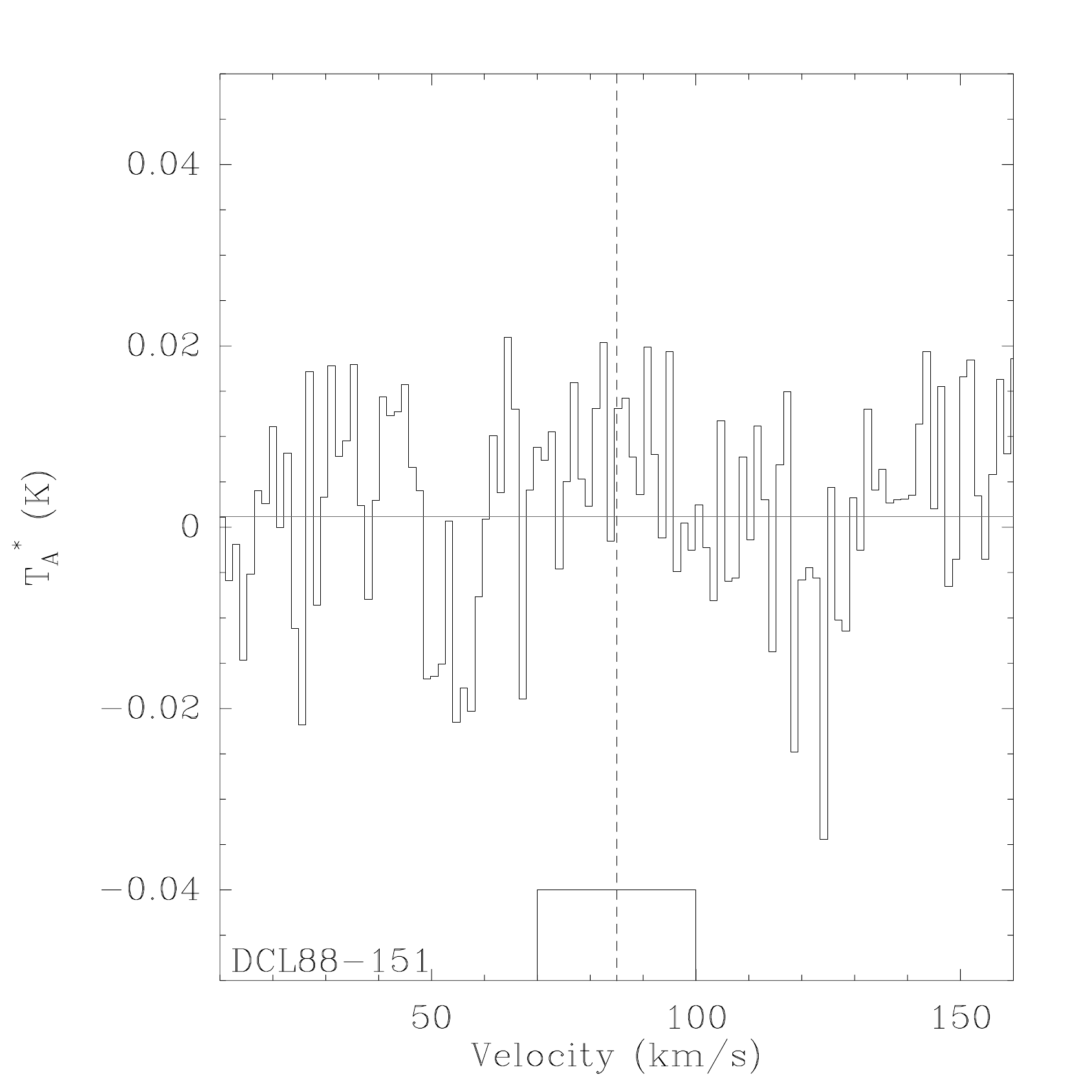}
\end{minipage}
\noindent\textbf{Figure~\ref{fig:stamps} -- continued.}
\label{fig:stamps15}
\end{figure*}

% end online only section

%--------------------------BIBLIOGRAPHY---------------------------

%\end{landscape}

\end{document}